%% file: LH_NewPhysics_Report.tex
\documentclass[12pt]{cernyrep}
\pdfoutput=1

\usepackage{subfigure}
\usepackage{multirow}
\usepackage{amsmath}
\usepackage{amssymb}
\usepackage{amsfonts}
\usepackage{amstext}
\usepackage[english]{babel}
\usepackage{epsfig}
\usepackage{cite}
\usepackage{graphicx}
\usepackage{color}
\usepackage{slashed}
\usepackage{bm}
\usepackage{xspace}

\input commands.tex

\input authors_commands.tex

\begin{document}

\setcounter{tocdepth}{0}
\thispagestyle{empty}


\vspace{1cm}

\begin{center}
{\Large {\bf NEW PHYSICS AT THE LHC: A LES HOUCHES REPORT\\[4mm]}}
{\large {\bf Physics at TeV Colliders 2009 -- New Physics Working Group}}
\end{center}

\vspace{1cm}
\input authors.tex

 
 \vspace{2cm}
\begin{center}
{\large {\bf Abstract}}\\[.2cm]
\end{center}
We present a collection of signatures for physics beyond the 
standard model that need to be explored at the LHC. 
First, are presented various tools developed to measure new 
particle masses in scenarios where all decays include an unobservable particle.
Second, various aspects of supersymmetric models are discussed.
Third, some signatures of models of strong electroweak symmetry are discussed.
In the fourth part, a special attention is devoted to high mass resonances, as the ones appearing in models with warped extra dimensions. Finally, prospects for models with a hidden sector/valley are presented.
Our report, which includes brief experimental and theoretical reviews 
as well as original results, summarizes the activities of the 
``New Physics'' working group for the ``Physics at TeV Colliders"
workshop (Les Houches, France, 8--26 June, 2009).

\vspace{2cm}
\begin{center}
{\bf Acknowledgements}\\[.2cm]
\end{center}
We would like to heartily thank the funding bodies, the organisers 
(G.~B\'elanger, F.~Boudjema, L. di Ciaccio, P.A.~Delsart, S.~Gascon, C.~Grojean, J.P.~Guillet, S.~Kraml, R.~Lafaye, G.~Moreau, E.~Pilon, G.~Salam, P.~Slavich and D.~Zerwas), the staff and the other
participants of the Les Houches workshop for providing a stimulating and
lively environment in which to work.

\newpage

\vspace{1cm}

\thispagestyle{empty}
\setcounter{page}{2}

\begin{center}
\input addresses.tex

\end{center}

\vspace{3cm}

\newpage

\tableofcontents

\newpage


\input intro.tex
\addtocontents{toc}{\protect\contentsline{part}{\protect\numberline{} \hspace{-2cm}Introduction}{\thepage}}
\AddToContent{G.~Brooijmans,  C.~Grojean, G.D.~Kribs and C.~Shepherd-Themistocleous}

\superpart{Mass determination methods}

\input Robens/masses.tex

\AddToContent{L.~Basso et al.}
%
\renewcommand{\thesection}{\arabic{section}}

\input Gripaios/gripaios.tex

\AddToContent{B.~Gripaios}
%

\superpart{Supersymmetry}

\input RSW/RSW.tex
\AddToContent{A.R.~Raklev, G.P.~Salam and J.G.~Wacker}
%

\input Muehlleitner/Muehlleitner.tex

\AddToContent{M.~M\"uhlleitner, H.~Rzehak and M.~Spira}
%

\input Fichet-Kraml/fichet-kraml.tex
\AddToContent{S.~Fichet and S.~Kraml}

\input Kraml/cpvmcmc.tex

\AddToContent{G. Belanger, S. Kraml, A. Pukhov and R.K. Singh}
\clearpage
%

\superpart{Strong EWSB}

\input Espinosa/composite.tex

\AddToContent{J.~Espinosa, C.~Grojean and M.~M\"uhlleitner}

\input Lane/Lane.tex

\AddToContent{K.~Black et al.}
%

\superpart{High mass resonances}

\input Karagoz/LH_warped.tex

\AddToContent{K.~Agashe et al.}

\input Basso/Zp.tex

\AddToContent{L.~Basso et al.}
%

\input Moreau/Moreau.tex

\AddToContent{S.~Gopalakrishna, G.~Moreau and R.K.~Singh}
\renewcommand{\thesection}{\arabic{section}}


\input Servant/Servant.tex

\AddToContent{G.~Servant, M.~Vos, L.~Gauthier and A.-I.~Etienvre}

\input Brooijmans/Brooijmans.tex
\AddToContent{G.~Brooijmans, G.~Moreau and R.K.~Singh}
%

\input Deandrea/resonances.tex

\AddToContent{G.~Cacciapaglia, A.~Deandrea and S.~De~Curtis}
%

\superpart{Hidden sectors}


\input Henderson/Henderson.tex
\AddToContent{C.~Henderson}
%

\input TomalinStrassler/TomalinStrassler.tex

\AddToContent{M.J.~Strassler and I.~Tomalin}
%

\input Morrissey/Morrissey.tex

\AddToContent{D.E.~Morrissey, D.~Poland and K.M.~Zurek}
%


\newpage

\input LH_NewPhysics_Report_biblio.tex

\end{document}

%% file: commands.tex
\addto\captionsenglish{}

\makeatletter
\def\myaddcontentsline@toc#1#2#3{%
   \addtocontents{#1}{\protect\thispagestyle{empty}}%
   \addtocontents{#1}{\protect \vspace{-.5cm}}
   \addtocontents{#1}{\protect\contentsline{#2}{#3}{}}
   \addtocontents{#1}{\protect \vspace{-.3cm}}}
\def\myaddcontentsline#1#2#3{%
  \expandafter\@ifundefined{addcontentsline@#1}%
  {\addtocontents{#1}{\protect \vspace{-.5cm}}
  \addtocontents{#1}{\protect\contentsline{#2}{#3}{}}
  \addtocontents{#1}{\protect \vspace{-.3cm}}}
  {\csname addcontentsline@#1\endcsname{#1}{#2}{#3}}}
\makeatother

\makeatletter
\def\mysuperaddcontentsline@toc#1#2#3{%
   \addtocontents{#1}{\protect\thispagestyle{empty}}%
   \addtocontents{#1}{\protect \vspace{.1cm}}
   \addtocontents{#1}{\protect\contentsline{#2}{#3}{\thepage}}
   \addtocontents{#1}{\protect \vspace{-.1cm}}}
\def\mysuperaddcontentsline#1#2#3{%
  \expandafter\@ifundefined{addcontentsline@#1}%
  {\addtocontents{#1}{\protect \vspace{.1cm}}
  \addtocontents{#1}{\protect\contentsline{#2}{#3}{\thepage}}
  \addtocontents{#1}{\protect \vspace{-.1cm}}}
  {\csname addcontentsline@#1\endcsname{#1}{#2}{#3}}}
\makeatother

\newcommand{\superpart}[1]{\clearpage \thispagestyle{empty}  \vphantom{blabla}\vspace{5cm} \centerline{\huge #1} 
\mysuperaddcontentsline{toc}{part}{\protect\numberline{} \hspace{-2cm}#1}%
\clearpage}
   
\newcommand{\AddToContent}[1]{\myaddcontentsline{toc}{chapter}{\protect  \textit{\hspace{1.5cm} \small#1}}}


%% file: authors_commands.tex
\newcommand{\sumet}{Sum-$E_T$}

\newcommand{\mP}{{\bar M}_{P}}

\newcommand{\pslash}{\mbox{$\not \hspace{-0.08cm} p$ }}

\newcommand{\pb}{\,\rm{pb}}
\newcommand{\wt}{\widetilde}

\newcommand{\eqn}{equation}
\newcommand{\fb}{{\ensuremath\rm fb}\xspace}
\newcommand{\TeV}{{\ensuremath\rm TeV}\xspace}
\newcommand{\mrm}[1]{\mathrm{#1}}
\newcommand{\Pt}{\ensuremath{p_{\mrm{T}}}\xspace}
\newcommand{\EtMiss}{\ensuremath{E_{\mrm{T}}^{\mrm{miss}}}\xspace}
\newcommand{\GeV}{{\ensuremath\rm GeV}}
\newcommand{\Meff}{\ensuremath{M_{\mrm{eff}}}\xspace}
\newcommand{\Msusy}{\ensuremath{M_{\mrm{SUSY}}}\xspace}
\newcommand{\Mtt}{\ensuremath{M_{T2}} }

\newcommand{\Smin}{\ensuremath{\hat{s}^{1/2}_{\mrm{min}}}\xspace}
\newcommand{\Minv}{\ensuremath{M_{\mrm{inv}}}\xspace}

\newcommand{\lsim}{\raisebox{-0.13cm}{~\shortstack{$<$ \\[-0.07cm] $\sim$}}~}
\newcommand{\gsim}{\raisebox{-0.13cm}{~\shortstack{$>$ \\[-0.07cm] $\sim$}}~}
\newcommand{\beq}{\begin{eqnarray}}
\newcommand{\eeq}{\end{eqnarray}}

\newcommand\ltap{\
  \raise.3ex\hbox{$<$\kern-.75em\lower1ex\hbox{$\sim$}}\ }
\newcommand\gtap{\
  \raise.3ex\hbox{$>$\kern-.75em\lower1ex\hbox{$\sim$}}\ }

\newcommand\simge{\mathrel{%
   \rlap{\raise 0.511ex \hbox{$>$}}{\lower 0.511ex \hbox{$\sim$}}}}
\newcommand\simle{\mathrel{
   \rlap{\raise 0.511ex \hbox{$<$}}{\lower 0.511ex \hbox{$\sim$}}}}

\newcommand{\slashchar}[1]%
        {\kern .25em\raise.18ex\hbox{$/$}\kern-.75em #1}
\def\lsim{\mathrel{\raise.3ex\hbox{$<$\kern-.75em\lower1ex\hbox{$\sim$}}}}
\def\gsim{\mathrel{\raise.3ex\hbox{$>$\kern-.75em\lower1ex\hbox{$\sim$}}}}
\renewcommand{\bs}{\boldsymbol}

\newcommand\CG{{\cal G}}

\newcommand\CL{{\cal L}}

\newcommand\CO{{\cal O}}

\newcommand\be{\begin{equation}}
\newcommand\ee{\end{equation}}
\newcommand\bea{\begin{eqnarray}}
\newcommand\eea{\end{eqnarray}}
\newcommand\ba{\begin{array}}
\newcommand\ea{\end{array}}
\newcommand\nn{\nonumber}

\newcommand\Tr{{\rm Tr}}

\newcommand{\half}{\ensuremath{\frac{1}{2}}}

\newcommand{\tfourth}{\ensuremath{\textstyle{\frac{1}{4}}}}

\newcommand\dagg{\dagger}
\newcommand\ts{\thinspace}
\newcommand\ra{\rightarrow}

\newcommand\ol{\bar}
\newcommand\mev{{\rm MeV}}
\newcommand\gev{{\rm GeV}}
\newcommand\tev{{\rm TeV}}
\newcommand\MeV{{\rm MeV}}
\newcommand\ipb{{\rm pb}^{-1}}
\newcommand\ifb{{\rm fb}^{-1}}
\newcommand\ecm{\sqrt{s}}

\newcommand\shat{\hat s}

\newcommand\etmiss{\slashchar{E}_T}

\newcommand\cm{{\rm cm}}
\newcommand\ellm{\ell^-}

\newcommand\ellp{\ell^+}

\newcommand\mm{{\rm mm}}

\newcommand\suc{SU(3)_C}
\newcommand\Ntc{N_{TC}}

\newcommand\atc{\alpha_{TC}}

\newcommand\Leff{{\cal L}_{\rm eff}}
\newcommand\Lsig{{\cal L}_{\Sigma}}
\newcommand\LFF{{\cal L}_{\rm gauge}}
\newcommand\LWZW{{\cal L}_{\rm WZW}}
\newcommand\Lff{{\cal L}_{\bar f f}}
\newcommand\Lpifbf{{\cal L}_{\tpi \bar f f}}
\newcommand\grpp{g_{\rho_T\pi_T\pi_T}}
\newcommand\condtc{{\langle \ol T T \rangle}_{TC}}
\newcommand\condetc{{\langle \ol T T \rangle}_{ETC}}

\newcommand\tom{\omega_{T}}
\newcommand\tro{\rho_{T}}

\newcommand\trho{\rho_T}
\newcommand\ta{a_T}

\newcommand\taz{a_T^0}
\newcommand\tapm{a_T^\pm}

\newcommand\tropm{\rho_{T}^\pm}

\newcommand\troz{\rho_{T}^0}

\newcommand\tpi{\pi_T}

\newcommand\tpiz{\pi_T^0}
\newcommand\tpipr{\pi_T^{0 \prime}}

\newcommand\jets{{\rm jets}}

\newenvironment{changemargin}[2]{\begin{list}{}{
        \setlength{\topsep}{0pt}\setlength{\leftmargin}{0pt}
        \setlength{\rightmargin}{0pt}
        \setlength{\listparindent}{\parindent}
        \setlength{\itemindent}{\parindent}
        \setlength{\parsep}{0pt plus 1pt}
        \addtolength{\leftmargin}{#1}\addtolength{\rightmargin}{#2}
        }\item }{\end{list}}

%% file: authors.tex
\begin{center}
\textbf{G.~Brooijmans}$^{1}$, 
\textbf{C.~Grojean}$^{2,3}$, 
\textbf{G.D.~Kribs}$^{4}$ 
\textbf{and}
\textbf{C.~Shepherd-Themistocleous}$^{5}$ 
\textbf{(convenors)}\\
%
K.~Agashe$^{6}$, 
L.~Basso$^{5, 7}$, 
G.~Belanger$^{8}$, 
A.~Belyaev$^{5, 7}$, 
K.~Black$^{9}$, 
T.~Bose$^{10}$, 
R.~Bruneli\`ere$^{11}$, 
G.~Cacciapaglia$^{12, 13}$, 
E.~Carrera$^{10, 14}$, 
S.P.~Das$^{15}$, 
A.~Deandrea$^{12, 13}$, 
S.~De~Curtis$^{16}$, 
A.-I.~Etienvre$^{17}$, 
J.R.~Espinosa$^{18}$, 
S.~Fichet$^{19}$, 
L.~Gauthier$^{17}$, 
S.~Gopalakrishna$^{20}$, 
H.~Gray$^{21}$, 
B.~Gripaios$^{2}$, 
M.~Guchait$^{22}$,  
S.J.~Harper$^{5}$, 
C.~Henderson$^{23}$, 
J.~Jackson$^{5, 24}$, 
M.~Karag\"oz$^{25}$, 
S.~Kraml$^{19}$, 
K.~Lane$^{10}$, 
T.~Lari$^{26}$, 
S.J.~Lee$^{27}$, 
J.R.~Lessard$^{28}$, 
Y.~Maravin$^{29}$, 
A.~Martin$^{30}$, 
B.~McElrath$^{31}$, 
G.~Moreau$^{32}$, 
S.~Moretti$^{5, 7, 33}$, 
D.E.~Morrissey$^{34}$, 
M.~M\"uhlleitner$^{35}$, 
D.~Poland$^{36}$, 
G.M.~Pruna$^{5, 7}$, 
A.~Pukhov$^{37}$, 
A.R.~Raklev$^{38}$, 
T.~Robens$^{39}$, 
R.~Rosenfeld$^{40}$, 
H.~Rzehak$^{35}$, 
G.P.~Salam$^{41}$, 
S.~Sekmen$^{42}$, 
G.~Servant$^{2, 3}$, 
R.K.~Singh$^{43}$, 
B.C.~Smith$^{9}$, 
M~Spira$^{44}$, 
M.J.~Strassler$^{45}$, 
I.~Tomalin$^{5}$, 
M.~Tytgat$^{46}$, 
M.~Vos$^{47}$, 
J.G.~Wacker$^{48}$, 
P.~v.~Weitershausen$^{39}$, 
and~K.M.~Zurek$^{49}$ 
\end{center}

%% file: addresses.tex
{\small
$^{1}$ Physics Department, Columbia University, New York , USA\\
$^{2}$ CERN, Physics Departement, Theory Unit, Geneva, Switzerland\\ 
$^{3}$ IPhT, CEA--Saclay, Gif-sur-Yvette, France\\
$^{4}$Department of Physics, University of Oregon, Eugene, USA \\
$^{5}$Particle Physics Department, STFC, Rutherford Appleton Laboratory, Didcot, UK\\
$^{6}$ Maryland Center for Fundamental Physics, Department of Physics, U. of Maryland, USA\\ 
$^{7}$School of Physics \& Astronomy, University of Southampton, Highfield, Southampton, UK\\
$^{8}$  LAPTH, Universit\'e de Savoie, CNRS,  Annecy-le-Vieux, France\\
$^{9}$ Laboratory for Particle Physics and Cosmology, Harvard University, Cambridge, USA\\ 
$^{10}$ Department of Physics, Boston University, Boston, USA\\  
$^{11}$ Fakult\"at f\"ur Mathematik und Physik, Albert-Ludwigs-Universit\"at, Freiburg, Germany\\
$^{12}$ Universit\'e Lyon 1, Villeurbanne, France\\
$^{13}$ Institut de Physique Nucl\'eaire de Lyon, CNRS/IN2P3, UMR5822, Villeurbanne, France\\
$^{14}$ Department of Physics, Boston University, Boston, USA\\ 
$^{15}$ AHEP Group, Institut de F\'{\i}sica Corpuscular -- C.S.I.C./Universitat de Val{\`e}ncia, Spain\\
$^{16}$INFN, Sesto Fiorentino, Firenze, Italy\\
$^{17}$ IRFU/Service de physique des particules, CEA--Saclay, Gif-sur-Yvette, France\\
$^{18}$ ICREA and IFAE, Universitat Aut{\`o}noma de Barcelona, Barcelona, Spain \\
$^{19}$ LPSC, UJF Grenoble 1, CNRS/IN2P3, Grenoble, France\\
$^{20}$ The Institute of Mathematical Sciences, C.I.T. Campus, Taramani, Chennai, India\\
$^{21}$ Physics Department, California Institute of Technology, Pasadena, USA\\
$^{22}$ Department of High Energy Physics, Tata Institute of Fundamental Research, Mumbai, India\\
$^{23}$ CERN, Geneva, Switzerland\\
$^{24}$ H.H.~Wills Physics Laboratory, University of Bristol, UK\\
$^{25}$ University of Oxford, Subdepartment of Particle Physics, Oxford, UK\\
$^{26}$ INFN, Sezione di Milano, Milano, Italy\\
$^{27}$ Department of Particle Physics, Weizmann Institute of Science, Rehovot, Israel\\
$^{28}$ Department of Physics and Astronomy, University of Victoria, Victoria, Canada\\
$^{29}$ Kansas State University, Manhattan, USA\\
$^{30}$ Fermi National Accelerator Laboratory, Batavia, USA\\
$^{31}$ Heidelberg University, Heidelberg, Germany \\
$^{32}$ Laboratoire de Physique Th\'{e}orique, Universit\'{e} Paris XI, Orsay, France\\
$^{33}$ Dipartimento di Fisica Teorica, Universit\`a di Torino, Torino, Italy\\
$^{34}$ TRIUMF, Vancouver, Canada\\
$^{35}$ Institute for Theoretical Physics, Karlsruhe Institute of Technology,  Karlsruhe, Germany\\
$^{36}$ Jefferson Physical Laboratory, Harvard University, Cambridge, USA\\
$^{37}$ Skobeltsyn Inst. of Nuclear Physics, Moscow State University, Moscow, Russia\\
$^{38}$ Oskar Klein Centre, Department of Physics, Stockholm University, Stockholm, Sweden\\
$^{39}$ Department of Physics and Astronomy, University of Glasgow, Glasgow, UK\\
$^{40}$ Instituto de Fisica Teorica -- UNESP,  Sao Paulo, Brazil\\
$^{41}$LPTHE, UPMC Univ. Paris 6, CNRS UMR 7589, Paris, France\\
$^{42}$ Department of Physics, Florida State University, Tallahassee, FL 32306, USA\\
$^{43}$ Institut f\"ur Theoretische Physik und Astrophysik, W\"urzburg, Germany\\
$^{44}$ Paul Scherrer Institute, Villigen PSI, Switzerland\\
$^{45}$ Department of Physics and Astronomy, Rutgers University, Piscataway, USA\\
$^{46}$ Department of Physics and Astronomy, University of Gent,  Gent, Belgium\\
$^{47}$ IFIC --- centre mixte Univ. Val\`encia/CSIC, Valencia, Spain\\
$^{48}$ Theory Group, SLAC, Menlo Park, USA\\
$^{49}$ University of Michigan, Ann Arbor, USA
}

%% file: intro.tex
\noindent {\Large {\bf Introduction}}
\vspace{.5cm}

{\it G.~Brooijmans,  C.~Grojean, G.D.~Kribs and C.~Shepherd-Themistocleous}
\vspace{.5cm}

The LHC has started colliding proton beams at a center of mass energy of 7~TeV,
ushering in a new era of physics at the energy frontier.  The exploration of physics in the 
multi-TeV energy domain will take another major step forward in~2013 when the 
LHC will run at a center of mass energy close to 14~TeV.

The minimal discovery scenario for the LHC is the Higgs boson, but it is
likely that there will be a lot more.  In the case of observation of the Higgs 
boson, the mechanism responsible for stabilizing its mass at the electroweak 
scale should manifest itself.  If the Standard Model Higgs boson is shown not to exist,
other new particles or interactions fulfilling its role in regulating the massive
vector boson scattering cross-section should be observed.  The LHC's discovery
potential spans a broad spectrum, including the direct production of the
dark matter components in the universe and the manifestation of new degrees of freedom 
in space-time.  In this report a wide variety of new physics signals are studied, 
exploring mostly areas that have emerged recently.  

The first two contributions examine the various tools developed to measure new 
particle masses in scenarios where all decays include an unobservable particle,
usually a dark matter candidate.  The performance of these tools is evaluated
and compared for different new physics scenarios, illustrating their complementary
strengths.

A second group of studies use Supersymmetry as a working model.  These evaluate
multiple aspects: how to observe R-parity violating decays of gluinos, the
impact of SUSY-QCD corrections to MSSM Higgs production, how to distinguish 
Supersymmetry from Gauge-Higgs Unification, and the allowed region for CP-violating
phases in the MSSM.

This is followed by two contributions in the area of strong electroweak symmetry 
breaking, one on the impact of a composite nature of the Higgs boson on the LHC
Higgs discovery potential, and the second a study of the LHC discovery reach for techni-vector
mesons in their decays to electroweak vector bosons.

Another possibility is that new high mass resonances will produced at the LHC.  A
first article on that topic reviews processes inspired by models of warped extra 
dimensions and shows the results of some applications of the special techniques needed in their 
discovery.  In other contributions to this section, the LHC discovery potential at multiple 
center of 
mass energies for the specific case of a $Z'_{B-L}$ boson is revisited, the production
of heavy Kaluza-Klein quarks and four-top final states are studied, the LHC sensitivity 
to very wide high mass $t\bar{t}$ resonances is examined, and the effects of nearby 
resonances are considered.

The final set of studies included in this report tackle the novel signatures introduced
in recent hidden sector models.  One of these proposes a new trigger scheme for signatures
with multitudes of low energy photons, another defines a set of trigger, reconstruction and
analysis benchmarks that should be appropriate for early LHC searches, and a third
explores what would happen if the MSSM were coupled to a new gauged hidden sector with 
characteristic energy scale in the GeV region.

This report does not attempt to present an exhaustive picture of new physics at LHC.
However, in presenting a wide variety of signatures motivated by very different 
models and exploring the performance of sets of techniques to observe and/or 
measure these new physics scenarios, it will hopefully serve as a useful resource 
for the exploitation of the LHC physics potential. 

%% file: Robens/masses.tex
\chapter{Comparison of mass determination methods at the LHC}

{\it L.~Basso, R.~Bruneli\`ere, T.~Lari, J.-R.~Lessard, B.~McElrath, T.~Robens,\\
 S.~Sekmen, M.~Tytgat and P.~v.~Weitershausen} 

\begin{abstract}
\input{Robens/TR_abstract}
\end{abstract}

\section{Introduction}
\input{Robens/TR_intro}

\section{Event generation and detector simulation}
\label{massd_sec:generation}
\input{Robens/Generation}

\section{Mass variables}
\input{Robens/Meff_Description}
\input{Robens/Smin_Description}

\input{Robens/MT_Description}
\input{Robens/MT2_Description}

\input{Robens/Edge_Description}

\input{Robens/Polynomial_Description}


\section{Results}
\label{massd_sec:results}
All analyses use object definitions and cuts as given in~\ref{massd_sec:delphesdef} if not stated otherwise.
\input{Robens/Meff_Results}
\input{Robens/Smin_Results}
\input{Robens/MT_Results}
\input{Robens/MT2_Results}

\input{Robens/Edge_Results}

\input{Robens/Polynomial_Results}

\section{Conclusions}\label{massd_sec:conclusion}
\input{Robens/TR_conclude}


\input{Robens/susy_spec}
\input{Robens/delphes_def}

\section*{Acknowledgements}
We thank Benjamin Fuks, Claude Duhr, and Priscila de Aquino for their help during the setup of Feynrules during an earlier version of this study. We also thank Xavier Rouby for clarifying some Delphes-related issues. Furthermore, TR and PvW are indebted to David Miller for helpful discussions and careful reading of the edge-related sections of this report. We finally want to congratulate the Les Houches organizers for a great and fruitful workshop atmosphere, good food, and the amazing scenery of the French
alps.

%% file: Robens/TR_abstract.tex
For any BSM theory, the underlying particle mass spectrum will be
among the first available information on the new physics involved. A
multitude of techniques is currently available to determine the masses of
new particles in these models from measured data. Here, we report on an
ongoing study in which different mass determination methods are applied to
a common SUSY event sample, generated including a generic collider
detector simulation. The event sample was produced with and without the
explicit generation of an additional hard jet by the hard matrix element
to investigate possible effects of extra hard jet radiation. We report on
first results of this study for several of the more commonly used mass
determination methods.

%% file: Robens/TR_intro.tex
The start of data taking at CERN's Large Hadron Collider at the end of
2009~\cite{twitter} 
promises the beginning of an exciting era for both
Standard Model physics and beyond the Standard Model searches. Currently,
many BSM models are on the market, which promise to solve some SM inherent
puzzles such as the hierarchy problem, or the absence of dark matter
candidates. These models typically introduce additional massive particles,
where coupling and mass exclusion limits are obtained from past and
current BSM collider searches~\cite{Amsler:2008zzb}. For the center of
mass energies at the LHC, many allowed scenarios exist where the new
particles are produced at a relatively high rate, and typically decay
through long decay chains containing both SM and BSM decay products. The
measurement of these BSM masses at the LHC will be among the first available
information about the structure of the underlying theory.

In the past years, a large number of methods, widely varying in
applicability and accuracy, have been proposed for measuring the masses of
the new particles at colliders (for a comprehensive
review, see Ref.~\cite{Barr:2010zj}).  Many of the well established methods have
already been tested to high accuracy in realistic experimental setups, where parton showers, hadronization, and detector simulation are all included.
However, similar studies for many of the more recently proposed variables, as
well as a consistent comparison of the existing methods are still lacking.
Here, we initiate a comparative investigation of various mass determination methods.
For this, we use common Monte Carlo samples for the mSUGRA scenario
SPS1a~\cite{Allanach:2002nj}\footnote{The superpartner masses and production cross-sections
for this scenario are given in~\ref{massd_app_spectrum}.}, where parton shower, decays and
hadronization were included and a generic LHC detector response was
modelled with a fast detector simulator. We also produced event samples
where one hard jet is explicitely generated by the hard matrix element,
and matched with the parton shower using the MLM matching algorithm
\cite{Alwall:2008qv}. For methods relying on jet spectra, the effect
of this more accurate description of the jet energy distribution needs to
be taken into account in a realistic application of the respective
variable. First results on these are presented in this study as well.\\

Throughout our work, we have used a center of mass energy of $14\,\TeV$. We
focused on a luminosity of $10\,\fb^{-1}$. Our results should apply for
the first stage analyses at the nominal LHC energy.

%% file: Robens/Generation.tex

\subsection{Event generation}

In this project, we have generated events for supersymmetric processes and the two main background processes.
In order to include a better description of initial state radiation with high $P_{T}$ and three-body decays which could affect mass determination, event generation of the supersymmetric signal has been performed in three steps:
\begin{enumerate}
\item Matrix element generation has been done with Madgraph 4.4.24~\cite{Stelzer:1994ta,Maltoni:2002qb}. Samples were divided according to the different final states $\tilde{g}\tilde{g}$, $\tilde{g}\tilde{q}$, $\tilde{q}\tilde{q}$ and $\chi\chi$ and the number of QCD radiations (0 or 1). $\tilde{g}$ means a gluino, $\tilde{q}$ is a squark and $\chi$ is either a chargino or a neutralino. Samples generated with no or one additional QCD radiation will be named in the following $2\rightarrow2$ and $2\rightarrow3$ processes, respectively.
\item As a second step, the particles produced during matrix element event generation are provided to BRIDGE. BRIDGE v1.8~\cite{Meade:2007js} is used to decay supersymmetric particles according to its own decay rates using all possible 2 and 3-body decays.
\item Finally, decayed events are passed to Pythia~\cite{pythia} version 6.420 for parton showering and hadronization. The merging of samples with different parton multiplicity is also performed during this last step using the MLM matching scheme as explained in~\cite{Alwall:2007fs,Alwall:2008qv}. The main matching parameter $Q_{match}$, used to determine whether a jet after showering is matched to one of the initial partons, is set to $40\ \mathrm{GeV}$. In order to avoid double counting between e.g. $\tilde{g}\tilde{g}$ and $\tilde{g}\tilde{q}\bar{q}$~\cite{Alwall:2008qv}, events from the latter process including an intermediate gluino resonance are excluded.
\end{enumerate}
The $W+jets$ and $t\bar{t}+jets$ backgrounds have been generated with Alpgen~\cite{Mangano:2002ea} plus Pythia~\cite{pythia} generators using the standard MLM matching procedure.\newline


\subsection{Detector simulation}

Simulation for a multipurpose LHC detector response was implemented using the fast simulation package Delphes 1.8~\cite{Ovyn:2009tx}.  Simulation includes a tracking system embedded into a magnetic field, calorimeters, a muon system, and very forward detectors arranged along the beamline, and takes into account the effect of magnetic field, the granularity of the calorimeters and subdetector resolutions.  We have used the default detector configuration.  Definitions of objects used in the analysis are given 
in~\ref{massd_sec:delphesdef}.

All generated signal and background samples were stored in the Monte Carlo Database MCDB~\cite{Belov:2007qg}.


%% file: Robens/Meff_Description.tex
\subsection{Effective mass}
The effective mass (\Meff) is used to estimate the SUSY mass scale (\Msusy). For hadronic processes, \Msusy usually refers to the masses of the strongly interacting SUSY particles. Authors in \cite{Hinchliffe:1996iu} use \Msusy as the lowest of these masses, while the author in \cite{Tovey:2000wk} defines it to be their average. Similarly, there is no universal way to define \Meff. The most widely used is described in equation (\ref{massd_sec:masses;Meff_eq}).

\begin{equation}
\Meff = p_{\mrm{T},1} + p_{\mrm{T},2} + p_{\mrm{T},3} + p_{\mrm{T},4} + \EtMiss
\label{massd_sec:masses;Meff_eq}
\end{equation}

It is mainly used in the 4 jets + \EtMiss channel. Nevertheless, the 2 jets + 2 leptons + \EtMiss channel will also be studied in this note. In this latter case, \ensuremath{p_{\mrm{T},3}} and \ensuremath{p_{\mrm{T},4}} of equation (\ref{massd_sec:masses;Meff_eq}) correspond to the leptons transverse momenta instead of the jets transverse momenta.

Independently of the definition of \Meff and \Msusy, the strategy is always the same. The correlation between these two is determined by simulating many points in SUSY parameter space. This correlation has been shown to be linear, although the correlation coefficient varies significantly depending on the exact definition of these two variables and the SUSY model used \cite{Tovey:2000wk}.

%% file: Robens/Smin_Description.tex
\subsection{Square root of s-hat min (\Smin)}
The \Smin variable, equation (\ref{massd_sec:masses;Smin_eq}), is another variable used to establish the SUSY energy scale. It is designed to have its distribution peak at the threshold center of mass energy ($\hat{s}^{1/2}$) of the studied process \cite{Konar:2008ei}. In the context of SUSY processes produced in hadron colliders, the threshold value of $\hat{s}^{1/2}$ corresponds to about twice the mass of the lightest gluino or squark.

\begin{equation}
\Smin(\Minv) \equiv \sqrt{E^{2} - P_{z}^{2}} + \sqrt{(\EtMiss)^{2} + \Minv^{2}}
\label{massd_sec:masses;Smin_eq} 
\end{equation}

The total visible energy is $E = \sum_{i} E_{i}$ and the corresponding longitudinal momentum is $P_{z} = \sum_{i} E_{i}\mrm{cos}\theta_{i}$, where index $i$ labels the calorimeter towers. \Minv is the sum of the masses of all the particles that cannot be detected (invisible). When muons are present in the event, their energy is added to $E$ and their longitudinal momentum is added to $P_{z}$.

%% file: Robens/MT_Description.tex
\subsection{Transverse Mass}
When particles in the final state escape detection, their momentum can only be constrained from the missing momentum in the transverse plane. Therefore, a simple variable that can extract the absolute masses of intermediate particles is the transverse mass. Such a variable does not rely on the event topology. The only requirement for the variable to work properly is that all the missing energy comes from the same particle (which mass is to be reconstructed). If such a requirement is not matched, it gets harder to extract information.
A typical suitable event is of the form:
\begin{equation}\label{massd_MTdef}
A + X \rightarrow B\mrm{(vis)} + C\mrm{(inv)} + X\, ,
\end{equation}
where the particle $A$ decays into some visible ($B$) and some invisible ($C$) particles. We use $X$ to identify anything else taking part in the event and not being important here.

Several definitions of the transverse mass exist in the literature: we quote here Barger's Transverse Mass \cite{Barger:1987du}:
\begin{equation}\label{massd_M_T}
M^2_T = \left( \sqrt{M^2\mrm{(vis)}+\vec{\Pt}^2\mrm{(vis)}}+\left|{\slash\!\!\!\!\!\:\Pt} \right| \right) ^2
	- \left( \vec{\Pt}\mrm{(vis)} + {\slash\!\!\!\!\!\:\vec{\Pt}} \right) ^2\, ,
\end{equation}
where $\mrm{(vis)}$ means the sum over the visible particles one wants to consider. A general feature of this variable is the prominent peak\footnote{Notice that this is not the case for chiral bosons, for which the Jacobian peak is absent \cite{Chizhov:2006nw}.} and sharp edge at the absolute mass of the parent particle $A$.

However, this variable assumes that all the missing energy comes from one particle, which is not generally true for BSM models with pair production and a stable particle, such as in the MSSM considered here.

%% file: Robens/MT2_Description.tex
\subsection{\Mtt Stransverse Mass and \Mtt Kink}
The \Mtt stransverse mass collider observable first introduced 
in~\cite{LesterSummers:1999} is
useful to measure masses of pair produced particles, with each of them
decaying to one or more directly visible particles and one invisible particle
leading to missing transverse momentum. It was shown that the endpoint of the \Mtt 
distribution yields an estimate
of the mass of the decaying particle, provided that the mass of the
invisible daughter is known. The method is especially suited for R-parity
conserving SUSY models, where superparticles are pair produced and the LSP at
the end of the decay chains is
stable and undetectable.
As an example, we give the expression for \Mtt as originally derived
in~\cite{LesterSummers:1999} for slepton pair production $pp\rightarrow
X+\tilde{l}_1\tilde{l}_2\rightarrow X+l_1\tilde{\chi}^0_1 l_2\tilde{\chi}^0_1$
\begin{equation}
M_{T2}\equiv
\underset{\not\text{p}_1+\not\text{p}_2=\not\text{p}_T}{\text{min}}\left[\text{max}\left\{m_T^2(\text{p}_T^{l_1},\not\!\text{p}_1),m_T^2(\text{p}_T^{l_2},\not\!\text{p}_2)\right\}\right],
\end{equation}
with
$m_T^2(\text{p}_T^{l_i},\not\!\text{p}_i)=m_{l_i}^2+m_{\tilde{\chi}}^2+2(E_{Tl_{i}}E_{Ti}-\text{p}_{Tl_i}\not\!\text{p}_{i})$
and $E_T=\sqrt{\text{p}_T^2+m^2}$
and where 
the minimization runs over all possible 2-momenta,
$\not\!\text{p}_{1,2}$ (corresponding to the unknown  
2-momenta of the two neutralinos), such
that their sum equals the total missing
transverse momentum, $\not\!\text{p}_T$, observed in the event. 
The condition that the mass of the invisible daughter is known beforehand is of course a
problem since none of
the SUSY particle masses have been determined yet. However, this problem can be avoided
with the
\Mtt kink method introduced in~\cite{ChoChoi:2007}, in which the \Mtt endpoint 
distribution considered as 
function of a trial mass for the invisible particle may
reveal a kink yielding the exact two unknown particle
masses separately. In the example of slepton pair production given above,
where the decay of the mother particle contains 1 visible particle, the
strength of the kink depends on $p_T(X)$ (or the total $p_T$ of the slepton
pair system) and the kink is expected to disappear for 
$p_T\rightarrow 0$~\cite{ChoChoi:2008, BurnsKong:2009}.

%% file: Robens/Edge_Description.tex
\subsection{Edges}
In contrast to other methods, mass determination from edges does not rely on a specific event topology. The method is typically used for long decay chains of the form
\begin{\eqn}
A\,\rightarrow\,B+C\,\rightarrow\,B+D+E\,\rightarrow\,...
\end{\eqn}
where the intermediate decay chain particles are taken onshell; in general, it can be used to extract masses from decay chains of arbitrary length\footnote{For $1\,\rightarrow\,2$ and $1\,\rightarrow\,3$ decays, only relative mass differences can be determined.}. From the four-momenta of the outgoing visible particles, invariant masses
\begin{\eqn}
m_{ab...n}^{2}\,=\,(p_{a}+p_{b}+...+p_{n})^{2}
\end{\eqn}
are constructed.
The  minimal and maximal allowed values of these variables, which are visible as ``edges'' in the respective distributions,
are completely determined by phase space and given in terms of the decay-chain masses only, therefore being independent of the total energy of the process. The explicit analytic form of the distribution endpoints depends on relative mass hierarchies between the intermediate onshell particles; in case of no a priori knowledge, all possible sets of inversion relations need to tested. Studies of edges have been presented in e.g. \cite{Hinchliffe:1996iu,Bachacou:1999zb,:1999fr,Allanach:2000kt}, and these (and similar) variables have found wide applications. \\
In our present study, we focus on the decay chain
\begin{\eqn}\label{massd_eq:edge_signal}
 \tilde{q} \to
\tilde{\chi}_2^0 q \to \tilde{l}lq \to \tilde{\chi}_1^0 llq,
\end{\eqn}
 where we consider the following variables
\begin{eqnarray*}
 m^2_{ll}\,=\, (p_{l_1} + p_{l_2})^2,&&
  m^2_{qll} \,=\, (p_{l_1} + p_{l_2} + p_q)^2,\\ m^2_{ql{\rm (low)}} \,=\, \min\{ (p_{l_1} + p_q)^2, (p_{l_2} + p_q)^2 \},&&
  m^2_{ql{\rm (high)}} \,=\, \max\{ (p_{l_1} + p_q)^2, (p_{l_2} + p_q)^2 \}
\end{eqnarray*}
The endpoints in the distributions of these variables are denoted by
$m_{ll}^{\rm max}$, $m_{qll}^{\rm max}$, $m_{ql{\rm (low)}}^{\rm
  max}$, $m_{ql{\rm (high)}}^{\rm max}$.

%% file: Robens/Polynomial_Description.tex
\subsection{Polynomial Intersection}

Unlike the preceding methods, the estimator in the case of Polynomial methods
is the mass itself, and not an auxiliary variable.  They work by hypothesizing a
kinematic topology consistent with the particles in the final state, and for
each assignment of visible objects to external legs, deriving a polynomial
equation for the event.  This polynomial is a function of unknown kinematic
quantities and masses.  One may then consider different ways to solve these
polynomials by making further assumptions.  Applications of these ideas were
pursued in Ref.\cite{Kawagoe:2004rz} for a single decay chain with some masses
known. Considering both decay chains simultaneously can potentially give us
more information and allow a better determination of the
masses~\cite{Cheng:2007xv,ChoChoi:2007,Nojiri:2007pq}.  Ref.\cite{Cheng:2007xv}
considered symmetric decay chains with two intermediate resonances on each
side, Ref.\cite{Cheng:2008mg,Cheng:2009fw} considered symmetric chains
with three intermediate resonances on each side, and Ref.\cite{Webber:2009vm}
used this same symmetric 3-resonance topology but omitted the quadratic missing
mass shell condition and instead used a likelihood to achieve similar results.
The relationship between these variables and the $M_{T2}$ "kink" observable was
explored in Ref.\cite{Cheng:2008hk}.

\begin{figure}[h]
\begin{center}
 \includegraphics[width=0.27\textwidth]{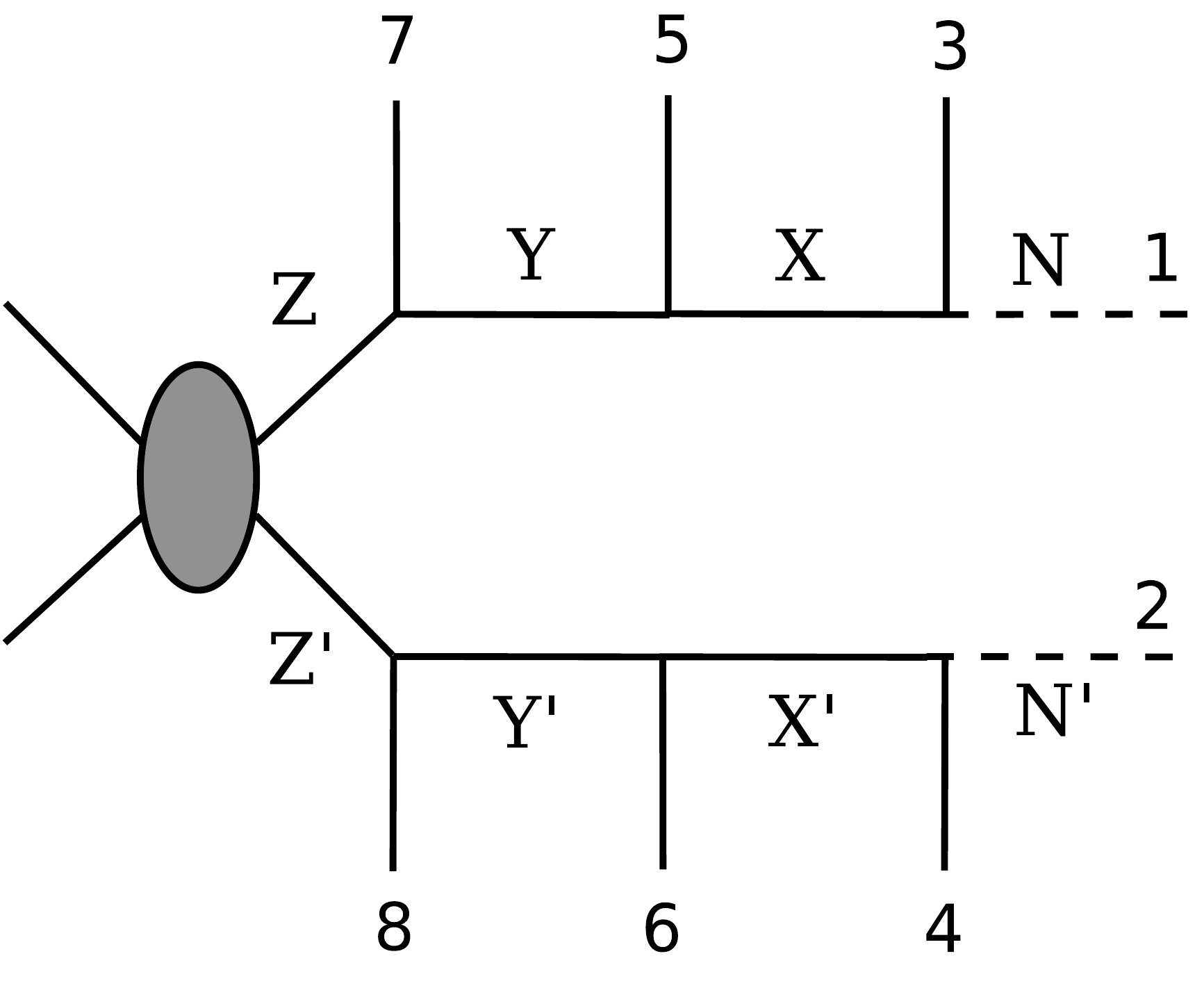}
\vspace*{-.1in}
\caption{\label{massd_fig:topology} The event topology considered.}
\vspace*{-.3in}
\end{center}
\end{figure}

For this study we examined events with resonances as shown in
Fig.\ref{massd_fig:topology}.  The equations for a single event, assuming the two
sides of the decay have the same masses are
\begin{equation}
\label{massd_massshell}
\begin{array}{lcrcl}
(M_Z^2 &=)& (p_1+p_3+p_5+p_7)^2&=&(p_2+p_4+p_6+p_8)^2,\\
( M_Y^2 &=)& (p_1+p_3+p_5)^2&=&(p_2+p_4+p_6)^2,\\
(M_X^2 &=)& (p_1+p_3)^2&=&(p_2+p_4)^2,\\
(M_N^2 &=)& p_1^2&=&p_2^2.
\end{array}
\end{equation}
where $p_i$ is the 4-momentum for particle $i$ $(i=1\ldots 8)$.
Since the only invisible particles are $1$ and $2$ and since we can
measure the missing transverse energy,
there are two more constraints:
\begin{eqnarray}
p_1^x+p_2^x=p_{miss}^x,\quad 
p_1^y+p_2^y=p_{miss}^y.\label{massd_misspt}
\end{eqnarray}
Given the 6 constraints in Eqs.~(\ref{massd_massshell}) and (\ref{massd_misspt}) 
and 8 unknowns from the 4-momenta of the missing particles, there remain two 
unknowns per event.  The system is under-constrained and cannot be
solved.
This situation changes if we use a second event with the same decay
chains, under the assumption that the invariant masses are the same in the two
events. Denoting the 4-momenta in the second event as $q_i$
$(i=1\ldots 8)$, we have 8 more unknowns, $q_1$ and $q_2$, but 10 more
equations,
\begin{eqnarray}
&&\begin{array}{rcccl}
q_1^2&=&q_2^2&=&p_2^2,\\
(q_1+q_3)^2&=&(q_2+q_4)^2&=&(p_2+p_4)^2,\\
(q_1+q_3+q_5)^2&=&(q_2+q_4+q_6)^2&=&(p_2+p_4+p_6)^2,\\
\multicolumn{5}{c}{
\begin{array}{rcl}
(q_1+q_3+q_5+q_7)^2&=&(q_2+q_4+q_6+q_8)^2\\
    &=&(p_2+p_4+p_6+p_8)^2,\\
\end{array}
} 
\end{array}
\label{massd_misspt2}
\nonumber \\ 
&&q_1^x+q_2^x=q_{miss}^x,\quad
q_1^y+q_2^y=q_{miss}^y.
\end{eqnarray} 
Altogether, we have 16 unknowns and 16 equations. The system can be
solved numerically and we obtain discrete solutions for $p_1$, $p_2$,
$q_1$, $q_2$ and thus the masses $m_N$, $m_X$, $m_Y$, and $m_Z$. Note
that the equations always have 8 complex solutions, but we will keep
only the real and positive ones which we henceforth call ``solutions''.
The code used to solve the polynomials is publicly available in
Ref.\cite{wimpmass}.

%% file: Robens/Meff_Results.tex
\subsection{Effective mass}
\begin{figure} [htbp]
\includegraphics[width=0.49\textwidth]{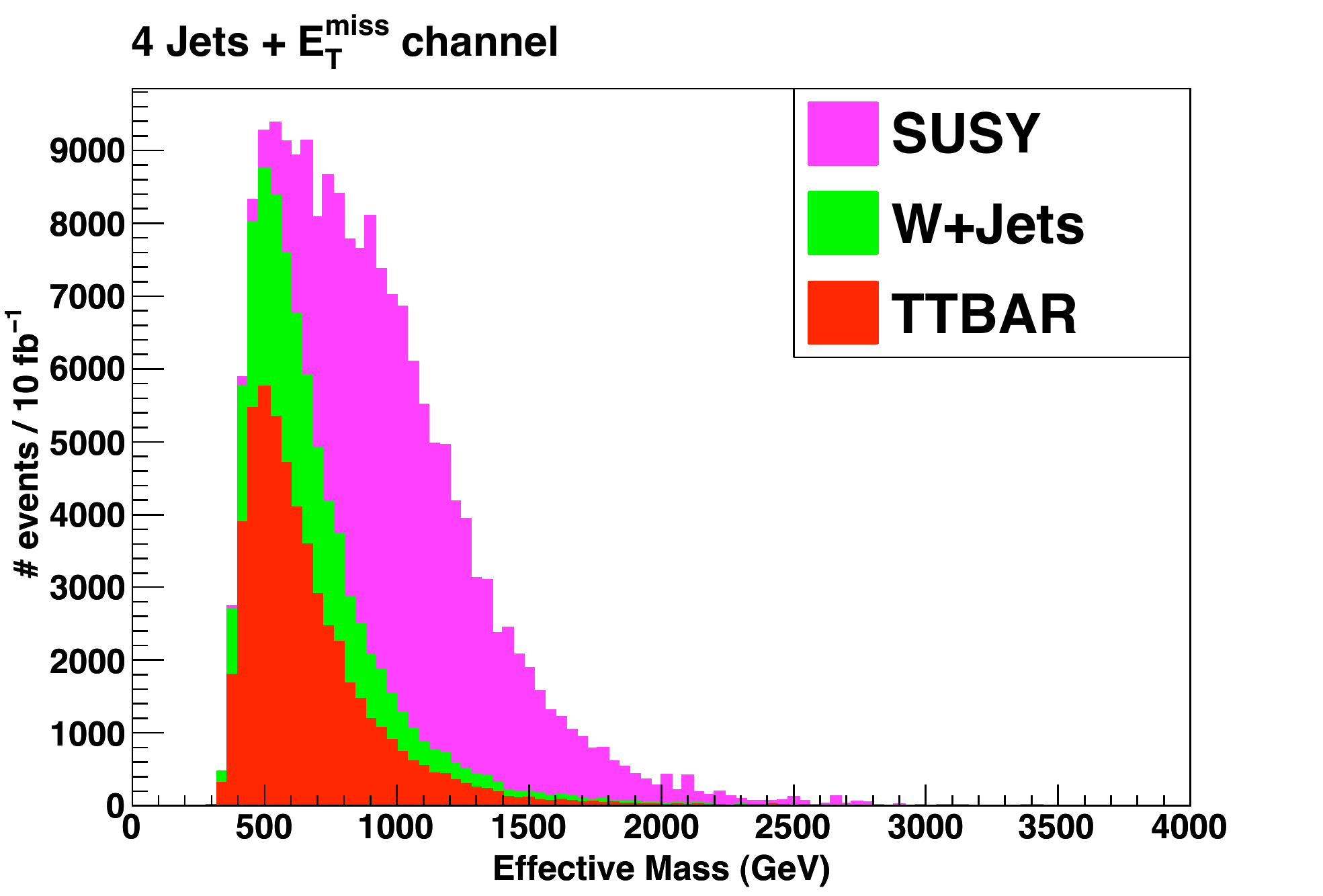}
\hfill
\includegraphics[width=0.49\textwidth]{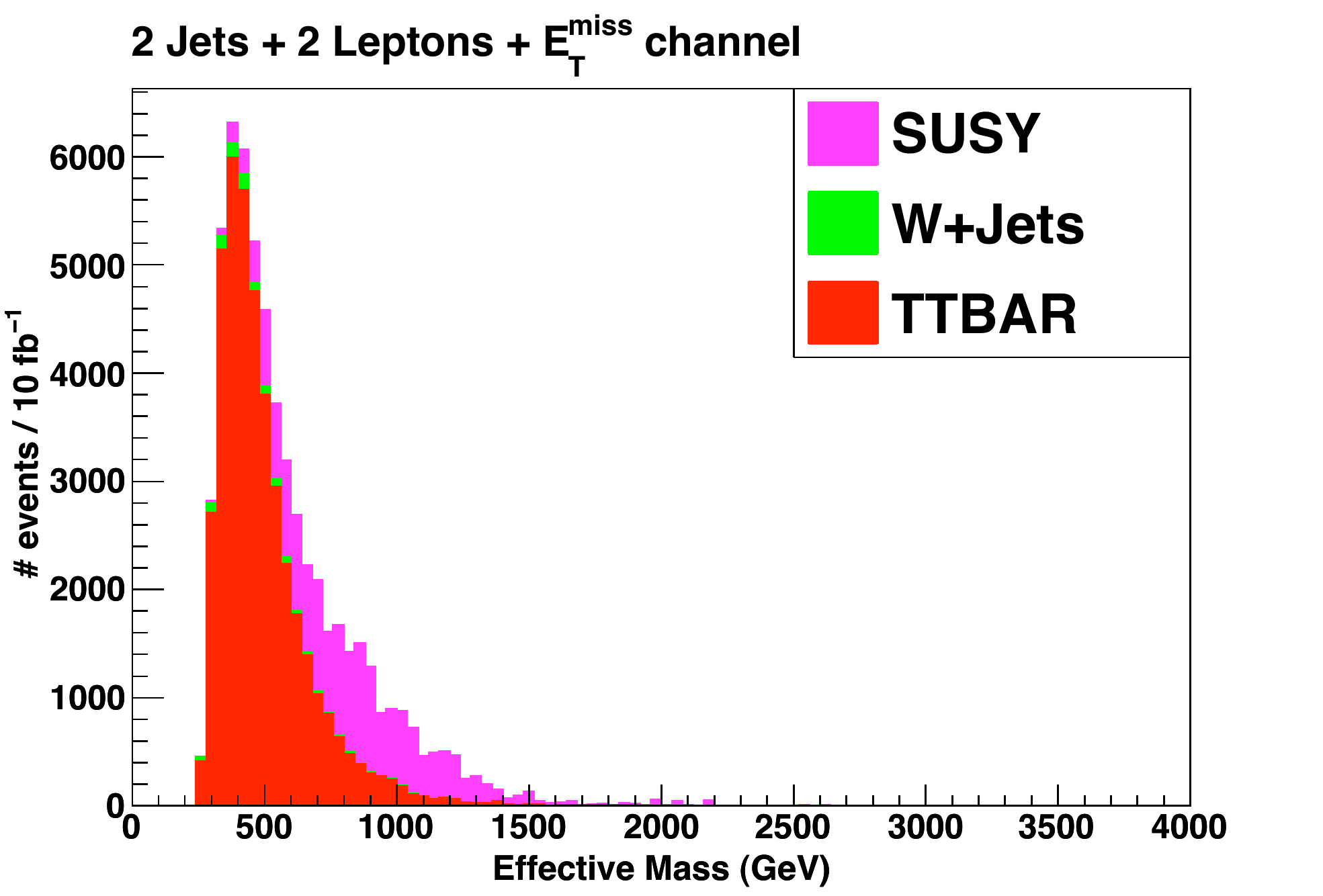}
\caption{The distribution of the effective mass for the 4 jets + \EtMiss channel (left) and the 2 jets + 2 leptons + \EtMiss channel (right). The SUSY events are in purple, while the background from top-antitop and W+jets events are in red and green respectively.}
\label{massd_sec:masses;Meff_SUSYwithBackground}
\end{figure}

The effective mass distribution is shown in Fig. \ref{massd_sec:masses;Meff_SUSYwithBackground}. 

\begin{figure} [htbp]
\begin{center}
\includegraphics[width=0.49\textwidth]{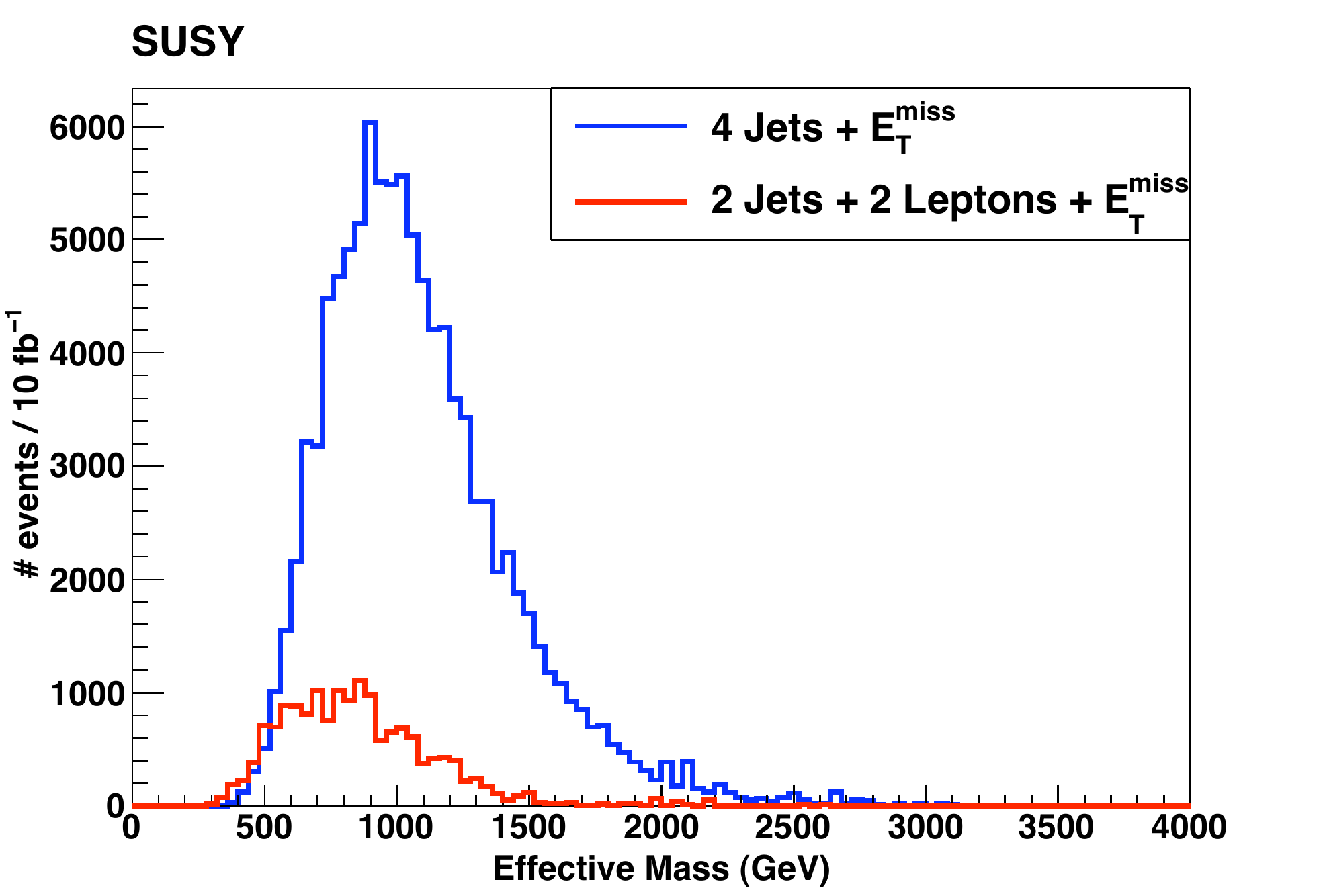}
\end{center}
\caption{The distribution of the effective mass for the 4 jets + \EtMiss channel (blue) and the 2 jets + 2 leptons + \EtMiss channel (red) when only SUSY events are present.}
\label{massd_sec:masses;Meff_2channels}
\end{figure}

The \Meff from SUSY events can be clearly distinguished from the backgrounds
considered\footnote{Due to computing constraints, multi-jets from QCD have not
  been simulated. They could be a significant source of background for the 4
  jets + \EtMiss channel although we are confident that requiring \EtMiss $>
  100$ GeV in the analysis would keep this type of background under
  control.}. This makes \Meff a good variable for early SUSY
  discovery. Moreover, given a good understanding of the backgrounds,
  the \Meff distribution from SUSY events could be deduced. From
  Fig. \ref{massd_sec:masses;Meff_2channels}, the peak of the distribution
  could be established with a precision of 10 to 100 GeV. Nevertheless, to estimate \Msusy from the distribution, the
corresponding SUSY model needs to be known. This is needed to find the
correlation between \Meff and \Msusy via MC simulation. However, an
effective mass analysis cannot discriminate between different SUSY
models. Consequently, external input from other studies is needed to
estimate \Msusy when using \Meff.

%% file: Robens/Smin_Results.tex
\subsection{Square root of s-hat min (\Smin)}
\begin{figure} [htbp]
\includegraphics[width=0.49\textwidth]{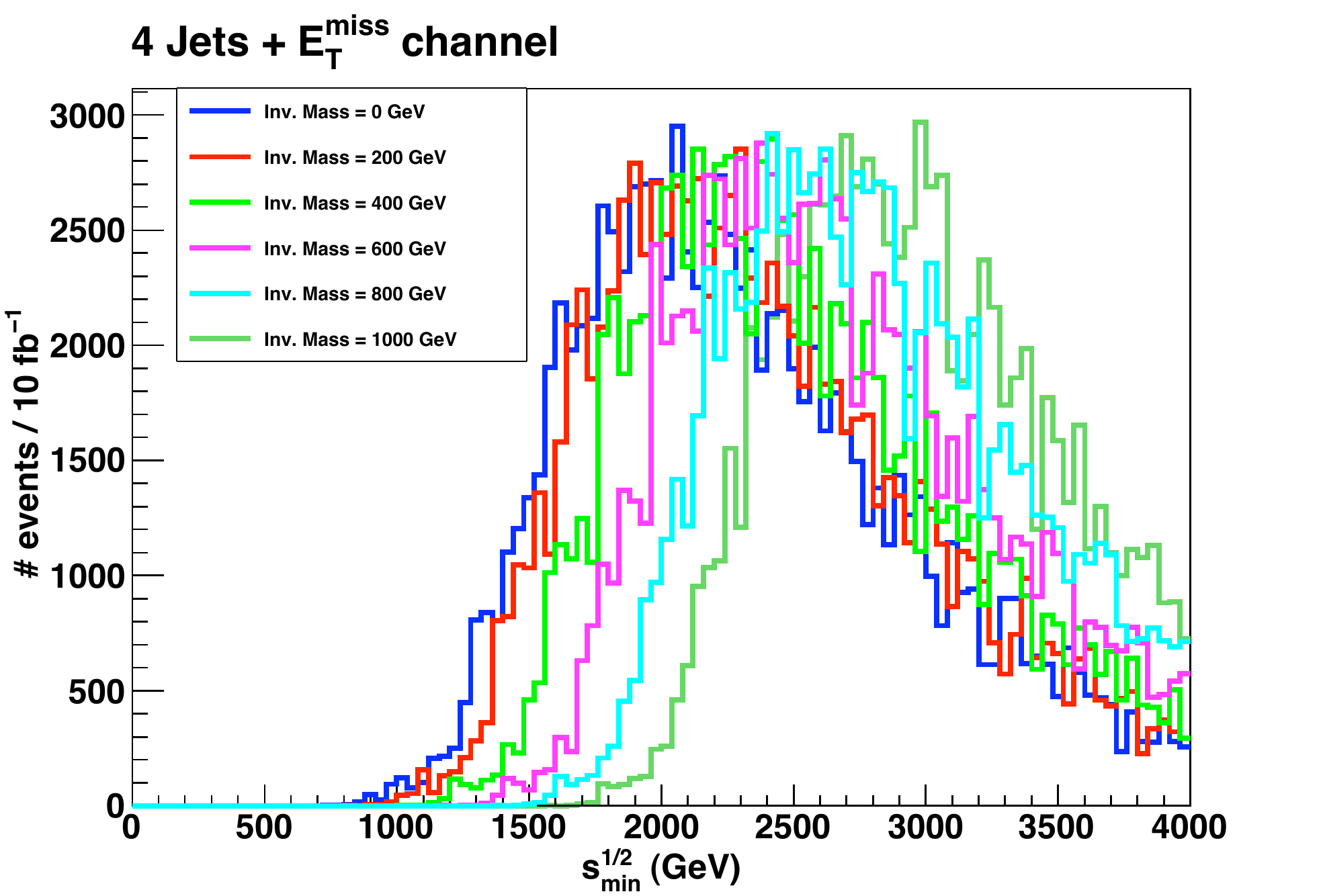}
\hfill
\includegraphics[width=0.49\textwidth]{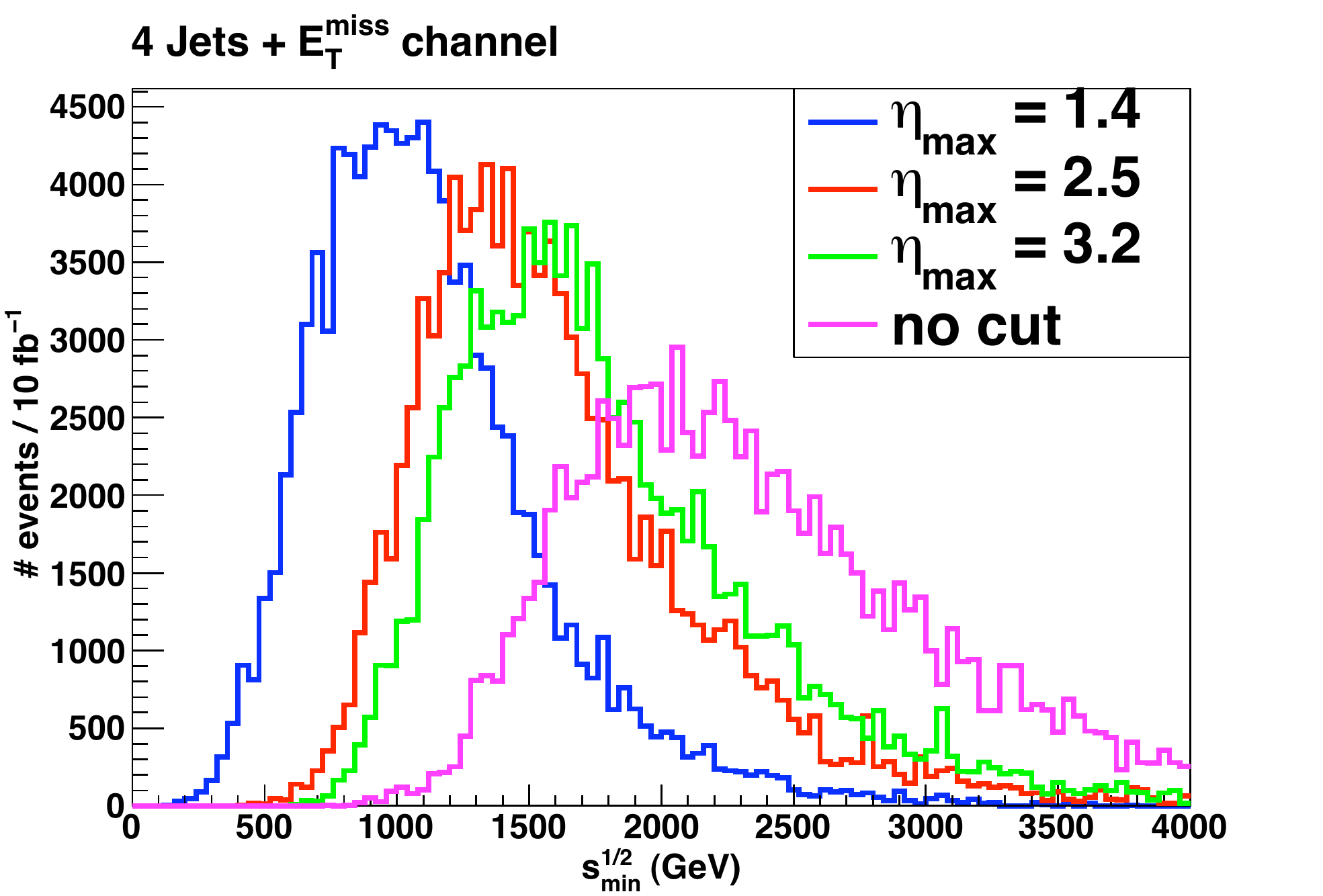}
\caption{The left hand side plot shows \Smin for six different values of \Minv, 0 (blue), 200 (red) , 400 (light green), 600 (purple), 800 (cyan) and 1000 (dark green) GeV. The right hand side figure shows the $\Smin(0)$ distribution for different $\eta_{\mrm{max}}$ cuts: $\eta_{\mrm{max}} = 1.4$ (blue), $\eta_{\mrm{max}} = 2.5$ (red), $\eta_{\mrm{max}} = 3.2$ (green) and no $\eta_{\mrm{max}}$ cut (purple). Both plots are using the 4 jets + \EtMiss channel.}
\label{massd_sec:masses;Smin_Minv}
\end{figure}

Although \Smin is model independent, it needs \Minv as input. In the SUSY context, it means that the neutralino mass needs to be known. The dependence of \Smin on \Minv is shown in Fig. \ref{massd_sec:masses;Smin_Minv}. Another issue with the \Smin variable is that it is very sensitive to initial state radiation (ISR). The solution proposed by the authors in \cite{Konar:2008ei} is to use only calorimeter towers with $|\eta|$ smaller than $\eta_{\mrm{max}}$ in the calculation of $E$ and $P_{z}$, equation (\ref{massd_sec:masses;Smin_eq}). They choose $\eta_{\mrm{max}} = 1.4$ based on the fact that this is where the CMS barrel ends. The effect of using different $\eta_{\mrm{max}}$ is shown in Fig.\ref{massd_sec:masses;Smin_Minv}. 

\begin{figure} [htbp]
\includegraphics[width=0.49\textwidth]{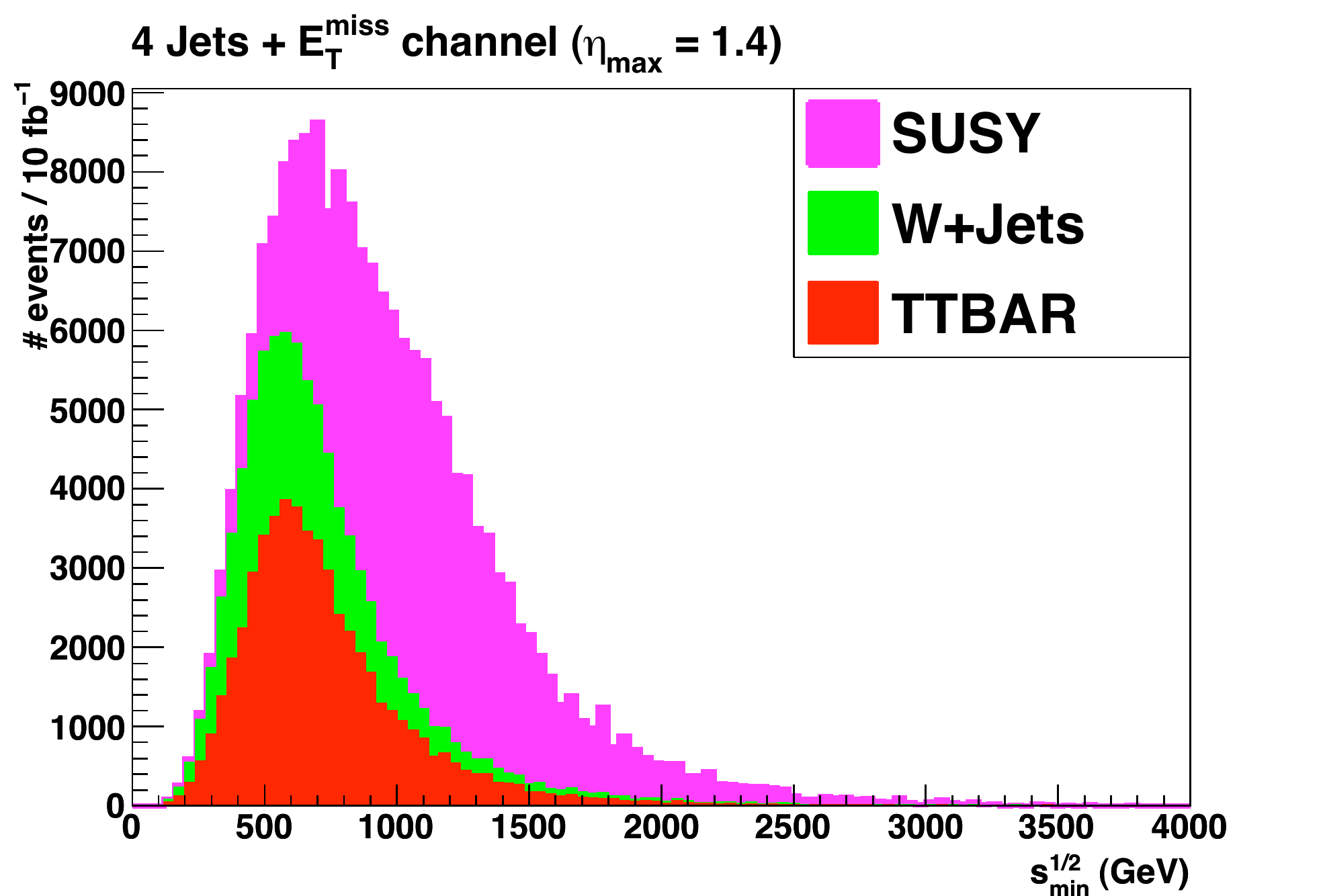}
\hfill
\includegraphics[width=0.49\textwidth]{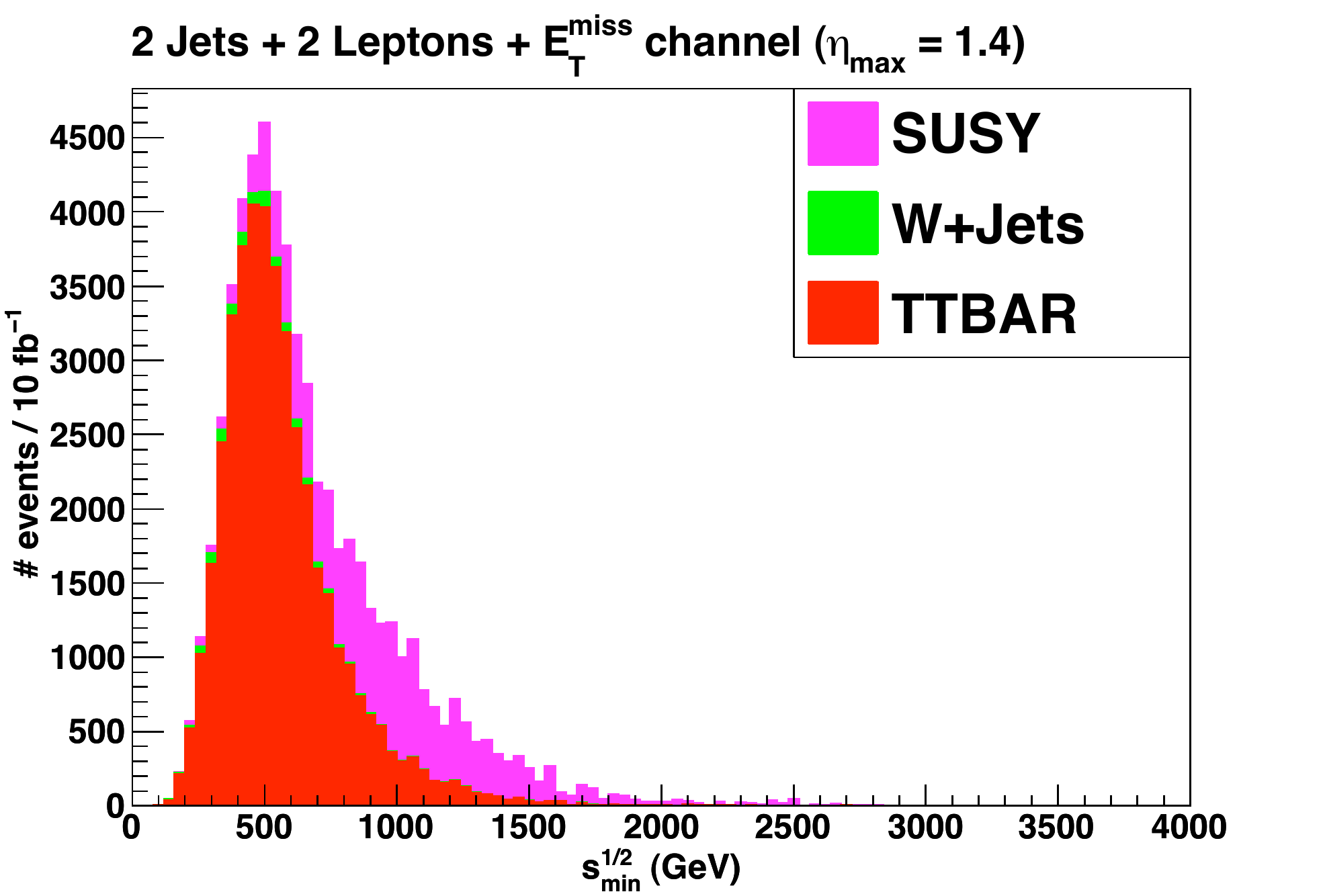}
\caption{The \Smin{}(0) distribution for the 4 jets + \EtMiss (left) and 2 jets + 2 leptons + \EtMiss (right) channels. $\eta_{\mrm{max}} = 1.4$ is used as described in the text.}
\label{massd_sec:masses;Smin_Distributions}
\end{figure}

The \Smin distribution of SUSY and background events (without QCD multi-jets) in the 4 jets + \EtMiss and 2 jets + 2 leptons + \EtMiss channels can be seen in Fig. \ref{massd_sec:masses;Smin_Distributions}. The \Smin SUSY distribution peaks at a different position than the SM model background making \Smin a good variable for early SUSY discovery. Nevertheless, it is doubtful that it will be useful in establishing the SUSY scale. First, Fig.\ref{massd_sec:masses;Smin_Minv} shows that \Smin(0) peaks at about 2000 GeV while \Smin(1000) peaks around 3000 GeV. It means that a bias of about twice the lightest SUSY particle mass would be introduced. Second, the choice of $\eta_{\mrm{max}}$ cut can induce another significant deviation. For example, the $\Smin(0)$ distribution of the top-antitop background should peak at twice the top mass ($\sim350$ GeV). However, from Fig. \ref{massd_sec:masses;Smin_Distributions}, it peaks at 500 GeV in the 2 jets + 2 leptons + \EtMiss channel and at 600 GeV for the 4 jets + \EtMiss channel. It is also worrisome that the variable \Smin is channel dependent. These observations lead to conclude that \Smin requires a very trustworthy MC to properly understand the effect of the $\eta_{\mrm{max}}$ cut.

%% file: Robens/MT_Results.tex
\subsection{Transverse Mass}
\begin{figure} [!htbp]
\centering
\includegraphics[width=0.49\textwidth]{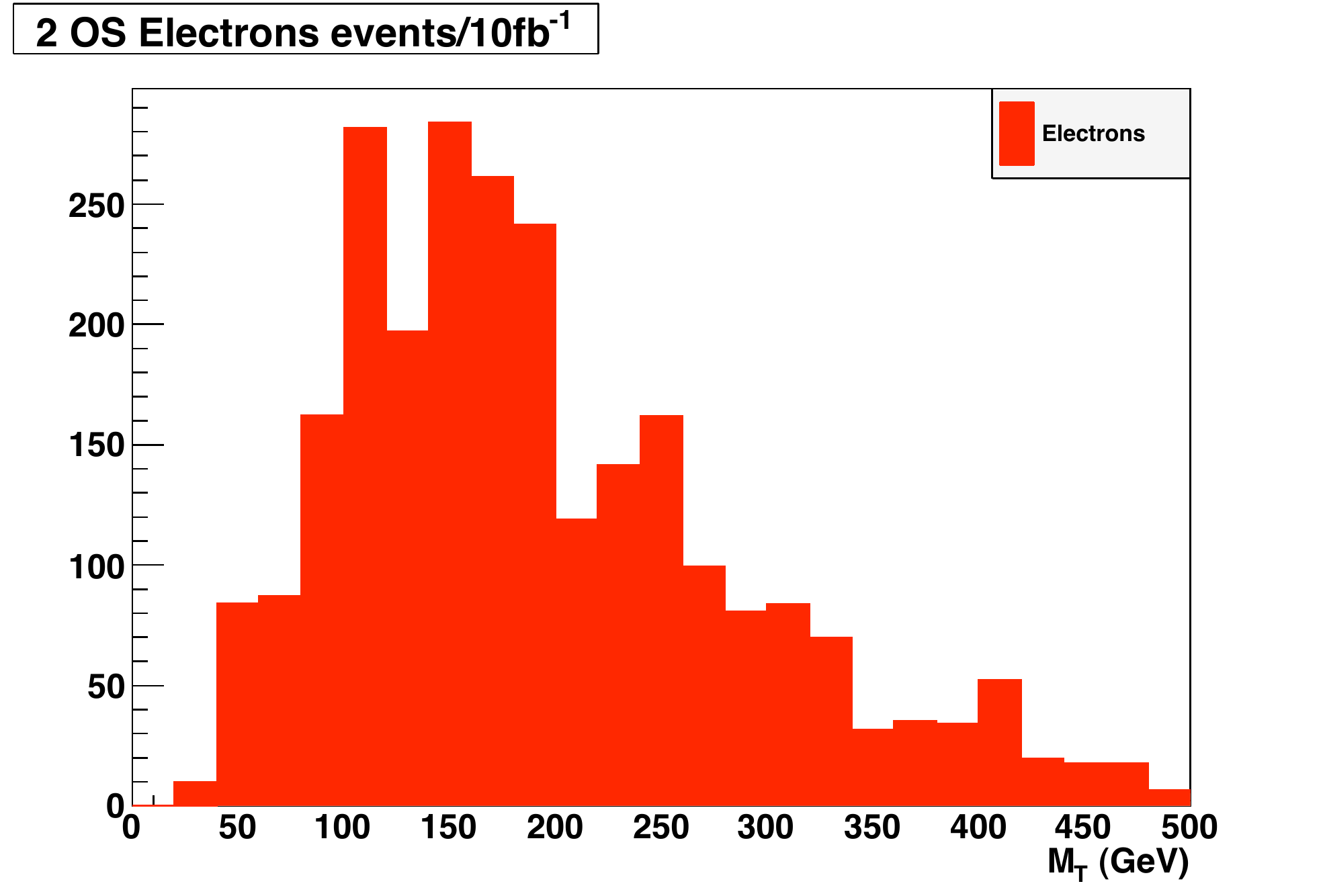}
\hfill
\includegraphics[width=0.49\textwidth]{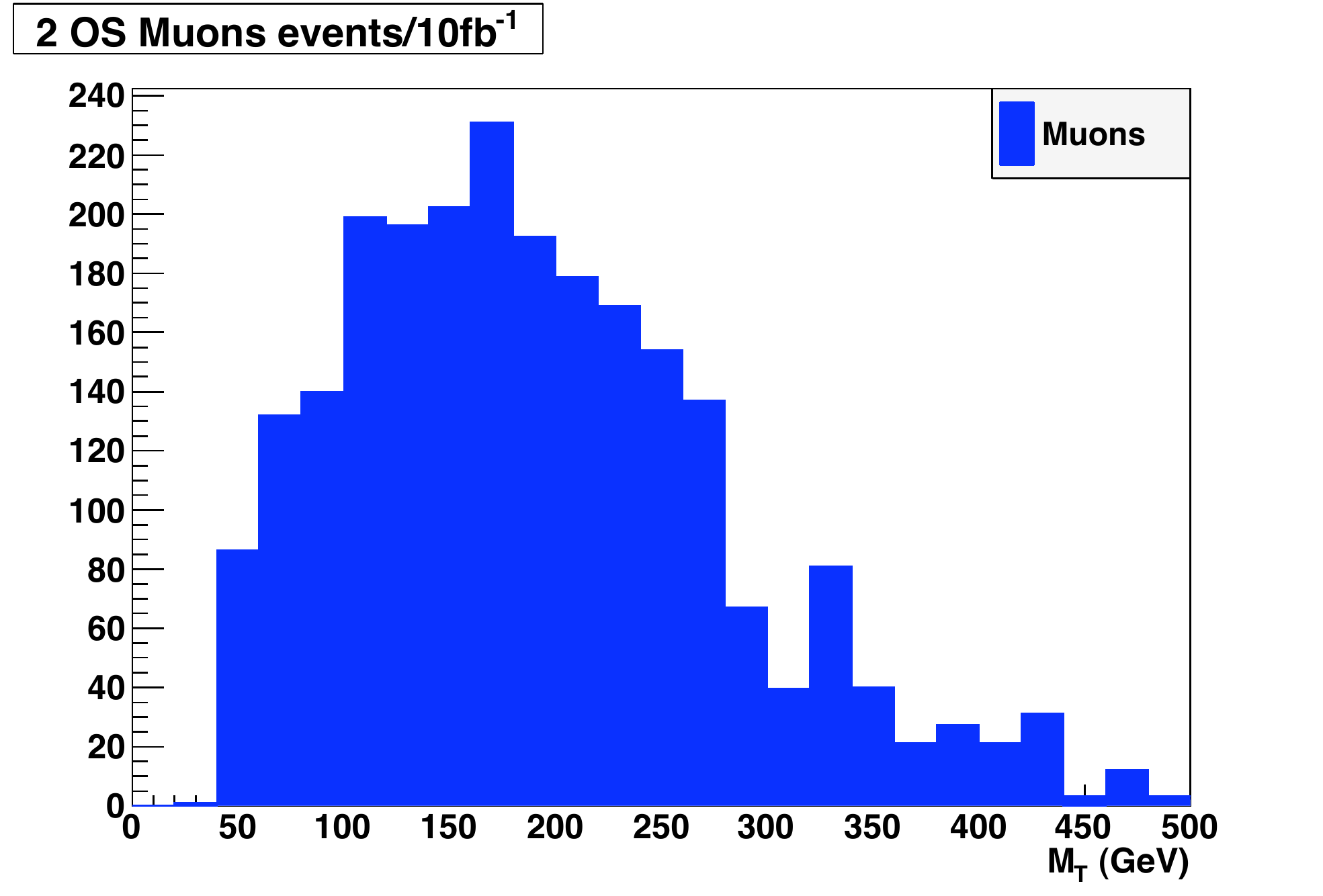}
\caption{Transverse mass distribution for $e^+e^-$ pairs (left) and $\mu ^+\mu ^-$ pairs (right) for the SUSY $2\rightarrow 2$ scenario. \label{massd_MT_Distributions}}
\end{figure}

The general assumption for this variable to work properly is to have only one source of missing energy. The presence of more than one source in an event (both real particles and detector leaks) generally spoils the results. In fact, the fraction of events for which the missing energy is effectively coming from just one source, matching the definition of the variable, is small. Therefore, instead of a well defined peak with a sharp edge on a flat distribution, we can expect a smooth distribution peaking at the correct mass value. This is indeed what we see. Figure \ref{massd_MT_Distributions} shows the transverse mass distributions for two opposite sign leptons: $e^+e^-$ (left) and $\mu ^+\mu ^-$ (right) for the SUSY $2\rightarrow 2$ scenario. The distributions for the SUSY $2\rightarrow 3$ scenario are very similar.

As suggested previously, looking at figure \ref{massd_MT_Distributions} we see a continuous distribution peaking at the correct $\widetilde{\chi} ^0_2$ mass value ($m_{\chi ^0_2}=181.1$ GeV). Since the peak is not very prominent, more detailed analysis of the background is required for a quantitative statement. Also, the shape of the distribution is not really characteristic: similar studies in the literature \cite{Basso:2008iv} showed that typical SM backgrounds can lead to the same shape. 


Notice that, given the low statistics, fluctuations may be misunderstood as peaks in the distribution. Both electron and muon distributions show possible secondary peaks at $250$ GeV and $330$ GeV respectively, none of which corresponding to actual particles in the spectrum giving rise to pairs of (opposite sign) charged leptons.

The conclusion from this study is that the application of the transverse mass to processes with more than one missing energy source may yield some information, but ultimately, the (simple) transverse mass is not a suitable variable for SUSY or UED events, since more than one particle is escaping the detection and the definition of eq. (\ref{massd_MTdef}) is not matched.

When applied to the proper events instead, this variable is very powerful in addressing quantitatively the intermediate particle's mass, as shown in \cite{Basso:2008iv} for the $B-L$ model \footnote{Only at the parton-level. The  detector level analysis is still on going.}.

%% file: Robens/MT2_Results.tex
\subsection{\Mtt Stransverse Mass and \Mtt Kink}

\begin{figure}[htbp]
\begin{center}
\includegraphics[width=0.49\textwidth]{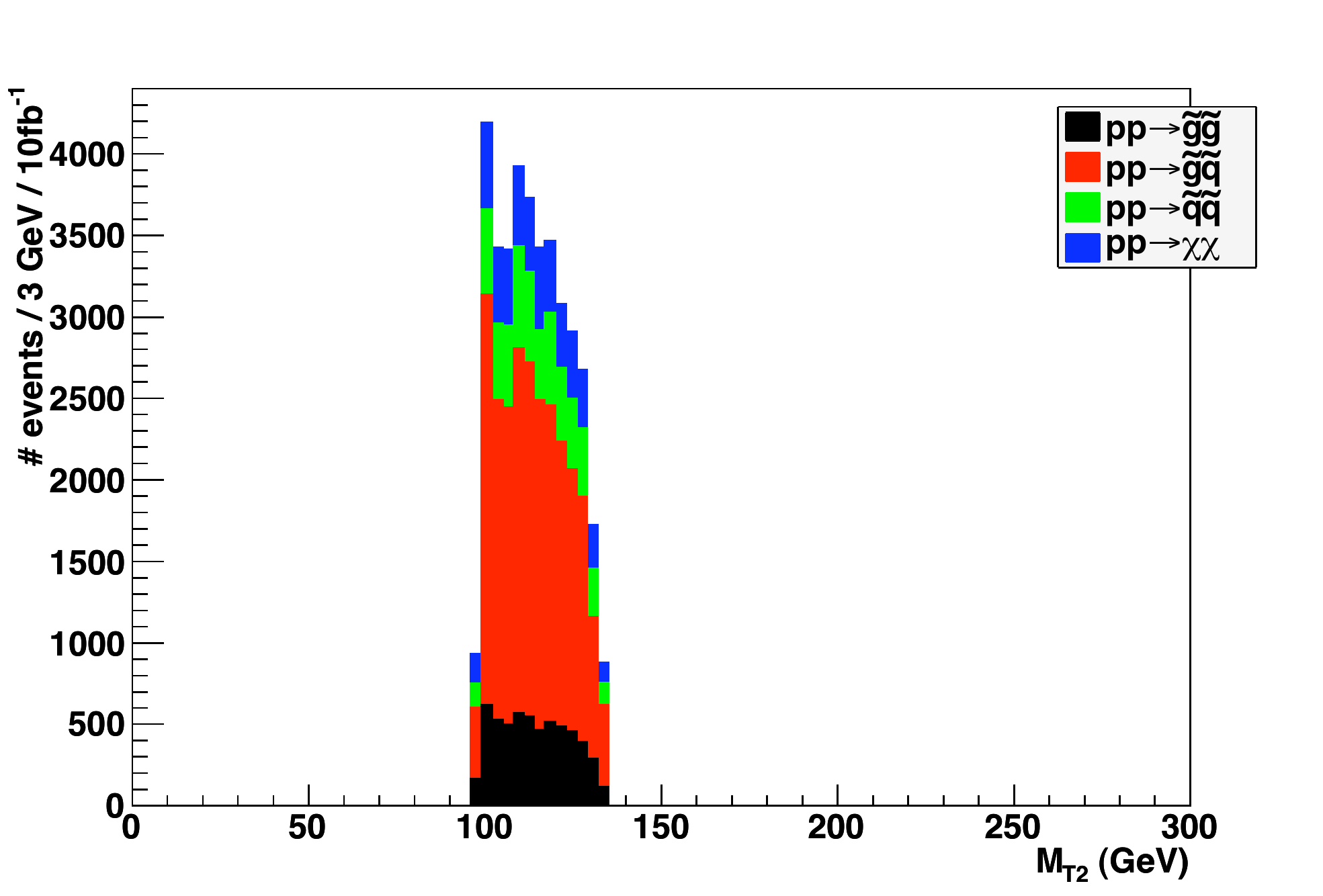}
\includegraphics[width=0.49\textwidth]{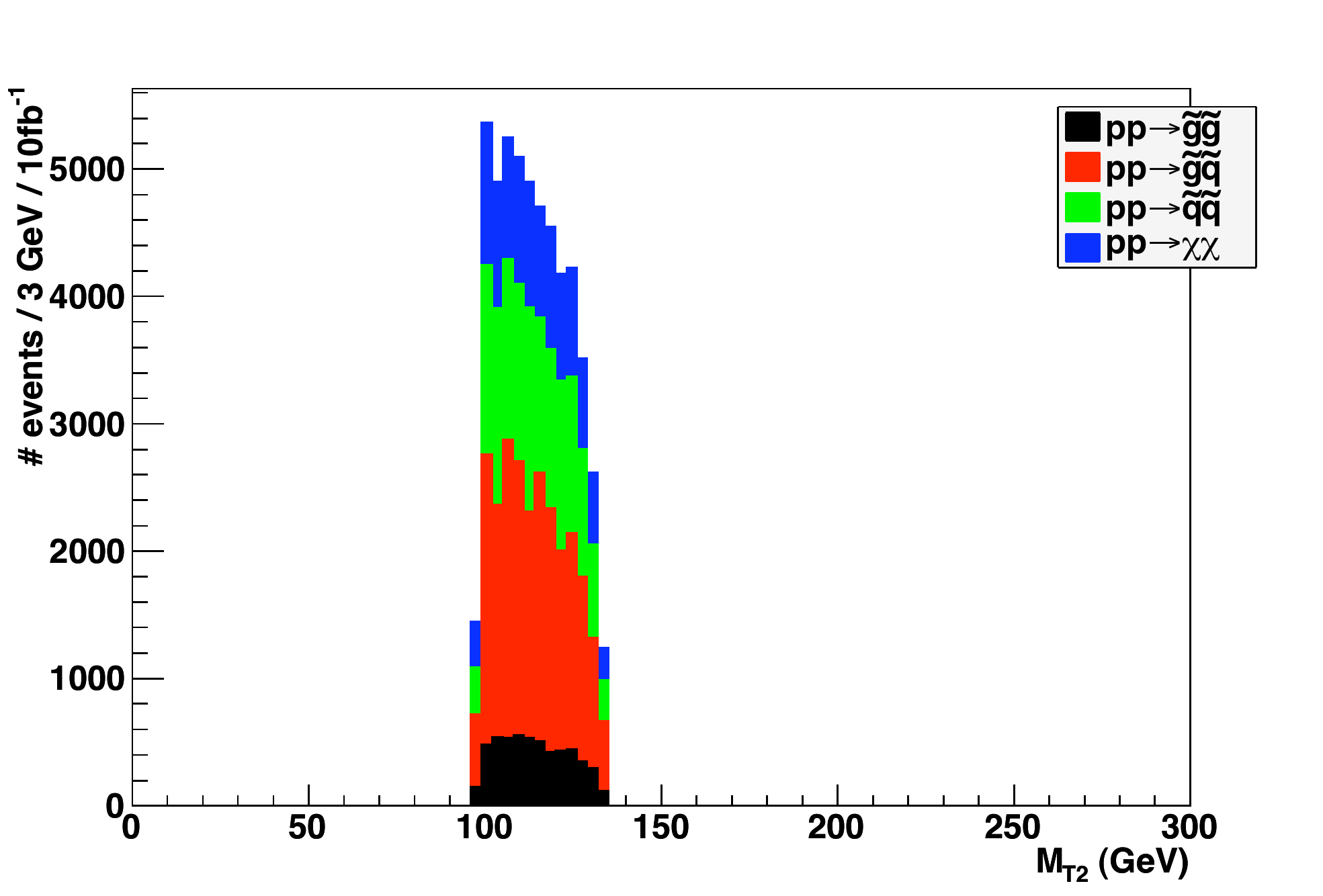}
\end{center}
\caption{\Mtt distribution for same sign (left) and opposite sign (right) SUSY
  $\tilde{\tau}_1$ pair production on parton level.}
\label{massd_fig:gen_stau1}
\end{figure}

\begin{figure}[htbp]
\begin{center}
\includegraphics[width=0.49\textwidth]{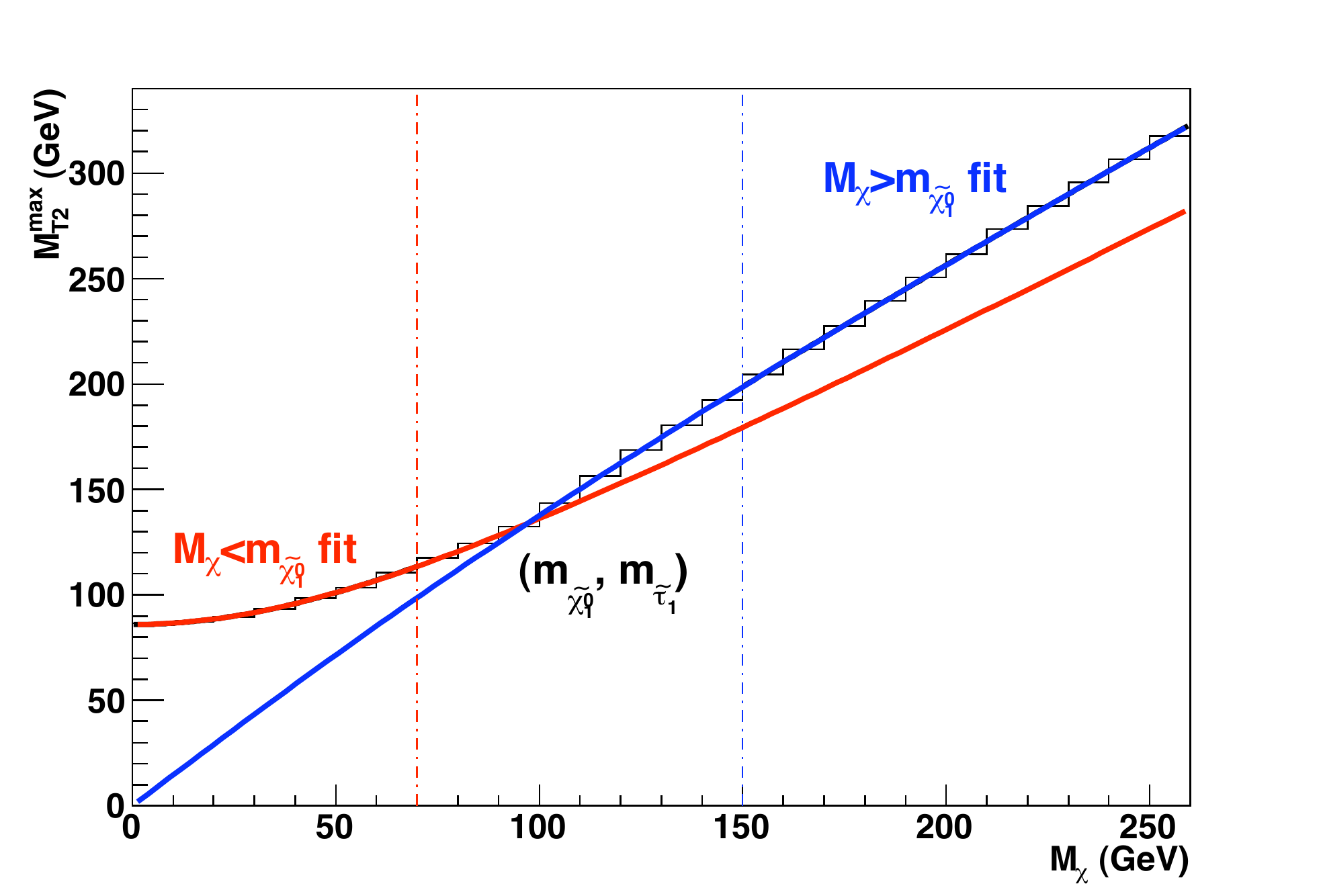}
\includegraphics[width=0.49\textwidth]{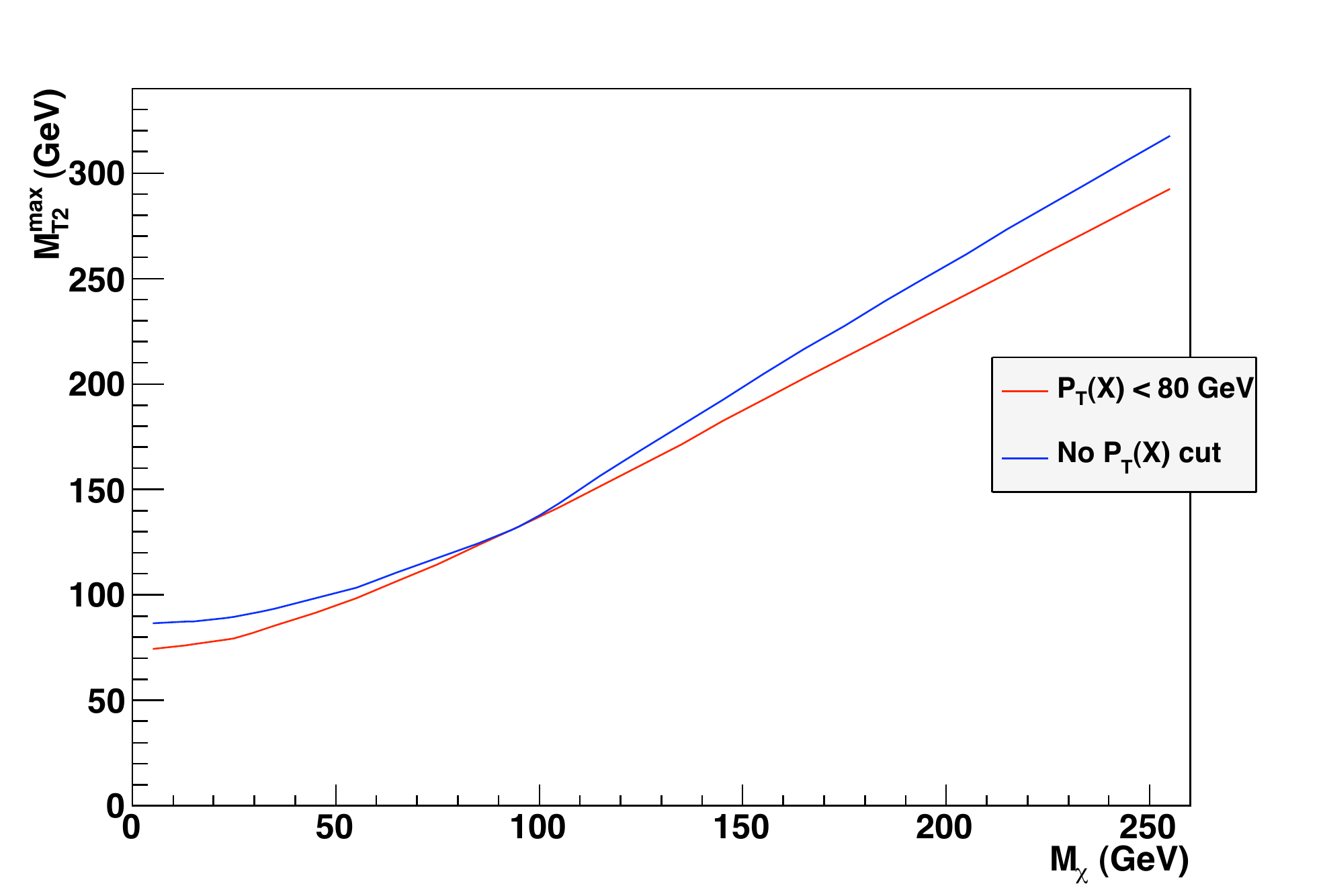}
\end{center}
\caption{$M_{T2}^{\text{max}}$ distribution as function of the trial LSP mass,
  $M_{\chi}$, for same sign SUSY $\tilde{\tau}_1$ pair production in the
  $\tilde{g}\tilde{q}$ production channel, with the fit revealing the kink at
  ($m_{\tilde{\chi}_1^0}, m_{\tilde{\tau}_1}$) as described in the text
  (left) and with different $p_T(X)$ cuts (right).}
\label{massd_fig:Mt2Kink}
\end{figure}

At the SPS1a point used here, the most abundantly produced slepton is the
$\tilde{\tau}_1$. Fig.~\ref{massd_fig:gen_stau1} shows the \Mtt distribution
obtained for the SUSY $2\rightarrow 2$ sample for
both same sign and opposite sign $\tilde{\tau}_1$ pair production
$pp\rightarrow X+ \tilde{\tau}_1\tilde{\tau}_1\rightarrow X + \tau\tilde{\chi}^0_1
\tau\tilde{\chi}^0_1$ using parton level information and using the exact
mass of the invisible LSP
$\tilde{\chi}^0_1$,  96.7~GeV (see Table~\ref{tab:sps1amasses}).
For the computation of \Mtt the Oxbridge \Mtt / Stransverse Mass
Library~\cite{oxbridge} was used. As by construction
$m_{\tilde{\tau}_1}\geq\Mtt$, the endpoints
of these distributions are expected to be a good estimate of the $\tilde{\tau}_1$
mass, 134.5~GeV, which is clearly the case here. For this particular channel,
where each
of the $\tilde{\tau}_1$ decays to 1 visible and 1 invisible
daughter, the
endpoint of the \Mtt distribution, $M_{T2}^{\text{max}}(M_{\chi},p_T)$, considered
as function of the trial LSP
mass, $M_{\chi}$, 
is expected to exhibit a kink at $M_{\chi}=m_{\tilde{\chi}_1^0}$,
only when the $\tau\tilde{\chi}^0_1\tau\tilde{\chi}^0_1$ system is recoiling
with significant $p_T$ against $X$~\cite{ChoChoi:2008, BurnsKong:2009}, as
will be demonstrated below.  
As an example, Fig.~\ref{massd_fig:Mt2Kink} (left) shows the
$M_{T2}^{\text{max}}$ 
distribution as function of the trial LSP mass for the same sign
$\tilde{\tau}_1$ pair 
production events in the $\tilde{g}\tilde{q}$ production channel.   
The distribution was fitted in the low ($0<M_{\chi}<60$~GeV) and high
($140<M_{\chi}<260$~GeV) $M_{\chi}$ range with
the functional forms $M_{T2}^{\text{max}}(M_{\chi},p_T)$ 
taken from~\cite{BurnsKong:2009}, which describe the region
below $(M_{\chi}<m_{\tilde{\chi}^0_1})$ and above
$(M_{\chi}>m_{\tilde{\chi}^0_1})$ the kink\footnote{See Eqn.~(4.10)-(4.13) of
that reference.}. The latter functions implicitly also depend on $m_{\tilde{\chi}_1^0}$ and
$m_{\tilde{\tau}_1}$. Instead
of doing a 2-dimensional fit we performed a 1-dimensional fit of the
$M_{\chi}$ dependence only, leaving the
average  $p_T$ of the sample as a free parameter in the fit together with the two
unknown particle masses, $m_{\tilde{\chi}_1^0}$ and
$m_{\tilde{\tau}_1}$. Here,
the average $p_T(X)$ of the event sample was determined to be 234~GeV.
As can be seen, both fits describe
the $M_{T2}^{\text{max}}$ perfectly in their respective regions and the two
curves cross each
other at the 
expected kink position at $(m_{\tilde{\chi}_1^0}, m_{\tilde{\tau}_1})$. The
fitted mass values were $m_{\tilde{\chi}_1^0}=96\pm4 \, (97\pm 2)$~GeV and
$m_{\tilde{\tau}_1}=133\pm 4 \, (133\pm 3)$~GeV, whereas
$\left<p_T\right>=205\pm58 \, (370\pm 28)$~GeV for the low (high) $M_{\chi}$
fit respectively, which is in very good agreement with the actual mass 
and $\left<p_T\right>$ values.
The observed kink is expected to disappear for $p_T(X)$ going to zero. This
effect can indeed be seen in Fig.~\ref{massd_fig:Mt2Kink} (right) where the $\Mtt$ endpoint
distribution is plotted for events with $p_T(X)<80$~GeV and compared to the
case without any $p_T(X)$ cut. The position of the kink is determined by the
two particle masses only and therefore independent of
$p_T(X)$, but the kink itself becomes less pronounced for small $p_T(X)$ values.
For $p_T(X)\rightarrow 0$ the entire 
$\Mtt$ endpoint distribution
can be described by one single function of the two unknown 
particle masses~\cite{BurnsKong:2009}.  
From the above observations, we conclude that the $\Mtt$ kink method 
is at least in principle applicable to the $\tilde{\tau}_1$ pair production channel considered
here and may yield an accurate determination of both the LSP and
$\tilde{\tau}_1$ mass.  

\begin{figure}[htbp]
\begin{center}
\includegraphics[width=0.49\textwidth]{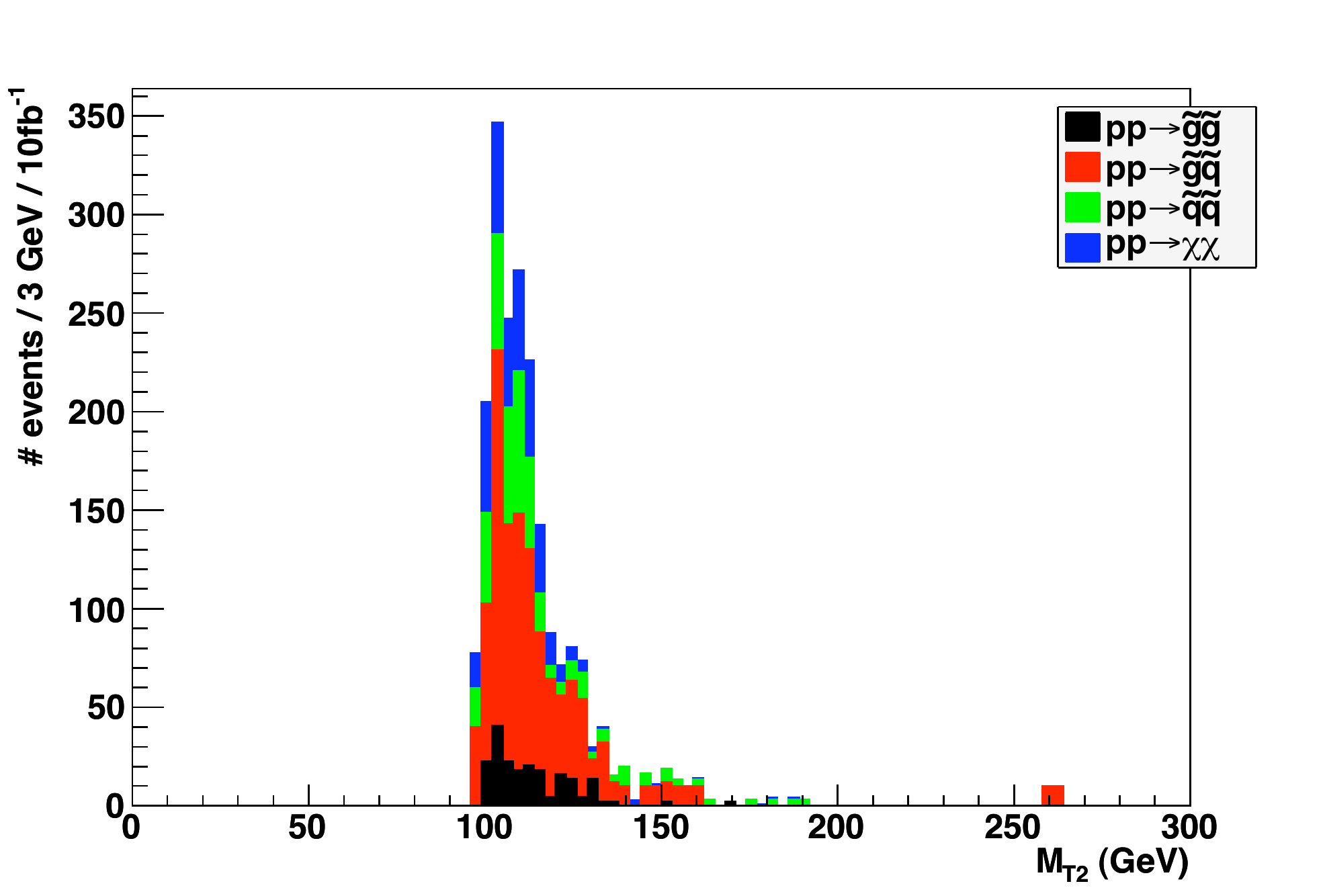}
\includegraphics[width=0.49\textwidth]{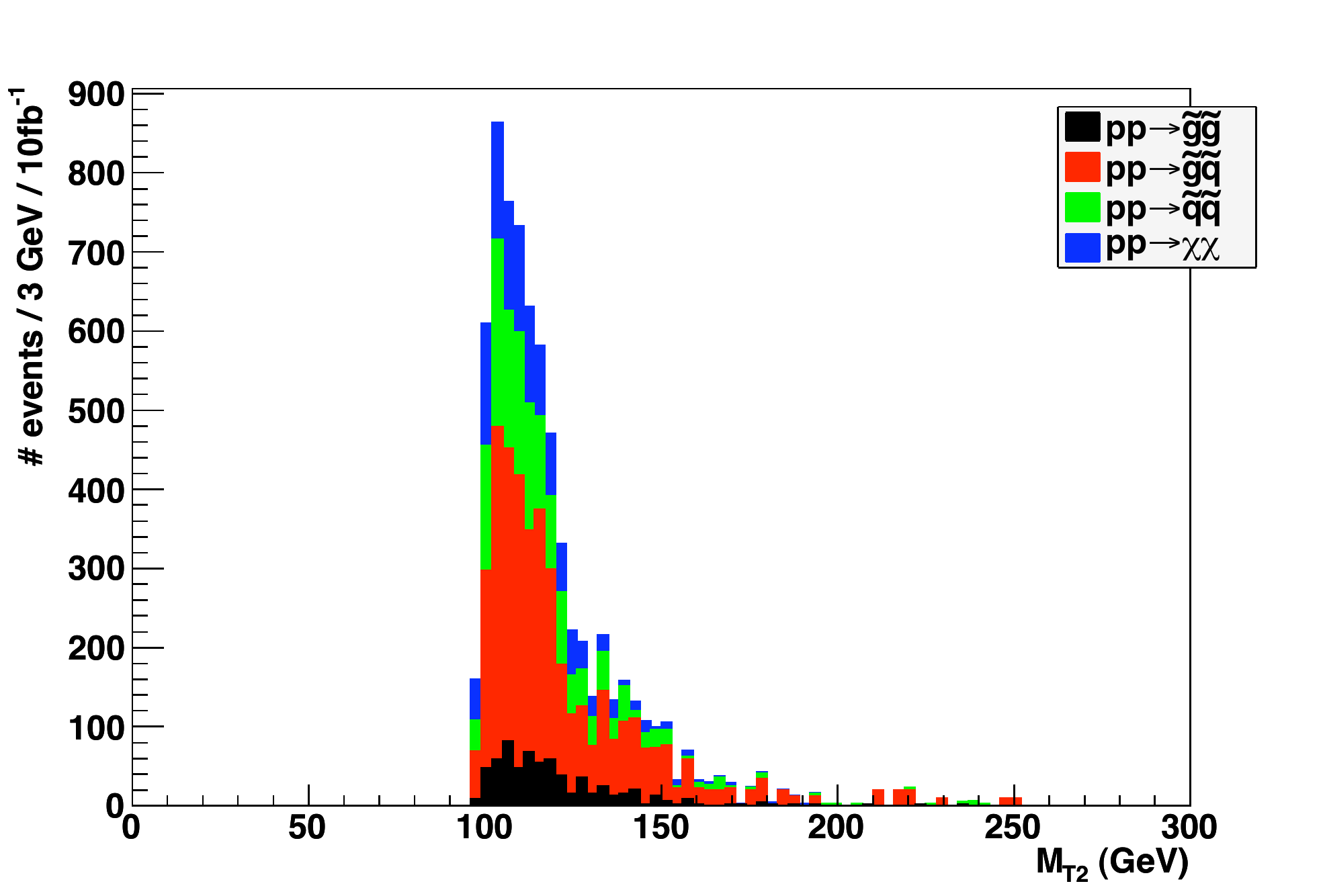}
\end{center}
\caption{\Mtt distribution for same sign (left) and opposite sign (right)
  $\tau$ pair production for the SUSY sample.}
\label{massd_fig:rec_stau1}
\end{figure}

Turning to an analysis at detector level, in Fig.~\ref{massd_fig:rec_stau1} the $\Mtt$ 
distributions are shown for $\tau$
pair production for the SUSY $2\rightarrow 2$ sample, where events with
exactly 2 
same sign or opposite sign $\tau$ jets
were selected and where the exact LSP mass was used for the $\Mtt$
computation. 
Due to experimental resolution, possible $\tau$ misidentification and due to
$\tau$ jets not
originating from $\tilde{\tau}_1$ decays, the
sharp edges of the
distributions are blurred compared to the corresponding parton level distributions in 
Fig.~\ref{massd_fig:gen_stau1}. The precise position of the endpoints will also
be affected by e.g. the tau jet calibration. Note also that a study of the
different backgrounds for this particular channel was not yet included
here. The extraction of the LSP and $\tilde{\tau}_1$ mass values from actual measured
data therefore requires further investigation and will definitely be more challenging.

%% file: Robens/Edge_Results.tex
\subsection{Edges}

\begin{center}
\begin{figure}
\begin{center}
  \includegraphics[width=0.45\textwidth]{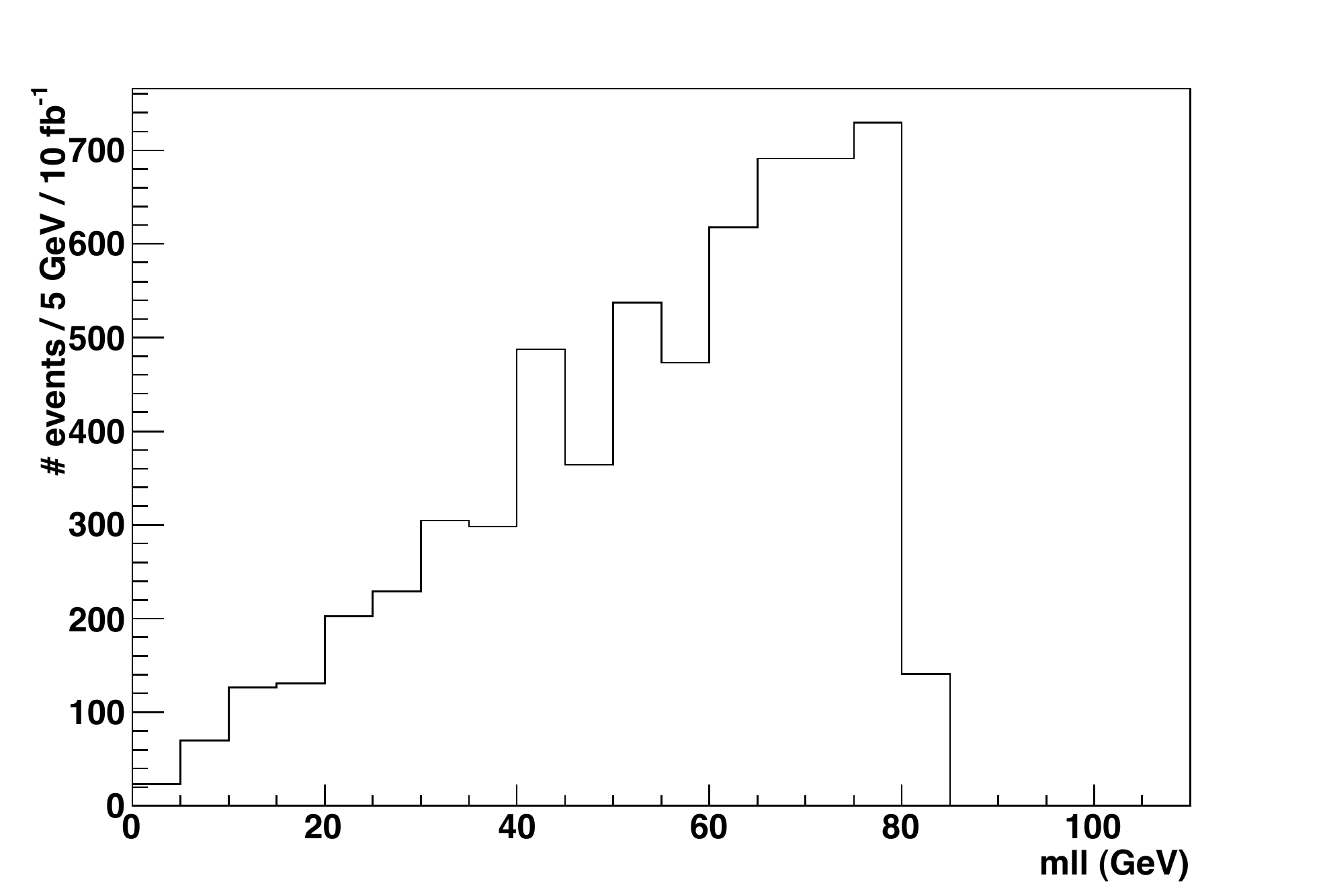}\hspace{5mm}
  \includegraphics[width=0.45\textwidth]{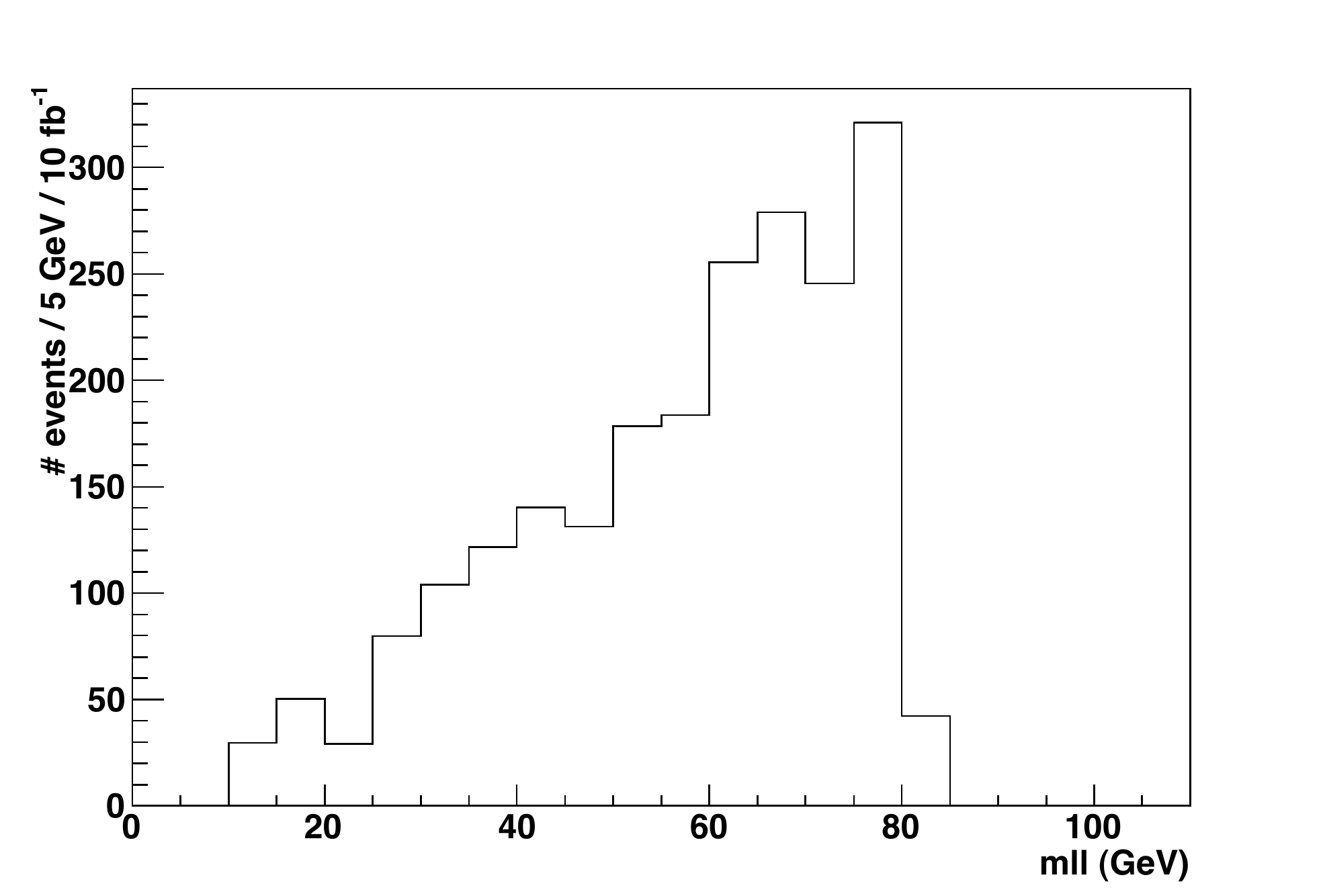}
  \caption{ \label{massd_Fig:Edge_hmll_gen_smur} The $m_{ll}$ distribution for events exhibiting a
    $\mu^+\mu^-$ signature as well as involving the $\tilde{\chi}_2^0
    \to \mu_L \tilde{\mu}_R$ decay, both on generator level (left) and
    detector level (right), for $2\,\rightarrow\,2$ sample. The left (right) plot corresponds to 6049 (2173) events. The expected edge at $81\,\GeV$ is clearly visible in both samples. }
\end{center}
\end{figure}
\end{center}

We have analysed the signal chain as given in (\ref{massd_eq:edge_signal}), for both $2\,\rightarrow\,3$ and $2\,\rightarrow\,2$ SUSY samples, considering $l=\mu$ only.  For the
parameter point considered here, the theoretical
values for the endpoints are given by
\begin{eqnarray*}
  m_{ll}^{\rm max} \,=\, 81\ {\rm GeV},&&
  m_{qll}^{\rm max} \,=\, 455\,/ 448\ {\rm GeV}\\         
  m_{ql{\rm (low)}}^{\rm max} \,=\, 320\,/315\ {\rm GeV},&&
  m_{ql{\rm (high)}}^{\rm max} \,=\, 398\,/392\ {\rm GeV}   
\end{eqnarray*}
for initial ($\tilde{d},\,\tilde{s}$) and  ($\tilde{u},\tilde{c}$) squarks respectively.
Our experimental signature is exactly one
$\mu^+\mu^-$ pair, both at generator and detector level. In addition to this "dimuon" signal, we also investigate the behaviour of the "pure" signal, which was selected by additionally requiring the existence of the
$\tilde{\chi}_2^0 \to \mu_L \tilde{\mu}_R$ decay on generator level. Note that this sample also includes events where the neutralino was not produced according to (\ref{massd_eq:edge_signal}); the majority of these additional events comes from direct $\wt{\chi}^{0} \wt{\chi}^{0}$ production and subsequent decays.  In our analysis, we applied standard Delphes cuts and detector level object definitions\footnote{We used the Delphes lepton isolation criteria with no track with $p_T\,>\,2\,\GeV$ in the $dR\,=\,0.5$ cone.} (cf appendix), as well as lepton isolation for all leptons considered. In addition, we cut out the $Z$ peak as well as all invariant masses below $10\,\GeV$ in $m_{ll}$ for all variables.
The overall pure (dimuon) signal cross sections on detector level, which take the above mentioned cuts as well as object definitions into account, are $0.22\,\pb \,(0.35\,\pb) $ for the $2\,\rightarrow\,2$ and $0.24\,\pb \,(0.40\,\pb) $  for the $2\,\rightarrow\,3$ sample\footnote{The relative contributions to the pure signal on detector level for the $\tilde{g}\tilde{g}$/ $\tilde{g}\tilde{q}$/ $\tilde{q}\tilde{q}$ , and $\wt{\chi}\wt{\chi}$ samples are $12\%,\,53\%,\,20\%,$ and $15\%$ for the $2\,\rightarrow\,3$ and $11\%,\,44\%,\,18\%$, and $27\%,$ for the $2\,\rightarrow\,2$ sample respectively.}. \\ 
Considering the pure signal only, the characteristic triangle-shaped distribution of the $m_{ll}$
variable \cite{Gjelsten:2004ki}  can easily be reproduced on generator level and persists on the detector level, cf. 
Fig.~\ref{massd_Fig:Edge_hmll_gen_smur}. The dimuon signal contains additional background which peaks at lower energies.
About two thirds of the background can be
attributed to stau pairpoduction, with subsequent leptonic tau decays.  Since the tau does not distinguish between first and second family leptons, this background can be nearly completely reduced by
subtracting the $m_{ll}$ distribution for events with the $e^-\mu^+$
and $e^+\mu^-$ signatures, respectively (see Fig.~\ref{massd_Fig:Edge_hmll_ana_diff}). After the subtraction, the expected triangular shape is recovered and the edge is clearly visible.\\

Unlike $m_{ll}$, the $m_{qll}, m_{ql{\rm (low)}}, m_{ql{\rm (high)}}$
variables involve identifying the correct quark jet. As an example, we here discuss $m_{qll}$, where similar results were obtained for the other variables. First, we consider the behaviour of the pure signal without additional background, where we now additionally require a squark parent for the $\wt{\chi}^{0}_{2}$, such that events stem from the decay chain (\ref{massd_eq:edge_signal}) only. As for $m_{ll}$, the distribution shape doesn't change much when moving from
generator to detector level, given the correct identification of the
jet\footnote{We here used a $\chi^{2}$ minimalization in order to identify the "proper" jet at detector level, in order to test detector effects on the pure signal distribution.}, cf. Fig.~\ref{massd_Fig:Edge_hmqll_gen_smur_proper}. In general, however, combinations with either one of the two hardest jets in the event have to be considered, and each variable will then inevitably include
misidentified jets. In Ref.~\cite{Gjelsten:2004ki}, a subtraction method similar to the opposite sign opposite flavour subtraction as described above was used. The background resulting from incorrectly identified jets is eliminated by
subtracting a mass distribution
with a random uncorrelated hard jet, for instance the hardest or
second hardest jet from a previous event candidate. However, for the low luminosity considered here, this subtraction method does not immediately result in the expected shape distributions, and further investigation is needed\footnote{In Ref.~\cite{Wienemann:2008jk}, a more dedicated study results in percent-range errors for distributions including jets, for a slighlty different point in SUSY parameter space.}.  
\begin{center}
\begin{figure}
\begin{center}
  \includegraphics[width=0.5\textwidth]{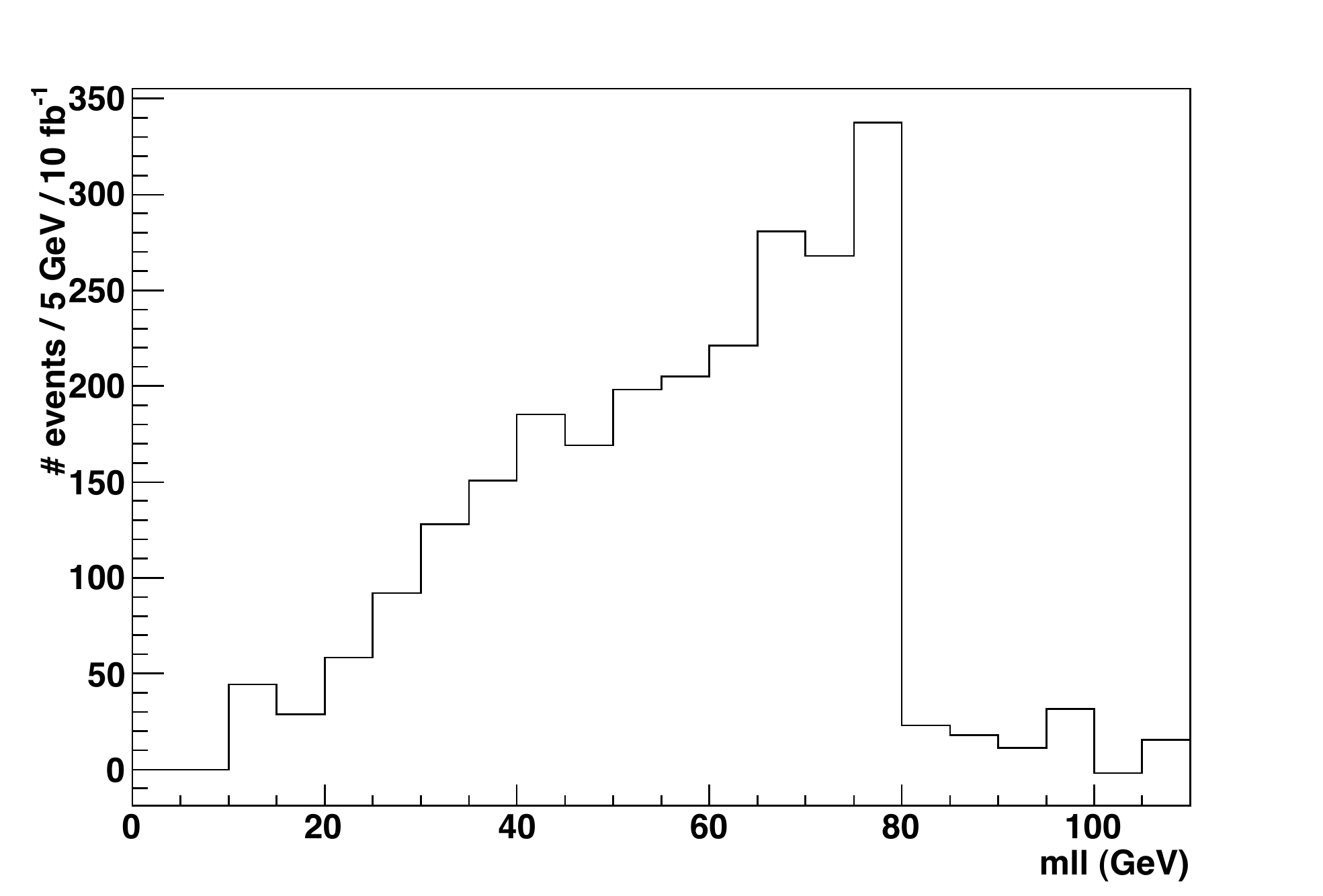}
  \caption{ \label{massd_Fig:Edge_hmll_ana_diff} The $m_{ll}$ distribution for all $\mu^+\mu^-$ events after
    subtracting the uncorrelated lepton background. The plot corresponds to 2435 events after subtraction. The expected edge at $81\,\GeV$ is clearly visible. }
\end{center}
\end{figure}
\begin{figure}
  \includegraphics[width=0.5\textwidth]{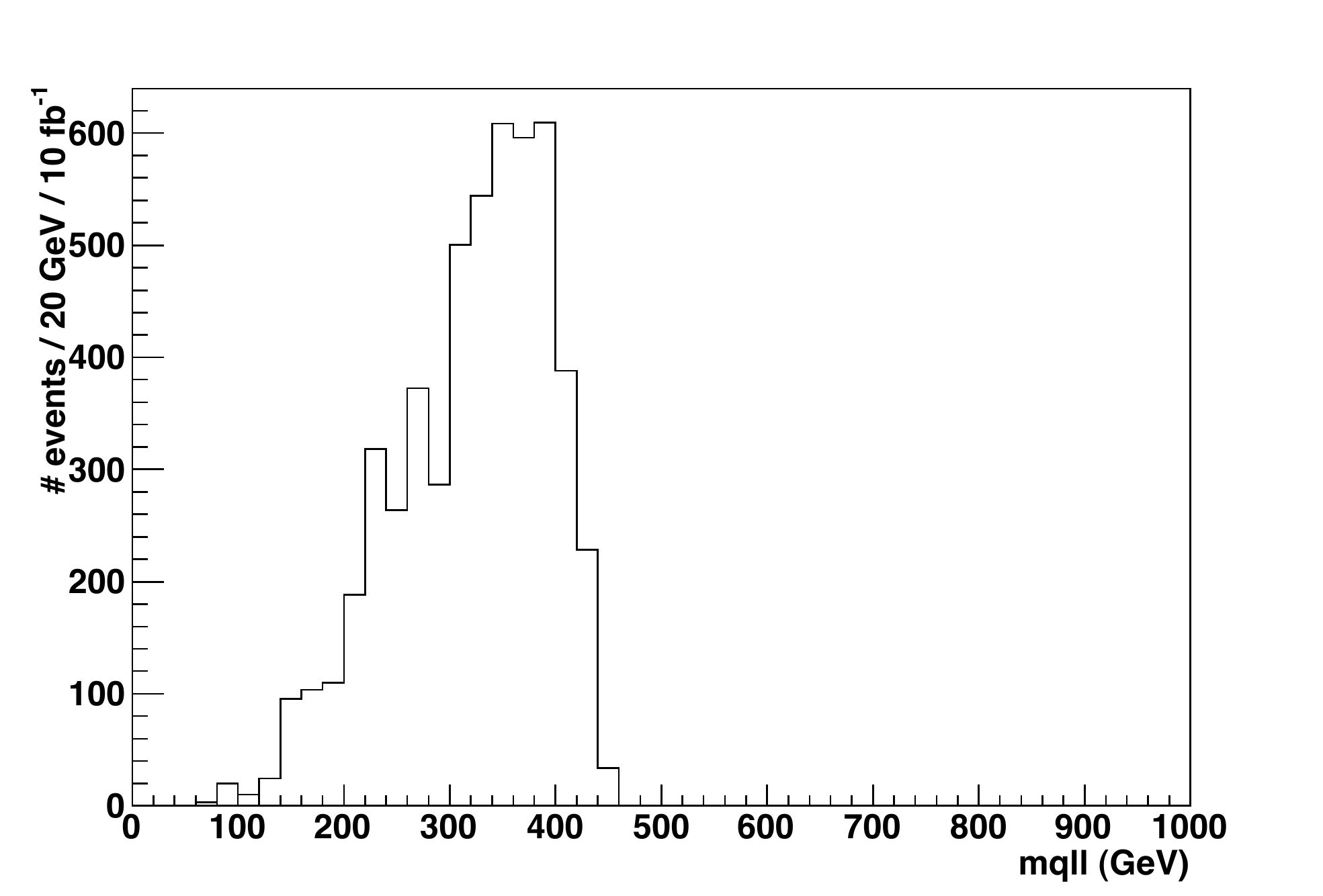}
  \includegraphics[width=0.5\textwidth]{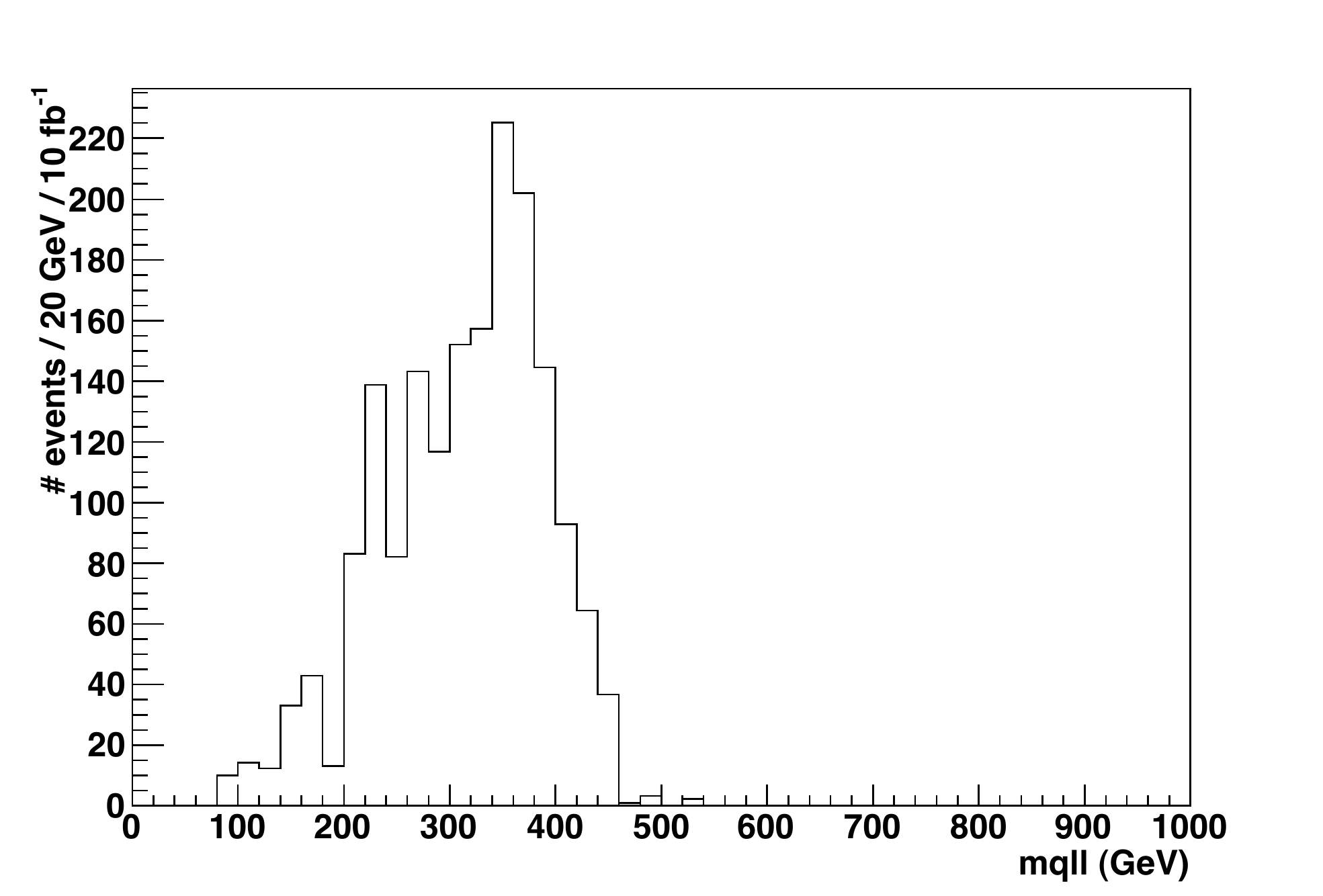}
  \caption{ \label{massd_Fig:Edge_hmqll_gen_smur_proper} The $m_{qll}$ distribution for $\mu^+\mu^-$ events
    involving the $\tilde{\chi}_2^0 \to \mu_L \tilde{\mu}_R$ decay,
    using the proper quark on generator level (left) and corresponding
    jet on detector level (right). This is from the $2\,\rightarrow\,2$ sample, corresponding to 5241 and 1803 events respectively.}
\end{figure}
\end{center}
Finally, we want to comment on the inclusion of additional backgrounds. Specifically, the preliminary results for the edge mass method presented here did not take SM background into account. In the high luminosity study \cite{Gjelsten:2004ki}, however, this background was well under control after applying similar suppression techniques as discussed for the SUSY induced background above. Summarizing, we can say that, given that the SM background is under control, the $m_{ll}$ edge, including all SUSY induced background, is clearly visible even at an early stage of data taking, and can be used to constrain the number of unknown masses by one. However, full knowledge of the relative mass spectrum includes edge measurements involving jets. These have proven to be more challenging, and further studies are needed in order to obtain the correct jet assignment for these variables on detector level.

%% file: Robens/Polynomial_Results.tex
\subsection{Polynomial Intersection}
Here we considered the topology of
Refs.\cite{Cheng:2008mg,Cheng:2009fw,Webber:2009vm}, shown in Fig.~\ref{massd_fig:topology}.
This occurs in SPS1a in large numbers with taus on the external legs, because
the stau is the NLSP.  This is generic across SUSY models in the "coannihilation
region", in which the correct relic density is achieved by the enhanced
annihilation cross section due to the near degeneracy of the stau and
neutralino.  To achieve precise results one can restrict to only events with
smuons or selectrons instead of staus, but the statistics are much lower.
Instead here we tried using the taus themselves.  We define a tau as either an
isolated muon, electron, or hadronic tau candidate as defined by Delphes.  There
is inherently missing energy in the tau decays, so we expect resonances to be
smeared compared to refs.\cite{Cheng:2008mg,Cheng:2009fw,Webber:2009vm}.
Additionally we require:
\begin{itemize}
    \item 2 or more jets with $p_T > 50$ GeV (only $p_T > 50$ GeV jets are considered)
    \item all possible combinations of jets and tau's are considered.
\end{itemize} 

To solve the system of equations presented previously, one must choose two
events.  Refs.\cite{Cheng:2008mg,Cheng:2009fw} computed all $N(N-1)$ possible
pairs for $N$ events to avoid questions about subset size, which is very CPU
intensive, but in principle one can Monte Carlo over pair choice (with
replacement) and as the number of pairs approaches infinity this is
mathematically equivalent to taking all possible pairs.  In practice the error
on mass determination is fundamentally set by the number of events, therefore
one should not need very many more solutions than the number of events before
the errors from pair choice are sub-dominant.  Therefore instead of plotting the
solutions from all $N(N-1)$ pairs of events, we Monte Carlo'ed over pair choice,
plotting all solutions from each pair, until the number of entries in the
histograms were 10 times the number of events.  Future work should quantify the
errors on mass determination as a function of the number of pairs chosen.

There are several possible particles which can appear at each point in the
chain, generally with similar masses, so that no double-peak structures are
seen.  We have used the entire SPS1a $2\to 3$ dataset, so the heaviest particle
$M_Z$ is always a squark (possibly with an upstream gluon) with masses from
$513-568$ GeV.  The second heaviest $M_Y$ is the  $\chi_2^0$ at 181 GeV.  The
third heaviest $M_X$ is dominantly $\tilde \tau_1$ at 135 GeV, and the lightest
$M_N$ is the $\tilde \chi_1^0$ at 97 GeV.

\begin{figure}[h]
\begin{center}
 \includegraphics[width=0.6\textwidth]{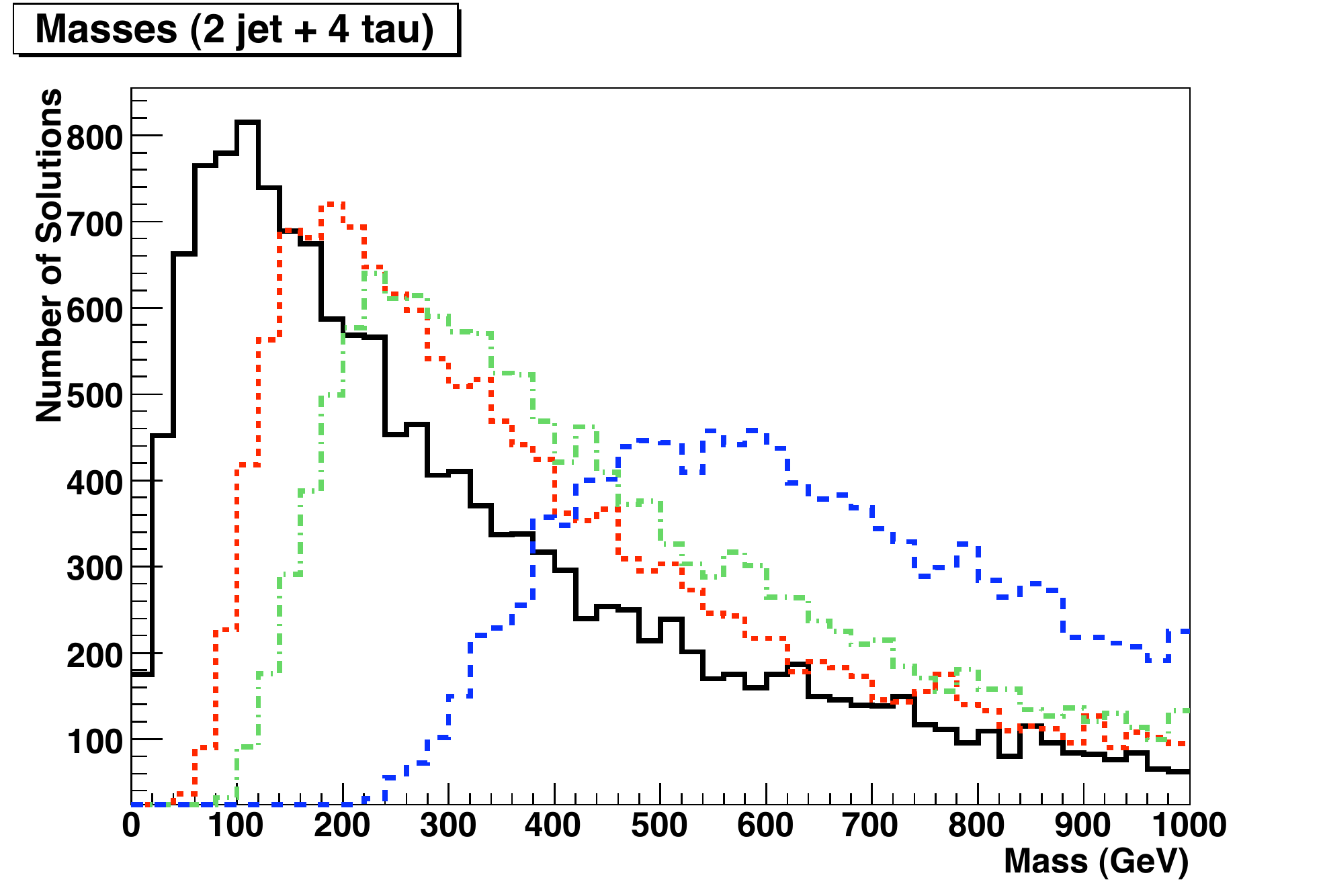}
\vspace*{-.1in}
\caption{\label{massd_fig:polynomial_results} $M_N$ (black solid), $M_X$ (red dashed), $M_Y$ (green dot-dash), and $M_Z$ (blue dashed) polynomial solutions.}
\vspace*{-.3in}
\end{center}
\end{figure}

Our results are shown in Fig.\ref{massd_fig:polynomial_results}.  We find
unsurprisingly that the slepton and neutralino mass peaks are broader than in
Refs.\cite{Cheng:2008mg,Cheng:2009fw}, due to the extra missing energy from
neutrinos.


%% file: Robens/TR_conclude.tex
In this writeup, we reported on the first results of an ongoing comparative study of different mass determination methods. We used a common Monte Carlo data sample, which was generated for the MSSM mSugra point SPS1a, for a proton-proton collider with a c.m. energy of $14\,\TeV$ and an integrated luminosity of $10\,\fb^{-1}$. Our sample includes parton shower, hadronization, and detector simulation.  We investigated several mass determination variables. Most of these were specifically designed for a scenario with long decay chains and missing energy from one or more invisible final state particles. At this stage of the study, comparative statements cannot yet be made. Therefore, we only comment on the status of the analyses and point to directions which need to be taken in further investigation. 
\begin{itemize}
\item{\bf Effective mass} The effective mass variable is designed to determine the lowest or average BSM mass scale in the considered process. It only uses transverse information of the involved particles and does not rely on additional mass assumptions. In this study, we found that, assuming the background to be under control, the distribution peaks at the expected values. However, for a thorough investigation of any BSM model, a parameter scan needs to be done which establishes the relation between $M_{\rm{eff}}$ and $M_{\text{BSM}}$; therefore, the final interpretation of the result is highly model dependent. 
\item{\bf Square of shat-min} Similar to the effective mass, the
square of shat-min tries to determine an overall scale of the BSM
process by exploration of the threshold region of BSM particle
pair-production. In contrast to $M_{\rm{eff}}$, this variable directly
relies on an additional input of the LSP mass, which needs to be determined elsewhere. Furthermore, this variable is highly sensitive to initial state radiation. Cutting out ISR events with a rapidity cut leads to a high cut dependence of the result. We therefore conclude that, although in principle applicable, the effects of different rapidity cuts need to be further under control before this variable can be used to determine a mass scale for new physics.
\item{\bf Transverse mass} In contrast to the other variables considered in this report, the transverse mass is not applicable for scenarios where the missing energy stems from more than one particle; in a way, our results can be seen as a test of an a priori false assumption. As expected, we do not obtain a distinct peak in the $M_{T}$ distributions, but rather a broad spectrum which however peaks at the expected value. This is caused by a small number of events which have effectively one source of missing energy. The distribution can furthermore be polluted by additional background; therefore, the use of the transverse mass is quite limited in the scenario considered here. 
\item{\bf \Mtt Stransverse Mass and \Mtt kink}
The stransverse mass, \Mtt, has the advantage over the transverse
mass that the missing energy in the process can come from more than one
particle. At the parton level, the endpoint of the \Mtt distribution of the
event sample considered here could effectively be used to estimate the mass
of the decaying particle. However, the mass of the invisible particle is
required as input and needs to be determined elsewhere. The latter problem
could be overcome by the \Mtt Kink method in which the \Mtt endpoint, $\Mtt^{\rm{max}}$,
is considered as function of a trial mass corresponding to the invisible
particle. A kink effectively appears at a position which depends only on
the mass of the LSP and the considered decaying particle. The strength of
the kink was seen to depend on the total $P_T$ of the decaying particle
pair. Both the mass of the decaying particle and the LSP could be
extracted quite well from a fit to the $\Mtt^{\rm{max}}$ trial mass dependence.
However, further investigation at detector level, including the
consideration of the SM and BSM background as well as reconstruction
inefficiencies, is needed before a definite statement can be made about
the use of this method in our present study.
\item{\bf Edges} Using the information of edges of invariant mass distributions is one of the more classical methods for BSM mass determination. It is in principle applicable to any event topology which involves on-shell decays of (B)SM particles. In our study, we found that edge measurements which only rely on the leptonic information of the event can easily be determined, especially after a simple background reduction. However, edge measurements involving jets are much more challenging, and for the low luminosity considered here, we did not manage to efficiently subtract the background stemming from wrong combinatorics. This point needs further investigation before any statement about jet-related quantities can be made. From the measurement of the dilepton mass only, the number of unknown masses can be reduced by one. Note however that, depending on relative mass hierarchies within the decay chains,  different inversion relations hold for extraction of the correct mass assignments. This can in principle lead to further misinterpretations, even when the complete edge information is available.
\item{\bf Polynomial Intersection} The polynomial intersection method uses exact solutions for the kinematic configurations of long decay chains with intermediate on-shell particles, and in general can only be applied to specific topologies, as it relies on the overall number of unknowns and constraints in the considered system. In this study, we investigated a topology with eight external legs, assuming symmetric decay chains. This allows for an exact solution of the polynomial equation system if any two events of the same topology are combined. Pair assignment in $N$ events as well as the related error determination proved to pose the biggest challenge in our study. Instead of combining all possible pair choices, we used a random Monte Carlo pair assignment. The resulting distributions for the masses peak at the expected values, where peaks are broad mainly due to extra energy losses in the tau decays. A big advantage of this method is that all intermediate masses can be determined and fitted simulaneously. Future investigation concerns error estimation as well as the inclusion of all backgrounds.
\end{itemize}
We consider this report as a starting point for a more thorough investigation. More detailed studies adressing the issues mentioned above, as well as the inclusion of all background, are needed before we can compare different variables in a quantitative way. However, our results already point to advantages and drawbacks of the variables considered here, and further investigation and eventual synergies of different determination methods will hopefully lead to promising results in the near future.

%% file: Robens/susy_spec.tex
\section*{Appendices}
\setcounter{section}{0}
\renewcommand{\thesection}{App.~\Alph{section}}

\section{SPS1a spectrum}
\label{massd_app_spectrum}
The SPS1a spectrum use here was generated using SOFTSUSY \cite{Allanach:2001kg} version 2.0.5, with $m_{t}\,=\,175\,\GeV$. We give the mass values for particles relevant in this study in Table \ref{tab:sps1amasses}, and total cross sections for the $2\,\rightarrow\,3$ and $2\,\rightarrow\,2$ samples in Table \ref{tab:sps1axsections}.
\begin{table}
\begin{\eqn*}
\begin{array}{|cc|cc|cc|cc|cc|cc|cc|cc|}\hline
\tilde{d}_{L}&568.4&\tilde{d}_{R}&545.2&\tilde{u}_{L}&561.1&\tilde{u}_{R}&549.3&
\tilde{b}_{1}&513.1&{\tilde{b}_{2}}&543.7&
{\tilde{t}_{1}}&399.7&{\tilde{t}_{2}}&585.8\\ 
\tilde{l}_{L}&202.9&\tilde{l}_{R}&144.1&\tilde{\tau}_{1}&134.5&\tilde{\tau}_{2}&
206.9&\tilde{\nu}_{l}&185.3&\tilde{\nu}_{\tau}&184.7&\tilde{g}&607.7&& \\
\wt{\chi}^{-}_{1}&181.7&\wt{\chi}^{-}_{2}&380.0&\wt{\chi}^{0}_{1}&96.7&\wt{\chi}^{0}_{2}&181.1&|\wt{\chi}^{0}_{3}|&363.8&\wt{\chi}^{0}_{4}&381.7&&&&\\ \hline
\end{array}
\end{\eqn*}
\caption{Relevant masses for SPS1a in \GeV. $u\,=\,(u,c),\,d\,=\,(d,s),\,l\,=\,(e,\mu)$.}
\label{tab:sps1amasses}
\end{table}
\begin{table}
\begin{\eqn*}
\begin{array}{l|c|c}
X_{1}X_{2}&2\,\rightarrow\,2&2\,\rightarrow\,3\\ \hline
\tilde{q}\tilde{q}\,(j)&6.56&7.83\\
\tilde{q}\tilde{g}\,(j)&19.52&21.75\\
\tilde{g}\tilde{g}\,(j)&4.53&5.47\\
\wt{\chi}\wt{\chi}\,(j)&1.97 &4.89
\end{array}
\end{\eqn*}
\caption{Production cross sections in $\pb$ for $p\,p\,\rightarrow\,X_{1}\,X_{2}$, for a cm energy of $14\,\TeV$. CTEQ6L1 PDFs were used. $2\,\rightarrow\,3$ sample includes an explicitly generated hard jet, where hard is defined by $p_{T,\rm{jet}}\,>\,40\,\GeV$. }
\label{tab:sps1axsections}
\end{table} 
Branching ratios have been calculated using BRIDGE \cite{Meade:2007js} and are
available upon request\footnote{With slight numerical variations, we reproduce the decay tables in appendix D of \cite{Porod:2003um}.}.

%% file: Robens/delphes_def.tex
\section{Delphes precuts and object definitions}
\label{massd_sec:delphesdef}
In all detector level analyses, a minimal set of cuts was used, corresponding to the Delphes \cite{Ovyn:2009tx} pre set cuts. We also list the object definitions on detector level used in all analyses. Additional cuts might have been applied for different variables; cf the respective subsections for further details.\\
\vspace{2mm}\\
{\bf Delphes pre cuts}
\begin{itemize}
\item{}electron/ positron definition: $|\eta|\,<\,2.5$ in the tracker, $p_T\,>\,10\,\GeV$
\item{}muon definition: $|\eta|\,<\,2.4$ in the tracker, $p_T\,>\,10\,\GeV$
\item{}taujet definition\footnote{For a more detailed description of the reconstruction algorithm see \cite{Ovyn:2009tx}.}: $p_T\,>\,10\,\GeV$
\item{}jet definition: $p_T\,>\,20\,\GeV$; CDF jet cluster algorithm \cite{Abe:1991ui} was used, with $R\,=\,0.7$  
\item{}lepton isolation criteria (if applied): no track with
$p_T > 2\, \GeV$ in a cone with $dR=0.5$ around the considered lepton
\end{itemize}
\vspace{5mm}
{\bf Analysis object definitions}
\begin{itemize}
\item{} Missing transverse energy: requires $E_T^{\rm{miss}}\,>\,100\,\GeV$.
\item{} jet criteria: $p_{T,\rm{jet}}\,>\,50\,\GeV$, $|\eta|_{\rm{jet}}\,<\,3$
\item{} electron/ muon: isolated; no track with
$p_T > 6\, \GeV$ in a cone with $dR=0.5$ around the considered lepton
\item{} any signal involing $n$ leptons: exactly $n$ isolated leptons at detector level
\end{itemize}

%% file: Gripaios/gripaios.tex
\chapter{LHC mass measurement, algebraic singularities, and the transverse mass}
{\it B.~Gripaios}

\begin{abstract}
I consider the recently-proposed `algebraic singularity method' for measuring masses of invisible particles produced at the LHC in arbitrary decay topologies.
I apply the method to the simplest case of a single parent particle decaying to an invisible daughter particle and a visible daughter particle,
and show that it gives a local approximation to the usual transverse mass variable. In doing so, I identify some issues that may need to be taken into consideration in generalizing the algebraic singularity method to more complicated decay topologies. One is that, in order to measure masses unambiguously with this method, one may need to identify not only the presence, but also the nature, of singularities in experimental distributions.
\end{abstract}
\section{Introduction}
Invisible particles will be a fact of life at the LHC, {\em nolens volens}. Whether they take the form of neutrinos, or dark matter candidates, or even visible particles that escape into dead regions of the detector, invisible particles will be omnipresent. The problem with invisible particles is that they carry away kinematic information in events in which they are present, making the reconstruction of events, and hence particle mass measurements, a non-trivial exercise. In the presence of a concrete dynamical model, missing information is not necessarily a problem, in that one can simply marginalize the likelihood that comes from the matrix element. But if we profess to be ignorant of dynamics (which is certainly the case if we are looking for new physics), then we must address the question of what can be learnt from the residual kinematic information alone.

We have known for a long time that the situation is not hopeless. Indeed, in the canonical example of a $W$-boson undergoing a decay to a charged lepton and an invisible neutrino, the transverse mass variable was exploited long ago in UAs 1 and 2 to measure the mass of the $W$~\cite{Arnison:1983rp,Banner:1983jy}, and even today it provides the best individual measurement~\cite{Aaltonen:2007ps}. A more modern example is the top quark, pair produced at the Tevatron and undergoing a decay in the di-leptonic channel: $2t \rightarrow 2b 2l 2\nu$. Here there is an extra complication, in that each decay in the pair produces an invisible particle, and even more information is lost.\footnote{More precisely, the measured missing transverse momentum in an event constrains only the sum of the transverse momenta of the two neutrinos.} Nevertheless, the so-called $m_{T2}$ variable~\cite{Lester:1999tx,Barr:2003rg}, has recently successfully been used to measure the top mass in this channel~\cite{Cho:2008cu,cdfmt2}. At the LHC, we can expect (or at least hope) to encounter even more complicated scenarios. For example, a heavy Higgs boson may decay in a di-leptonic channel via two $W$s, resulting in a topology in which a single particle decays to two invisible neutrinos: $h\rightarrow 2W \rightarrow 2l2\nu$~\cite{Barr:2009mx}. Even worse (or better, depending on one's perspective), the LHC may produce an invisible dark-matter candidate, whose unknown (and, in contrast to the neutrino, non-negligible) mass further increases the number of unknowns. Dark-matter candidates may also be multiply produced, if there is a discrete symmetry that guarantees their stability. A final complication is that theories of physics beyond the Standard Model, such as supersymmetry, typically predict a plethora of new states clustered around the TeV scale. Given the presence of light SM states, these are likely to undergo cascade decays, resulting in sizable combinatoric ambiguities in observed final (and initial) states. 

In recent years, a large number of methods have been proposed for measuring the masses of particles produced in these topologies;  reviews may be found in~\cite{DeRoeck:2009id, Barr:2010zj} and elsewhere in these proceeedings (along with a complete set of references). Although most of these methods are, to a large extent, {\em ad hoc}, in that they focus on a particular decay topology, a 
number of results of a more general nature have been obtained along the way. Among these is the observation 
that in longer decay chains, the system of kinematic equations from one or more events may be sufficient to solve directly for the masses~\cite{Nojiri:2003tu}. Even for shorter decays chains, we now know that all masses can be measured, given sufficiently many events. Indeed, even in the decay topology with the fewest constraints, namely one (or more) parent particle(s) undergoing a two-body decay to a visible and an invisible particle, the kinematics allow both the masses of the parent and the invisible daughter to be measured~\cite{gripaios:2007is}. As a corollary, one has the result that all masses can be measured in any set of decays where each decay contains only one invisible daughter particle, no matter how many visible particles are involved. A third result is that for variables, such as the transverse mass, that enjoy boundedness properties, issues of combinatorics (which arise from assignments of particles to decays~\cite{Lester:2007fq} or of radiation to the initial or final state) can be solved by extremization with respect to assignments. 

Furthermore, we have also begun to arrive at a deeper understanding of kinematics itself. The breakthrough in this direction came from Cheng and Han~\cite{Cheng:2008hk}, who observed that the $m_{T2}$ variable mentioned above has an interpretation as `the' natural kinematic function for the topology of pair-produced particles undergoing identical two-body decays, in the following sense. Imagine writing down the kinematic constraints, corresponding to conservation of (energy-)momentum, and the mass-shell conditions for some assumed decay topology. Now, for a given event, in which some of the energy-momenta are measured and some are not, one may ask whether the measured momenta impose any constraint on the unknown masses that appear in the kinematic constraints. Apparently, the answer is negative, because the constraints are just a set of underconstrained polynomials in the unmeasured momenta and masses. In fact the answer is affirmative, essentially because the masses and energies are restricted to take values in $\mathbb{R}^+$, whereas the solutions of polynomial equations generally take values in $\mathbb{C}$. The upshot is that, each event divides the space of unknown mass parameters into an allowed region and a disallowed region. The boundary of the two regions is defined by the function $m_{T2}$. 

From this we learn not only that the {\em ad hoc} variable $m_{T2}$ is a natural kinematic object, but we also learn that $m_{T2}$ encodes all of the information about particle masses that is contained in an event of this topology. This means that, absent dynamic information or other assumptions, there seems to be little point in searching for an alternative variable to measure masses in this topology. 

An obvious follow-up question is: what is the function that defines the kinematic boundary for other decay topologies? For simple cases, this is easily answered: For a single parent, two-body decay, it is the transverse mass~\cite{Barr:2009jv}, whereas for asymmetric pair decays (where either parents or daughters (or both) are different) one is led to a generalized version of $m_{T2}$~\cite{Barr:2009jv}. Unfortunately, addressing this question on a case-by-case basis becomes increasingly difficult as the the decay topology, and the set of kinematic constraints, become increasingly complex.   

Very recently, I.-W. Kim has proposed~\cite{Kim:2009si}
a related method, which, although only approximate (in a sense to be defined below) allows an elegant, and more importantly general, algorithm for mass measurement to be defined, for any decay topology.

The starting point for the method is to note that the full phase space (defined by the various momenta, subject to the kinematic constraints) is smooth (modulo singularities arising from soft and collinear divergences of massless particles, which do not concern us here). But when some of the particles are invisible, we must
project out the kinematic variables that go unmeasured; the resulting {\em observable} phase space is a singular manifold. The singularities in observable phase space give rise to singularities in the distributions of functions on observable phase space, which in turn give rise to sharp features that can be easily identified in experimental data, notwithstanding the presence of smoothly-varying backgrounds or detector acceptances. These singularities generalize the well-known edges that appear in invariant- or transverse-mass distributions. 

With this observation in hand, one can define an algorithm for measuring masses, schematically described as follows.\footnote{For a fuller description, see~\cite{Kim:2009si}.} Firstly, assume some decay topology, and write down the corresponding kinematic constraints. Secondly, identify the locations of the singular points in observable phase space. Thirdly, construct a co-ordinate, called the singularity co-ordinate, in the vicinity of a given singular point, which: (i) vanishes at the singularity; (ii) corresponds to a direction normal to the singular phase space; and (iii) is normalized such that every event has the same significance. Fourthly, for each event, find the nearest singularity, and the value of the associated singularity co-ordinate for that event. Fifthly, plot the distribution over events of the singularity co-ordinate for all possible guesses for the unknown mass values. When the mass guesses are correct, the distribution will feature a singularity of the origin.

The reader may already have noted three potential thorns in the side of the algorithm. Firstly, the notion of `nearest singularity' needs an explicit definition if there is more than one. Secondly, one might fear that the local approximation made will be of limited use for a sample of experimental events that are spread roughly uniformly in phase space~\cite{Lester}. 
Thirdly, the method only guarantees the presence of a singularity at the origin of the singularity co-ordinate when the hypothesized masses are the correct ones. It does not guarantee the converse, namely the absence of a singularity at the origin when the hypothesized masses are incorrect.

In what follows, I hope to shed some light on these issues by applying the method, {\em verbatim}, to the simplest decay topology, namely a single parent particle undergoing a two-body decay to visible particle and an invisible particle. In doing so, we will see that the singularity co-ordinate is just a local approximation to the usual transverse mass variable. We will also see in these examples that the algorithm, as it stands, does not identify the correct masses uniquely; to do so, one needs to identify not just the presence of a singularity at the origin of the singularity co-ordinate, but also its nature. 

I start, in the next Section, by considering as a special case the subset of events in which the visible daughter particle is produced at rest. I treat the general case of moving visible daughters in the subsequent Section.   

\section{A special case}
Consider a single parent particle $Y$, of mass $m_Y$, undergoing a two-body decay to an invisible particle $X$, of mass $m_X$,  and a visible particle $V$, of mass $m_V$. Consider, for now, the restricted subset of events in which the momentum of the visible system $V$ vanishes. The four-momenta of $V$, $X$ and $Y$ may, therefore, be written as
\begin{align}
V_\mu &= (m_V , \mathbf{0},0 ), \\
X_\mu &= ( p_0, \mathbf{p}, p_3), \\
Y_\mu &= ( p_0 + m_V,\mathbf{p},p_3 ),
\end{align}
and the three mass-shell constraints may be written as $V_\mu V^\mu = m_V^2$, $X_\mu X^\mu = m_X^2$, and $Y_\mu Y^\mu = m_Y^2$. To make the analysis as straightforward as possible, I solve the constraint $X_\mu X^\mu = m_X^2$ for $p_0$, such that the remaining constraint is
\begin{gather} \label{gripaios_sphere}
g \equiv \mathbf{p}^2 + p_3^2 - M^2 = 0,
\end{gather}
where I set
\begin{gather} 
\Bigg(\frac{m_Y^2 - m_X^2 - m_V^2}{2m_V}\Bigg)^2 - m_X^2 \equiv M^2.
\end{gather}

It is now very simple to apply the method of~\cite{Kim:2009si}. The full phase space is defined by three momenta, namely $\mathbf{p}$ and $p_3$, subject to the constraint~(\ref{gripaios_sphere}). Geometrically, phase space is a two-sphere embedded in $\mathbb{R}^3$. According to~\cite{Kim:2009si}, we should now split the momentum variables into those momenta which are measured in an experiment (or `known 
unknowns'~\cite{Rumsfeld}), {\em viz.}\ $\mathbf{p}$, and those `unknown 
unknowns' which are not measured,  {\em viz. } $p_3$. The observable phase space is then obtained by projecting out the unmeasured $p_3$, and is given by the disk
\begin{gather} \label{gripaios_disk}
 \mathbf{p}^2 \leq M^2 ,
\end{gather}
 in  $\mathbb{R}^2$. The full phase space is clearly a non-singular manifold (it is, after all, just a two-sphere), but the observable phase space, obtained by projection, exhibits singularities whenever the tangent space to the full phase space is parallel to the direction of projection. In our simple example, this is equivalent to the simple algebraic condition $0 = \frac{\partial g}{\partial p_3} = 2p_3$. But when $p_3 = 0$, Eq.~(\ref{gripaios_sphere}) implies that $\mathbf{p}^2 = M^2$, so the singular points of the observable phase space correspond to the boundary of the disk in~(\ref{gripaios_disk}).

Now let us build the singularity co-ordinate in the vicinity of a given singular point. Since the disk is rotationally symmetric, we may, without loss of generality, choose the singular point to be at $(\mathbf{p}, p_3) = ((M,0),0)$. Following the rubric of~\cite{Kim:2009si}, the first step is to choose a system of orthonormal co-ordinates, $(n,t_1,t_2)$ in the neighbourhood of the singularity, corresponding to directions normal and tangent to the full phase space (the two-sphere in the case at hand). Thus I 
write
\begin{gather}
(\mathbf{p} , p_3)=((M+ n , t_1),t_2) 
\end{gather}
In these co-ordinates, the constraint~(\ref{gripaios_sphere}) may be written as
\begin{gather} 
 (M + n)^2 + t_1^2+ t_2^2 - M^2 = 0 \implies n = \frac{-1}{2M}(t_1^2+ t_2^2) + O(n^2)
\end{gather}
The un-normalized singularity co-ordinate is just $n$; to normalize it, we restrict the second fundamental form $II = \frac{-1}{2M}(t_1^2+ t_2^2)$ to the invisible direction $t_2$. The volume of phase space in the invisible direction thus scales as $(2M II)^{\frac{1}{2}}$ and the normalized singularity co-ordinate is $\Sigma =  2 M n = 2 M \delta p_1$. For a general singular point on the boundary of the disk, we find $\Sigma = 2\mathbf{p} \cdot \delta \mathbf{p} = \delta (\mathbf{p}^2 - M^2) $. So the singularity co-ordinate is just the observable $(\mathbf{p}^2 - M^2)$, linearized about a point where $p^2 = M^2$.
Note that the singularity co-ordinate depends on the event (through $\mathbf{p}$), on the chosen singularity (which defines $\delta \mathbf{p}$, and on the hypothesis for the masses (through $M$).

Having computed explicitly the singularity co-ordinate, we are now in a position to answer several questions. 
Firstly, what is the relation to the usual transverse mass variable? The transverse mass variable is defined by
\begin{gather} \label{gripaios_mt}
m_T^2 \equiv  m_X^2 + m_V^2 + 2(ef - \mathbf{p} \cdot \mathbf{q})
\end{gather}
where $\mathbf{p}$ and $\mathbf{q}$ are the transverse momenta of $X$ and $V$, respectively, and $e \equiv \sqrt{\mathbf{p}^2 + m_X^2} $ and $f \equiv \sqrt{\mathbf{q}^2 + m_V^2}$ are their transverse energies. For the special case of $\mathbf{q}= \mathbf{0}$, this reduces to 
\begin{gather} \label{gripaios_mTspecial}
m_T^2 \equiv m_V^2 + m_X^2 + 2m_V\sqrt{\mathbf{p}^2 + m_X^2}.
\end{gather}
Now we know that when we hypothesize the correct value $m_X$ for the {\em a priori} unknown mass of $X$, the distribution of $m_T$ has its maximum at $m_T = m_Y$. That is to say, the $m_T$ distribution has a singularity (an edge, in fact) at $m_T = m_Y$. Equivalently, we can say that the distribution of $m_T^2 - m_Y^2$
will be singular at the origin (when the correct hypothesis of the masses $m_X$ and $m_Y$ is chosen), or indeed that the distribution of the observable $(\mathbf{p}^2 - M^2)$ has a singularity at the origin. But as we saw above, $(\mathbf{p}^2 - M^2)$, when linearized about a singularity, is precisely the singularity co-ordinate constructed according to the recipe of~\cite{Kim:2009si}. Note that it is not correct to say that the transverse mass and the singularity co-ordinate are equivalent, because the latter is linearized about a singular point, whereas the former is not. But it is correct to say that the transverse mass and the singularity co-ordinate are equivalent in the neighbourhood of a given singular point, modulo an overall scale factor. Nevertheless, away from the singular point, the transverse mass and singularity co-ordinate distributions will disagree.

Secondly, since there is a whole $S^1$ of singular points, given by the boundary of the disk, which one should we choose to construct the singularity co-ordinate for a given event?
Na\"{\i}vely, the rotational invariance tells us that any one is as good as any other. But if we choose just a single point, most events (assuming they are spread uniformly over the disk) will be a long way away from the singular point. To counteract this, it is suggested in~\cite{Kim:2009si} that for any event, we should choose the `nearest' singular point to compute the singularity co-ordinate for any event. This then raises the question of what metric defines the concept of nearness, and of whether a singular point thus defined is unique. One answer might be to define the nearest point with respect to the metric on observable phase space induced by the Euclidean embedding; in that case the nearest singularity is obtained by drawing a radius through the event and finding its intersection with the disk's boundary. Another solution might be to minimize the singularity co-ordinate itself, constructed with respect to all singularities. At least in the example here, this alternative definition yields the same singular point at the nearest one, for a given event.

Thirdly, what masses can we actually measure with the singularity co-ordinate in this case? We know on the basis of general kinematic arguments that to measure both 
the invisible masses $m_X$ and $m_Y$, one needs events in which the parent particle $Y$ has variable transverse boosts with respect to the laboratory frame. But here, we restricted events to the subset with $\mathbf{q}= 0 $, corresponding to a fixed transverse boost of the parent. In this case we know that we should only be able to measure the combination of masses given by $M$ using kinematic methods alone. 
To see that this is what happens here, we need to recall how the algorithm of~\cite{Kim:2009si} is defined. The algorithm instructs us to construct the singularity co-ordinate for all possible hypothetical values of the unknown masses. It then tells us that for true values of the masses, we will observe a singularity at the origin.

Now, since the singularity co-ordinate is just $\delta (\mathbf{p}^2 - M^2)$, it is clear that if we instead choose wrong values for the masses, the singularity will, in general, have a singularity that is translated away from the origin. But if one makes a {\em wrong} guess for $m_X$ and $m_Y$ individually that yields the {\em right} value of $M$ in combination, then the singularity will be at the origin. 

This is, of course, fully consistent with the kinematic observation that in such a subset of events, one can do no better than measure the combination $M$, and the singularity method does no worse than any other method in this respect. But it does raise the worry that, in other cases, more than one set of mass values will give rise to a singularity at the origin, even when general kinematic arguments tell us that the masses can be measured unambiguously. 
Indeed, this is exactly what will happen when we consider, in the next Section, the general case of visible particles with arbitrary momentum. 

Lastly, can we get rid of the linearization? In this simple case, we can simply take the `known unknown' to be $\mathbf{p}^2$ (taking values in $\mathbb{R}^+$).
Phase space is then a parabola in $\mathbb{R} \times \mathbb{R}^+$ and the projected phase space is the interval $\mathbf{p}^2 \in [0,M^2]$. Then the singularity co-ordinate is just $\mathbf{p}^2 - M^2$. So in this case, because the constraint is a single quadratic function, linearization is an unnecessary simplification.  
\subsection{The general case}
Now let me proceed to the general case, where $V$ is produced with arbitrary four-momentum. In this case, we have events in which the parent may have an arbitrary boost with respect to the laboratory frame, and know on general kinematic grounds that it should be possible to measure both of the unknown masses $m_X$ and $m_Y$. We would like to see, explicitly, whether (and if so, how) this may be achieved using the algebraic singularity method.

To prevent the proliferation of unknowns, let me consider the case of $2+1$-dimensional spacetime. The energy-momentum vectors then become
\begin{align}
V_\mu &= (q_0 , q,q_3 ), \\
X_\mu &= ( p_0, p, p_3), \\
Y_\mu &= ( p_0 + q_0,p+q,p_3 + q_3 ).
\end{align}

To render the analysis straightforward, I first use the two constraints $V_\mu V^\mu = m_V^2$ and $X_\mu X^\mu = m_X^2$ to solve for $p_0$ and $q_0$. To wit,
\begin{align}  \label{gripaios_pq}
q_0 &= \sqrt{ q^2 + q_3^2 + m_V^2} \\
p_0 &= \sqrt{ p^2 + p_3^2 + m_X^2}.
\end{align}
This leaves a set of  three `known unknowns', namely $\{ p,q,q_3\}$, and one `unknown unknown', $p_3$, subject to the single constraint
\begin{gather}  \label{gripaios_g2}
g \equiv 2p_0 q_0 - 2 pq - 2p_3 q_3 + m_X^2 + m_V^2 -m_Y^2 = 0,
\end{gather}
where $p_0,q_0$ are, of course, given by Eq.~(\ref{gripaios_pq}). Geometrically, the full phase space is a three-dimensional hypersurface in $\mathbb{R}^4$, defined by the quartic constraint~(\ref{gripaios_g2}). The observable phase space is obtained by projecting with respect to the co-ordinate $p_3$, and is singular when
\begin{gather}
0 = \frac{\partial g}{\partial p_3} = 2(\frac{q_0}{p_0} p_3 - q_3) \implies q_0p_3- p_0 q_3 = 0.
\end{gather}
Substituting into~(\ref{gripaios_g2}), it is easy enough to show that, at the singularities, the observable momenta satisfy
\begin{gather}
m_Y^2 = m_X^2 + m_V^2 + 2(ef - pq).
\end{gather}
Perhaps unsurprisingly, this is just the condition that the transverse mass variable~(\ref{gripaios_mt}) be at its maximum.

At a singularity, the normal vector to phase space has direction
\begin{gather}
\begin{pmatrix} 0& 0& q_0&-p_0 \end{pmatrix}^T,
\end{gather}
so that the tangent space may be defined by the three vectors
\begin{gather}
\begin{pmatrix} 1&0& 0&0 \end{pmatrix}^T,\; \begin{pmatrix} 0&1& 0&0 \end{pmatrix}^T,\; \begin{pmatrix} 0&0& p_0&q_0 \end{pmatrix}^T
\end{gather}
Using these vectors to define the directions of the orthonormal co-ordinates in the neighbourhood of the singularity, we have
\begin{gather}
\begin{pmatrix} 
n \\ t_1 \\ t_2 \\ t_3
\end{pmatrix}  =
\begin{pmatrix} 
\sin \theta & \cos \theta & 0 & 0\\
\cos \theta & -\sin \theta & 0 & 0\\
0 & 0 & 1&0\\
0 & 0 & 0&1
\end{pmatrix} 
\begin{pmatrix} 
\delta p \\ \delta q  \\ \delta q_3 \\ \delta p_3
\end{pmatrix} ,
\end{gather}
where I defined
\begin{gather}
 \tan \theta \equiv - \frac{q_0}{p_0} = - \frac{q_3}{p_3} = - \frac{f}{e}.
\end{gather}
We are now in a position to compute the singularity co-ordinate. Going through the normalization procedure given in~\cite{Kim:2009si}, we end up with
\begin{align}
\Sigma &= \frac{\partial g}{\partial n}n, \\
&= 2\frac{pq_0 - qp_0}{p_0q_0}( q_0 \sin \theta + p_0 \cos \theta) (\delta p \sin \theta  + \delta q \cos \theta ), \\
&= 2 \frac{pq_0 - qp_0}{p_0q_0} (q_0 \delta p  - p_0 \delta q ), \\
&= 2 \frac{pf - qe}{ef} (f \delta p  - e \delta q ).
\end{align}

To show the relation with the transverse mass, linearize~(\ref{gripaios_mt}) about a point with $m_T = m_Y$. One obtains
\begin{gather}
m_Y^2 - m_T^2 =  2 \frac{pf - qe}{ef} (f \delta p  - e \delta q ).
\end{gather}
So the singularity co-ordinate is equivalent to the transverse mass variable, expanded linearly about its maximum.

Let me again make some remarks. Firstly, I pointed out at the beginning of this section that, in this general case, kinematics tells us that it should be possible to measure both invisible masses, $m_X$ and $m_Y$, in this case. How is this achieved using the singularity co-ordinate? For the transverse mass variable, this may be achieved, at least in principle, in the following way~\cite{gripaios:2007is}. Guess a value for the mass $m_X$ and compute the resulting distribution over events of the transverse mass~(\ref{gripaios_mt}). Extract the endpoint of the distribution. Now plot the endpoint as a function of the guessed mass $m_X$. Kinematics tells us that the function has a `kink'~\cite{Cho:2007qv} (that is, is $C^0$, but not $C^1$), at the point where $m_X$ takes its true value.

\begin{figure}
\begin{center}
\includegraphics[width=0.5\textwidth]{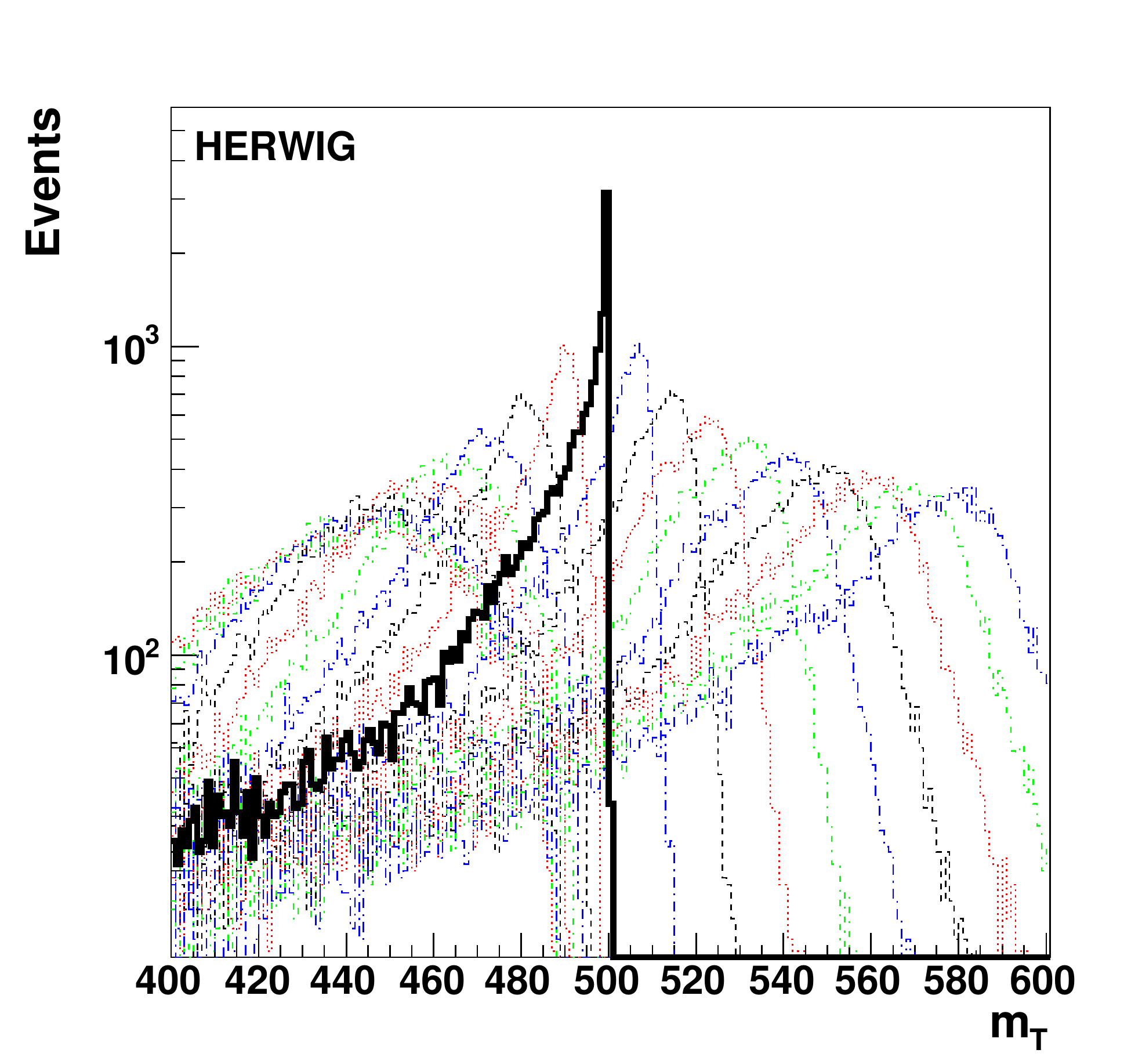}
 \caption{Simulation of the transverse mass variable for decay of a single gluino, with different curves corresponding to different hypothesized values of the invsible daughter mass $m_X$. The distribution
 has a singular endpoint for all values of $m_X$, which translates into a singularity in the singularity co-ordinate for all $m_X$. Reproduced from~\cite{Barr:2007hy}.
}
\label{gripaios_fig}
\end{center}
\end{figure}

For the singularity co-ordinate, we are instructed to compute the value of the co-ordinate for all events and for all possible hypothesized values of the masses. We should then look for values of the masses that give rise to a singularity at the origin in the distribution over events of the singularity co-ordinate. Now, just as for the special case considered in the last section, although it is true that the set of mass values giving rise to a singularity at the origin contains the point corresponding to the true mass values, it is not true that the set contains {\em only} this point. This is easily seen by considering a plot of the transverse mass distribution for various values of $m_X$, an example of which is reproduced in Fig.~\ref{gripaios_fig}. The point is simply that the $m_T$ distribution has an endpoint for all values of $m_X$, and since the distribution
is $C^0$, but not $C^1$ at the endpoint, it seems reasonable to describe the endpoint as singular. Therefore, the singularity co-ordinate, which is simply a linearized version of the transverse mass variable, will feature a singularity in this sense for all values of $m_X$. And for each value of $m_X$, there exists a value of $m_Y$ that will map this singularity to the origin in the singularity co-ordinate. 

We conclude that the set of masses, $\{(m_X,m_Y)\}$ that give rise to a singularity at the origin of the singularity co-ordinate does not consist of the single point corresponding to the true mass values. But rather consists of a curve in the space $(m_X,m_Y)$. How then are we to find the true masses? Two methods suggest themselves. One is to note that this curve is nothing but the kink curve discussed above, and the location of the kink gives the true masses. A second is to observe in Fig.~\ref{gripaios_fig} that the nature of the singularity qualitatively changes as one approaches the true value of $m_X$. That is to say, the distribution becomes discontinuous.

A second remark is that, in order to compute the singularity co-ordinate according to the prescription given in~\cite{Kim:2009si}, one needs a definition of what is the `nearest
singularity to a given event'. In the special case considered in the previous Section, where the observable phase space is a disk with the singularities living on its boundary, it seems easy enough to make an unambiguous definition. But in the more general case considered in this Section, the observable phase space is a three-volume in $(p,q,q_3)$, whose singular boundary is the two-surface defined by the quartic $2(p_0 q_0 - pq -p_3q_3) = M^2$. (In the case of massless daughter particles, this describes a hyperbola.) It is now unclear, at least from a geometer's viewpoint, what the nearest singularity should be.
\section{Conclusions}
The algebraic singularity method of~\cite{Kim:2009si} constitutes an elegant general method for measuring masses, 
exploiting singularities that arise in projecting the smooth phase space of events involving invisible particles onto the observable phase space.
To better understand the method, I applied it to the simple example of a single parent particle undergoing a point-like decay to an invisible daughter particle and a visible daughter particle. For this decay topology, it is known that the kinematic properties are completely captured by the transverse mass variable, and I showed that the variable that arises from the algebraic singularity method
is nothing but a linearized version of the usual transverse mass. 

The algorithm works by looking for a singularity at the origin in the distribution of a certain co-ordinate, the so-called singularity co-ordinate. Such a singularity is shown to arise when the hypothesis for the masses is the correct one. I argued that the method suffers from the problem that many mass hypotheses (including incorrect ones) give rise to such a singularity, even in cases where general kinematic arguments tell us that it ought to be possible to determine the masses unambiguously.
To do so using the singularity method seems to require supplementing the algorithm with a means to identify the nature of a singularity, as well as its mere presence.

I also discussed the issue of the definition of the nearest singularity, required to implement the algorithm. This issue will need to be borne in mind when applying the method to more complicated decay topologies, such as cascade decays.

\section*{Acknowledgements}
I thank A. J. Barr, I.-W. Kim and C. G. Lester for discussions.

%% file: RSW/RSW.tex
\chapter{Light gluinos in hiding: reconstructing R-parity violating decays at the Tevatron}
{\it A.R.~Raklev, G.P.~Salam and J.G.~Wacker}

\begin{abstract} 
If gluinos exist with masses less than 200 GeV, they are 
copiously produced at the Tevatron, but still may not have been discovered if they decay through baryon number violating operators. We show that using cuts on jet substructure can enable a discovery with existing data and even determine the gluino's mass.  \end{abstract}

\section{Introduction}
The TeV scale is the high energy frontier and New Physics (NP) is currently searched for in a multitude of channels at hadron colliders.  Searches for NP with final state jets have yielded bounds on new resonances decaying into pairs of jets through di-jet searches~\cite{Aaltonen:2008dn}, or into jets and missing energy \cite{Abazov:2007ww,Aaltonen:2008rv}.  These resonance searches reach up to masses at the TeV scale for resonant production and 500~GeV for pair-produced particles. However, other NP possibilities can appear in exclusively hadronic final states, with no missing energy, that will not be discoverable with the di-jet search or jets and missing energy searches. These NP possibilities fail to stand out over background because hadronic final states are challenging to calibrate and analyse.  Even some relatively basic searches have not been performed at the Tevatron.  For instance, \cite{Kilic:2008pm} showed that a resonance that decays into two other new particles that subsequently decay to jets has not been explicitly searched for, even though the backgrounds are manageable.  In light of this dearth of searches, it is possible that NP candidates could escape detection at much lower masses than 1~TeV, not because of small couplings to Standard Model (SM) particles and low cross sections, but because of detector signatures that are sufficiently different from the searches  performed so far at the Tevatron.

This contribution considers one such possibility: a relatively light gluino, copiously pair-produced at the Tevatron, in a model with trilinear R-parity violating (RPV) operators that violate baryon number. As a result, the gluino decays into three quarks, $\tilde g\to qqq$, and the event contains six final state partons.  It is conceivable that these six partons will produce a six-jet signature if the gluinos are produced near threshold, however, in the busy hadronic environment of the Tevatron, with potential multiple jet overlaps and unreconstructed jets, it is not evident that this will show up over SM backgrounds in searches requiring high jet-multiplicity, {\it e.g.}\ the all-hadronic $t\bar t$ channel~\cite{RSW:Essig}. The reconstruction of a gluino mass peak is similarly very hard due to the large combinatorics.  Taken to its other extreme, the signature of gluino pair-production far above threshold, the six final state partons can merge into two back-to-back jets, each consisting of three collimated sub-jets from the individual final-state quarks. Here di-jet searches along the same lines as~\cite{Aaltonen:2008dn} may be effective, but the non-resonant origin of the gluino pair makes discovery through simple di-jet invariant mass distributions very difficult.

The current limit on gluino masses in these scenarios comes from event shapes at LEP, which give a model independent gluino mass bound of 51.0~GeV~\cite{Kaplan:2008pt}.  In this contribution, we will show how jet-substructure information for hard jets yields much better expected sensitivity to gluinos, and how a discovery may be lurking in Tevatron data. Our analysis will follow a pattern similar to those suggested for reconstructing RPV neutralino decays to three quarks in~\cite{Butterworth:2009qa}, and benefits by lessons learned from the recent large interest in reconstructing the hadronic decays of other massive particle species, such as gauge bosons, top quarks and the Higgs~\cite{Butterworth:2002tt,Butterworth:2007ke,Butterworth:2008iy,Thaler:2008ju,Almeida:2008yp,Kaplan:2008ie,Krohn:2009zg,Ellis:2009su,Plehn:2009rk,Krohn:2009th}.

\section{The light gluino}

Searches for supersymmetry necessarily make assumptions on the sparticle spectrum, and usually it is assumed that color neutral particles (neutralinos, charginos, and sleptons) are lighter than the colored particles (gluinos and squarks), however, this choice is motivated by top-down model building considerations. See \cite{Raby:1997bpa,Raby:1998xr} for examples of models with gluinos as the lightest supersymmetric particle (LSP) and \cite{Berger:2008cq,Alwall:2008va} for studies without top-down prejudices. If a colored LSP decays relatively quickly it avoids the standard cosmological constraints on stable charged particles.

In this contribution, the gluino is assumed to be the lightest supersymmetric particle produced in a collider and it decays via the $R$-parity violating superpotential operator
\begin{eqnarray}
\mathcal{L}_{\text{RPV}} = \int\!\!d^2\theta \; \lambda''_{ijk} U^c_i D^c_j D^c_k +\text{ h.c. },
\end{eqnarray}
where $\lambda''_{ijk}$ are flavor dependent coupling constants that are antisymmetric in the $jk$ indices.  It is likely that there are strong flavor hierarchies in these coupling constants which lessen the constraints on the operator. For instance, a minimal flavor violation scenario gives $\lambda''_{ijk} \propto y_{u\, i}^{1/2}y_{d\, j}^{1/2}y_{d\, k}^{1/2} $. If this is the case, the gluino decays dominantly to the heavy flavor combination of $\tilde{g} \rightarrow cbs$, assuming that it is beneath the top quark threshold. However, to be conservative, we will not make any assumptions of heavy flavours in the final state, and we will use $\lambda''_{112}=0.001$, leading to the gluino decay $\tilde g\to uds$ and its charge conjugate. The size of the coupling avoids both resonant single-sparticle production and displaced vertices/metastable charged sparticles, which have signatures that should be more easily detectable. 

\begin{figure}[ht]
\begin{center}
\includegraphics[width=2.5in]{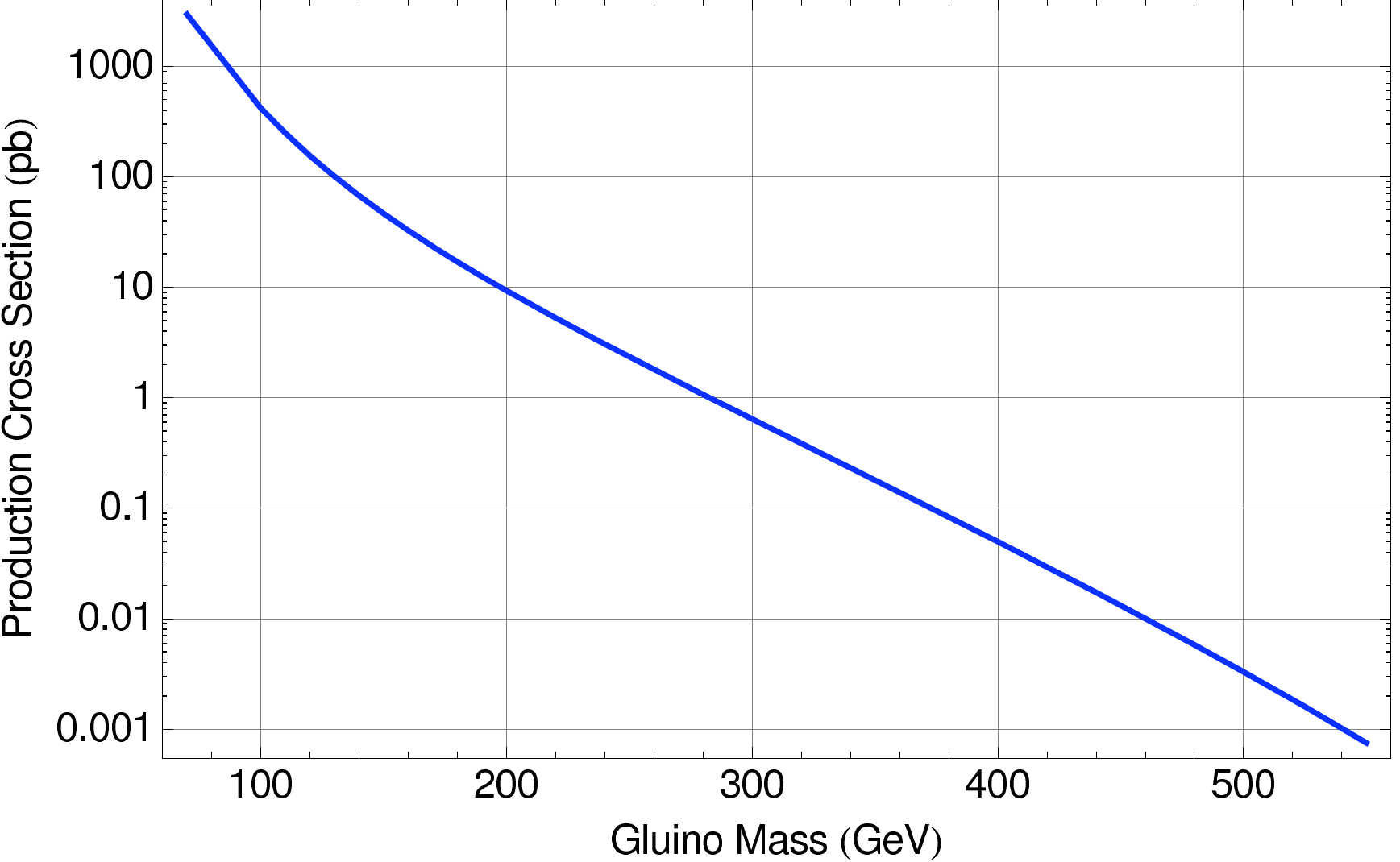}
\caption{Leading order gluino pair-production cross section versus mass at the Tevatron.
}
\label{RSW:Fig:GluinoCrossSection}
\end{center}
\end{figure}

A gluino LSP that decays via the baryon number violating RPV operator completely changes the search strategy for supersymmetry at a hadron collider.  Pair-production of light gluinos has a huge cross section, as shown in Fig. \ref{RSW:Fig:GluinoCrossSection}.   If the gluino is light and undiscovered, the rest of the susy spectrum is possibly not much heavier and we give a rough outline of the generic phenomenology here.   

Any squarks produced decay quickly into the gluino, giving rise to short cascades that rarely contain secondary leptons or missing energy, even if charginos are kinematically accessible in the decay chain.
The best chance for a ``spectacular'' leptonic event at the Tevatron is then if the direct production of charginos and neutralinos is kinematically accessible, and they decay to a lighter slepton.  The slepton will in turn decay via a prompt four-body decay, $\tilde{\ell} \rightarrow \ell \tilde{g} \bar{q}q$.  Thus the entire event would be $\chi\chi' \rightarrow 2 \tilde{g} 4 \ell 4 j$, where some of the $\ell$ may be neutrinos, giving rise to a modest amount of missing energy.  A tri-lepton search might be effective with this channel, however, most current searches place jet multiplicity cuts to reduce the $t\bar{t}$ background \cite{Aaltonen:2008pv}.  The searches for tri-leptons with jet vetoes set limits of $m_{\chi^\pm} \lower.7ex\hbox{$\;\stackrel{\textstyle>}{\sim}\;$} 150$~GeV. Given that a gluino LSP gives a more challenging environment, this is an optimistic estimate of the Tevatron's reach for charginos and neutralinos in these scenarios.

Associated production of a squark and gluino can be an important contribution to the boosted gluino spectrum because the gluino's $p_T$ can arise through the decay of the squark; however, the final state decay products of $\tilde{g} \tilde{q} \rightarrow q \tilde{g} \tilde{g}$ are not more visible than gluino pair-production. The squark may also appear as a di-jet resonance, however, the production cross sections are much smaller.  For instance, a spectrum with $(m_{\tilde{g}}, m_{\tilde{q}} ) = (51\text{ GeV}, 400\text{ GeV})$ has a production cross section of $\sigma_{\tilde{g} \tilde{q}} \simeq 6$~pb, beneath the 10 pb bound for a 400~GeV resonance in the di-jet search \cite{Aaltonen:2008dn}.  Associated production could become an effective search channel for the squark with the jet substructure methods described below.

\subsection{Monte Carlo simulation}

For concreteness, we consider a ``worse case'' scenario with only gluinos light, keeping all other sparticle masses at 1 TeV. Our conservative approach means that the discovery potential should be independent of the RPV coupling flavour and value, and to first order only depend on the gluino mass. We simulate gluino pair-production with RPV decays at the $\sqrt{s}=1.96$~TeV Tevatron using the {\sc Herwig~6.510} Monte Carlo event generator~\cite{Corcella:2000bw,Dreiner:1999qz,Moretti:2002eu,Corcella:2002jc} with {\sc CTEQ 6L}~\cite{Pumplin:2002vw} PDFs, and using {\sc Jimmy~4.31}~\cite{Butterworth:1996zw} for the simulation of multiple interactions. The {\sc Herwig} event generator includes spin correlations in the gluino decays. The leading-logarithmic parton shower approximation used in {\sc Herwig} has been shown to model jet substructure well in a wide variety of processes~\cite{Buskulic:1995sw,Abazov:2001yp,Abbiendi:2003cn,Abbiendi:2004pr,Chekanov:2004kz,Acosta:2005ix}.

The resulting events are interfaced to the {\sc FastJet~2.4.0}~\cite{Cacciari:2005hq,RSW:FastJetWeb} jet-finder package using the {\sc Rivet}~\cite{Waugh:2006ip} framework, with some minor modifications to cope with the simulation of sparticles. Our background sample, consisting of QCD $2\to 2$ events, $t\bar t$, $W$+jet, $Z$+jet and $WW/WZ/ZZ$ production is simulated with the same setup. To explore the ultimate reach of the Tevatron, we use statistics comparable to 10~fb$^{-1}$ of integrated luminosity. 
A cut of $|\eta|<1.1$ is imposed upon all jets as a realistic geometrical acceptance; however, no other
detector effects are included.

\subsection{Analysis}
The signal is isolated by searching for jets that contain all three quarks from the gluino decay. These jets are necessarily very hard; indeed, the distance $\Delta R = \sqrt{\Delta\phi^2+\Delta\eta^2}$ between the decay products of a massive particle with mass $m$ and transverse momentum $p_T$ should be $\Delta R\;\raisebox{-0.9ex} {$\textstyle\stackrel{\textstyle >}{\sim}$}\;2m/p_T$ ~\cite{Butterworth:2008iy}.  The sensitivity to gluinos is strongly dependent upon the jet-algorithm size parameter, and the momentum cut and these parameters should be optimised for any specific search.  This contribution uses values that have good sensitivity over a large gluino mass range, but no optimisation is performed.  In the following we will use the $k_T$ jet-algorithm~\cite{Catani:1993hr,Ellis:1993tq} in the inclusive mode, as currently used at the Tevatron \cite{Abulencia:2007ez}, with the following choices for jet size: $R=0.5,0.7,1.0,1.5$.
 
Gluino candidates are identified by requirements on jet-substructure of the hardest jets in the event. For each merging $i$ of sub-jets $k$ and $l$ in the jet clustering we define
\begin{equation} 
y_i = \frac{\min{(p_{Tk}^{2},p_{Tl}^{2})}}{m_j^2} \Delta R_{kl}^2,
\end{equation}
where $p_{Tk}^2$ and $p_{Tl}^2$ are the transverse momenta of the sub-jets and $m_j$ is the mass of the final jet after all mergings.  The expectation is that $y_1$ and $y_2$, from the last two mergings of the jet, are distributed very differently from ordinary QCD jets because of the three-parton structure of the gluino jet.  This turns out to be the case, see Fig.~1 of~\cite{Butterworth:2009qa} for the similar case of a three-quark neutralino decay.

Since we have pair-production of gluinos, there is a choice of whether to perform an inclusive analysis searching for at least one gluino candidate jet in each event, or an exclusive analysis, reconstructing both gluinos. This is a balance between signal efficiency and SM background rejection that should be optimized to maximize discovery potential.  In order to trigger and collimate the decay products of the gluino, at least one high $p_T$ jet is required, however, this also implies another back-to-back high $p_T$ gluino. To arrive at as clean a signal sample as possible we choose an exclusive analysis.

\begin{figure}[ht]
\begin{center}
\includegraphics[width=4.5in]{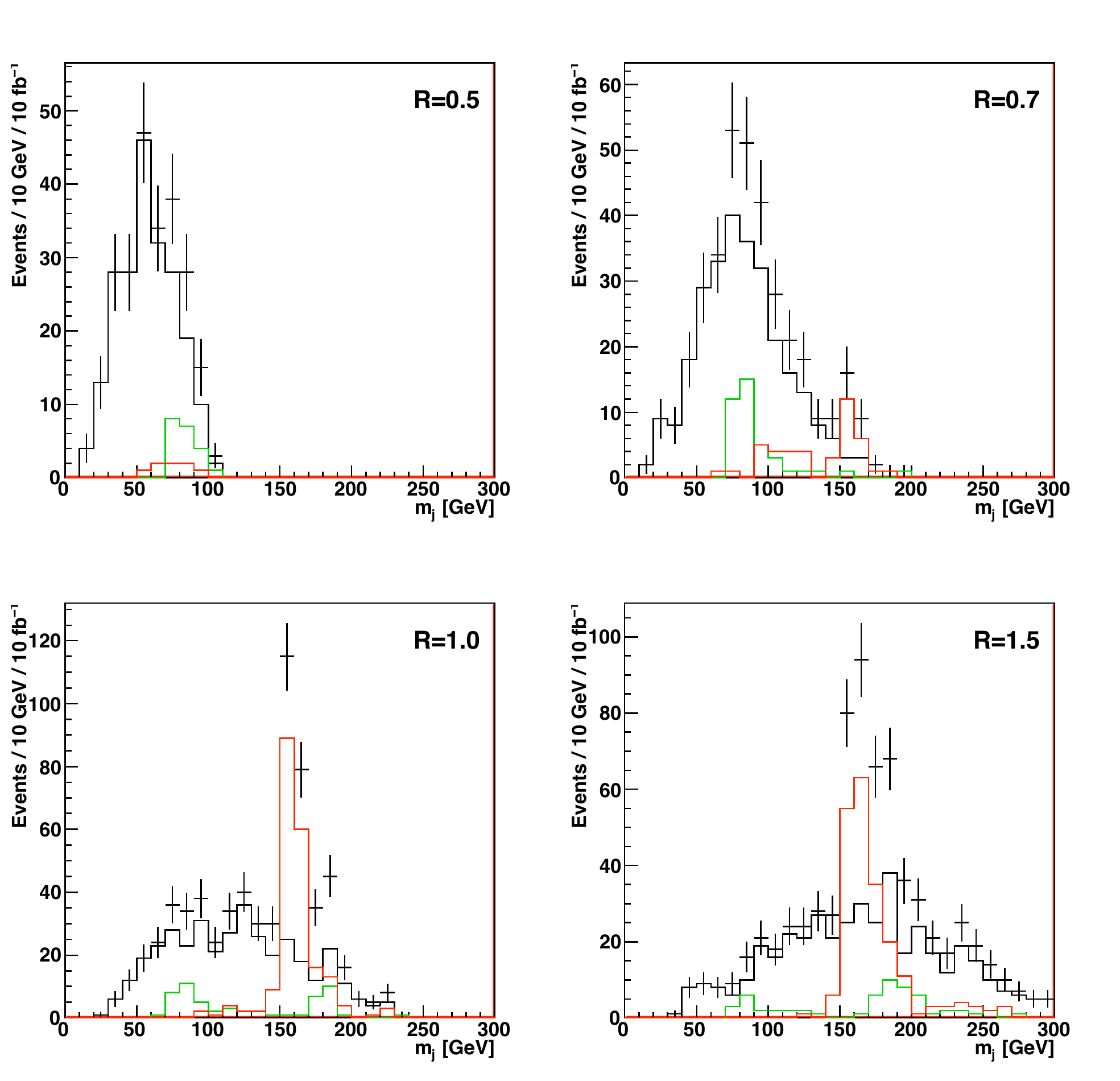}
\caption{Jet mass distribution after cuts for 10~fb$^{-1}$ integrated luminosity (black with error bars). Also shown is contribution from signal events for $m_{\tilde{g}} = 150$ GeV (red), QCD (black) and other high $p_T$ SM background events ($W,Z$ and $t\bar{t}$) (green). 
}
\label{RSW:Fig:mj}
\end{center}
\end{figure}

The following cuts are used: i) we require two hard jets with $p_T>350$~GeV, 
ii) both must be candidate gluino jets satisfying the substructure constraint $y_1>0.1$, and iii) their masses must be within 20\% of each other. The resulting jet mass spectrum is shown in Fig.~\ref{RSW:Fig:mj} for a gluino mass of $m_{\tilde g}=150$~GeV and different values of the jet-size parameter $R$.  A gluino mass peak is clearly observable with a small background, consisting mostly of QCD events, when the jet-size $R$ is large enough to contain the complete gluino jets. This is generically true for gluino masses up to around 200~GeV with our $p_T$ cut, where we start to loose containment of the gluino for the largest jet-size considered.

There is some systematic bias in signal events, visible in Fig.~\ref{RSW:Fig:mj}, towards jet masses larger than the nominal. This arises from the jet-algorithm sweeping up extra energy from initial state radiation and the underlying event, however, it should be possible to calibrate this with known particle masses, {\it e.g.}\ the top quark, and to limit the bias with filtering \cite{Butterworth:2008iy} and related techniques \cite{Ellis:2009su,Krohn:2009th}. The final achievable precision on mass seems likely to be limited by statistics and the jet mass resolution of the experiment.

\begin{figure}[ht]
\begin{center}
\includegraphics[width=4.0in]{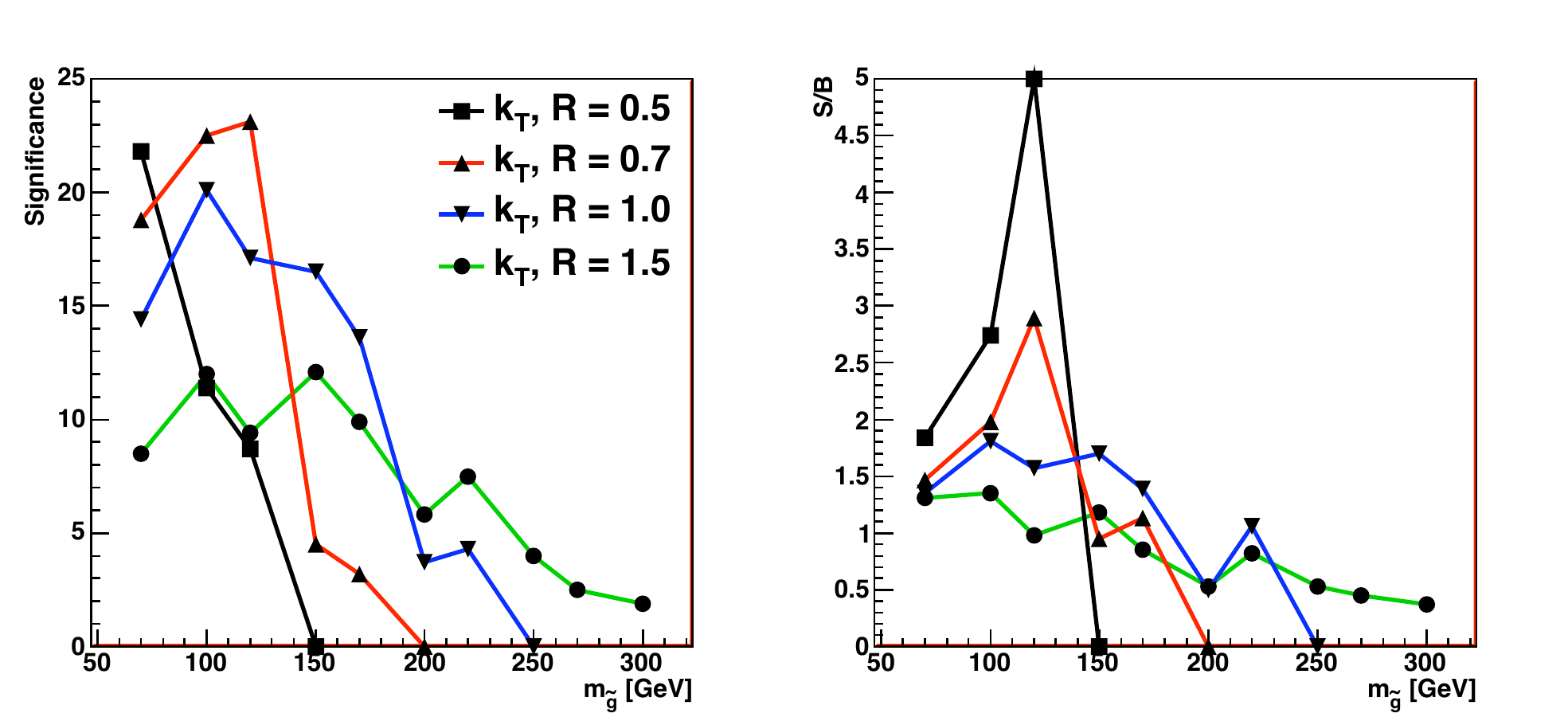}
\caption{Significance (left) and signal over background (right) as a function of gluino mass for various values of the jet-size $R$.
}
\label{RSW:fig:significance}
\end{center}
\end{figure}

In Fig.~\ref{RSW:fig:significance} we show the resulting signal significance, $S/\sqrt{B}$, as a function of gluino mass for all four jet-sizes. The significance is estimated by the number of events in a 40~GeV interval around the nominal mass, which corresponds to a semi-realistic experimental jet mass resolution. This ignores the ``looking-elsewhere'' problem, but should serve as a first estimate of the possible reach.

We can see that even above the point where we start to loose containment of the gluino jet the signal stands out over the background. This is in part caused by the $k_T$ algorithm's efficiency in sweeping up soft radiation somewhat outside of its ``radius'' $R$, and in part by the jet substructure cut not requiring two significant structures, allowing partially reconstructed gluinos. The latter effect is visible in the mass tails for the signal distribution in Fig.~\ref{RSW:Fig:mj}, and also in the presence of vector bosons which should only have one significant structure. For low mass gluinos this is clearly not desirable as it could obfuscate a signal, however, the vector bosons can serve as a jet mass calibration tool along with the top.   

Figure \ref{RSW:fig:significance} also shows the signal to background ratio as a function of the gluino mass.
For light gluino masses the lowest jet-sizes allows discovery and large $S/B$ for gluino masses from 70~GeV to 150~GeV, above which containment of the gluino is lost. For higher masses progressively larger jet-sizes must be used.  For gluino masses significantly above 200~GeV, the rate for producing boosted gluinos is too low; the requirement of 10 signal events with gluino transverse momentum of $p_T>1.5m_{\tilde g}$ sets an upper limit to the reach of $m_{\tilde g}=280$~GeV with 10~fb$^{-1}$ of integrated luminosity.  However, these heavier gluinos are sufficiently massive that their decay products are multiple hard jets at the Tevatron.

\section{Conclusions}

This contribution has demonstrated that the Tevatron can discover new light colored particles that decay into complicated hadronic final states, by using events where these particles are produced at high $p_T$ and their boosted decay products are collimated.  The recently developed techniques using jet substructure can effectively separate signal from background, allowing the new particle to appear as a jet-mass resonance. This work is a proof-of-principle, but there is significantly more work to be done.  Neither of the Tevatron's two detectors have calorimetry that is as finely segmented as the LHC detectors, where most studies have been done so far, and this may reduce the jet mass resolution and restrict the Tevatron reach.  Some of this resolution loss may be recovered by using tracking information, particularly at CDF \cite{Abulencia:2007iy}.

On the theoretical side, the optimal cuts and jet-size for a given mass have not been determined, nor have other jet algorithms such as Cambridge/Aachen been studied. Switching to a more inclusive analysis may also improve sensitivity.  The primary challenge is to keep signal efficiency high due to the low number of gluinos in the high $p_T$ tail, {\it e.g.}\ with $m_{\tilde g}=150$~GeV we see 18\% of gluinos with $p_T>350$~GeV reconstructed using $R=1.0$.  Therefore it may be beneficial to search for one narrow gluino candidate jet with tight constraints recoiling against another wide jet-size gluino candidate jet. The application of loose substructure cuts to the jets of the ordinary di-jet search is also interesting.

If the squarks become light enough that associated squark-gluino production or squark pair production becomes sizable, then the dominant source of boosted gluinos may come from these secondary processes.  This is qualitatively similar to the models studied in \cite{Butterworth:2009qa} where the hadronically decaying LSP was being produced in cascade decays of squarks and gluinos.

\section*{Acknowledgements}
ARR thanks the Swedish Research Council (VR) for financial support through the Oskar Klein Centre. GPS acknowledges support from the French ANR under grant ANR-09-BLAN-0060.  JGW is supported by the US DOE under contract number DE-AC02-76SF00515 and receives partial support from the Stanford Institute for Theoretical Physics and the US DOE Outstanding Junior Investigator Award.

%% file: Muehlleitner/Muehlleitner.tex
\chapter{SUSY-QCD corrections to MSSM Higgs boson production via gluon fusion}

{\it M.~M\"uhlleitner, H.~Rzehak and M.~Spira}

\begin{abstract}
 In the MSSM scalar $h,H$ production is mediated by heavy
 quark and squark loops. The higher order QCD corrections have been 
 obtained some time ago and turned out to be large. The full SUSY-QCD
 corrections have been obtained recently including the full mass
 dependence of the loop particles. We describe our calculation and 
 present first numerical results. We also address the question of the
 proper treatment of the large gluino mass limit, {\it i.e.} the
 consistent decoupling of heavy gluino effects, and present the
 effective Lagrangian for decoupled gluinos.
\end{abstract}

\section{Introduction}
One of the major goals at the LHC is the detection of Higgs
boson(s)
\cite{Higgs:1964ia,Higgs:1964pj,Higgs:1966ev,Englert:1964et,Guralnik:1964eu}. In
the Minimal Supersymmetric Extension of the
Standard Model (MSSM) two complex Higgs doublets are introduced to 
give masses to up- and down-type fermions
\cite{Fayet:1974pd,Fayet:1976et,Fayet:1977yc,Dimopoulos:1981zb,Sakai:1981gr,Inoue:1982ej,Inoue:1982pi,Inoue:1983pp}. After electroweak symmetry breaking there are five
physical Higgs states, two CP-even neutral Higgs bosons $h,H$, one
neutral CP-odd Higgs state $A$ and two charged Higgs bosons
$H^\pm$. At tree level, the Higgs sector can be parameterized by two
independent parameters, the pseudoscalar Higgs boson mass $M_A$
and the ratio of the two vacuum expectation values (VEV) of the two
complex Higgs doublets, $\tan\beta=v_2/v_1$.  The Higgs couplings to
quarks and gauge bosons are modified with $\sin$ and $\cos$ of the
mixing angles $\alpha$ and $\beta$ with respect to the Standard Model (SM)
couplings, where $\alpha$ denotes the $h,H$ mixing angle. 
The bottom (top) Yukawa couplings are enhanced (suppressed) 
for large values of $\tan\beta$, so that top Yukawa
couplings play a dominant role at small and moderate values of
$\tan\beta$. 

At the LHC and Tevatron neutral Higgs bosons are copiously produced
via gluon fusion $gg \to h,H,A$, which is mediated in the case of
$h,H$ by (s)top and (s)bottom loops
\cite{Spira:1997dg,Djouadi:2005gi,Djouadi:2005gj}. The pure
QCD corrections to the (s)quark loops have been obtained including the
full Higgs and (s)quark mass dependencies and increase the cross
sections by $\sim 100$\% \cite{Graudenz:1992pv,Spira:1993bb,Spira:1995rr,Muhlleitner:2006wx,Anastasiou:2006hc,Aglietti:2006tp,Bonciani:2007ex}. This result can be
approximated by very heavy top (s)quarks with $\sim 20-30$\% accuracy
for $\tan\beta \lsim 5$ \cite{Kramer:1996iq}. In this limit the
next-to-leading order (NLO) QCD
\cite{Djouadi:1991tka,Dawson:1990zj,Kauffman:1993nv,Dawson:1993qf,Dawson:1996xz}
and later the
next-to-next-to-leading order (NNLO) QCD corrections
\cite{Harlander:2002wh,Harlander:2002vv,Anastasiou:2002yz,Anastasiou:2002wq,Ravindran:2003um} have been
obtained, the latter leading to a moderate increase of
20-30\%. Finite top mass effects  at NNLO have been discussed in
\cite{Harlander:2009bw,Pak:2009bx,Harlander:2009mq,Pak:2009dg,Harlander:2009my}.
Finally, the estimate of the next-to-next-to-next-to-leading
order effects \cite{Catani:2003zt,Moch:2005ky,Ravindran:2005vv,Ravindran:2006cg} indicates improved perturbative convergence. The
full supersymmetric (SUSY) QCD corrections have been obtained in the limit of heavy
SUSY particle masses
\cite{Harlander:2003bb,Harlander:2003kf,Harlander:2004tp,Harlander:2005if,Degrassi:2008zj} 
and more
recently including the full mass dependence \cite{Anastasiou:2008rm}. 
The electroweak loop effects have been calculated in
\cite{Degrassi:2004mx,Aglietti:2006yd,Actis:2008ug,Anastasiou:2008tj}. 
In this article we will describe in Section 2 
the calculation of the full SUSY-QCD
corrections in gluon fusion to $h,H$, and we will
present for the first time numerical results. 
In Section 3 we will discuss the consistent derivation of the
effective Lagrangian for the scalar Higgs couplings to gluons after the
gluino decoupling.

\section{Gluon fusion\label{gghsusyqcd_label1}}
\begin{figure}[t]
\begin{center}
\includegraphics[width=0.8\textwidth]{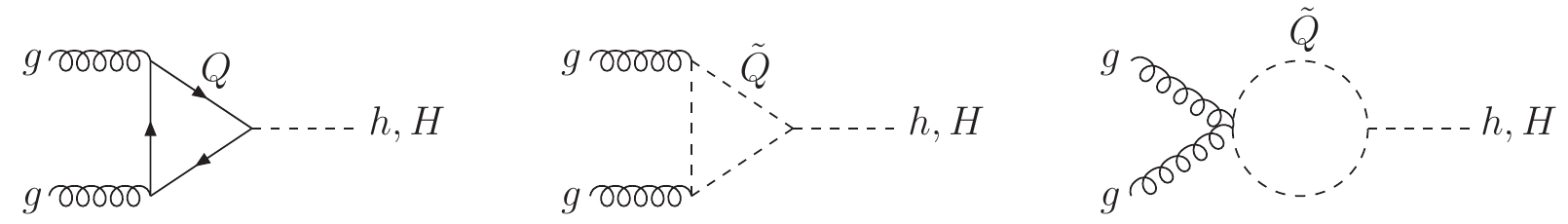}
\caption{Diagrams contributing to $gg\to h,H$ at leading order.}
\label{gghsusyqcd_label3}
\end{center}
\end{figure}

\begin{figure}[hbtp]
\begin{center}
\includegraphics[width=0.9\textwidth]{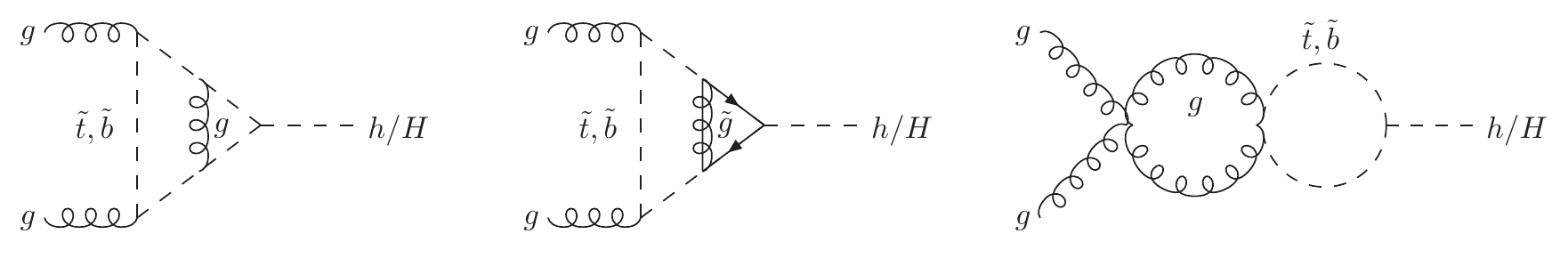}
\\
\caption{Some generic diagrams for the virtual NLO SUSY-QCD
  corrections to the gluonic Higgs couplings.}
\label{gghsusyqcd_label4}
\end{center}
\end{figure}

\begin{figure}[hbtp]
\begin{center}
\includegraphics[width=0.8\textwidth]{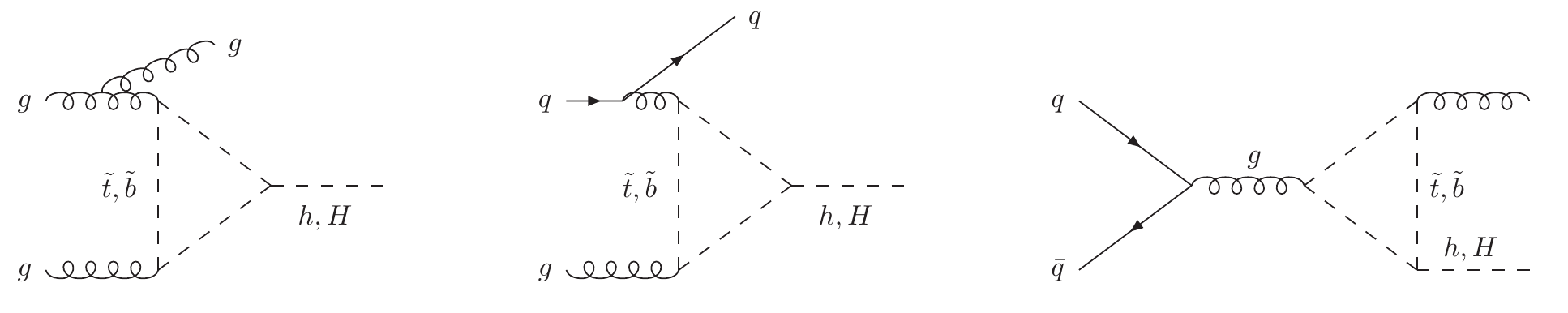}
\\
\caption{Typical diagrams for the real NLO QCD corrections to the squark contributions to the gluon fusion processes.}
\label{gghsusyqcd_label6}
\end{center}
\end{figure}

\begin{figure}[ht]
\begin{center}
\begin{picture}(150,160)(0,0)
\put(-25,-10){\includegraphics[width=0.5\textwidth]{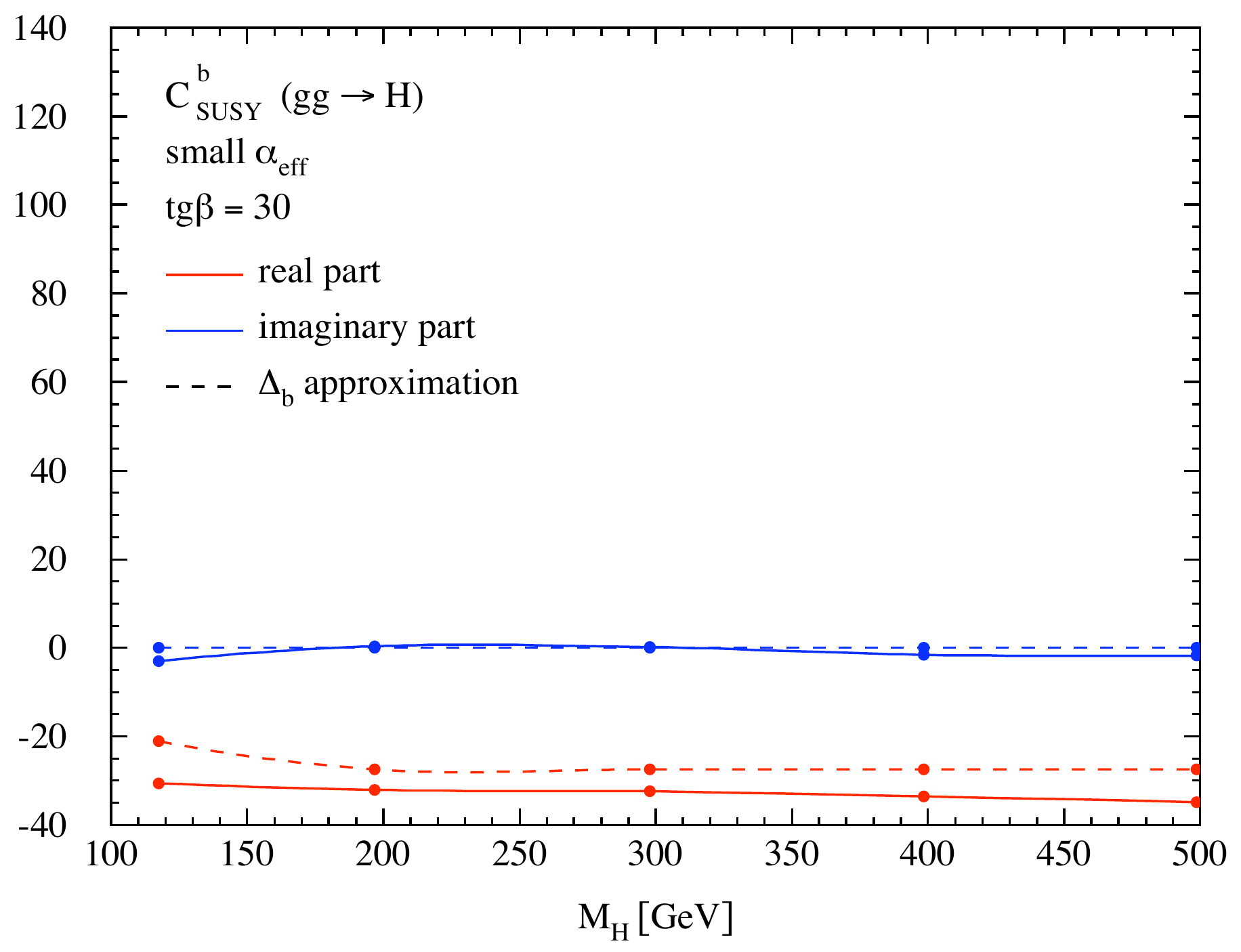}
                  }
\put(125.0,140.0){\small Preliminary}
\end{picture}
\caption{The genuine SUSY-QCD corrections 
normalized to the  LO bottom quark form factor. Real corrections: red
(light gray), virtual corrections: blue (dark gray), compared to the
$\Delta_b$ approximation (dashed lines). $A_b$ has been renormalized
in the $\overline{\mbox{MS}}$ scheme.}
\label{gghsusyqcd_label7}
\end{center}
\end{figure}

At leading order (LO) the gluon fusion processes $gg\to h/H$ are mediated
by heavy quark and squark triangle loops, {\it cf.}
Fig.\ref{gghsusyqcd_label3}, the latter contributing significantly for 
squark masses $\lsim 400$~GeV. The LO cross section in the
narrow-width approximation can be obtained from the $h/H$ gluonic decay
widths, \cite{Spira:1997dg,Djouadi:2005gi,Djouadi:2005gj ,Georgi:1977gs} 
\begin{eqnarray}
\sigma_{LO}(pp\to h/H) & = & \sigma^{h/H}_0 \tau_{h/H}\frac{d{\cal
L}^{gg}}{d\tau_{h/H}} \\
\sigma^{h/H}_0 & = & \frac{\pi^2}{8M_{h/H}^3}\Gamma_{LO}(h/H\to gg)
\nonumber \\
\sigma^{h/H}_0 & = & \frac{G_{F}\alpha_{s}^{2}(\mu_R)}{288 \sqrt{2}\pi} \
\left| \sum_{Q} g_Q^{h/H} A_Q^{h/H} (\tau_{Q})
+ \sum_{\widetilde{Q}} g_{\widetilde{Q}}^{h/H} A_{\widetilde{Q}}^{h/H}
(\tau_{\widetilde{Q}}) \right|^{2} \, , 
\end{eqnarray}
where $\tau_{h/H} = M_{h/H}^2/s$ with $s$ being the squared hadronic
c.m.\ energy and $\tau_{Q/\tilde{Q}}=4 m_{Q/\tilde{Q}}^2/M_{h/H}^2$. 
The LO form factors are given by
\beq
A_Q^{h/H}(\tau) & = & \frac{3}{2} \tau [1+(1-\tau)f(\tau)] \nonumber \\
A_{\tilde Q}^{h/H} (\tau) & = & -\frac{3}{4} \tau[1-\tau
f(\tau)] \\
f(\tau) & = & \left\{ \begin{array}{ll}
\displaystyle \arcsin^2 \frac{1}{\sqrt{\tau}} & \tau \ge 1 \\
\displaystyle - \frac{1}{4} \left[ \log \frac{1+\sqrt{1-\tau}}
{1-\sqrt{1-\tau}} - i\pi \right]^2 & \tau < 1
\end{array} \right. \nonumber\;.
\eeq
And the gluon luminosity at the factorization scale $\mu_F$ is defined as
\begin{displaymath}
\frac{d{\cal L}^{gg}}{d\tau} = \int_\tau^1 \frac{dx}{x}~g(x,\mu_F^2)
g(\tau /x,\mu_F^2) \, ,
\end{displaymath}
where $g(x,\mu_F^2)$ denotes the gluon parton density of the proton.
The NLO SUSY-QCD corrections consist of the virtual two-loop
corrections, {\it cf.} Fig.\ref{gghsusyqcd_label4}, and the real
corrections due to the radiation processes $gg\to gh/H, gq\to qh/H$
and $q\bar{q}\to gh/H$, {\it cf.} Fig.\ref{gghsusyqcd_label6}.
The final result for the total hadronic cross sections can be split
accordingly into five parts,
\beq
\sigma(pp \rightarrow h/H+X) = \sigma^{h/H}_{0} \left[ 1+ C^{h/H}
\frac{\alpha_{s}}{\pi} \right] \tau_{h/H} \frac{d{\cal
L}^{gg}}{d\tau_{h/H}} + \Delta \sigma^{h/H}_{gg} + \Delta
\sigma^{h/H}_{gq} + \Delta \sigma^{h/H}_{q\bar{q}} \, .
\label{gghsusyqcd_label5}
\eeq
The strong coupling constant is renormalized in the $\overline{\rm MS}$
scheme, with the top quark, gluino and squark contributions decoupled from the
scale dependence. The quark and squark masses are renormalized
on-shell. The parton densities are defined in the $\overline{\rm MS}$
scheme with five active flavors, i.e. the top quark, the gluino and the squarks are
not included in the factorization scale dependence. After
renormalization we are left with collinear divergences in the sum of
the virtual and real corrections which are absorbed in the
renormalization of the parton density functions, so that the result
Eq.~(\ref{gghsusyqcd_label5}) is finite and depends on the
renormalization and factorization scales $\mu_R$ and $\mu_F$,
respectively. The  natural scale choices turn out to be
$\mu_R=\mu_F\sim M_{h/H}$. 
The numerical results are presented for the modified small
$\alpha_{eff}$ scenario \cite{Carena:2002qg}, defined by the
following choices of MSSM parameters [$m_t=172.6$~GeV],
\beq
\begin{array}{llllll}
M_{\tilde{Q}} &=& 800\;\mbox{GeV} & \qquad \tan\beta &=& 30 \\
M_{\tilde{g}} &=& 1000\;\mbox{GeV} & \qquad \mu &=& 2\;\mbox{TeV} \\
M_2 &=& 500\;\mbox{GeV} & \qquad A_b = A_t &=& -1.133\;\mbox{TeV} \;.
\end{array}
\eeq
In this scenario the squark masses amount to
\beq
\begin{array}{llllll}
m_{\tilde{t}_1} &=& 679\;\mbox{GeV} & \qquad m_{\tilde{t}_2} &=&
935\;\mbox{GeV} \\
m_{\tilde{b}_1} &=& 601\;\mbox{GeV} & \qquad m_{\tilde{b}_2} &=&
961\;\mbox{GeV} \;.
\end{array}
\eeq
Fig. \ref{gghsusyqcd_label7} displays the genuine SUSY-QCD corrections 
normalized to the  LO bottom quark form factor, {\it i.e.}
$A_b^{h/H} (\tau_b) \to A_b^{h/H} (\tau_b) (1+C^b_{SUSY}
\frac{\alpha_s}{\pi})$. The corrections can be sizeable, but can be
described reasonably with the usual $\Delta_b$ approximation
\cite{Carena:1999py,Guasch:2003cv}, if $A_b$ is renormalized in the $\overline{\mbox{MS}}$ scheme.

\section{Decoupling of the gluinos\label{gghsusyqcd_label2}}
In this section we will address the limit of heavy quark, squark and
gluino masses, where in addition the gluinos are much heavier than the quarks and
squarks. For the derivation of the effective Lagrangian
for the scalar Higgs couplings to gluons we analyze the relation
between the quark Yukawa coupling $\lambda_Q$ and the Higgs coupling
to squarks $\lambda_{\tilde{Q}}$ in the limit of large gluino
masses. We define these couplings at leading order in the case of
vanishing mixing,
\beq
\lambda_Q = g_Q^{\cal H} \frac{m_Q}{v}\; , \qquad \lambda_{\tilde{Q}} = 2
g_Q^{\cal H} \frac{m_Q^2}{v} = \kappa \lambda_Q^2\; , \qquad 
\mbox{with }
\kappa = 2 \frac{v}{g_Q^{\cal H}} \;,
\label{gghsusqcd_label8}
\eeq
where $g_Q^{\cal H}$ denotes the normalization factor of the MSSM
Higgs couplings to quark pairs with respect to the SM. In the
following we will sketch how the modified relation between these
couplings for scales {\it below} the gluino mass $M_{\tilde{g}}$ is
derived. For details, see Ref.~\cite{Muhlleitner:2008yw}. We
start with the unbroken relation between the running
$\overline{\mbox{MS}}$ couplings of Eq.~(\ref{gghsusqcd_label8}) and the
corresponding renormalization group equations (RGE) for scales
{\it above} $M_{\tilde{g}}$. If the scales decrease {\it below}
$M_{\tilde{g}}$ the gluino decouples from the RGEs leading to modified
RGEs which are different for the two couplings $\lambda_{\tilde{Q}}$
and $\kappa \lambda_Q^2$ so that the two couplings deviate for scales
below $M_{\tilde{g}}$. The proper matching at the gluino mass scale
yields a finite threshold contribution for the evolution from the
gluino mass scale to smaller scales, while the logarithmic structure of
the matching relation is given by the solution of the RGEs {\it below}
$M_{\tilde{g}}$. 
In order to decouple consistently the gluino from the RGE for gluino
mass scales large compared to the chosen renormalization scale, a
momentum substraction of the gluino contribution for vanishing
momentum transfer has to be performed \cite{Collins:1978wz}. We refer the reader
to \cite{Muhlleitner:2008yw} for details and give here directly the result for
the modified relation between the quark Yukawa coupling and the
effective Higgs coupling to squarks taking into account the proper gluino
decoupling:
\beq
2 g_Q^{\cal H} \frac{m_Q^2}{v} = \bar \lambda_{\tilde Q,MO}(m_{\tilde Q}) \left\{ 1 +
C_F\frac{\alpha_s}{\pi} \left(\log\frac{M_{\tilde{g}}^2}{m_{\tilde Q}^2} +
\frac{3}{2}\log\frac{m_{\tilde Q}^2}{m_Q^2} + \frac{1}{2} \right)
\right\} \;,
\eeq
where $m_Q$ is the pole mass and $MO$ denotes the momentum substracted
coupling, which is taken at the squark mass scale, which is the
proper scale choice of 
the effective Higgs coupling to squarks and which is relevant for an
additional large gap between the quark and squark masses. 

Taking into account the radiative corrections to the relation between the effective
couplings after decoupling the gluinos leads to the following
effective Lagrangian in the limit of heavy squarks and quarks,
\begin{equation}
{\cal L}_{eff} = \frac{\alpha_s}{12\pi} G^{a\mu\nu} G^a_{\mu\nu}
\frac{\cal H}{v} \left\{\sum_Q g_Q^{\cal H}
\left[1+\frac{11}{4}\frac{\alpha_s}{\pi}\right] + \sum_{\tilde Q}
\frac{g_{\tilde Q}^{\cal H}}{4} \left[1 + C_{SQCD}
\frac{\alpha_s}{\pi}\right] + {\cal O}(\alpha_s^2) \right\} \; ,
\end{equation}
where $g_{\tilde Q}^{\cal H}=v\bar \lambda_{\tilde Q,MO}(m_{\tilde Q}) /
m_{\tilde Q}^2$. The cofficient $C_{SQCD}$ is given by
\beq
C_{SQCD} = \frac{37}{6} \;.
\eeq
It is well-defined in the limit of large gluino masses and thus
fulfills the constraint of the Appelquist--Carazzone decoupling theorem
\cite{Appelquist:1974tg}. 

\section{Conclusions}
We have presented first results for the NLO SUSY-QCD corrections to
gluon fusion into CP-even MSSM Higgs bosons, including the full mass
dependence of the loop particles. The genuine SUSY-QCD corrections can
be sizeable. We furthermore demonstrated that the gluino
contributions can be decoupled in the large $M_{\tilde{g}}$ limit in
accordance with the Appelquist--Carazzone theorem.

\section*{Acknowledgements}
We thank the organizers of the 2009 Les Houches workshop for the
organization of this very interesting and fruitful workshop, in which
to participate is always a big pleasure.

%% file: Fichet-Kraml/fichet-kraml.tex

\chapter[Discriminating SUSY models at the LHC: gauge-Higgs unification vs mSUGRA]{Discriminating SUSY models at the LHC: a case study of gauge-Higgs unification versus mSUGRA}

{\it S. Fichet and S. Kraml}

\begin{abstract}
We investigate whether a sparticle spectrum arising from supersymmetric gauge-Higgs 
unification (SGHU) can be discriminated against the minimal supergravity (mSUGRA) 
model by LHC measurements. 
To this aim we assume that a realistic part of the mass spectrum has been 
measured with a reasonable accuracy and perform Markov Chain Monte Carlo 
fits of the two models, SGHU and mSUGRA, to the expected data.  
\end{abstract}

\section{Introduction}

The model of supersymmetric gauge-Higgs unification (SGHU) we published 
recently in~\cite{Brummer:2009ug} features light selectrons and smuons, which are 
systematically lighter than the second-lightest neutralino $\tilde\chi^0_2$. 
Same-flavour opposite-sign (SFOS) dileptons stemming from 
$\tilde\chi^0_2 \to \ell^\pm\tilde\ell^\mp\to \ell^\pm\ell^\mp\tilde\chi^0_1$ (with $\ell=e,\mu$) 
in cascade decays of squarks and gluinos are hence expected to have a large rate in 
this model. 

The SFOS dilepton signature arising from on-shell decays of $\tilde\chi^0_2$ to 
sleptons is also prominant in the minimal supergravity (mSUGRA) model with small 
$M_0$~\cite{Baer:1995va}. Indeed, most benchmark studies are performed within 
the mSUGRA model, see e.g.~\cite{:1999fr,Ball:2007zza,Aad:2009wy}. 
Top-down fits of the model to expected LHC measurements look quite promising; 
as shown in \cite{Allanach:2006fy} they are, however, largely 
dominated by the gaugino and slepton masses (or mass differences).

In this contribution, we investigate whether SGHU can be discriminated 
against mSUGRA based on LHC measurements. 
To this aim, we perform a case study for the SGHU point D of~\cite{Brummer:2009ug}. 
We assume that a realistic part of the mass spectrum has been measured 
with a reasonable accuracy, and perform Markov Chain Monte Carlo (MCMC)
fits of the two models to the expected measurements.\footnote{For details 
on the MCMC method see, e.g., \cite{Allanach:2005kz,deAustri:2006pe,Allanach:2007qk,Belanger:2009ti} and references therein. MCMC fits of mSUGRA parameters to expected LHC data 
(at mSUGRA benchmark points) were recently done in \cite{Aad:2009wy,Roszkowski:2009ye}.} 
These measurements, although not sufficient
to do a Lagrangian reconstruction, may permit to exclude models
of supersymmetry breaking, or to conclude that one model is more likely than
another from a Bayesian point of view. 
If it is not the case, the posterior distributions may help identify additional observables 
with better discriminating power.

In general, depending on the measurements available, there are two ways 
of comparing the agreement between models and data. If the models are 
overconstrained by measurements, one can simply compare the maxima 
of their likelihoods.
On the other hand, if the models are underconstrained, continous sets
of points reach the maximum of likelihood, and it becomes relevant
to compare the average of the likelihood on the whole parameter space
allowed (Bayes factor). Although the likelihood functions remain the
same in the two approaches, the first is the Frequentist approach, 
whereas the second corresponds to Bayesian statistics. 
In this contribution, we will consider both points of view.

After explaining details of the analysis in Section~\ref{sec:fichet-analysis}, we 
present in Section~\ref{sec:fichet-results} the results of the MCMC fits. 
In particular we show the marginalized likelihood distributions
of the model parameters, sparticle masses and other observables,
as well as the Bayes factors to compare the two models.

\section{Setup of the analysis}\label{sec:fichet-analysis}

We use a modified version of SUSPECT2.41 \cite{Djouadi:2002ze} as spectrum generator, 
and MICROMEGAS \cite{Belanger:2004yn,Belanger:2006is} for computing additional 
observables. 
While there exist specialized fitting tools like SFITTER \cite{Lafaye:2004cn} 
and FITTINO~\cite{Bechtle:2004pc}, these are not directly applicable to the 
SGHU model for various reasons. We have therefore programed our own 
MCMC analysis with a Metropolis algorithm, largely following the procedure 
of \cite{Belanger:2009ti}. Below we explain some details which are specific to 
our analysis. 

\subsection{Reference scenario and assumed measurements}

As reference scenario we use the SGHU point D of~\cite{Brummer:2009ug}. 
The (s)particle masses that are accessible to LHC measurements are shown in 
Table~\ref{tab:susyghuscenario}. 
Since at present no experimental simulation is available 
for this scenario, we simply assume that the sparticle masses can be extracted 
from invariant-mass distributions (following, e.g.,~\cite{Cheng:2008mg,Cheng:2009fw})
with 3\% accuracy. This is in agreement with the discussions in the ``Spins and Masses'' 
subgroup at this Workshop. We keep the input $m_t$ used in \cite{Brummer:2009ug}
and assume that it will be measured to 1~GeV accuracy at the LHC. The error on the 
top quark mass feeds into a parametric uncertainty on $m_h$; we therefore take  
$m_t=172.4\pm 1$~GeV and $m_h=117.3\pm 1$~GeV in our fits, assuming that other theoretical 
uncertainties on $m_h$ will be under control by the time these measurements become available.
Finally, we consider two cases: hypothesis H0 without measurement of 
the heavy Higgs sector and hypothesis H1 with measurement of the heavy Higgs 
sector. Throughout the analysis, we demand that the ${\tilde\chi^0_1}$ is the lightest SUSY particle, LSP.

Some more comments are in order. First, $m_{\tilde q}$ in Table~\ref{tab:susyghuscenario} 
is the average mass of the 1st and 2nd generation squarks, 
$m_{\tilde{q}}=\frac{1}{4}(m_{\tilde{u}_{L}}+m_{\tilde{u}_{R}}+m_{\tilde{d}_{L}}+m_{\tilde{d}_{R}})$.
Second, a priori we cannot know the chirality of the slepton in the 
$\tilde\chi^0_2 \to \ell^\pm\tilde\ell^\mp\to \ell^\pm\ell^\mp\tilde\chi^0_1$ decay chain:
the extracted slepton mass $m_{\tilde\ell}\equiv m_{\tilde e}=m_{\tilde\mu}$  
is the mass of either the left- or the right-chiral selectron/smuon, depending on the 
mass ordering with respect to $\tilde{\chi}^0_2$. 
If $m_{\tilde\ell_L}<m_{\tilde{\chi}_{2}^{0}}$ the wino-like $\tilde{\chi}_{2}^{0}$ decays 
mainly into $\ell\tilde\ell_{L}$ even if $m_{\tilde\ell_R}<m_{\tilde\ell_L}$, and it is 
$m_{\tilde\ell_L}$ that is measured. This is in fact the case at our reference 
point, which has $m_{\tilde\ell_R}=217$~GeV and $m_{\tilde\ell_L}=327$~GeV. 
If, however, $m_{\tilde\ell_R}<m_{\tilde\chi^0_2}<m_{\tilde\ell_L}$, then 
$\tilde\chi^0_2\to\ell\tilde\ell_R$, and what is measured is  $m_{\tilde\ell_R}$. 
This is typically the case in mSUGRA.  
In the MCMC scans we therefore take $m_{\tilde\ell}=326.8\pm 9.8$~GeV as 
being $m_{\tilde\ell_L}$ or $m_{\tilde\ell_R}$ depending on the mass ordering 
at a particular parameter point. 
Third, we note that at point~D both staus are heavier than the $\tilde\chi^0_2$ and 
hence do not appear in the decay chains. This is neglected in this simple study; in a 
more sophisticated analysis, however, one should take the absence of a $\tau^+\tau^-$ 
edge into account.

\begin{table}
\begin{center}
\begin{tabular}{|c|c|c|c|c|c||c|}
\hline
$m_{\tilde\chi^0_1}$ & $m_{\tilde\ell}$ & $m_{\tilde\chi^0_2}$ & $m_{\tilde q}$ & $m_{\tilde g}$ & $m_h$ & $m_H$ \\ \hline
{\small $208.7\pm 6.3$} & {\small $326.8\pm9.8$} & {\small $400.4\pm12$} & {\small $1022\pm30.7$} & {\small $1155\pm34.7$} & {\small $117.3\pm1$} & {\small $637.4\pm19.1$} \\
\hline
\end{tabular}
\end{center}
\caption{Masses (in GeV) accessible to LHC measurements at the SGHU point~D, 
and assumed experimental errors. We consider two cases: case H0 without measurement of 
$m_H$, and case H1 with measurement of $m_H$.}
\label{tab:susyghuscenario}
\end{table}

We do not include constraints from B-physics observables nor the dark matter relic density 
in the fit, but use them only {\it a posteriori}. The nominal values at point D are 
BR$(b\to s\gamma)=2.89\times 10^{-4}$, 
BR$(B_s\to \mu^+\mu^-)=5.76\times 10^{-9}$, 
and $\Omega h^2=0.108$. 

\subsection{Model parameters}

The familiar mSUGRA model depends on four continuous 
parameters ---$\tan\beta$,  
the universal gaugino mass $M_{1/2}$, 
the universal scalar mass parameter $M_0$ and 
the universal trilinear coupling $A_0$ (the latter three being input at $M_{\rm GUT}$)--- 
and the sign of $\mu$.

The SGHU model also depends on $\tan\beta$, $M_{1/2}$ and sign($\mu$). 
The boundary conditions for the Higgs and scalar sectors are, however,  
considerably different from the mSUGRA case. 
First of all, the Higgs 
soft terms are fixed by the SGHU relation 
\begin{equation}
  m_{H_{1,2}}^2=\epsilon_H B\mu-|\mu|^2
  \label{eq:susyghucondition}
\end{equation}
at $M_{\rm GUT}$, with $\epsilon_H=\pm1$; this is computed iteratively in our 
modified SUSPECT version~\cite{Brummer:2009ug}.
Moreover, the soft terms of the first and second generation sfermions vanish at $M_{\rm GUT}$, 
while those of the third generation are non-zero and non-universal. In the full model developed in~\cite{Brummer:2009ug}, 
the third generation soft terms depend on the GUT-scale Yukawa couplings and two bulk mixing angles,   
and are computed in our modified SUSPECT version using an additional level of iteration. 
This procedure being very time consuming, we do not consider the complete model here, but simply let 
the third-generation scalar soft-terms vary independently.
The cost of this is a larger number of free parameters, which will have repercussions on the Bayes factor, 
as explained in Section~\ref{sec:bayes_factor}. On the other hand, this approach 
is less dependent on the model building of the matter sector.

The parameters to be fitted to the data are hence:
\begin{eqnarray}
  {\rm mSUGRA:} && \tan\beta,~M_{1/2},~M_0,~A_0\\
  {\rm SGHU:} && \tan\beta,~M_{1/2},~M_{Q_3},~M_{U_3},~M_{D_3},~A_t,~A_b, 
  ~M_{L_3},~M_{E_3},~A_\tau
\end{eqnarray}
We take $\mu>0$ throughout, and $\epsilon_H=-1$ in the SGHU case. 
Generally, both signs of $\mu$ and all sign combinations of $\mu$ and $\epsilon_H$ should be investigated,  
but this is not possible here because of CPU limitations. The choice of $\epsilon_H=-1$ is, however, justified
because, as we will see, in the mSUGRA case we find large negative $A_0$, dominated by the effect of $A_t$. 
In the SGHU case, we know from \cite{Brummer:2009ug} that only one sign combination of $\epsilon_H$ and 
$A_t$ gives acceptable phenomenology.

An important difference between mSUGRA and SGHU lies  
in the gaugino and slepton mass ratios. The gaugino masses are determined by 
$M_{1/2}$ in both models. The slepton masses, however, are driven by $M_0$ 
in the mSUGRA case, while in the SGHU case they are driven by $M_{1/2}$ 
and the U(1)$_Y$ D-term contribution from the $S$ parameter, 
$S=(m_{H_2}^2-m_{H_1}^2) + {\rm Tr}(m_{Q}^2 - 2m_{U}^2 + m_{D}^2+m_{R}^2-m_{L}^2)$. 
Roughly,  
$m_{\tilde\chi^0_1}\approx 0.43\,M_{1/2}$, 
$m_{\tilde\chi^0_2}\approx 0.83\,M_{1/2}$, 
$m_{\tilde e_R}^2\approx M_0^2+(0.39\,M_{1/2})^2 - 0.052\,S_{\rm GUT}$, and 
$m_{\tilde e_L}^2\approx M_0^2+ (0.68\,M_{1/2})^2 + 0.026\,S_{\rm GUT}$, 
where $S_{\rm GUT}$ is the value of $S$ at $M_{\rm GUT}$. 
Note that $S_{\rm GUT}\equiv 0$ in mSUGRA, while $M_0\equiv 0$ in SGHU. 
From this we can already estimate $M_{1/2}\approx 500$~GeV in both models, 
$M_0\approx 260$~GeV in the mSUGRA case, and 
$S_{\rm GUT}\approx -(280~\rm GeV)^2$ in the SGHU case.
Moreover, from these considerations we expect the mass ordering 
$m_{\tilde\ell_R}<m_{\tilde\chi^0_2}<m_{\tilde\ell_L}$ in mSUGRA, but 
$m_{\tilde\ell_R}<m_{\tilde\ell_L}<m_{\tilde\chi^0_2}$ in SGHU.
 
Another important difference lies in the higgsino and heavy Higgs masses. 
Since $\mu\sim 1300$~GeV at point D, the higgsino states are not accessible 
at LHC. The heavy Higgs masses, however, are around $640$~GeV, which 
might be within reach.  
In order to test the discriminating power of the heavy Higgs sector, 
we perform fits without and with including a measurement of one of the 
heavy Higgs masses. We here use the mass of $H^0$, but taking instead $m_A$ or $m_{H^\pm}$ 
is completely equivalent.  

Regarding parameter ranges, since $m_{\tilde g}\simeq1150$~GeV, 
we vary $M_{1/2}$ in $\left[0,\,1000\right]$~GeV only.  
In SGHU, the scalar mass parameters are allowed to vary within $\left[0,\,2000\right]$~GeV.
The $A$ terms are allowed to vary within specific ranges, which contain the parameter space of the full SGHU model:
$A_t=\left[-2600,1400\right]$~GeV, $A_b=\left[-3200,200\right]$~GeV, $A_\tau=\left[-3200,1200\right]$~GeV. 
In the mSUGRA case, the scalar masses and $A_0\equiv A_t=A_b=A_\tau$ are allowed to vary without bounds. 
We do not constrain $\tan\beta$. A posteriori, it does
not exceed $60$ due to theoretical constraints from tachions and color or charge breaking. 
Last but not least, we use flat priors for all model parameters. For a discussion of prior (in)dependence 
in the presence of LHC data, see \cite{Roszkowski:2009ye}.

\subsection{Likelihoods}

In the likelihood function, all measurements are taken into account
as gaussians proportional to $\exp\left(-\left(x_{th}- x_{exp} \right)^{2}/\sigma_{exp}^2\right)$. 
Here $x_{exp}$ and $\sigma_{exp}$ are the nominal value and assumed experimental 
error as given in Table~\ref{tab:susyghuscenario}, and $x_{th}$ is the prediction 
at a given parameter point.
The global level of convergence of the Markov chains is evaluated using the
procedure described in \cite{Allanach:2005kz}. 
For the parameters which give the maximum likelihood in each case, we evaluate the 
68\% and 95\% Bayesian Credibility (BC) intervals, using the full likelihood. 
If the maximum likelihood is constrained by gaussian measurements, these 
correspond to the usual $1\sigma$ and $3\sigma$ confidence intervals. 
We also evaluate the 68\% and 95\% BC regions from the 2D marginalized distributions.

\subsection{Bayes factor\label{sec:bayes_factor}}

The Bayes factor is defined as the ratio of the posterior probability
of two models given a set of data: 
\begin{equation}
   \mathcal{K}=P(\mathcal{M}_{1}|\textrm{data})/P(\mathcal{M}_{2}|\textrm{data}).
\end{equation}
Assuming that both models have the same global probability, $P(\mathcal{M}_{1})=P(\mathcal{M}_{2})$, 
to describe reality, this ratio is reduced to the ratio of global
likelihoods: $\mathcal{K}=P(\textrm{data}|\mathcal{M}_{1})/P(\textrm{data}|\mathcal{M}_{2})$.

There is, however, a subtlety: assuming that a set $M$ of data is measured
implies that the discovery $D$ is already done: $\textrm{data}=M\cap D$.
This implies that $\mathcal{K}=P(M|\mathcal{M}_{1}\cap D_{1})P(D_{1}|\mathcal{M}_{1})/P(M|\mathcal{M}_{2}\cap D_{2})P(D_{2}|\mathcal{M}_{2})$.
Here $P(D|\mathcal{M})$ is the probability to make a discovery assuming
the model $\mathcal{M}$, i.e.\ the potential of discovery of $\mathcal{M}$.
For a supersymmetric model at the LHC, we can consider this is roughly
equal to $P(M_{1/2}<1-2~\textrm{TeV})$. In the particular case
we study, as we compare two supersymmetric models, this ratio cancels.
The likelihood $P(M|\mathcal{M}\bigcap D)$ becomes equal to 
$\int\mathcal{L}(M,\theta_{i})P(\theta_{i}\cap D)d\theta_{i}$
where the $\theta_{i}$ are the parameters of the model. By taking
flat internal priors on the parameters, this reduces to the integral
of the likelihood over the volume of the parameter space $V_{D}$ allowing
the discovery: $P(M|\mathcal{M}\cap D)=\intop^{V_{D}}\mathcal{L}(M,\theta_{j})d\theta_{j}$.
Outside of this volume, the likelihood must be considered as null.
In our case, the Bayes factor is therefore simply reduced to the ratio of the
two average likelihoods, computed on the discovery volume: 
\begin{equation}
    \mathcal{K}=\left\langle \mathcal{L}_{1}\right\rangle /\left\langle \mathcal{L}_{2}\right\rangle =\overset{N_{1}}{\underset{i=1}{\sum}}\mathcal{L}_{1}(x_{1}^{(i)})/\overset{N_{2}}{\underset{i=1}{\sum}}\mathcal{L}_{2}(x_{2}^{(i)}).
\end{equation}
where the sums are over the points of Markov Chains. For two models to be discriminated, 
the Bayes factor should be at least around 3 (30) to constitute a weak (strong) evidence. 
A Bayes factor larger than 100 is considered as a decisive evidence.   

It is important to note that the Bayes factor favorizes models with small number of parameters.
This implies that the SGHU model with independant scalar soft terms we consider here should 
be less favored than the complete one with only two mixing angles.
A detailed discussion of the Bayes factor can be found in, e.g., \cite{AbdusSalam:2009tr}.

\section{Results}\label{sec:fichet-results}

\begin{figure}[ht!]\centering
\includegraphics[width=\textwidth]{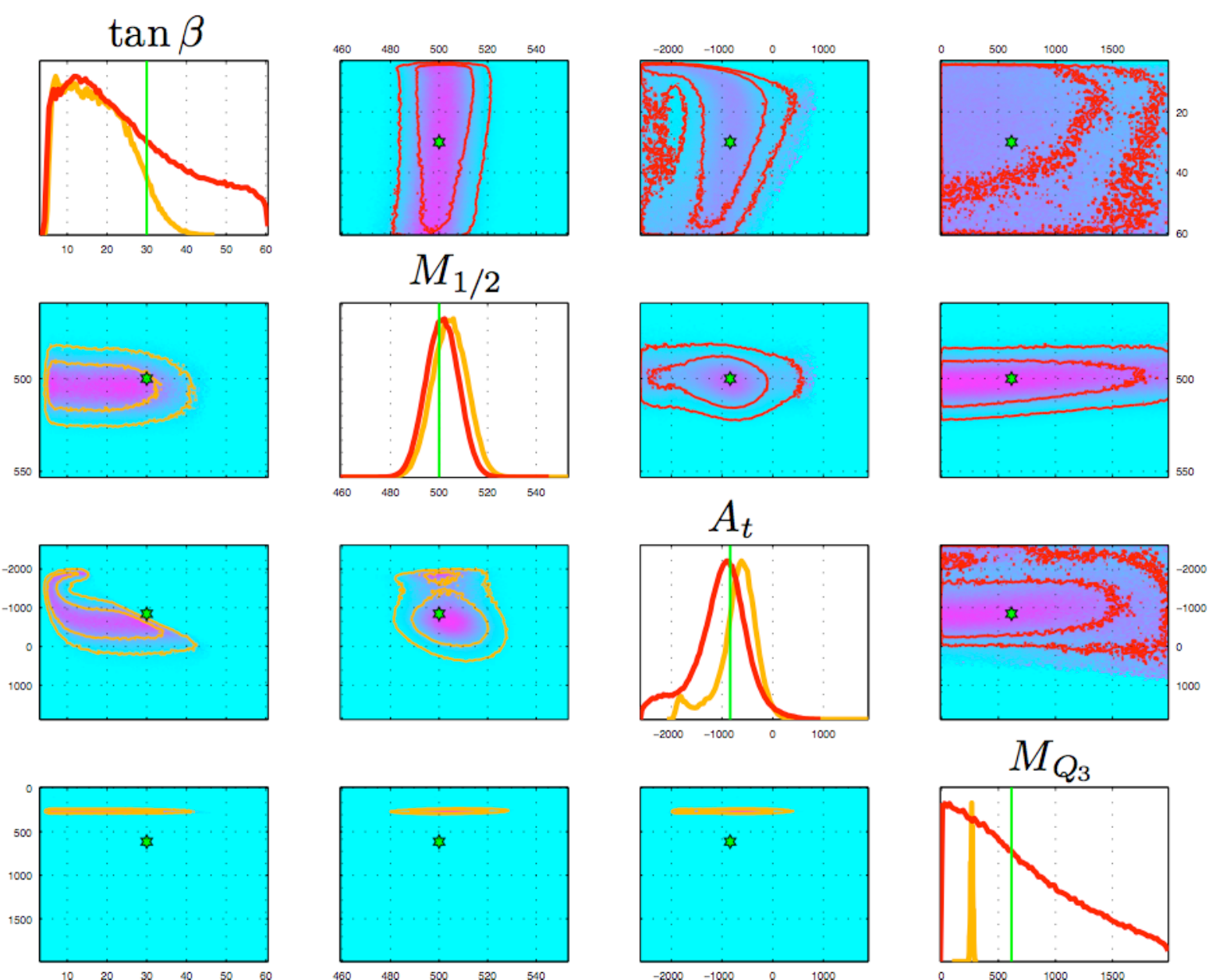}
\caption{Marginalized likelihood distributions in 1D and 2D for the 
mSUGRA (orange) and SGHU (red) models in the H0 hypothesis. 
In the mSUGRA case, $A_0\equiv A_t$ and $M_0\equiv M_{Q_3}$. 
The plots on the diagonal show the 1D likelihoods of both models, normalized 
to have the same maximum. The off-diagonal plots show iso-contours of  
68\% and 95\% BC, computed within the 2D marginalized likelihood. 
The upper triangle of 2D plots is the SGHU case, while the lower triangle 
is the mSUGRA case. The color maps indicate the empirically averaged likelihoods. 
The axes of the 2D plots are shown on the outer boundary of the figure.
The green lines/stars indicate the nominal values of point~D.}
\label{fig:susyghu-dist-H0}
\end{figure}

\begin{figure}[ht!]\centering
\includegraphics[width=\textwidth]{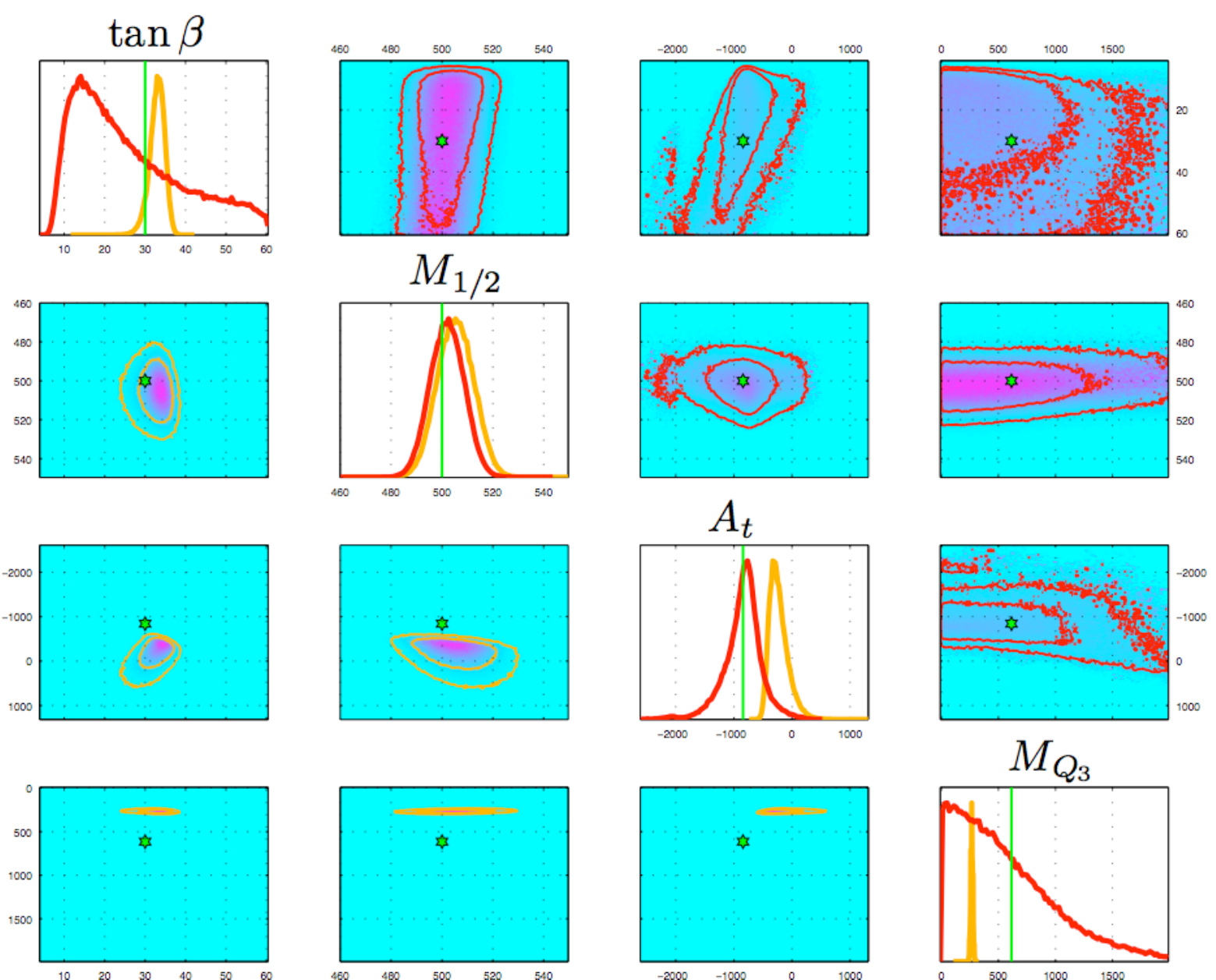}
\caption{Same as Fig.~\ref{fig:susyghu-dist-H0} but for the H1 hypothesis.}
\label{fig:susyghu-dist-H1}
\end{figure}

In this section, we present the results of MCMC scans which collected around $10^6$ 
points for each case, i.e.\ for each of the two hypotheses in the two models.
Figure~\ref{fig:susyghu-dist-H0} shows 1D and 2D marginalized likelihoods 
for the mSUGRA and SGHU model parameters under the H0 hypothesis 
(no measurement of heavy Higgses). The marginalized likelihoods for the 
H1 hypothesis (assumed measurement of $m_H$) are shown in Fig.~\ref{fig:susyghu-dist-H1}.
In both figures, the 2D marginalized likelihoods are plotted as isolines corresponding 
to 68\% and 95\% BC regions. The colored 2D maps correspond
to the empirical averages of the sampled likelihoods. They have only 
indicative value, to show what the zones of high likelihood are,
independent of the volume effect which is taken into account in the true
marginalization. 
We recall that the 68\% (95\%) BC intervals are defined by the hypersurface enclosing
68\% (95\%) of the integral likelihood around the maximum.
When this limit is identical to the boundary of the scan, this means that the distribution is too flat
to give a prefered value with 68\% (95\%) credibility.

We see from Figs.~\ref{fig:susyghu-dist-H0} and~\ref{fig:susyghu-dist-H1} that in the mSUGRA 
case $M_{1/2}$, $A_0$ and $M_0$
and in the SGHU case $M_{1/2}$ and $A_t$ are well constrained, but the other parameters 
are not. We also note a considerable tightening of the correlations between 
$\tan\beta$, $M_{1/2}$ and $A_{0(t)}$ when information on the heavy Higgs sector is added.
In particular, a measurement of $m_H$ very much constrains $\tan\beta$ in the mSUGRA case, 
with the fitted value being in fact quite close to the ``true'' one, see Fig.~\ref{fig:susyghu-dist-H1}.
In the SGHU case, on the other hand, $\tan\beta$ is much less constrained. 

The values of maximal and averaged likelihoods and convergence parameter $r$ are 
given in Table~\ref{tab:susyghu-likes}.
In the H0 hypothesis, both models fit the data very well without preference for the one or the other, 
the maximum likelihoods as well as the Bayes factor being close to one.
This is in fact only little different in the H1 hypothesis: the mSUGRA fit still gives a high 
${\cal L}_{max}\simeq 0.7$, and the Bayes factor is of order 2, {\it i.e.}\ 
not sufficiently large to constitute an evidence.
In order to separate the effect of the ``pure'' SGHU condition eq.~(\ref{eq:susyghucondition}) 
from that of the non-universal sfermion soft terms,    
we also performed a fit for a SGHU model variant with universal $M_0$ and $A_0$
for all three generations (in other words, mSUGRA supplemented  by  eq.~(\ref{eq:susyghucondition})).
In this case, we find ${\cal L}_{max}= 0.992$ and $\langle {\cal L}\rangle=0.182$ in the H1 hypothesis, 
that means a Bayes factor of ${\cal K}\approx 2$ with respect to strict mSUGRA, and 
${\cal K}\approx 1$ w.r.t.\ SGHU with 10 free parameters. So the small preference of 
SGHU over mSUGRA in the H1 case comes indeed from the degeneracy of  
the Higgs soft terms, $m_1^2=m_2^2=m_3^2$ at $M_{\rm GUT}$ 
($m_{1,2}^2=m_{H_{1,2}}^2+|\mu|^2$, $m_3^2=|B\mu|$). 

The 68\% and 95\% Bayesian credibility intervals (BCIs) for the model parameters 
are given explicitly in Table~\ref{tab:susyghu-msugrafit} for mSUGRA 
and in Table~\ref{tab:susyghu-ghufit} for SGHU. For comparison, 
the input values at point~D are: $\tan\beta=30$, $M_{1/2}=500$~GeV, 
$M_{Q_3}=614$~GeV, $M_{U_3}=635$~GeV, $M_{D_3}=414$~GeV, 
$A_t=-842$~GeV, $A_b=-966$~GeV, $M_{L_3}=408$~GeV, $M_{E_3}=433$~GeV, $A_\tau=-1070$~GeV.

\begin{figure}[t]\centering
\includegraphics[width=0.8\textwidth]{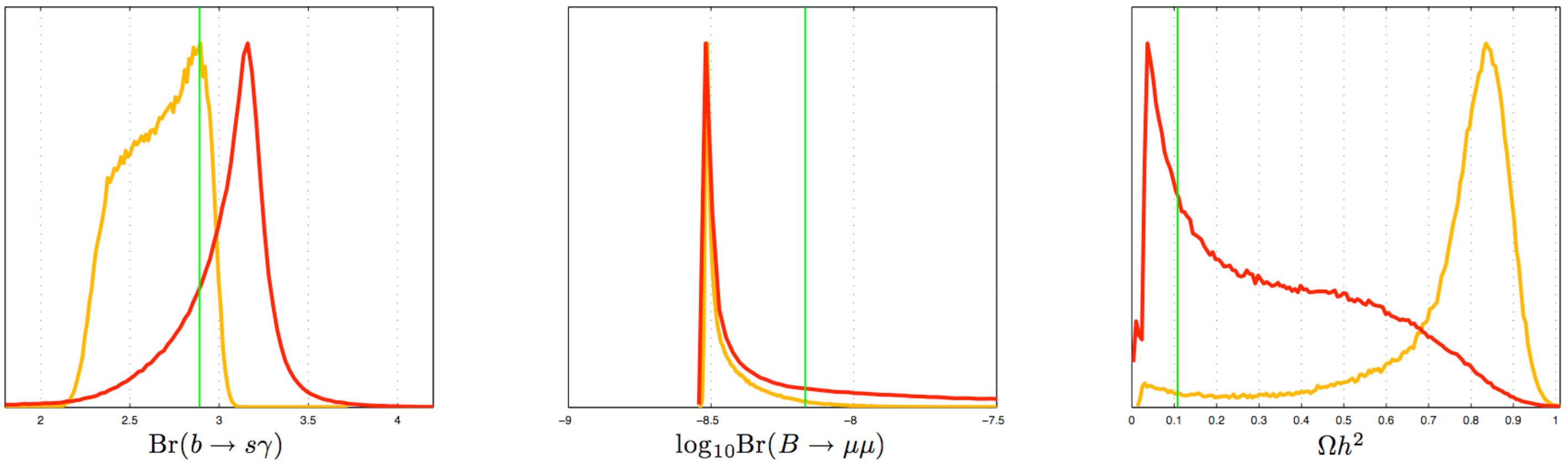}
\includegraphics[width=0.8\textwidth]{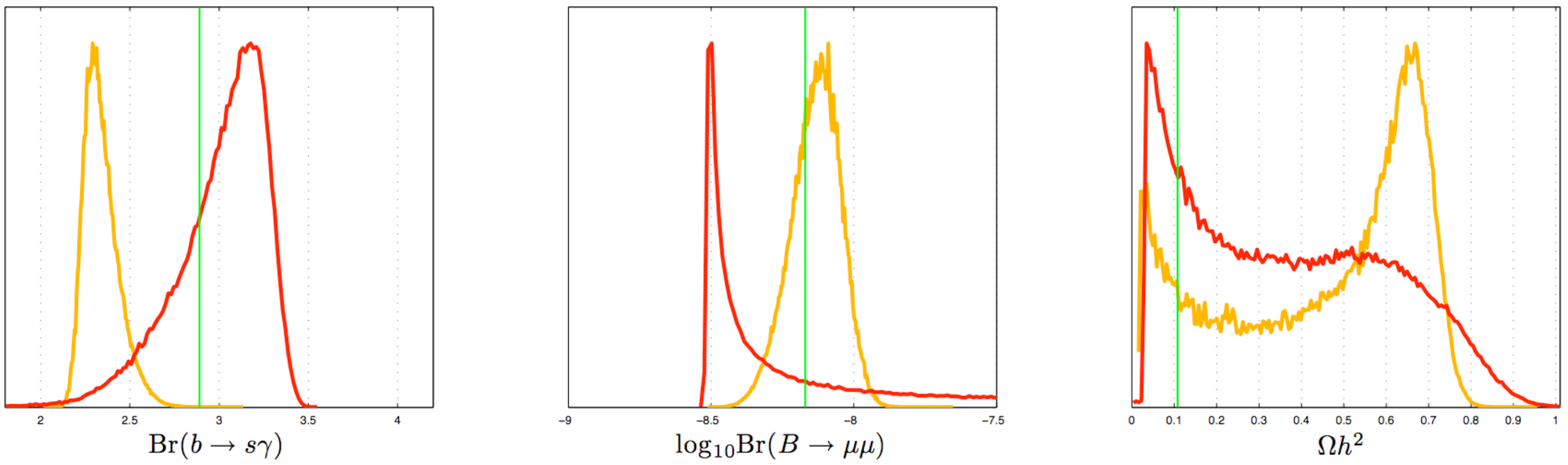}
\caption{Marginalized likelihood distributions in 1D for indirect observables predicted in the
mSUGRA (orange) and SGHU (red) models; the upper row of plots is for the H0, 
and the lower row for the H1 hypothesis.
The green lines indicate the nominal values at the reference point~D.}
\label{fig:susyghu-indirect}
\end{figure}

We next ask whether indirect observables can help discriminate the two models. 
To this aim, Fig.~\ref{fig:susyghu-indirect} shows the 1D marginalized distributions  
for $Br(b\to s\gamma)$, $Br(B_s\to \mu^+\mu^-)$, and the neutralino relic density $\Omega h^2$
as obtained from the mSUGRA and SGHU fits. 
The 68\% and 95\% BCIs are given explicitly in Table~\ref{tab:susyghu-indirect}. 
We see that  the B-physics observables have a good discriminating power in case 
the heavy Higgs sector is known (H1 hypothesis), but not so in the H0 hypothesis.
Regarding the relic density, we note that the mSUGRA model predicts a much too 
large $\Omega h^2\sim 0.6$--$0.9$ at 68\% BC if the heavy Higgs sector is unconstrained. 
In the H1 case, when $m_H$ (and hence $m_A$ and $\tan\beta$) are fixed, then 
the $\Omega h^2$ prediction within mSUGRA also gives smaller values in agreement 
with WMAP observations. This is different for the SGHU model, for which 
$\Omega h^2$ peaks towards values smaller than $\sim 0.1$. However, the distribution 
is rather flat and when considering the 68\% or 95\% BCIs, no definite conclusion can be 
obtained, see Table~\ref{tab:susyghu-indirect}. 

\begin{table}[h!]\centering
\begin{tabular}{|c|c|c|c|}
\cline{2-2} \cline{3-3} \cline{4-4} 
\multicolumn{1}{c|}{} & $\mathcal{L}_{max}$ & $\langle{\mathcal{L}}\rangle$ & $r$ \tabularnewline
\cline{2-2} \cline{3-3} \cline{4-4} 
\hline 
mSUGRA H0 & 0.984 & 0.200 & 1.0037\tabularnewline
\hline 
mSUGRA H1 & 0.742 & 0.080 & 1.0058\tabularnewline
\hline
\end{tabular}\quad
\begin{tabular}{|c|c|c|c|}
\cline{2-2} \cline{3-3} \cline{4-4} 
\multicolumn{1}{c|}{} & $\mathcal{L}_{max}$ & $\langle{\mathcal{L}}\rangle$ & $r$ \tabularnewline
\cline{2-2} \cline{3-3} \cline{4-4} 
\hline 
SGHU H0 & 0.995 & 0.221 & 1.0064 \tabularnewline
\hline 
SGHU H1 & 0.995 & 0.166 & 1.0065 \tabularnewline
\hline
\end{tabular}
\caption{Values of the maximum and averaged likelihoods, and of the convergence
parameter $r$.}\label{tab:susyghu-likes}
\end{table}
  
\begin{table}[h!]\centering
\begin{tabular}{|c|cc|cc|}
\cline{2-2} \cline{3-3} \cline{4-5} 
\multicolumn{1}{c|}{}  & \multicolumn{2}{c|}{mSUGRA H0} & \multicolumn{2}{c|}{mSUGRA H1} \\
\multicolumn{1}{c|}{}  & 68\% BCI & 95\% BCI & 68\% BCI & 95\% BCI  \\
\hline       
$\tan\beta$ & [9, 27] & [6, 36] & [29, 35] & [25,37]  \\
$M_{1/2}$ & [495, 515] & [485, 525] & [496, 516] & [487, 526]  \\
$M_0$ & [252, 280] & [239, 292] &  [252, 280] & [239, 292]  \\
$A_0$ & [$-$1065, $-$197] & [$-$1065, 200] &  [$-$338, 145] &  [$-$468, 500]  \\
\hline
\end{tabular}
\caption{68\% and 95\% Bayesian credibility intervals (BCIs) for the mSUGRA parameters 
in the H0 and H1 hypotheses.}\label{tab:susyghu-msugrafit}
\end{table}
 
\begin{table}[h!]\centering
\begin{tabular}{|c|cc|cc|}
\cline{2-2} \cline{3-3} \cline{4-5} 
\multicolumn{1}{c|}{}  & \multicolumn{2}{c|}{SGHU H0} & \multicolumn{2}{c|}{SGHU H1} \\
\multicolumn{1}{c|}{}  & 68\% BCI & 95\% BCI & 68\% BCI & 95\% BCI  \\
\hline       
$\tan\beta$ & [4, 43] & [4, 57] & [13,42] & [8,56]  \\
$M_{1/2}$ & [493, 512] & [484, 520] & [494, 512] & [485,521]  \\
$M_{Q_3}$ & [1, 1341] & [1, 1837] &  [0, 1093] & [0, 1689]  \\
$M_{U_3}$ & [3, 1413] & [3, 1766] & [2, 1257] & [2, 1626] \\
$A_t$ & [$-$1309, $-$773] & [$-$2215, $-$120] & [$-$975, $-$687] & [$-$1522, $-$267] \\
\hline
\end{tabular}
\caption{68\% and 95\% BCIs for SGHU parameters in the H0 and H1 hypotheses.
The limits for $M_{D_3,L_3,E_3}$ are very similar to those for $M_{Q_3,U_3}$.  
There are no reasonable limits for $A_{b,\tau}$. }\label{tab:susyghu-ghufit}
\end{table}

\begin{table}[h!]\centering
\begin{tabular}{|c|c|c|c|}
\cline{2-2} \cline{3-3} \cline{4-4} 
\multicolumn{1}{c|}{} & $Br(b\to s\gamma)\times10^{4}$ & $\log_{10}(Br(B_s\to \mu^{-}\mu^{+}))$ & $\Omega h^{2}$\tabularnewline
\hline 
mSUGRA H0 & {\small $[2.42,\,2.90],[2.26,\,3.02]$ } & {\small $[-8.5,\,-8.3],[-8.5,\,-8.0]$} & {\small $[0.60,\,0.89],[0.11,\,0.96]$} \tabularnewline
\hline 
mSUGRA H1 & {\small $[2.28,\,2.56],[2.19,\,2.75]$} & {\small $[-8.3\,,-8.1],[-8.3,\,-8.0]$} & {\small $[0.02,\,0.72],[0.02,\,0.79]$}\tabularnewline
\hline
SGHU H0 & {\small $[2.78,\,3.26],[2.24,\,3.55]$} & {\small $[-8.5,\,-7.6],[-8.5,\,-6.5]$} & {\small $[0.01,\,0.71],[0.01,\,0.90]$}
\tabularnewline
\hline 
SGHU H1 & {\small $[2.72,\,3.27],[2.26,\,3.39]$} & {\small $[-8.5,\,-7.7],[-8.5,\,-7.0]$} & {\small $[0.03,\,0.75],[0.03,\,0.91]$}
\tabularnewline
\hline 
\end{tabular}\caption{68\% and 95\% BCIs intervals of predicted indirect observables: 
$Br(b\to s\gamma)$, $Br(B_s\to \mu^+\mu^-)$, and  $\Omega h^2$.}\label{tab:susyghu-indirect}
\end{table}

\begin{figure}[h!]
\centering
\includegraphics[width=\textwidth]{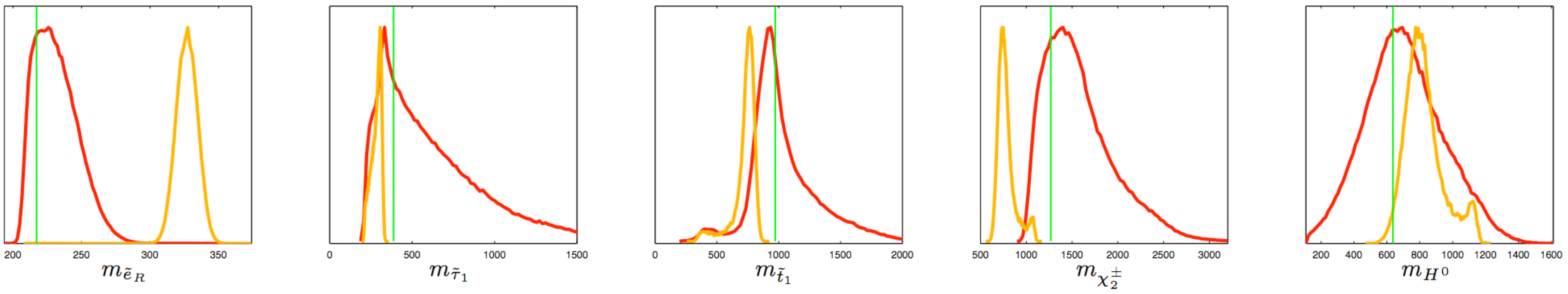}
\includegraphics[width=0.8\textwidth]{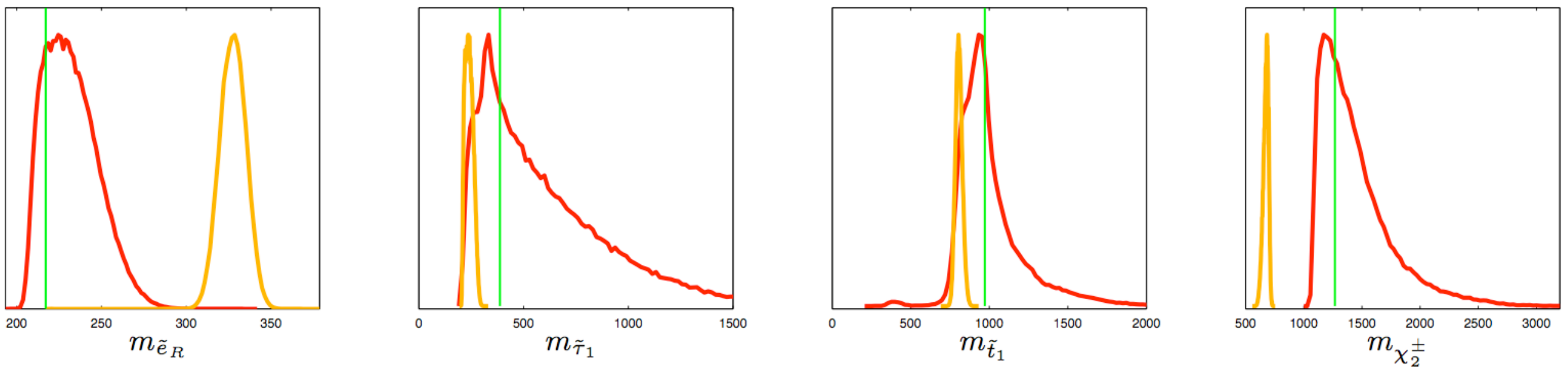}
\caption{Marginalized likelihood distributions in 1D for some predicted masses for
mSUGRA (orange) and SGHU (red) models in the H0 (upper row) and H1 (lower row) hypotheses.
The green lines indicate the nominal values at the reference point~D.}
\label{fig:susyghu-masses}
\end{figure}

Obviously, improving the model discrimination requires the measurement of additional parts 
of the mass spectrum. 
To this end, we show in Fig.~\ref{fig:susyghu-masses} 
the 1D marginalized likelihood distributions for some predicted masses, in particular the  masses of 
$\tilde e_R$, $\tilde\tau_1$, $\tilde t_1$, and $\tilde\chi^\pm_2$. 
As expected, a very good discrimination would be obtained by measuring the $\tilde e_R$  
mass (note that the posterior distributions for $m_{\tilde e_R}$ do not overlap). 
Measurement of the $\tilde\tau_1$ and/or $\tilde t_1$ masses would help reveal the
non-universality of the scalar soft terms.  A powerful test in particular of the SGHU 
condition eq.~(\ref{eq:susyghucondition})) would be the determination of the $\mu$ parameter 
through a measurement 
of the higgsino sector: the distributions for $m_{\tilde\chi^\pm_2}$ hardly overlap 
in the H0 case and do not overlap at all in the H1 case. 
All this may best be done at an $e^+e^-$ linear collider with high enough centre-of-mass energy.
Nevertheless, at the LHC a first hint for a non-universal structure may be obtained from 
the absence of a kinematic endpoint in the $\tau^+\tau^-$ invariant-mass distribution, since 
in the mSUGRA case we typically have 
$m_{\tilde\tau_1}<m_{\tilde e_{R}}<m_{\tilde\chi^0_2}<m_{\tilde e_{L}}$. 
Indeed, in the mSUGRA fit, 
$\tilde\chi^0_2\to \tau^\pm\tilde\tau_1^\mp$ 
typically has about 80--90\% branching ratio, followed by $\tilde\chi^0_2\to h^0\tilde\chi^0_1$ 
as the next-important channel, while $\tilde\chi^0_2\to e^\pm\tilde e_R^\mp$ often has a 
branching ratio below 1\%. 

Before concluding, we recall that in the complete SGHU model in \cite{Brummer:2009ug}, 
where the third generation soft terms are computed from two bulk mixing angles, 
$M_{Q_3}$, $M_{U_3}$, $M_{D_3}$, $A_t$ and $A_b$  are not independent of  
each other. Therefore the SGHU distributions in Fig.~\ref{fig:susyghu-masses} 
will be a bit narrower in the complete model than in the more general version 
presented here.

\section{Conclusions}

We investigated whether a sparticle spectrum arising from SGHU 
can be discriminated against the mSUGRA model by LHC measurements. 
To this end we performed MCMC fits of the two models to assumed LHC data for 
a particular SGHU benchmark point, which is characterized by GUT-scale degenerate 
Higgs mass parameters and non-universal third-generation soft terms. 

It turned out that the mSUGRA model can fit the anticipated LHC data well; 
a measurement of the $\tilde\chi^0_1$, $\tilde\chi^0_2$, $\tilde e$, $\tilde g$ and $h^0$ masses 
(with percent-level precision) is not sufficient to discriminate the  structure of 
the underlying model. Also the Bayes factor does not allow to favour the SGHU 
model over mSUGRA. This does not change significantly if information on the heavy Higgs 
sector is included. However, information on the heavy Higgs sector in combination with improved 
B-physics constraints would significantly influence the fits.

A decisive model discrimination would be possible through a measurement of 
the $\tilde e_R$ mass in $e^+e^-$ collisions (together with refined measurements of the 
rest of the spectrum). Besides, a measurement of the higgsino mass 
should provide a test of the SGHU condition $m_1^2=m_2^2=m_3^2$ at $M_{\rm GUT}$.  
Accurate  measurements of the sparticle spectrum in $e^+e^-$ should also allow to determine 
the neutralino relic density with good precision.

Last but not least we note that our analysis is based on assumed LHC measurements 
of absolute masses. It should be possible to improve the fits by including more information, 
{\it e.g.}\ the positions of kinematic endpoints and event rates. Moreover, a lower limit on 
the $\tilde\tau_1$ mass from the absence of a $\tau^+\tau^-$ signal would considerably 
impact the results obtained here.
How well this can be done should be subject to further investigation.

\section*{Acknowledgements}
We are grateful to Ritesh~K.~Singh for inspiring discussions about Markov Chains, 
which triggered this analysis. We also thank Michael Rauch for comparisons 
of the mSUGRA case with SFITTER. 

%% file: Kraml/cpvmcmc.tex
\chapter{MCMC Analysis of the MSSM with arbitrary CP phases}

{\it G.~Belanger, S.~Kraml, A.~Pukhov and R.K.~Singh}

\begin{abstract}
We explore the parameter space of the MSSM with explicit CP-violating (CPV)
phases by means of a Markov Chain Monte Carlo analysis, 
imposing constraints from direct Higgs and SUSY searches at colliders,  
B-physics, EDM measurements, and the relic density of dark matter.
We find that over most of the parameter space, large phases are 
compatible with experimental data. We present likelihood maps 
of the CPV-MSSM, concentrating in particular on 
quantities relevant for the neutralino relic density.
\end{abstract}

\section{Introduction}

It was noted early on~\cite{Goldberg:1983nd,Ellis:1983ew} that a neutralino LSP in the MSSM 
with conserved R-parity is an excellent cold dark matter candidate. 
Detailed studies showed that in the MSSM, or constrained versions thereof, there are 
several mechanisms that provide the correct rate of neutralino
annihilation, such that $\Omega h^2\simeq 0.1$: 
annihilation of a bino LSP into fermion pairs through $t$-channel
sfermion exchange in case of very light sparticles;
annihilation of a mixed bino-higgsino or bino-wino LSP into gauge
boson pairs through $t$-channel chargino and neutralino exchange,
and into top-quark pairs through $s$-channel $Z$ exchange;
and finally annihilation near a Higgs resonance (the so-called Higgs funnel).
Furthermore, coannihilation processes with sparticles that are close
in mass with the LSP may bring $\Omega h^2$ in the desired range.
This way, the measured relic density of dark matter is often used to severely 
constrain the MSSM parameter space. 

In \cite{Belanger:2009ti}, some of us explored the parameter space 
of the phenomenological MSSM that is allowed when requiring that the 
neutralino LSP constitutes all the dark matter by means of a Markov Chain 
Monte Carlo (MCMC) scan.This was done for the 
case of seven free parameters, where it was assumed  that there 
are no new sources of CP violation beyond the CKM.\footnote{An analogous 
analysis of the phenomenological MSSM with 25 free parameters was performed 
in \cite{AbdusSalam:2009qd} employing a MultiNest algorithm.}  
Here we go a step further and perform a MCMC analysis of the 
MSSM parameter space allowing for arbitrary CP phases. 

The parameters that can have CP phases in the MSSM are the gaugino
and higgsino mass parameters and the trilinear sfermion-Higgs couplings.
Although constrained by electric dipole moments (EDMs), nonzero phases can
significantly influence the phenomenology of SUSY particles, see 
e.g.~\cite{Kraml:2007pr} and references therein.
They can also have a strong impact on the Higgs sector, inducing
scalar-pseudoscalar mixing through loop 
effects~\cite{Pilaftsis:1998dd,Demir:1998dp,Pilaftsis:1999qt}.
Moreover, CP phases can have a potentially dramatic effect on the relic density of the 
neutralino~\cite{Falk:1995fk,Gondolo:1999gu,Nihei:2005va,Gomez:2005nr,Belanger:2006qa}. 
This is true not only in the Higgs funnel region: since the couplings of the LSP to other 
sparticles depend on the phases, so will all the annihilation and coannihilation
cross sections, even though this is not a CP-violating (CP-odd) effect.
Therefore also the phenomenology of a ``well-tempered'' neutralino 
LSP~\cite{ArkaniHamed:2006mb} 
sensitively depends on possible CP phases~\cite{Belanger:2006qa}.
For the same reasons, CP phases can also significantly modify 
the cross sections for direct and indirect dark matter detection.
 
It is therefore interesting to explore the parameter space of neutralino dark matter 
in the presence of CP phases. The advantage of the MCMC approach (or related 
scanning techniques) is that it provides a way to regard the full volume of the 
parameter space rather than just taking slices through it. This is what we do 
in this contribution for the CPV-MSSM.

\section{Setup of the MCMC scans}

Table~\ref{tab:cpvmssm-para} lists the free parameters of the CPV-MSSM 
together with ranges within which they are allowed to vary in our scan. 
We take common masses for the first and second generation of sfermions to avoid FCNC constraints and 
assume universality of the gaugino masses at the GUT scale as motivated in 
the context of models defined at the GUT scale. 
The trilinear couplings of the first and second generation are taken to be zero.
For the third generation, mass parameters and trilinear soft terms are 
treated as independent parameters. 
In addition, we allow for arbitrary phases of all the gaugino mass parameters 
and trilinear couplings of the third generation. The higgsino mass parameter 
$\mu$, on the other hand, is taken to be real. 
This can be done without loss of generality because the physically relevant 
phases are ${\rm arg}(M_i\mu)$ and ${\rm arg}(A_f\mu)$. 

\begin{table}[h]
\begin{center}
\begin{tabular}{||c|l|l||}\hline\hline
Symbol &  stands for & General range\\ \hline
$m_{H^\pm}$ &  mass of $H^\pm$ &  $[100, 2000]$ GeV\\ \hline
$\tan\beta$ & $\tan\beta$ & $[2.5, 50]$ \\ \hline
$\mu$ &  $\mu$ parameter &  $[-3000, 3000]$ GeV\\ \hline
$A_t$ & Trilinear stop coupling  & $[0, 5000]$ GeV\\ \hline
$\Phi_{t}$ & Phase of $A_t$ & $[0, 2\pi]$ \\ \hline
$A_b$ & Trilinear sbottom coupling & $[0, 5000]$ GeV\\ \hline
$\Phi_{b}$ & Phase of $A_b$ & $[0, 2\pi]$ \\ \hline
$M_1$ &  Gaugino mass, $2 M_1 = M_2 = M_3/3$ & $[50, 1000]$ GeV\\\hline
$\Phi_{1}$ & Phase of $M_1$ & $[0, 2\pi]$ \\ \hline
$\Phi_{2}$ & Phase of $M_2$ & $[-\pi, \pi]$ \\ \hline
$\Phi_{3}$ & Phase of $M_3$ & $[0, 2\pi]$ \\ \hline
$M_l$&Common slepton mass for first two generations&$[500, 10000]$ GeV \\\hline 
$M_{l3} $& Mass of left stau &$[100, 5000]$ GeV \\\hline 
$M_{r3} $& Mass of right stau &$[100, 5000]$ GeV \\\hline 
$M_q$&Common squark  mass for first two generations&$[500, 10000]$ GeV \\\hline 
$M_{Q3} $& Mass of left stop--sbottom doublet &$[100, 5000]$ GeV \\\hline 
$M_{u3} $& Mass of right stop  &$[100, 5000]$ GeV \\\hline 
$M_{d3} $& Mass of right sbottom &$[100, 5000]$ GeV \\\hline 
$m_t$ & Top quark mass & $173.1 \pm 1.3$ GeV \cite{Group:2009qk}\\ \hline
\hline
\end{tabular}\\[0.5cm]
\begin{tabular}{||c|l|l||}\hline\hline
Symbol &  stands for & General range\\ \hline
$\Phi_\mu$ &  Phase of $\mu$ parameter & $0$ or $\pi$ for $\pm$ve value of $\mu$
\\ \hline
$A_l$ & Trilinear coupling of 1st\& 2nd gen.\ sleptons & $0$ GeV\\ \hline
$A_q$ & Trilinear coupling of 1st\& 2nd gen.\ squarks& $0$ GeV\\ \hline
\hline
\end{tabular}
\end{center}
\caption{\label{tab:cpvmssm-para} Model parameters and their ranges used in the scan.}
\end{table}

For the numerical analysis, we use 
{\tt micrOMEGAs2.2} \cite{Belanger:2006is,Belanger:2008sj} 
linked to {\tt CPsuperH2} \cite{Lee:2007gn}. The latter gives the CPV Higgs 
sector, B-physics observables and EDMs. We use the thallium, mercury and 
electron EDMs $d(Tl)$, $d(Hg)$ and $d(e^-)$; the neutron EDM is not used 
because of its big uncertainty stemming from the quark model~\cite{Lee:2007gn}. 
To evaluate the limits on the light Higgs mass, we make use of 
the {\tt HiggsBounds}~\cite{Bechtle:2008jh} program. 
For the scan we use the directed random search MCMC method as 
described in detail in Ref.~\cite{Belanger:2009ti} (see also references therein).

We compute the likelihood of a parameter point as the product of likelihoods of 
all the observables under consideration. The observables considered in our 
analysis are listed in Table~\ref{tab:cpvmssm-obs} along with the shapes of 
the likelihood functions used. These probability distribution funtions 
(PDFs) are given as:
\begin{equation}
  G(x,x_0,\sigma_x) = \exp\left[\frac{-(x-x_0)^2}{2 \ \sigma_x^2}\,\right], \qquad
  F(x,x_0,\sigma_x) = \frac{1}{1+\exp[-(x-x_0)/\sigma_x]}.
\end{equation}  
We use the Gaussian function $G$ for 
observables for which a measurement is available, 
and function $F$ when there is only an upper or lower bound. 
Last but not least, we use flat priors for all input parameters, 
and base the analysis on ten chains with $10^6$ points each. 

\begin{table}[t]
\begin{center}
\begin{tabular}{||l|c|l|l||}\hline\hline
Observable & Limit & Likelihood function&Ref. \\ 
\hline
$\Omega h^2$ & $0.1099\pm 0.0062$ & $\mathbf{G}(x,0.1099,0.0062)$ & \cite{Dunkley:2008ie} \\ 
\hline
${\rm BR}(b\to s\gamma)$ & $(3.52 \pm 0.34) \times 10^{-4}$ &  $\mathbf{G} 
(x,3.52 \times 10^{-4}, 0.34 \times 10^{-4})$& \cite{Barberio:2008fa,Misiak:2006zs} \\ \hline
$A_{CP}(b\to s\gamma)$ & $(1.0 \pm 4.0) \times 10^{-2}$ &  $\mathbf{G} 
(x,1.0 \times 10^{-2}, 4.0 \times 10^{-2})$& \cite{Amsler:2008zzb} \\ 
\hline
${\rm BR}(B_s\to \mu^+\mu^-)$ & $\le 5.8 \times 10^{-8}$ & $\mathbf{F}(x,5.8 \times
10^{-8}, -5.8 \times 10^{-10})$ & \cite{Aaltonen:2007kv} \\ 
\hline
$R(B_u\to \tau\nu_\tau)$ & $1.28\pm0.38$ & $\mathbf{G}(x,1.28,0.38)$ & \cite{Barberio:2008fa} \\
\hline
${\rm BR}(B_d\to \tau^+\tau^-)$ & $\le 4.1 \times 10^{-3}$ & $\mathbf{F}(x,4.1 \times
10^{-3}, -8.2 \times 10^{-5})$ & \cite{Amsler:2008zzb} \\ 
\hline
$d(Tl)$ e\,cm & $\le 9.0 \times 10^{-25}$ & $\mathbf{F}(x,9.0 \times 10^{-25},
-1.8 \times 10^{-25})$ & \cite{Regan:2002ta} \\ \hline
$d(Hg)$ e\,cm & $\le 2.0 \times 10^{-28}$ & $\mathbf{F}(x,2.0 \times 10^{-28},
-2.0 \times 10^{-29})$ & \cite{Romalis:2000mg} \\ \hline
$d(e^-)$ e\,cm & $\le 1.6 \times 10^{-27}$ & $\mathbf{F}(x,1.6 \times 10^{-27},
-3.2 \times 10^{-28})$ & \cite{Regan:2002ta} \\ \hline
$R_{H_1}$ (Higgs mass) & $\le 1.00$ & $\mathbf{F}(x,1.00,0.01)$ & \cite{Bechtle:2008jh} \\ \hline
Mass limits & LEP limits & $1$ or $10^{-9}$ & \cite{lepsusy} \\\hline
\hline
\end{tabular}
\end{center}
\caption{\label{tab:cpvmssm-obs} Observables used in the likelihood calculation.}
\end{table}

\section{Results}

\begin{figure}[t]\centering
\includegraphics[width=0.6\textwidth]{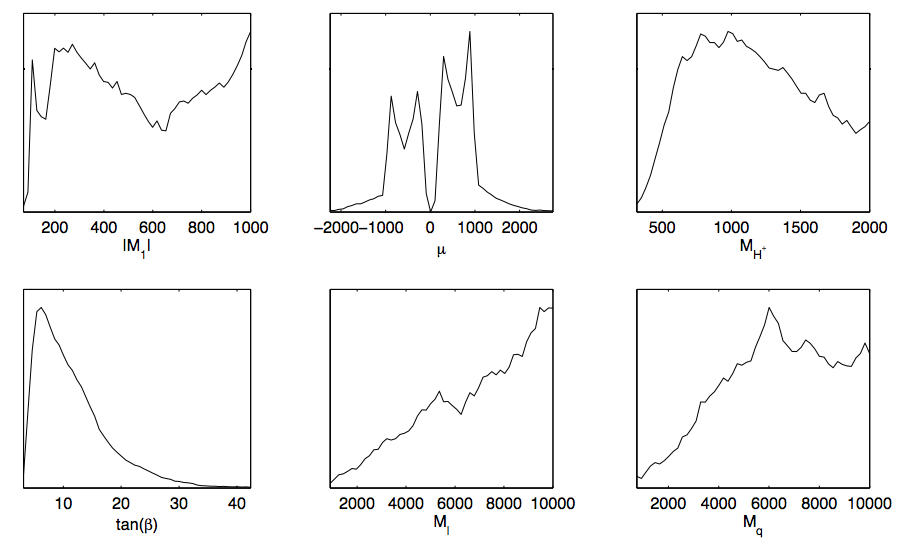}
\caption{1D posterior PDFs  for some model parameters; 
from left to right: $|M_1|$, $\mu$, $m_{H^+}$ (top row) and 
$\tan\beta$, $M_l$, $M_q$ (bottom row). Dimensionful parameters 
are in GeV.}
\label{fig:cpv1D-pars}
\end{figure}

Figure~\ref{fig:cpv1D-pars} shows the 1D posterior 
PDFs for some of the most important model parameters like 
$|M_1|$, $\mu$, $m_{H^+}$, $\tan\beta$, $M_l$, $M_q$.  
(Here and in the following, dimensionful parameters are in GeV.) 
Some explanatory comments are in order. 
First, we observe a slight preference for positive $\mu$, at the level of 
40\% minus versus 60\% plus sign. A priori this seems in agreement 
with the preference of ${\rm sign}(M_2\mu)=+1$ found in \cite{AbdusSalam:2009qd} 
caused by the $b\to s\gamma$ constraint (we do not use any  
constraint on the muon $(g-2)$).  
In our case it is, however, mostly due to the fact that we have six chains that 
converged in the $\mu>0$ subspace but only four in the $\mu<0$ one.  
Either way, the preference of one sign over the other is not significant.
Second, the heavy Higgs sector is pushed to masses above ca.\ 500~GeV 
by B-physics constraints, while EDM constraints push the masses of the first 
and second generation sfermions to the multi-TeV range. 
Third, regarding $\tan\beta$,
we observe a preference for small values, caused again by EDM constraints.

\begin{figure}[t]\centering
\includegraphics[width=\textwidth]{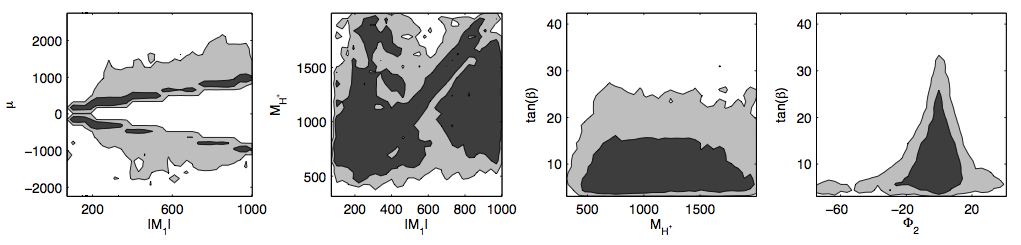}
\caption{Regions of 68\% BC (dark grey) and 95\% BC (light grey) 
in the plane  $\mu$ vs.\ $|M_1|$, $m_{H^+}$ vs.\ $|M_1|$,
$\tan\beta$ vs.\ $m_{H^+}$ and $\tan\beta$ vs.\ $\Phi_2$.}
\label{fig:cpv2D-pars}
\end{figure}

Correlations between the input parameters can be seen in Fig.~\ref{fig:cpv2D-pars},  
which shows the 2D 68\% and 95\% Bayesian Credibility (BC) regions in the 
$(\mu,|M_1|)$, $(m_{H^+},|M_1|)$, $(\tan\beta,m_{H^+})$ and 
$(\Phi_2,\tan\beta)$ planes. 
CP-conserving (CPC) analogs of the 
first two plots can be seen in Fig.~3 of Ref.~\cite{Belanger:2009ti}. 
The correlations between $|M_1|$--$\mu$ and $|M_1|$--$m_{H^+}$ 
are dominantly driven by  the relic density constraint. 
The CPV and CPC cases show the same basic features, 
favouring the mixed bino-higgsino ($|M_1|\approx \mu$) or the Higgs-funnel regions $|M_1|\approx M_{H^+}/2$. 
It is, however, apparent that allowing for nonzero phases considerably 
enlarges the parameter space that is compatible with a relic density 
within WMAP bounds. For example, the 68\% BC range includes a region 
far from the Higgs funnel where $|M_1|\approx \mu\approx {\cal O}(1)$~TeV 
and the $\tilde\chi^\pm_1$ and $\tilde\chi^0_2$ have a small mass difference 
with the LSP. 
This region occurs with much smaller likelihood in the CPC case~\cite{Belanger:2009ti}.
The impact of the EDM constraints on $\tan\beta$ is apparent from the fourth panel 
in Fig.~\ref{fig:cpv2D-pars}: when $\Phi_2$ is nonzero, $\tan\beta$ is constrained 
to very small values, while the large values of $\tan\beta$ are allowed only for very 
small values of $\Phi_2$.

\begin{figure}[htbp]\centering
\includegraphics[width=\textwidth]{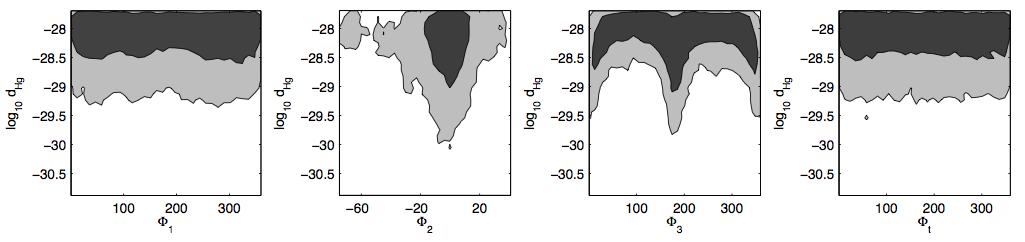}
\caption{Regions of 68\% BC (dark grey) and 95\% BC (light grey) of 
the mercury EDM versus phases (in degrees).}
\label{fig:cpv2D-edm}
\end{figure}

One advantage of the MCMC is that it lets us explore the constraints on the phases 
in a general way, by marginalization over parameters. As expected, we find that 
the phase that is most constrained by the EDMs is the relative phase between 
$M_2$ and $\mu$. 
Since we take $\mu$ to be real without loss of generality, this means severe 
constraints on  $\Phi_2$, as illustrated in Fig.~\ref{fig:cpv2D-edm}. 
The other phases are much less constrained. In particular the phases of 
$M_1$ and of the trilinear soft terms
can vary over the full range,  $\Phi_{1,t,b}=[0,\,2\pi]$, if the sfermions 
of the first two generations have masses of few TeV. Only for $\Phi_3$ 
there is also some preference for the near-CPC case. 

\begin{figure}[htbp]\centering
\includegraphics[width=0.8\textwidth]{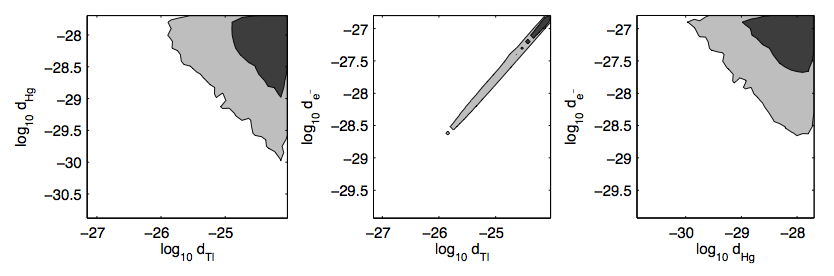}
\caption{Regions of 68\% BC (dark grey) and 95\% BC (light grey) showing the 
correlations between the EDMs.}
\label{fig:cpv2D-edm-edm}
\end{figure}

Overall, with five phases to vary, the CPC case becomes a point in a 5D
parameter space. This has important consequences for the EDMs, 
since they will be near zero only when all the dominant phases go to 
zero simultaneously. This means that the EDMs dominantly saturate the present 
bounds: they are predicted to be large and potentially observable over most 
of the allowed parameter space. 
This is illustrated in  Fig.~\ref{fig:cpv2D-edm-edm}, which shows 
the 2D marginalized distributions of EDMs at 68\% and 95\% BC. 
We see that (i) the EDMs are highly correlated and (ii) the CPC case 
is just a small corner of the large parameter space we are considering.

\begin{figure}[htbp]\centering
\includegraphics[width=0.95\textwidth]{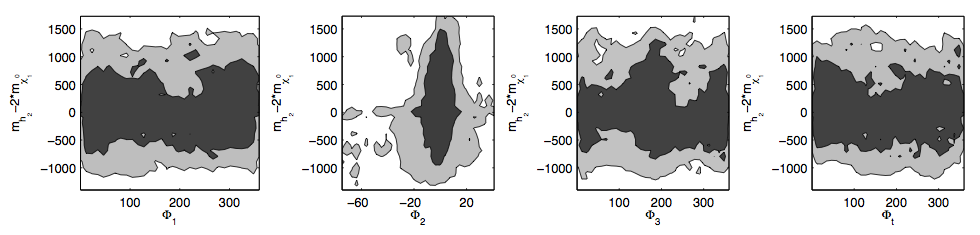}
\caption{Regions of 68\% BC (dark grey) and 95\% BC (light grey) showing the 
correlation between distance from the $h_2$ pole ($m_{h_2}-2m_{\tilde\chi^0_1}$) 
and the various phases (in degrees).}
\label{fig:cpv2D-mh}
\end{figure}

\begin{figure}[htbp]\centering
\includegraphics[width=0.98\textwidth]{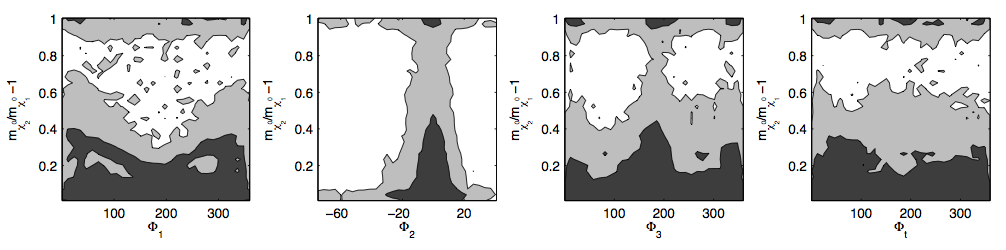}
\caption{Regions of 68\% BC (dark grey) and 95\% BC (light grey) showing the 
correlation between the relative $\tilde\chi^0_2$--$\tilde\chi^0_1$ mass difference 
and the various phases  (in degrees).}
\label{fig:cpv2D-mchi}
\end{figure}

Let us now turn to two key quantities determining the dominant annihilation 
channel of the neutralino LSP: the distance from the (mostly pseudoscalar) 
Higgs pole, $\delta m_H\equiv m_{h_2}-2m_{\tilde\chi^0_1}$, and the relative 
mass difference 
between the lightest and second-lightest neutralino, 
$\Delta_\chi \equiv (m_{\tilde\chi^0_2}-m_{\tilde\chi^0_1})/m_{\tilde\chi^0_1}$. 
In the CPC case with gaugino mass universality, the latter quantity is a direct 
measure of the higgsino fraction of the LSP. 
The 2D likelihood functions for $\delta m_H$ versus 
$\Phi_i$ (with $i=1,2,3,t$) are shown in Fig.~\ref{fig:cpv2D-mh}.
The analogous distributions for $\Delta_\chi$
are shown in Fig.~\ref{fig:cpv2D-mchi}.
We see that for nonzero phases the preferred values of both 
$\delta m_H$ and $\Delta_\chi$ can considerably differ from those 
in the CPC case.
This was already noted in \cite{Belanger:2006qa} and is confirmed
here in a more general way. 

Finally, Table~\ref{tab:cpvmssm-bci} explicitly lists the 68\% and 95\% BC 
intervals for CPV-MSSM parameters, Higgs and sparticle masses, 
and several low-energy observables. Note, for instance, that the squark and 
slepton masses of the first two generations are above 2 (4) TeV at 95\% (68\%)
BC. The third generation can be much lighter, with a 95\% (68\%)  lower limit of 
300--400 GeV (around 800 GeV) for the lighter mass eigenstates $\tilde t_1$, 
$\tilde b_1$, and $\tilde\tau_1$. Neutralinos and charginos cover a large 
mass range, from about 100 GeV up to ca.\ 1 TeV. This also holds for the 
LSP. In turn, the gluino can be rather light, leading to a large pair production 
cross section at the LHC followed dominantly by decays into third generation 
quarks---or very heavy, beyond the reach of the LHC. Gaugino--higgsino 
mixing is sizable over a large part of the parameter space; whether this can 
lead to observable rates of electroweak $\tilde\chi^0_2\tilde\chi^\pm_1$ 
production at the LHC depends, however, on the neutralino/chargino 
mass scale, which as said above spans a wide range. 
All these issues will be considered in detail elsewhere~\cite{Belanger:inprep}.

\begin{table}
\begin{center}
\begin{tabular}{|l|r r|r r|l|}\hline
Parameters, masses, & 68\% BCI & 68\% BCI & 95\% BCI & 95\% BCI & Remarks\\
observables & min & max & min & max & \\ \hline
 $M_{H^+}                                   $&$      704    $&$    1999     $&$     473    $&$    1999     $&Upper limit saturated \\
 $\tan\beta                                $&$        5.82   $&$      17.2    $&$       4.10   $&$      27.8    $& \\
 $|M_1|                                     $&$      235    $&$    1002     $&$     113    $&$    1002     $& Upper limit saturated\\
 $\mu                                       $&$     -750    $&$     881    $&$   -1364     $&$    1567     $& \\
 $|A_t|                                     $&$     1683     $&$    5008     $&$     511    $&$    5008     $&Upper limit saturated \\
 $\Phi_2                                    $&$      -12.7    $&$       7.03  $&$     -50.17   $&$      23.68   $& \\
 $M_l                                       $&$     4379     $&$   10000      $&$    2121     $&$   10000      $& Upper limit saturated\\
 $M_q                                       $&$     3960    $&$   10000      $&$    1912     $&$   10000      $& Upper limit saturated\\
 $M_{l3}                                    $&$     1077     $&$    5002     $&$     365    $&$    5002     $& Upper limit saturated\\
 $M_{r3}                                    $&$     1289     $&$    5006     $&$     578    $&$    5006     $& Upper limit saturated\\
 $M_{q3}                                    $&$     1116     $&$    3570     $&$     618    $&$    4569    $& \\
 $M_{u3}                                    $&$     1132    $&$    3613     $&$     552    $&$    4669    $& \\
 $M_{d3}                                    $&$      989    $&$    5008     $&$     389    $&$    5008     $& Upper limit saturated\\
\hline
 $m_{h_1}                                   $&$      114    $&$     120    $&$     114    $&$     123    $& Lower limit saturated\\
 $m_{h_2}                                   $&$      700   $&$    1997     $&$     466    $&$    1998     $& Upper limit saturated\\
 $m_{h_3}                                   $&$      700   $&$    1998     $&$     466    $&$    1998     $& Upper limit saturated\\
 $m_{\tilde\chi^0_1}                              $&$       70   $&$     841    $&$      70   $&$     935     $& \\
 $m_{\tilde\chi^0_2}                              $&$      107    $&$     925    $&$     107    $&$    1440     $& Lower limit saturated \\
 $m_{\tilde\chi^0_3}                              $&$      139    $&$     996    $&$     139    $&$    1758    $& Lower limit saturated \\
 $m_{\tilde\chi^\pm_1}                              $&$      104    $&$     923     $&$     104    $&$    1440    $& Lower limit saturated\\
 $m_{\tilde\chi^\pm_2}                              $&$      232    $&$    1789     $&$     232    $&$    1989    $& Lower limit saturated\\
 $m_{\tilde e_{L,R}}                           $&$     4379     $&$   10000      $&$    2122     $&$   10000      $& Upper limit saturated\\
$m_{\tilde\tau_1}                                $&$      868    $&$    3236     $&$     335    $&$    4297    $& \\
 $m_{\tilde\tau_2}                                $&$     2157     $&$    5012     $&$     876    $&$    5012     $& Upper limit saturated\\
 $m_{\tilde q_{L,R}}                   $&$     3960     $&$   10000      $&$    1911     $&$   10000      $& Upper limit saturated\\
 $m_{\tilde t_1}                                   $&$      867    $&$    2444     $&$     463    $&$    3471    $& \\
 $m_{\tilde t_2}                                   $&$     1941     $&$    4999     $&$    1167     $&$    4999    $& Upper limit saturated\\
 $m_{\tilde b_1}                                   $&$      815    $&$    2782     $&$     378    $&$    3973     $& \\
 $m_{\tilde b_2}                                   $&$     1883     $&$    5010     $&$     987    $&$    5010     $& Upper limit saturated\\
\hline
$\Omega h^2                                $&$        0.1032 $&$       0.1156 $&$       0.0972 $&$       0.1216 $& Post-diction\\
 BR$(B \rightarrow X_s \gamma)\times10^4     $&$3.44           $&$3.78          $&$3.23          $&$   4.05       $& Post-diction\\
 $A_{CP} [B \rightarrow X_s \gamma](\%)     $&$       -0.0835 $&$       0.0883 $&$      -0.3210 $&$       0.3871 $& Post-diction\\
 $R(B_u \rightarrow \tau\nu)                $&$        0.9828 $&$       0.9998 $&$       0.9385 $&$       0.9998 $& Post-diction\\
 BR$(B_s \rightarrow \mu\mu)\times10^9       $&$    3.63       $&$   3.72       $&$   3.31       $&$   4.10       $& Pre-diction\\
 BR$(B_d \rightarrow \tau\tau)\times10^8     $&$    2.25       $&$   2.30       $&$   2.04       $&$   2.53       $& Pre-diction\\
$\log_{10} |d_{Tl}|                          $&$      -24.88 $&$     -24.03   $&$     -25.71   $&$     -24.03   $& \\
$\log_{10} |d_{Hg}|                          $&$      -28.69 $&$     -27.69   $&$     -29.51   $&$     -27.69   $& \\
$\log_{10} |d_{e^-}|                         $&$      -27.64 $&$     -26.80   $&$     -28.48   $&$     -26.80   $& \\
\hline
 $R_{Higgs}                                 $&$        0.3872 $&$       0.7239 $&$       0.1766 $&$       0.9529 $& c.f. {\tt HiggsBounds} \\
 $          f_H                             $&$    0.0133    $&$       0.6083 $&$   0.0008     $&$       0.8054 $& LSP higgsino fraction\\
 $m_{h_2}-2m_{\chi^0_1}                    $&$     -527    $&$     775    $&$   -1041     $&$    1355     $& \\
  $m_{\tilde\chi^0_2}/m_{\tilde\chi^0_1} -1              $&$0.008          $&$   1.005      $&$ 0.008        $&$   1.005      $& Peaks at both ends\\
\hline
\end{tabular}
\end{center}
\caption{\label{tab:cpvmssm-bci} The min/max limits of the 68\% and 95\% BC 
intervals (BCI) for CPV-MSSM parameters, Higgs and sparticle masses, and various observables.}
\end{table}

\section{Conclusions}

We have presented a first Bayesian analysis of the CPV-MSSM model with
parameters defined at the electroweak scale, taking into account constraints 
from collider searches, B-physics and EDMs and requiring that the neutralino 
LSP be the dark matter of the Universe with a relic density in agreement 
with WMAP observations. 
We find that phases can be large 
if the first two generations of sfermions are above 2 (4) TeV at 95\% (68\%) BC.
In fact only one phase, $\Phi_2$, is strongly constrained. 
A large fraction of the parameter space features heavy sparticles
that are beyond the reach of the LHC. This is just a reflection of the fact 
that,  apart from  $\Omega h^2$, other measurements do not  require a supersymmetric contribution. 
Clearly improvements on the experimental determination of the EDMs will play a crucial role in revealing 
or further constraining phases.
The implications of the phases for LHC phenomenology as well as for dark matter direct and indirect
detection will be presented in an expanded and updated version \cite{Belanger:inprep} 
of this analysis.\\

\noindent
{\bf Note added:}
On completion of this work, we became aware of an improved limit on the mercury 
EDM of $d(Hg)<3.1\times 10^{-29}$~e\,cm~\cite{Griffith:2009zz}. This new limit leads 
to stronger constraints on the parameter space, especially on $\Phi_2$, and will be
included in the more detailed report.

\section*{Acknoledgements}
We thank Oliver Brein for discussions on {\tt HiggsBounds}.

%% file: Espinosa/composite.tex
\chapter{Composite Higgs boson search at the LHC}

{\it J.R.~Espinosa, C.~Grojean and M.~M\"uhlleitner}

\begin{abstract}
 In composite Higgs models the Higgs boson emerges as a
 pseudo-Goldstone boson from a strongly-interacting sector. While in
 the Standard Model the Higgs sector is uniquely determined by the
 mass of the Higgs boson, in composite Higgs models additional
 parameters control the Higgs properties. In consequence the LEP and
 Tevatron exclusion bounds are modified and the Higgs boson searches
 at the LHC are significantly affected. The consequences
 for the LHC Higgs boson search in the composite model will be discussed.
\end{abstract}

\section{Introduction}
The massive nature of the weak gauge bosons $W,Z$ requires new degrees
of freedom and/or new dynamics around the TeV scale to ensure
unitarity in the scattering of longitudinal gauge bosons $W_L,Z_L$. In
the Standard Model (SM) unitarity is assured by the introduction of an
elementary Higgs boson. The SM Higgs couplings are proportional to the
mass of the particle to which it couples, and the only unknown parameter
in the SM is the mass of the Higgs boson. Furthermore, the electroweak 
precision observables and the
absence of large flavor-changing neutral currents strongly constrain
departures from this minimal Higgs mechanism and rather call for
smooth deformations, at least at low energy (see Ref.~\cite{Grojean:2009fd} for a general discussion).
This supports the idea of a light Higgs boson emerging as a
pseudo-Goldstone boson from a strongly-coupled sector, the so-called
Strongly Interacting Light Higgs (SILH) scenario~\cite{Giudice:2007fh,Contino:2010mh}. The low-energy content is
identical to the SM with a light, narrow Higgs-like scalar, which
appears, however, as a bound state from some strong dynamics~\cite{Dimopoulos:1981xc,Banks:1984gj,Kaplan:1983fs,Kaplan:1983sm,Georgi:1984ef,Georgi:1984af,Dugan:1984hq}.
A mass gap separates the Higgs boson from the other usual
resonances of the strong sector as a result of the Goldstone nature of
the Higgs boson. Since the rates for production and decay, however, can
differ significantly from the SM results we study in the present work
how the LHC Higgs boson search channels are affected by the
modifications of the composite Higgs boson couplings. 
We estimate the experimental sensitivities in the main search channels studied by
ATLAS and CMS as well as the luminosities needed for discovery.

The effective Lagrangian constructed in~\cite{Giudice:2007fh}, which
involves higher dimensional operators for the low-energy degrees of
freedom, should be seen as an expansion in $\xi=(v/f)^2$ where $v=
1/\sqrt{\sqrt{2} G_F} \approx 246$
GeV and $f$ is the typical scale of the Goldstone bosons of the strong
sector. It can therefore be used in the vicinity of the SM limit
($\xi\to 0$), whereas  the technicolor limit ($\xi \to 1$) requires a
resummation of the full series in $\xi$. Explicit models, built in 5D
warped models, provide concrete examples of such a resummation. Here,
we will rely on two representative 5D models exhibiting different behaviours of the
Higgs couplings. In these models the deviations from the SM Higgs
couplings are controlled by the parameter $\xi= (v/f)^2$ which varies
from 0 to 1. The two extra parameters which generically control the
composite Higgs couplings are thus related and our analysis is hence
an exploration of the parameter space along some special
directions. On the other hand, the technicolor limit can be approached.

In Section~\ref{composite_label1} we give the general parameterization
of the composite Higgs couplings derived from the SILH Lagrangian of
Ref.~\cite{Giudice:2007fh}. For the two explicit 5D composite models
we give the exact form of these couplings. The LEP and Tevatron limits are
studied in Section~\ref{composite_label3}. The Higgs decay rates are
discussed in Section~\ref{composite_label2}. 
Section~\ref{composite_label4} presents the Higgs 
boson production cross sections, before in Section~\ref{composite_label12}  the
modifications of the significances with respect to the SM search
channels are discussed. Furthermore, the luminosities needed for
discovery will be presented. Section~\ref{composite_label13} contains our conclusions.

\section{Parameterization of the Higgs
  couplings \label{composite_label1}}
The effective SILH Lagrangian involves two classes of higher
dimensional operators: (i) operators genuinely sensitive to the new
strong force, which will affect qualitatively the Higgs boson physics
and (ii) operators sensitive to the spectrum of the resonances only,
which will simply act as form factors. The effective Lagrangian
generically takes the form
\begin{eqnarray}
&&\mathcal{L}_{\rm SILH} = \frac{c_H}{2f^2} \left( \partial_\mu |H|^2 \right)^2
+ \frac{c_T}{2f^2}  \left(   H^\dagger{\overleftrightarrow D}_\mu H\right)^2 
- \frac{c_6\lambda}{f^2} |H|^6
+ \left( \frac{c_yy_f}{f^2} |H|^2 {\bar f}_L Hf_R +{\rm h.c.}\right) \nonumber \\ 
&&
+\frac{ic_Wg}{2m_\rho^2}\left( H^\dagger  \sigma^i \overleftrightarrow {D^\mu} H \right )( D^\nu  W_{\mu \nu})^i
+\frac{ic_Bg'}{2m_\rho^2}\left( H^\dagger  \overleftrightarrow {D^\mu}
  H \right )( \partial^\nu  B_{\mu \nu})  +\ldots 
\label{composite_label5}
\end{eqnarray}
where $g,g'$ denote the SM electroweak (EW) gauge couplings, $\lambda$
the SM Higgs quartic coupling and $y_f$ the SM Yukawa coupling to the
fermions $f_{L,R}$. The coefficients $c_H,c_T,...$ appearing in
Eq.~\ref{composite_label5} are expected to be of order 1 unless
protected by some symmetry. The operator $c_H$ gives a corrections to
the Higgs kinetic term. After rescaling the Higgs field, in order
to bring the kinetic term back to its canonical form, the Yukawa
interactions read (see Ref.~\cite{Giudice:2007fh} for details)
\begin{eqnarray}
&&g_{hf\bar{f}}^{\xi} = g_{hf\bar{f}}^\textrm{\tiny SM}\times (1-(c_y + c_H/2) \xi),\\
&&g_{hVV}^{\xi} = g_{hVV}^\textrm{\tiny SM} \times (1-c_H\, \xi/2),
\quad
g_{hhVV}^{\xi} = g_{hhVV}^\textrm{\tiny SM} \times (1-2 c_H\, \xi/2) \;,
\end{eqnarray}
where $V=W,Z$, $g_{hf\bar{f}}^\textrm{\tiny SM}=m_f/v$,
$g_{hW^+W^-}^{\xi} = gM_W$, $g_{hZZ}^{\xi} = \sqrt{g^2+g^{'2}}M_Z$,
$g_{hhW^+W^-}^\textrm{\tiny SM} = g^2$ and $g_{hhZZ}^\textrm{\tiny SM}
= (g^2+ g^{'2})$ and $m_f,M_W,M_Z$ denote the fermion, $W$ and $Z$
boson masses.  
The dominant corrections controlled by the strong operators preserve
the Lorentz structure of the SM interactions, while the form factor
operators will also introduce couplings with a different Lorentz
structure.

For our two concrete models studied hereafter we refer to the
Holographic Higgs models of
Refs.~\cite{Contino:2003ve,Agashe:2004rs,Contino:2006qr}, which are
based on a five-dimensional theory in Anti-de-Sitter (AdS) space-time. The bulk
gauge symmetry $SO(5)\times U(1)_X \times SU(3)$ is broken down to the
SM gauge group on the UV boundary and to $SO(4)\times U(1)_X \times
SU(3)$ on the IR. In the unitary gauge this leads to the following
Higgs couplings to the gauge fields ($V=W,Z$) in terms of the
parameter $\xi = (v/f)^2$ 
\begin{equation}
g_{hVV}=g_{hVV}^{SM}\ \sqrt{1-\xi}\ , \hspace{1cm}
g_{hhVV}=g_{hhVV}^{SM}\ (1-2\xi)\ .
\label{composite_label6}
\end{equation}
The Higgs couplings to the fermions will depend on the way the SM
fermions are embedded into representations of the bulk symmetry. In
the MCHM4 model~\cite{Agashe:2004rs} with SM fermions transforming as
spinorial representations of $SO(5)$, the Higgs fermion interactions are given by
\begin{equation}
\label{composite_label7}
\textrm{MCHM4:} \hspace{1cm} g_{hff}=g_{hff}^{SM}\ \sqrt{1-\xi}\ .
\end{equation}
In the MCHM5 model~\cite{Contino:2006qr} with SM fermions transforming
as fundamental representations of $SO(5)$, the Higgs fermion
couplings take the form
\begin{equation}
\label{composite_label8}
\textrm{MCHM5:} \hspace{1cm} g_{hff}=g_{hff}^{SM}\ \frac{1-2\xi}{\sqrt{1-\xi}}\ .
\end{equation}
While the Higgs gauge couplings are always reduced compared to the SM,
the Higgs couplings to fermions behave differently in the two
models. In the vicinity of the SM the couplings are reduced, with the
reduction being more important for the MCHM5 than for the MCHM4
model. For larger values of $\xi$, the MCHM5 Higgs fermion couplings
raise again and can even become larger than in the SM, leading to
enhanced gluon fusion Higgs production cross sections. The latter will
significantly affect the Higgs searches.

\section{Constraints from LEP and Tevatron and EW precision data\label{composite_label3}}
\begin{figure}[t]
\begin{center}
\includegraphics[width=5.cm]{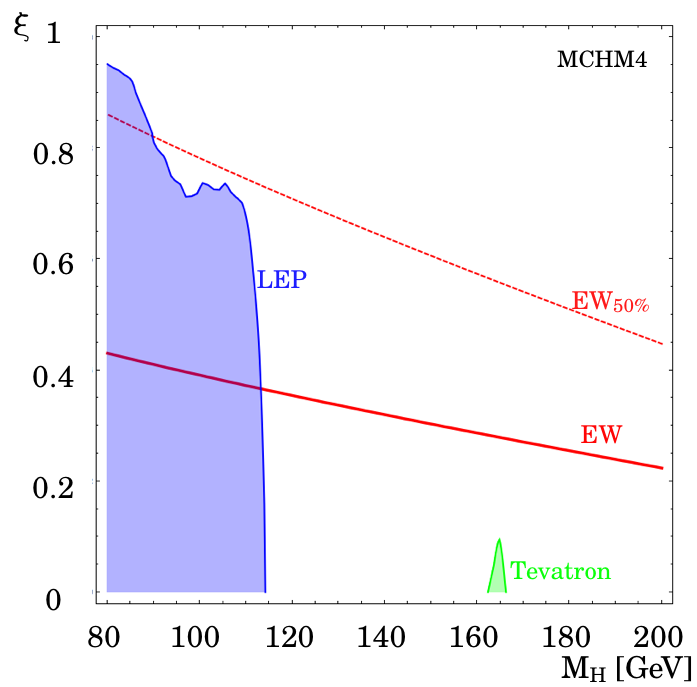} 
\hspace*{1cm}
\includegraphics[width=5.cm]{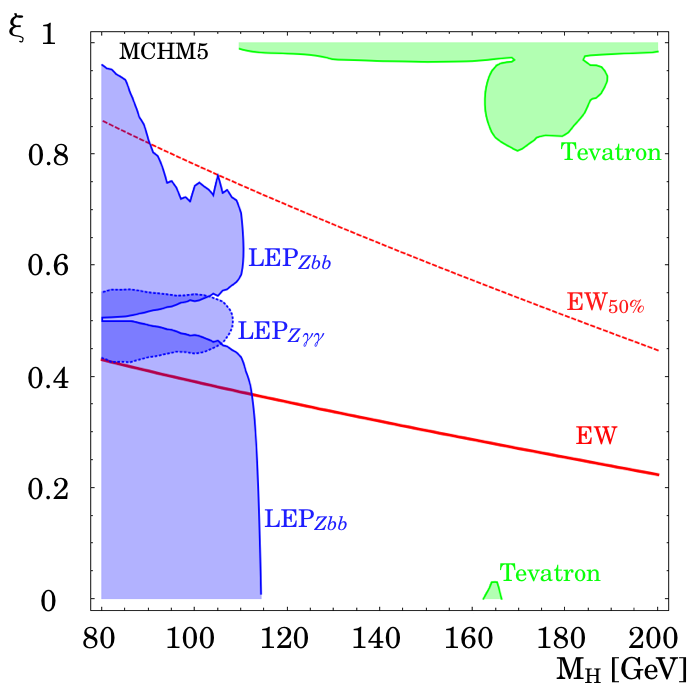} 
\caption{\label{composite_label10}
Experimental limits from Higgs searches at LEP (blue/dark gray) and the Tevatron 
(green/light gray) in the plane $(M_H,\xi)$ for MCHM4~(left) and
MCHM5~(right). EW  
precision data prefer low value of $\xi$: the red continuous line 
delineates the region favoured at 99\% CL (with a cutoff scale fixed at 
2.5~TeV) while the region below the red dashed line 
survives if there is an additional 50\% cancellation of the oblique 
parameters.}
\end{center}
\end{figure}

The $(M_H,\xi)$ parameter region is constrained from the Higgs
searches at LEP and Tevatron. The excluded regions are shown in
Fig.~\ref{composite_label10}. For the generation of the plots the
program HiggsBounds~\cite{Bechtle:2008jh} has been used, modified to
take into account the latest Tevatron limits.

In both composite models the SM Higgs mass LEP limit $M_H\gsim 114.4$ GeV is
lowered, since at LEP the most relevant search channel is
Higgs-strahlung with subsequent decay into $b\bar{b}$~\cite{Barate:2003sz,Schael:2006cr}. In both models 
the production process is suppressed compared to the SM. Since in
MCHM5 at $\xi=0.5$
the Higgs fermion coupling vanishes, this channel cannot be used in
the area around this $\xi$ value. Constraints are set by
Higgs-strahlung production with subsequent decay into $\gamma \gamma$
instead~\cite{LEPHWG}.  

At Tevatron, low $\xi$ values are excluded by the Higgs decay into a $W$
pair for Higgs masses around 160 GeV\footnote{Tevatron searches in
  $H\to WW$ decays exclude the SM Higgs boson in the mass range 162
  GeV$\le M_H \le$ 166 GeV~\cite{Aaltonen:2010yv}.}. The exclusion region
quickly shrinks to 0, since the relevant Higgs-strahlung production is
suppressed compared to the SM for non-vanishing $\xi$
values. In MCHM5, an additional region $M_H \sim 165-185$ GeV can be excluded for
$\xi \gsim 0.8$ through $H\to WW$~\cite{Aaltonen:2010yv} where the
enhanced Yukawa coupling 
increases the production in gluon fusion and the $WW$ branching ratio
is still high, before fermionic decays take over close to
$\xi=1$. The exclusion is then set by $H\to \tau \tau$ decays~\cite{tautaulim}. These results should be regarded, however, as rough
guidelines. The Tevatron searches combine several search channels from
both experiments in a sophisticated way. We cannot perform such an
analysis at the same level of sophistication. 

Further constraints arise from the electroweak precision (EWP) data. The
oblique parameters are logarithmically sensitive to the Higgs boson
mass~\cite{Peskin:1991sw}. The EWP limits are also shown in
Fig.~\ref{composite_label10}. In our 
set-up they are due to the incomplete cancellation between the Higgs
and gauge boson contributions to $S$ and $T$ and low $\xi$ values are
preferred. The upper bound on $\xi$ is relaxed by a factor of $\sim 2$
if one allows for a partial cancellation of the order of 50\%.

\section{Branching ratios \label{composite_label2}}
The partial widths in the composite Higgs models are obtained by
rescaling the corresponding Higgs couplings involved in the decay. In
the MCHM4 model all couplings are multiplied by the same factor
$\sqrt{1-\xi}$ so that the branching ratios are the same as in the
SM. 
In the MCHM5 model the partial decay width into fermions can be obtained from
the corresponding SM width by, {\it cf.} Eq.~(\ref{composite_label8}),
\beq
\Gamma(H\to f\bar{f}) = \frac{(1-2\xi)^2}{(1-\xi)} \;
\Gamma^{SM} (H\to f\bar{f}) \; .
\eeq
The Higgs decay width into gluons, mediated by heavy quark loops, reads
\beq
\Gamma(H\to gg) = \frac{(1-2\xi)^2}{(1-\xi)} \;
\Gamma^{SM} (H\to gg) \; .
\eeq
The Higgs decay width into massive gauge bosons $V=W,Z$ is given by
\beq
\Gamma(H\to VV) = (1-\xi) \; \Gamma^{SM}
(H\to VV) \; .
\eeq
The Higgs decay into photons proceeds dominantly via $W$-boson and
top and bottom loops. Since the couplings to gauge bosons and
fermions scale differently in MCHM5, the various
loop contributions have to be multiplied with the corresponding Higgs
coupling modification factor. As QCD corrections do not involve the
\begin{figure}[h]
\begin{center}
\includegraphics[width=6.cm]{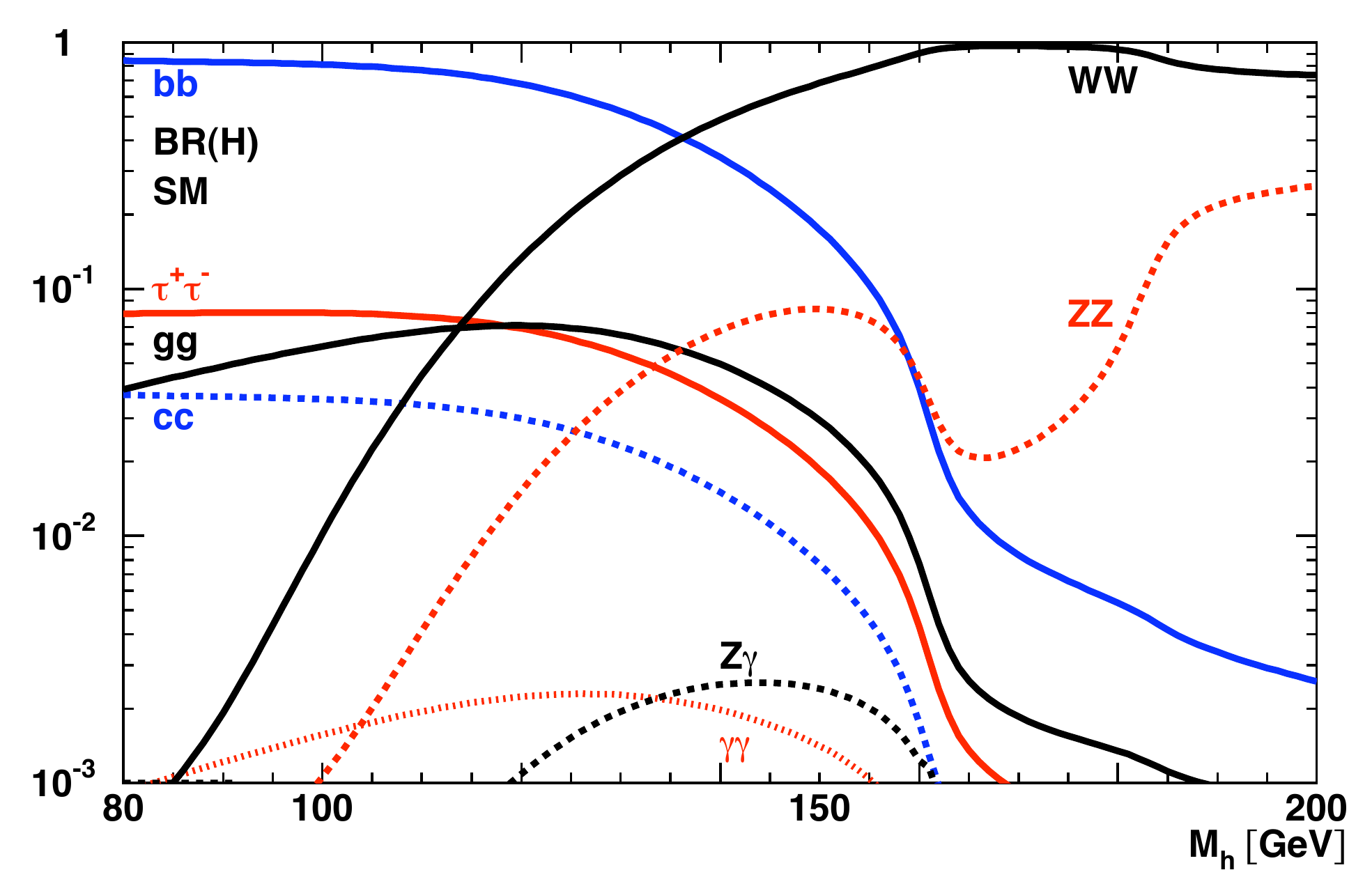} 
\hspace*{0.5cm}
\includegraphics[width=6.cm]{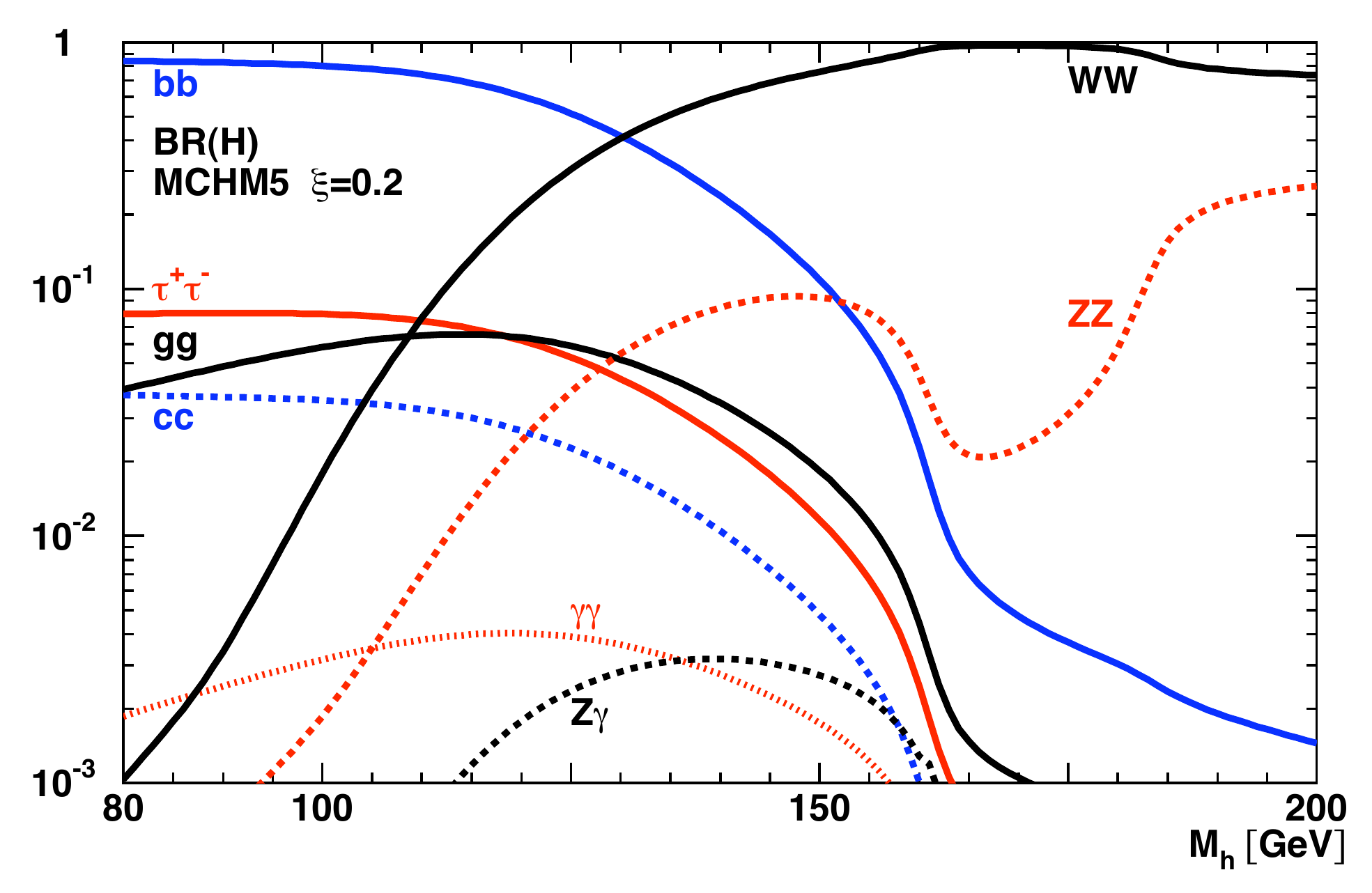} 
\vskip 0.5cm
\includegraphics[width=6.cm]{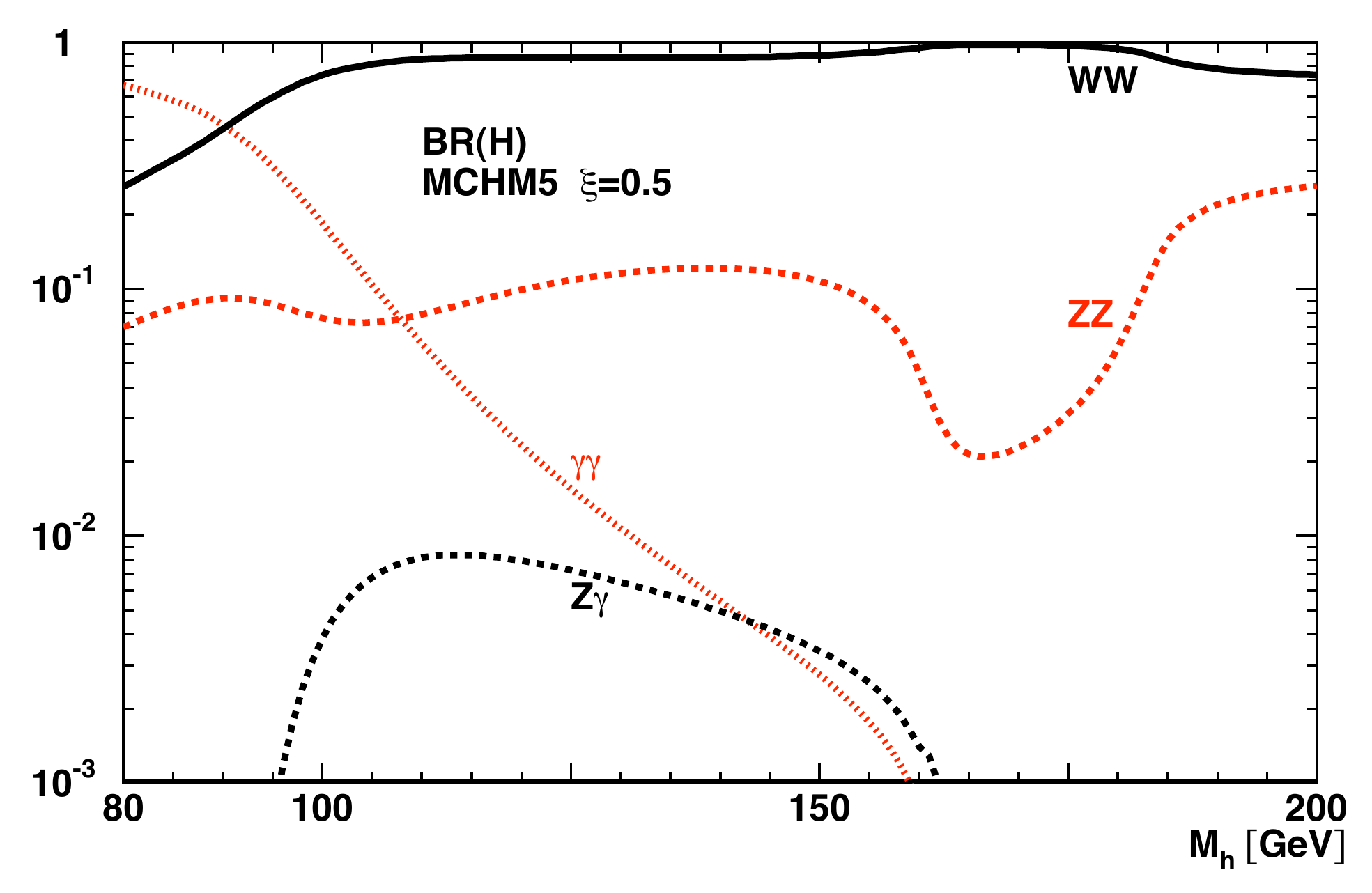} 
\hspace*{0.5cm}
\includegraphics[width=6.cm]{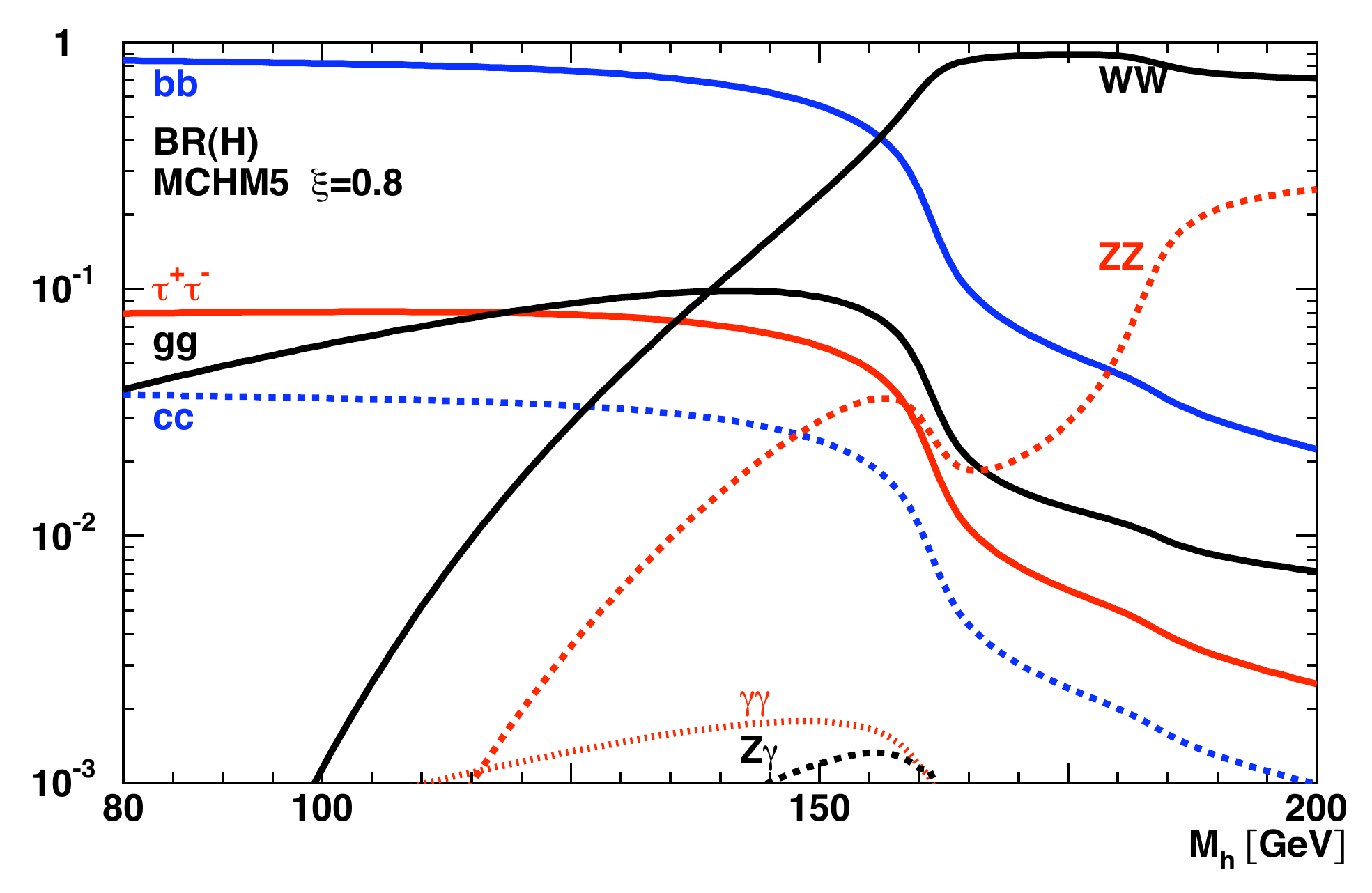} 
\caption{\label{composite_label9}
Higgs branching ratios as a function of the Higgs boson mass in
the SM ($\xi=0$, upper left) and MCHM5 with $\xi=0.2$ (upper right), 0.5 
(bottom left) and 0.8 (bottom right).}
\end{center}
\end{figure}
Higgs couplings the higher order QCD corrections to the decays are
unaffected and can readily be taken over from the
SM. 

Fig.~\ref{composite_label9} shows the SM branching ratios and the
composite Higgs branching ratios of MCHM5 for three representative
values of $\xi=0.2,0.5,0.8$ in the mass range favoured by composite
Higgs models between 80 and 200 GeV, which has not been completely
excluded by the LEP bounds yet (see Section~\ref{composite_label3}). 
The branching ratios have been obtained with the program
HDECAY~\cite{Djouadi:1997yw}, where the modifications due to the
composite nature of the Higgs boson have been implemented.
For $\xi=0.2$ the behaviour is almost the same as in the SM, with the
Higgs below $\sim 2M_Z$ decaying dominantly into $b\bar{b}$ and a pair of
massive gauge bosons, one or two of them being virtual. Above the gauge
boson threshold, it almost exclusively decays into $WW,ZZ$. The decays
into $\gamma \gamma$ and $Z\gamma$ are slightly enhanced compared to
the SM though, behaviour which culminates at $\xi = 0.5$. Here, due to the
specific Higgs fermion coupling in MCHM5, see Eq.~(\ref{composite_label8}),
the decays into fermions and fermion-loop mediated decays into gluons
are closed and the branching ratio into $\gamma\gamma$ dominates in
the low Higgs mass region. This cannot be exploited for the LHC
searches, however, which rely on this search channel in the low mass
region, since the gluon fusion production is absent for the same
reason and the vector boson fusion process is suppressed by a factor
two compared to the SM. At $\xi=0.8$ the branching ratios into
fermions dominate at low-Higgs mass and are enhanced compared to the
SM above the gauge boson threshold, which is due to the enhancement
factor in the Higgs fermion coupling, while the Higgs couplings to
massive gauge bosons are suppressed.

\section{LHC production cross sections \label{composite_label4}}
The Higgs boson search channels at the LHC can be significantly
changed in composite Higgs models due to the modified production cross
sections and branching ratios. The main characteristics of the
production cross sections shall be presented here. At the LHC the relevant
production channels are 

\underline{\it Gluon fusion:} The gluon fusion process $gg\to H$~\cite{Georgi:1977gs}
 constitutes the dominant production mechanism in the SM. At
leading order it is mediated by heavy quark loops. The next-to-leading
order QCD corrections~\cite{Spira:1995rr}, which enhance the cross section by
50-100\%, do not involve Higgs couplings and thus are unaffected by the
composite nature of the Higgs boson in our specific
parameterization. The NLO gluon fusion cross section
in the composite model can hence be obtained from the SM by
\beq
\begin{array}{llll}
\sigma_{NLO} (gg\to H) &=& (1-\xi) \,\, \sigma^{SM}_{NLO} (gg \to H) &
\qquad \mbox{MCHM4} \\[0.1cm]
\sigma_{NLO} (gg\to H) &=& \frac{(1-2\xi)^2}{(1-\xi)} \, 
\sigma^{SM}_{NLO} (gg \to H) &
\qquad \mbox{MCHM5} \;.
\end{array} 
\eeq

\underline{\it $W/Z$ boson fusion:} Weak boson fusion $qq\to qq+
W^*W^*/Z^* Z^* \to qqH$~\cite{Cahn:1983ip,Hikasa:1985ee,Altarelli:1987ue} is the
next important SM Higgs production process. Due to the additional forward
jets, which allow for a strong background reduction, it plays an
important role for the Higgs boson search. NLO QCD corrections~\cite{Spira:1997dg,Han:1992hr}, 
accounting for a 10\% correction, are
unaffected by the modified composite Higgs couplings, so that for our
models it is given by 
\beq
\sigma_{NLO} (qqH) = (1-\xi) \,\, \sigma^{SM}_{NLO} (qqH)
\qquad \mbox{for MCHM4 and MCHM5} \;. 
\eeq

\underline{\it Higgs-strahlung:} In the intermediate mass range $M_H \lsim
2 M_Z$ Higgs-strahlung off $W,Z$ bosons $q\bar{q} \to Z^*/W^* \to H+
Z/W$  provides another production mechanism~\cite{Glashow:1978ab,Kunszt:1991xk}. The cross section including NLO
QCD corrections, which add $\sim 30$\% in the SM~\cite{Spira:1997dg,Han:1991ia}, is given by 
\beq
\sigma_{NLO} (VH) = (1-\xi) \,\, \sigma^{SM}_{NLO} (VH)
\qquad \mbox{for MCHM4 and MCHM5} \;.
\eeq

\underline{\it Higgs radiation off top quarks:} This production
mechanism~\cite{Raitio:1978pt,Ng:1983jm,Kunszt:1984ri,Gunion:1991kg,Marciano:1991qq} only plays a role for Higgs masses $\lsim 150$ GeV. NLO QCD
corrections increase the cross section at the LHC by $\sim 20$\%~\cite{Beenakker:2001rj,Beenakker:2002nc,Dawson:2002tg}, 
and in the composite Higgs models studied here it is given by
\beq
\begin{array}{llll}
\sigma_{NLO} (Ht\bar{t}) &=& (1-\xi) \,\, \sigma^{SM}_{NLO}
(Ht\bar{t}) & \qquad \mbox{MCHM4} \\[0.1cm]
\sigma_{NLO} (Ht\bar{t}) &=& \frac{(1-2\xi)^2}{(1-\xi)} \,
\sigma^{SM}_{NLO} (Ht\bar{t}) &  \qquad \mbox{MCHM5} \; .
\end{array}
\eeq  
While being excluded as discovery channel due to the large background
and related uncertainties, in MCHM5 it may provide an interesting
search channel for large values of $\xi$ near one due to a significant
enhancement factor.

\begin{figure}[h]
\begin{center}
\includegraphics[width=6.cm]{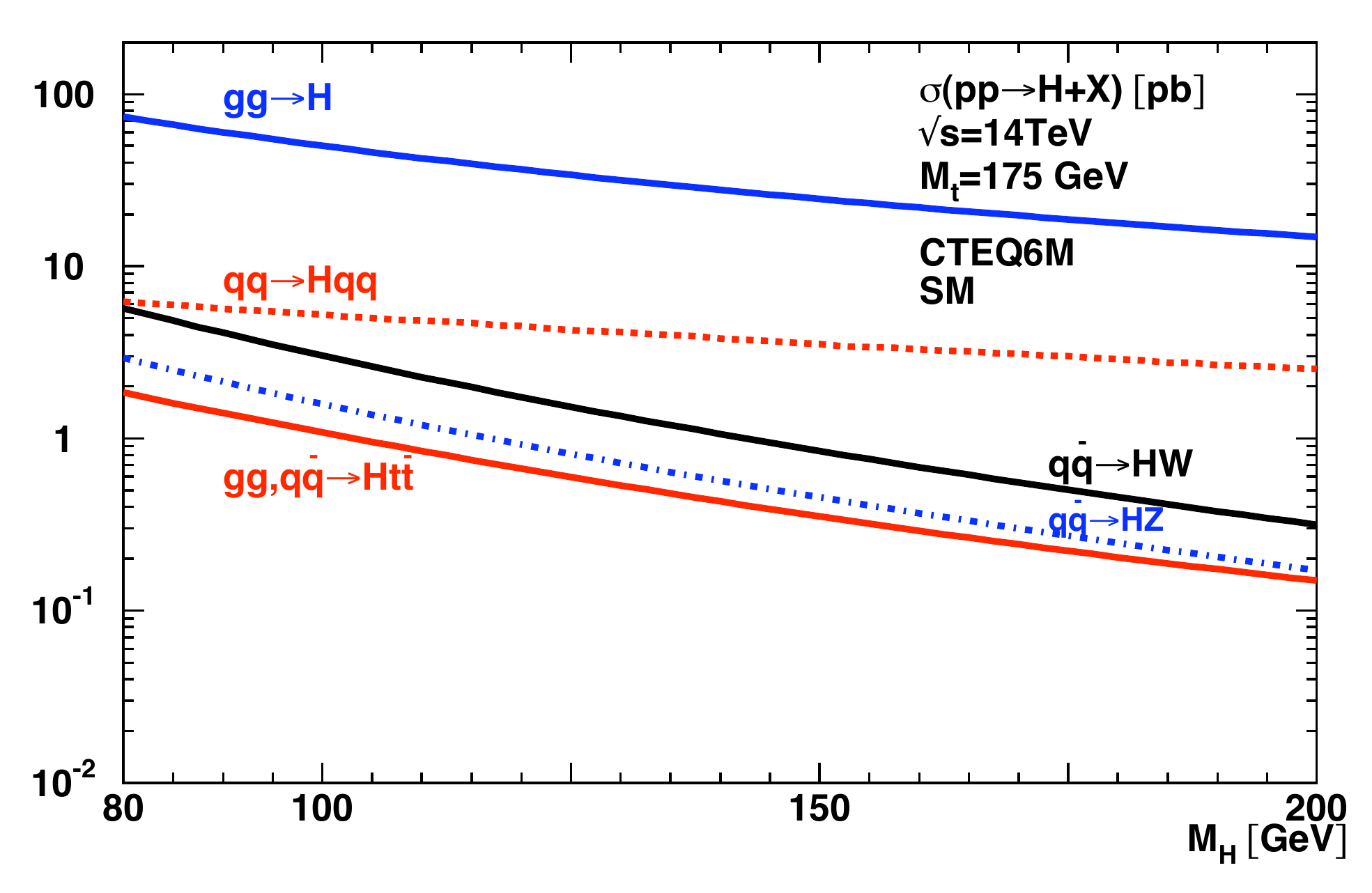} 
\hspace*{0.5cm}
\includegraphics[width=6.cm]{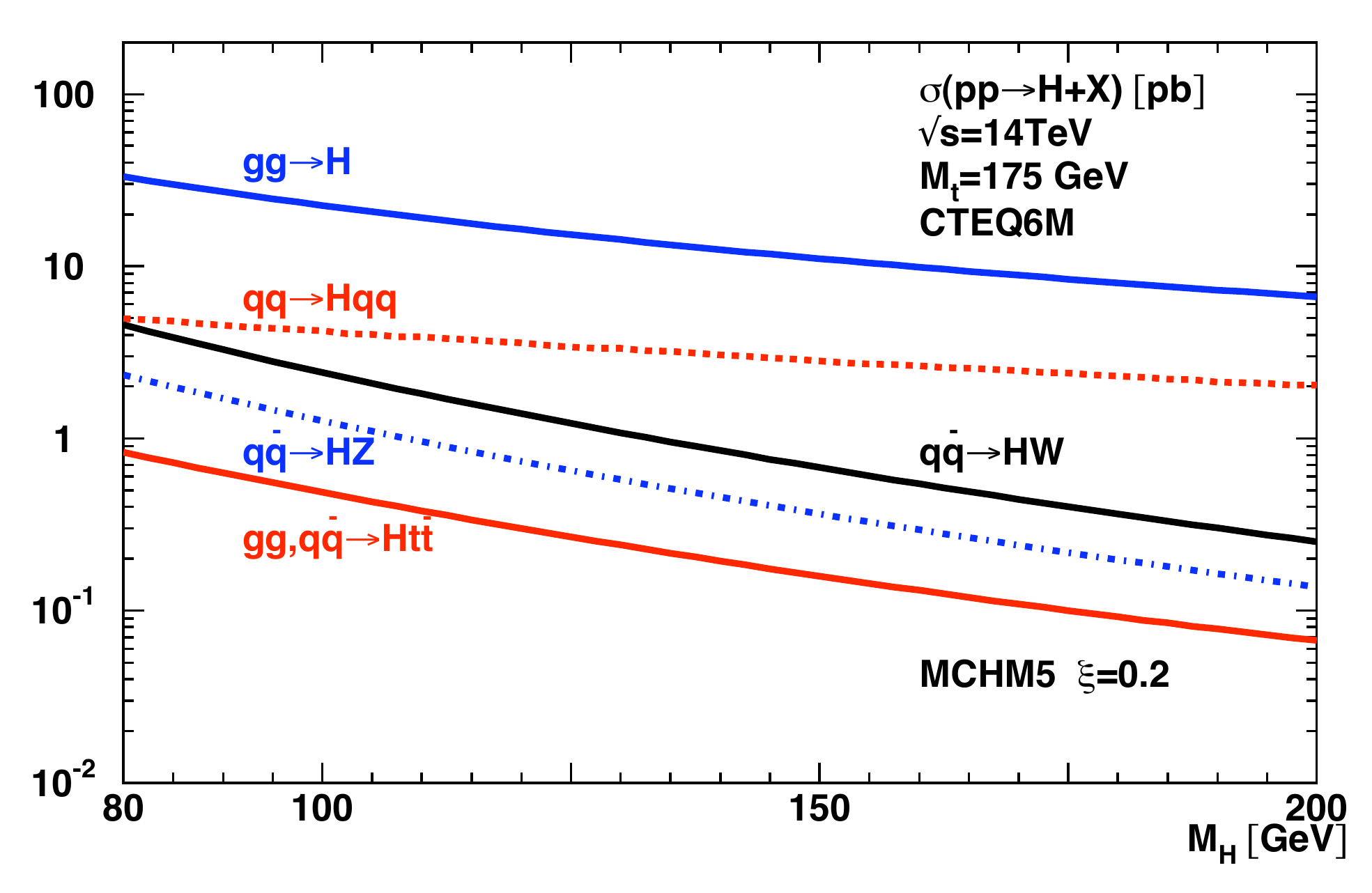} 
\vskip 0.5cm
\includegraphics[width=6.cm]{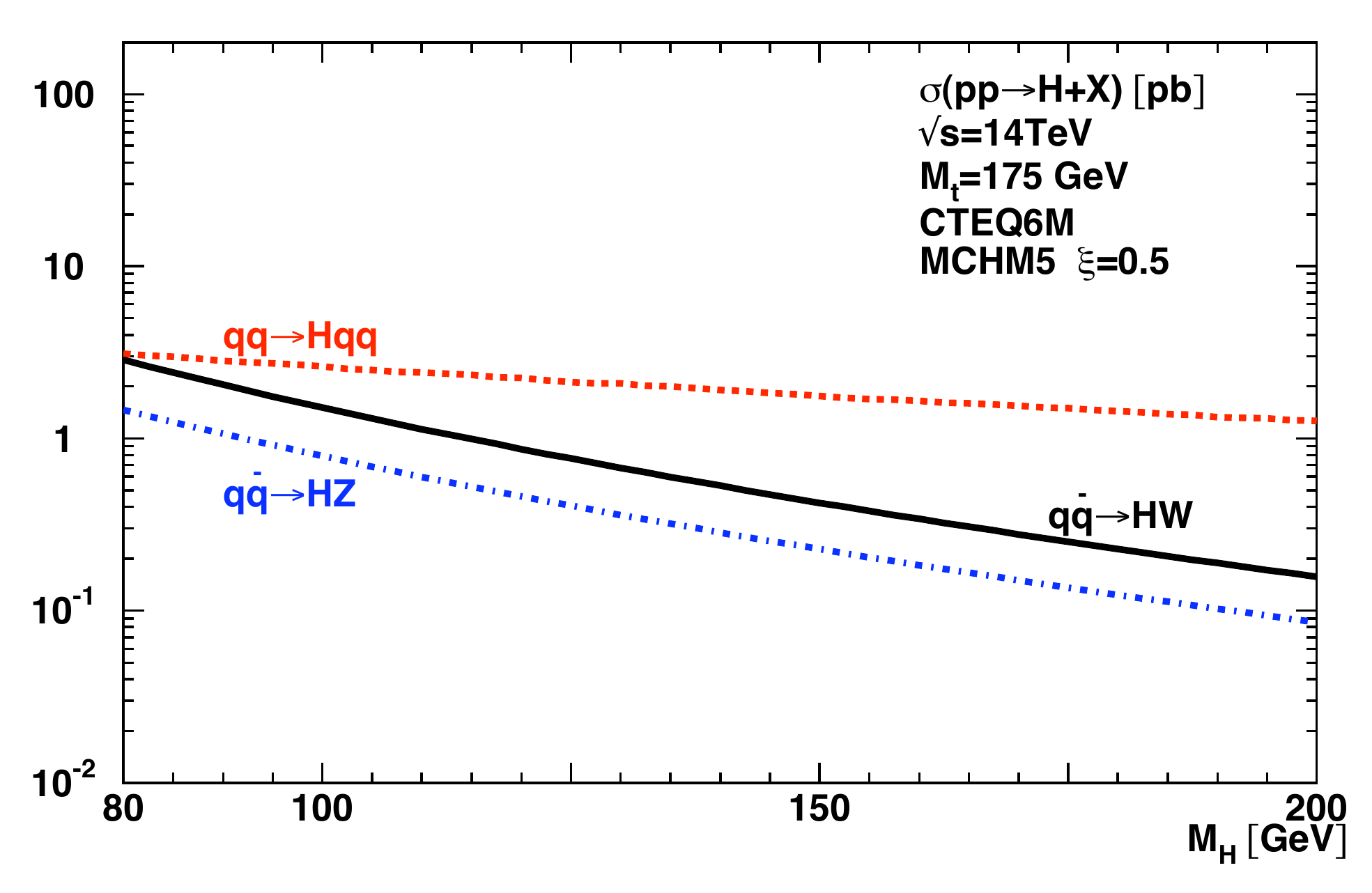} 
\hspace*{0.5cm}
\includegraphics[width=6.cm]{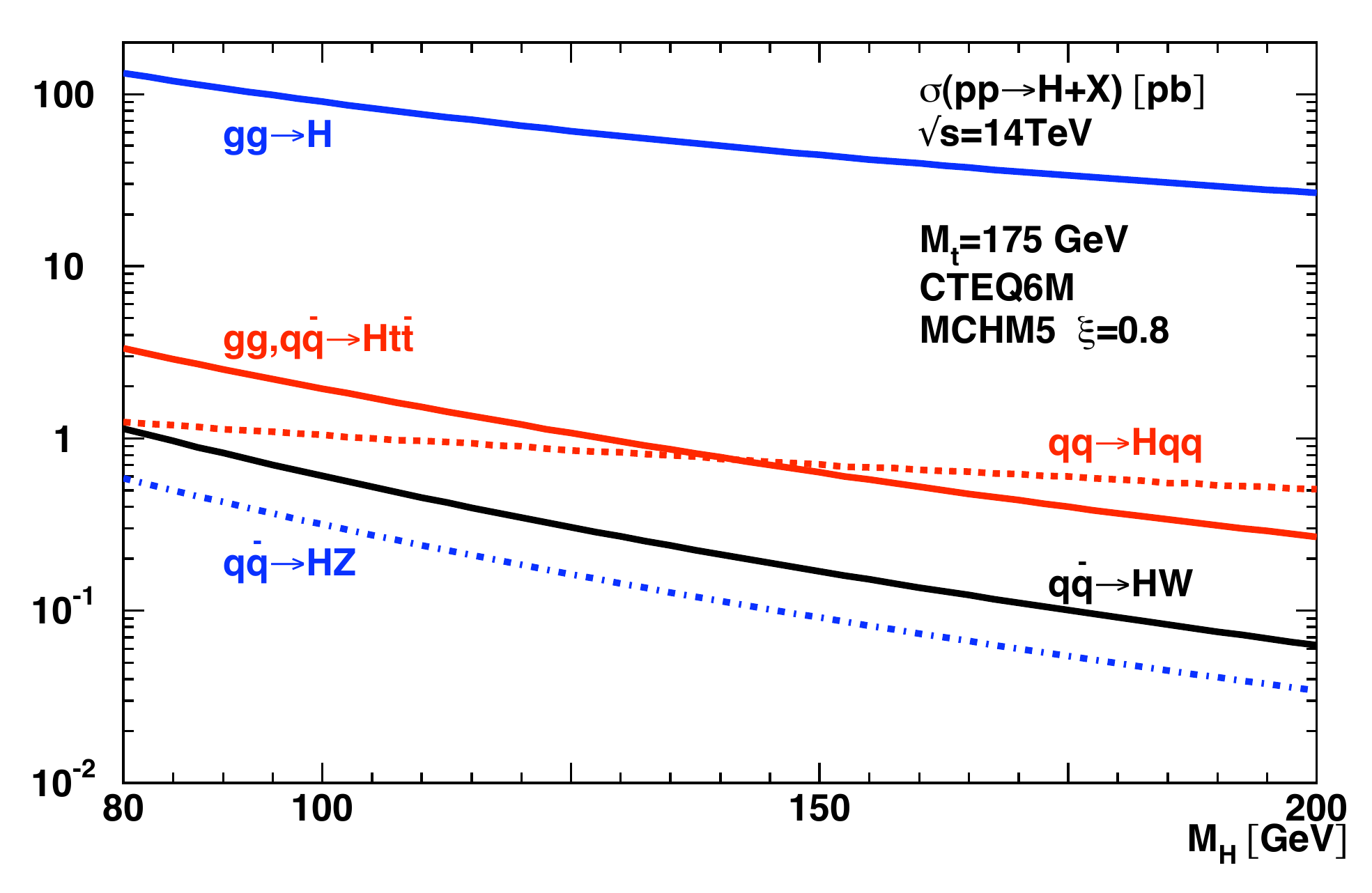} 
\caption{\label{composite_label11}
The LHC Higgs boson production cross-sections as a function
of the Higgs boson mass in the SM ($\xi=0$, upper left) and for MCHM5 
with $\xi=0.2$ (upper right), 0.5 (bottom left) and 0.8 (bottom right). The cross
  sections include NLO QCD corrections and have been obtained by use
  of the programs HIGLU~\protect\cite{Spira:1995mt}, VV2H~\protect\cite{Spira:Prog}, V2HV~\protect\cite{Spira:Prog}, 
  HQQ~\protect\cite{Spira:Prog}.}
\end{center}
\end{figure}

Fig.~\ref{composite_label11} shows the production cross sections as
function of $M_H=80...200$ GeV in the SM and MCHM5 for
$\xi=0.2,0.5$ and $0.8$. For $\xi=0.2$ the inclusive cross section is
considerably reduced due to reduced couplings in the production cross sections,
situation which is even worse for $\xi=0.5$ where the gluon fusion and
$Ht\bar{t}$ cross sections vanish and the others are reduced. For
$\xi=0.8$ the situation is reversed due to the significantly enlarged
gluon fusion process. The cross sections for MCHM4 are not shown
separately. They can be obtained from the SM ones by multiplying each
with $1-\xi$.

\section{Statistical significances\label{composite_label12}}
In order to study how the Higgs prospects of discovery will be changed
in composite models, we evaluated the statistical significances for
the different search channels at the LHC. We referred to the CMS
analyses~\cite{Ball:2007zza}. The results presented hereafter are not
significantly changed when applying the ATLAS analyses~\cite{Aad:2009wy}. 
Assuming that only the signal rates are changed but not the
backgrounds rates, since only Higgs couplings are affected in our
models, the significances in MCHM4 and MCHM5 can be obtained by
applying a rescaling factor $\varkappa$ to the number of signal
events. Referring to a specific search channel, it is given by
taking into account the change in the production process $p$ and in the
subsequent decay into a final state $X$ with respect to the SM, hence
\beq
\varkappa = \frac{\sigma_{p} \, BR(H\to X)}{\sigma_{p}^{SM}
  BR(H^{SM} \to X)} \;.
\eeq
The number of signal events $s$ is obtained from the SM events
$s^{SM}$ by
\beq
s = \varkappa\, \cdot \, s^{SM} \; ,
\eeq
where $s^{SM}$ after application of all cuts is taken from the
experimental analyses. The signal events $s$ and the  background
events after cuts, i.e. $b \equiv b^{SM}$, are used to calculate  the
corresponding significances in the composite Higgs model. 
The various channels studied are

\underline{$H\to \gamma\gamma$:} This channel is crucial for Higgs
searches at low masses $M_H \lsim 150$ GeV. Despite the clean signal,
the channel is challenging due to small signal and large background
rates. The production is given by the inclusive cross section composed
of gluon fusion, vector boson fusion, Higgs-strahlung and $Ht\bar{t}$
production.

\underline{$H\to ZZ \to 2l 2l'$:} The gold-plated channel for Higgs
masses above $\sim 130$ GeV with the Higgs decaying through $ZZ^{(*)}$
in the clean $4e,2e2\mu$ and $4\mu$ final states is based on
gluon fusion and vector boson fusion in the production. Since the production
cross section is large as well as the branching ratio into $ZZ^{(*)}$ it allows for a
precise determination of the Higgs boson mass and cross section.

\underline{$H\to WW \to 2l 2\nu$:} Higgs decay into $WW$ with
subsequent decay in leptons is the main discovery channel in the
intermediate region $2M_W \lsim M_H \lsim 2M_Z$. Spin correlations can
be exploited to extract the signal from the background. The CMS
analyses use gluon and vector boson fusion to get the signal rates.

\underline{$H\to WW \to l\nu jj$:} Higgs production in vector boson
fusion with subsequent decay $H\to WW \to l\nu jj$ covers the mass
region $160$ GeV$\lsim M_H \lsim$180 GeV, where the $H\to ZZ^{(*)}$
branching ratio is largely suppressed. The event topology with two
energetic forward jets and suppressed hadronic activity in the central
region can be exploited to extract the signal from the background.

\underline{$H\to \tau \tau \to l+j+E_T^{miss}$:} This channel with the
Higgs produced in vector boson fusion, adds to the difficult Higgs
search in the low mass region $M_H \lsim 140$ GeV. The specific
signature of vector boson fusion production (see above) helps for the
extraction of the signal.

\begin{figure}[ht]
\begin{center}
\includegraphics[width=7.cm]{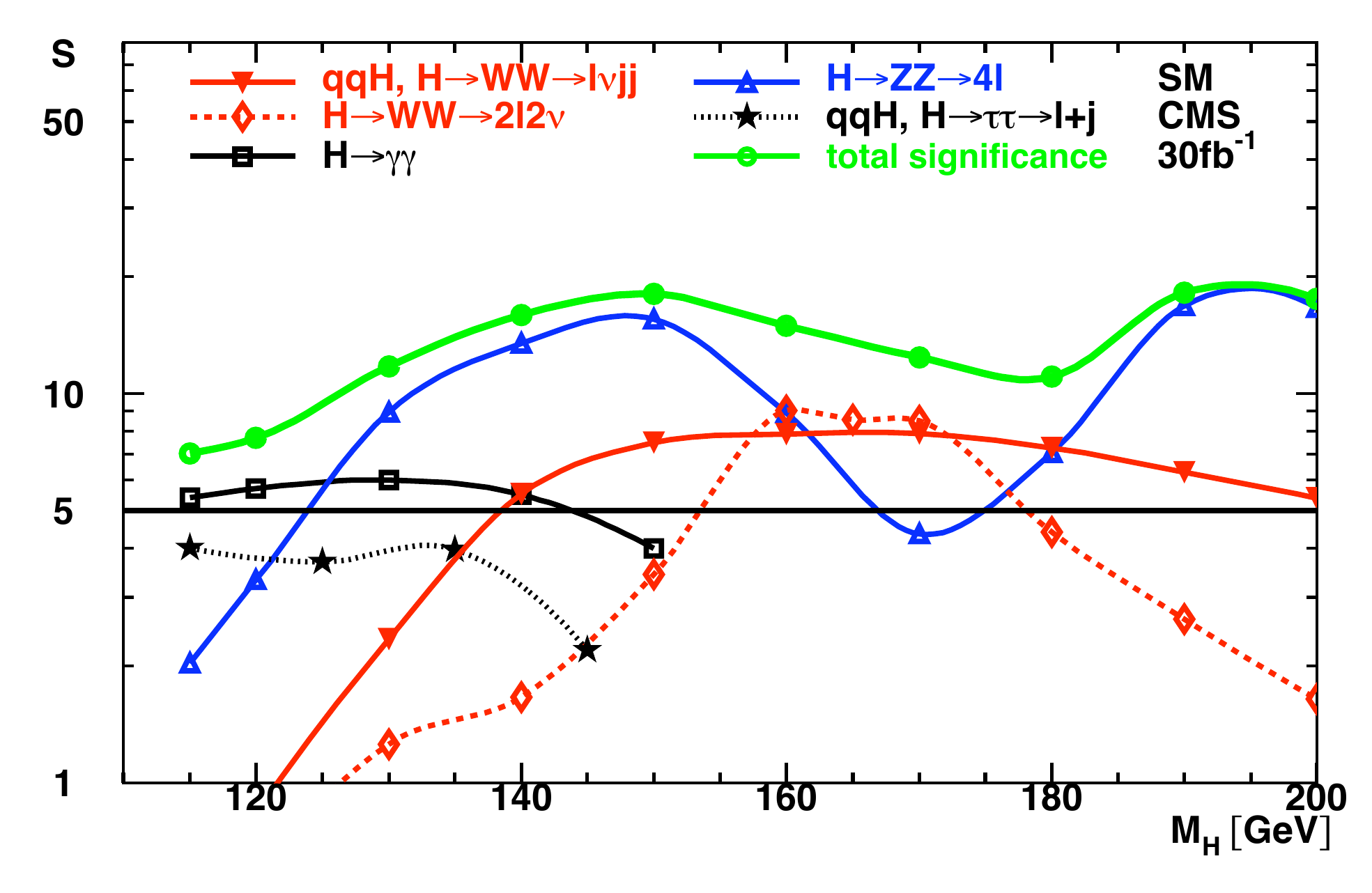} 
\hspace*{0.5cm}
\includegraphics[width=7.cm]{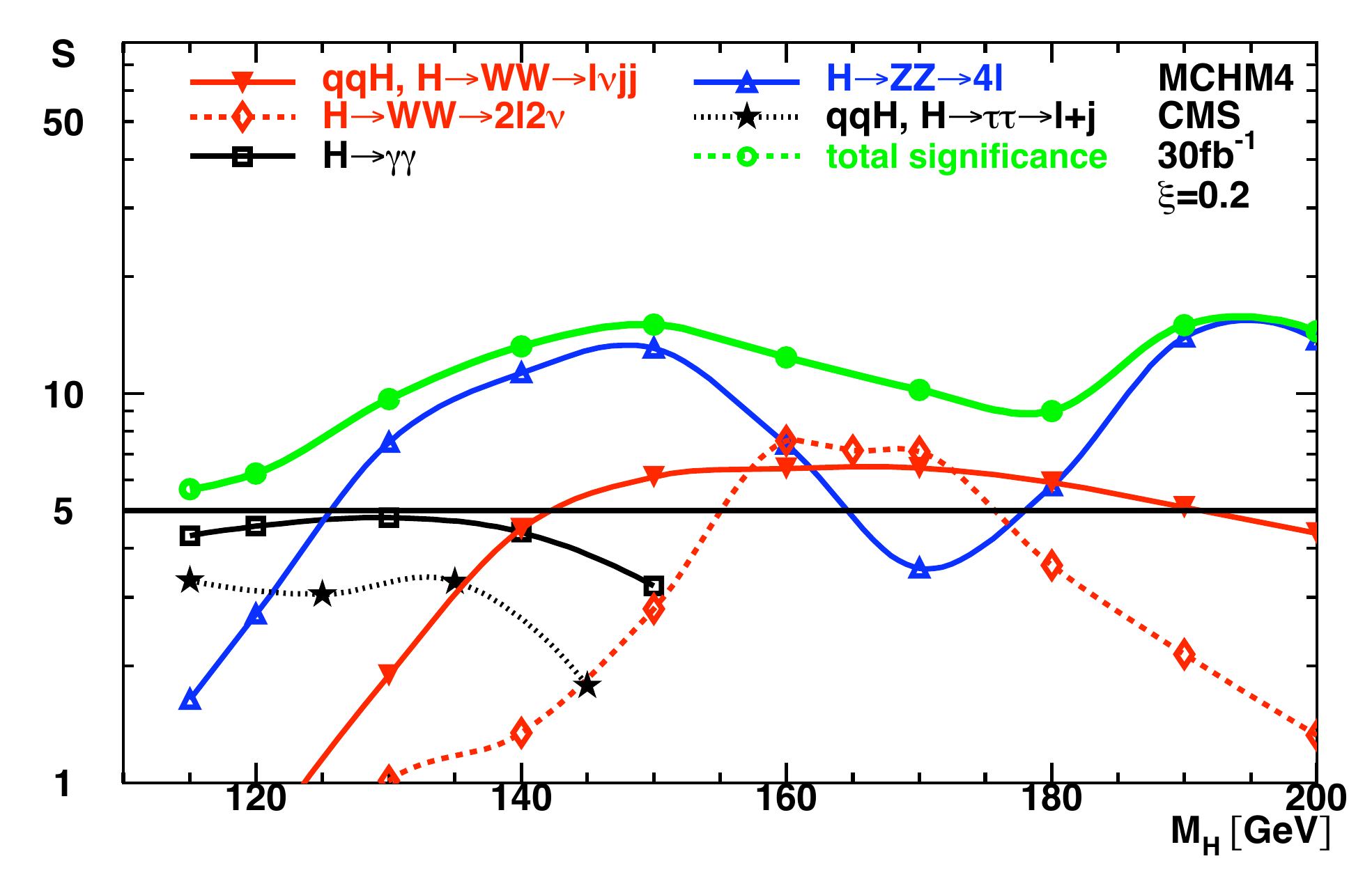} 
\vskip 0.5 cm
\includegraphics[width=7.cm]{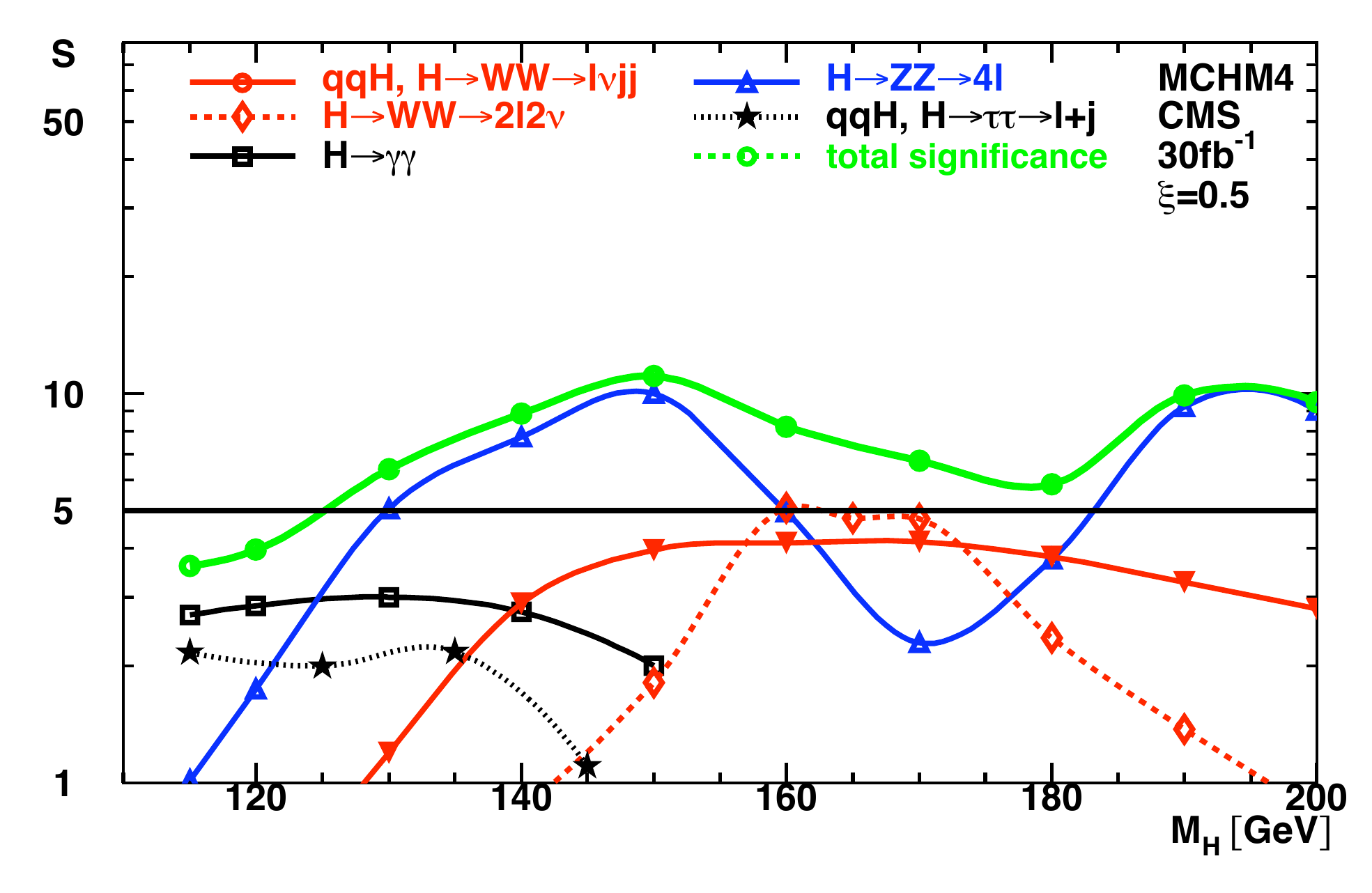} 
\hspace*{0.5cm}
\includegraphics[width=7.cm]{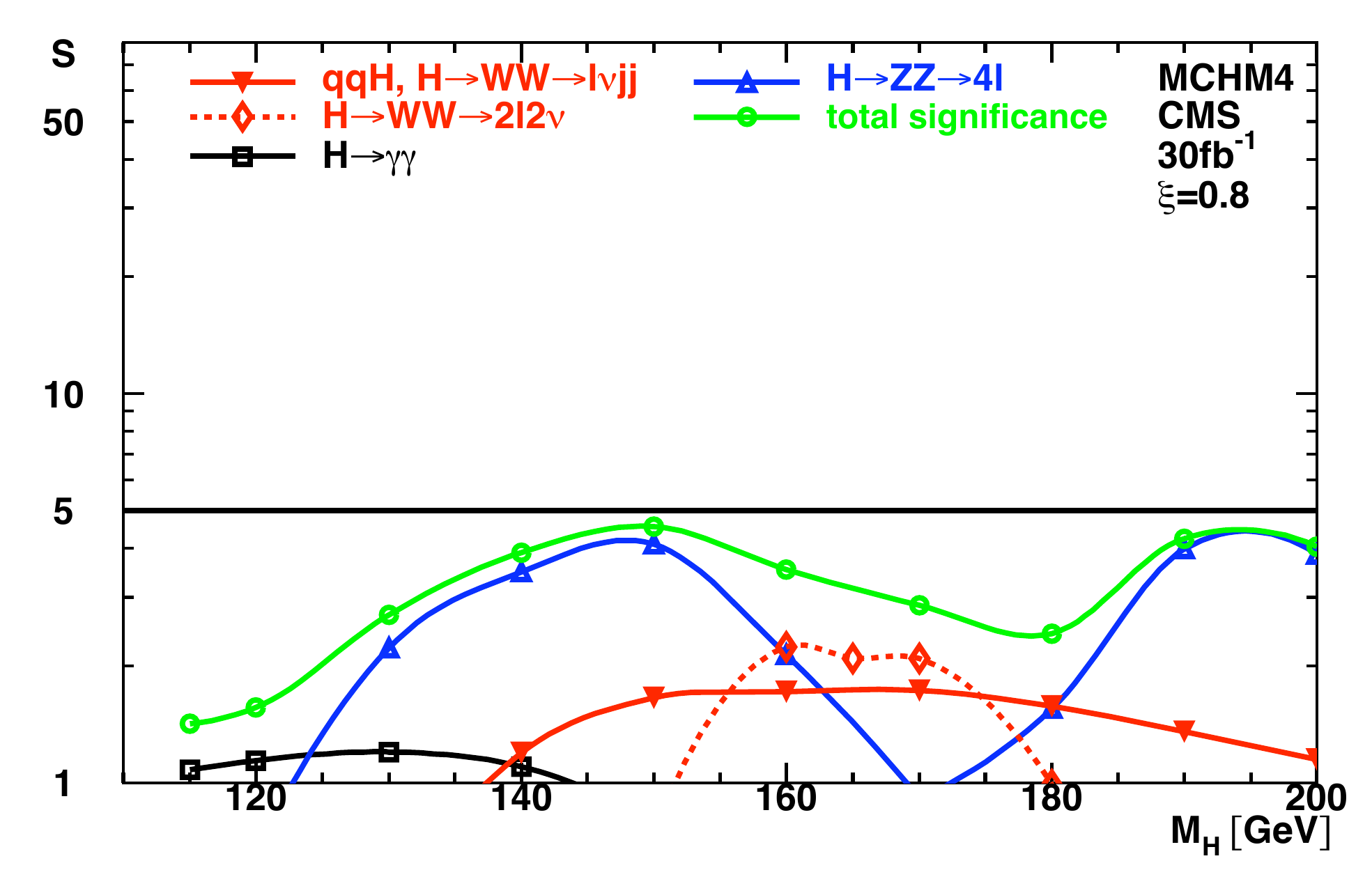} 
\caption{\label{composite_label14}
The significances in different channels as a function of the Higgs boson 
mass in the SM ($\xi=0$, upper left) and for 
MCHM4 with $\xi=0.2$ (upper right), 0.5 (bottom left) and 0.8 (bottom 
right).} \end{center}
\end{figure}
\begin{figure}[ht]
\begin{center}
\includegraphics[width=7.cm]{Espinosa/figures/composite_fig9.pdf} 
\hspace*{0.5cm}
\includegraphics[width=7.cm]{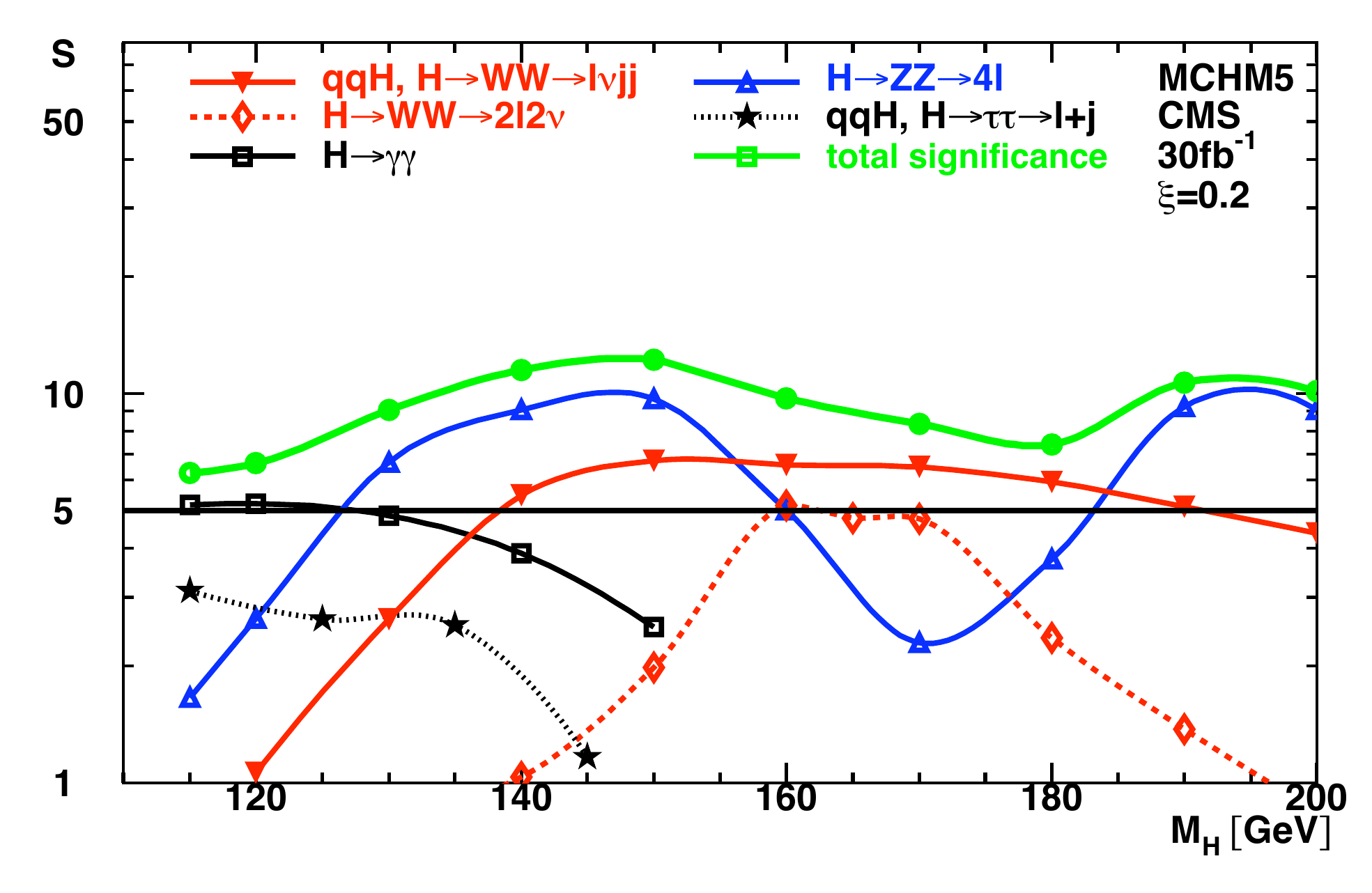} 
\vskip 0.5 cm
\includegraphics[width=7.cm]{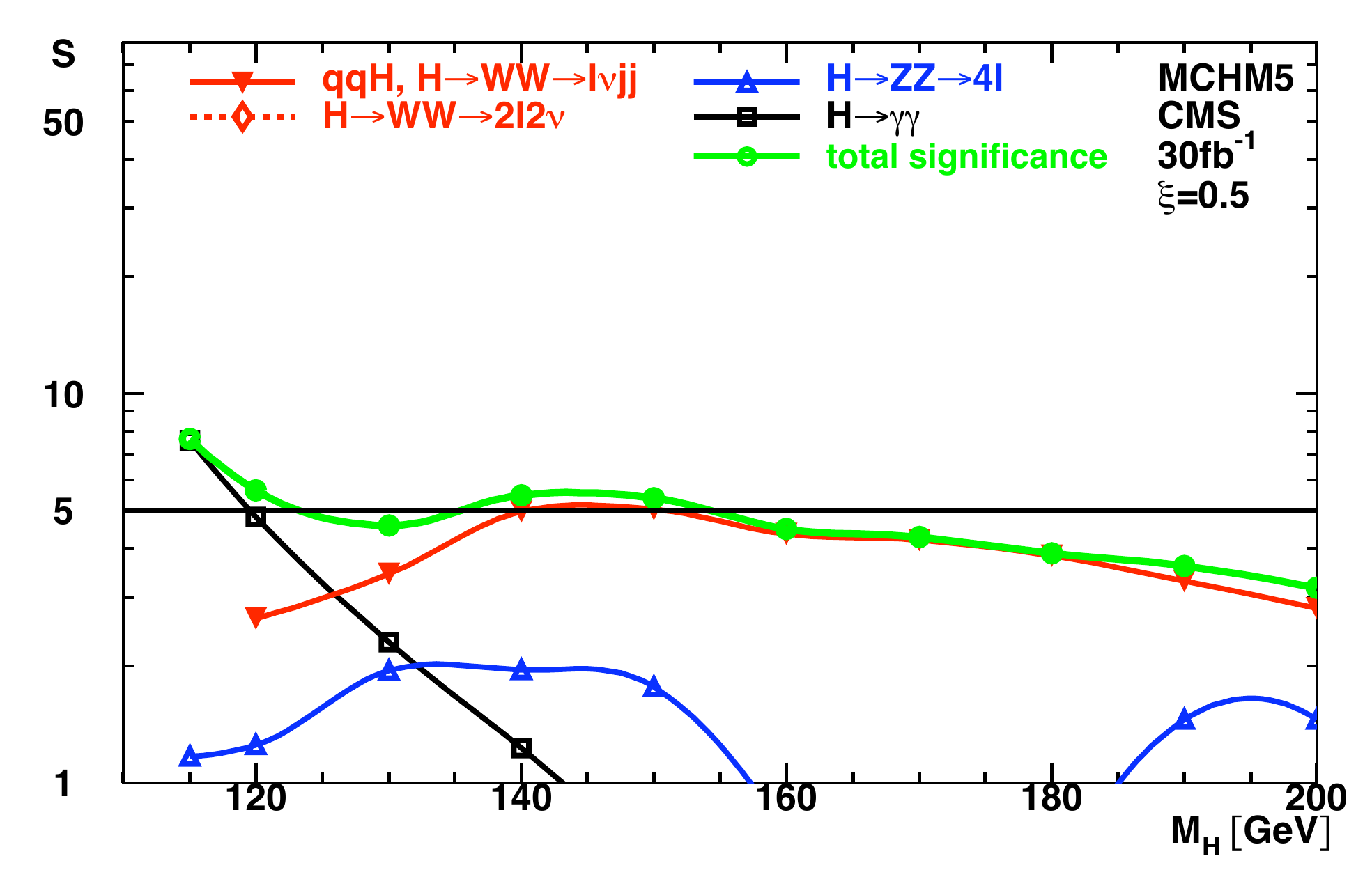} 
\hspace*{0.5cm}
\includegraphics[width=7.cm]{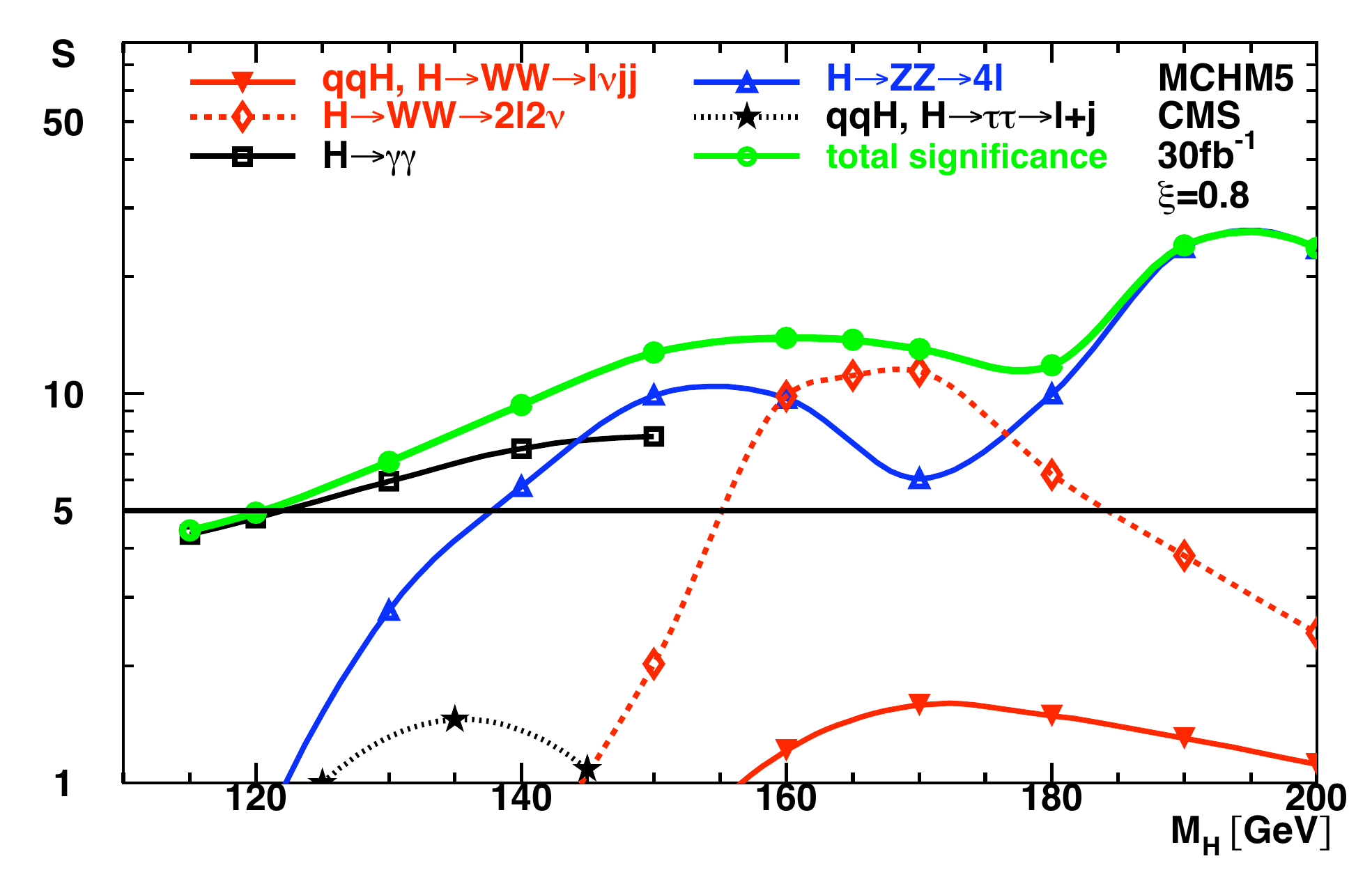} 
\caption{\label{composite_label15}
The significances in different channels as a function
  of the Higgs boson mass in the SM ($\xi=0$, upper left) and for 
MCHM5 with $\xi=0.2$ (upper right), 0.5 (bottom left) and 0.8 (bottom 
right).} \end{center}
\end{figure}

For more details on each search channel and on the significance
estimators we used we
refer the reader to~\cite{Espinosa:2010vn}. In Figs.~\ref{composite_label14} 
and~\ref{composite_label15} we present the SM significance 
(for comparison) and the MCHM4 and MCHM5 significances for
$\xi=0.2,0.5,0.8$. The results should be understood as estimates. They
cannot replace experimental analyses. But they can serve as a
guideline of what is changed in composite models and where to be
careful when it comes to interpretation of experimental results.
As can be inferred from Figs.~\ref{composite_label14}, in MCHM4 in all
search channels the significance is always below the corresponding
significance in the SM. With the branching ratios being unchanged,
this is due to the production cross sections which are all suppressed
by the universal factor $1-\xi$. The Higgs search will hence be much
more difficult. For $\xi=0.8$ the total significance even drops
below 5.

In MCHM5 the behaviour of the significances is more involved due to
the interplay of modified production and decay channels. For $\xi=0.2$
the reduction in production channels cannot be compensated by the
enhancement in the branching ratios into $\gamma\gamma$ and massive
gauge bosons, so that the significances are below the SM ones. In
total the significance is also below the total MCHM4 significance,
as gluon fusion production which contributes to the main search
channels, is more strongly reduced in MCHM5. The situation looks even
worse for $\xi=0.5$ where gluon fusion (and also $Ht\bar{t}$
production) is completely erased from the list of production
channels. Only for low Higgs masses the strong enhancement in the
$\gamma\gamma$ branching ratio can rise the significance above 5,
even for $M_H$ below the LEP limit, although that has to be  confirmed by detailed experimental analyses
though. For higher Higgs masses one has to rely on weak boson fusion
with $H\to WW$ decay. For $\xi=0.8$ the picture is totally different
from MCHM4. The production is completely taken over by gluon fusion
and leads to large significances in the massive gauge boson final
states. Also $\gamma\gamma$ final states contribute for $M_H \gsim
120$ GeV, and probably for $M_H>150$ GeV, although this also has to be confirmed by experimental analyses though.

\section{Conclusions \label{composite_label13}}
We have shown by focusing on two particular directions in the
parameter space of the composite Higgs model, that the search modes
and significances can deviate significantly from the SM
expectations. In the MCHM4 model all couplings are reduced compared to
the SM values and hence the Higgs searches deteriorate. In the MCHM5
model, however, the production in gluon fusion is enhanced if the
composite scale is low enough. The significances can then be larger
than in the SM case. Once the Higgs boson will show up in the LHC
experiments, the study of the relative importance of the various
production and decay channels will thus provide us to a certain extent with
information on the dynamics of the Higgs sector and tell us whether the
electroweak symmetry breaking is weak or strong.

%% file: Lane/Lane.tex
\chapter{Low-Scale Technicolor at the 10 TeV LHC}

{\it K.~Black, T.~Bose, E.~Carrera, S.~J.~Harper, K.~Lane, Y.~Maravin, A.~Martin and B.C.~Smith}\\

\begin{abstract}
  This report summarizes Low-Scale Technicolor (LSTC) and the work done by
  the LSTC-at-LHC group for Les Houches~2009. We study the reach of the LHC
  with $\sqrt{s} = 10\,\tev$ for the lightest $\tro,\tom,\ta$ technivectors
  decaying to $WZ$, $\gamma W$, $\gamma Z$ followed by leptonic decays of the
  weak bosons, and to $e^+e^-$. For the most part, we restrict ourselves to
  luminosities of $\CO(1\,\ifb)$. The revised $7\,\tev$ LHC run schedule for
  2010--11 was established as this report was being completed.

\end{abstract}

\section{Introduction}

\begin{figure}[!Hht]
 \begin{center}
\includegraphics[width=3.00in, height=2.35in, trim=60 60 60
60]{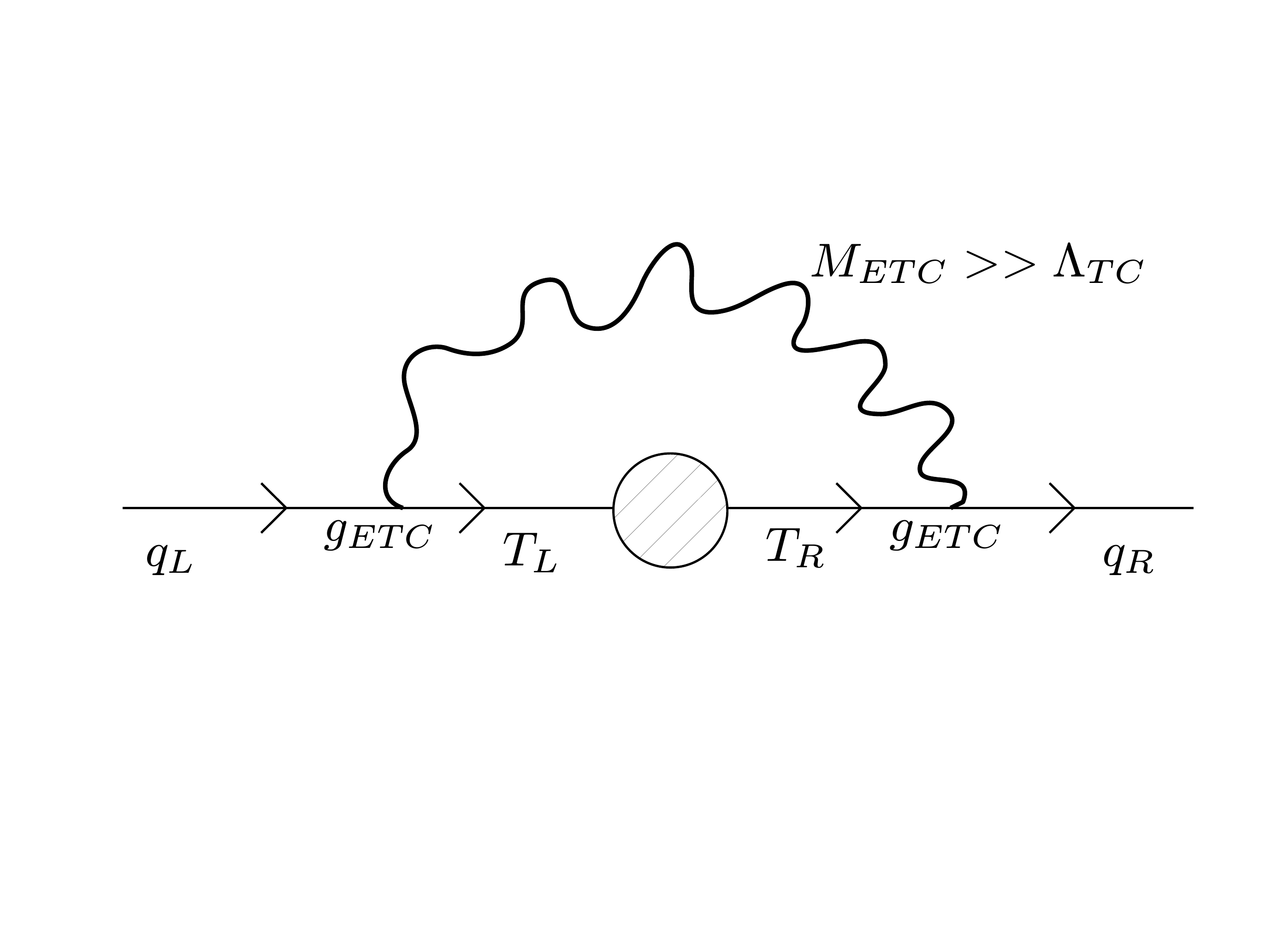}
\includegraphics[width=3.00in, height=2.35in, trim=60 60 60 60]{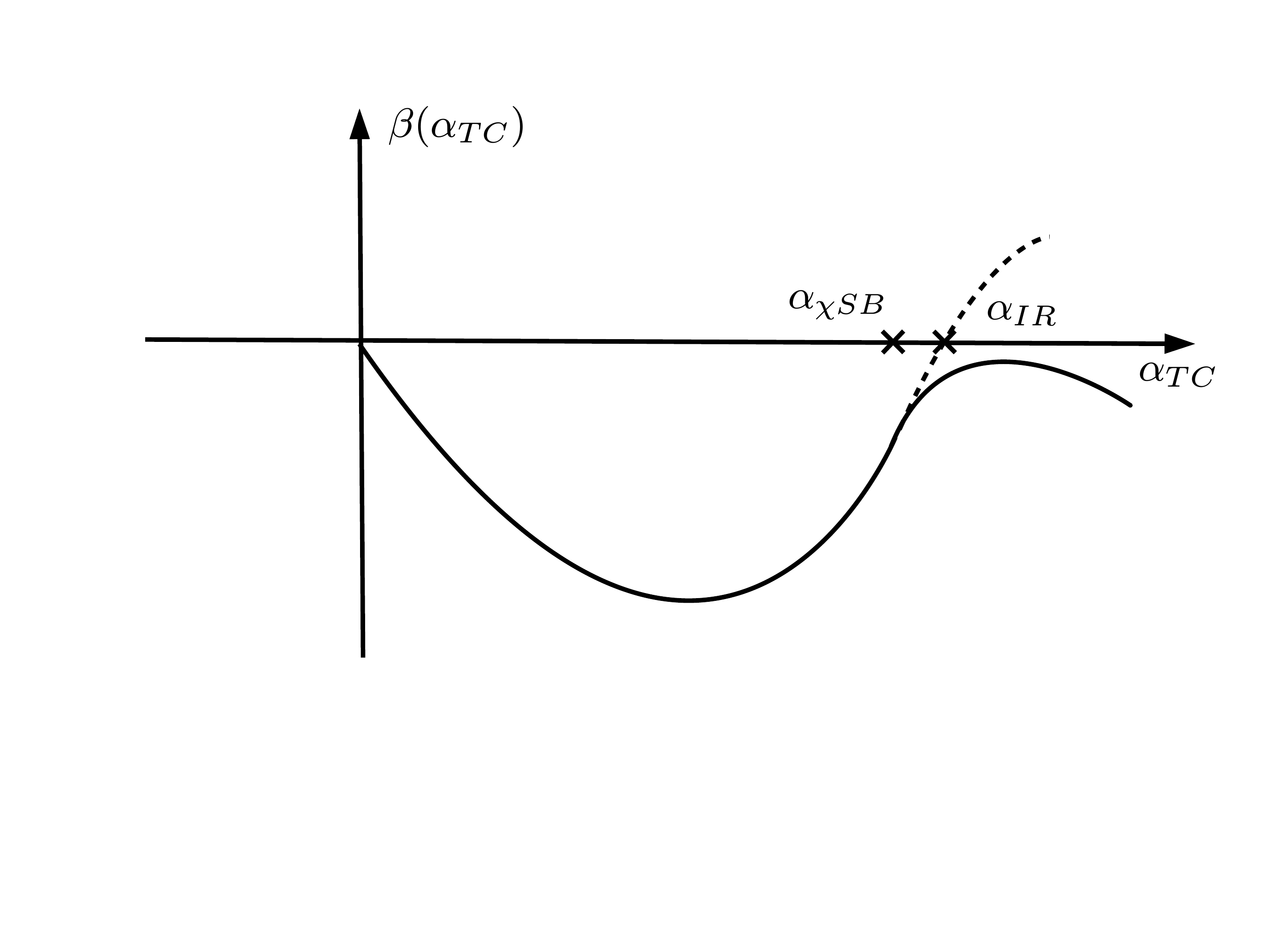}
\vskip -0.5in
 \caption{The quark and lepton mass generating mechanism in ETC (left).
   The $\beta$-function in walking technicolor, with the chiral symmetry
   breaking value of $\atc$ just below an approximate infrared fixed point
   (right).\label{fig:Livia}}
 \end{center}
 \end{figure}

Technicolor (TC)~\cite{Weinberg:1979bn,Susskind:1978ms,Lane:2002wv,
  Hill:2002ap} was invented to provide a natural and consistent
quantum-field-theoretic description of electroweak (EW) symmetry breaking ---
{\em without} elementary scalar fields. Extended technicolor
(ETC)~\cite{Eichten:1979ah} was invented to complete that description by
including quark and lepton flavors and their chiral symmetry breaking as
interactions of fermions and gauge bosons alone. In particular, from
Fig.~\ref{fig:Livia}, $m_{q,\ell} \simeq g_{ETC}^2\condetc/M_{ETC}^2$, where
$\condetc$ is the technifermion condensate renormalized at $M_{ETC}$. From
the beginning, ETC was recognized to have a problem with flavor-changing
neutral current interactions, especially those inducing $K^0$--$\bar K^0$
mixing. Masses $M_{ETC}$ of several 100, possibly 1000, TeV are required to
suppress these interactions to an acceptable level. The problem is that this
implies $m_{q,\ell}$ of at most a few~MeV if one assumes that, as in QCD, (1)
asymptotic freedom sets in quickly above the TC scale of a few $100\,\gev$ so
that $\condetc \simeq \condtc$ and (2) $\condtc$ can be estimated by scaling
from the quark condensates of QCD.  Walking technicolor~\cite{Holdom:1981rm,
  Appelquist:1986an,Yamawaki:1986zg, Akiba:1986rr} was invented to cure this
problem. The cure is that the QCD-based assumptions may not apply to
technicolor after all. In particular, in walking TC the gauge coupling
decreases very slowly, staying large for 100s, perhaps 1000s, of TeV and
remaining near its critical value for spontaneous chiral symmetry breaking.
Then, the $\bar T T$ anomalous dimension $\gamma_m \simeq 1$ over this large
energy range~\cite{Cohen:1988sq}, so that $\condetc \gg \condtc$ and
reasonable fermion masses result.\footnote{Except for the top quark, which
  needs an interaction such as topcolor to explain its large
  mass~\cite{Hill:1994hp}.}  The important lesson of walking technicolor is
that QCD-based assumptions for technicolor must, at best, be viewed with
suspicion and used with caution. In particular, all estimates of the
precision electroweak parameter $S$ for TC models~\cite{Peskin:1990zt,
  Golden:1990ig,Holdom:1990tc,Altarelli:1991fk} are based on scaling from QCD
and, as such, are untrustworthy~\cite{Lane:1993wz,Lane:1994pg}. Lattice
gauge-theoretic techniques appear to be a promising way to test the ideas of
walking technicolor in a nonperturbative way.

 A walking TC gauge coupling with $\gamma_m \simeq 1$ for a large energy
 range occurs if, as in Fig.~\ref{fig:Livia}, the critical coupling for chiral
 symmetry breaking lies just {\em below} a value at which the TC
 $\beta$-function vanishes (an infrared fixed point)~\cite{Lane:1991qh,
   Appelquist:1997fp}. This requires a {\em large} number of technifermions,
 which may be achieved by having $N_D \gg 1$ doublets in the fundamental
 representation $\bs{N}_{TC}$ of the TC gauge group, $SU(N_{TC})$, or by
 having a few doublets in higher-dimensional
 representations~\cite{Lane:1989ej,Dietrich:2005wk}. In the latter case,
 constraints on ETC representations~\cite{Eichten:1979ah} almost always imply
 other technifermions in the fundamental representation as well. In either
 case, then, there generally are technifermions whose technipion ($\tpi$)
 bound states have a decay constant $F_1^2 \ll F_\pi^2 = (246\,\gev)^2$. This
 low scale implies there are, in addition to the $\tpi$, technihadrons
 $\tro$, $\tom$ and $\ta$ with masses well below a TeV. We refer to this
 situation as low-scale technicolor (LSTC)~\cite{Lane:1989ej, Eichten:1996dx,
   Eichten:1997yq}. These technivector mesons can be produced as $s$-channel
 resonances in $q\bar q$ annihilation at the LHC. As we discuss next, they
 will be extremely narrow, with striking signatures visible above manageable
 backgrounds.

\vskip 0.50cm

There are two important consequences of this picture of walking TC. First, to
restate what we just said, $N_D > 1$ technifermion doublets implies the
existence of physical technipions, some of which couple to the lightest
technivector mesons. Second, since $M_{\tpi}^2 \propto \langle\ol T T \ol T
T\rangle_{ETC} \approx (\condetc)^2$, walking TC enhances the masses of
technipions much more than it does other technihadron masses. Thus, it is
very likely that the lightest $M_{\tro} < 2M_{\tpi}$ and that the two and
three-$\tpi$ decay channels of the light technivectors are
closed~\cite{Lane:1989ej}. This further implies that these technivectors are
{\em very} narrow, a few~GeV or less, because their decay rates are
suppressed by phase space and/or small
couplings (see below). 

A simple phenomenology of LSTC is provided by the Technicolor Straw-Man Model
(TCSM)~\cite{Lane:1999uh,Lane:2002sm,Eichten:2007sx}. The TCSM's ground rules
and major parameters are these:

\begin{enumerate}
  
\item The lightest doublet of technifermions $(T_U,T_D)$ are color-$\suc$
  singlets.\footnote{Colored technifermions get a large contribution to
    their mass from $\suc$ gluon exchange. We also assume implicitly that, in
    the case of $N_D$ fundamentals, ETC interactions split the doublets
    substantially.}
  
\item The decay constant of the lightest doublet's technipions is $F_1 =
  (F_\pi = 246\,\gev)\cdot \sin\chi$. In the case of $N_D$ fundamentals,
  $\sin^2\chi \cong 1/N_D \ll 1$. In the case of two-scale TC, $F_\pi =
  \sqrt{F_1^2 + F_2^2} = 246\,\gev$ with $F_1^2/F_2^2 \cong \tan^2\chi \ll
  1$.
  
\item The isospin breaking of $(T_U,T_D)$ is small. Their electric charges
  are $Q_U$ and $Q_D = Q_U - 1$. In the TCSM, the rates for several decay
  modes of the technivectors to transversely-polarized electroweak gauge
  bosons ($\gamma,W_\perp^\pm,Z_\perp^0$) plus a technipion or longitudinal
  weak boson ($W_L^{\pm,0} \equiv W_L^\pm,Z_L^0$) and for decays to a
  fermion-antifermion pair depend sensitively on $Q_U+Q_D$.
  
\item The lightest technihadrons are the pseudoscalars $\pi_{T1}^{\pm,0}(I =
  1)$ and the vectors $\tro^{\pm,0}(I=1)$, $\tom(I=0)$ and axial vectors
  $a_T^{\pm,0}(I=1)$, $f_T(I=0)$. Isospin symmetry and quark-model experience
  strongly suggest $M_{\tro} \cong M_{\tom}$ and $M_{\ta} \cong
  M_{f_T}$.\footnote{We assume that the isosinglet $\pi_{T1}^{0\prime}$ is
    too heavy to play a part in LSTC phenomenology. The $f_T$ doesn't either
    because it cannot be produced as an $s$-channel resonance in $q \bar q$
    collisions.}
  
\item Since $W_L^{\pm,0}$ are superpositions of all the isovector
  technipions, the $\pi_{T1}$ are not mass eigenstates. This is parameterized
  in the TCSM as a simple two-state admixture of $W_L$ and the lightest
  mass-eigenstate $\tpi$:
\be\label{eq:pistates}
 \vert\pi_{T1}\rangle = \sin\chi \ts \vert
W_L\rangle + \cos\chi \ts \vert\tpi\rangle\ts.
\ee
Thus, technivector decays involving $W_L$, while nominally, strong
interactions, are suppressed by powers of $\sin\chi$.

\item The lightest technihadrons, $\tpi$, $\tro$, $\tom$ and $\ta$, may be
  studied {\em in isolation}, without significant mixing or other
  interference from higher-mass states. This is the most important of the
  TCSM's assumptions. It is made to avoid a forest of parameters.
    
\item In addition to these technihadrons and $W_L^\pm$, $Z^0_L$, the TCSM
  involves the transversely-polarized $\gamma$, $W^\pm_\perp$ and
  $Z^0_\perp$. The principal production process of the technivector mesons at
  hadron and lepton colliders is Drell-Yan, e.g, $\ol q q \ra \gamma,Z^0 \ra
  \troz, \tom, \taz \ra X$. This gives strikingly narrow $s$-channel
  resonances at $M_X = M_{\troz,\tom,\ta}$ {\em if} $M_X$ can be
  reconstructed.
  
\item Technipion decays are mediated by ETC interactions and, therefore, are
  expected to be Higgs-like, i.e., $\tpi$ preferentially decay to the
  heaviest fermion pairs they can. There is one exception. Something like
  topcolor-assisted technicolor~\cite{Hill:1994hp} is required to give the
  top quark its large mass. Then, the coupling of $\tpi$ to top quarks is not
  proportional to $m_t$, but more likely to $\CO(m_b)$~\cite{Hill:1994hp}.

\end{enumerate}

This TCSM phenomenology was tested at LEP (see, e.g.,
Refs.~\cite{Abdallah:2001ft,Schael:2004tq}) and the
Tevatron~\cite{Abazov:2006iq,Abazov:2009eu,Aaltonen:2009jb} for some generic
values of the parameters. So far there is no compelling evidence for TC, but
there are also no significant restrictions on the masses and couplings
commonly used in the TCSM search analyses carried out so far: For $\tro \ra W
\tpi$, the limits are $M_{\tro} \simge 210$--$250\,\gev$, $M_{\tpi} \simge
125$--$145\,\gev$ when $M_W + M_{\tpi} < M_{\tro} <
2M_{\tpi}$~\cite{Aaltonen:2009jb}; for $\tropm \ra W Z$, they are $M_{\tro} >
400\,\gev$, $M_{\tpi} > 350\,\gev$ when $M_{\tro} < M_W +
M_{\tpi}$~\cite{Abazov:2009eu}. Both sets of limits use the {\sc Pythia}
defaults~\cite{Sjostrand:2006za}: $\sin\chi = 1/3$, $Q_U \simeq 1$, $\Ntc =
4$, and the $\tro \ra \tpi \tpi$ coupling scaled from QCD, $\grpp =
\sqrt{4\pi(2.16)(3/\Ntc)}$.\footnote{See Sect.~5 for a discussion of this
  assumption on $\grpp$.} On the other hand, the more general idea of LSTC
makes little sense if the limit on $M_{\tro}$ is pushed past $\sim
700\,\gev$. Therefore, we believe that the LHC can discover it or certainly
rule it out.

In the June 2007 Les Houches summary report~\cite{Brooijmans:2008se}, several
of the current authors used {\sc Pythia}~\cite{Sjostrand:2006za} together
with the the PGS detector simulator~\cite{LSTCPGS} to study the reach of the
LHC with $\sqrt{s} = 14\,\tev$ for the LSTC processes
\be\label{eq:VTdecays}
q \bar q \ra \tropm \ra W^\pm Z^0,\,\, a_T^\pm \ra
\gamma W^\pm,\,\, \tom \ra \gamma Z^0\,.
\ee
In all cases, the $W$ and $Z$ decay to $e$ or $\mu$-type leptons. These decay
modes were chosen because they are not overwhelmed by backgrounds (as is
$\tro \ra W \tpi$ which is swamped by $t\bar t$ at the LHC). Thus, we
expected that they are the most likely LSTC discovery channels. We shall see
in Sect.~4 that neutral technivector decays to $\ellp\ellm$ are also quite
promising discovery modes.

For Les Houches 2007, we concentrated on three TCSM mass points that cover
most of the reasonable range of LSTC scales; they are listed in
Table~\ref{tab:LH07}. In all cases, we assumed isospin symmetry, together
with $M_{\tro} = M_{\tom}$ and $M_{a_T} = 1.1 M_{\tro}$. The near degeneracy
of $\tro$ and $\ta$ was motivated by the argument that it makes the low-scale
TC contribution to the $S$-parameter small (see Ref.~\cite{Lane:2009ct} and
references therein).  The {\sc Pythia} defaults listed above were used as
well as $M_{V_{1,2}} = M_{A_{1,2}} = M_{\tro}$ for the LSTC mass parameters
controlling the strength of $\tro$, $\tom$, $\ta$ decays to a transverse
electroweak boson plus $\tpi$ or $W_L$~\cite{Lane:2002sm, Eichten:2007sx}.
The table also lists the signal cross sections at 14~TeV and, in parentheses,
the minimum luminosities for a $5\,\sigma = S/\sqrt{S+B}$ discovery.

 \begin{table}[!ht]
     \begin{center}{
  \begin{tabular}{|c|c|c|c|c|c|c|c|}
  \hline
 Case & $M_{\tro} = M_{\tom}$ & $M_{a_T}$ & $M_{\tpi}$ & &
 $\sigma(W^\pm Z^0)$ & $\sigma(\gamma W^\pm)$ & $\sigma(\gamma Z^0)$ \\
  \hline\hline
 A & 300 & 330 & 200 & & 110 (2.4)&  168  (2.3)& 19.2 (17)\\
 B & 400 & 440 & 275 & & 36.2 (7.2)& 64.7 (4.5)&  6.2 (46)\\
 C & 500 & 550 & 350 & & 16.0 (15)&  30.7 (7.8)&  2.8 (97)\\
   \hline\hline
 \end{tabular}}
 \caption{The LH~2007 study's TCSM masses (in GeV) and signal cross sections
 times $e,\mu$ branching ratios (in fb) for $pp$ collisions at $\sqrt{s} =
   14\,\tev$ producing the lightest technihadrons. Numbers in parentheses are
   the luminosities (in $\ifb$) needed or a $5\,\sigma$
   discovery~\cite{Brooijmans:2008se}. \label{tab:LH07}}
 \end{center}
 \end{table}
 
 In addition to discovering the narrow resonances in these channels, the
 angular distributions of the two-body final states in the technivector rest
 frame provide compelling evidence of their underlying technicolor origin.
 Because all the modes involve at least one longitudinally-polarized weak
 boson, the distributions are
\bea\label{eq:angdist}
&& \frac{d\sigma(\bar q q \ra \tropm \ra W_L^\pm Z_L^0)}{d\cos\theta}
\propto \sin^2\theta, \\
&& \frac{d\sigma(\bar q q \ra a_T^\pm,\tropm \ra \gamma
  W_L^\pm)}{d\cos\theta},\quad
\frac{d\sigma(\bar q q \ra \tom,\troz \ra \gamma Z_L^0)}{d\cos\theta}
  \propto 1+ \cos^2\theta \,.
\eea
Simulations were presented in the LH~2007 report. While these studies were
very preliminary, they indicated that the $\tropm \ra W^\pm Z$ and $\tapm \ra
\gamma W^\pm$ distributions easily could be distinguished from background for
$M \simeq 300\,\gev$ with $10\,\ifb$ of data and $M \simeq 400\,\gev$ with
20--$40\,\ifb$. The smaller $\tom \ra \gamma Z$ signal rates require much
more luminosity, e.g., $40\,\ifb$ for $M_{\tom} \simeq 300\,\gev$.

There are three motivations for the present study. First, for some time to
come, the main operating c.m.~energy of the LHC will, with some luck, be 10,
not 14, TeV. This requires that our studies be repeated and the reach for
LSTC signals be estimated for the lower energy --- and lower luminosities ---
expected for the next several years.\footnote{Our luck did not hold. As this
  document was being completed, a new LHC run plan was adopted in which the
  machine would begin an 18--24 month run in 2010 run at $\sqrt{s} =
  7\,\tev$, followed by a long shutdown in which it would be prepared for
  running at the design c.m.~energy of $14\,\tev$. See {\tt
    http://indico.cern.ch/conferenceDisplay.py?confId=83135}. Some
  justification for our studies may be derived from the fact that $10\,\tev =
  \sqrt{(7\,\tev)(14\,\tev)}$.} Second, as noted above, most of the 2007 work
was carried out using the PGS detector simulator. While adequate for a first
look at LSTC for the LHC, one really wants more substantial studies using the
ATLAS and CMS detector simulations and, where possible, more reliable
estimates of backgrounds.\footnote{This motivation was thwarted to some extent
  by the collaborations' requirements for publishing analyses made with their
  software and simulation tools.} Finally, two of us have developed an
effective Lagrangian for LSTC~\cite{Lane:2009ct}. This can be interfaced
with such tools as MadGraph and CalcHEP to generate cross sections for
particle production and decay using {\sc Pythia} or HERWIG. We present here a
selection of first results comparing the parton-level cross sections
generated by our Lagrangian with the TCSM as implemented in {\sc Pythia}.

In this paper we report on several more-in-depth studies for some of the
classic LSTC discovery channels at the LHC, and we add some new
ones. The LSTC processes investigated in this report and the principal
results are the following:

\begin{enumerate}

\item A CMS study of $\tropm \ra W^\pm Z$ (Bose, Carrera, Maravin).
  
\item A PGS-based study of $\tom \ra \gamma Z \gamma \ellp\ellm$ (Black,
  Smith).

\item A CMS-based study of $\tom, \troz, \taz \ra e^+e^-$ (Harper).
  
\item Comparisons of an effective Lagrangian, $\Leff$, for LSTC with the TCSM
  in {\sc Pythia}, including an investigation of the accuracy of the
  longitudinal gauge boson approximation for technivector decays (Martin and
  Lane). The effective Lagrangian implies some striking differences with the
  TCSM defined in Refs.~\cite{Lane:1999uh,Lane:2002sm,Eichten:2007sx} and
  implemented in {\sc Pythia}. In particular, the value of $\grpp$ is
  predicted by $\Leff$ and it is considerably smaller than the value
  $\sqrt{4\pi(2.16)(3/\Ntc)}$ obtained by scaling from QCD. Thus, the rate
  for $\tro \ra WZ$ predicted by $\Leff$ is much smaller than in the TCSM,
  while the rate for $\tro \ra \gamma W$ can be much larger. This is a new
  result. It is unclear whether it is more or less credible than the TCSM,
  but experiment can decide.

\end{enumerate}
The mass points and signal cross sections at $\sqrt{s} = 10\,\tev$ (computed
from the TCSM in {\sc Pythia}) are listed in Table~\ref{tab:cases}. Note that
$\tro \ra W\tpi$ is forbidden in Case~1a, enhancing the $\tropm \ra WZ$
branching ratio.

\begin{changemargin}{-3.0cm}{-3.0cm}
%
%
\begin{table}[!ht]
\begin{center}{
  \begin{tabular}{|c|c|c|c|c|c|c|c|c|c|}
  \hline
 Case & $M_{\tro,\tom}$ & $M_{a_T}$ & $M_{\tpi}$ &
 $M_{V_1,\dots,A_2}$ && $\sigma(W^\pm Z^0)$ & $\sigma(\gamma
 W^\pm)$ & $\sigma(\gamma Z^0)$ & $\sigma(e^+e^-)$ \\
  \hline\hline
 1a & 225 & 250 & 150 & 225 && 230 & 330 & 60 & 1655 (980)\\
 1b & 225 & 250 & 140 & 225 && 205 & 285 & 45 & 1485 (980)\\
 2a & 300 & 330 & 200 & 300 &&  75 & 105 & 11 &  425 (290)\\
 2b & 300 & 330 & 180 & 300 &&  45 &  85 &  7 &  380 (290)\\
 3a & 400 & 440 & 275 & 400 &&  22 &  40 &  4 &  130  (90)\\
 3b & 400 & 440 & 250 & 400 &&  14 &  35 &  3 &  120  (90)\\
   \hline\hline
 \end{tabular}}
 \caption{Technihadron masses, LSTC mass parameters (in GeV) and {\em
     approximate} signal cross sections for $pp$ collisions at $\sqrt{s} =
   10\,\tev$ (in $\fb$) for the 2009 Les Houches study. Isospin symmetry is
   assumed. Other TCSM parameters are $\sin\chi = 1/3$, $\Ntc = 4$, $Q_U =
   Q_D + 1 = 1$, $\grpp = \sqrt{4\pi(2.16)(3/\Ntc)} = 4.512$, and CTEQ5L
   parton distribution functions were used. Branching ratios of $W$ and $Z$
   to electrons and muons are included.  $\sigma(e^+ e^-)$ includes signal
   plus standard-model production integrated over approximately
   $M_{\tro,\tom} - 25\,\gev$ to $M_{\ta} + 25\,\gev$; the standard model
     cross section for this range is in parentheses.  
\label{tab:cases}}
 \end{center}
 \end{table}
\end{changemargin}

\section{$\tropm \ra W^\pm Z^0$} 

This section summarizes a CMS study of the detector's reach for $\tropm \ra
W^\pm Z^0 \ra \ell^\pm \nu_\ell \ell^+ \ell^-$ for $\ell = e$ and/or $\mu$ as
described in the TCSM and encoded in {\sc
  Pythia}~\cite{LSTCCMS}.\footnote{While {\sc Pythia} shows the $\ta \ra WZ$
  resonance, the $\etmiss$ resolution in the detector simulation results in
  its coalescing with the larger $\tro$ peak.} This study updates one carried
out for Les Houches~2007~\cite{LSTCbose}, with $pp$ collisions at $\sqrt{s} =
10\,\tev$ and concentrating on four TCSM mass points not excluded by other
experiments and covering a range accessible with an integrated luminosity
$\simle 5\,\ifb$, namely, the three cases of Table~\ref{tab:cases} plus
$M_{\tro} = 500\,\gev$. This analysis uses the detailed {\sc geant4}
simulation of the CMS detector, improved object identification algorithms,
and formulates methods for data-driven background estimation.

\subsection{Analysis Strategy}

Sources of background are the standard model $WZ$ production, plus $ZZ$ and
$WW$, $Z+\gamma$, $W+{\rm jets}$ and $Z+{\rm jets}$ production ($W$ or $Z$
boson production in association with a pair of heavy quark jets, referred to
as $VQQ$, is treated separately), and $t\bar{t}$ production. The
statistically significant instrumental backgrounds come from $Z + {\rm jets}$
and $t \bar t$ production. For instance, in an energetic $Z+{\rm jet}$ event,
the footprint of a jet in the detector can mimic the leptonic decay of
a $W$ boson, making it a perfect technicolor candidate event. Massive top
quark pair events also populate the invariant mass peaks. To overcome these
backgrounds, the analysis puts stringent identification requirements on final
state leptons, enforces constraints on the particle transverse momenta and on
invariant quantities such as the mass of the $Z$ boson, making using of the
aforementioned data-driven techniques known to have worked in previous
experiments.

Signal samples are produced with {\sc Pythia} and processed using a detector
simulation based on CMS {\sc geant4}. To simulate next-to-leading order
predictions, a $K$-factor of $1.35 \pm 0.27$ is applied to all signal cross
section values. Most backgrounds are produced with {\sc Pythia} (although,
for some processes, MadGraph was used in the generation) and the same
selection criteria are applied to signal and background simulation samples.
Whenever fast simulation is used for the backgrounds, a cross-check with the
full detector simulation is performed to ensure proper description of
detector effects. Next-to-leading order background cross sections and
$K$-factors used in the study can be found in~\cite{LSTCCMS}.

\subsection{Signal and Event Selection}

Events are pre-selected using single muon and electron triggers which are
$99\%$ efficient and at least $3$ leptons with $p_T > 10\,\gev$ are required.
The pair of like-flavored, opposite charge leptons with invariant mass
$M_{\ell\ell}$ closest to the $Z$ nominal mass are assigned as $Z$ decay
products. To reject $ZZ$ background, events with two non-overlapping $Z$
candidates that are found within $50\ {\rm GeV}<M_{\ell\ell}<120\,\gev$ are
eliminated.  The most energetic lepton in the remaining pool is assigned to
the $W$ boson, and the corresponding neutrino assigned transverse energy
equal $-\etmiss$, the event missing transverse energy. The $WZ$ candidate
invariant mass is determined by forcing the known $W$ invariant mass to the
lepton-neutrino system while choosing the smaller solution in the calculation
of the longitudinal momentum of the neutrino.

Electron candidates, which are reconstructed as energy clusters in the
electromagnetic calorimeter with a matched pixel track, are required to have
$p_T>15\ {\rm GeV}$, to be consistent with shape and energy deposition of an
electron shower, and to be isolated in order to suppress misidentified
jets.  Muons are reconstructed using information from the muon detectors and
the silicon tracker.  Those assigned to a $Z$-boson must have $p_T>10\,\gev$,
with no track or isolation requirement due to the low misidentification rate.
Tighter selection criteria ($p_T> 20\,\gev$ and isolation) are applied to
muons from $W$ candidates since a higher misidentification rate is expected.
In addition, a quality cut on the impact parameter significance of the muons
is applied.

To enhance the signal to background ratio, two sets of further requirements
are used in this study.  The first one optimized for early conditions (or for
$M_{\tro} = 225\,\gev$), and another one optimized for higher luminosity
scenarios (or for $M_{\tro} > 300\,\gev$.  These requirements for early
(late) conditions are: $p_T(Z) > 50\, (90)\,\gev$, $p_T(W) > 50\,
(90)\,\gev$, and $H_T > 130\, (160)\,\gev$, where $H_T$ is the
scalar sum of the transverse momentum of the three charged leptons in the
final state.

Figure~\ref{fig:WZ-mass} shows, the $WZ$ invariant mass distributions for the
various mass points for $1\,\ifb$ of integrated luminosity.
Table~\ref{main-eff} lists the number of signal events expected with
$200\,\ipb$ of data within a mass window of $1.4$ Gaussian standard
deviations around the $\tro$ mass peak.

\begin{figure}
 \begin{center}
\includegraphics[width=0.45\textwidth]{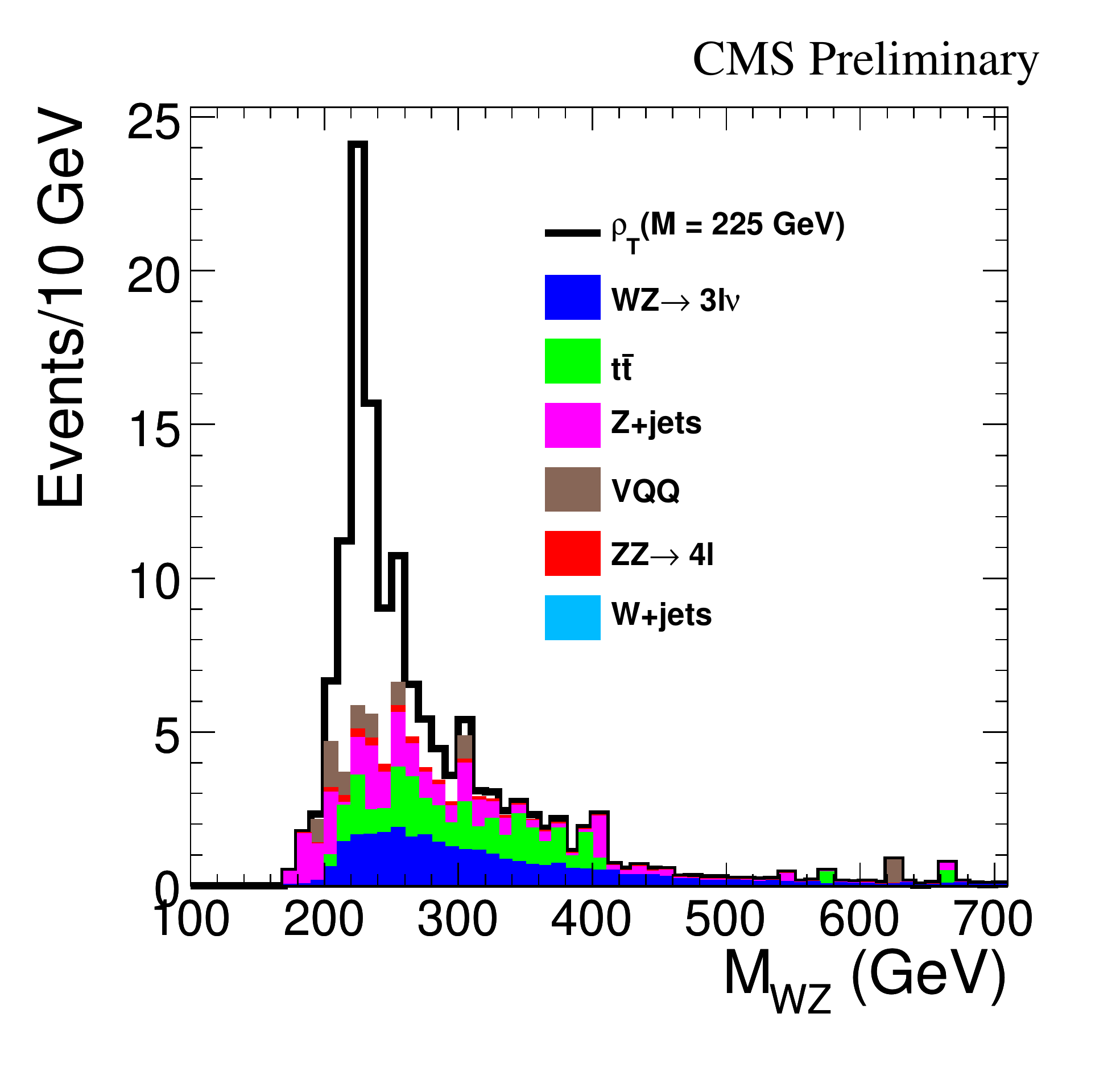}
\includegraphics[width=0.45\textwidth]{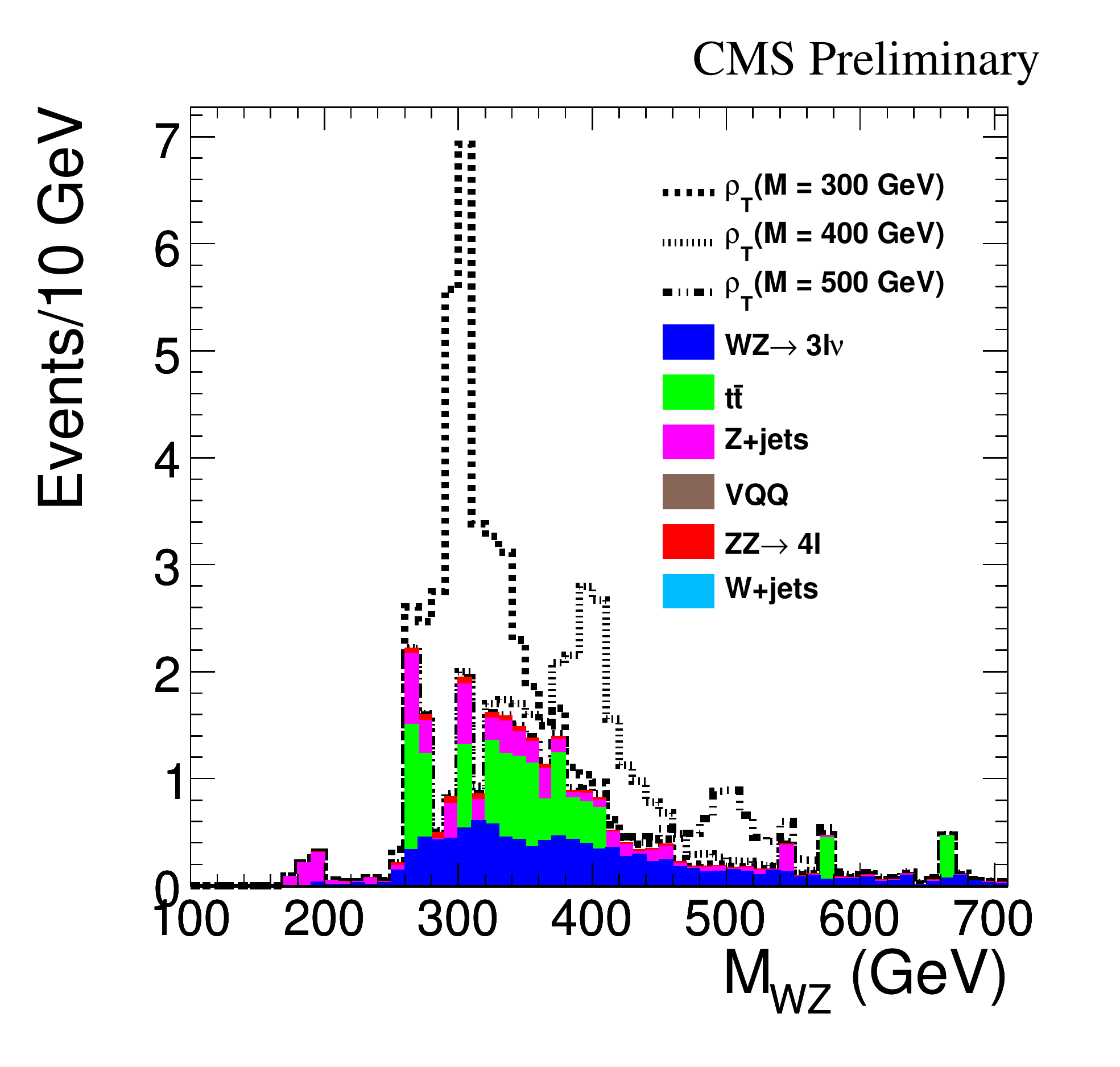}
 \caption{$WZ$ invariant mass distributions for the Case~1a signal
   ($M_{\tro} = 225\,\gev$) and background samples (left). $WZ$ invariant
   mass distributions for signal ($M_{\tro}$ in the range 300--500~GeV) and
   background samples (right). The distributions are normalized to an
   integrated luminosity of $1\,\ifb$.\label{fig:WZ-mass}}
 \end{center}
 \end{figure}

\begin{table}[!ht]
\begin{center}
\begin{tabular}{|l||c|c|c|}
\hline
    Process               & Efficiencies   &   Expected
signal & Expected background \\ 
&Signal ($\epsilon_{main}$) & events per 200 pb$^{-1}$& events per 200
    pb$^{-1}$  \\
\hline \hline
 $\rho_{\rm T}$ (M=225 GeV) &$0.137\pm 0.037$ & $8.60\pm 3.17$ &$4.75\pm
    0.95$           \\ 
 $\rho_{\rm T}$ (M=300 GeV) &$0.186\pm 0.034$ & $3.71\pm 1.15$ &$1.79\pm
    0.39$           \\  
   $\rho_{\rm T}$ (M=400 GeV) &$0.251\pm 0.046$ &$1.62\pm 0.50$  &$1.05\pm
    0.27$            \\ 
   $\rho_{\rm T}$ (M=500 GeV) &$0.254\pm 0.047$&$0.65\pm 0.20$  &$0.24\pm
    0.06$         \\ \hline \hline 
\end{tabular}
\caption{Final efficiencies and number of events for the various selection
  criteria for $200\,\ipb$ of data at $\sqrt{s} = 10\,\gev$. The first three
  cases are 1a, 2a, 3a in Table~\ref{tab:cases}; in the last case $M_{\tpi} =
  350\,\gev$ and $M_{V_i} = M_{A_i} = 500\,\gev$. The quoted uncertainties
  include statistical and systematic uncertainties (purely from simulation),
  the latter described later in the text.  \label{main-eff}}
\end{center}
\end{table}

\subsection{Background Estimation}

The physics backgrounds, $WZ$ and $ZZ$, are estimated from Monte Carlo
simulation.  The instrumental backgrounds fall into two groups, one that
includes a genuine $Z$-boson and one that does not.  The $Z+{\rm jets}$
background dominates the first group, which also includes $Z\gamma$
production (found to be negligible), and $Zbb$ production. In the second
group $t\bar{t}$ production dominates, followed by $W+{\rm jets}$, and QCD
multi-jet production (found to be negligible).

The $Z+\jets$ background (including $VQQ$) is estimated using a data-driven
technique, the ``matrix method'', used successfully in previous experiments.
This method makes use of two samples, a ``tight-cut'' sample with events
passing all the signal selection criteria, and a ``loose-cut'' sample where
events pass all the signal selection requirements except the isolation cuts
on the $W$'s charged lepton.  Hence, the number of events in each sample are
given by $N_{loose} = N_{lep}+N_{jet}$ and $N_{tight} = \epsilon_{tight}
N_{lep}+P_{fake}N_{jet}$.  Here, $N_{lep}$ and $N_{jet}$ is the number of
events with the $W$ candidates reconstructed from true leptons and the fake
ones from misidentified jets, respectively; $\epsilon_{tight}$ is the
efficiency for true leptons to pass the isolation cuts and $P_{fake}$ is the
corresponding efficiency for fake leptons.  These efficiencies will be
extracted from data using the standard ``tag and probe'' method, thus
minimizing systematic errors due to simulation. Using Monte Carlo simulation,
the efficiencies $\epsilon_{tight}$ for muons and electrons are estimated to
be $(93.9\pm 0.8)\%$ and $(96.5\pm 1.3)\%$, respectively, while the rates
$P_{fake}$ for misidentified jets are $0.30\pm 0.04$ for electrons and
$0.33\pm 0.03$ for muons. The signal and background contributions are
estimated with these measured efficiencies.

The $t\bar{t}$ and other backgrounds without a genuine $Z$-boson, which are
assumed to dominate the tails of the $Z$-boson mass distribution, are
estimated using the sideband subtraction method. The final $Z$-mass
distribution, for an integrated luminosity of $200\,\ipb$, is fit to a linear
sum of a histogram and a quadratic function.  The ``$Z$-shaped'' histogram is
extracted from a combination of $Z+{\rm jets}$ and $WZ$ samples with much
looser requirements, and the quadratic contribution from a combination of
$t\bar{t}$ and $W+{\rm jets}$ samples (which are expected to be rather flat).

Table~\ref{tab:data_back} presents a summary of the number of background
events expected with $200\,\ipb$ for the $1.4\sigma$ mass window used above
for the signal.  The uncertainties in the $Z+{\rm jets}$, $VQQ$, $t\bar{t}$,
and $W+{\rm jets}$ backgrounds are taken from the data-driven techniques.

\begin{table}[!Hhtb]
\begin{center}
{\scriptsize
\begin{tabular}{|l||c|c|c|c|}
\hline
    Process               & $\rho_{\rm T}$ (M=225 GeV) &  $\rho_{\rm T}$
    (M=300 GeV) & $\rho_{\rm T}$ (M=400 GeV)   & $\rho_{\rm T}$ (M=500 GeV)
    \\ \hline \hline
   $WZ$ & $1.416\pm 0.043\pm 0.502$ & $0.699\pm 0.030\pm 0.214$ & $0.508\pm
0.026 \pm 0.156$ & $0.190\pm 0.016\pm 0.058$\\
   $ZZ$ & $0.236\pm 0.004\pm 0.084$ & $0.079\pm 0.003\pm 0.024$ & $0.032\pm
0.002\pm 0.010$ & $0.015\pm 0.001\pm 0.005$\\
   $Z$+jets and $VQQ$ & $2.082\pm 2.663\pm 0.506$   & $0.384\pm 1.521\pm 0.064$ &
$0.121\pm 0.479\pm 0.020$ & $0.034\pm 0.135\pm 0.006$\\
   $t\bar{t}$ and $W$+jets &$1.014\pm 1.016\pm 0.247$ & $0.624\pm 0.101\pm 0.104$
& $0.390\pm 0.063\pm 0.065$ &$0.000\pm 0.000\pm 0.000$\\ 
   Total&$4.76\pm 2.85\pm 0.76$ & $1.79\pm 1.52\pm 0.25$
& $1.05\pm 0.48\pm 0.17$ &$0.24\pm 0.14\pm 0.06 $\\ \hline 
\hline
\end{tabular}
}
\caption{Summary of final number of background events for 200 pb$^{-1}$ of
  data at $\sqrt{s} = 10\,\gev$. Statistical and systematic uncertainties (in
  this order) are also given. Statistical uncertainties include those from
  data-driven methods for this low luminosity. \label{tab:data_back}}
\end{center}
\end{table}

\begin{figure}[!Hht]
\begin{center}
\includegraphics[width=0.7\textwidth]{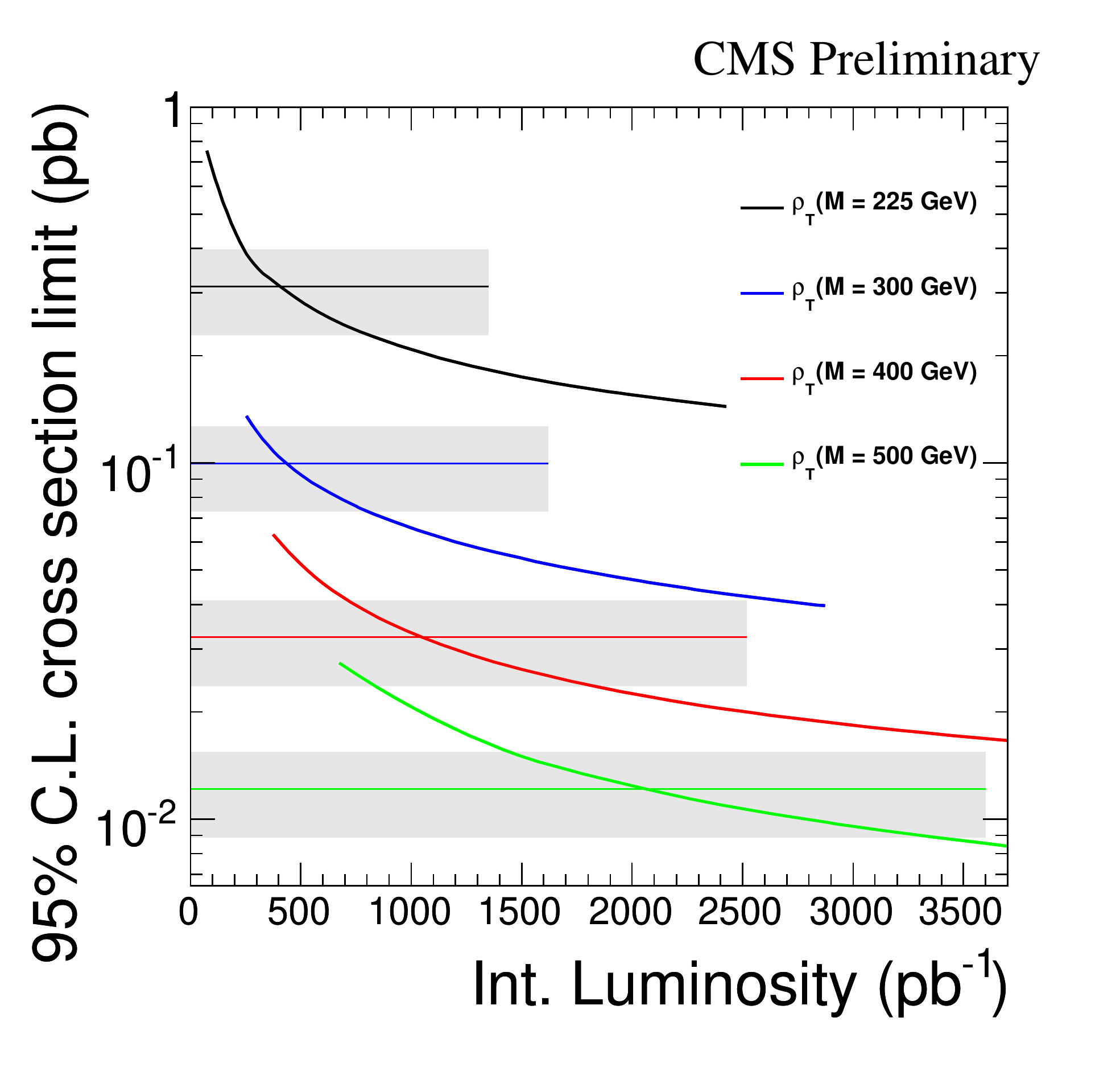}
\caption{95\% C.L. limits for $\sigma(\tro \ra WZ)$ as a function of integrated
  luminosity for $pp$ collisions at $\sqrt{s} = 10\,\gev$. The cross sections
  include the branching ratio to electrons and muons. The horizontal bands,
  which indicate the theoretical cross sections (and their associated $27\%$
  uncertainty), intersect the limit curves at approximately the values given
  in Table~\ref{tab:sensitivity}.
  \label{fig:xseclim}}
\end{center}
\end{figure}

\begin{table}[!Ht]
\begin{center}
{\small
\begin{tabular}{|l||c|c|c|}
\hline
    Mass values   &Int. luminosity &Int. luminosity &Int. luminosity \\           
& for $95\%$ C.L limit&  for $95\%$ C.L limit  &for $95\%$ C.L limit \\ 
&(pb$^{-1}$) & (+ theoretical& (- theoretical  \\
& &uncertainty) (pb$^{-1}$)&uncertainty) (pb$^{-1}$)\\
\hline \hline
$M_{\tro} = 225\,\gev$, $M_{\tpi} = 150\,\gev$  &$400$ & $240$ &$790$   \\ 
$M_{\tro} = 300\,\gev$, $M_{\tpi} = 200\,\gev$  &$440$ & $290$ &$790$   \\ 
$M_{\tro} = 400\,\gev$, $M_{\tpi} = 275\,\gev$  &$1040$ &$710$  &$1800$ \\
$M_{\tro} = 500\,\gev$, $M_{\tpi} = 350\,\gev$  &$2050$&$1450$  &$3310$
\\ \hline
$M_{\tro} = 225\,\gev$, $M_{\tpi} = 140\,\gev$  &$540$&$300$  &$1060$\\ 
$M_{\tro} = 300\,\gev$, $M_{\tpi} = 180\,\gev$  &$1300$&$800$  &$2550$\\
\hline \hline
\end{tabular}
}
\caption{Integrated luminosity at $\sqrt{s} = 10\,\gev$ needed for exclusion
  at $95\%$ C.L. The last two columns indicate the values of integrated
  luminosity (in $\ifb$) needed if the theoretical uncertainty in the signal
  is taken into account.  The last two rows show results for different
  parameter sets for the mass points $\rho_{\rm T}=225$~GeV and $\rho_{\rm
    T}= 300$~GeV.\label{tab:sensitivity}}
\end{center}
\end{table}

\subsection{Results and Conclusions}

In the absence of an excess of signal events, $95\%$ C.L.~upper limits can be
set on the cross sections. These limits, as functions of integrated
luminosity, are summarized in Fig.~\ref{fig:xseclim}.  The final results are
presented in Table~\ref{tab:sensitivity}, which include a second set of
technicolor parameters that use lower values for $M_{\tpi}$ from cases 1b and
2b in Table~\ref{tab:cases}. These limits use the results for $200\,\ipb$
given in Table~\ref{tab:data_back}. The statistical uncertainty in the total
background is scaled with luminosity while the relative systematic
uncertainty is kept constant throughout.

As expected from Table~\ref{tab:LH07} (constructed for $\sqrt{s} =
14\,\tev$), a $5\sigma$ discovery of technicolor particles via the $\rho_{\rm
  T} \rightarrow WZ \ra {\rm leptons}$ process will require well over
$1\,\ifb$ of data.

\section{$\tom \ra \gamma Z^0 \ra \gamma \ellp\ellm$}

\subsection{Introduction}

The decay $\tom \ra \gamma Z^0 \ra \gamma \ellp\ellm$ may be the discovery
channel for $\tom$ at the LHC. This is especially true if $Q_U + Q_D = 0$, in
which case $\tom \ra \ellp\ellm$ is forbidden (just as in QCD!). This
section presents a simplified study of $\tom \ra \gamma Z^0 \ra
\gamma\mu^+\mu^-$ using the PGS detector simulator~\cite{LSTCPGS}. A more
in-depth analysis using ATLAS simulation tools for $\tom \ra \gamma Z^0 \ra
\gamma e^+ e^-$ could not receive collaboration approval for its release in
time for this document's submission.  The present PGS-based analysis should
be a plausible feasibility study.  Another very important feature of the
$\gamma Z$ mode is its angular distribution. In the approximation that the
$Z$ is longitudinally polarized, as expected in LSTC, it is $1+\cos^2\theta$.

Signal and background cross sections were calculated using {\sc Pythia}.  The
$\gamma\mu^+\mu^-$ signal rates are half those in the $\sigma(\gamma Z)$
column of Table~\ref{tab:cases}. The two principal backgrounds are the
standard-model production of $\gamma Z$ and $Z+\jets$ where a jet fakes a
photon; see the 2007 Les Houches study of LSTC in
Ref.~\cite{Brooijmans:2008se}. The cross sections for the standard
$\gamma\mu^+\mu^-$ and $Z + \jets$ cross sections are $7.3\,\pb$ and
$1144\,\pb$, respectively.

\subsection{Analysis}

A parameterized detector simulation with PGS was used to give an estimate of
an LHC detector's response. The parameterization was chosen to correspond to
the approximate behavior of ATLAS and CMS. Most notably we assumed a muon
identification efficiency of 95\%, a photon efficiency of 80\%, and a jet to
photon misidentification rate of $10^{-4}$.

The most significant backgrounds are expected from $Z$ events with (1) a
photon radiated from the initial $\bar q q$ or from the $Z$'s decay leptons
or (2) a quark or gluon jet misidentified as a photon. To reduce these
backgrounds we take advantage of two aspects that differ in signal and
background kinematics.

\begin{enumerate}
  
\item The signal $Z$-boson will be centrally produced and with typically
  large transverse momenta. In contrast, $p_T(Z) = 0$ in lowest order and
  nonzero $p_T$ comes from parton or photon radiation processes having
  rapidly falling cross-sections.
  
\item The signal photons should be isolated from the $Z$ or its decay
  products whereas the radiated photons and gluons tend to follow the object
  which produced them.

\end{enumerate}

\noindent Therefore, we required the following:

\begin{enumerate}
  
\item Two muons of opposite sign, each with $p_T > 15\,\gev$ and $\eta < 2.5$
  reconstructing a $Z$-boson within $15\,\gev$ of the nominal $Z$-mass of
  $91.2\,\gev$.

\item A photon with $p_{T} > 35$ GeV and $\eta < 2.5$.

\item The photon and muons have $\Delta \phi > 1$.

\item The photon and $Z$ have $\Delta \phi > 2$.

\end{enumerate}
The efficiencies on the signal and background samples are displayed in
Tables~\ref{tab:signal_eff} and~\ref{tab:back_eff}

\begin{table}[ht]
\centering
\begin{tabular}{| c  | c  |  c  |  c  | c  |  c  | c | }
\hline\hline
Case  & $Z$-boson selection  & photon selection  &  $\Delta \phi(\gamma\mu) >
1$ & $\Delta \phi(\gamma Z) > 2$ \\ 
\hline\hline
1a & 0.45 $\pm$ 0.01 & 0.43 $\pm$ 0.01  & 0.33 $\pm$ 0.02 & 0.31 $\pm$ 0.02  \\
\hline
1b & 0.45 $\pm$ 0.01 & 0.43 $\pm$ 0.01  & 0.32 $\pm$ 0.02 & 0.31 $\pm$ 0.02  \\
\hline
2a & 0.49 $\pm$ 0.01 & 0.48 $\pm$ 0.01  & 0.39 $\pm$ 0.02 & 0.37 $\pm$ 0.02  \\
\hline
2b & 0.49 $\pm$ 0.01 & 0.47 $\pm$ 0.01  & 0.39 $\pm$ 0.02 & 0.36 $\pm$ 0.02  \\
\hline
3a & 0.55 $\pm$ 0.01 & 0.55 $\pm$ 0.01  & 0.47 $\pm$ 0.01 & 0.45 $\pm$ 0.01  \\
\hline
3b & 0.54 $\pm$ 0.01 & 0.53  $\pm$ 0.01&  0.47 $\pm$ 0.01 & 0.44 $\pm$ 0.01   \\
\hline \hline

\end{tabular}
\caption{Cumulative efficiencies for signal event selection in $pp \ra \tom +
  X$, $\tom \ra \gamma Z \ra \gamma \mu^+ \mu^-$ at $\sqrt{s} =
  10\,\tev$. \label{tab:signal_eff}} 
\end{table}

\begin{table}[ht]
\centering
\begin{tabular}{| c  | c  |  c  |  c  | c  |  c  | c | }
\hline\hline
Background &  { $Z$-boson election } &  { photon selection }  & { $\Delta \phi(\gamma\mu) >
1$ } &  {\small $\Delta \phi(\gamma Z) > 2$ } \\ 
\hline\hline
$Z \gamma$ & {\small 0.074 $\pm$ 0.01} &  {\small 0.043 $\pm$ 0.029}  &  {\small 0.005 $\pm$ 0.001} &  {\small 0.028 $\pm$ 0.005}  \\
$Z$ + jets &  {\small 0.003 $\pm$ 0.001} &  {\small 0.00011 $\pm$ 0.00005 } &  {\small $ ( 7 \pm 1) \times 10^{-5}$} &  
{\small $ (5.5 \pm 1) \times 10^{-5}$ }\\ 
\hline \hline

\end{tabular}
\caption{Cumulative efficiencies for background event selection  in $pp \ra
  \tom + X$, $\tom \ra \gamma Z \ra \gamma \mu^+ \mu^-$ at $\sqrt{s} =
  10\,\tev$. \label{tab:back_eff}}
\end{table}

\begin{figure}[!ht]
\begin{center}
\includegraphics[width=3.00in, height=3.00in]{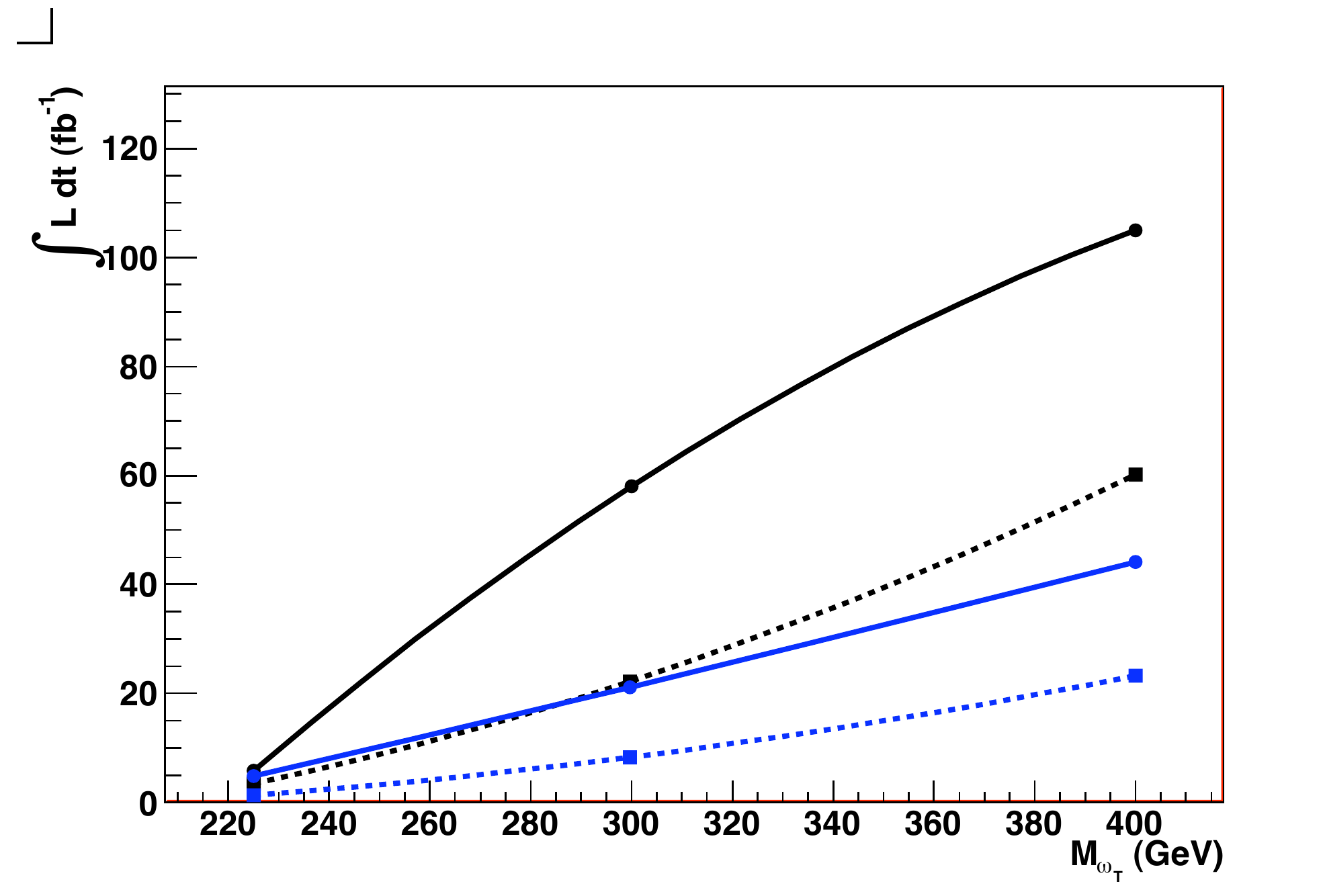}
\includegraphics[width=3.00in, height=3.00in]{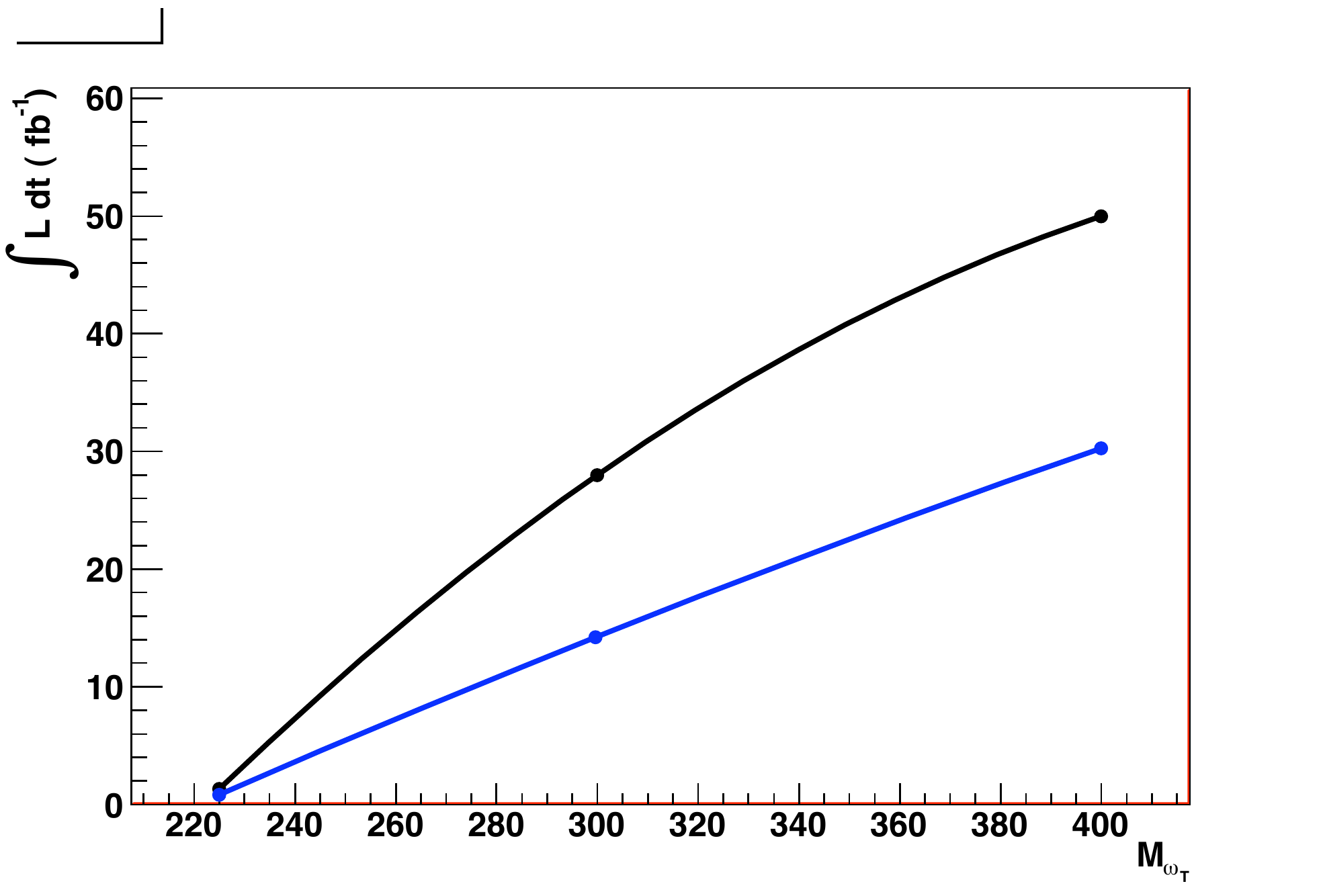}
\caption{Left: Integrated luminosity of $pp$ collisions at $\ecm = 10\,\tev$
  required for $3\sigma$ evidence (dashed) and $5\sigma$ observation (solid)
  of $\tom \ra \gamma Z^0 \ra \gamma\mu^+\mu^-$ as a function of $M_{\tom}$
  for LSTC Cases~a (blue) and~b (black). Right: Integrated luminosity required
  for 95\% C.L. exclusion of Cases~a (blue) and~b (black).
  \label{fig:Zgamma_lumi}}
\end{center}
\end{figure}

The low branching ratio for $\tom \ra \gamma Z$ makes this analysis channel
significantly more challenging than the other LSTC processes considered in
this report. To evaluate the channel's discovery potential we computed two
quantities by counting the events within a $20\,\gev$ window of the assumed
signal mass window: (1) the discovery potential by evaluating the 3~and
$5\sigma$ luminosity contours by a simple event counting method; (2) the
luminosity required for 95\% C.L.~exclusion of the signal if none is found.
The results are shown in Fig.~\ref{fig:Zgamma_lumi}. Depending on the masses
the luminosity for $5\sigma$ discovery ranges from a few to $100\,\ifb$. The
exclusion contours are approximate because the rate of $Z + \jets$ passing
the selection cuts is only approximately known.

\section{$\tom,\, \troz,\, \taz  \ra e^+e^-$}
 
The neutral states $\troz$, $\tom$ and $\taz$ all decay to $\ellp\ellm$
(unless $Q_U + Q_D = 0$ in which case $\tom \ra \ellp\ellm$ vanishes.) In the
TCSM as implemented in {\sc Pythia}, the $\omega_T$ signal is generally much
greater that the $\troz$ one because of the latter's larger rates into
$W\tpi$ and $WW$. In this section we present an estimate of LHC reach for
these technivectors decaying to $e^+e^-$ based on a CMS study of the
Drell-Yan process at $\sqrt{s} = 10\,\tev$~\cite{LSTCcmsZPrime}. As we shall
see, the $\taz$ may be visible in this mode with only moderate luminosity at
$M_{\taz} \simle 330\,\gev$. The presence of the nearby second resonance
distinguishes this LSTC signal from $Z'$ or $G_{RS}$ searches. (An ATLAS
study of $\tom \ra \mu^+\mu^-$ at $\sqrt{s} = 14\,\tev$ may be found in
Ref.~\cite{Aad:2009wy}.)

The CMS Collaboration has released public results showing the expected result
of an $e^+e^-$ mass spectrum from $50\,\gev$ to $2\,\tev$ for $pp$ collisions
at $10\,\tev$~\cite{LSTCcmsZPrime}; this is an update of a previous study for
$\sqrt{s} = 14\,\tev$~\cite{LSTCcmsZPrime14TeV}. This result is
re-interpreted in this report to estimate the sensitivity of the LHC to
technicolor using CMS. This is a private interpretation using information the
CMS collaboration has made public and is not an official approved result of
CMS collaboration.

\subsection{Method}

The $e^+e^-$ mass spectrum measurement along with the estimated systematic
uncertainties is taken from a preliminary CMS summer 09
result~\cite{LSTCcmsZPrime}. The parameters for this study are the following:
the electron ID efficiency is $89\pm 4$\%; $e^+e^-$ mass resolution is 2\%;
the uncertainty in the standard-model Drell-Yan is 11\%; the $t\bar t$
background uncertainty is 16\%; the jet background uncertainty is 50\%; and a
$K$-factor of 1.35 is used for the Drell-Yan signal and background. The
systematic uncertainties on the backgrounds are conservative and
approximately twice as large as a similar CDF analysis~\cite{:2007sb}.
Therefore, the possibility that the systematic uncertainties are half as
large is also considered here. While the CMS Collaboration has made no
statement on whether this reduction is possible, experience at the Tevatron
suggests that it will be. The technicolor signal sample is generated using
{\sc Pythia}. Both generator level electrons are required to satisfy
$E_{T}>50\,\gev$ and $|\eta|<1.442$ or $1.56<|\eta|<2.5$ corresponding to the
kinematic and geometric acceptance of the CMS analysis. As can be seen from
Fig.~\ref{tc:sh:fig:sigMass} for Case~2a in Table~\ref{tab:LH07}, the 2\%
mass resolution is sufficient to resolve the $\tom$ and $\taz$ resonances at
300 and $330\,\gev$. While the two peaks are distinguishable, the
interference effect between the standard model and TC signal below the first
peak is not visible with this resolution.  Figure~\ref{tc:sh:fig:sigMass}
also shows a sample pseudo-experiment in the presence of technicolor with the
predicted standard model backgrounds.

The technique used to estimate the significance of a technicolor signal is a
$p$-value method used in the CDF $e^+e^-$ search described in~\cite{:2007sb}.
This method addresses the ``look-elsewhere'' effect resulting from the fact
that the mass of a new boson resulting from new physics is not known. First a
pseudo-experiment is generated from the expected standard model background
mass distribution using a Poisson distributed random number for each bin.
Then in a mass window of $\pm 1.5$ times the mass resolution, the Poisson
probability, or $p$-value, of observing the number of observed events or
greater in the absence of new physics is calculated. The uncertainty on the
number of background events is included by averaging the $p$-values for all
possible background values weighted by a Gaussian with mean and sigma equal
to the expected background and its uncertainty. This is done in 1~GeV steps
for masses between 200 and 1000~GeV. This process is repeated for $2 \times
10^8$ pseudo-experiments per luminosity point and the two smallest $p$-values
in each pseudo-experiment are recorded. The mass windows used to calculate
the $p$-values are not allowed to overlap to ensure that they do not share any
events. Then the process is repeated in the presence of the technicolor
signal and the median $p$-value is obtained for the signal bins. The fraction
of standard-model-only pseudo-experiments which observe this $p$-value or
greater is then obtained to determine how often a similar sized signal
can be produced from chance alone.

The advantage of this search technique is that it uses very few assumptions
and is generic to all new physics types. As there are two peaks, the
$p$-values for both peaks are calculated. Then the fraction of
pseudo-experiments generated with standard-model-only templates that have a
$p$-value $< p_{\tom}$ and another $p$-value $< p_{\taz} $ is determined,
where $p_{\tom}$ and $p_{\taz}$ are the $p$-values of the two peaks.  This
offers some increase in sensitivity compared to using only the leading peak.

Limits are then set via a simple Bayesian likelihood method using Poisson
statistics. The $\pm 1.5$ mass resolution region around each peak correspond
to the two bins of the likelihood. The background uncertainty is assumed to
be modeled by a truncated Gaussian and that background uncertainty is 100\%
correlated between the two bins.

\begin{figure}
\begin{center}
\includegraphics[width=3.00in, height=3.00in, angle=90]{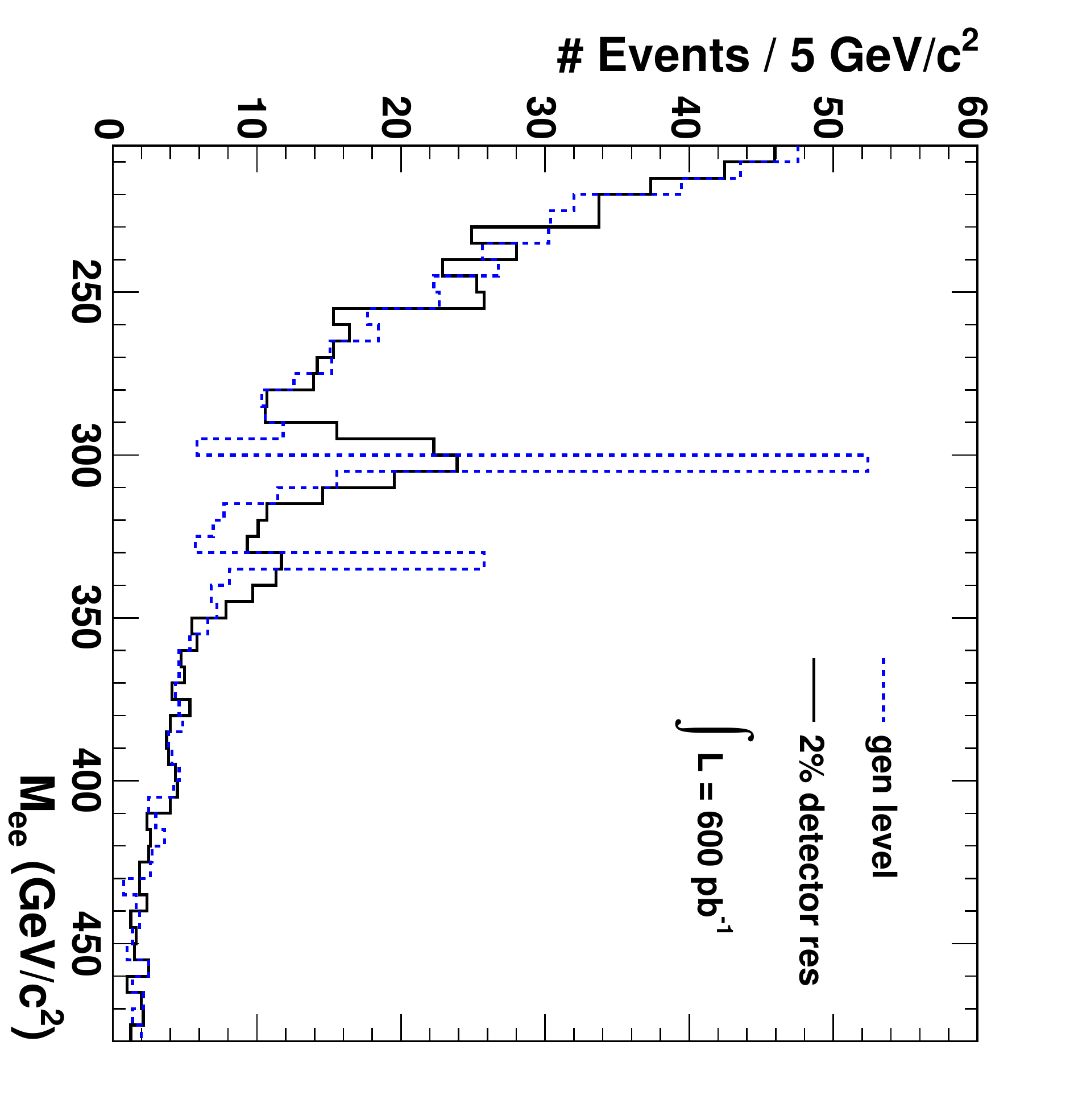}
\includegraphics[width=3.00in, height=3.00in,angle=90]{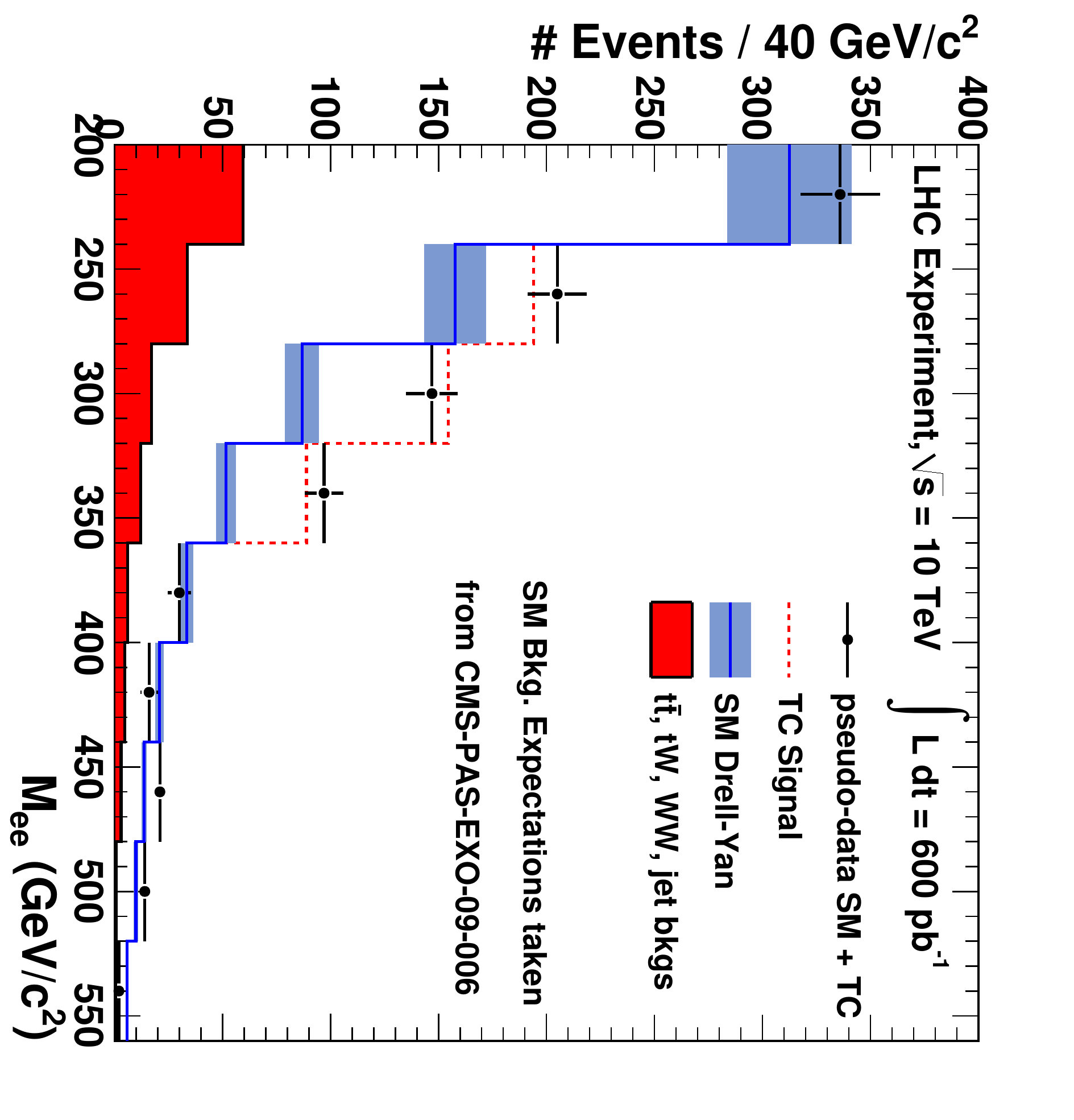}
\caption{A comparison of the LSTC signal at generator level and after detector
  resolution (left) and a pseudo-experiment for Case~2a together with the
  standard model backgrounds (right); $E_T(e^\pm) > 50\,\gev$. The standard
  model backgrounds are taken from Fig.~2a of~\cite{LSTCcmsZPrime}, scaled by
  a factor of six to account for the luminosity difference~\cite{:2007sb}.}
\label{tc:sh:fig:sigMass}
\end{center}
\end{figure}

\begin{table}[!ht]
\begin{center} 
{\scriptsize
\begin{tabular}{|c|c|c|c|c|} \hline 
$\int\CL dt$ ($\ipb$)  &  Case 1a  &  Case 1a (imp. syst.)  &  Case 1b  &
Case 1b (imp. syst.) \\
\hline 
50  &  0.022 (5.5$\times 10^{-4}$)  &  0.017 (3.9$\times 10^{-4}$)  &  0.24
(6.1$\times 10^{-3}$)  &  0.20 (5.0$\times 10^{-3}$)\\ 
100  & 1.0$\times 10^{-4}$ (3.2$\times 10^{-6}$)  &  5.5$\times 10^{-5}$ (9.7$\times
10^{-7}$)  &  0.017 (4.8$\times 10^{-4}$)  &7.0$\times 10^{-3}$ (1.2$\times 10^{-4}$)\\ 
150  &  1.4$\times 10^{-6}$ (7.1$\times 10^{-8}$)  &  2.7$\times 10^{-7}$
(8.0$\times 10^{-9}$)  &  9.7$\times 10^{-4}$ (3.0$\times 10^{-5}$)  & 
3.6$\times 10^{-4}$ (8.0$\times 10^{-6}$) \\ 
200  &  $<$1.5$\times 10^{-8}$ (5.2$\times 10^{-10}$)  &  $<$1.5$\times 10^{-8}$
(3.5$\times 10^{-11}$)  &  3.5$\times 10^{-5}$ (2.4$\times 10^{-6}$) 
&  7.3$\times 10^{-6}$ (2.9$\times 10^{-7}$) \\ 
250  &  $<$1.5$\times 10^{-8}$ (9.3$\times 10^{-12}$)  &  $<$1.5$\times 10^{-8}$
(9.2$\times 10^{-14}$)  &  2.1$\times 10^{-6}$ (2.3$\times 10^{-7}$) &  2.1$\times
10^{-7}$ (1.2$\times 10^{-8}$) \\\hline  
\end{tabular} 
}
\end{center}
\caption{The fraction of standard-model-only pseudo-experiments which observe
  a $p$-value equal to or lower than the median $p$-value of the first peak
  (shown in parentheses) and a second $p$-value equal to or lower than the
  median $p$-value of the second peak of LSTC Cases~1a and~1b. The improved
  systematics correspond to a reduction of background uncertainty by 50\%
  which is a level comparable to that in a similar CDF
  analysis~\cite{:2007sb}.}
\label{tc:sh:tab:pvaluesVsLumiCase1}  
\end{table} 

\subsection{Results and Conclusion}

Tables~\ref{tc:sh:tab:pvaluesVsLumiCase1}
and~\ref{tc:sh:tab:pvaluesVsLumiCase2} show the fraction of
standard-model-only pseudo-experiments that have have two $p$-values
somewhere in the mass spectrum larger than the median $p$-value of each peak
in the presence of technicolor for Cases~1a,b and~2a,b respectively. An
$\tom$ with mass $225\,\gev$ and $M_{\tpi} = 150\,\gev$ (Case~1a) is
discoverable at the $5\sigma$-level with $200\,\ipb$, while Case~1b requires
$\sim 300\,\ipb$. Strong evidence can be obtained for $100\,\ipb$. This puts
discovery of such a model well within the expected reach of the first run of
the LHC. For $\sqrt{s} = 10\,\tev$ and luminosities of 600--$800\,\ipb$
strong evidence can be obtained for Cases 2a,b.  Cases~3a,b can not be
distinguished from background at luminosities less than $1\,\ifb$. Improving
the systematic uncertainties gives on average a factor of five increase in
significance.

In the absence of a signal, limits can be set on the technicolor models.
Table~\ref{tc:sh:tab:lumiFor95Exclude} shows the luminosity required at
$10\,\tev$ to exclude the various cases considered. Cases~1a and~1b can be
excluded very quickly, requiring~20 and $31\,\ipb$, respectively.  Cases~2a
and~2b can be excluded with~170 and $360\,\ipb$ of data.  Cases~1a,b and 2a
and, possibly, 2b could therefore be excluded by an LHC experiment in
2010--11, however Cases~3a,b require significantly more data, on the order of
an inverse femtobarn. Reducing the systematic uncertainties would reduce the
luminosity required to exclude the LSTC models considered here by 10-15\%
which could be important in excluding Cases~2a and~2b by the end of 2011.

\begin{table}[!Hhtb]
\begin{center} 
{\scriptsize
\begin{tabular}{|c|c|c|c|c|} \hline 
$\int\CL dt$ ($\ipb$)  &  Case 2a  &  Case 2a (imp. syst.)  &  Case 2b  &  Case
2b (imp. syst.) \\\hline 
400  &  0.064 (1.4$\times 10^{-3}$)  &  0.042 (6.1$\times 10^{-4}$)  & 
0.72 (0.068)  &  0.72 (0.053) 
\\ 
600  &  8.1$\times 10^{-3}$ (2.2$\times 10^{-2}$)  &  3.0$\times 10^{-3}$
(4.4$\times 10^{-5}$)  &  0.36 (0.039)   &  0.33 (0.024) \\ 
800  &  2.0$\times 10^{-3}$ (6.5$\times 10^{-5}$)  &  3.6$\times 10^{-4}$
(6.2$\times 10^{-6}$)  &  0.089 (0.025)   &  0.066 (0.012) \\ 
1000  &  2.0$\times 10^{-4}$ (1.0$\times 10^{-5}$)  &  1.3$\times 10^{-5}$
(3.8$\times 10^{-7}$)  &  0.067 (0.017)  &  0.033 (0.0063) \\ 
1500  &  3.1$\times 10^{-6}$ (5.1$\times 10^{-7}$)  &  3.5$\times 10^{-8}$
(2.2$\times 10^{-9}$)  &  4.7$\times 10^{-3}$ (9.6$\times 10^{-3}$)  &
7.1$\times 10^{-4}$ (1.9$\times 10^{-3}$) \\\hline
\end{tabular} 
}
\caption{The fraction of standard-model-only pseudo-experiments which observe
  a $p$-value equal to or lower than the median $p$-value of the first peak
  (shown in parentheses) and a second $p$-value equal to or lower than the
  median $p$-value of the second peak of LSTC Cases~2a and~2b. The improved
  systematics correspond to a reduction of background uncertainty by 50\%
  which is a level comparable to that in a similar CDF
  analysis~\cite{:2007sb}.}
\label{tc:sh:tab:pvaluesVsLumiCase2}  
\end{center} 
\end{table} 

\begin{table}[!Hhtb]
\begin{center}
\begin{tabular}{|c|c|c|} \hline 
Model  & nominal syst. & improved syst. \\\hline
1a & 20 & 20  \\
1b & 31 & 31   \\
2a & 170 & 150  \\
2b & 360 & 320  \\
3a & 610 & 560  \\
3b & 1120 & 930  \\\hline
\end{tabular} 
\caption{Luminosities (in $\ipb$) needed at $\sqrt{s} = 10\,\tev$ to exclude
  $\tom \ra e^+e^-$ in the LSTC models in Table~\ref{tab:LH07} at 95\%~C.L.
  Nominal systematics are on the left. Improved systematics on the right
  correspond to a reduction of background uncertainty by 50\% which is a
  level comparable to that in a similar CDF analysis~\cite{:2007sb}.}
\label{tc:sh:tab:lumiFor95Exclude}  
\end{center} 
\end{table}

\section{$\Leff$ for low-scale technicolor}

There are three motivations for an effective Lagrangian for
LSTC~\cite{Lane:2009ct}. First, longitudinally polarized electroweak bosons,
$W_L^\pm$ and $Z_L^0$, play an important role in the TCSM described in
Sect.~1 and are expected to appear in many of the technivector decays
accessible at a hadron collider. They are treated in the TCSM in the
approximation $W^\pm_{L\mu} = \partial_\mu \Pi_T^\pm/M_W$ and $Z^0_{L\mu} =
\partial\mu \Pi_T^0/M_Z$. This is valid when $p_W^2 \gg M_W^2$, but that is
not always the case, especially when $\tro \ra WZ$ for the lightest $\tro$ we
consider here.  Therefore, we want a consistent mathematical treatment of
longitudinal {\em and} transverse weak bosons that a Lagrangian can furnish.
This will also allow us to assess the transverse weak boson contribution to
the angular distributions in Eq.~(\ref{eq:angdist}). Second, a Lagrangian
makes available the versatility of such programs as MadGraph and CalcHEP for
generating amplitudes to be used in {\sc Pythia} and HERWIG.  Finally, the
TCSM describes a phenomenology of LSTC expected to be valid only in the
limited energy $\sqrt{\shat} \simle M_{\tro}$, where the lightest
technihadrons may be treated in isolation. An effective Lagrangian, $\Leff$,
is well-suited for this description because it gives warning of its
limitation.

The hidden local symmetry (HLS) formalism of Bando, {\it et
  al.}~\cite{Bando:1984ej, Bando:1987br} was adopted to construct an $\Leff$
describing the technivector mesons, electroweak bosons and technipions of
LSTC. Such an $\Leff$ guarantees that production of $W_L,\,Z_L$ via
annihilation of massless fermions is well-behaved at all energies {\em in
  tree
  approximation.}
Elastic $W_L W_L$ scattering still behaves at high energy as it does in the
standard model without a Higgs boson, i.e., the amplitude $\sim
\shat/F_\pi^2$ at $\shat \gg M_{\tro}^2$. Of course, this violation of
perturbative unitarity signals the strong interactions of the underlying
technicolor theory. The HLS method also guarantees that the photon is
massless and the electromagnetic current conserved.

The gauge symmetry group of $\Leff$ is $\CG = SU(2)_W \otimes U(1)_Y \otimes
U(2)_L \otimes U(2)_R$. The first two groups are the standard electroweak
gauge symmetries, with primordial couplings $g$ and $g'$ and gauge bosons
$\bs{W} = (W^1,W^2,W^3)$ and $B$. The latter two are the ``hidden local
symmetry'' groups. The underlying TC interactions are parity-invariant, so
that their zeroth-order couplings are equal, $g_L = g_R = g_T$. The assumed
equality of the $SU(2)_{L,R}$ and $U(1)_{L,R}$ couplings reflect the isospin
symmetry of TC interactions and the expectation that $M_{\tro} \cong
M_{\tom}$ and $M_{\ta} \cong M_{f_T}$.  This symmetry must be broken
explicitly if $\Leff$ is to allow an appreciable $\tro$--$\tom$ splitting. We
have not done that.\footnote{Mixing between $\troz$ and $\tom$ is limited by
  the smallness of the $T$-parameter.} The gauge bosons $(\bs{L},L^0)$ and
$(\bs{R},R^0)$ contain the primordial technivector mesons,
$\bs{V},V_0,\bs{A},A_0 \cong \bs{\rho}_T,\tom,\bs{a}_T,f_T$.

To describe the lightest $\tpi$ and to mock up the heavier TC states that
contribute most to electroweak symmetry breaking (see Sect.~1), and to break
all the gauge symmetries down to electromagnetic $U(1)$, the nonlinear
$\Sigma$-model fields in $\Leff$ are $\Sigma_2$, $\xi_L$, $\xi_R$ and
$\xi_M$, transforming under $\CG$ as
\bea\label{eq:sigmatransforms}
\Sigma_2 &\ra& U_W\Sigma_2U_Y^\dagg,\qquad
\xi_L \ra U_W U_Y \xi_L U_L^\dagg ,\nn \\
\xi_M &\ra& U_L \xi_M U_R^\dagg,\qquad
\xi_R \ra U_R \xi_R U_Y^\dagg\,.
\eea
The covariant derivatives describing their coupling to the gauge fields are
\bea\label{eq:covderivs}
D_\mu \Sigma_2 &=& \partial_\mu\Sigma_2 -ig \bs{t}\cdot\bs{W}_\mu \Sigma_2
+ ig' \Sigma_2 t_3 B_\mu, \nn \\
D_\mu \xi_L &=& \partial_\mu\xi_L -i(g\bs{t}\cdot\bs{W}_\mu +g' y_1 t_0
B_\mu) \xi_L + ig_T \xi_L\, t\cdot L_\mu, \nn\\ 
D_\mu \xi_M &=& \partial_\mu\xi_M  -ig_T (t \cdot L_\mu\, \xi_M - \xi_M\,  t
\cdot R_\mu), \nn\\ 
D_\mu \xi_R &=& \partial_\mu\xi_R -ig_T t \cdot R_\mu\, \xi_R + ig' \xi_R (t_3
+ y_1 t_0)B_\mu \,,
\eea
where $t\cdot L_\mu = \sum_{\alpha=0}^3 t_\alpha L_\mu^\alpha$ and $\bs{t} =
{\half}\bs{\tau}$, $t_0 = {\half}\bs{1}$. The hypercharge $y_1 = Q_U + Q_D$
of the TCSM. The field $\Sigma_2$ contains the technipions that get absorbed
by the $W$ and $Z$ bosons. They are an isotriplet of $F_2$-scale Goldstone
bosons, where $F_2 = F_\pi \cos\chi \gg F_1$, and $\chi$ was introduced in
Sect.~1. It is parameterized as $\Sigma_2(x) =
\exp{(2i\bs{t}\cdot\bs{\pi}_2(x)/F_2)}$. It is convenient to define $\Sigma_1
= \xi_L \xi_M \xi_R$; then
\bea\label{eq:sigone}
\Sigma_1 &\ra& U_W \Sigma_1 U_Y^\dagg \nn\\
D_\mu \Sigma_1 &=& \partial_\mu\Sigma_1 -ig \bs{t}\cdot\bs{W}_\mu \Sigma_1
+ ig' \Sigma_1 t_3 B_\mu\,.
\eea
In the unitary gauge ($\Sigma_2,\, \xi_L,\,\xi_R\ra 1$) this field will be
parameterized as $\Sigma_1 = \exp{(2it\cdot\tilde\pi/F_1)}$, where $\tilde
\pi$ are the isovector and isoscalar technipions ${\bs \pi}_T, \tpipr$ up to
a normalization constant.

The complete effective Lagrangian is
\be\label{eq:Leff}
\Leff = \Lsig + \LWZW + \LFF + \Lff + \CL_{M_\pi^2} + \Lpifbf\,.
\ee
Here,
\bea\label{eq:Lsigone}
\Lsig = && {\tfourth} F_2^2 \Tr|D_\mu\Sigma_2|^2 
      + {\tfourth} F_1^2 \Bigl\{a \Tr|D_\mu\Sigma_1|^2 + 
        b\Bigl[\Tr|D_\mu\xi_L|^2 + \Tr|D_\mu\xi_R|^2\Bigr] \nn\\
        && + c\, \Tr|D_\mu\xi_M|^2 
           + d\, \Tr(\xi_L^\dagg D_\mu \xi_L D_\mu \xi_M \xi_M^\dagg + 
                   \xi_R D_\mu \xi_R^\dagg D_\mu \xi_M^\dagg
                   \xi_M)\nn\\
        && - \frac{if}{2g_T}\,\Tr(D_\mu\xi_M \xi_M^\dagg D_\nu\xi_M
        \xi_M^\dagg\, t\cdot L_{\mu\nu} + 
                    \xi_M^\dagg D_\mu\xi_M \xi_M^\dagg D_\nu\xi_M\,
                    t\cdot R_{\mu\nu})\Bigr\}\,.
\eea
The dimensionless constants $a,b,c,d,f$ are expected to be $\CO(1)$. The
first four terms are those involving only two derivatives and/or gauge fields
that are consistent with the symmetries of TC interactions. The $f$-term is
needed to describe decays of $\ta$. It is one of several possibilities and,
to minimize the number of free parameters, only one such term is used. The
$\LWZW$ interaction includes the Wess-Zumino-Witten (WZW)
terms~\cite{Wess:1971yu, Witten:1983tw} implementing the effects of
anomalously nonconserved symmetries of the underlying TC theory. They are
essential for describing the radiative decays of $\tro$ and $\tom$ as well as
$\tpiz\ra \gamma \gamma$. They are described in more detail in
Refs.~\cite{Lane:2009ct} and~\cite{Harvey:2007ca}. The remaining terms in
$\Leff$ are the gauge kinetic terms, couplings of quarks and leptons to
$(SU(2)\otimes U(1))_{EW}$ gauge bosons, $\tpi$ mass terms, and the couplings
of $\tpi$ to quarks and leptons.

This Lagrangian describes production and decay of the technivector mesons. In
this section we concentrate on the modes $\tropm,\,\tapm \ra W^\pm Z^0$ and
$\gamma W^\pm$. The operators describing the on-shell decays $\tropm,\,\tapm
\ra WZ$ are rather complicated and they are given in Ref.~\cite{Lane:2009ct},
Eqs.~(47) and~(56). The purely longitudinal process $\tropm \ra W_L Z_L$ is
controlled by the coupling $\grpp$ and, as we discuss below, $\Leff$ predicts
a considerably smaller value of this parameter than was used in the TCSM.
This and the small $W,Z$ momenta make the transverse $W$ and $Z$
contributions to this decay at least as important as the longitudinal ones.
The longitudinal-$W$ approximation is accurate for the radiative decays with
their larger momenta. The effective Lagrangians for these decays are
\bea\label{eq:raddecays}
\CL(\tropm \ra \gamma W^\pm) &=& \frac{eg y_1 F_\pi\sin\chi}
  {2M_{V_1}}[\rho^+_{T\mu}W^-_\nu + \rho^-_{T\mu}W^+_\nu]
  \widetilde F^{\mu\nu} \nn\\ 
&\simeq& 
\frac{ey_1\sin\chi}{2M_{V_1}}\,
 [\rho_{T\mu\nu}^+ \Pi_T^- + \rho_{T\mu\nu}^- \Pi_T^+]\widetilde F^{\mu\nu}
\,;\\
\CL(\tapm \ra \gamma W^\pm) &=& -\frac{ieg F_\pi\sin\chi}
  {2M_{A_2}}\,(a_{T\nu}^+ W_\mu^- - a_{T\nu}^- W_\mu^+)F^{\mu\nu} \nn\\
&\simeq& \frac{ie\sin\chi}{2M_{A_2}}\, (a_{T\mu\nu}^+ \Pi_T^-
- a_{T\mu\nu}^- \Pi_T^+) F^{\mu\nu} \,.
\eea
Here, $F_{\mu\nu}$ is the electromagnetic field strength and $\widetilde
F_{\mu\nu} = \half \epsilon_{\mu\nu\lambda\rho} F^{\lambda\rho}$ is its dual.
The mass parameters $M_{V_1}$ and $M_{A_2}$ are set equal $M_{\tro}$ in this
study.\footnote{In QCD, the parameter $M_V$ controlling $\rho^0 \ra
  \gamma\pi^0$ is $700\,\mev$, very close to $M_{\rho}$.}

The coupling $\grpp$ and the TCSM mass parameters $M_{V_i}$ and $M_{A_i}$ are
functions of the $\Leff$ couplings $a,\dots,f$ and of $F_\pi$, $\sin\chi$ and
$\Ntc$. It is both possible and natural to choose as inputs $F_\pi$,
$\sin\chi$, $\Ntc$, $M_{\tro} = M_{\tom}$, $M_{\ta}$ and the mass parameters
$M_{V_1}$, $M_{A_1}$ and $M_{A_2}$ (only these enter the technivector decays
we study) and to express $f$, $g_T$ and $\grpp$ in terms of them. This is
what was done in the TCSM in {\sc Pythia} {\em except} that there $g_T$ {\em
  is} the $\trho \ra \tpi\tpi$ coupling and was chosen to be
$(g_T^2/4\pi)_{TCSM} = 2.16(3/\Ntc)$.  We obtain:
\bea\label{eq:params}
&& g_T = \frac{16\sqrt{2}\,\pi^2 M_{A_1} F_\pi\sin\chi}{\Ntc M_{V_1}(M_{A_1}
  + M_{A_2})} \,,\nn\\
&& f = \frac{(4\pi M_{A_1} F_\pi\sin\chi)^2}{\Ntc M_{V_1} M_{A_2}^2
(M_{A_1} + M_{A_2})} \,,\nn\\
&& \grpp = \frac{M_{\tro}^2}{\sqrt{2}g_T (F_\pi \sin\chi)^2}\left[1 +
  (f-1)\frac{M_{A_2}^2}{M_{A_1}^2}\right]\,.
\eea
In the present study we set $M_{V_i} = M_{A_i} = M_{\tro}$.\footnote{The
  $F_1$-scale contribution to the $s$-parameter vanishes in this limit.} In
this case,
\be\label{eq:ourparams}
g_T = \frac{8\sqrt{2}\,\pi^2 F_\pi\sin\chi}{\Ntc M_{\tro}}\,,\qquad
\grpp = \frac{M_{\tro}}{2F_\pi\sin\chi}\,.
\ee
This expression for $\grpp$ (but not for $g_T$) is what one would expect for
a Higgs mechanism origin for $M_{\tro}$ with gauge coupling $\simeq \grpp$
and Goldstone boson decay constant $F_\pi \sin\chi \simeq F_1$. It is
also reminiscent of the KSRF relation~\cite{Kawarabayashi:1966kd,
  Riazuddin:1966sw}.

\begin{table}[!Hhtb]
\begin{center}
  \begin{tabular}{|c|c|c|c|c|c|c|c|}
  \hline
 Case & $\grpp$ & $\Gamma(\tropm)$ & $B(WZ)_{\tro}$ &
 $B(\gamma W)_{\tro}$ & $\Gamma(\tapm)$ & $B(WZ)_{\ta}$ & $B(\gamma W)_{\ta}$ \\
  \hline\hline
 1a & 1.372 & 46 & 0.349 & 0.133 &  93  & 0.103  & 0.095 \\
 1b & 1.372 & 84 & 0.191 & 0.072 & 113 & 0.085   & 0.078 \\
 2a & 1.829 & 282 & 0.221 & 0.033& 146 & 0.124   & 0.087 \\
   \hline\hline
\end{tabular}
 \caption{The $\tro\ra\tpi\tpi$ decay constant $\grpp$ and total widths (in
 MeV) and branching ratios for $\tropm$ and $\tapm$ decays 
 to $W^\pm Z^0$ and $\gamma W^\pm$ for cases 1a,b and 2a. Note that the
 QCD-inspired value of $\grpp$ used in {\sc Pythia} is
 $\sqrt{4\pi(2.16)(3/\Ntc)} = 4.512$ for $\Ntc = 4$. Other TCSM parameters
 used are $\sin\chi = 1/3$, $\Ntc = 4$ and $Q_U = Q_D + 1 = 1$ (i.e., $y_1 =
 1$).
     \label{tab:widths}}
\end{center}
 \end{table}
\begin{table}[!Hhtb]
\begin{center}
  \begin{tabular}{|c|c|c|c|c|}
  \hline
 Case & $\sigma(WZ)_{\tro}$ & $\sigma(WZ)_{\ta}$ & $\sigma(\gamma W)_{\tro}$
 & $\sigma(\gamma W)_{\ta}$ \\ 
  \hline\hline
 1a & 45 (35) & 4.3 (30) & 1765 (905) & 860 (555) \\
 1b & 25 (35) & 3.4 (30) & 920 (905)  & 695 (555) \\
 2a & 17 (20) & 3.7 (17) & 280 (245)  & 575 (160) \\
   \hline\hline
\end{tabular}
 \caption{Parton-level $\tropm$, $\tapm$ signal cross sections (in $\fb$) for
   $pp$ collisions at $\sqrt{s} = 10\,\tev$ for cases 1a,b and 2a.  Cross
   sections were calculated using $\Leff$ and by integrating over $\pm
   20\,\gev$ about the resonances. Cross sections in parentheses are the
   underlying standard model rates. Branching ratios of $W$ and $Z$ to
   electrons and muons are included. Other TCSM parameters used are
   $\sin\chi = 1/3$, $\Ntc = 4$ and $Q_U = Q_D + 1 = 1$ (i.e., $y_1 = 1$).
     \label{tab:xsections}}
\end{center}
\end{table}
\vfil\eject

The important consequence of Eq.~(\ref{eq:params}) is that
$\alpha_{\tro\tpi\tpi} = \grpp^2/4\pi$ is proportional to $M_{\tro}^2$.  For
the $M_{\tro}$ of low-scale technicolor, $\alpha_{\tro\tpi\tpi}$ is
considerably smaller than the default value $2.16(3/\Ntc)$ used in the {\sc
  Pythia} implementation of the TCSM. This greatly reduces the branching
ratios $B(\tro \ra W\tpi,\, WZ)$ and, so long as $y_1$ is not small,
correspondingly enhances $B(\tro \ra \gamma \tpi,\,\gamma W)$; see
Table~\ref{tab:widths}. We do not know which value of $\grpp$ is more
reliable. The KSRF relation $g_{\rho\pi\pi} = M_\rho/\sqrt{2} f_\pi$ works
well in QCD. If HLS is more than an accidental description of the low-energy
QCD spectrum (see Ref.~\cite{Georgi:1989xy} for a contrary view), that may
lend credence to using the smaller value of $\grpp$ here. Still, we must
remember the admonition to rely with suspicion on QCD for describing walking
technicolor.  Only experiment can decide.

\begin{figure}
\begin{center}
\includegraphics[width=3.00in, height=2.95in, angle=0]{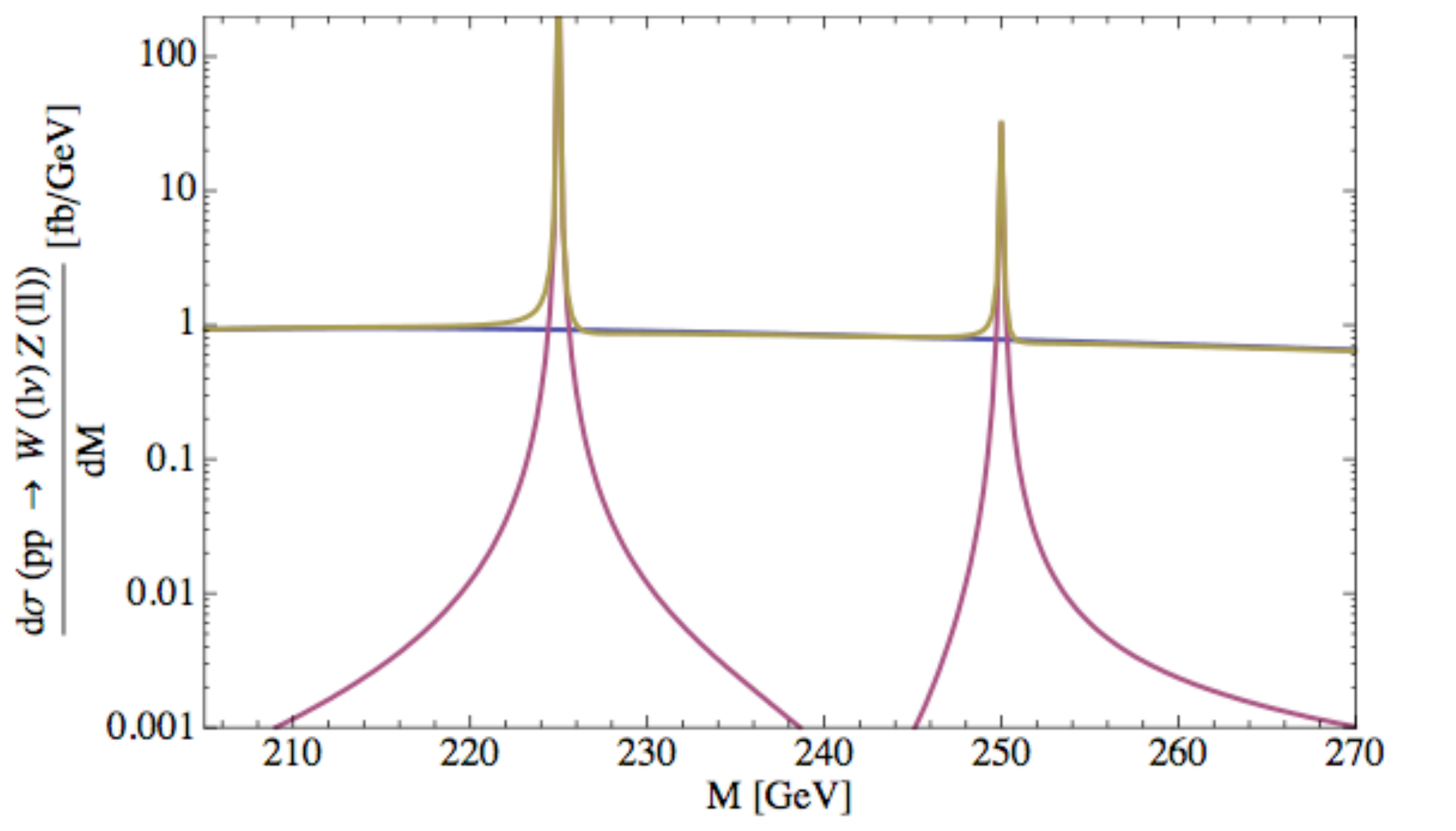}
\includegraphics[width=3.00in, height=2.80in,angle=0]{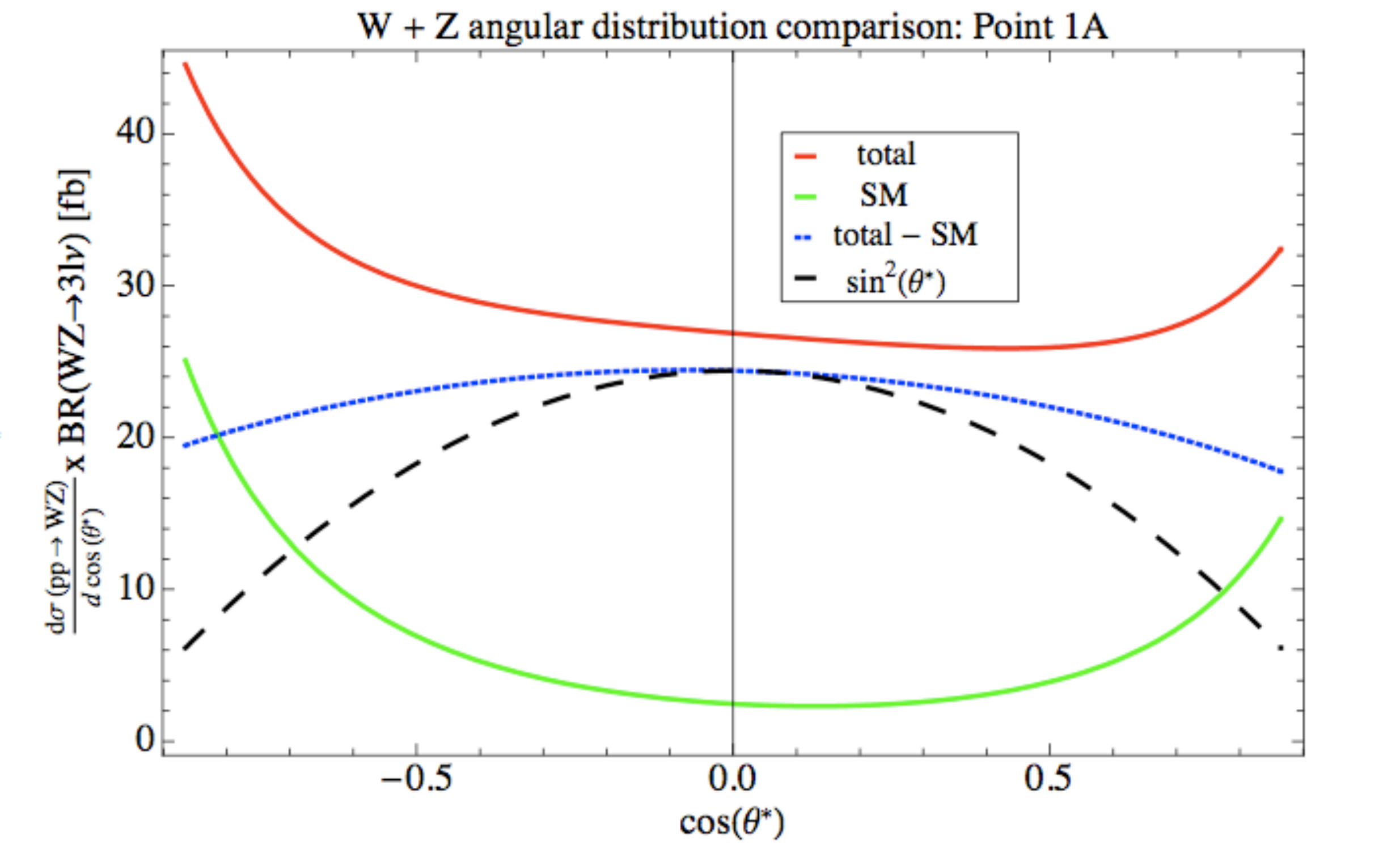}
\caption{The $W^\pm Z^0$ invariant mass (left) and angular (right) distributions
  calculated from $\Leff$ for Case~1a with $\sqrt{s} = 10\,\tev$, $M_{\tro} =
  225\,\gev$, $M_{\ta} = 250\,\gev$ and $M_{\tpi} = 150\,\gev$. The angular
  distributions are for the $\tropm \ra W^\pm Z^0$ region, and the total
  (red), standard-model (green), total - SM (blue dashed) and pure
  $\sin^2\theta$ (black dashed) are shown. The standard-model contribution is
  calculated over the resonance region.}
\label{fig:WZ_1a}
\end{center}
\end{figure}

\begin{figure}
\begin{center}
\includegraphics[width=3.00in, height=2.95in, angle=0]{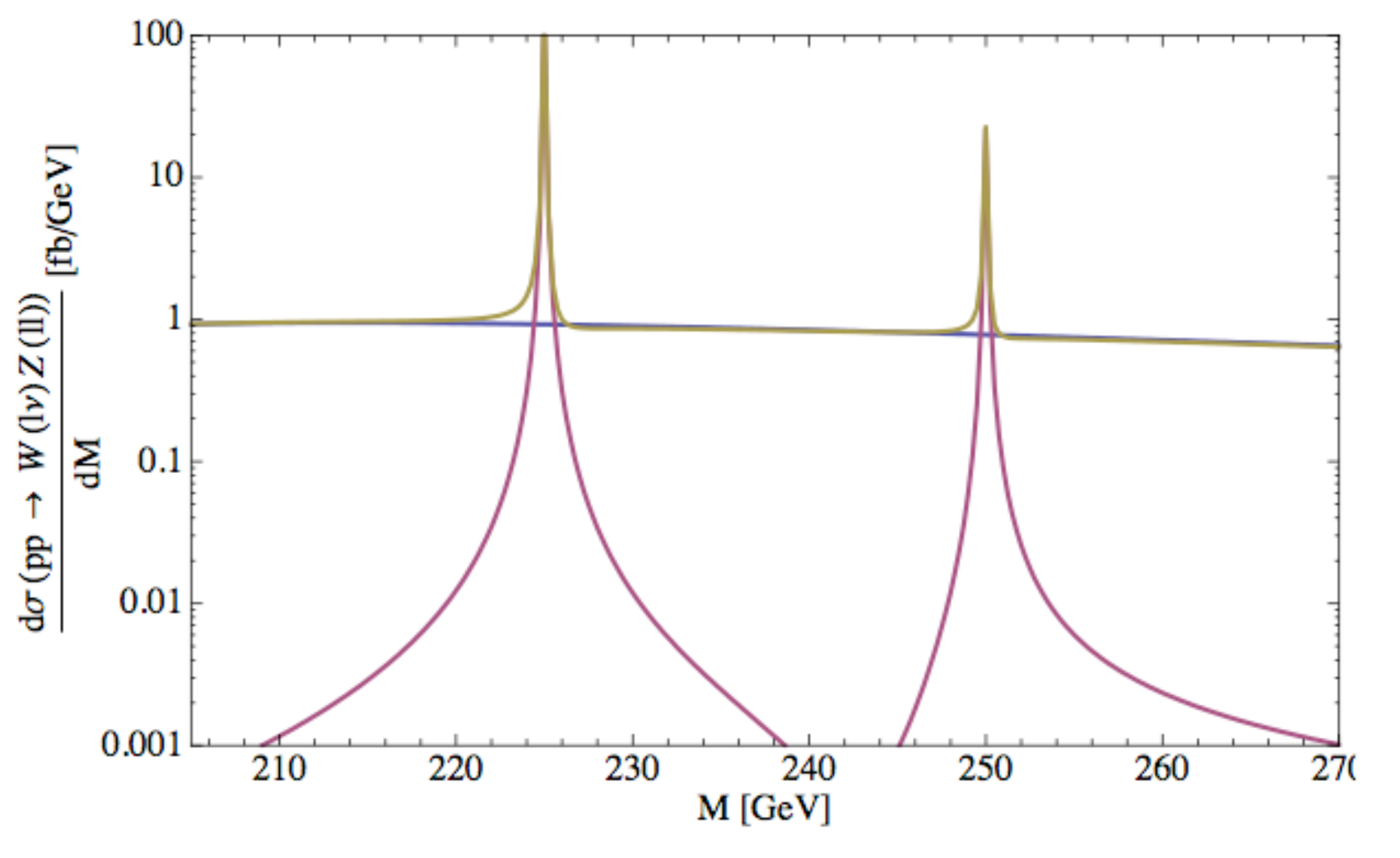}
\includegraphics[width=3.00in, height=2.80in,angle=0]{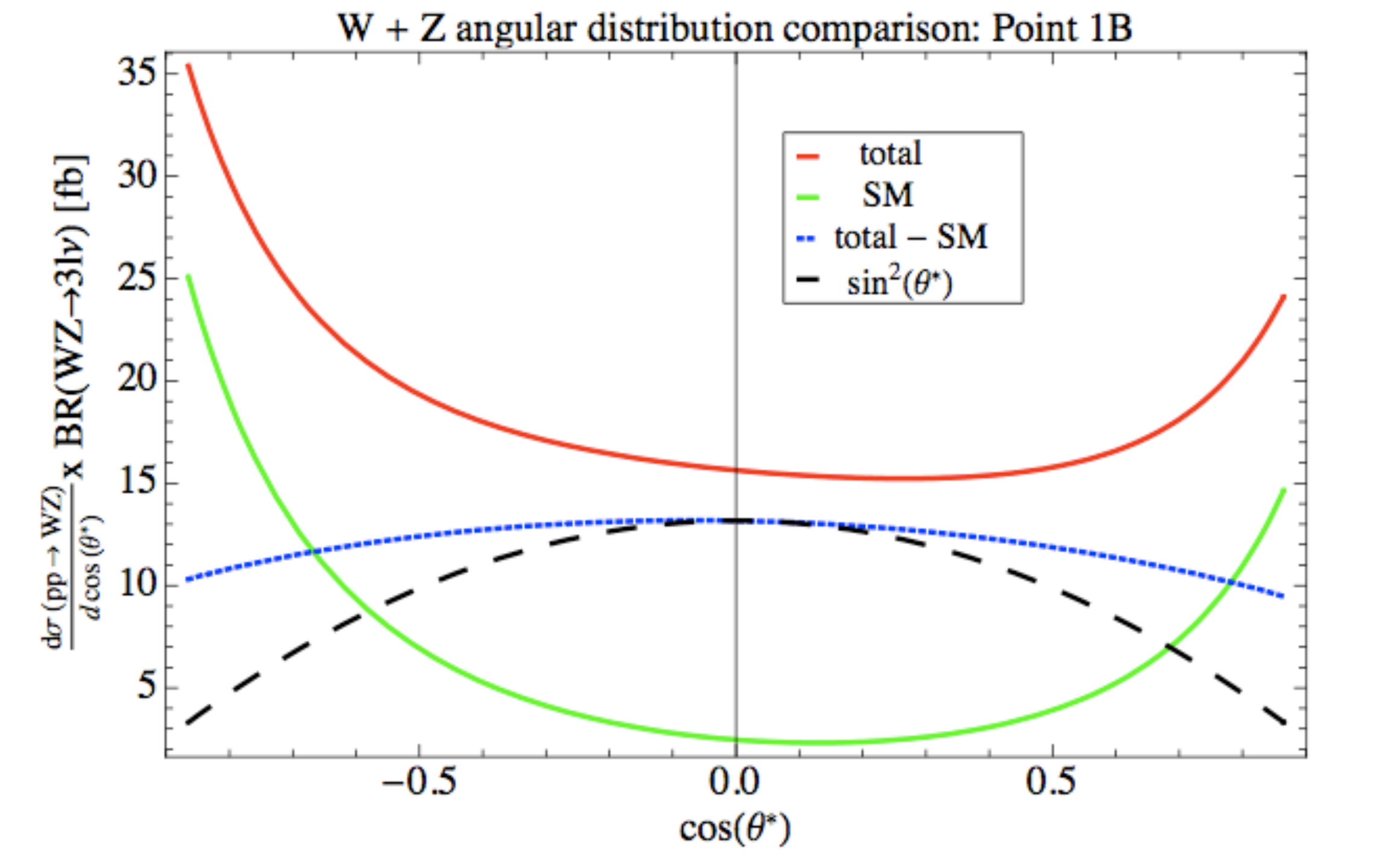}
\caption{The $W^\pm Z^0$ invariant mass (left) and angular (right) distributions
  calculated from $\Leff$ for Case~1b with $\sqrt{s} = 10\,\tev$, $M_{\tro} =
  225\,\gev$, $M_{\ta} = 250\,\gev$ and $M_{\tpi} = 140\,\gev$. The angular
  distribution is for $\tropm \ra W^\pm Z^0$.}
\label{fig:WZ_1b}
\end{center}
\end{figure}

\begin{figure}
\begin{center}
\includegraphics[width=3.00in, height=2.95in, angle=0]{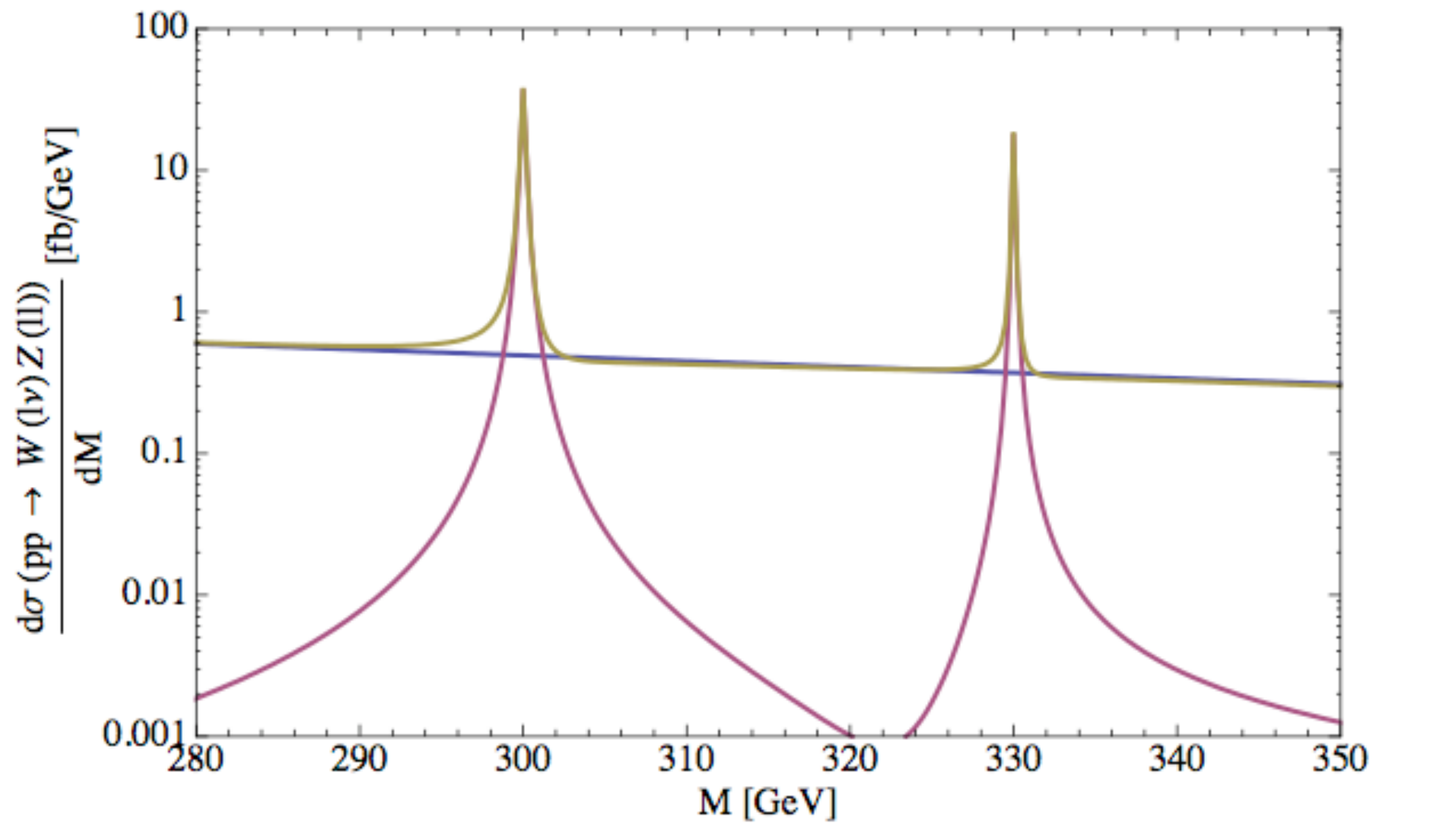}
\includegraphics[width=3.00in, height=2.80in,angle=0]{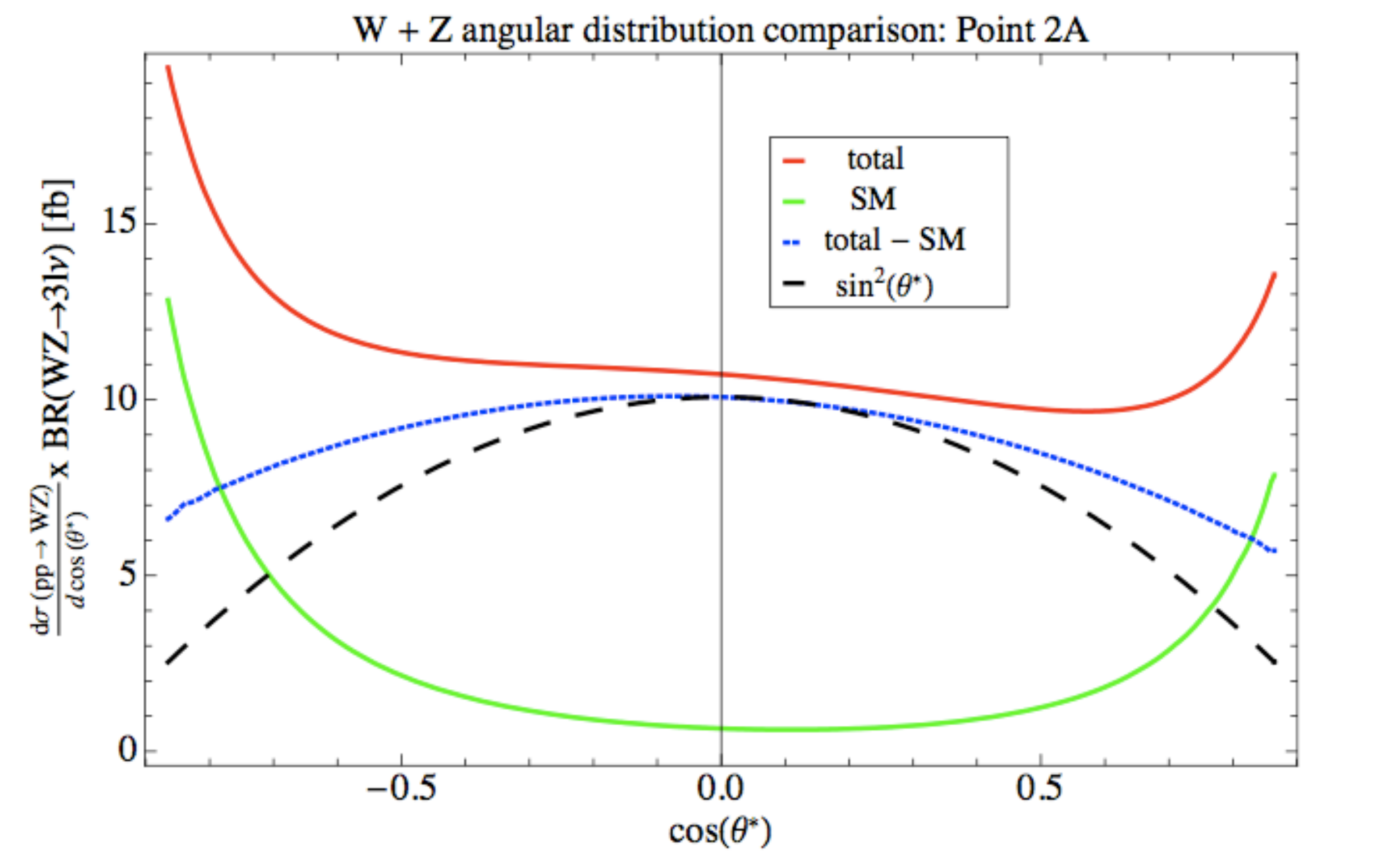}
\caption{The $W^\pm Z^0$ invariant mass (left) and angular (right) distributions
  calculated from $\Leff$ for Case~2a with $\sqrt{s} = 10\,\tev$, $M_{\tro} =
  300\,\gev$, $M_{\ta} = 330\,\gev$ and $M_{\tpi} = 200\,\gev$. The angular
  distribution is for $\tropm \ra W^\pm Z^0$.}
\label{fig:WZ_2a}
\end{center}
\end{figure}

The cross sections for $\tropm,\,\tapm \ra W^\pm Z^0$ and $\gamma W^\pm$,
followed by $W$ and $Z$ decays to electrons and muons, for cases 1a (in which
$\tro \ra W\tpi$ is forbidden), 1b and 2a are listed in
Table~\ref{tab:xsections}. The effect of the small $\grpp$ on these cross
sections compared to the {\sc Pythia} rates in Table~\ref{tab:cases} is
dramatic.

The parton-level invariant mass and angular distributions for these three
cases of $\tropm,\,\tapm \ra W^\pm Z^0$ are shown in
Figs.~\ref{fig:WZ_1a}, \ref{fig:WZ_1b} and \ref{fig:WZ_2a}. CTEQ5l parton
distribution functions were used. Although no experimental realism was
included in these calculations, comparing with the results of the CMS study
in Sect.~3 (see Table~\ref{tab:cases} and Fig.~\ref{fig:WZ-mass}), it seems
unlikely that $\tropm \ra WZ$ with such small $\grpp$ could be discovered
with only 1--$2\,\ifb$ at $\sqrt{s} = 10\,\tev$. We won't speculate on what
it would take to observe the angular distributions and determine whether or
not they fit the LSTC expectation because no serious studies have been done.
However, it is noteworthy that the sideband-subtracted angular distribution
(calculated by integrating the standard-model contribution over the resonance
region and subtracting it from the total cross section) is considerably
larger than the standard-model one and that it looks much more like
$\sin^2\theta$ than the standard model does. It is also clear that, as
expected for small $\grpp$, there is substantial contribution to $\tropm \ra
WZ$ from transversely-polarized $W$ or $Z$, and that this flattens out the
angular distributions compared to $\sin^2\theta$. Figure~\ref{fig:WZ_2a}
shows that $\tropm \ra W_L Z_L$ becomes more important as $M_{\tro}$
increases.

The invariant mass and angular distributions of $\tropm,\,\tapm \ra \gamma
W^\pm$ for cases~1a,b and~2a are shown in
Figs.~\ref{fig:gW_1a}, \ref{fig:gW_1b} and \ref{fig:gW_2a}. Thanks to the
substantially larger branching ratio for $\tropm \ra \gamma W$ that $\Leff$
predicts (for $y_1 = \CO(1)$), both resonances can be seen with quite modest
luminosity.  Conversely, it appears that $1\,\ifb$ at $\sqrt{s} = 7\,\tev$
would be sufficient to exclude these cases. If the resonances are discovered
at the rates shown here, the angular distributions, shown for $\tapm \ra
\gamma W$, should be measurable as well. The sideband-subtracted
distributions are quite close to the $1 + \cos^2\theta$ expected for a
$\gamma W_L$ signal.

\begin{figure}
\begin{center}
\includegraphics[width=3.00in, height=2.95in, angle=0]{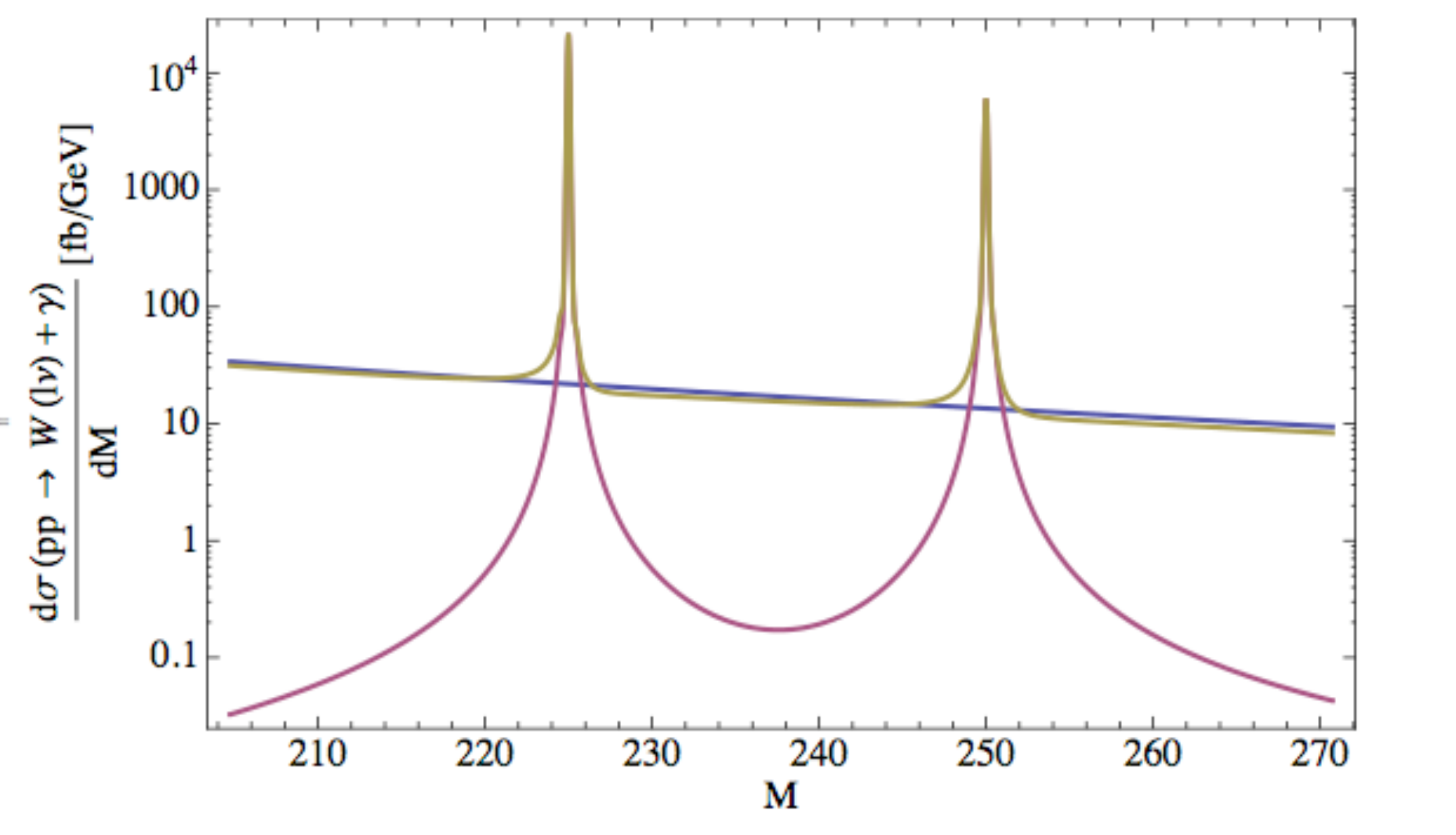}
\includegraphics[width=3.00in, height=2.80in,angle=0]{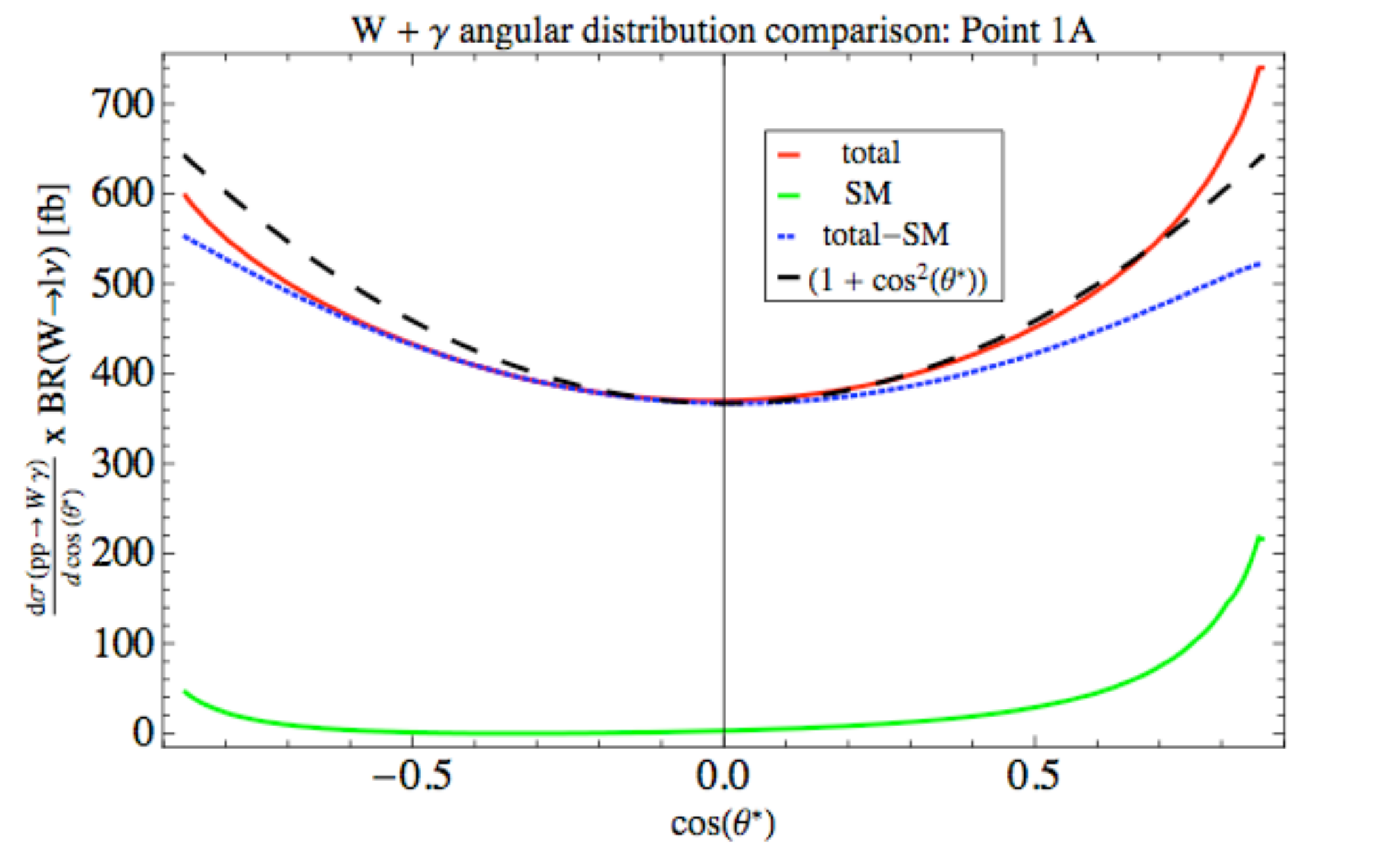}
\caption{The $\gamma W^\pm$ invariant mass (left) and angular (right)
  distributions calculated from $\Leff$ for Case~1a with $\sqrt{s} =
  10\,\tev$, $M_{\tro} = 225\,\gev$, $M_{\ta} = 250\,\gev$ and $M_{\tpi} =
  150\,\gev$. The angular distributions are for the $\tapm \ra \gamma W^\pm$
  region, and the total (red), standard-model (green), total - SM (blue
  dashed) and pure $\sin^2\theta$ (black dashed) are shown. The
  standard-model contribution is calculated over the resonance region.}
\label{fig:gW_1a}
\end{center}
\end{figure}

\begin{figure}
\begin{center}
\includegraphics[width=3.00in, height=2.95in, angle=0]{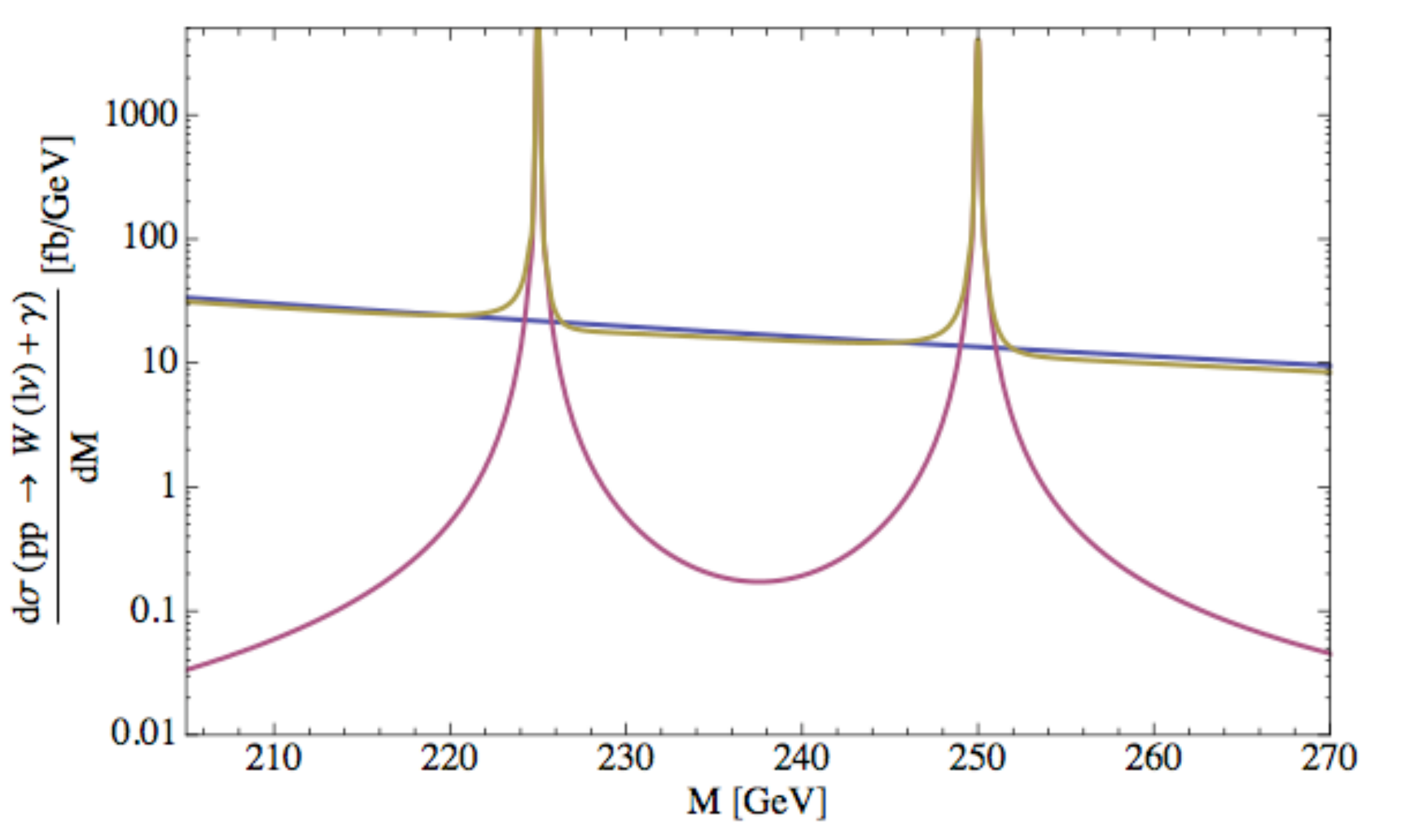}
\includegraphics[width=3.00in, height=2.80in,angle=0]{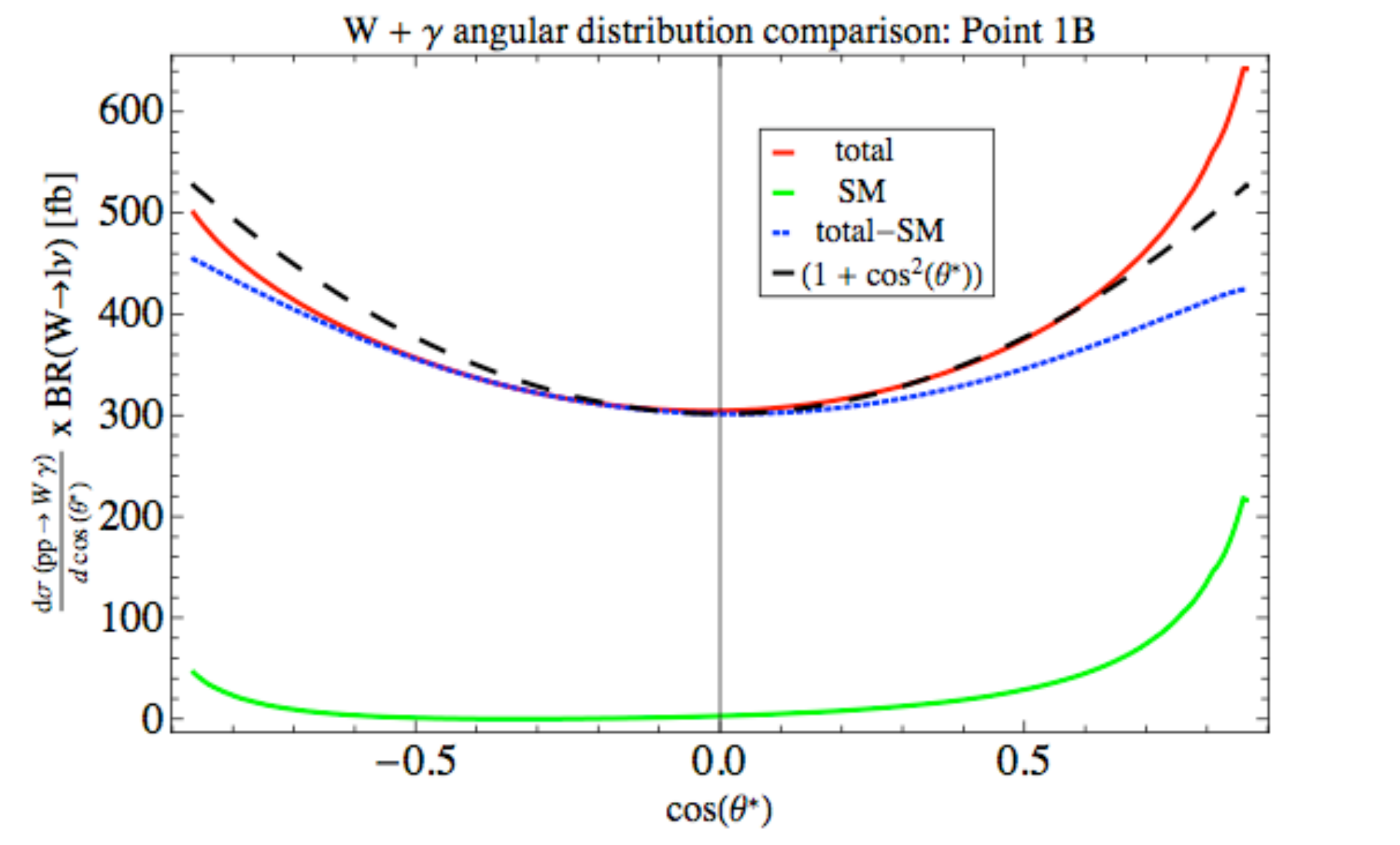}
\caption{The $\gamma W^\pm$ invariant mass (left) and angular (right) distributions
  calculated from $\Leff$ for Case~1b with $\sqrt{s} = 10\,\tev$, $M_{\tro} =
  225\,\gev$, $M_{\ta} = 250\,\gev$ and $M_{\tpi} = 140\,\gev$. The angular
  distribution is for $\tapm \ra \gamma W^\pm$.}
\label{fig:gW_1b}
\end{center}
\end{figure}

\begin{figure}
\begin{center}
\includegraphics[width=3.00in, height=2.95in, angle=0]{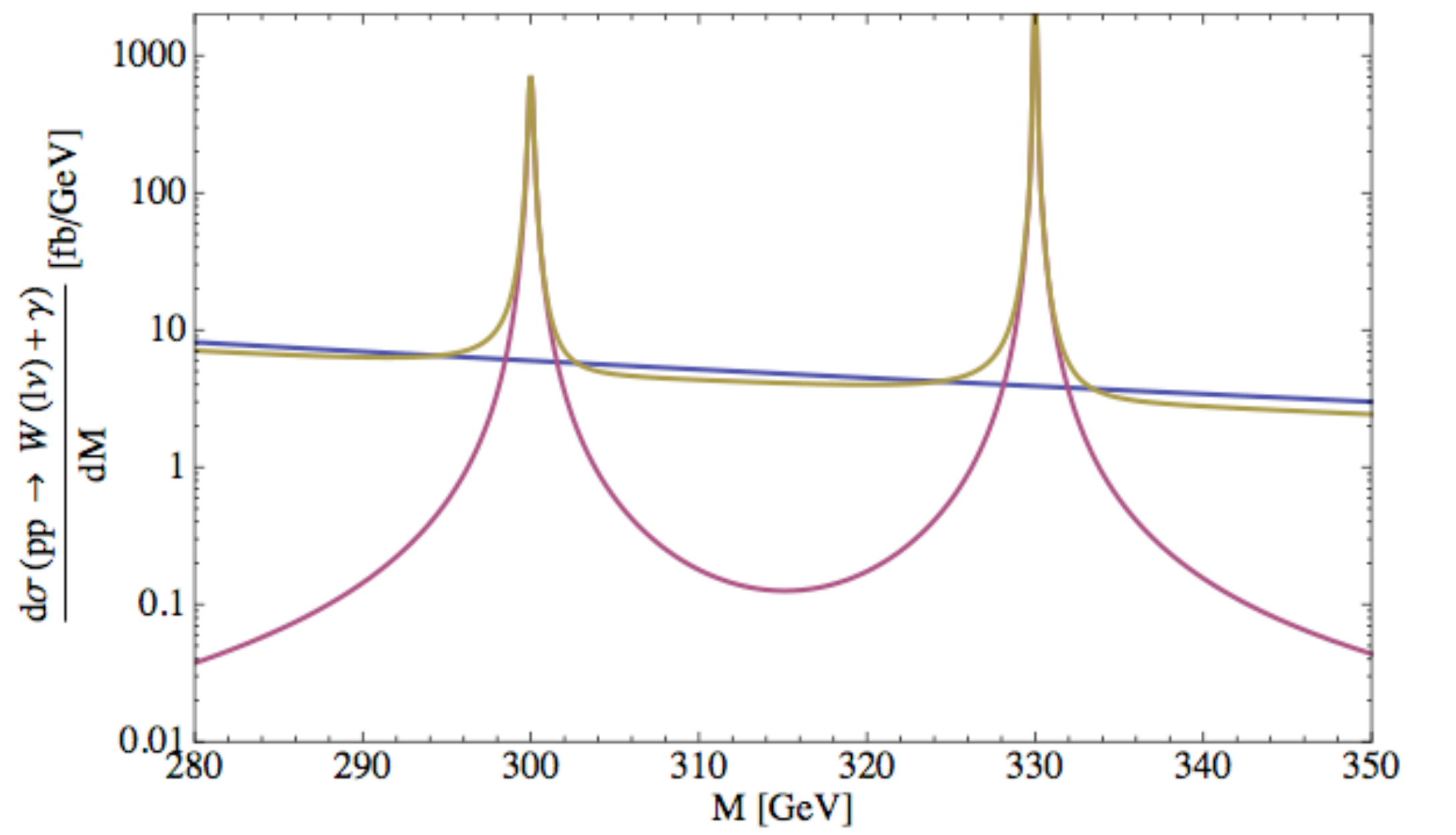}
\includegraphics[width=3.00in, height=2.80in,angle=0]{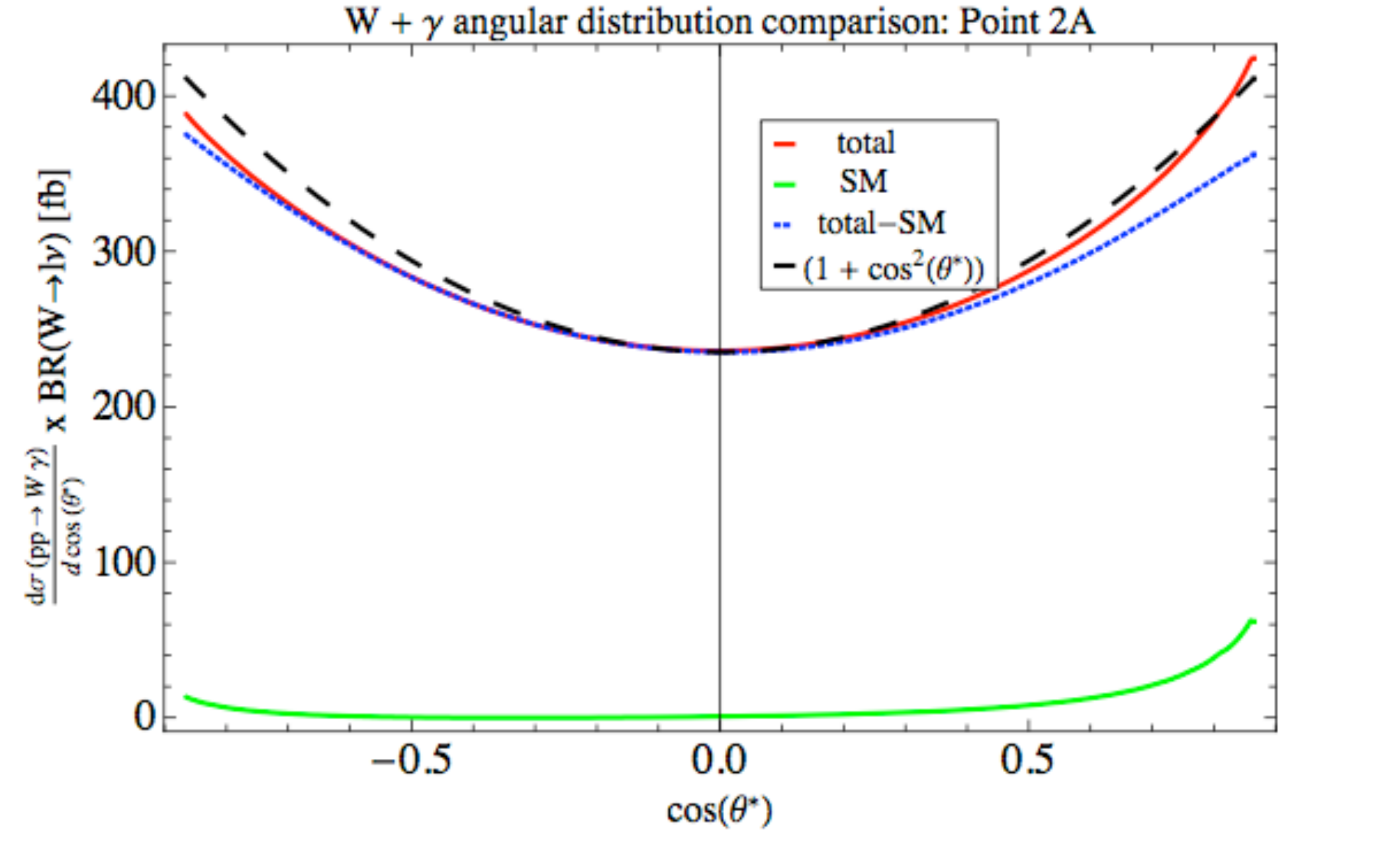}
\caption{The $\gamma W^\pm$ invariant mass (left) and angular (right) distributions
  calculated from $\Leff$ for Case~2a with $\sqrt{s} = 10\,\tev$, $M_{\tro} =
  300\,\gev$, $M_{\ta} = 330\,\gev$ and $M_{\tpi} = 200\,\gev$. The angular
  distribution is for $\tapm \ra \gamma W^\pm$.}
\label{fig:gW_2a}
\end{center}
\end{figure}

\section{Conclusions and outlook}

Low-scale technicolor remains a well-motivated scenario for strong
electroweak symmetry breaking with a walking TC gauge coupling. The
Technicolor Straw-Man framework outlined in Sect.~1 provides the simplest
phenomenology of this scenario by assuming that the lightest technihadrons
--- $\tro$, $\tom$, $\ta$ and $\tpi$ --- and the electroweak gauge bosons can
be treated in isolation. This framework is implemented in {\sc Pythia}. A new
effective Lagrangian approach allows direct quantitative tests of some the
assumptions on which the TCSM is based, in particular, the dominance of
longitudinally-polarized gauge bosons in technivector decay rates and
angular distributions.

In this report, we used {\sc Pythia} and various detector simulations, and
the effective Lagrangian (at the parton level) to study technivector decays
to $W^\pm Z^0$, $\gamma Z^0$, $\gamma W^\pm$ and $e^+e^-$. At the time of the
2009 Les Houches Summer Study, the initial LHC plan was to run at $\sqrt{s} =
10\,\tev$, and so all our studies were carried out for this energy and
luminosities of $\CO(1\,\ifb)$. As the report was being written, the LHC run
plan for 2010-11 changed to running at $7\,\tev$ with the aim of collecting
about $1\,\ifb$ of data. The reach of LHC experiments at $7\,\tev$ for the
resonant processes discussed here may be estimated from our results by using
the parton-parton luminosities and their ratios in Ref.~\cite{Quigg:2009gg}.
Overall, the first run of the LHC should be able to set some useful new
limits on low-scale technicolor. We reiterate what we said two years ago:
With sufficient luminosity, generally in the range of 5--$40\,\ifb$, the LHC
at its design energy of $14\,\tev$ can discover or rule out low-scale
technicolor in the channels discussed here; with more luminosity angular
distributions can be measured to determine whether technicolor is the
underlying dynamics of discovered resonances. Thus, by the time of the next
Les Houches Summer Study, we all hope that we can return to more in-depth
studies of LHC reach at $14\,\tev$. We conclude as we did two years ago: the
main goal of our Les Houches work, as it is for the other ``Beyond the
Standard Model'' scenarios investigated for Les Houches 2009, is to motivate
the ATLAS and CMS collaborations to broaden the scope of their searches for
the origin and dynamics of electroweak symmetry breaking.

\vskip0.15truein
\hskip0.44truein ``Faith'' is a fine invention

\hskip0.5truein    When Gentlemen can see ---

\hskip0.5truein    But  {\em Microscopes}  are prudent

\hskip0.5truein    In an Emergency.\hfil\break


\hskip1.0truein --- Emily Dickinson, 1860

\bigskip

\section*{Acknowledgements}

We thank the organizers and conveners of the Les Houches workshop, ``Physics
at TeV Colliders'', for a most stimulating meeting and for their
encouragement in preparing this work. We benefited from Conor Henderson's
participation in our group at Les Houches. We thank many other participants,
too numerous to name, for spirited discussions.  Lane is indebted to
Laboratoire d'Annecy-le-Vieux de Physique des Particules (LAPP) and
Laboratoire d'Annecy-le-Vieux de Physique Theorique (LAPTH) for generous
hospitality and support. He thanks Louis Helary and Nicolas Berger of LAPP
for many illuminating discussions. This research was supported by the
U.S.~Department of Energy under Grants DE-FG02-91ER40654 (Black and Smith),
DE-FG02-91ER40676 (Bose and Lane), and Fermilab operated by Fermi Research
Alliance, LLC under contract number DE-AC02-07CH11359 by the U.S.~Department
of Energy (Martin).

%% file: Karagoz/LH_warped.tex
\chapter{LHC studies inspired by warped extra dimensions}
\label{chapter:RS}

{\it 
K.~Agashe,
L.~ Basso, 
G.~Brooijmans,
S.P.~Das,
H.~Gray,
M.~Guchait,  
J.~Jackson, \\
M.~Karag\"oz,
S.J.~Lee,
R.~Rosenfeld,
C.~Shepherd-Themistocleous and
M.~Vos}

\begin{abstract}
The framework of a warped extra
dimension with the Standard Model (SM) fields propagating in it is a
very well-motivated extension of the SM since it can address both
the Planck--Weak and flavor hierarchy problems of the SM.
We consider processes at the large hadron collider (LHC) {\em inspired} by 
signals for new particles in
this framework. Our studies include
identification of boosted top quarks and $W/Z$, production of a particle called radion 
with 
Higgs-like properties
and effects of flavor violating $tc$Higgs coupling.
\end{abstract}


\section{Introduction}

The framework of a warped extra dimension \`a la Randall-Sundrum
(RS1) model~\cite{Randall:1999ee}, but 
with all the SM fields propagating
in it~\cite{Davoudiasl:1999tf,Pomarol:1999ad,Chang:1999nh,Grossman:1999ra,Gherghetta:2000qt}
is a very-well motivated extension of the Standard Model (SM):
for a review and further
references, see Ref.~\cite{Davoudiasl:2009cd}.
Such a framework
can address both the Planck--Weak and the
flavor hierarchy problems of the SM, the latter without resulting in 
(at least a severe) flavor problem.
The versions of this framework
with a grand unified gauge symmetry in the bulk 
can naturally lead to precision unification of the three
SM gauge couplings~\cite{Agashe:2005vg} and 
%
%
a candidate for the 
dark matter of the universe 
(the latter from
requiring longevity of the proton)~\cite{Agashe:2004ci,Agashe:2004bm}. 
The
new particles in this framework are Kaluza-Klein (KK) 
excitations of all SM fields with masses at $\sim {\rm TeV}$ scale.
In addition, there is a particle, denoted by the
``radion'', which is  roughly the degree of freedom 
corresponding to the fluctuations of the size of extra dimension, and typically has a mass
at the
weak scale.
In this write-up, we summarize some of the signals at the large hadron collider
(LHC) for these new particles.
Some of these studies can be useful in other contexts as well.

\section{Review of Warped Extra Dimension}
\label{review}

The framework 
consists of a slice of anti-de Sitter
space in five dimensions (AdS$_5$), where (due to the
warped geometry) the effective 4D mass scale is dependent
on position in the extra dimension.
The 4D
graviton, i.e., the zero-mode
of the 5D graviton, is automatically localized
at one end of the extra dimension (called the Planck/UV brane).
If the Higgs sector is localized at the other end (in fact
with SM Higgs originating as 5th
component of a 5D gauge field
($A_5$) it is automatically so~\cite{Contino:2003ve}), then the warped geometry
naturally generates the Planck--Weak hierarchy.
Specifically, TeV $\sim \mP
e^{ - k \pi r_c }$, where $\mP$ is
the reduced 4D Planck scale,
$k$ is the AdS$_5$ curvature scale and $r_c$ is the proper
size of the extra dimension. The crucial 
point is that the required
modest size of the radius (in units of the curvature radius), i.e.,
$k r_c \sim 1 / \pi \log \left( \mP / \hbox{TeV}
\right) \sim 10$ can be 
%
%
stabilized (i.e., the radion given a mass) with only a corresponding
modest tuning in the fundamental or 5D parameters of
the theory~\cite{Goldberger:1999uk,Garriga:2002vf}.
Remarkably, the correspondence between
AdS$_5$  
and 4D conformal field theories (CFT)~\cite{Maldacena:1997re,Gubser:1998bc,Witten:1998qj}
suggests that 
the scenario with warped extra dimension is 
dual to the idea of a composite Higgs in 4D~\cite{Contino:2003ve,ArkaniHamed:2000ds,Rattazzi:2000hs}.

\subsection{SM in warped bulk}

It was realized that with 
SM fermions propagating in the extra dimension, we can also
account for the hierarchy between quark 
and lepton masses and mixing angles (flavor hierarchy)
as follows~\cite{Grossman:1999ra,Gherghetta:2000qt}: 
the basic idea is that the 4D Yukawa coupling are given by the product of the 
5D Yukawa and the overlap of the profiles in the extra dimension of the SM 
fermions (which are the zero-modes of the 5D fermions) with that of the Higgs.
The light SM fermions can be localized 
near the Planck brane, resulting in a 
small overlap with the TeV-brane localized SM Higgs, while
the top quark is localized near the TeV brane with a large 
overlap with the Higgs.
The crucial point is that such vastly different profiles for zero-mode fermions
can be realized with small variations in the 
5D mass parameters of fermions.
Thus we can obtain hierarchical SM Yukawa couplings without any 
large hierarchies in the parameters of the 5D theory, i.e. the
5D Yukawas and the 5D masses.

With SM fermions emerging as zero-modes of 5D fermions,
so must the SM gauge fields. Hence, this scenario can be dubbed ``SM in the (warped) bulk''.
Due to the different profiles
of the SM fermions in the extra dimension, 
flavor changing neutral
currents (FCNC) are generated
by their non-universal couplings to gauge KK 
states.
However, 
these contributions to
the FCNC's are suppressed due to an analog of the Glashow--Iliopoulos--Maiani
(GIM) mechanism of the SM, i.e. RS--GIM, 
%
%
which is ``built-in''~\cite{Gherghetta:2000qt,Huber:2003tu,Agashe:2004cp}.
The point is that {\em all} KK modes
(whether gauge, graviton or fermion) are localized near the
TeV or IR brane (just like the Higgs) so that non-universalities
in their couplings to SM fermions are of
same size as couplings to the Higgs.
In spite of this RS--GIM
suppression, the lower limit on the KK mass scale can be
$5-10$ TeV~\cite{Csaki:2008zd,Blanke:2008zb,Bauer:2009cf}~\footnote{See Refs~\cite{Agashe:2009tu} and~\cite{Gedalia:2009ws} for ``latest''
constraints from
lepton and quark flavor violation, respectively, i.e.,including
variations of the minimal framework.} although these constraints can
be ameliorated by addition of 5D flavor symmetries~\cite{Fitzpatrick:2007sa,Chen:2008qg,Perez:2008ee,Csaki:2008qq,Santiago:2008vq,Csaki:2008eh,Csaki:2009wc,Chen:2009hr}.
Finally, various custodial symmetries~\cite{Agashe:2003zs, Agashe:2006at}
can be incorporated such that the constraints from the various 
(flavor-preserving) electroweak precision tests (EWPT)
can be satisfied for a few TeV KK scale~\cite{Carena:2006bn,Carena:2007ua}.
The 
bottom line is that a 
few TeV mass scale for the KK gauge bosons can be consistent with both
electroweak and flavor precision tests.

\subsection{Couplings of KK's}

Clearly, the light fermions have a small
couplings to all KK's (including graviton)
based simply on the overlaps
of the corresponding profiles, 
while the top quark and Higgs have a large coupling to the KK's.
To repeat, light SM fermions are localized near the Planck brane and photon, gluon and transverse $W/Z$ have flat profiles, whereas all KK's, Higgs (including longitudinal $W/Z$) and top quark are localized near the TeV brane. 
Schematically,
neglecting effects related to electroweak symmetry breaking (EWSB), we find the following ratio of
RS1 to SM
gauge couplings:
\begin{eqnarray}
{g_{\rm RS}^{q\bar q,l\bar l\, A^{ (1) }}\over g_{\rm SM}}
&\simeq&
- \zeta^{-1}\approx - {1\over5} \nonumber \\
{g_{\rm RS}^{Q^3\bar Q^3 A^{ (1) }}\over g_{\rm
    SM}},
{g_{\rm RS}^{t_R\bar t_R A^{ (1) }}\over g_{\rm
    SM}} 
& \simeq & 
1 \; \hbox{to} \; \zeta \; ( \approx 5 ) \nonumber \\
{g_{\rm RS}^{ HH A^{(1)}}\over g_{\rm
    SM}}  
& \simeq & 
\zeta \approx 5 \; \; \; \left( H = h, W_L, Z_L \right)
\nonumber \\
{g_{\rm RS}^{ A^{ (0) }A^{ (0) } A^{ (1) }}\over g_{\rm
    SM}}  
& \sim & 0
\label{RScouplings}
\end{eqnarray}
Here $q=u,d,s,c,b_R$, $l =$ all leptons, $Q^3= (t, b)_L$, 
and $A^{ (0) }$ ($A^{ (1) }$) correspond
to zero (first KK) states of the gauge fields. Also, 
$g_{\rm RS}^{xyz}, g_{\rm SM}$ stands for the RS1 and the three SM (i.e.,
4D) gauge couplings respectively.
Note that 
$H$ includes both the physical Higgs ($h$) and 
{\em un}physical Higgs, i.e., {\em longitudinal}
$W/Z$ by the equivalence theorem
(the derivative involved in this coupling is 
similar for RS1 and SM cases and hence is not shown for
simplicity). Finally, the parameter $\xi$ is
related to the Planck--Weak hierarchy: $\zeta \equiv \sqrt{ k \pi r_c }$.

We also present 
the couplings of the KK
graviton to the SM particles. 
These couplings involve derivatives
(for the case of {\em all} SM particles),
but (apart from a factor from the overlap
of the profiles) it turns out that 
this energy-momentum dependence is
compensated (or made dimensionless) by the $\mP e^{ - k \pi r_c }\sim$ 
TeV scale, instead of 
the $\mP$-suppressed coupling to the SM graviton. Again, schematically:
\begin{eqnarray}
g_{ \rm RS }^{ q\bar q,l\bar l\, G^{ (1) } } & \sim & 
\frac{E}{ \mP e^{ - k \pi r_c } } \times 4D \; \hbox{Yukawa}
\nonumber \\
g_{ \rm RS }^{ A^{ (0) }A^{ (0) } G^{ (1) } } &
\sim & \frac{1}{ k \pi r_c }  \frac{E^2}{ \mP e^{ - k \pi r_c } }
\nonumber \\
g_{ \rm RS }^{ Q^3\bar Q^3 A^{ (1) } }, g_{ \rm RS }^{ t_R\bar t_R G^{ (1) } }
& \sim & \left( \frac{1}{ k \pi r_c } 
\; \hbox{to} \; 1 \right) 
\frac{E}{ \mP e^{ - k \pi r_c } }\nonumber \\
g_{ \rm RS }^{ H H G^{ (1) } } & \sim  & 
\frac{E^2}{ \mP e^{ - k \pi r_c } }
\end{eqnarray}
Here, $G^{ (1) }$ is the KK graviton
and the 
tensor
structure of the couplings is not shown
for simplicity.

\subsection{Couplings of radion~\protect\cite{Giudice:2000av,Rizzo:2002pq,Toharia:2008tm,Csaki:2007ns} } 

The unperturbed metric is written as:
\begin{equation}
ds^2 = \left( \frac{R}{z} \right)^2 \left( \eta_{\mu\nu} dx^\mu dx^\nu - dz^2 \right),
\end{equation}
where $z$ refers to the coordinate in the 5th dimension restricted to $R<z<R'$, and $R$ is
the AdS curvature. The radion is related to the scalar perturbation of the metric, which at leading order is
given by:
\begin{equation}
\delta g_{M N} = -2 F \left( \frac{R}{z} \right)^2 \left( \begin{array}{cc}
\eta_{\mu\nu} & 0  \\
0 & 2   \end{array} \right)
\end{equation}
where $F(x,z)$ is the 5D radion field.

The linear radion couplings are determined by the modification
of the action due to the linear perturbation of the metric, which by the definition of the 
energy-momentum tensor is given by:
\begin{equation}
\delta S = -\frac{1}{2} \int d^5 x \sqrt{g} T^{M N} \delta g_{M N} = 
 \int d^5 x \sqrt{g} F ( Tr T^{M N} - g_{55} T^{55})
\end{equation}

The canonically normalized scalar radion field in 4D is related
to $F(x,z)$ by:
\begin{equation}
r(x) = \Lambda_r \left( \frac{R'}{z} \right)^2 F(x,z)
\end{equation}
where $\Lambda_r = \sqrt{6}/R'$.\footnote{$\Lambda_r\approx$KK scale, which can be varied by $\cal{O}$(1) number. But canonical value is given by the above equation.}
For fields that are strongly localized in the infrared brane, such as the Higgs boson and the top quark,
the coupling to the radion is given by
the usual term 
\begin{equation}
{\cal L} = \frac{r(x)}{\Lambda_r} T^{\mu}_{\mu}
\end{equation}

For the top quark one has
\begin{equation}
T_{\mu \nu}^{(t)} = i \bar{t} \gamma_\mu \partial_\nu t - 
\eta_{\mu \nu} \bar{t} \left( i\gamma_\alpha \partial^\alpha -m \right) t  
\end{equation}
which implies
\begin{equation}
{\cal L}_{rtt} =  \frac{r}{\Lambda_r} m_t \bar{t} t.
\end{equation}

However, for the Higgs boson the situation is complicated because of two factors:
spontaneous symmetry breaking and the fact that the
energy-momentum tensor of a scalar field must be modified in order for its trace to
vanish in the zero-mass limit, as it is required by conformal invariance~\cite{Callan:1970ze}.

For a Higgs lagrangian (after symmetry breaking)
\begin{equation}
{\cal L}_h = \frac{1}{2} (\partial_\mu h)^2 - \lambda \left( \frac{(h+v)^2}{2} - \frac{v^2}{2} \right)^2
\end{equation}
the modified energy-momentum tensor $\Theta_{\mu \nu}$ reads:
\begin{equation}
\Theta_{\mu \nu} = \partial_\mu h \partial_\nu h - \eta_{\mu \nu} {\cal L}_h + 
\xi \left(  \eta_{\mu \nu} \partial_\lambda \partial^\lambda - \partial_\mu \partial_\nu \right) 
\left( \frac{(h+v)^2}{2} \right)
\end{equation}
which leads to
\begin{equation}
\Theta^{\mu}_\mu = -(1-6\xi) (\partial_\mu h)^2 +  (1-6\xi) (\lambda h^4 + 4 \lambda v h^3) + (4-30 \xi) \lambda v^2 h^2
- 12 \xi \lambda v^3 h.
\end{equation}
Therefore, for $\xi = 1/6$, one gets
\begin{equation}
\Theta^{\mu}_\mu = - \lambda v^2 h^2 - \frac{1}{2} \lambda v^3 h
\end{equation}
where the first term of the trace of the modified energy-momentum tensor is proportional to the
Higgs mass whereas the second term will induce a mixing between the radion and the Higgs boson.

Radion phenomenology is very sensitive to the values of $\xi$. The $\xi$ term for a general scalar field $\phi$
can be written as a coupling to the Ricci scalar R as
\begin{equation}
{\cal L}_\xi = \xi R \phi^2
\end{equation}
and it breaks a shift symmetry in the scalar field. In models where the Higgs is a Goldstone boson,
one would expect the residual shift symmetry to forbid such a term, which corresponds to setting $\xi = 0$.
Even if the Higgs is an approximate Goldstone boson, $\xi$ should be small.
Since in this note we will be interested in the case where the radion mass is at least twice the Higgs boson mass,
we will neglect the possibility of Higgs--radion mixing. In this case it follows that
\begin{equation}
{\cal L}_{rhh} = \frac{r}{\Lambda_r} \left((\partial_\mu h)^2 - 2 m_h^2 h^2 \right)
\end{equation}
where the Higgs mass is $m_h^2 = 2 \lambda v^2$.

The leading contribution in the radion interaction with massive gauge bosons $W^\pm$ and $Z$ is given by
\begin{equation}
{\cal L}_{rVV} = -\frac{r}{\Lambda_r} \left( 2 M_W^2 W^2 + M_Z^2 Z^2  \right)
\end{equation}
but there are model dependent corrections that we include in our analyses.

Usually the coupling of the radion to massless gauge bosons vanishes at tree level. At 1-loop it arises due to two
contributions: the trace anomaly, which is related to the beta function, and the top quark triangle diagram.
However, in the warped scenario, there are two main differences: a tree level bulk contribution from radion and gauge
bosons wave functions and a modification in the beta function term to take into account that only particles in the 
infrared brane contribute to the running. The final result for this coupling is:
\begin{eqnarray}
{\cal L}_{rAA} &=& \frac{r}{4 \Lambda_r \ln (R'/R)} \left( 1- 4 \pi \alpha \left(\tau_{UV}^{(0)}+\tau_{IR}^{(0)} \right) + \right. \\ \nonumber
&&  \left. \frac{\alpha}{2 \pi} \left( -\frac{11}{3} - F_1(\tau_w) - \frac{4}{3} F_{1/2}(\tau_t) \right) \ln (R'/R) \right) 
F_{\mu \nu}  F^{\mu \nu}
\end{eqnarray}
for photons and 
\begin{eqnarray}
{\cal L}_{rgg} &=& \frac{r}{4 \Lambda_r \ln (R'/R)} \left( 1- 4 \pi \alpha_s \left(\tau_{UV}^{(0)}+\tau_{IR}^{(0)} \right) + \right. \\ \nonumber
&&  \left. \frac{\alpha_s}{2 \pi} \left(7 - \frac{1}{2} F_{1/2}(\tau_t) \right) \ln (R'/R) \right) 
G_{\mu \nu}^a  G^{\mu \nu}_a
\end{eqnarray}
for gluons where
$\tau_x = 4 m_x^2/m_r^2$ and the functions $F_{1,1/2}(\tau)$ vanishes when $\tau < 1$.
The parameters
$\tau_{UV}^{(0)}$ and $\tau_{UV}^{(0)}$ are related to the Planck and TeV-brane induced kinetic
terms.

\subsection{Masses}

As indicated above, masses below about $2$ TeV for gauge KK
particles are strongly disfavored by precision tests, whereas masses
for other KK particles are expected
(in the general framework) 
to be of similar size to gauge KK mass and hence are (in turn) also constrained to be above 
$2$ TeV. However, {\em direct} 
constraints on masses of other (than gauge) KK particles can be weaker.
Radion mass can vary from $\sim 100$ GeV to $\sim 2$ TeV.
In 
{\em minimal} models, KK graviton is actually about $1.5$ heavier than
gauge KK modes, i.e., at least $3$ TeV.

As far as KK fermions are concerned, in minimal models,
they have typically masses same as (or slightly heavier than) gauge KK and hence are constrained to be heavier than $2$ TeV (in turn, based on masses of gauge KK required to satisfy precision tests). 
However, the masses of 
the KK excitations of top/bottom (and their other gauge-group partners) in some 
non-minimal (but well-motivated) models
(where the 5D gauge symmetry is extended beyond that in the SM)
can be (much) smaller than gauge KK modes, 
possibly $\sim 500$ GeV.

\section{KK signals at the LHC}

Based on these KK couplings 
and masses, 
we are faced with the following challenges 
in obtaining signals at the LHC
from direct production of the KK modes, namely,
\begin{itemize}
\item[(i)]
Cross-section for production of these
states is suppressed to begin with
due to a small coupling to
the protons' constituents, and due to the large mass of the new particles; 
\item[(ii)]
Decays to ``golden'' channels (leptons, photons)
are suppressed. Instead, the decays are 
dominated by top quark and Higgs
(including longitudinal $W/Z$); 

\item[(iii)]
These resonances tend to be quite
broad due to the enhanced couplings to top quark/Higgs.

\item[(iv)]
The SM particles, namely, top quarks/Higgs/$W/Z$ gauge bosons, produced in the decays of the heavy
KK particles are highly boosted, 
resulting in a high degree of collimation
of the SM particles' decay products. 
Hence,
conventional methods for identifying top quark/Higgs/$W/Z$ might no longer work 
for such a situation.

\end{itemize}

However, such challenges also present research 
opportunities -- for example, several techniques to identify highly boosted top quark/Higgs/$W/Z$
have been developed~\cite{Thaler:2008ju,Kaplan:2008ie,Almeida:2008yp,Almeida:2008tp,
Skiba:2007fw,Holdom:2007nw,Holdom:2007ap,Seymour:1993mx,Butterworth:2007ke,
Butterworth:2008iy,Butterworth:2008sd,Ellis:2009su,Ellis:2009me,Plehn:2009rk,Benchekroun:2001je,Butterworth:2002tt}.

\section{Direct KK effects}

Next, we summarize decay channels and production cross-sections
for the KK particles: for more details, see corresponding references given in each title
and for an overview, see Ref.~\cite{Davoudiasl:2007wf}.
Based on the above discussion, note that the polarization of $W/Z$'s in these decay channels is dominantly {\em longitudinal}.

\subsection{KK gluon~\protect\cite{Agashe:2006hk,Lillie:2007ve,Djouadi:2007eg,Guchait:2007jd,Baur:2007ck,Baur:2008uv,Lillie:2007yh}}


Kaluza Klein partners of the gluon offer a particulary interesting phenomenology at the LHC. The cross-section of such coloured states can exceed that of typical electro-weak ($Z'$) resonances by one or even several orders of magnitude. However, these states cannot be observed through the {\em golden} di-lepton resonance searches and discovery is only possible in the more challenging hadronic final states.

 In this contribution, the focus is on the basic RS setup of Ref.~\cite{Lillie:2007yh}\footnote{In Ref.~\cite{Baur:2008uv} many different parameter sets for the KK gluon, each with a quite different phenomenology, are discussed.}. In this model the KK gluon displays strongly enhanced couplings to (right-handed) top quarks. The most promising signature of the KK gluon is resonant $ t \bar{t} $ production on top of the Standard Model $ t \bar{t} $ continuum. The LHC (14 TeV) production rate of the $ pp \rightarrow g_{KK} \rightarrow t \bar{t} $ process ranges from nearly 30 pb for a 1 TeV resonance to approximately 3 pb for a 3 TeV resonance.

\begin{figure}[h]
     \begin{center}
     \vspace{.2cm}
     \includegraphics[width=10cm]{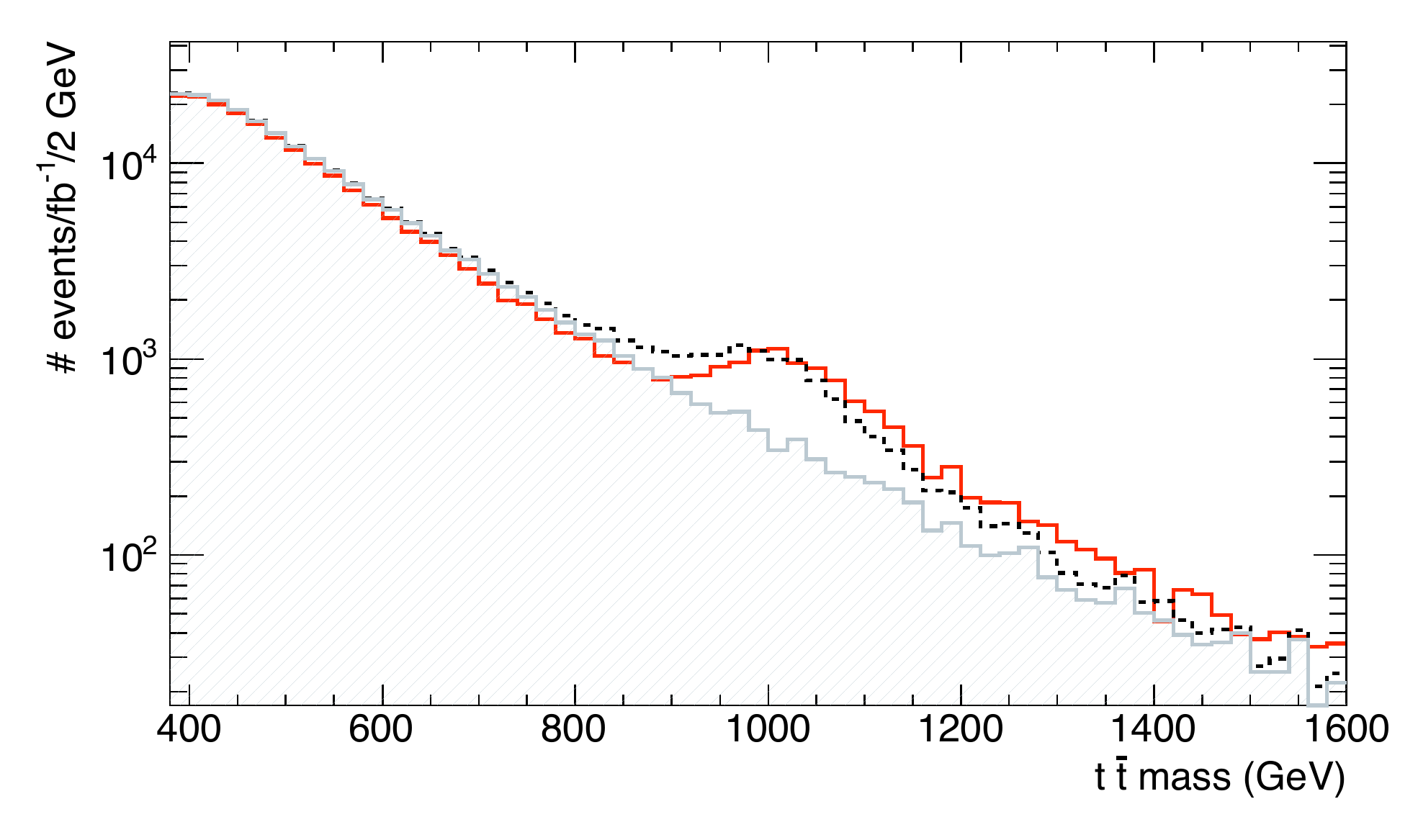}
     \end{center}
     \caption{The $ t \bar{t} $ invariant mass distribution: Standard Model continuum (shaded histogram), the sum of SM and resonant production (dashed line) and the full interference of SM and resonant production (continuous line).}
     \label{Fig:kkgluon_mass_distribution}
 \end{figure}

The KK gluon of the basic RS setup has a number of features that do not satisfy the usual assumptions of model-independent narrow resonance searches. A KK gluon search must take into acount the following:
\begin{itemize}
\item The width of the KK gluon, 17 \% of the mass in the basic RS setup, is not negligible compared to the experimental mass resolution. The model-independent limit for {\em narrow} resonances derived in the large majority of published $ t \bar{t} $ resonance searches therefore does not apply. An experimental strategy must be developed to deal with the width explicitly.
\item The interplay between the width of the resonance and the parton luminosity function leads to a significant skew of the mass distribution of the $ pp \rightarrow g_{KK} \rightarrow t \bar{t} $ process. Especially for large KK gluon mass a long tail towards lower mass develops. It is therefore non-trivial to relate an excess of events in a mass window to a total cross-section.
\item The interference between the resonant production and Standard Model $ t \bar{t} $ production can be significant. 
Figure~\ref{Fig:kkgluon_mass_distribution} shows the difference between the full interference (continuous line) and the sum of signal and background processes (dashed line) for a generic, spin-1 colour octet with a mass of 1 TeV and the couplings of the KK gluon implemented in MadGraph~\cite{Maltoni:2002qb}. The interference leads to a pronounced reduction of the production rate for $ M_{t \bar{t}} \sim M_{g_{KK}}/2$.
\end{itemize}

Therefore, while the KK gluon could be rather abundantly produced at the LHC, a complete experimental strategy for this type of broad coloured resonances is not yet fully developed (see, however, Contribution~\ref{chapter:KKgluons} in these proceedings).


\subsection{KK graviton~\protect\cite{Fitzpatrick:2007qr,Agashe:2007zd,Antipin:2007pi,Antipin:2008hj}}

The dominant decay channels are
into $t \bar{t}$, $WW$, $ZZ$, $hh$.
For a $2$ TeV  KK graviton, each of these cross-sections
can be $\sim \mathcal{O}(10~\textrm{fb})$ with a total decay width of 
$\sim \mathcal{O}(100~\textrm{GeV})$.

\subsection{$W^{ \prime }$~\protect\cite{Agashe:2008jb}}

It turns out that in addition to KK $W_L^+$, these models also have a KK $W_R^+$
(with no corresponding zero-mode), due to the custodial (i.e., extended 5D gauge) symmetry.
These two KK states mix after EWSB and the mass eigenstates
are generically denoted by $W^{ \prime }$.
The dominant decay modes for
$W^{ \prime }$ are into $WZ$ and $Wh$.
For each $W^\prime$ and with a mass of 2 TeV, the cross-section is $\sim \mathcal{O}(10~\mathrm{fb})$ with a total decay width of 
$\sim  \mathcal{O}(100~\mathrm{GeV})$.
In some models, 
$W^{ \prime }$ decays to $t \bar{b}$ --- giving boosted top and bottom ---
can also have similar cross-section. Interestingly, the process
KK gluon $\rightarrow t \bar{t}$ --- with KK
gluon mass being similar to $W^{ \prime }$ --- can be a significant background to this channel since a highly boosted top quark can fake a bottom quark: 
techniques similar to the ones used to identify highly boosted tops can now be applied to {\em veto} this possibility!

\subsection{$Z^{ \prime }$~\protect\cite{Agashe:2007ki}}

There are actually three neutral KK states:
KK $Z$, KK photon and a KK mode of an extra $U(1)$ (again, with no 
corresponding zero-mode).
These states mix after EWSB and  the mass eigenstates are generically denoted by
$Z^{ \prime }$.
The dominant decay modes are to $t \bar{t}$, $WW$ and $Zh$, each with a 
cross-section of 
$\sim \mathcal{O}(10~\mathrm{fb})$ for a
$2$ TeV $Z^{ \prime }$ with a total decay width of $\sim 100$ GeV.
However, the $t \bar{t}$ channel can be swamped by KK gluon
$\rightarrow t \bar{t}$ if the $Z^{ \prime }$ and KK gluon have similar mass.

\subsection{Heavier KK fermions~\protect\cite{Davoudiasl:2007wf}}

The KK fermions in the minimal model being $2$ TeV or heavier,
even single production of these particles can be very small (pair production is even smaller).

\subsection{Light KK fermions~\protect\cite{Dennis:2007tv,Contino:2008hi,Mrazek:2009yu}}

As mentioned above, in non-minimal models, KK partners of top/bottom
can be light so that their
production (both pair and single, the latter 
perhaps in association with SM particles) can be significant. 
As these particles are ``top-like" with respect to their production at the LHC, the yields can be sizeable. For example, the pair production cross-section of a KK
bottom with mass of $500$ GeV is $\sim 1$ pb at $\sqrt{s} = 10$ TeV.
These particles decay into $t/b + W/Z/h$, where the  $W/Z$ can be boosted at the LHC (even for fermionic KK
partners with masses as low as $\sim 500$ GeV). 
Some of these light KK fermions can have ``exotic" electric charges 
-- for example, $4/3$ and $5/3$. This makes them appealing with respect to a generic $b\prime/t\prime$ from, for example, a minimal extension to 
SM generations~\cite{delAguila:2008iz}. 
Recently, Tevatron experiments have placed limits on such KK fermions~\cite{Aaltonen:2009nr}. 
Various search strategies for KK fermions are being developed at the LHC~\cite{Dennis:2007tv,Contino:2008hi,Mrazek:2009yu} (see also Contribution~\ref{chapter:BrooijmansEtAl} in these proceedings).

In addition, the other heavier (spin-1 or 2) KK modes can decay
into these light KK fermions, resulting in perhaps more distinctive
final states for the heavy KK's than the pairs of $W/Z$ or top quarks
that have been studied so far -- for 
such a study for KK gluon, see Ref.~\cite{Carena:2007tn}.

\subsection{Radion~\protect\cite{Giudice:2000av,Rizzo:2002pq,Toharia:2008tm,Csaki:2007ns}}

Radion production at the LHC could be substantial due to the fact that the branching fraction of the radion to two gluons could be enhanced by as much as a factor of 10 (for $\Lambda_r = 1$ TeV) in comparison with the Higgs branching fraction to gluons. The enhancement is due to the fact that the radion couples to massless gauge bosons through the conformal anomaly, which is rather large for QCD.
As a bona fide dilaton, the radion couples to the energy-momentum tensor of the theory.
Hence, its couplings are proportional to masses of particles, in much the same way as the usual Higgs boson.
As mentioned above, radion mass is a free parameters of the theory, varying from $\sim 100$ GeV to $\sim 2$ TeV,
which means that dominant decay channels are determined by radion mass.
For radion mass lighter than $2M_{W}$, $r\rightarrow \gamma\gamma$ is a promising channel,
which can be also dramatically enhanced in the presence of Higgs--radion mixing. 
For larger radion mass, $WW$, $hh$, $ZZ$, $t \bar{t}$ channels are the dominant channels,
which can pose a challenge for detecting highly boosted signals.

\begin{figure}
\begin{center}
	\includegraphics[width=90mm]{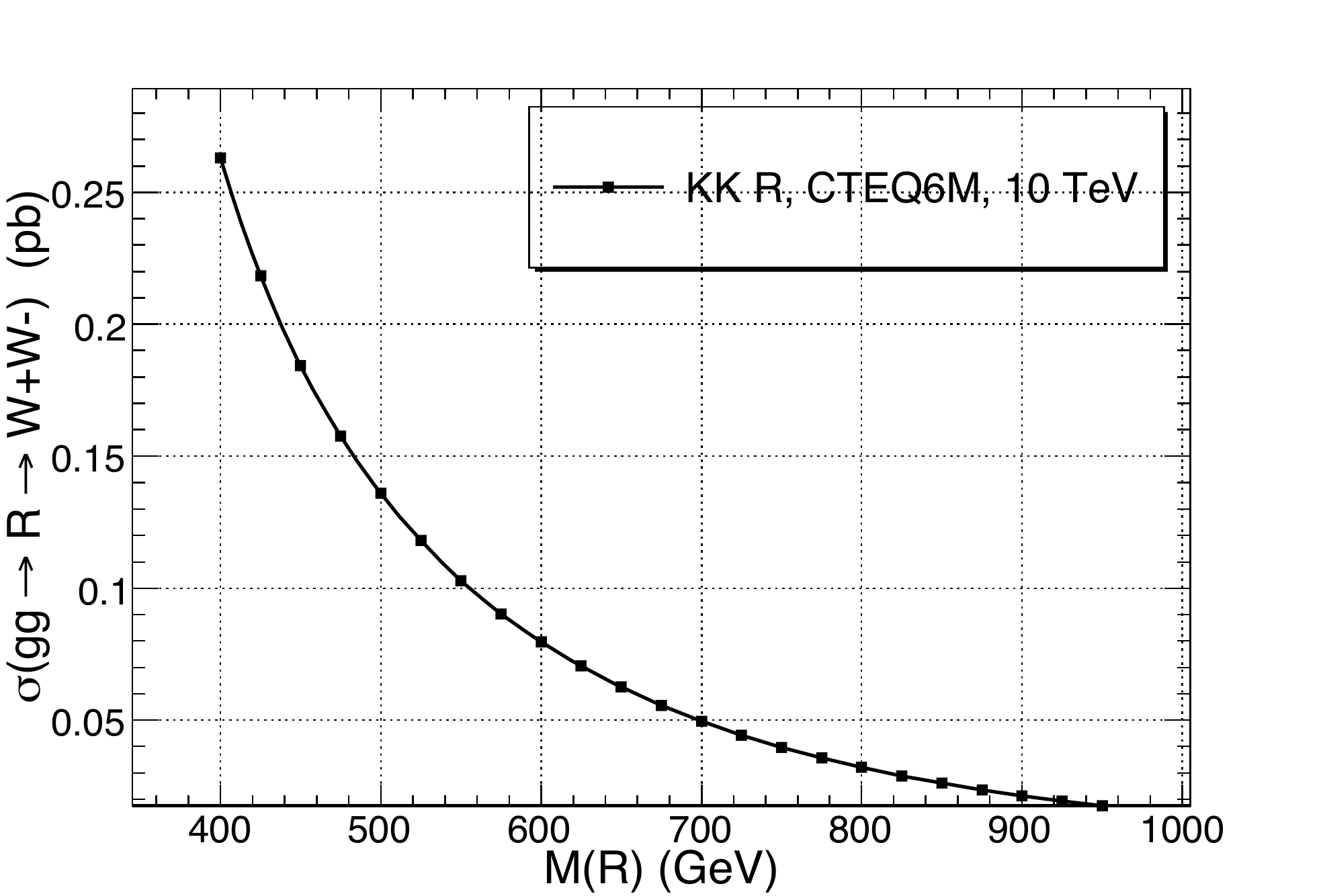}
	\end{center}
	\caption{$\sigma(gg\rightarrow r\rightarrow W W)$ cross section at the LHC for 10 TeV center of mass energy.}
	\label{fig:radionWWDecay}
\end{figure}

Above a radion mass of 400 GeV or so, where decay products of radions can start to be boosted, the branching fractions of radion into SM particles are reasonably flat. 
Depending on model parameters, the $WW$ channel can be the most dominant channel with a branching fraction of about 50\%. Figure~\ref{fig:radionWWDecay} shows $WW$ 
cross section of radions as a function of radion mass at 10 TeV LHC center of mass energy using the CalcHEP implementation of 
Ref.~\cite{Csaki:2007ns}. It can be seen that even for a high value of $\Lambda_r$ at 3 TeV, the cross sections can be as high as a fraction of a picobarn. 
Reach prospects improve when a value of 2 TeV for $\Lambda_r$ is chosen, as allowed by the EWPT results. 
The largest yield in $WW$ channel would come from fully hadronic decays of the $W$ boson, however, this channel may suffer largely from QCD dijet production at the LHC. 
Looking at the semi-leptonic channel, as was done for $WW$ scattering searches at ATLAS~\cite{Benchekroun:2001je,Butterworth:2002tt,Aad:2009wy}, may provide a way to 
observe radion production in $WW$ channel. For example, for a radion mass of 600 GeV, the $\sigma(r\rightarrow W_{had} W_{lep})$ is $\sim \mathcal{O}(10~\mathrm{fb})$, for $\Lambda_r = $ 3~TeV. 
This value would be comparable to that of a direct SM Higgs production at the LHC. 

\begin{figure}
\begin{center}
	\includegraphics[width=80mm, angle=0]{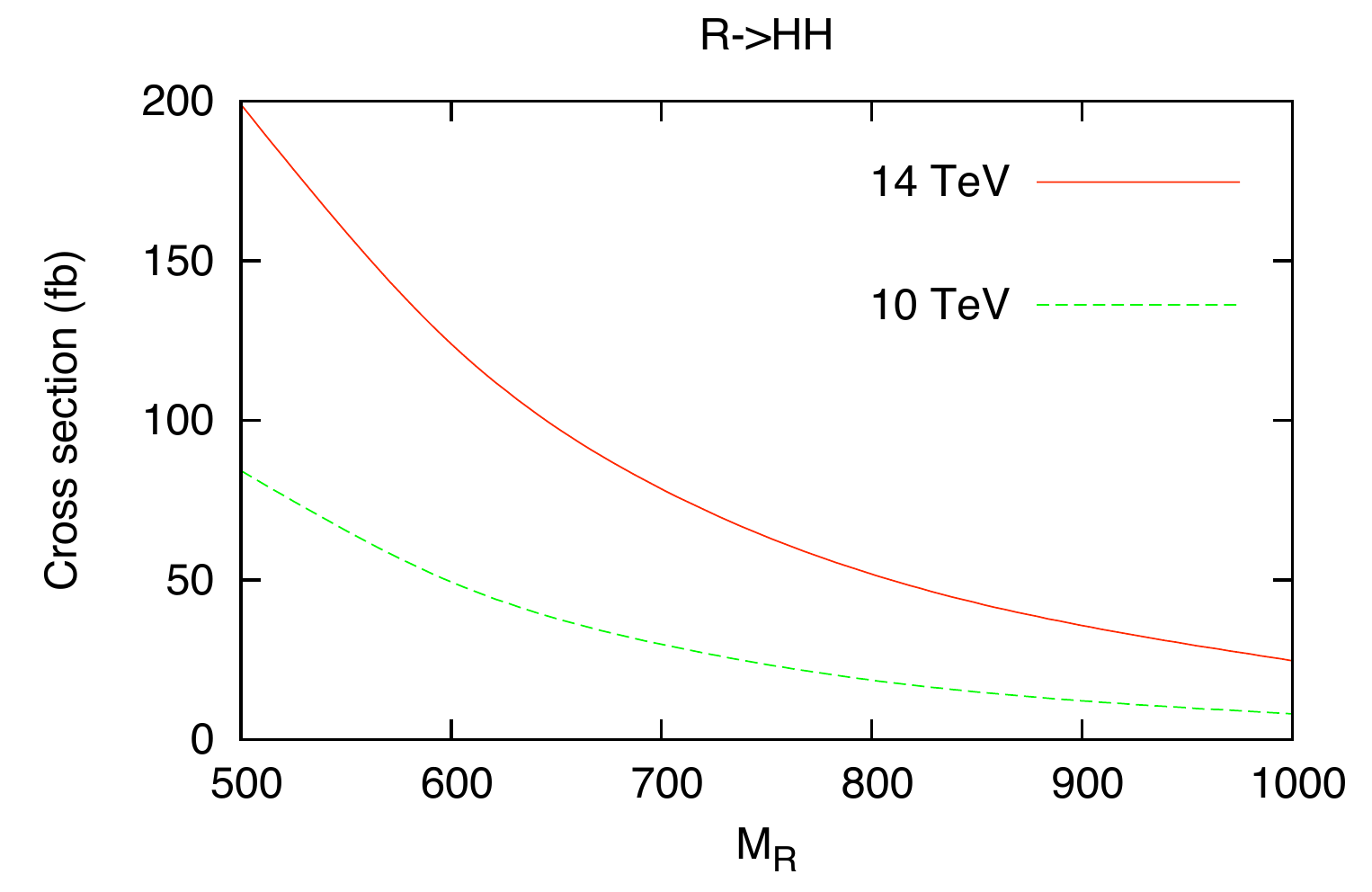}
	\end{center}
	\caption{$\sigma(pp\rightarrow r\rightarrow H H)$ cross section at the LHC for 14 and 10 TeV center of mass energies.}
	\label{fig:radionHHDecay}
\end{figure}

Figure~\ref{fig:radionHHDecay} shows the production cross section of radion in the $HH$ channel for the same settings as before in the $WW$ channel.

\section{Identification of boosted objects from KK direct production}

Motivated by above discussion of signals for KK particles in warped extra dimensional framework, we study in this section the identification of boosted SM particles which decay.

\subsection{Identification of boosted $W$ and $Z$ decay products}
The identification of $W$ and $Z$ decays products from the models discussed will be experimentally challenging due to the boosted nature of the decaying system. For available LHC energies, the angular separation in the lab frame of the decay products will be of the order $0.1\,\mathrm{rad}$.

For decays to $e$, $\mu$, $\nu$, this hampers traditional reconstruction techniques which rely on isolated leptons in order to reject jet backgrounds and to clean fake ${E\!\!\!\!/_T}$. For hadronic decays, the two decaying quarks will merge into one collimated jet. It is possible that by exploring jet substructure, these will be identifiable with backgrounds under control. Studies in that direction have already been performed and discussed elsewhere (see, e.g., Ref.~\cite{Aad:2009wy}), thus here we only concentrate on the leptonic decays.

\subsubsection{Leptonic $Z$}
The main challenge in identifying boosted $Z\to e^+e^-$ will be the merging of electromagnetic clusters. The granularity of typical LHC calorimetry is such that this will be an algorithmic rather than a physical issue. In particular, algorithms designed to recover energy lost due to Brehmstrahlung radiation may be detrimental to boosted $Z$ identification.

\begin{figure}[htbp]
\begin{center}
     \includegraphics[width=80mm]{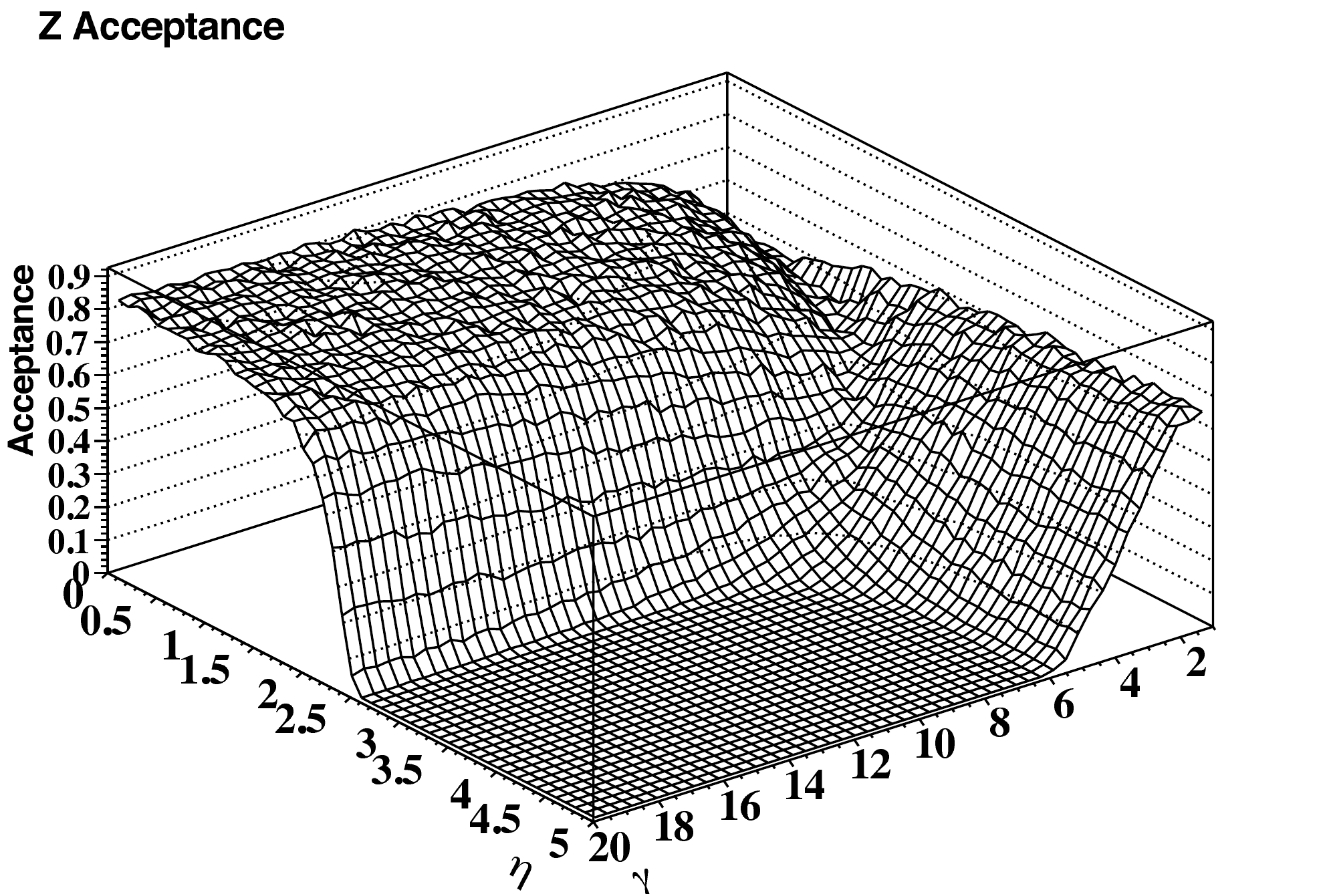}
     \end{center}
     \caption{Toy Monte Carlo simulation of boosted $Z\to e^+e^- identification$}
     \label{fig:boostedZee}
\end{figure}

The results of a toy Monte Carlo simulation of boosted $Z\to e^+e^-$, assuming a $90\%$ efficiency to identify a single electron, are shown in Fig.~\ref{fig:boostedZee}. Within typical LHC detector acceptance (electron acceptance is taken to be $100\%$ in the region $|\eta | < 2.5$), identification is possible for centrally boosted $Z$'s with high relativistic $\gamma$. At high energy, the energy resolution is dominated by the constant term, and as such resolutions of the order $1-5\%$ can be expected. Existing background rejection methods, such as the jet fake rate, developed for non-boosted decays of heavy neutral particles to di-lepton pairs will be equally applicable to the boosted reconstruction scenario.

In the $\mu^+\mu^-$ channel, angular separation is not an issue, however the momentum measurement will be affected by the low-curvature tracks. CMS and ATLAS expect a momentum resolution of order $10\%$ for TeV muons~\cite{:1999fq, Bayatian:2006zz}.

\subsubsection{Leptonic $W$}
Where a $W$ decays leptonically, $W \to e\nu_e$, $W \to \mu\nu_\mu$ is also of interest to the models discussed. Such a decay leads to significant ${E\!\!\!\!/_T}$, correlated with the electron (muon) direction. This allows the $W$ mass to be reconstructed in the collinear approximation, where the neutrino three-vector is defined as
\begin{equation}
\vec{p}_{\nu_e} = ({\not\!\!{E_x}}, {\not\!\!{E_y}}, \frac{\sqrt{{\not\!\!{E_x^2}} + {\not\!\!{E_y^2}}}}{\sqrt{p_{x,e}^2 + p_{y,e}^2}}p_{z,e}),
\end{equation}
where $\vec{p}_e$ is the electron momentum. The neutrino four-vector is defined as $p_\mu^{\nu_e} = (\vec{p}_{\nu_e}, |\vec{p}_{\nu_e}|)$.

\begin{figure}[htbp]
    \centering
    \begin{tabular}{cc}
    \includegraphics[angle=90,width=80mm]{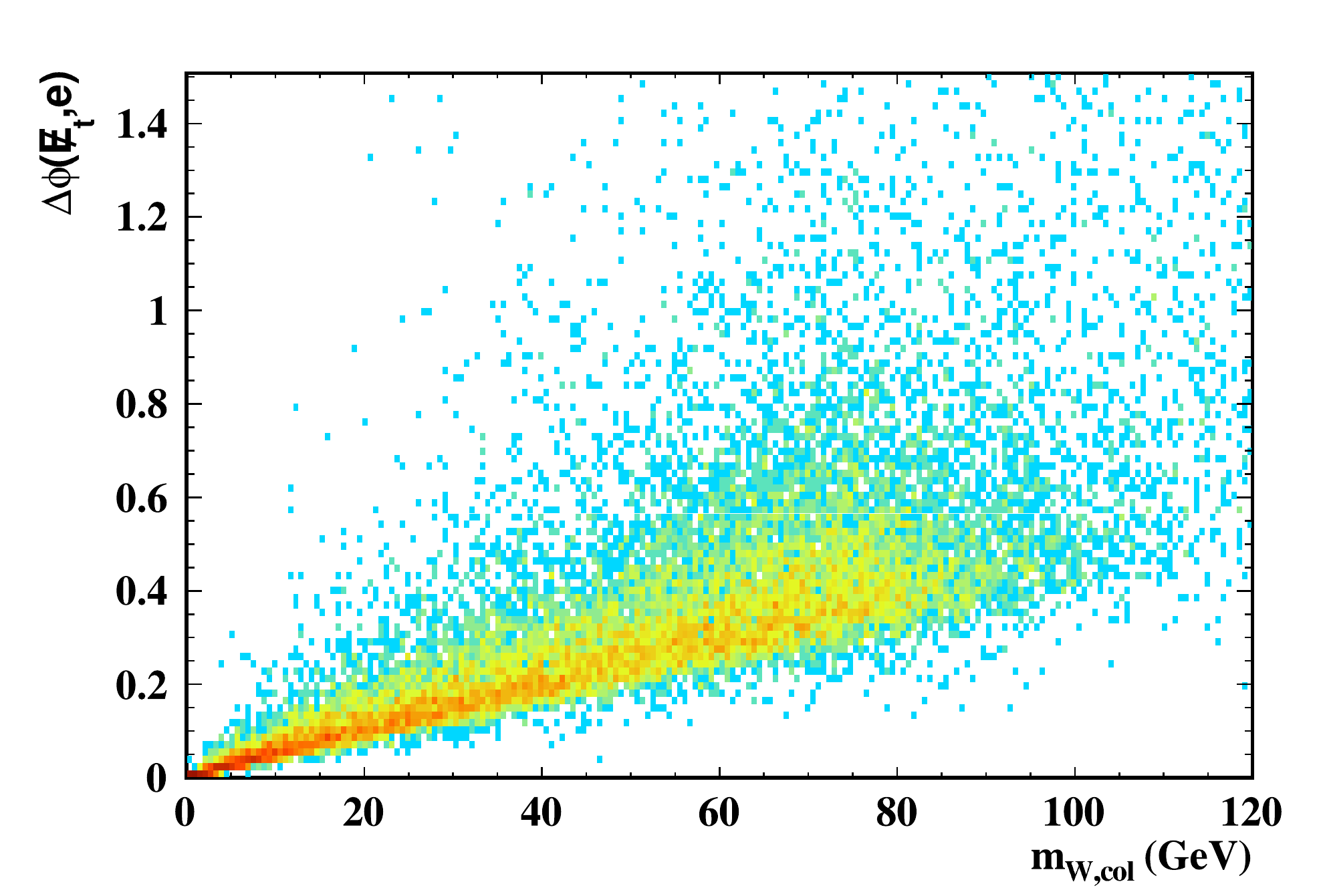}&
    \includegraphics[angle=90,width=80mm]{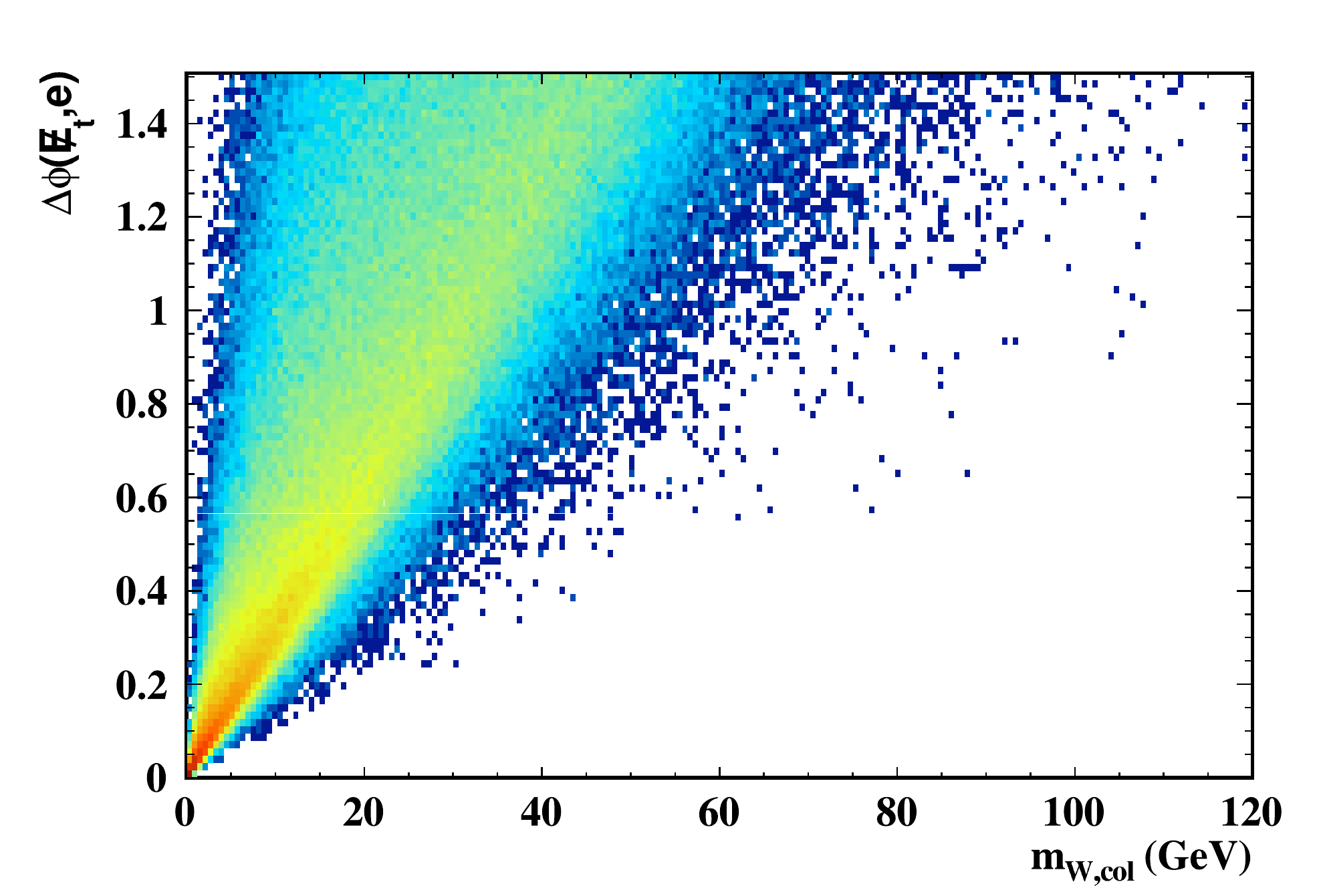}\\
    {\small (a)}&{\small (b)}\\
    \includegraphics[angle=90,width=80mm]{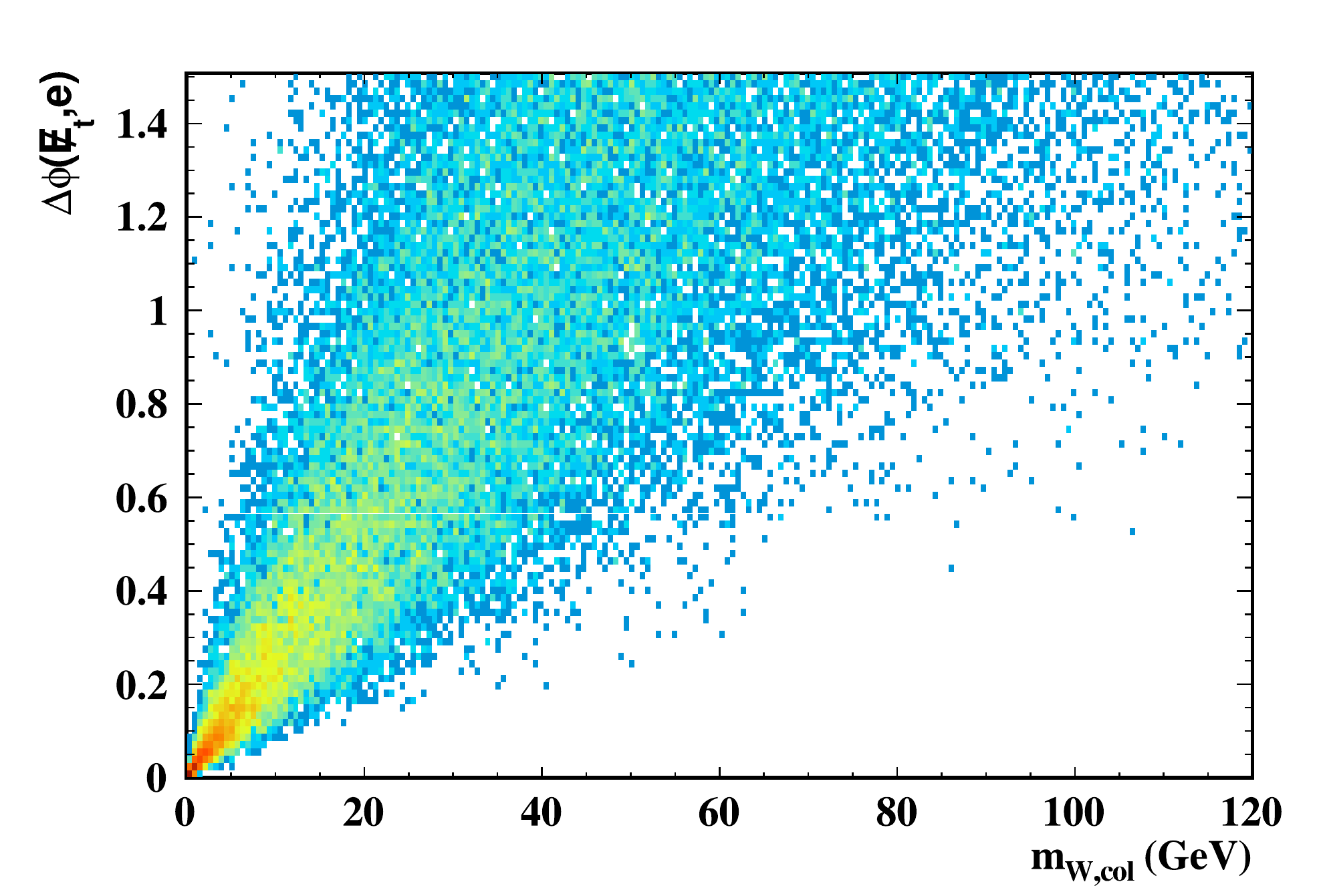}&
    \includegraphics[angle=90,width=80mm]{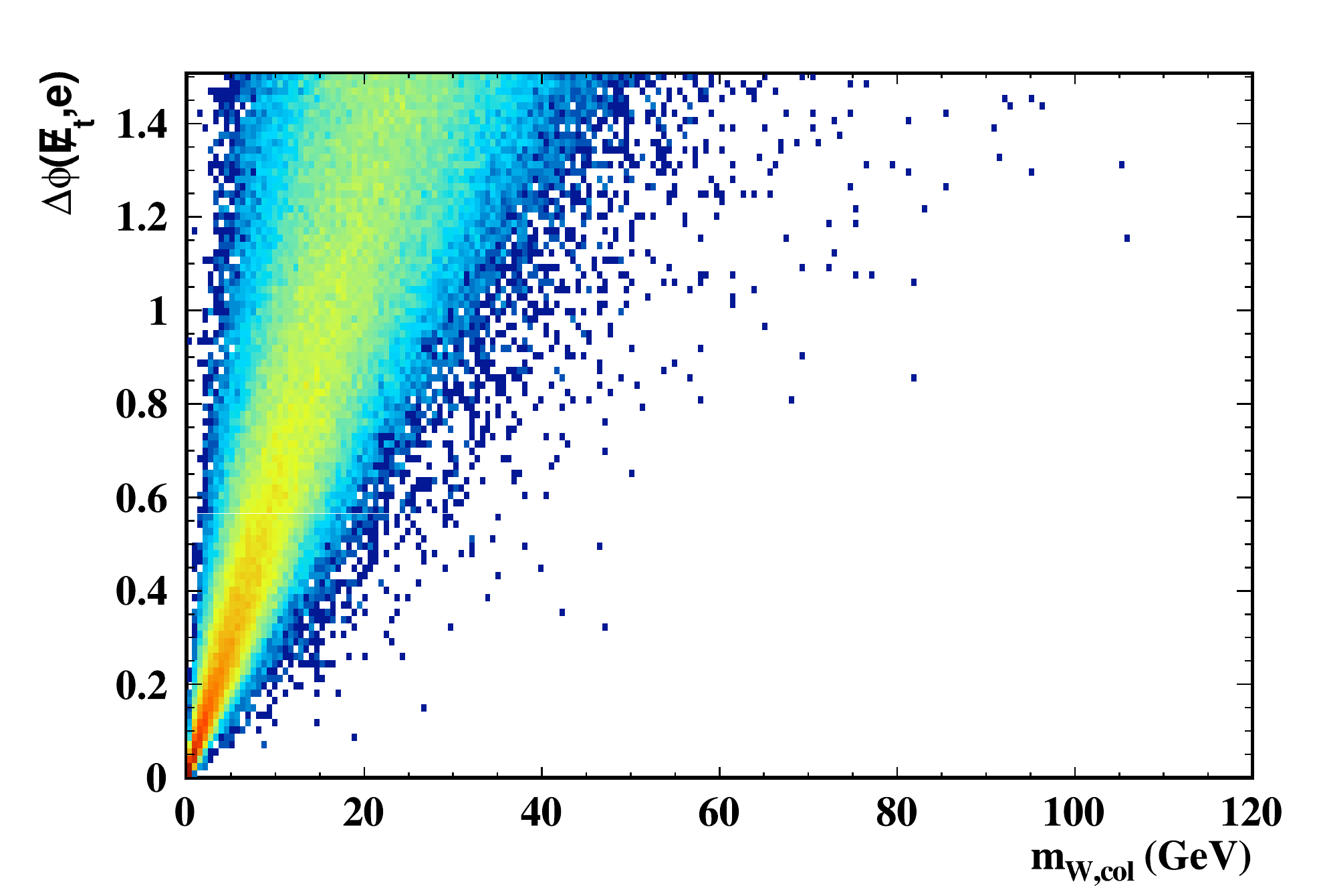}\\
    {\small (c)}&{\small (d)}
    \end{tabular}
    \caption[Discriminating boosted $W^\pm$s from background]{Discriminating boosted $W^\pm$s from background for signal (a), $W$ + Jets (b), $t\bar{t}$ (c) and $Z\to e^+e^-$ (d)}
    \label{fig:wdiscriminant}
\end{figure}

\begin{figure}[htbp]
  \begin{center}
  \includegraphics[width=80mm]{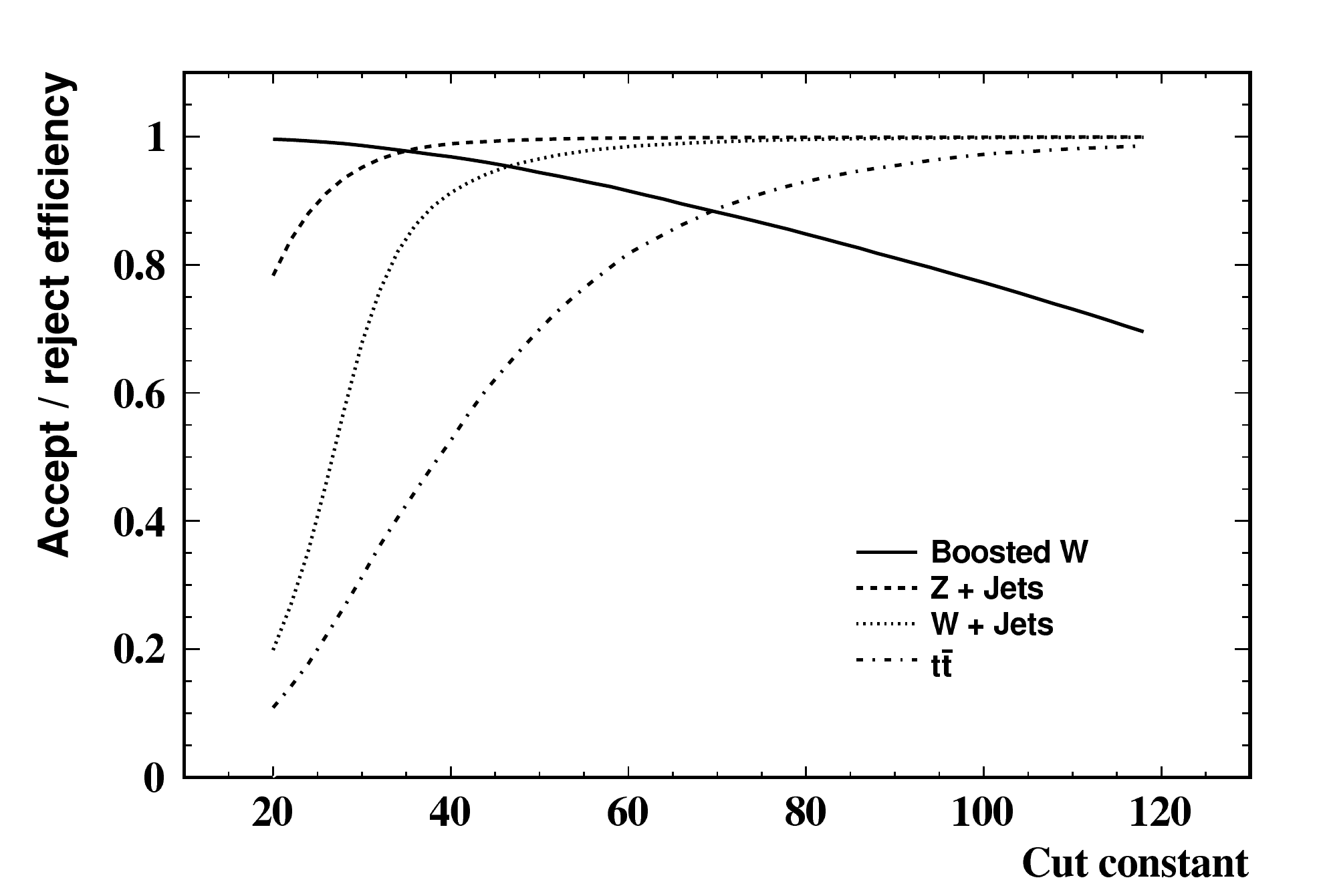}
  \end{center}
  \caption{Acception / rejection efficiencies for signal and background, varying the constant term $c$ in the 2D-plane cut.}
  \label{fig:boostedwcuteff}
\end{figure}

Plotting the electron-neutrino invariant mass against the angle in $\phi$ between the electron and ${E\!\!\!\!/_T}$ provides a powerful discriminant between signal and background, as shown in Fig.~\ref{fig:wdiscriminant} for events simulated with Pythia~\cite{Sjostrand:2006za} and PGS~\cite{PGS}. The signal is a $1\,\mathrm{TeV}$ excited quark, which can be taken as producing a generic boosted $W$ with momentum near $500\,\mathrm{GeV}$. A cut in the 2D plane of $\Delta\phi < M_{W,col} / c$ with $c=100$ yields a boosted $W$ identification efficiency of $77\%$ and a $t\bar{t}$ rejection of $97\%$. Further study and tuning is needed with full detector simulation, but it appears that powerful signal selection and background rejection is possible (Fig.~\ref{fig:boostedwcuteff}).

\label{ss:futurew}

\subsection{High $p_T$ top reconstruction}
CDF and D0 have performed extensive $t \bar{t} $ resonance searches~\cite{:2007dz,Abazov:2008ny} and a $ \frac{\delta \sigma}{\delta M_{t \bar{t} }}$ measurement~\cite{Aaltonen:2009iz}. No deviations from the Standard Model prediction have been observed and limits are derived for several models.

At the Tevatron, the large majority of $ t \bar{t} $ pairs are produced essentially at rest. The $ t \bar{t} $ pair with the largest invariant mass is registered with approximately 1 TeV. At 14 TeV, in 20 \% of $ t \bar{t}$ produced, one of the top quarks has a transverse momentum greater than 200 GeV~\footnote{Estimate obtained using MC@NLO~\cite{mcatnlo,mcatnlo2}}. The LHC will be able to explore the $ t \bar{t} $ mass spectrum into the several TeV regime.

The reconstruction of highly boosted top quarks is an experimental challenge. The top quark decay products are collimated in a narrow cone. The hadronic decay products often cannot be individually resolved by jet algorithms. The isolation of the leptons from $W$-decay is broken by the neighbouring b-jet. A number of references in the literature~\cite{Thaler:2008ju,Kaplan:2008ie,Almeida:2008yp,Almeida:2008tp,
Skiba:2007fw,Holdom:2007nw,Holdom:2007ap,Seymour:1993mx,Butterworth:2007ke,
Butterworth:2008iy,Butterworth:2008sd,Ellis:2009su,Ellis:2009me,Plehn:2009rk,Benchekroun:2001je,Butterworth:2002tt} 
have addressed this issue proposing a new approach, where top decays (and similarly W/Higgs decays) are reconstructed as a single jet. A number of techniques has been developed that allow to identify (tag) these top mono-jets.

Recent CMS~\cite{cms_ttbar_resonance_semi,cms_ttbar_resonance_allhad} and ATLAS~\cite{atlas_ttbar_resonance, Brooijmans:2008zz,brooijmans2,Vos:2008zzb} studies have implemented these ideas and established their performance on fully simulated signal and background events. These techniques are indeed found to offer greatly improved top quark reconstruction efficiency, while maintaining an adequate reduction of non-$ t \bar{t} $ backgrounds (primarily W+jets and QCD di-jet production). Thus, the sensitivity of $ t \bar{t} $ resonance searches is improved with respect to that obtained with classical reconstruction techniques.

 

\section{Indirect KK effects}

In addition to signals from the {\em direct} production of the KK particles
at the LHC, there can also be effects of
these KK particles on the properties of the SM
particles themselves.

\subsection{Flavor-violating Higgs/Radion couplings}

Higgs flavor-violation induced by KK particles in the warped extra-dimensional framework 
is discussed in Refs.~\cite{Agashe:2009di,Azatov:2009na}. Estimates are BR $\left( 
t \rightarrow c h \right) \sim 10^{ -4 }$
and BR $\left( h \rightarrow tc \right) \sim 5 \times 10^{-3}$: see 
Ref.~\cite{Azatov:2009na} for more detailed numbers.
Note that the radion also decays to $tc$ (very similarly to Higgs): see Ref.~\cite{Azatov:2008vm}.
Reference~\cite{AguilarSaavedra:2004wm} claims LHC sensitivity of $5 \times 10^{ -5 }$ for BR $\left( t \rightarrow
cH \right)$ (obviously for $m_h < m_t$) 
(see also Ref.~\cite{AguilarSaavedra:2000aj} for more details).
A reference for a study of $h \rightarrow tc$ --- obviously for $m_h > m_t$ --- 
at a similar level of detail could not be found.

We suggest performing
a detailed study of LHC sensitivity for $tch$ coupling for
the case $m_h > m_t$, i.e.,
when $t \rightarrow c h$ is not allowed.
One method is via Higgs {\em decays}: the Higgs can be produced via gluon fusion or by $WW$ fusion (in the latter case, we can tag forward jets).
See Refs.~\cite{Azatov:2009na} and~\cite{Azatov:2008vm} for first steps toward this goal (including some analysis of background).
[Reference~\cite{Bejar:2003em} studied flavor-violating Higgs decays to top in a different framework (2-Higgs doublet model), but without any analysis of background].

Another option is to use the $tch$ coupling to {\em produce} the Higgs, 
for example, $g c \rightarrow
t h$ (see Ref.~\cite{AguilarSaavedra:2000aj} for a study of this channel, but using $h 
\rightarrow b \bar{b}$, 
whereas here we would like to use $h \rightarrow WW/ZZ$ since we have $m_h > m_t$).

In both directions mentioned above, a starting point might be to use existing studies of related channels in SM (or its extensions) in order
to see how background was reduced -- for example, $g b \rightarrow t H^+$ in $2$-Higgs doublet models vs. $g c \rightarrow t h$ here or 
single top production in SM vs. 
$gg \rightarrow h
\rightarrow tc$ here.


%
%
%

Finally, a leptonic (and thus cleaner) channel: $BR \left( h \rightarrow \mu \tau \right)$ can be large in this framework 
(see Ref.~\cite{Azatov:2009na}) which might
be within the LHC reach (see Ref.~\cite{Han:2000jz}).


\subsubsection{Sensitivity study for $t \rightarrow c h$ at LHC }
\label{lhc_tcH}

Even though
the LHC sensitivity for observing the flavor violating decay of top quark,
$t \rightarrow c h$ (when $m_h < m_t$), has been studied in detail in 
Ref.~\cite{AguilarSaavedra:2000aj}, we think it is useful to re-visit this analysis
which is the goal of this section.
Specifically, we focus on the following {\em new} aspects:
i) optimizing cuts to improve the sensitivity, (ii) tagging charm quark, motivated
by the fact that 
since typically
$t \rightarrow c h$ dominates over $t \rightarrow u h$, the signal under
consideration
contains a charm quark and (iii)
considering $m_h = 160$ GeV so that $h \rightarrow b\bar{b}$ is very small
and $h \rightarrow WW$ dominates (note that only the cases
$m_h = 110, 130$ GeV were studied in Ref.~\cite{AguilarSaavedra:2000aj}
such that
the dominant decay mode $h \rightarrow b \bar{b}$ was used). 

The signal at the LHC arises from $pp \rightarrow  t \bar t \rightarrow b W 
c h \rightarrow bWcWW$ leading
to $\ell bc4j  {E\!\!\!\!/_T} $ events, where $\ell = e$ or $\mu$ and $j =u,d,c,s$ 
is from $W$ decays. We allowed all  the three $W$-bosons decay into all 
possible channel. The effective cross section for this signal topology 
can be expressed as, 
\begin{eqnarray}  \label{effcross}
C_{{\ell bc4j}} & = & \sigma_{SM}(pp \rightarrow  t \bar t)  
BR(t \rightarrow c h)   BR(t \rightarrow b W)  BR(h \rightarrow W W)\,,
\end{eqnarray}
where $W$ stands for $W^{\pm}$. We consider $m_{t}$=175 GeV, $m_{h}$= 160 GeV and BR($t \rightarrow c h$)=$10^{-4}$. 
The SM backgrounds with the similar signal topology arises 
from many reducible and irreducible sources. However, for the present study we considered  
the dominant two, namely, $t \bar t$ and $t \bar t b \bar b$.

In our signal simulation we used the {\tt PYTHIA v6.408} event generator~\cite{Sjostrand:2006za}. 
The {\tt SLHA}~\cite{Skands:2003cj} input is used to provide 
the flavor violating branching ratios of the top quark. 
 for generating parton level SM backgrounds, 
 we used {\tt MadGraph/MadEvent v4.4.15}~\cite{Maltoni:2002qb}, and we later fed them to {\tt PYTHIA} for showering. 
 The backgrounds 
events were generated with the following preselection kinematical 
cuts: $p_T^{j,b} \raisebox{-0.13cm}{~\shortstack{$>$ \\[-0.07cm] $\sim$}} 5$ GeV; $\eta^{j,b} \raisebox{-0.13cm}{~\shortstack{$<$ \\[-0.07cm] $\sim$}}  5.0$; $ \Delta R (jj,bb,bj) \raisebox{-0.13cm}{~\shortstack{$>$ \\[-0.07cm] $\sim$}} 0.3$.  
We set the renormalization and factorization
scale to $Q= \sqrt {\hat s}$ and used CTEQ5L for the parton distribution
functions (PDF). All the masses and mass parameters are given in GeV.

We simulate our signal and backgrounds at the LHC for 14 TeV center of mass energy based on the following assumptions: 

\begin{itemize}

\item The ATLAS~\cite{:1999fr}  calorimeter coverage is $\rm |\eta| < 5.0$; 
\item The segmentation is $\Delta \eta \times \Delta \phi$=$0.087 \times 0.10$ 
(i.e., approximately $\Delta R = 0.13$) which resembles the ATLAS detector;  

\item  The toy calorimeter, {\tt PYCELL}, provided in {\tt PYTHIA} for the jet reconstruction.
The total energy of jets and leptons are smeared according to Gaussian distribution. 
The energy resolution is taken as
\begin{eqnarray}
{\Delta E_{j,\ell} \over E_{j,\ell}} = {50\% \over \sqrt{E_{j,\ell}}} \oplus 3\% \quad ;
\end{eqnarray}

We reconstructed the missing energy (${E\!\!\!\!/_T}$) from smeared observed particles. We have 
not included any real detector effects in our simulations; 

\item The showering scales are the following:
for ISR and FSR we multiplied the hard scattering scale, $Q^2$,
which we set as $f \times \hat s$, where $f$=4.0 ;

\item A cone algorithm with
        $\rm\Delta R(j,j) = \sqrt{\Delta\eta^{2}+\Delta\phi^{2}} \ge 0.4$
        has been used for jet finding ;
\item The $\rm E_{T,min}^{cell} \ge 1.0$  is considered to be a potential candidate for jet initiator. The
cell with $\rm E_{T,min}^{cell} \ge 0.1 $  
is treated as a part of the would be jet and
minimum summed $\rm E_{T,min}^{j} \ge 15.0$  is accepted as a jet and the jets are ordered in $E_{T}$;
  \item Leptons ( $\rm \ell = e, ~\mu$ ) are selected with
        $\rm E_T^{\ell} \ge 20.0$  and $\rm |\eta^{\ell}| \le 2.5$ ;

  \item  We have implemented jet and lepton ($\rm \ell = e$ or $\rm \mu$)
isolation using the following criteria: if there is a jet within the 
vicinity of the 
partonic lepton ($\ell^{p}$) with   
$\rm \Delta R (j-\ell^{p}) \ge 0.4$ and $ 0.8 \le \rm  E_{T}^{j}/E_{T}^{\ell^{p}} \le 1.2 $, the jet
is removed from the list of jets and treated as a lepton, else the lepton is
removed from the list of leptons;

  \item {\it b-tagging:} A jet with $\rm |\eta^{j}| \le 2.5$ matched {\footnote{ Unlike jet-lepton 
matching we considered only the minimum $\Delta R(j,B)$ and not the $E_{T}$ ratios.}} with a $b-$flavored hadron $B$, i.e.,
with $\Delta R(j, B) < 0.2$, is considered to be {\em b-taggable}. We imposed the 
$b$ tagging in these taggable jets with probability $\epsilon_b$=0.50;

  \item {\it c-tagging:} A jet with $\rm |\eta^{j}| \le 2.5$ matched (similar to B-Hadron) with a 
$C-$flavored hadron $C-hadron$ (e.g., D-meson, $\Lambda_c$-baryons), i.e.,
with $\Delta R(j, C-hadron) < 0.2$, is considered to be {\em c-taggable}. We imposed the 
$c$ tagging in these taggable jets with probability $\epsilon_c$=0.10;

 \item{\it b-mis-tagging:} Jets other than $b$-taggable/tagged and $c$-taggable/tagged
are matched with the light flavor parton (q = u,d,s,g and $\tau$ with minimum $\Delta R (j-q)$
and $\le$ 0.4. If $E_{T}^{j} \ge 15$ and $\eta_j \le 2.5$ then
the jet is treated as a mis-taggable jets with the flavor similar to the
matched parton, $q$. If a jet does not match
with any parton in the event we consider this jet as a gluon-jet originating from the
secondary radiation. The jets are mis-tagged by generating
random numbers according to the flavors; we considered $\epsilon_{u,d,s,g}$=0.0025  
following the recent {\tt ATLAS} analysis~\cite{Aad:2009wy,Lehmacher:2008hs} and~\cite{mistaggcharm}  
{\footnote{The $\tau$-lepton is considered to be a parton in our analysis
with nearly zero mis-tagging probability.}}. It is important to note that the mis-tagging rate can be 
known precisely once we have the real LHC data.

\end{itemize}

We need to retain as many signal events as possible and at the same time suppress the
backgrounds to a large extent by applying different kinematical selection. 
In doing so we introduce the following kinematical selection:

\begin{itemize}
\item C1: $N_{\rm jet} \ge 6$, $E_{T}^{j=1-6} > 15.0$  and $ |\eta^{j=1-6}| < 5.0$;
\item C2: $N_{\rm lepton} \ge 1$, $E_{T}^{\ell} > 20.0$ and $|\eta^{\ell}| <
  2.5$; 
\item C3: ${E\!\!\!\!/_T} > 20$  where ${E\!\!\!\!/_T}$ is calculated from all visible
  particles;  
\item {C4a: $N_{b-tag}  \ge 1$}; $|\eta^{b-jet}| < 2.5$, $\Delta R(j,B)\le  0.2$;  
\item {C4b: $N_{c-tag}  \ge 1$}; $|\eta^{c-jet}| < 2.5$, $\Delta R(j,C-hadron)\le  0.2$;  
\item {C4(with Mis-tagging from the light quarks and gluon) : $N_{tot}$=$N_{{(b+c)}-tag+{q}-mistag }  \ge 2$}.
\end{itemize}

The individual efficiencies for $N_{b-tag}$,  $N_{c-tag}$ and $N_{tot}$ are given in Table~\ref{tab:tch_wht_table_1}. 
As expected, $N_{c-tag}$ efficiencies for Signal is larger than $t \bar t$ and $t \bar t b \bar b$.
We have also shown the individual efficiencies for number of jets, number of lepton and missing energy
(${E\!\!\!\!/_T}$) in Table~\ref{tab:tch_wht_table_2}.

To ensure the flavor violating decay of top quark we reconstructed the $W$-boson, Higgs boson and top quark masses.  
In order to suppress the huge backgrounds, before mass reconstruction, at the first step we applied the basic
acceptance cuts, JLM = C1 $\otimes$ C2 $\otimes$ C3. We required one $c$-tagged (C4b) events  
to suppress more background and can be seen from Table~\ref{tab:tch_wht_table_3}. Finally, 
we consider events with at least two tagged jet (C4) for the mass reconstruction. 

\begin{table}[t!]
\begin{center}
\begin{tabular}{|c|c||c|c|c||c|c|c||c|c|c|}
\hline
& &\multicolumn{3}{|c||}{$N_{b} \ge $} &
\multicolumn{3}{|c||}{ $N_{c} \ge$} & \multicolumn{3}{|c|}{$N_{tot} \ge $} \\
Process & EvtSim & 1 & 2&3&  1 & 2&3& 1 & 2& 3\\
\hline 
\hline
$m_{h}$=160& 100000 &.4267 &.0063 &.0007 &.1090 &.0049 &.0001 &.4984 &.0641 &.0051\\
\hline 
\hline
$t \bar t$ & 1000000 &.6610 &.1795 &.0027 &.0585 &.0011 &.0000 &.6853 &.2147 &.0168\\
\hline
$t \bar t b \bar b$& 125000 &.8037&.4024 &.1077 &.0612 &.0012 &.0000 &.8185&.4330 &.1314 \\
\hline
\hline
\end{tabular}
\caption{
The Individual efficiencies for purely $b$-tagged ($N_b$), purely $c$-tagged ($N_c$) and with the inclusion of 
low flavor mis-tagged ($N_{tot}$) at LHC. EvtSim stands for number of event simulated.}
\label{tab:tch_wht_table_1}
\end{center}
\end{table} 

\begin{table}[t!]
\begin{center}
\begin{tabular}{|c|c|c|c|c|c|c|c|}
\hline
Process& EvtSim&C1& C2 & C3 & C4a& C4b &C4 \\
\hline
\hline 
$m_{h}$=160&100000 &.719 &.357 &.790 &.427 &.109 &.064  \\
\hline 
\hline
$t \bar t$ & 1000000 &.655 &.330 &.746 &.661 &.058 &.215 \\
\hline
$t \bar t b \bar b$& 125000 &.838 &.340 &.790 &.804 &.061 &.433\\
\hline 
\hline 
\end{tabular}
\caption{
The individual efficiencies of various kinematical selections for signal and backgrounds at LHC.  
EvtSim stands for the number of event simulated. See text for the numerical values of the 
kinematical selections.}
\label{tab:tch_wht_table_2}
\end{center}
\end{table}

\begin{table}[t!] 
\begin{center}
\begin{tabular}{|c|c|c|c|c|c|c|c|}
\hline
Process& RawEvt&JLM& JLMNc& $W$-reco&$W70$&$h100$&$t100$\\
\hline
\hline
$m_{h}$=160&8000.  &   1190.& 138.5 & 66.6 &  63.2 &  47.4 &   6.93 \\
\hline 
\hline
$t \bar t$& 80000000.&8081880.& 376880.& 261456.& 225745.& 129452.&8927.\\
\hline
$t \bar t b \bar b$& 299147.&  50012.&2371.& 1383.&1227.& 571.& 52.6\\
\hline 
\hline
\end{tabular}
\caption{The cumulative events for Signal and Backgrounds survived after different combination of
selection criterion at the LHC for 100 $fb^{-1}$ integrated luminosity. 
RawEvt stands for the number of events produced in reality. 
$W$-reco stands number of events for the combined selections: 
C1 $\otimes$  C2  $\otimes$   C3 $\otimes$  C4b $\otimes$  C4. 
$W70$, $h100$ and $t100$ represent the selection on the reconstructed masses of 
$W$, Higgs boson and top quark, see text for details.}
\label{tab:tch_wht_table_3}
\end{center}
\end{table}

We show the cumulative number of events after applying some combined 
selections in Table~\ref{tab:tch_wht_table_3}. 
The combined selections are the following:

\begin{itemize}
\item JLM: C1 $\otimes$  C2  $\otimes$  C3;
\item JLMNc: JLM $\otimes$  C4b;
\item $W$-reco: JLMNc $\otimes$ C4;
\item $W70$: $W$-reco $\otimes$ $m_{W}$ $\pm$ 70 GeV; 
\item $h100$: $W70$ $\otimes$ $m_{h}$ $\pm$ 100 GeV;  
\item $t100$ : $h100$ $\otimes$ $m_{t}$ $\pm$ 100 GeV
\end{itemize}

\begin{figure}
\centering
        \includegraphics[width=80mm]{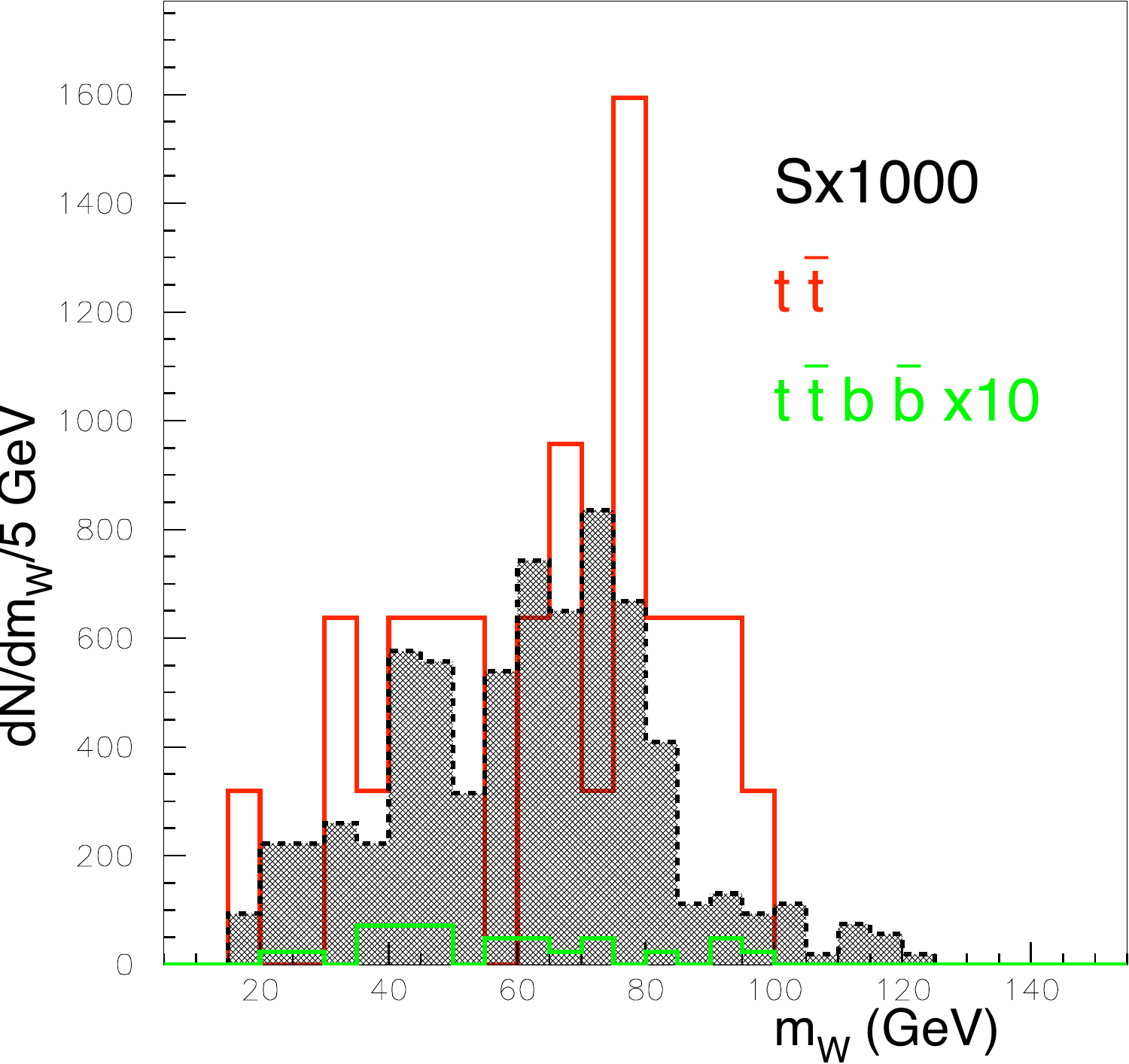}
        \caption{ The reconstructed W-boson mass ($m_W$) for signal and backgrounds.
The distribution is normalized to $t100$, see the last column in Table~\ref{tab:tch_wht_table_3}.
Signal and $t \bar t b \bar b$ are scaled with 1000 and 10 respectively.}
        \label{fig:tch_wht_fig_1}
\end{figure}

\begin{figure}
\centering
        \includegraphics[width=80mm]{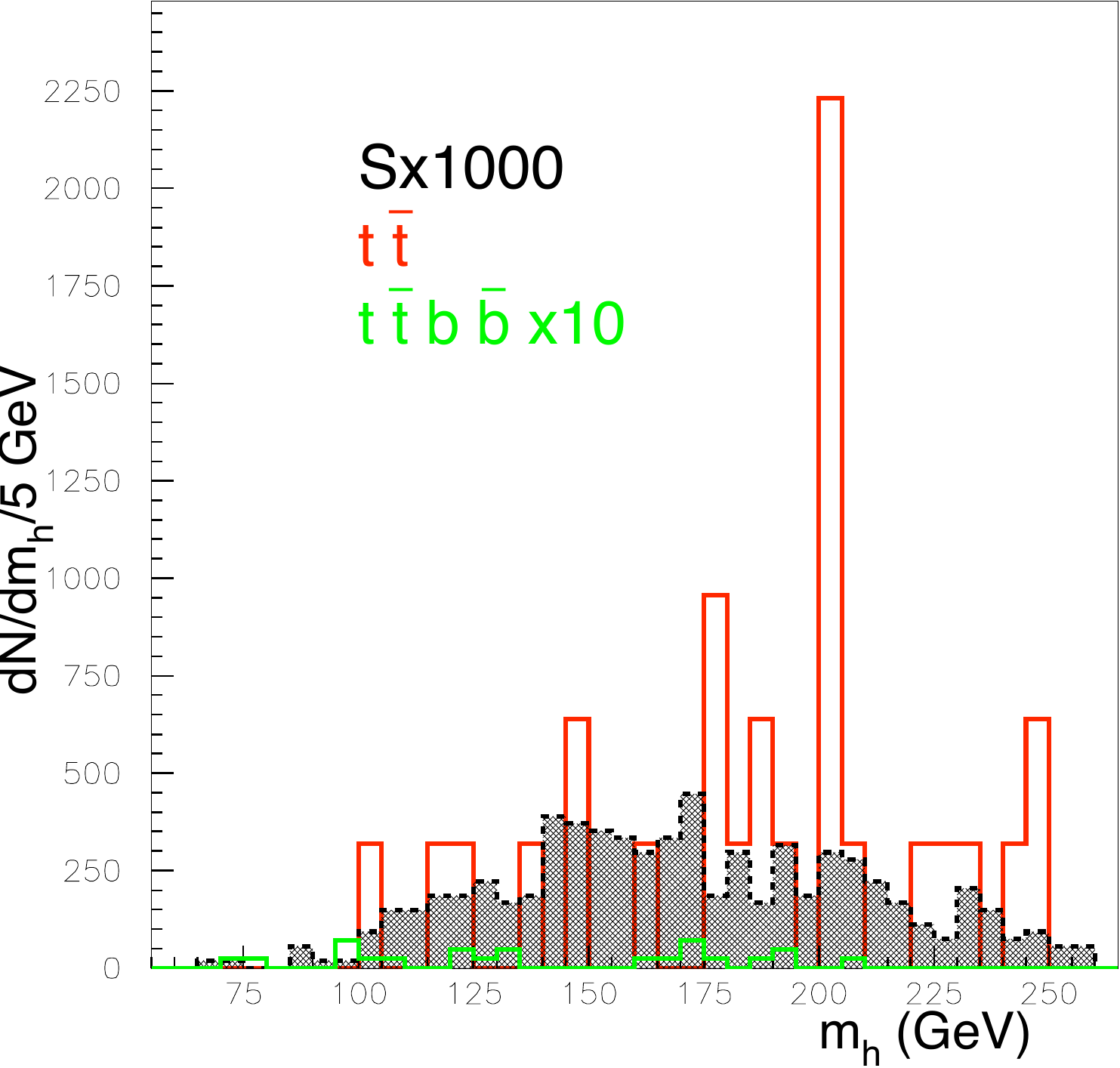}
        \caption{The reconstructed Higgs boson mass ($m_{h}$) for signal and backgrounds.
The distribution is normalized to $t100$, see the last column in Table~\ref{tab:tch_wht_table_3}.
Signal and $t \bar t b \bar b$ are scaled with 1000 and 10 respectively. }
        \label{fig:tch_wht_fig_2}
\end{figure}

\begin{figure}
\centering
        \includegraphics[width=80mm]{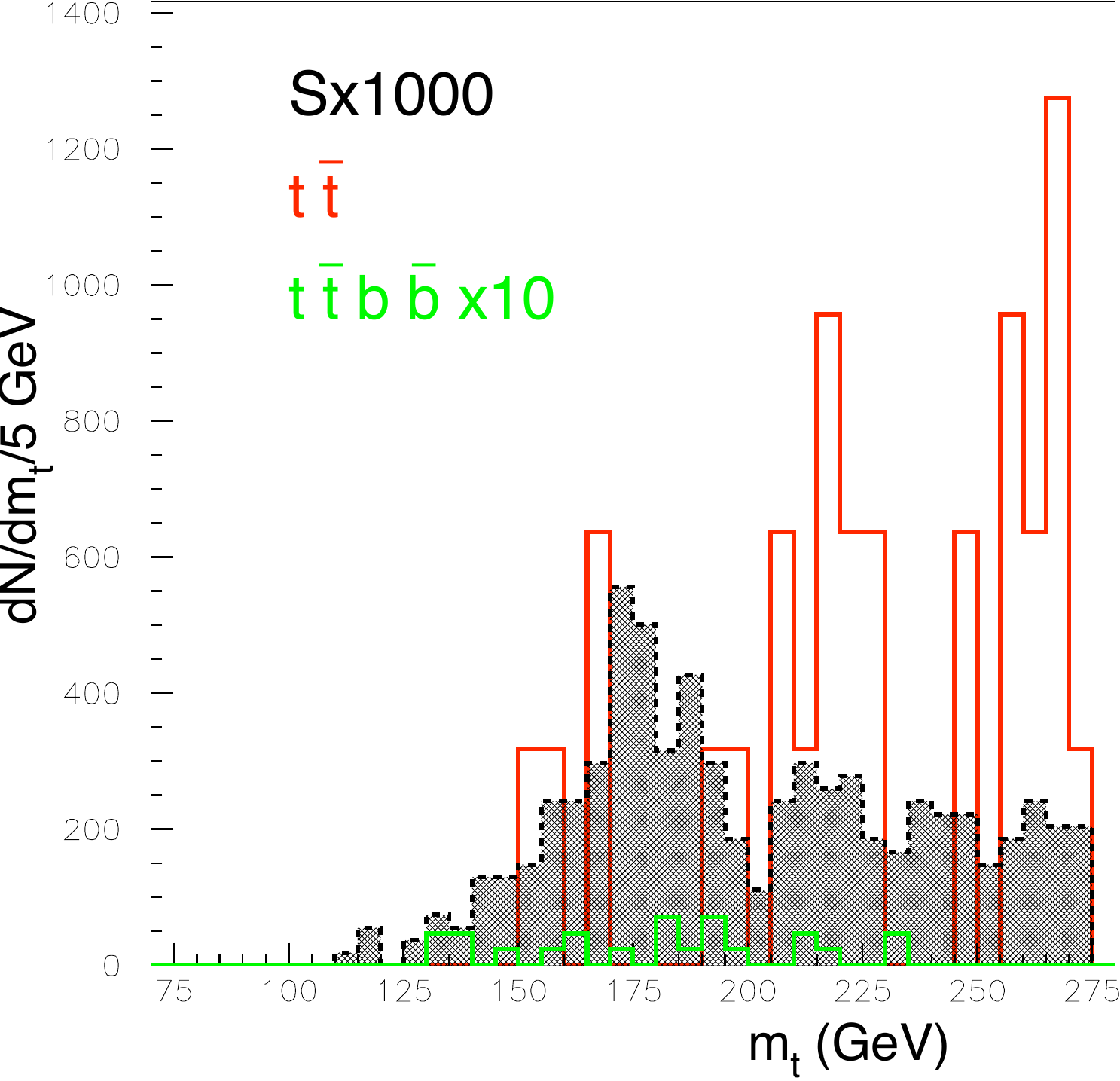}
        \caption{The reconstructed top quark mass ($m_{t}$) for signal and backgrounds.
The distribution is normalized to $t100$, see the last column in Table~\ref{tab:tch_wht_table_3}.
Signal and $t \bar t b \bar b$ are scaled with 1000 and 10 respectively.}
        \label{fig:tch_wht_fig_3} 
\end{figure}

We calculated all the possible di-jet invariant mass ($m_{jj}$) 
without considering the pure $b$-tagged jets (
since BR($W \rightarrow \bar b u(c)$ ) approximately $\cal O$ $(10^{-5(4)})$ ).  
The pair of jets for reconstructing each $m_{W}$
were selected by minimizing $|m_{{j_{1}}{j_{2}}} - m_{{j_{3}}{j_{4}}}|$. 
The reconstruction of $m_{h}$ is then straightforward, i.e., $m_{h} = m_{j_1j_2j_3j_4}$.  
Furthermore, we reconstruct the top quark mass to ensure the flavor violating decay.
We consider the remaining jets, without pure $b$-tagged jets   
(to ensure the flavor violating decay), combined with the selected four jets ($m_{h}$ candidates); by minimizing 
$|m_{j_1j_2j_3j_4j_5} - m_{t}|$. After mass reconstructions,  
we applied $t100$ selection and show the reconstructed masses 
for $W$, $H$ and top in Fig.~\ref{fig:tch_wht_fig_1}, \ref{fig:tch_wht_fig_2} and \ref{fig:tch_wht_fig_3} respectively
{\footnote{ We scaled the signal ($t \bar t b \bar b$) distribution by 1000 (10) in all the figures.}}. 
It can be seen from Table.\ref{tab:tch_wht_table_3} that the number of signal (total background) event is 
approximately 7(9000). 

Our preliminary analysis shows that the number of signal and total background events 
are approximately 7 and 9000. Thus the signal is very challenging to isolate 
from the SM backgrounds, mainly due to the very low branching ratio for $t \rightarrow c h$.  
With luminosity upgrade, one could get more signal events, however, the 
backgrounds will also be large. One has to design more clever selection 
to reject $t \bar t$ backgrounds. Intuitively, a slightly different 
approach while reconstructing the top mass might be useful, for example, 
considering explicit $c$-jet (i.e., $j_5$ candidate jet in our present analysis). 
Of course, an analysis by LHC {\em experimental} groups of sensitivity 
for $t \rightarrow c h$ is desirable. We are aware that ATLAS group is 
already undertaking such a study.


\section{Summary}

In this note, we have given an overview of LHC signals for the very well-motivated framework of SM particles propagating in a warped extra dimension.
We have also presented some {\em new} results of (or directions for) such studies, for example, identification of boosted $W/Z$'s and top quarks, $t \rightarrow c h$ and radion production.
It is worth pointing out that some of these studies might also be relevant in searching 
for other types of new physics, for example, other models beyond the
SM can also contain heavy particles decaying into
top quarks, $W/Z$ or can give rise to sizable flavor-violating $tc$Higgs coupling.

\section*{Acknowledgements}

The work of K.A. is supported by the NSF grant No. PHY-0652363.



%% file: Basso/Zp.tex
\def\met{{\slash\!\!\!\!\!\:E}_T}
\def\mpt{{\slash\!\!\!\!\!\:P}_T}
\def\metv{{\slash\!\!\!\!\!\:\vec{E}}_T}
\def\mptv{{\slash\!\!\!\!\!\:\vec{P}}_T}

\chapter{$\boldsymbol{Z'}$ discovery potential at the LHC in the minimal $\boldsymbol{B{\rm -}L}$ model}

{\it L.~Basso, A.~Belyaev, S.~Moretti, G.M.~Pruna and C.H.~Shepherd-Themistocleous}

\begin{abstract}
{\small \noindent We present the Large Hadron Collider (LHC) discovery
potential in the $Z'$ sector of a $U(1)_{B-L}$ enlarged Standard Model
for $\sqrt{s}=7$ and $14$ TeV centre-of-mass (CM) energies,
considering both the $Z'_{B-L}\rightarrow e^+e^-$ and
$Z'_{B-L}\rightarrow \mu^+\mu^-$ decay channels.  Electrons provide a
higher sensitivity to smaller couplings at small $Z'_{B-L}$ masses
than do muons. The resolutions achievable may allow 
the $Z'_{B-L}$ width to be measured at smaller masses in the case of
electrons in the final state.  The run of the LHC at $\sqrt{s}=7$ TeV,
assuming at most $\int \mathcal{L} \sim 1$ fb$^{-1}$, will be able to
give similar results to those that will be available soon at the
Tevatron in the lower mass region, and to extend them for a heavier
$M_{Z'}$. A run at $14$ TeV is needed to fully probe the parameter
space.  If no evidence is found in any energy configuration, $95\%$
C.L. limits can be determined, and, given their better resolution, the
limits from electrons will always be more stringent than those from
muons.  }
\end{abstract}

\section{Introduction}
The evidence for non vanishing (although very small) neutrino masses
is so far possibly the only hint for new physics beyond the Standard
Model (SM)~\cite{Fogli:2005cq,Fogli:2006yq}. It is noteworthy that the
accidental $U(1)_{B-L}$ global symmetry is not anomalous in the SM
with massless neutrinos though its origin is not well understood. It
thus becomes appealing to extend the SM to simultaneously explain the
existence of both (i.e., neutrino masses and the $B-L$ global
symmetry) by gauging the $U(1)_{B-L}$ group thereby generating a $Z'$
state.  This requires that the fermion and scalar spectra are enlarged
to account for gauge anomaly cancellations. The results of direct
searches constrain how this may be done~\cite{Carena:2004xs,Cacciapaglia:2006pk,Aaltonen:2008vx,Aaltonen:2008ah}. Minimally,
this requires the addition of a scalar singlet and three right-handed
neutrinos, one per generation~\cite{Buchmuller:1991ce,Khalil:2006yi,BL_master_thesis}, which could
trigger the see-saw mechanism explaining the smallness of the SM
neutrino masses~\cite{Minkowski:1977sc,VanNieuwenhuizen:1979hm,Yanagida:1979as,S.L.Glashow,Mohapatra:1979ia}.
Within this model, the masses of the heavy neutrinos are such that their
discovery falls within the reach of the LHC over a large portion of parameter space~\cite{Basso:2008iv,Huitu:2008gf}.

In general, studies of this model focus on a specific non-disfavoured
point in the parameter space and do not preform a systematic analysis
of the entire space. The $Z'_{B-L}$ boson is also not always
considered as a traditional benchmark for generic collider reach
studies~\cite{Langacker:1991pg,Rizzo:1996ce,Contino:2008xg,Gulov:2009tn,Erler:2009jh,Ball:2007zza}
or in data analyses~\cite{Aaltonen:2008vx,Aaltonen:2008ah}. We have
therefore performed a (parton level) discovery potential study for the
LHC in the $Z'$ sector of the $B-L$ model. In the light of the LHC
plan of action over the next few years~\cite{indico_chamonix}, we
consider the CM energies of $7$ and $14$ TeV, with integrated
luminosities up to $1$ fb$^{-1}$ for $7$ TeV, and up to $100$
fb$^{-1}$ for $14$ TeV. We also include
a comparison with the Tevatron reach for its expected $10$ fb$^{-1}$
of integrated luminosity. We chose to study the di-lepton channel
(both electrons and muons), the cleanest and most sensitive $Z'$ boson decay
channel in our model at colliders.

This work is organised as follows. Section~\ref{sect:model} describes
the $B-L$ model under consideration. Section~\ref{sect:comput}
illustrates the computational techniques adopted. The results are
presented in Section~\ref{sect:Zp_disc} for the $Z'$ boson
sector and finally the conclusions are given in Section~\ref{sect:conc}.

\section{The Model}\label{sect:model}
The model under study is the so-called ``pure'' or ``minimal''
$B-L$ model (see Ref.~\cite{BL_master_thesis,Basso:2008iv} for conventions and references) 
since it has vanishing mixing between the $U(1)_{Y}$ 
and $U(1)_{B-L}$ gauge groups.
In the rest of this paper we refer to this model simply as the ``$B-L$
model''.  This work focuses on the extended gauge sector of the model, whose Abelian Lagrangian can be written as follows:
\begin{equation}\label{La}
\mathscr{L}^{\rm Abel}_{YM} = 
-\frac{1}{4}F^{\mu\nu}F_{\mu\nu}-\frac{1}{4}F^{\prime\mu\nu}F^\prime _{\mu\nu}\, ,
\end{equation}
where
\begin{eqnarray}\label{new-fs3}
F_{\mu\nu}		&=&	\partial _{\mu}B_{\nu} - \partial _{\nu}B_{\mu} \, , \\ \label{new-fs4}
F^\prime_{\mu\nu}	&=&	\partial _{\mu}B^\prime_{\nu} - \partial _{\nu}B^\prime_{\mu} \, .
\end{eqnarray}
In this field basis, the covariant derivative is:
\begin{equation}\label{cov_der}
D_{\mu}\equiv \partial _{\mu} + ig_S T^{\alpha}G_{\mu}^{\phantom{o}\alpha} 
+ igT^aW_{\mu}^{\phantom{o}a} +ig_1YB_{\mu} +i(\widetilde{g}Y + g_1'Y_{B-L})B'_{\mu}\, .
\end{equation}
The ``pure'' or ``minimal'' $B-L$ model is defined by the condition $\widetilde{g} = 0$, that implies no mixing between the $Z'_{B-L}$ and SM $Z$ gauge bosons.

The fermionic Lagrangian (where $k$ is the
generation index) is given by
\begin{eqnarray} \nonumber
\mathscr{L}_f &=& \sum _{k=1}^3 \Big( i\overline {q_{kL}} \gamma _{\mu}D^{\mu} q_{kL} + i\overline {u_{kR}}
			\gamma _{\mu}D^{\mu} u_{kR} +i\overline {d_{kR}} \gamma _{\mu}D^{\mu} d_{kR} +\\
			  && + i\overline {l_{kL}} \gamma _{\mu}D^{\mu} l_{kL} + i\overline {e_{kR}}
			\gamma _{\mu}D^{\mu} e_{kR} +i\overline {\nu _{kR}} \gamma _{\mu}D^{\mu} \nu
			_{kR} \Big)  \, ,
\end{eqnarray}
 where the fields' charges are the usual SM and $B-L$ ones (in particular, $B-L = 1/3$ for quarks and $-1$ for leptons with no distinction between generations, hence ensuring universality).
  The  $B-L$ charge assignments of the fields
  as well as the introduction of new
  fermionic  right-handed heavy neutrinos ($\nu_R$'s) and a
  scalar Higgs field ($\chi$, with charge $+2$ under $B-L$)  
  are designed to eliminate the triangular $B-L$  gauge anomalies and to ensure the gauge invariance of the theory, respectively.
  Therefore, a $B-L$  gauge extension of the SM gauge group
  broken at the TeV scale requires
  at least one new scalar field and three new fermionic fields which are
  charged with respect to the $B-L$ group.

An important feature of the $Z'$ gauge boson in the $B-L$ model is the
chiral structure of its couplings to fermions: since the $B-L$ charges
do not distinguish between left-handed and right-handed fermions, the
$B-L$ neutral current is purely vector-like, with a vanishing axial
part\footnote{That is, $\displaystyle g_{Z'}^V =
\frac{g_{Z'}^L+g_{Z'}^R}{2}$, $\displaystyle g_{Z'}^A =
\frac{g_{Z'}^R-g_{Z'}^L}{2}=0$, hence $g_{Z'}^R=g_{Z'}^L$.}. As a
consequence, we do not study the asymmetries of the decay
products stemming from $Z'_{B-L}$ bosons, given that their distribution is trivial 
in the peak region which is studied here. However, asymmetries do become
important in the interference region, especially just before the $Z'$ boson 
peak, where the $Z-Z'$ interference will effectively provide an
asymmetric distribution somewhat milder than the case in which there
is no $Z'$ boson. This is a powerful method of discovery and identification
of a $Z'$ and it will be reported on separately~\cite{B-L_observ}.

The scalar and Yukawa sectors of the model play no relevant role in
this analysis, therefore we refer to Ref.~\cite{Basso:2010pe}
for a more detailed overview of the model\footnote{Although 
they do not modify the $Z'$ boson properties significantly, for completeness 
we state the chosen heavy neutrino and the scalar masses and the scalar
mixing angle: $m_{\nu ^1_h}=m_{\nu ^2_h}=m_{\nu ^3_h}=200$ GeV
(value that can lead to interesting phenomenology~\cite{Basso:2008iv}), $m_{h_1}=125$ GeV, $m_{h_2}=450$ GeV and
$\alpha=0.01$ (allowed by a preliminary study on the unitarity bound~\cite{Basso:2010jt}, as well as on the triviality bound~\cite{Basso:2010jm} of the scalar sector).}.

\section{Computational details}\label{sect:comput}

The study we present in this paper has been performed using the
CalcHEP package~\cite{Pukhov:2004ca}. The model under discussion has
previously been implemented in this package using the LanHEP tool~\cite{Semenov:1996es}, as discussed in Ref.~\cite{Basso:2008iv}.

The process we are interested in is di-lepton production. We define
our signal as~$pp\rightarrow \gamma,\,Z,\,Z'_{B-L}\rightarrow \ell^+
\ell^-$ ($\ell=e,\,\mu$), i.e., all possible sources together with
their mutual interferences, and the background as $pp\rightarrow
\gamma,\,Z\rightarrow \ell^+ \ell^-$ ($\ell=e,\,\mu$), i.e., SM
Drell-Yan production (including interference). No other sources of
background, such as $WW$, $ZZ$, $WZ$ or $t\overline{t}$,
have been taken into account. These can be suppressed or/and are insignificant~\cite{Ball:2007zza}.  For both the signal and background, we have
assumed standard acceptance cuts (for both electrons and muons) at the
LHC:
\begin{equation}\label{LHC_cut}
p_T^l > 10~{\rm GeV},\qquad |\eta^l|<2.5\qquad (l=e,\,\mu),
\end{equation} 
and we apply the following requirements on the di-lepton invariant
mass, $M_{ll}$, depending on whether we are considering electrons or
muons.  We distinguish two different scenarios: an ``early'' one (for
$\sqrt{s}=7$ TeV) and an ``improved'' one (for $\sqrt{s}=14$ TeV),
and, in computing the signal significances, we will select a 
window as large as either the width of the $Z'_{B-L}$ boson or twice the
di-lepton mass resolution\footnote{We take the CMS di-electron and
di-muon mass resolutions~\cite{Bayatian:2006zz} as representative of a typical LHC
environment. ATLAS resolutions~\cite{:1999fq} do not differ substantially.},
whichever is the largest. The windows in the invariant mass
distributions respectively are, for the ``early scenario''
\begin{eqnarray}\label{LHC_ris_el}
\mbox{electrons: }\; |M_{ee}-M_{Z'}| &<& 
\mbox{max} \left( \frac{\Gamma_{Z'}}{2},\; \left( 0.02\frac{M_{Z'}}{\rm GeV} \right) {\rm GeV}\; \right),\\ \label{LHC_ris_mu}
\mbox{muons: }\; |M_{\mu\mu}-M_{Z'}| &<& 
\mbox{max} \left( \frac{\Gamma_{Z'}}{2},\; \left( 0.08\frac{M_{Z'}}{\rm GeV} \right) {\rm GeV}\; \right),
\end{eqnarray}
and for the ``improved scenario''
\begin{eqnarray}\label{LHC_ris_el_imp}
\mbox{electrons: }\; |M_{ee}-M_{Z'}| &<& 
\mbox{max} \left( \frac{\Gamma_{Z'}}{2},\; \left( 0.005\frac{M_{Z'}}{\rm GeV} \right) {\rm GeV}\; \right),\\ \label{LHC_ris_mu_imp}
\mbox{muons: }\; |M_{\mu\mu}-M_{Z'}| &<& 
\mbox{max} \left( \frac{\Gamma_{Z'}}{2},\; \left( 0.04\frac{M_{Z'}}{\rm GeV} \right) {\rm GeV}\; \right).
\end{eqnarray}
Our choice reflects the fact that what we will observe is in fact the
convolution between the Gaussian detector resolution and
the Breit-Wigner shape of the peak, and such a convolution will be
dominated by the largest of the two. Our approach is to take the
convolution width exactly equal to the resolution width or to the peak
width, whichever is largest, and to count all the events within this
window.

In the next section we will compare the LHC and Tevatron discovery
reach. For the latter, we have considered typical acceptance cuts
(for both electrons and muons):
\begin{equation}\label{Tev_cut}
p_T^l > 18~{\rm GeV},\qquad |\eta^l|<1\qquad (l=e,\,\mu),
\end{equation}
and the following requirements on the di-lepton invariant mass,
$M_{ll}$, depending on whether we are considering electrons or
muons\footnote{We take the CDF di-electron and di-muon mass
resolution~\cite{Balka:1987ty} as respresentative of a typical Tevatron environment.}:
\begin{eqnarray}\label{Tev_ris}
\mbox{electron: }\; |M_{ee}-M_{Z'}| &<& 
\mbox{max} \left( \frac{\Gamma_{Z'}}{2},\; \left( 0.135 \sqrt{\frac{M_{Z'}}{\rm GeV}}{\rm GeV}+ 0.02\frac{M_{Z'}}{\rm GeV} \right) {\rm GeV}\; \right),\\
\mbox{muons: }\; |M_{\mu\mu}-M_{Z'}| &<& 
\mbox{max} \left( \frac{\Gamma_{Z'}}{2},\; \left( 0.0005\left(\frac{M_{Z'}}{2\, \rm GeV}\right) ^2 \right) {\rm GeV}\; \right).
\end{eqnarray}

In our analysis we also use a definition of the signal
significance $\sigma$, as follows.  In the region
where the number of both signal ($s$) and background ($b$) events is
``large'' (here taken to be bigger than 20), we use a definition of
significance based on Gaussian statistics:
\begin{equation}
{\sigma} \equiv {\it s}/{\sqrt{\it b}}.
\end{equation}
Otherwise, in case of smaller statistics,
we used the Bityukov algorithm~\cite{Bityukov:2000tt}, which basically uses the 
Poisson `true' distribution instead of the approximate Gaussian one.

Finally, as in~\cite{Basso:2008iv,Basso:2009hf}, we used CTEQ6L~\cite{CTEQ_website} as the default Parton Distribution Functions (PDFs),
evaluated at the scale $Q^2=M_{ll}^2$. Only the irreducible SM Drell-Yan
background has been considered. Reducible backgrounds, ISR,
photon-to-electron conversion etc.\ were neglected.

\section{$Z'$ Boson Sector: Results}\label{sect:Zp_disc}
In this section we determine the discovery potential and we present
exclusion plots for the LHC. We use centre-of-mass (CM) energies
of $7$ and $14$ TeV and relevant integrated luminosities.

The experimental constraints come from LEP and the Tevatron.  For the
$B-L$ model, the most recent limit from LEP~\cite{Cacciapaglia:2006pk} is:
\begin{equation}\label{LEP_bound}
\frac{M_{Z'}}{g'_1} \geq 7\; \rm{TeV}\, .
\end{equation}
The most recent limits from the Tevatron for the $Z'_{B-L}$ boson
(from the CDF analyses of Ref.~\cite{Aaltonen:2008vx,Aaltonen:2008ah}
using $2.5\,\mbox{fb}^{-1}$ and $2.3\,\mbox{fb}^{-1}$ of data for
electrons and muons in the final state, respectively), are shown in
table~\ref{mzp-low_bound} (for selected masses and couplings).
\begin{table}[h]
\begin{center}
\begin{tabular}{|c|c||c|c|}
\hline
\multicolumn{2}{|c||}{$p\overline{p}\rightarrow e^+ e^-$} & \multicolumn{2}{|c|}{$p\overline{p}\rightarrow \mu^+ \mu^-$} \\
\hline
 $g_1'$         & $M_{Z'}$ (GeV) & $g_1'$         & $M_{Z'}$ (GeV)\\
\hline
0.042 & 600 & 0.06	& 600        \\  
0.086 & 700 & 0.1	& 750        \\ 
0.115 & 800  & 0.123	& 800        \\ 
0.19 & 900  & 0.2	& 900        \\ 
0.3 & 1000  & 0.3	& 1000        \\
- & - & 0.5	& 1195       \\
\hline
\end{tabular}
\end{center}
\vskip -0.5cm
\caption{Lower bounds on the $Z'$ mass for selected $g_1'$ values in the 
$B-L$ model, at $95\%$ C.L.,
by comparing the collected data of Ref.~\cite{Aaltonen:2008vx,Aaltonen:2008ah} with our theoretical prediction for $p\overline{p}\rightarrow Z'_{B-L} \rightarrow e^+e^-(\mu ^+\mu ^-)$ at the Tevatron. 
\label{mzp-low_bound}}
\end{table}

\begin{figure}[!h]
\centering  
  \includegraphics[angle=0,width=0.55\textwidth ]{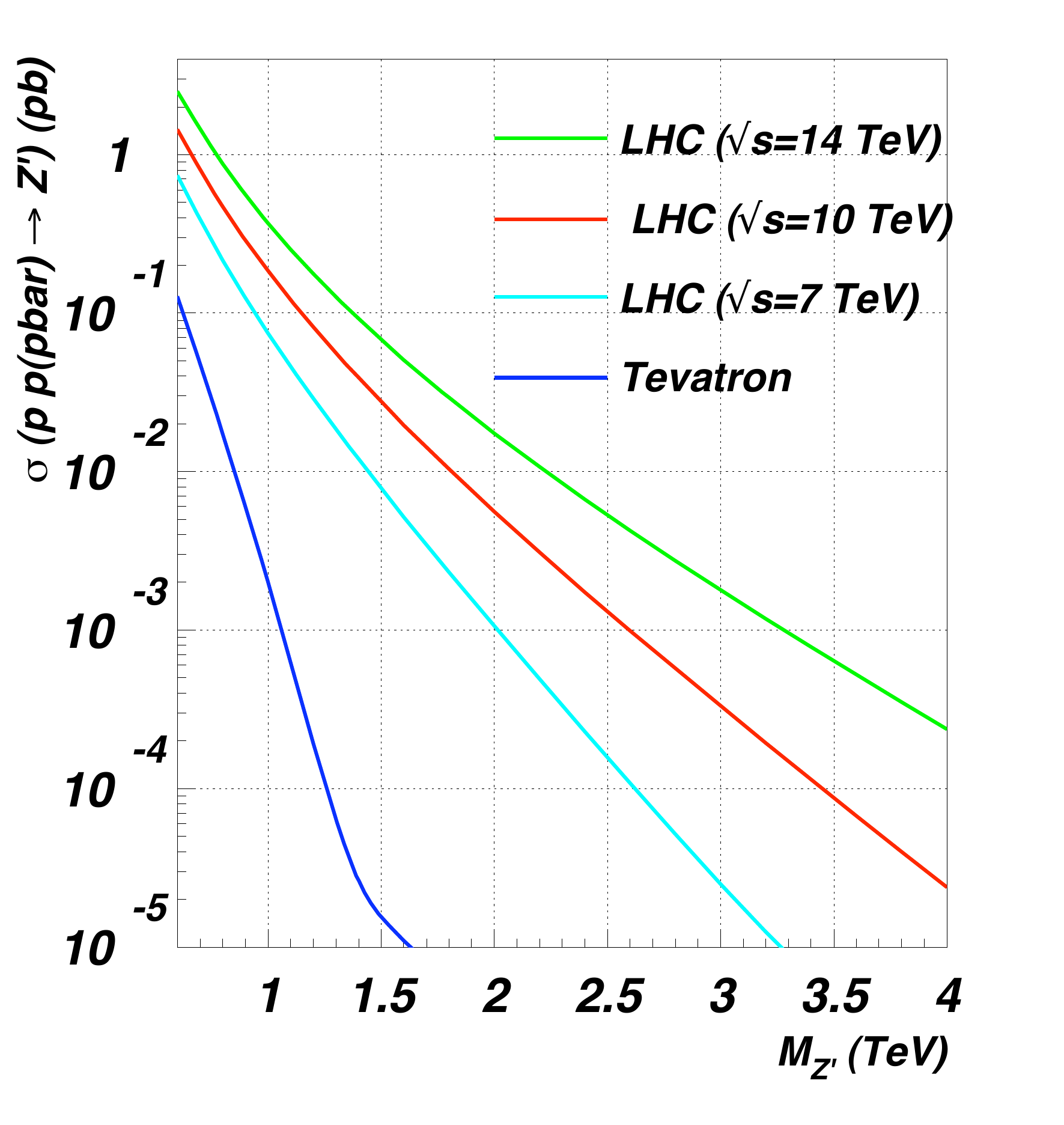}
  \caption{Cross sections for $pp(\overline{p}) \rightarrow
  Z'_{B-L}$ at the Tevatron and at the LHC (for $\sqrt{s}=7,10$ and
  $14$ TeV) for $g'_1=0.1$.}
  \label{Zp_xs}
\end{figure}

The production cross sections for the process $pp(\overline{p})
\rightarrow Z'_{B-L}$ for $g'_1=0.1$ are shown in Fig.~\ref{Zp_xs}.
Note that although at the Tevatron the production cross section is
smaller than at the LHC, the integrated luminosity 
considered here for the LHC at $\sqrt{s}=7$ TeV
(i.e. 1 fb$^{-1}$) is smaller than for the Tevatron
(i.e. 10 fb$^{-1}$).

\subsection{LHC at $\boldsymbol{\sqrt{s}=7}$ TeV}
\begin{figure}[!h]
  \includegraphics[angle=0,width=0.48\textwidth ]{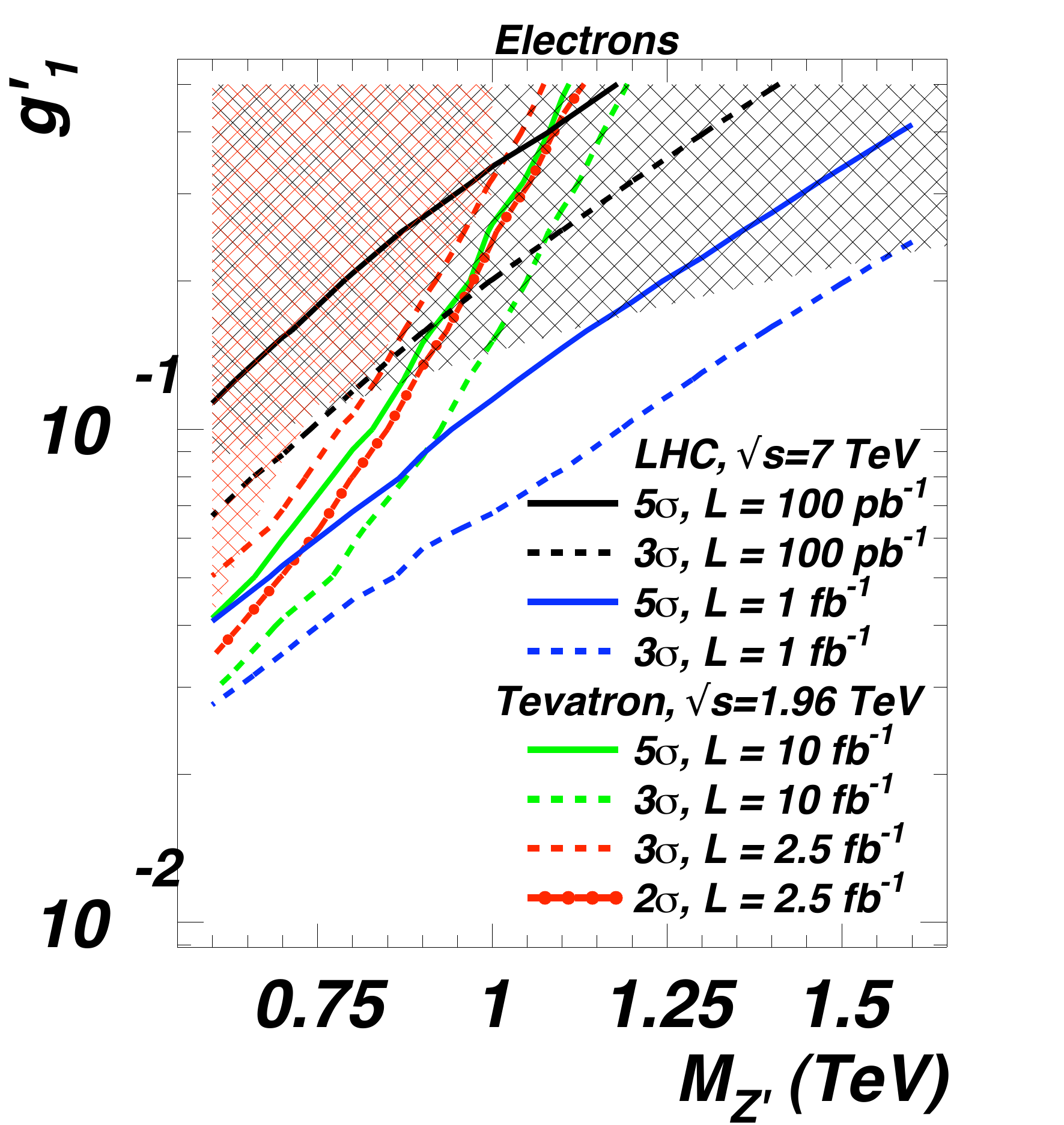}
  \includegraphics[angle=0,width=0.48\textwidth ]{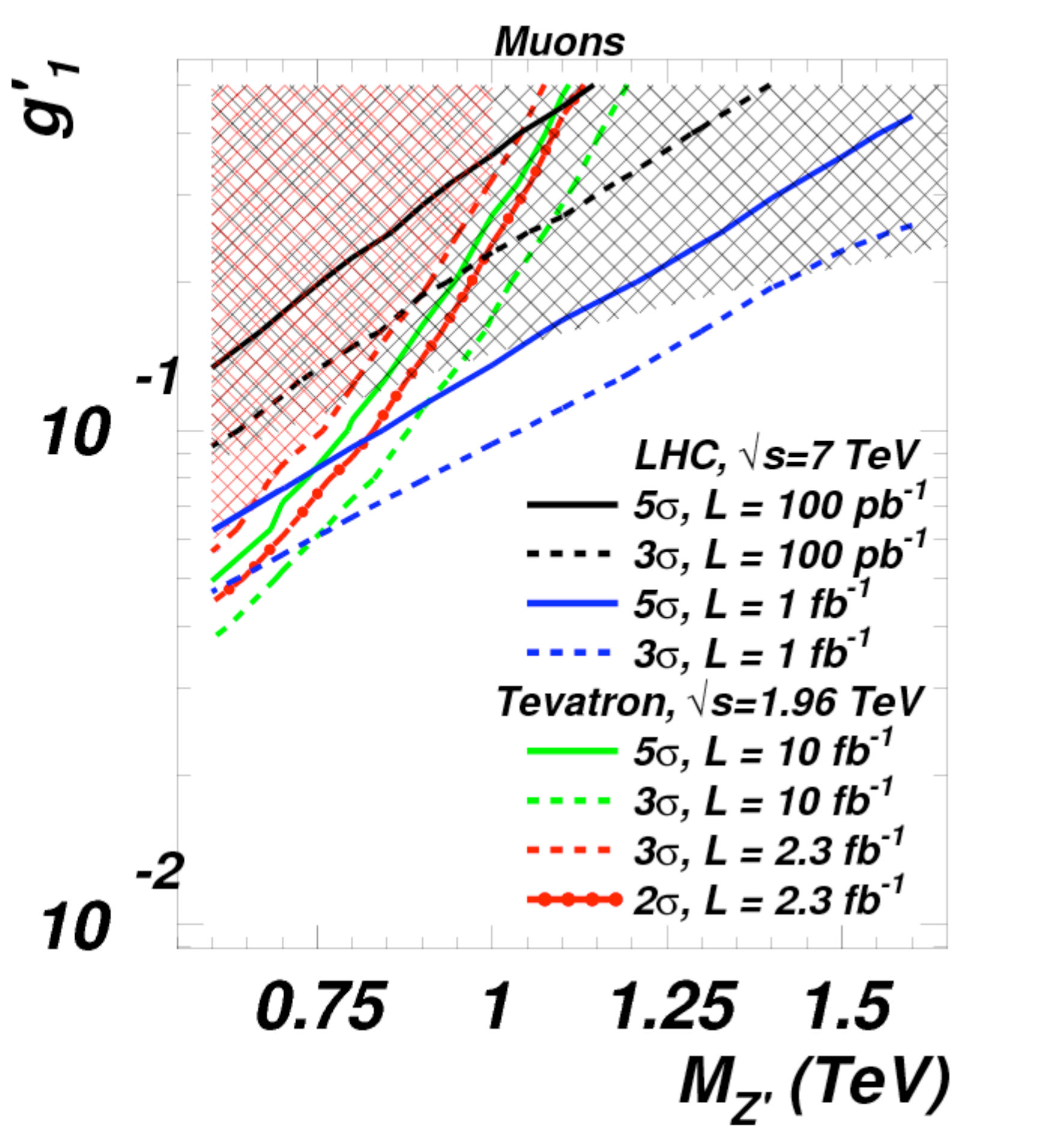}
  \caption{Significance contour levels plotted against $g_1'$ and
$M_{Z'}$ at the LHC for $\sqrt{s}=7$ TeV and $0.1-1$ fb$^{-1}$ and at
the Tevatron ($\sqrt{s}=1.96$ TeV) for (left,
electrons) $2.5-10\,\mbox{fb}^{-1}$ and (right, muons)
$2.3-10\,\mbox{fb}^{-1}$ of integrated luminosity. The shaded areas
correspond to the region of parameter space excluded experimentally
in accordance with Eq.~(\ref{LEP_bound}) (LEP bounds, in black) and
table~\ref{mzp-low_bound} (Tevatron bounds, in red).}
  \label{contour7}
\end{figure}

Initial LHC running will be at a CM energy of $7$ TeV, where the total
integrated luminosity is likely to be of the order of $1$
fb$^{-1}$. Figure~\ref{contour7} shows the discovery potential under
these conditions.  In the same figure we also include for comparison
the Tevatron discovery potential at the integrated luminosities used
for the latest published analyses ($2.5\,\mbox{fb}^{-1}$~\cite{Aaltonen:2008vx} and $2.3\,\mbox{fb}^{-1}$
\cite{Aaltonen:2008ah} for electrons and muons, respectively) as well
as the expected reaches at $\mathcal{L}=10\,\mbox{fb}^{-1}$. Ref.~\cite{Basso:2010pe}
where a comparision to Tevatron data is shown, one can see that our parton level simulation
reproduces experimental conditions reasonably well.

At this stage of the LHC, the Tevatron will still be competitive,
especially in the lower mass region where the LHC requires $1$
fb$^{-1}$ to be sensitive to the same couplings as the Tevatron.
The LHC will be able to probe the $Z'_{B-L}$ for values of the
coupling down to $4-6 \cdot 10^{-2}$ (for electrons and muons
respectively), while the Tevatron can be sensitive down to $4-5 \cdot
10^{-2}$.

\begin{figure}[!h]
  \includegraphics[angle=0,width=0.48\textwidth ]{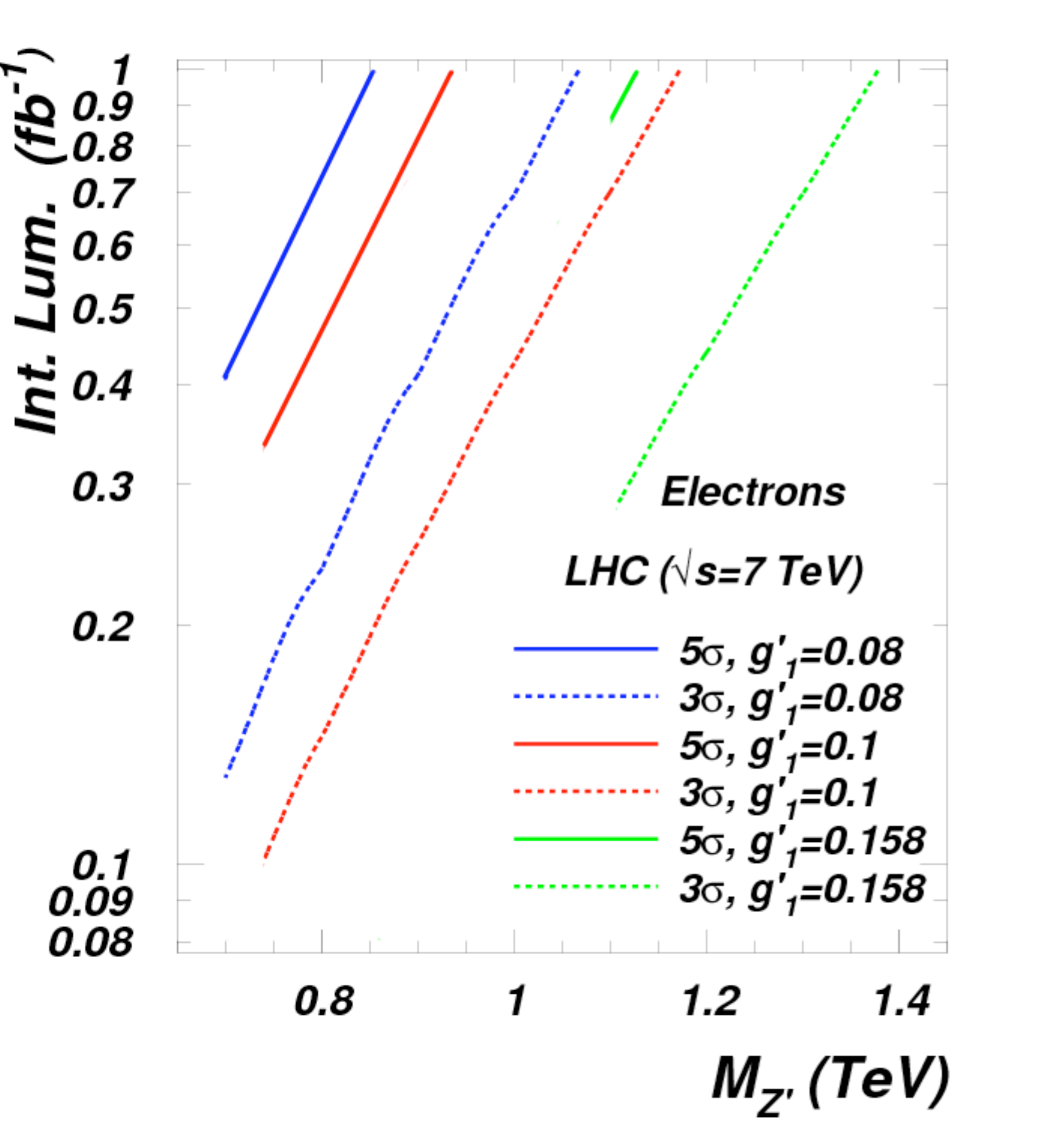}
  \includegraphics[angle=0,width=0.48\textwidth ]{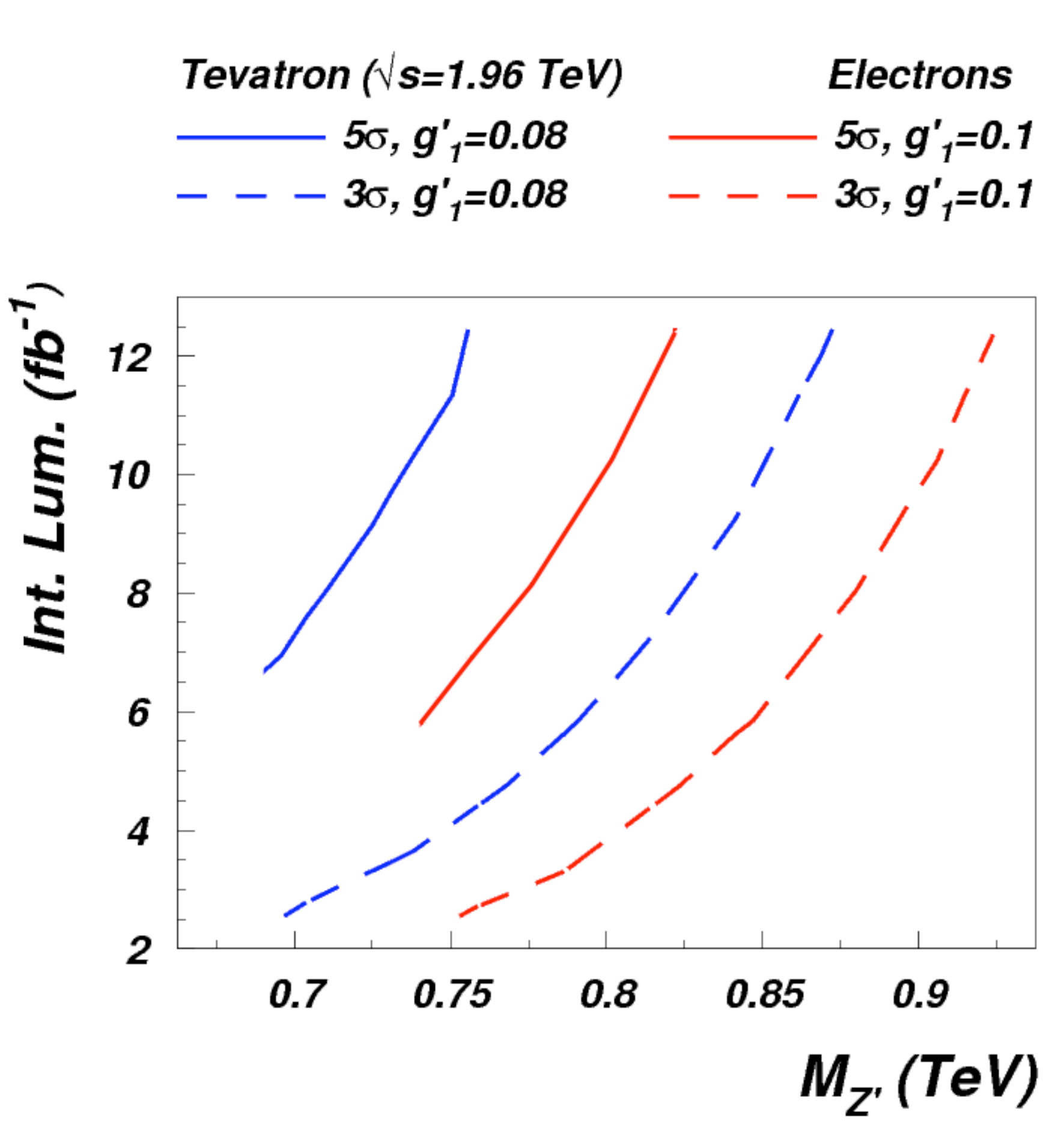}
  \caption{Integrated luminosity required for observations at
  3$\sigma$ and 5$\sigma$ vs $M_{Z'}$ for selected values of
  $g_1'$ for electrons (left) at the LHC for
  $\sqrt{s}=7$ TeV and (right) at the Tevatron
  ($\sqrt{s}=1.96$ TeV). Only combinations of masses and couplings not
  yet excluded are shown. Similar plots for muons in the final state
  are in Ref.~\cite{Basso:2010pe}.}
  \label{lumi_vs_mzp_7TeV}
\end{figure}

Figure~\ref{lumi_vs_mzp_7TeV} shows the integrated luminosity required
for $3\sigma$ evidence and $5\sigma$ discovery as a
function of the $Z'_{B-L}$ boson mass for selected values of the coupling
for the electron final state, both at the LHC and at the Tevatron.  We now
fix some values for the coupling ($g'_1=0.158,\, 0.1,\, 0.08$ for the
LHC analysis, $g'_1=0.1,\, 0.08$ for the Tevatron) and we see what
luminosity is required for discovery at each machine in the case
of electrons in the final state. For muons in the final state, see
Ref.~\cite{Basso:2010pe} for a similar analysis.  For $g'_1=0.1$ the
LHC requires $0.35$ fb$^{-1}$ to be sensitive at $5\sigma$, while
the Tevatron requires $6$ fb$^{-1}$. For the same value of the coupling,
the Tevatron can discover the $Z'_{B-L}$ boson up to $M_{Z'}=825$ GeV,
with $12$ fb$^{-1}$ of data. The LHC can extend the Tevatron
reach up to $M_{Z'}=925$ GeV for $g'_1=0.1$. For $g'_1=0.08$, a
discovery can be made, chiefly with electrons, requiring $0.4(7)$
fb$^{-1}$, for masses up to $850(750)$ GeV at the LHC(Tevatron). Both
machines will be sensitive at $3\sigma$ with much lower integrated
luminosities, requiring roughly $0.15-0.1$ fb$^{-1}$ to probe the
$Z'_{B-L}$ at the LHC and $2.5$ fb$^{-1}$ at the Tevatron, for
$g'_1=0.08-0.1$. Finally, larger values of the coupling, such as
$g'_1=0.158$, can be probed only at the LHC, which provides sensitivity at
$3\sigma$ for masses up to $1.4$ TeV. The lower masses kinematically accessible at 
the Tevatron limit the coupling that may be probed while satisfying the LEP 
constraints.

\begin{figure}[!h]
\centering
  \label{contour7_excl}
  \includegraphics[angle=0,width=0.48\textwidth ]{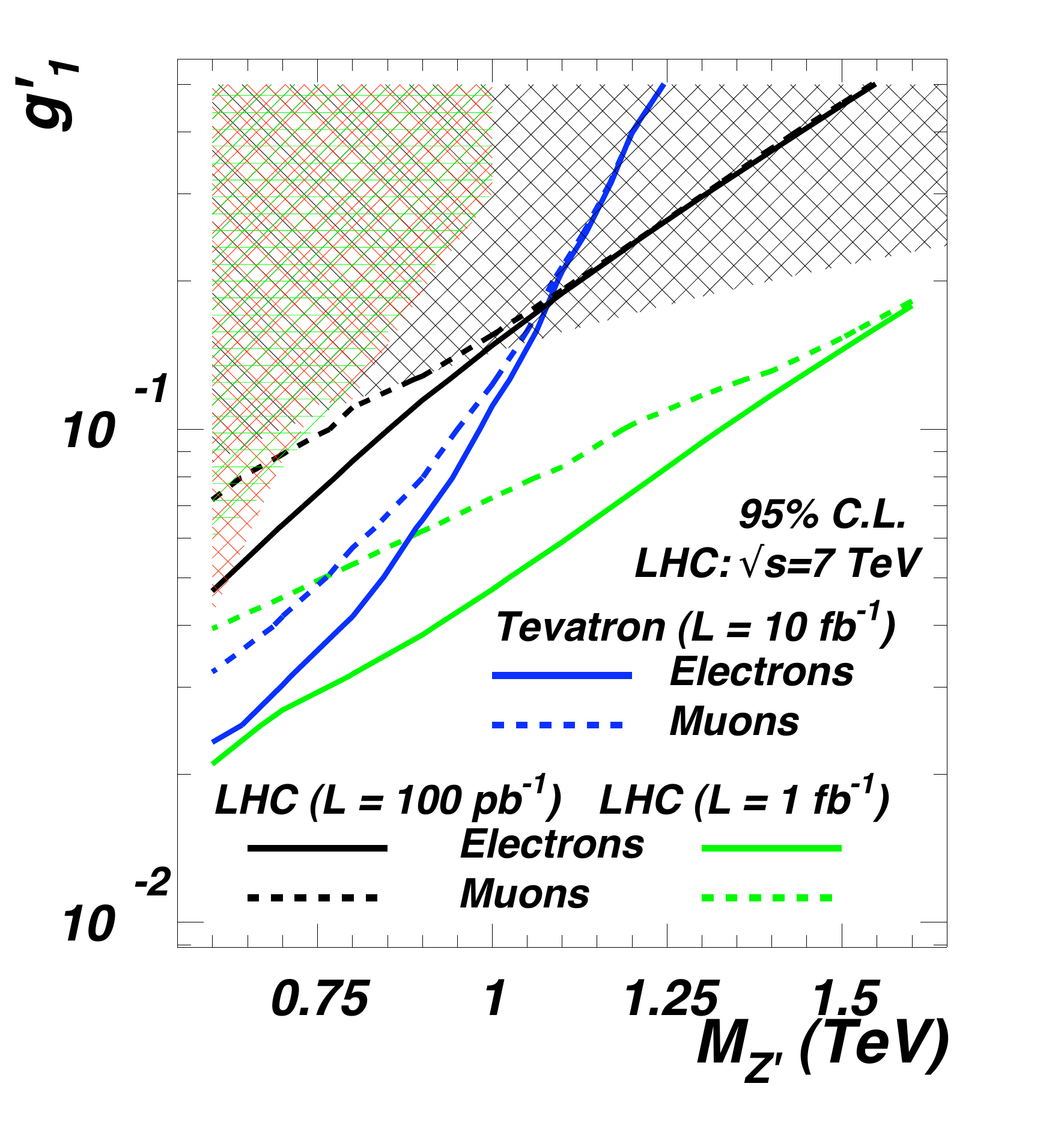}\\
\centering
  \label{lumi7LHC_excl}
  \includegraphics[angle=0,width=0.48\textwidth ]{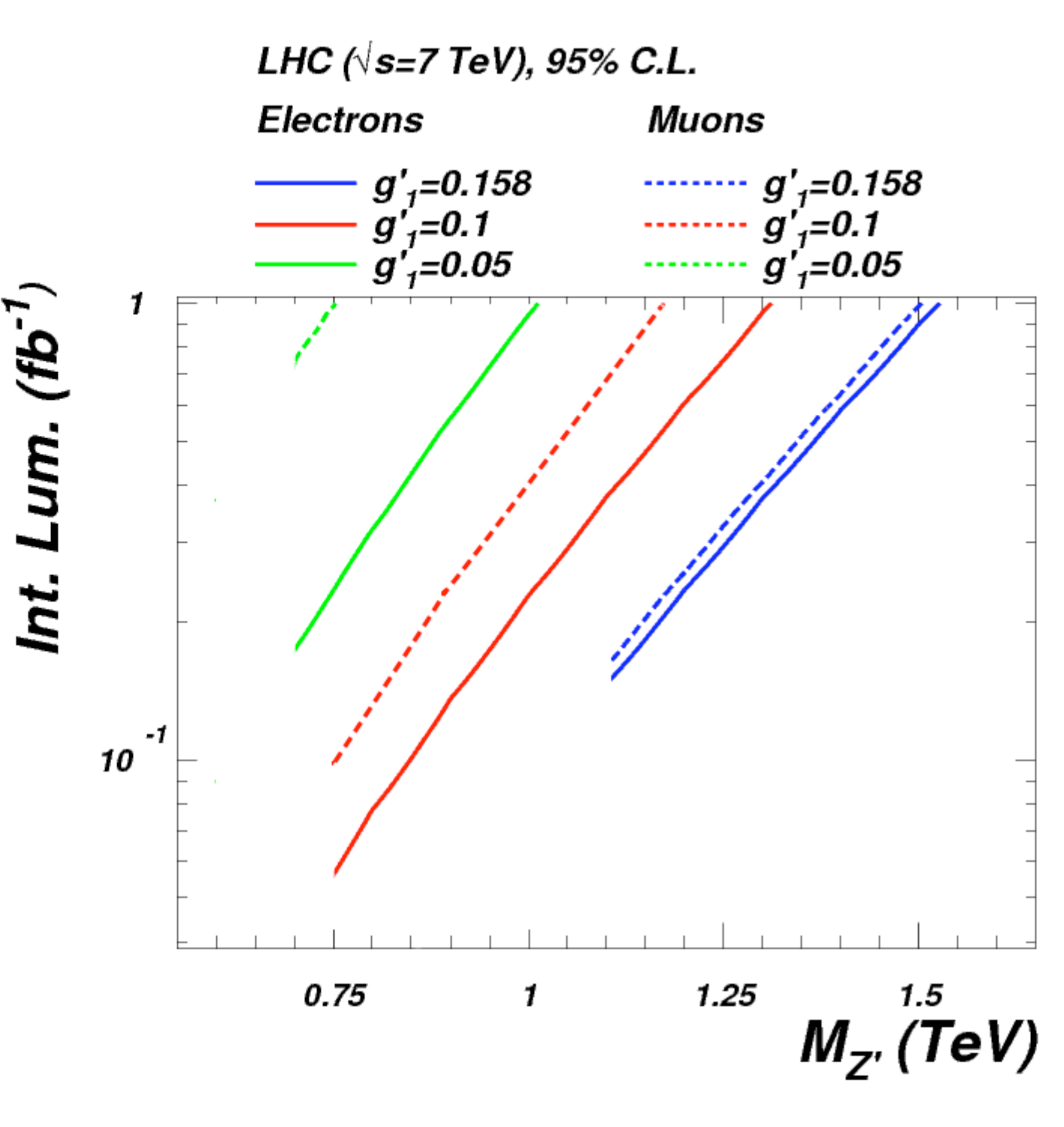}
  \label{lumi7Tev_excl}
  \includegraphics[angle=0,width=0.48\textwidth ]{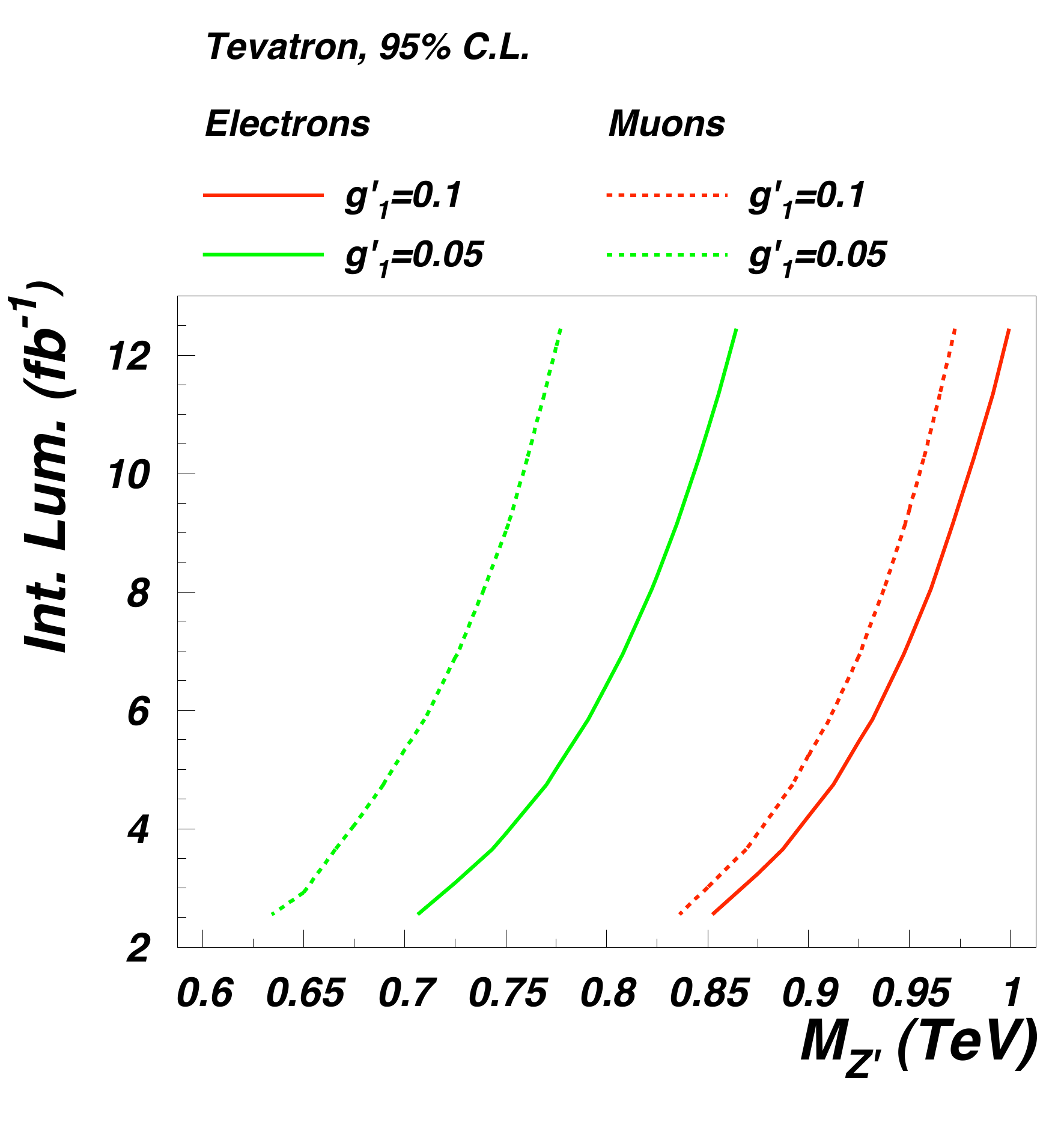}
  \caption{(top) Contour levels for $95\%$
C.L. exclusion in the ($g_1'$,$M_{Z'}$) plane at the LHC for selected
integrated luminosities, and in the (integrated luminosity, $M_{Z'}$) plane
for selected values of $g_1'$ (in which only the allowed
combination of masses and couplings are shown), for
(bottom left) the LHC at $\sqrt{s}=7$ TeV and
(bottom right) the Tevatron ($\sqrt{s}=1.96$ TeV), for both
electrons and muons.  The shaded areas and the allowed $(M_{Z'},g'_1)$
shown are in accordance with Eq.~(\ref{LEP_bound}) (LEP bounds, in
black) and table~\ref{mzp-low_bound} (Tevatron bounds, in red for
electrons and in green for muons).}
  \label{excl_7}
\end{figure}

If no evidence for a signal is found at this energy and luminosity
configuration of the LHC, $95\%$~C.L. exclusion limits can be
derived. We present here exclusion plots for the LHC as well as the
expected exclusions at the Tevatron for $\int{\mathcal{L}}=10$
fb$^{-1}$.  We start by looking at the $95\%$ C.L. limits presented in
Fig.~\ref{excl_7} for the Tevatron and for this stage of the LHC
(for $10$ fb$^{-1}$ and $1$ fb$^{-1}$ of integrated luminosities,
respectively).

One can see that the different resolutions imply that the limits
derived using electrons are always more stringent than those derived
using muons in excluding the $Z'_{B-L}$ boson.  As for the discovery reach,
the Tevatron is also competitive in setting limits, especially in the
lower mass region.  In particular, using electrons at the Tevatron for
$10$ fb$^{-1}$, the $Z'_{B-L}$ can be excluded for values of the
coupling down to $0.02$ ($0.03$ for muons) for
$M_{Z'}=600$ GeV. For the LHC to set the same exclusion limit the same mass, 
$1$ fb$^{-1}$ of integrated luminosity is
required, allowing the exclusion of $g'_1>0.02(0.04)$ using
electron(muons) in the final state. For the same integrated
luminosity, the LHC has much more scope in excluding a heavier
$Z'_{B-L}$ boson, for $M_{Z'}>750$ GeV.

For a coupling of $0.1$, the $Z'_{B-L}$ boson can be excluded up to
$1.3(1.15)$ TeV at the LHC considering electrons(muons) for $1$
fb$^{-1}$, and up to $975(900)$ GeV at the Tevatron for $10$ fb$^{-1}$
of data. For $g'_1=0.05$, the LHC when looking at muons will require
$800$ pb$^{-1}$ to start improving the current available limits, while
with $150$ pb$^{-1}$ it can set limits on $g'_1=0.158$, out of the
reach of Tevatron. It will ultimately be able to exclude $Z'_{B-L}$ up to
$M_{Z'}=1.5$ TeV for $1$ fb$^{-1}$ (both with electrons and muons).

\subsection{LHC at $\boldsymbol{\sqrt{s}=14}$ TeV}
We consider here the performance at the design centre of mass energy of
$\sqrt{s}=14$ TeV for an integrated luminosity $\int
\mathcal{L}=100$ fb$^{-1}$. We will present plots for the discovery
potential only for the $Z'_{B-L}\rightarrow e^+e^-$ channel. Similar
plots for the muon channel can be found in
Ref.~\cite{Basso:2010pe}. As before, exclusion plots will be
presented for both electrons and muons in the final state.

\begin{figure}[!h]
\centering
  \includegraphics[angle=0,width=0.48\textwidth ]{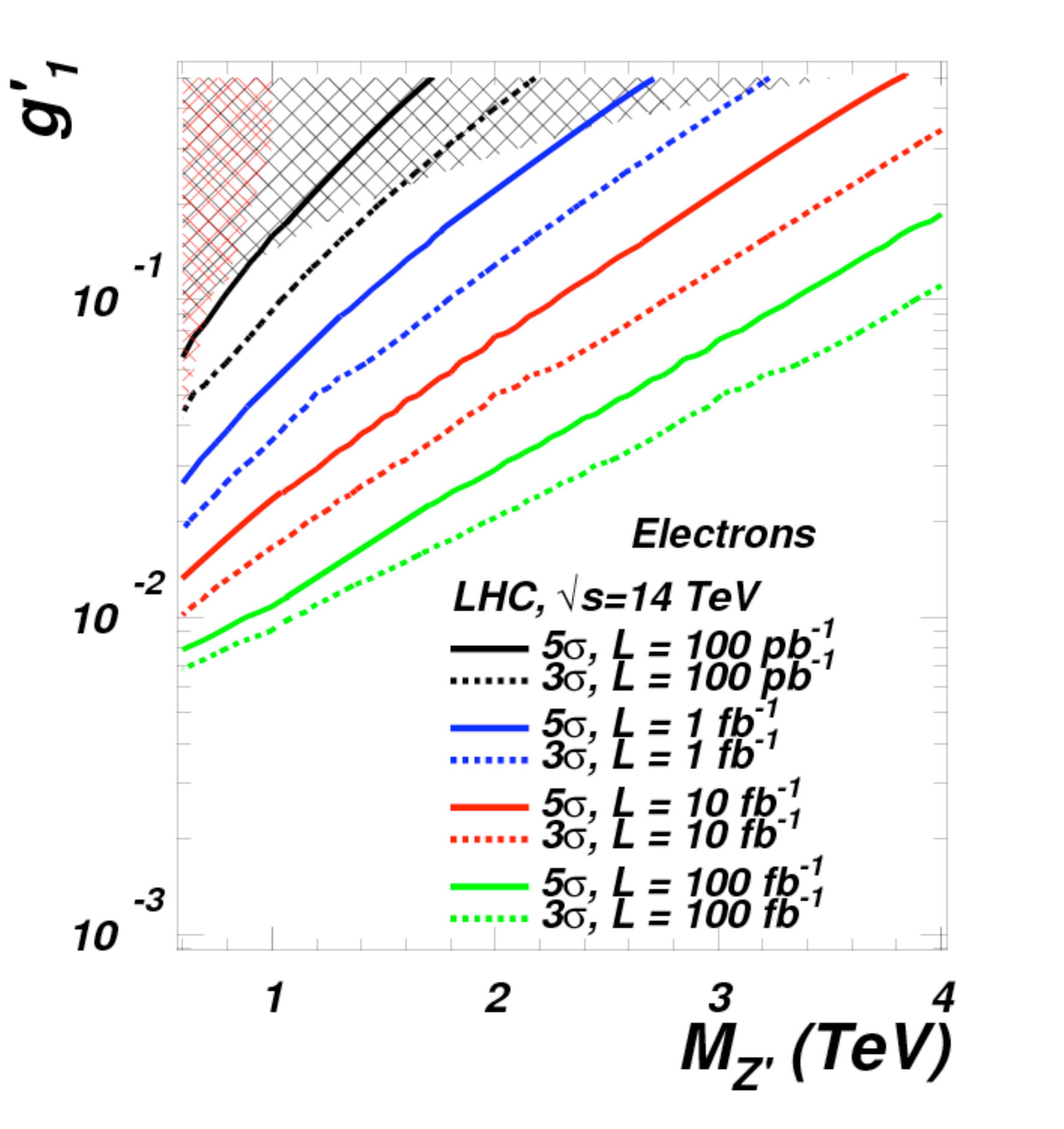}
  \includegraphics[angle=0,width=0.48\textwidth ]{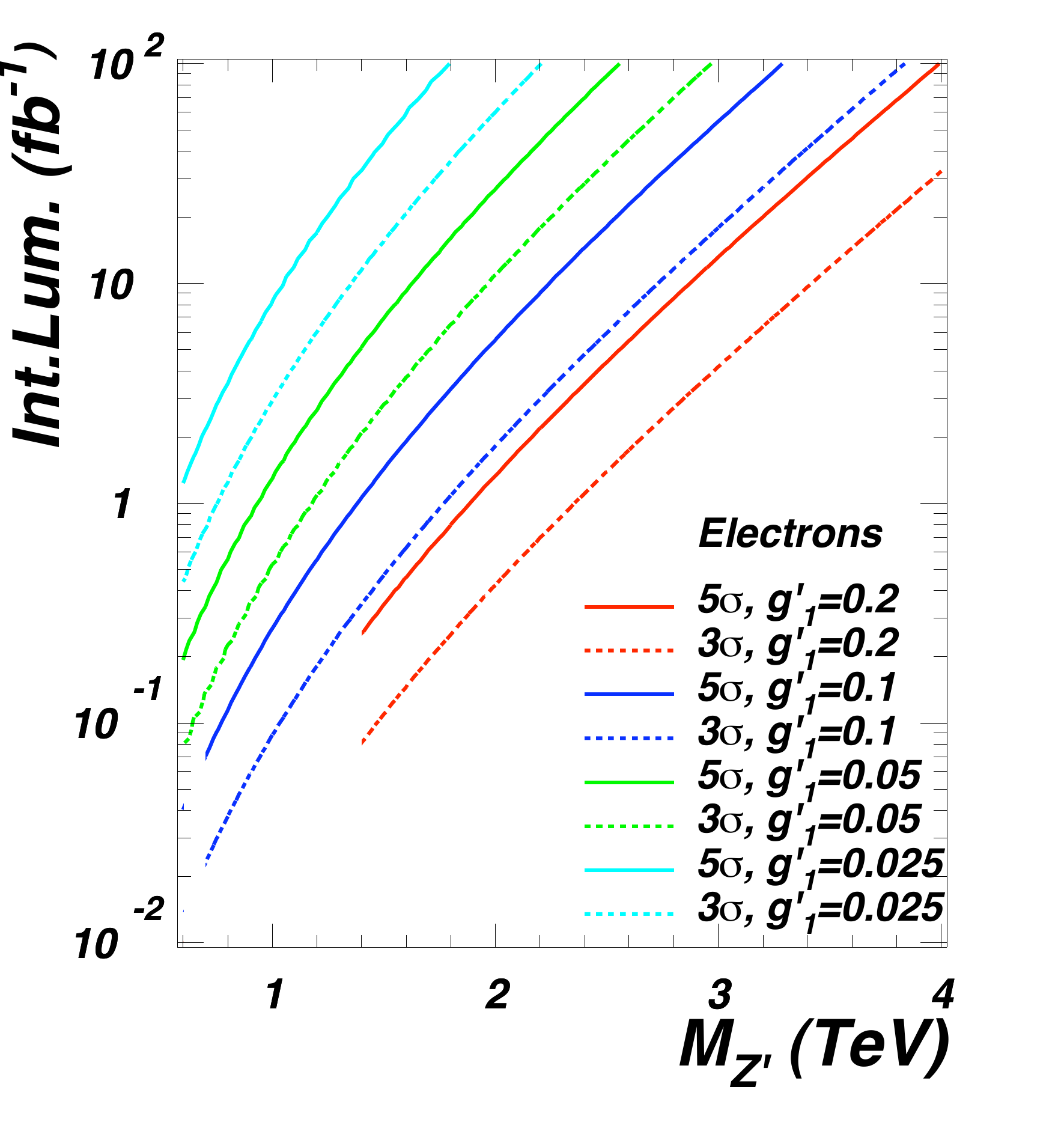}
  \caption{(left) Significance contour levels
  in the ($g_1'$,$M_{Z'}$) plane for several integrated
  luminosities and (right) in the (integrated luminosity,
  $M_{Z'}$) plane for selected values of $g_1'$ at the LHC
  for $\sqrt{s}=14$ TeV for electrons. The shaded areas and the
  allowed $(M_{Z'},g'_1)$ shown are in accordance with
  Eq.~(\ref{LEP_bound}) (LEP bounds, in black) and
  table~\ref{mzp-low_bound} (Tevatron bounds, in red). Similar plots
  for muons can be found in~\cite{Basso:2010pe}.}
  \label{contour14}
\end{figure}

Figure~\ref{contour14}~(left) shows the discovery potential for the
$Z'_{B-L}$ boson under these conditions, while
Fig.~\ref{contour14}~(right) shows the integrated luminosity
required for $3\sigma$ evidence as well as for $5\sigma$ discovery
as a function of the $Z'_{B-L}$ boson mass for selected values of the
coupling at $\sqrt{s}=14$ TeV. We consider the integrated luminosity
in the range between $10$ pb$^{-1}$ up to $100$ fb$^{-1}$.
After a number of years of data analysis, the performance of the detector will be well
understood. We therefore use the resolutions for both electrons and
muons quoted in Eqs.~(\ref{LHC_ris_el_imp}) and
(\ref{LHC_ris_mu_imp}), respectively.

From Fig.~\ref{contour14}~(left), we can see that the LHC at
$\sqrt{s}=14$ TeV will start probing a completely new region of the
parameter space for $\int \mathcal{L} \geq 1$ fb$^{-1}$.  For $\int
\mathcal{L} \geq 10$ fb$^{-1}$ $Z'_{B-L}$ gauge boson can be
discovered up to masses of $4$ TeV and for couplings as small as
$0.01$.  At $\int \mathcal{L} = 100$ fb$^{-1}$, the coupling can be
probed down to values of $8\,\cdot 10^{-3}$. The mass region that
can be covered extends towards $5$ TeV.

As before, Fig.~\ref{contour14}~(right) shows the integrated
luminosity required for $3(5)\sigma$ evidence(discovery) of the
$Z'_{B-L}$ boson as a function of its mass, for selected values of the
coupling. We explore luminosities in the range from $10$ pb$^{-1}$ to
$100$ fb$^{-1}$. However, only the configuration with $g'_1=0.1$ can
be probed with very low luminosity, and $80(200)$ pb$^{-1}$ is
required to enable sensitivity (at $3(5)\sigma$) to both $0.05$ and
$0.2$ values of the coupling. It is worth emphasising here  that the
first couplings that will start to be probed at the LHC are those
around $g'_1=0.1$, since the current experimental constraints are
looser for this value of the coupling.

At a given mass, the superior resolution in the electron w.r.t.\ the
muon channel results in greater sensitivity to smaller couplings.
For $M_{Z'}=600$ GeV, the LHC at $\sqrt{s}=14$ TeV requires
$0.2(1.2)$ fb$^{-1}$ to be sensitive at $5\sigma$ to a value of the
coupling of $0.05(0.025)$, in the electron channel. A measure of the
$Z'_{B-L}$ boson width is also possible over a range of masses. For a
comparison to the case with muons in the final state, we refer to
Ref.~\cite{Basso:2010pe}.

\begin{figure}[!h]
 \centering
   \includegraphics[angle=0,width=0.48\textwidth ]{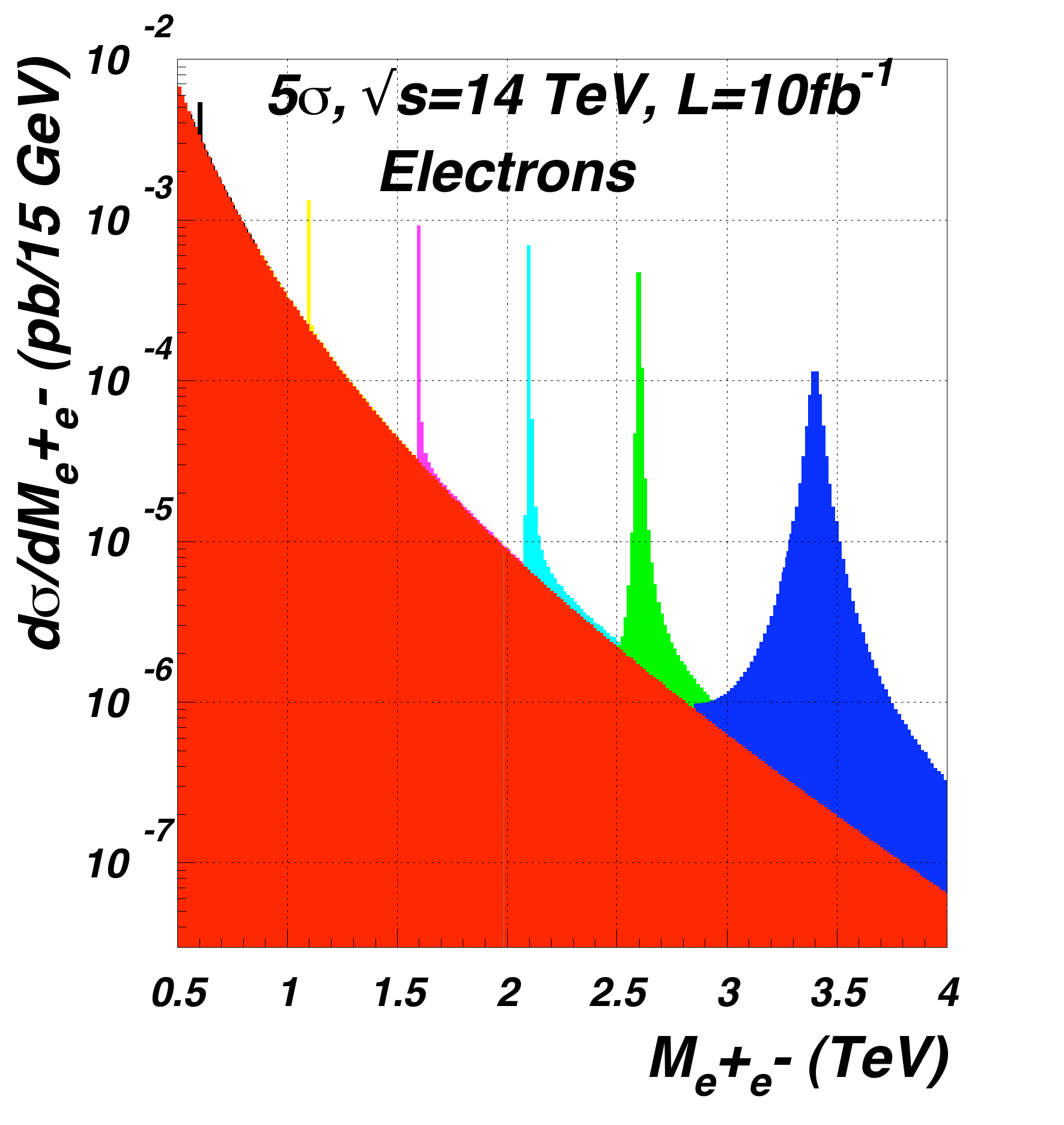}
  \caption{ \label{14TeV_5sigma_el}
$\frac{d\sigma}{dM_{ll}}(pp\rightarrow \gamma
  ,Z,Z'_{B-L} \rightarrow e ^+e ^-)$ for several masses and couplings
  ($M_{Z'}/$TeV, $g'_1$, $\Gamma _{Z'}/$GeV): ($0.6$, $0.009$,
  $0.009$), ($1.1$, $0.02$, $0.09$), ($1.6$, $0.04$, $0.53$), ($2.1$,
  $0.07$, $2.2$), ($2.6$, $0.12$, $7.9$) and ($3.4$, $0.3$, $61$),
($\sqrt{s}=14$ TeV), using 15 GeV
  binning. Notice that the asymmetry of the peaks is the result of our
  choice to consider here the full interference structure.}
\end{figure}

Figure~\ref{14TeV_5sigma_el} shows a pictorial representation of the
$Z'_{B-L}$ properties (widths and cross sections) for selected
benchmark points on the $5\sigma$ lines for $10$ fb$^{-1}$ of data at
$\sqrt{s}=14$ TeV, plotting the di-electron invariant mass to which
just the cuts of Eq.~(\ref{LHC_cut}) have been applied (without
selecting any mass window).

\begin{figure}[!h]
\centering
  \includegraphics[angle=0,width=0.48\textwidth ]{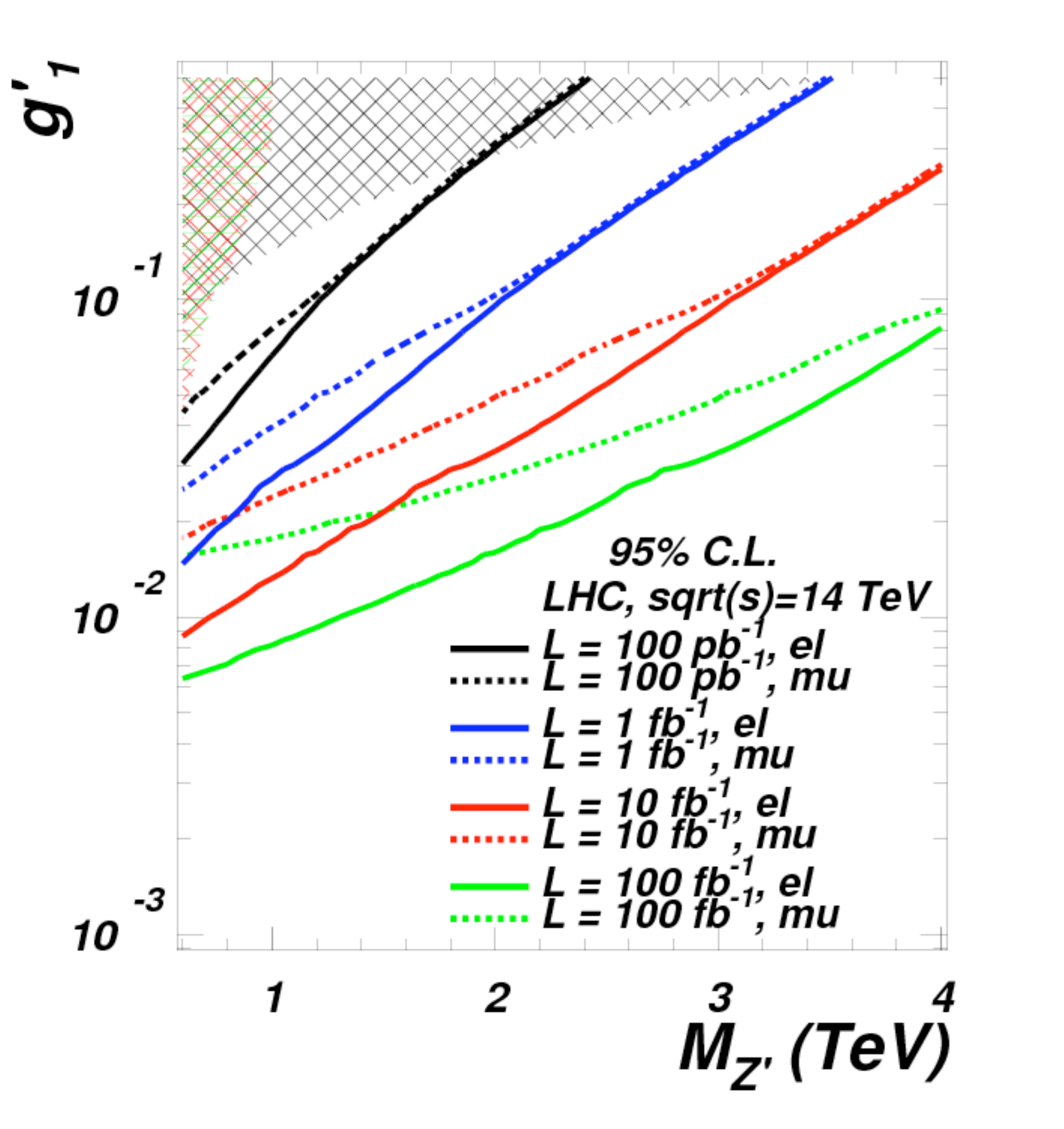}
  \includegraphics[angle=0,width=0.48\textwidth ]{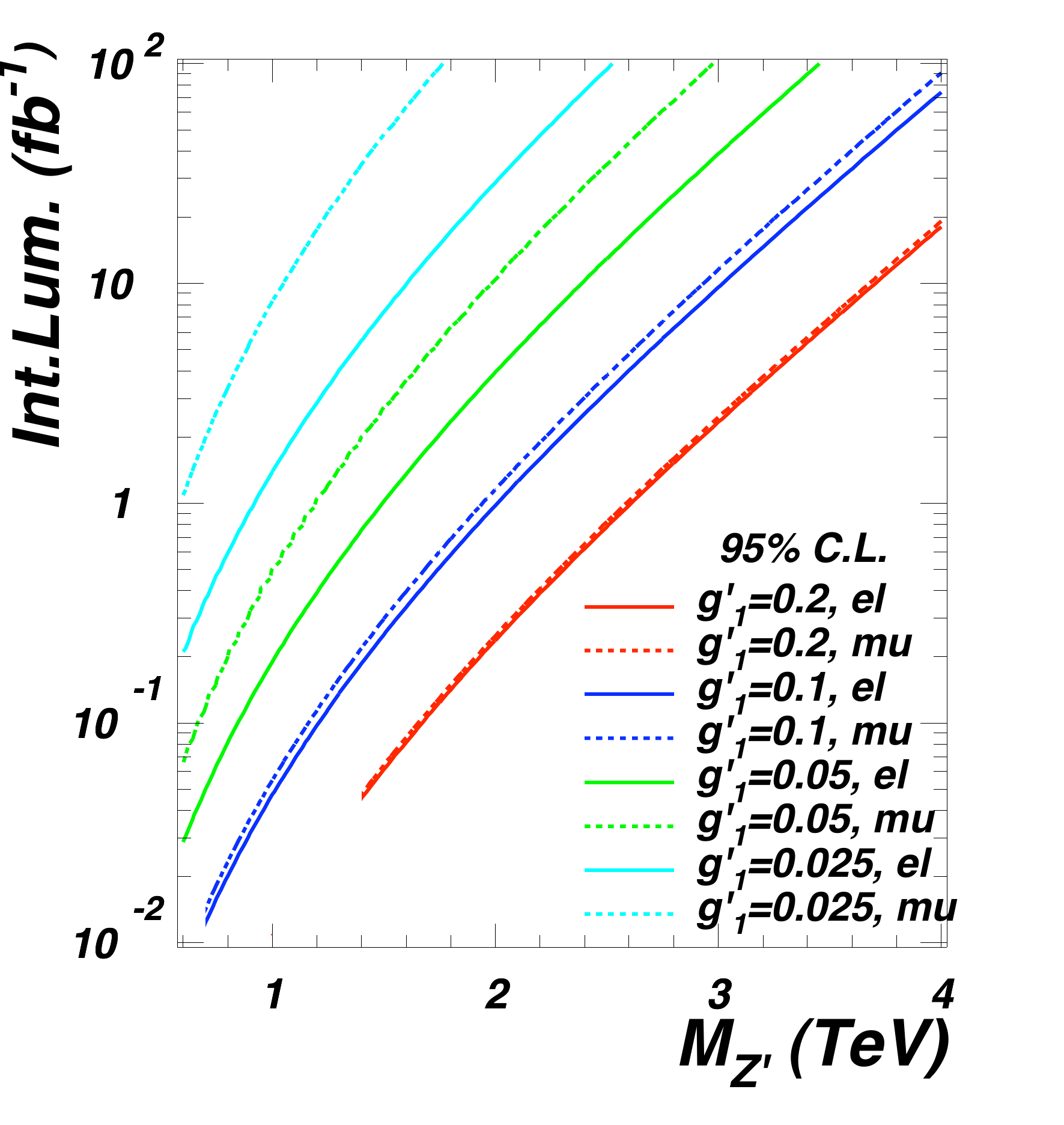}
  \caption{  \label{excl_14}
  (left) Contour levels for
$95\%$~C.L. exclusion in the ($g'_1$, $M_{Z'}$) plane at the LHC for
selected integrated luminosities and (right) in the (integrated
luminosity, $M_{Z'}$) plane for selected values of $g_1'$ (in
which only the allowed combination of masses and couplings are shown),
for $\sqrt{s}=14$ TeV, for both electrons and muons.  The shaded areas
and the allowed $(M_{Z'},g'_1)$ shown are in accordance with
Eq.~(\ref{LEP_bound}) (LEP bounds, in black) and
table~\ref{mzp-low_bound} (Tevatron bounds, in red for electrons and
in green for muons).}
\end{figure}

As before, if no evidence for a signal is found at this energy and
luminosity configuration of the LHC, $95\%$~C.L. exclusion limits can
be derived.  Due to the improved resolutions for both electrons and
muons, they have very similar exclusion powers for couplings
$g'_1\gtrsim 0.1$, therefore setting similar constraints (for details,
see Ref.~\cite{Basso:2010pe}). Depending on the amount of data that is
collected, several maximum bounds can be set (see
Fig.~\ref{excl_14})~(left): e.g.\ for $10$ fb$^{-1}$ of data, the
LHC at $14$ TeV can exclude masses, at the $95\%$~C.L., up to roughly
$5$ TeV for a value of the coupling\footnote{This is the largest
allowed value for the consistency of the model up to a scale
$Q=10^{16}$ GeV, from a Renormalisation Group (RG) analysis of the
gauge sector of the model~\cite{BL_master_thesis,Basso:2010jm}.}
$g'_1=0.5$. For $100$ fb$^{-1}$ and for the same value of the
coupling, the LHC can exclude masses, at the $95\%$ C.L., up to
roughly $6$ TeV. For $10$ fb$^{-1}$ it will be possible to exclude a
$Z'_{B-L}$ boson of $M_{Z'}=600$ GeV if the coupling is greater than
$1.8\,\cdot 10^{-2}(9\,\cdot 10^{-3})$ for muons(electrons) and values
of the coupling greater than $1.5\,\cdot 10^{-2}(7\,\cdot 10^{-3})$ 
for an integrated luminosity of $100$ fb$^{-1}$.
Figure~\ref{excl_14}~(right) shows the integrated luminosity that is
required to excluded a certain $Z'_{B-L}$ mass for fixed values of the
coupling. For $g'_1 \geq 0.1$ the same limits are obtained for
electrons and muons. An integrated luminosity of $10$ fb$^{-1}$ is
required to exclude a $Z'_{B-L}$ boson mass up to $3.6$ TeV for $g'_1=0.2$
and $40$ fb$^{-1}$ reduces this to $g'_1=0.1$.  For an integrated
luminosity of $10$ fb$^{-1}$ the LHC experiments will be able to exclude a $Z'_{B-L}$
for masses up to $3.0$ TeV for $g'_1=0.1$, $2.4(2.0)$ TeV for
$g'_1=0.05$ and $1.6(1.0)$ TeV for $g'_1=0.025$, when considering the
decay into electrons(muons). With and integrated luminosity of 
$100$ fb$^{-1}$ of data more stringent bounds can be derived: for
$g'_1=0.05(0.025)$ the $Z'_{B-L}$ boson can be excluded for masses up to
$3.4(2.5)$ TeV in the electron channel, and up to $3.0(1.8)$ TeV in the muons channel.

\section{Conclusions}\label{sect:conc}

We have presented the discovery potential for the $Z'$ gauge
boson of the $B-L$ minimal extension of the SM at the LHC for CM
energies of $\sqrt{s}=7$ and $14$ TeV, using the integrated
luminosities expected at each stage. This has been done for both the
$Z'_{B-L}\rightarrow e^+e^-$ and $Z'_{B-L}\rightarrow \mu ^+\mu ^-$
decay modes, and includes the most up-to-date constraints from LEP
and the Tevatron.

A general feature is that greater sensitivity to the $Z'_{B-L}$
resonance is provided by the electron channel. At the LHC this has
better energy resolution than the muon channel. A further consequence
of the better resolution of electrons is that an estimate of the gauge
boson width would eventually be possible for smaller values of the
$Z'_{B-L}$ mass than in the muon channel.  Limits from existing data
imply that the first couplings that will start to be probed at the LHC
are those around $g'_1=0.1$. Increased luminosity will enable both
larger and smaller couplings to be probed.

Our comparison shows that, for an integrated luminosity of $10$
fb$^{-1}$, the Tevatron is still competitive with the LHC in the small
mass region, being able to probe the coupling at the level of
$5\sigma$ down to a value of $0.04(0.05)$ using electrons(muons). The
LHC will start to be competitive in such a region only for integrated
luminosities close to $1$ fb$^{-1}$ at $\sqrt{s}=7$ TeV. At
$\sqrt{s}=7$ TeV the mass reach will be extended from the Tevatron
value of $M_{Z'}=750$ GeV up to $1.2(1)$ TeV for electrons(muons).

When the data from the high energy runs at the LHC becomes available,
the discovery reach of $Z'_{B-L}$ boson will be extended towards very
high masses and small couplings in regions of parameter space well
beyond the reach of the Tevatron and comparable in scope with those
accessible at a future LC~\cite{Basso:2009hf}.

If no evidence is found at any energies, $95\%$~C.L. limits can be
derived, and, given their better resolution, the bounds from electrons
will be more stringent than those from muons, especially at smaller
masses.

While this work was in progress, other papers dealing with the
discovery power at the LHC for the $Z'_{B-L}$ boson appeared, for CM
energies of $7$ TeV~\cite{Salvioni:2009mt} and $14$ TeV~\cite{Emam:2008zz}, as well as for other popular $Z'$ boson models. Our
results broadly agree with those therein.


\section*{Acknowledgements} 

LB thanks Muge Karagoz Unel and Ian Tomalin for useful discussions.
SM is financially supported in part by 
the scheme `Visiting Professor - Azione D - Atto Integrativo tra la 
Regione Piemonte e gli Atenei Piemontesi'.

%% file: Moreau/Moreau.tex
\chapter{Single custodian production in warped extra dimensional models}
\label{chapter:BrooijmansEtAl}

{\it S.~Gopalakrishna, G.~Moreau and R.K.~Singh}

\begin{abstract}
We examine the single production of heavy fermions at the LHC and show in particular the possible 
importance of some new types of processes,
namely single production of the heavy fermion in association with a standard model gauge boson
or a Higgs boson. 
The theoretical framework 
is the Randall-Sundrum scenario, motivated by the gauge hierarchy problem, with a bulk custodial 
symmetry. In this context, the heavy fermion considered in the Kaluza-Klein (KK) excitation 
of a fermion that does not have a zero mode. The location 
along the fifth dimension (affecting the single production) are selected such that they generate 
the masses of the third generation quarks.
From both the theoretical and phenomenological sides, the studied KK quarks can be lighter
than the KK excitations of gauge bosons which makes their potential discovery at the LHC easier,
as illustrated by the cross sections we obtain numerically.
\end{abstract}

\section{Introduction}

Recent alternatives to supersymmetric scenarios, like extra dimension theories, composite Higgs
and little Higgs models (as well as twin Higgs and fourth generation models), predict the existence 
of additional heavy fermions. Such fermions, e.g. exotic quarks, could be directly produced at the LHC providing
a clear discovery of new physics underlying the Standard Model (SM). In particular, the single 
production of such a heavy fermion is favored w.r.t. pair production from the point of view of the phase space 
\footnote{The {\it pair} production of exotic quarks has been studied in the
warped extra dimension scenario with a custodial symmetry~\cite{Dennis:2007tv,deSandes:2008yx} 
(within the gauge-Higgs unification context~\cite{Carena:2007tn}) or similarly within
their dual composite Higgs description~\cite{Brooijmans:2008se}. See also 
Ref.~\cite{Burdman:2008qh} for the case of a strongly coupled fourth generation and
Ref.~\cite{AguilarSaavedra:2005pv,AguilarSaavedra:2009es} for more general approaches (using jet mass~\cite{Skiba:2007fw}).}.

In the present work, we propose a systematic study of the various channels of single $b'$ (new quark
with an electric charge of $-1/3$) production at LHC. 
The exhaustive list of possible elementary processes of type $2\to 2$ body and $2\to 3$ body is: 
$qq \to tb'$, $qq \to bb'$, $qg \to qtb'$, $qg \to qbb'$, $bg \to b'Z$, $bg \to b'h$, $QQ/gg \to bb'Z$, $QQ/gg \to bb'h$,
$qb \to qb'Z$, $qb \to qb'h$, $qq/gg  \to tb'W$
and $qb  \to qb'W$, where $q$ stands for any SM quark except for the bottom and top quarks which are denoted $b$ and $t$
respectively ($QQ$ denotes either the initial state $qq$ or $bb$)
\footnote{The possible $b'$ decay channels, whose branching ratios depend on the considered model, are $b' \to b Z$ and $b' \to t W$.}. 
The theoretical framework we consider is precisely defined: it is the Randall-Sundrum (RS) scenario \cite{Randall:1999ee}
of warped extra dimensions with an extended bulk custodial symmetry. 
This symmetry gives rise to new fermions like the $b'$, 
called the custodians, which appear in the extended gauge multiplets. These custodians are pure Kaluza-Klein (KK)
excitations without zero-modes, due to specific boundary conditions. 

The well-known RS scenario is motivated by the
gauge hierarchy problem and the custodial symmetry allows to satisfy the ElectroWeak Precision Test (EWPT) constraints
for KK gauge boson masses in the vicinity of the TeV scale. 
The RS framework is also attractive as a flavor model
and we will carefully consider quark locations, reproducing quite precisely the $b$ and $t$ masses,
which are crucial for the single $b'$ production. In particular, in this flavor framework, the 
$b'$ quark, whose mass is controlled by the location of the right-handed $t_R$ (being in the same multiplet under
the custodial extension), tends to be particularly light due to the large top mass $m_t$
which also explains the motivation for analyzing its single production. 

The above type of single heavy fermion 
production processes with an EW gauge boson or Higgs field in the final state were never studied before. 
In contrast, the other single processes already considered have been within 
generic approaches~\cite{Atre:2008iu} or
within the different theoretical contexts of the
composite Higgs models~\cite{Contino:2008hi,Mrazek:2009yu}, the little Higgs scenario~\cite{Azuelos:2004dm} 
as well as the twin Higgs mechanism~\cite{Yue:2009cq}. 
The obtained production cross sections depend on the model considered. 
There were even NLO estimations of this second class of single heavy fermion production 
(with only fermions in the final state) 
in the fourth generation context~\cite{Berger:2009qy,Campbell:2009gj}. Nevertheless, to our knowledge, the contribution of 
KK excitations of gauge bosons, or even their mixing effect with SM bosons -- both of which we will 
consider here --
for this class of single heavy fermion production have never been studied previously.

\section{Theoretical framework}

\subsection{The RS scenario}

We consider the RS scenario under the theoretical assumption of a bulk gauge custodial symmetry
${\rm SU(2)_L\! \times\! SU(2)_R\! \times\! U(1)_X}$ which allows for a reduction of the final EWPT bound on 
the mass of the first KK gauge boson
excitation $M_{KK}$ (strictly speaking, the KK photon mass) 
from $\sim 10$ TeV down to a few TeV~\cite{Agashe:2003zs,Bouchart:2009vq},
improving then the situation with regard to the little hierarchy problem (fine tuning of the Higgs boson mass due to its loop level corrections sensitive to new physics). 
For simplicity, we take the minimal quark representations under the custodial symmetry~\cite{Agashe:2003zs}:
the corresponding multiplets under the custodial symmetry ${\rm SU(2)_L\! \times\! SU(2)_R\! \times\! U(1)_X}$ 
for the bottom and top quarks are the three doublets 
$$
Q_{L} \equiv ({\bf 2},{\bf 1})_{1/6} =(t_L,b_L), \ \ \ \ 
Q_{b_R} \equiv ({\bf 1},{\bf 2})_{1/6} =(t'_R,b_R), \ \ \ \
Q_{t_R} \equiv ({\bf 1},{\bf 2})_{1/6} =(t_R,b'_R)
$$
whereas the representation for the Higgs field, responsible for the EW symmetry breaking, is
$$
\Sigma \equiv ({\bf 2},{\bf 2})_{0}.
$$ 
Although for simplicity the quark representations just above are the only ones shown here,
the effects we present in this paper are also qualitatively relevant for 
the model~\cite{Agashe:2006at} where the $Z\bar{b}_L b_L$ coupling is protected by a 
custodial symmetry under which the quark doublet is enlarged to $Q_{L} \equiv ({\bf 2},{\bf 2})_{2/3}$
and e.g. $Q_{t_R} \equiv ({\bf 1},{\bf 3})_{2/3} \oplus ({\bf 3},{\bf 1})_{2/3}$.
We will address these details in Ref.~\cite{Gopalakrishna:2009xxx} 
(see also the discussion in next section). 

We will focus in this work on the production of $b'_R$ at the LHC. For notational ease we will
denote the $b'_R$ simply as $b'$.

\subsection{EW Precision Tests}

The bulk custodial symmetry ensures that the global EW fit can reach a better goodness-of-fit than for the SM case, in the 
EW gauge bosons and light fermions sector (not including the bottom and top quarks) as long as $M_{KK} > {\cal O}(3)$ TeV,
$m_h \simeq 115$ GeV and light fermions are localized towards the Planck-brane ($c_{\rm light} > 0.5$)~\cite{Bouchart:2008vp}. 
The $c$ parameters are the dimensionless quantities parameterizing the fermion five-dimensional (5$D$) masses which fix the 
profiles along the fifth dimension (see e.g. Ref.~\cite{Agashe:2003zs}).

In the $b$, $t$ sector, the $R_b$ observable is protected from excessively large deviations in the $Zbb$ coupling (induced by
$b-b'$ mixings) by taking $m_{b'} > {\cal O}(1.5)$ TeV~\cite{Agashe:2004bm}. We will {\it also} consider some masses
below ${\cal O}(1.5)$ TeV but then the $Z\bar b_Lb_L$ vertex can be protected by a subgroup of the custodial symmetry
${\rm O(3)}$~\cite{Agashe:2006at}
\footnote{The corrections to $Z\bar b_Rb_R$ from the mixing with KK excitations of the $Z$ boson are less dangerous 
than the $Z\bar b_Lb_L$ deviations
due to the vanishing ${\rm SU(2)_L}$ isospin of $b_R$ leading to a smaller coupling to $Z$ excitations. 
Moreover, we consider here only the domain $c_{b_R}>0.5$ which tends to minimize the effective four-dimensional 
(4$D$) $b_R$ couplings to KK $Z(')$ excitations.
In a more precise analysis~\cite{Gopalakrishna:2009xxx}, we will give the
numerical results for the corrections to the $Z\bar b_Rb_R$ vertex in the ${\rm O(3)}$ context.}.
We thus assume such a symmetry for this low mass spectrum, keeping in mind that this symmetry corresponds to
quark representations different from those given above but these new representations are not expected to modify significantly the $b'$ couplings involved in its
single production, and in turn our illustrative numerical results.
Given the present simple theoretical context, we are forced to 
assume that the anomalies on the forward-backward asymmetries $A^b_{FB}$ and $A^t_{FB}$
are due respectively to underestimated uncertainties and too preliminary data, so that we do not have
to interpret them in terms of new physics effects (see Ref.~\cite{Djouadi:2006rk} and Ref.~\cite{Djouadi:2009nb}, 
respectively, for interpretations based on KK contributions).

\section{$b'$ at the LHC}

\begin{figure}
\begin{center}
\includegraphics[width=0.9\textwidth]{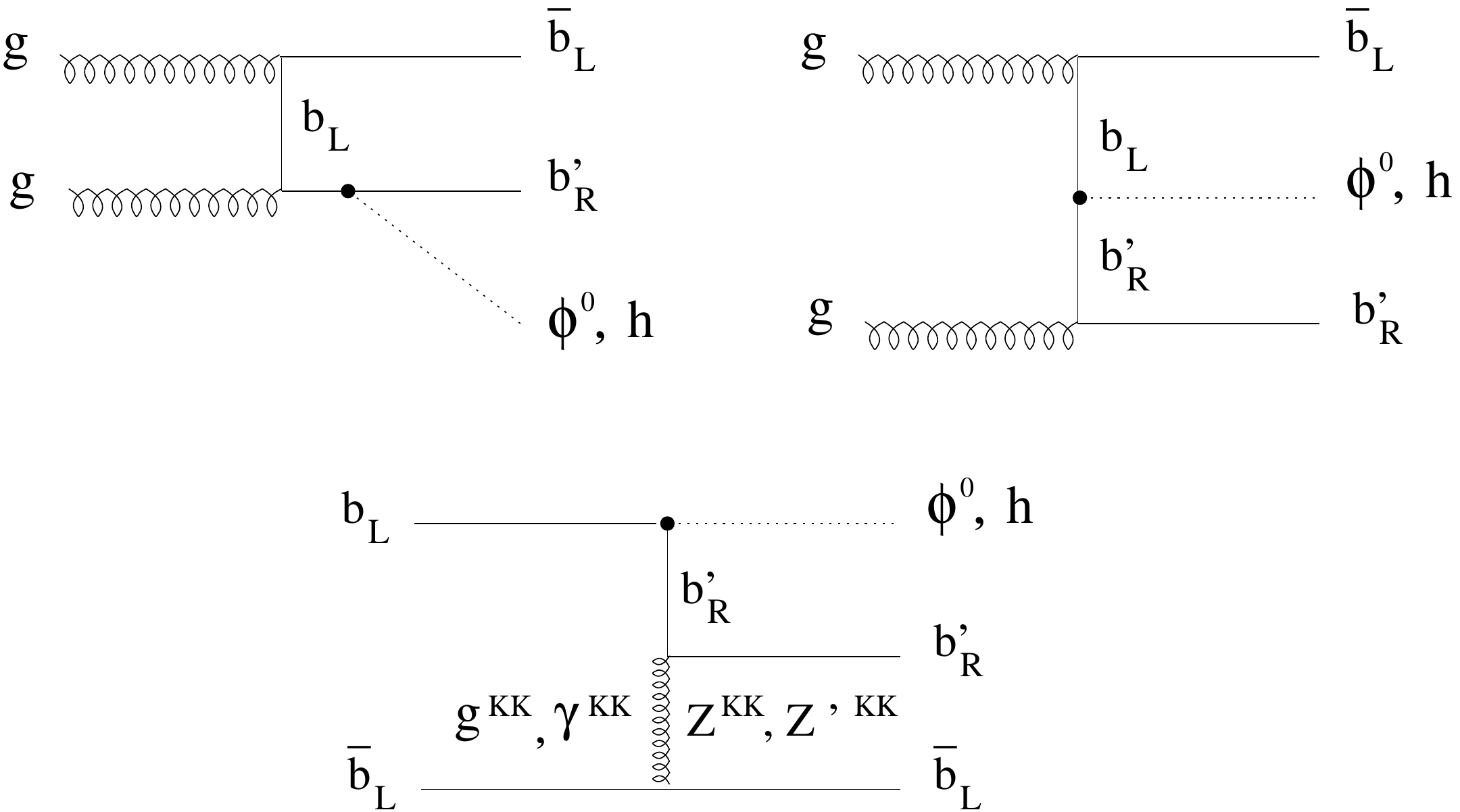}
\caption{Examples of Feynman diagram contributing to the single custodian productions
$\bar b_L b_L \to \bar b_L b'_R \phi^0, \ \bar b_L b'_R h$ and $gg \to \bar b_L b'_R \phi^0, \ \bar b_L b'_R h$ at LHC.
We indicate here the fermion fields in interaction basis (those are not the mass eigenstates).
$g$ stands for the gluon, $\phi^0$ for the neutral Goldstone boson (equivalently the longitudinal $Z$) 
and $h$ for the Higgs field. 
$Z^{KK}$ represents the full KK tower for the $Z$ boson (including the zero-mode), $\gamma^{KK}$ the full KK tower for the
photon, $Z^{\prime KK}$ is the KK tower for the additional neutral gauge boson and similarly $g^{KK}$ stands for the (KK) gluons.}
\label{SingleCusto_graph}
\end{center}
\end{figure}

We discuss here the production of the $b'$ in association with a (longitudinal) vector boson
at the LHC. 
In particular, we are interested in the processes: 
$gg \to \bar b_L b' _R Z_L/h$ and $\bar b_L b_L \to \bar b_L b' _R Z_L/h$,
where $Z_L$ is the longitudinal polarization. 
Owing to the Goldstone boson equivalence theorem, the longitudinal polarization of the 
vector boson is nothing but the corresponding Goldstone boson, and we have the correspondence
$V^\mu_L \leftrightarrow \partial^\mu \phi/M_V$ where $M_V$ is the vector boson mass. 
We show in Fig.~(\ref{SingleCusto_graph}) example Feynman diagrams for these processes. 
The couplings $\phi^\pm tb'$, $\phi^0bb'$ and $hbb'$, involved in each single $b'$ production, are important due to a large
wave function overlap near the TeV-brane owing to these fermion masses being sizable.  

\subsection{$b'$ couplings}
For computing the $b'$ couplings, we will need the Yukawa couplings given, in terms of the 5$D$ 
Yukawa coupling constants, by
\beq
{\cal L}_{5D}  \supset - \lambda_t \bar{Q}_L \Sigma Q_{t_R} 
                 - \lambda_b \bar{Q}_L \Sigma Q_{b_R} \ .
\label{yuk.EQ}
\eeq

Electroweak symmetry is broken by $\left< \Sigma \right> = {\rm diag}(v, v)$
(the Higgs boson vev is $v \approx 174$ GeV). 
The Goldstone bosons of electroweak symmetry breaking ($\phi$) are contained in 
$\Sigma = v e^{2i\phi^a T^a/v}$, written in the nonlinear realization,
where $T^a$ are the generators of $SU(2)_L$. 
We work here in the unitary gauge (as for our numerical calculations)
for which we absorb the Goldstone bosons as the longitudinal polarization of the
gauge bosons. Nevertheless, for completeness and to have a clear understanding of the involved 
couplings, a derivation of the couplings using Goldstone boson equivalence is
presented in~\ref{GBEq.SEC}.

Electroweak symmetry breaking also gives rise to fermion masses.
Taking the case of the $b_L$, 
for example, the off-diagonal terms in the bottom mass matrix resulting from Eq.~(\ref{yuk.EQ}) 
lead to the mixing of the fields 
$(b_L, {b'^{(n)}_R}_L, {b^{(n)}_L}_L, {b^{(n)}_R}_L)$, where when two subscripts are present, 
the first $L,R$ denote the gauge-group representation while the second subscript
denotes Lorentz chirality, and $(n)$ denotes the $n^{\rm th}$ KK state. 
Similarly, a mixing is induced amongst the corresponding Lorentz $R$ fields. 
Focusing on the dominant of these mixing terms for simplifying the present discussion
 \footnote{Numerically, we take into account {\it also} the $b$ mixing with the second KK excitations of the $b'$ custodian as well as with
the first KK excitations of the $b_L$ and $b_R$ fields.} we have
\begin{eqnarray}
{\cal L}_{4D} \supset - \left ( \begin{array}{cc} \bar{b}_L & \bar{b'}_L  \end{array} \right ) 
\left ( \begin{array}{cc} 
\frac{\lambda_b v}{\pi R_c}  f_{Q_L}({\scriptsize \pi R_c}) f_{b_R}({\scriptsize \pi R_c}) & \frac{\lambda_t v}{\pi R_c} 
f_{Q_L}({\scriptsize \pi R_c}) f_{b'_R}({\scriptsize \pi R_c}) \\
                    0             &         m_{b'}            \end{array} \right ) 
\left ( \begin{array}{c}  b_R \\ b'_R \end{array} \right ) + {\rm h.c.} 
\label{eq:SC_matrix}
\end{eqnarray}
where $L,R$ above (and in the text that follows) denote Lorentz chirality, 
$f_{i}$ are the wavefunction values of the corresponding fields at the IR-brane location. 
The above mass matrix is diagonalized by biorthogonal rotations, and we denote 
the sine (cosine) of the mixing angles by $s_\theta^{L,R}$ ($c_\theta^{L,R}$).
The sine of mixing angle $s_\theta^{L}$ is given in Appendix~\ref{GBEq.SEC}. 
We denote the corresponding mass eigenstates as $(b_1\ b_2)$. 


Due to these mixings, the $Z \bar{b}_{1L} b_{2L}$ and $W \bar{t}_{1L} b_{2L}$ couplings are
induced, where we similarly define $t_{1,2}$ as the top mass eigenstates and $c_\alpha$ as the cosine of mixing angle
between the top quark and $t'$ field.
The contribution to these couplings due e.g. to $W \leftrightarrow W^{KK}$ mixings are 
higher order in $v/M_{KK}$, which we have not shown here, but included in our numerical results.
%
%
The couplings of the heavier mass eigenstate $b_2$ in unitary gauge are given by
\bea
{\cal L}_{4D} \supset 
&-&\frac{e}{3} \bar{b}_{2L/R} \gamma^\mu b_{2L/R} A_\mu 
+ g_s \bar{b}_{2L/R}  \gamma^\mu T^\alpha b_{2L/R} g^\alpha_\mu 
- \left( 
\frac{g s^L_\theta c^L_\alpha}{\sqrt{2}} \bar{t}_{1L}  \gamma^\mu {b_2}_L W_\mu^+ + {\rm h.c.} \right) \nonumber \\
&+& g_Z \left( -\frac{1}{2}{s_\theta^L}^2 + \frac{1}{3} s_W^2 \right) \bar{b}_{2L} \gamma^\mu b_{2L} Z_\mu
+ \left[ g_Z c_\theta^L s_\theta^L \left(\frac{1}{2}\right) \bar{b}_{1L}  \gamma^\mu b_{2L}  Z_\mu + {\rm h.c.} \right] \nonumber \\
&+& g_Z \left(\frac{1}{3}s_W^2\right) \bar{b}_{2R}  \gamma^\mu b_{2R}  Z_\mu \ ,
\label{b'uni.EQ}
\eea
where $g_Z \equiv \sqrt{g^2+g'^2}$. 
The photon and gluon couplings are diagonal in the $(b_1\ b_2)$ basis and are identical to the 
b-quark couplings. 

To derive the Higgs couplings to fermions, we start with Eq.~(\ref{eq:SC_matrix}) and 
apply the vev replacement $v\rightarrow h/\sqrt{2}$. 
Writing in terms of the fermion mass eigenstates, we obtain
\bea
{\cal L}_{4D} \supset -\frac{h}{\sqrt{2}} \left[ 
\bar{b}_{1L} {b}_{1R} (c^L_\theta c^R_\theta \lambda^{4D}_b + c^L_\theta s^R_\theta \lambda^{4D}_{b'}) + 
\bar{b}_{2L} {b}_{2R} (s^L_\theta s^R_\theta \lambda^{4D}_b - s^L_\theta c^R_\theta \lambda^{4D}_{b'})
\right.
 \nonumber \\
\left.
+ \bar{b}_{1L} {b_2}_R (-c^L_\theta s^R_\theta \lambda^{4D}_b + c^L_\theta c^R_\theta \lambda^{4D}_{b'}) +
\bar{b}_{2L} {b_1}_R (-s^L_\theta c^R_\theta \lambda^{4D}_b - s^L_\theta s^R_\theta \lambda^{4D}_{b'}) \right] + {\rm h.c.} 
\eea
where $\lambda^{4D}_{b, b'} = \frac{\lambda_b}{\pi R_c} f_{Q_L}(\pi R_c) f_{b_R, b'_R}(\pi R_c)$.
For instance, $f_{b'_R}( \pi R_c)$ is the value of the wave function for $b'_R$
(controlled by the $c_{t_R}$ parameter of the $Q_{t_R}$ doublet) taken at the TeV-brane 
--- as induced by the overlap with the peaked profile of the Higgs field.

\subsection{Parameter space}

The motivation for the single $b'$ production resides in both its possibly low mass 
and its rather large coupling to $\phi$ and $h$ compared to lighter quark generations.

Let us first consider the most severe experimental lower bound on a fourth generation $b_4$ quark mass in order to get a rough idea
({\it a priori} the $b'$ couplings differ from the $b_4$ ones) of the realistic $m_{b'}$ range: $m_{b_4}>199$ GeV
at $95\% C.L.$~\cite{Amsler:2008zzb}. 
Now there might appear more severe lower bounds on $m_{b'}$ from FCNC considerations. 
Indeed, generally the new heavy fermions can contribute to $b \to s \gamma$ at the one-loop level
(see~\cite{Agashe:2008uz} for the two-site approach to 5$D$ AdS models with bulk Higgs, a framework that differs
from the present one where e.g. the Higgs field is confined on the TeV-brane). However such indirect constraints rely on the
whole set of 5$D$ Yukawa couplings and light fermion locations along the extra dimension that we do not specify
here. A complete three-flavor study, beyond our scope, should also reproduce the CKM angles for the quark mixing~\cite{Amsler:2008zzb}.

We consider a set of parameters, $M_{KK}=2.7$ TeV, $m_h \simeq 115$ GeV and $g_{Z'}=1.57$, justified by considerations
on the global EW fit~\cite{Bouchart:2008vp}. Then we choose various values of $c_{t_R}$ 
($c_{t_R}$ is the main parameter determining the single $b'$ production cross section) fixing the right-handed top
quark location and also mainly $m_{b'}$: $c_{t_R}=-0.55$ ($m_{b'}=225$ GeV), $c_{t_R}=-0.4$ ($m_{b'}=756$ GeV) and $c_{t_R}=-0.1$ ($m_{b'}=1574$ GeV).
Strictly speaking, the physical state is the $b_2$ state, namely the second lightest bottom quark mass 
eigenstate composed in general mainly of the first $b'$ custodian excitation but also partially of the pure SM $b$ field (due to the $b-b'$ mixing effect). 
So the lightest $b_1$ eigenstate is associated to the measured bottom quark mass $m_b$, while the $b_2$ eigenstate is associated to $m_{b'}$
(that we should write $m_{b_2}$ but leave as $m_{b'}$ for simplification reasons).

For each $c_{t_R}$ value, examples of $c_{Q_L}$ and the top quark 5$D$ Yukawa coupling constant $\lambda_t$ are chosen 
so that $m_{t} \sim 175$ GeV.
We note that the exact top mass value can only be fitted precisely after a complete three-flavor treatment of the full 
quark mass matrix. Then order one corrections can
bring the top mass obtained here exactly to the measured value.
Finally, we select some values of $c_{b_R}$ and $\lambda_b$ reproducing $m_{b} \simeq 4$ GeV.

\subsection{Numerical results and discussion}

Including the main contributions of the first KK gauge boson and fermion excitations, 
we obtain the cross sections for the single $b'$ production at LHC given in Table \ref{SingleCusto_table}.
Strictly speaking, the calculated amplitudes correspond to a final state with a unique $b_2$ state, 
namely the second lightest bottom quark mass 
eigenstate.
\begin{table}[!ht]
\begin{center}
\begin{tabular}{|c|c|c|c|}
\hline
&  {\bf  A)}\, $\begin{array}{l}  c_{Q_L} = -0.2 \\ c_{t_R}= -0.55 \\ c_{b_R}=+0.58 \end{array}$
& {\bf  B)}\, $\begin{array}{l}  c_{Q_L} = -0.2 \\ c_{t_R}= -0.40 \\ c_{b_R}=+0.62 \end{array}$
& {\bf  C)}\, $\begin{array}{l}  c_{Q_L} = -0.2 \\ c_{t_R}= -0.10 \\ c_{b_R}=+0.62 \end{array}$\\[.1cm] \hline
 {\vrule height 12pt depth 0pt width 0pt}  $m_{b'} = $ & $225$ GeV & $756$ GeV  & $1574$ GeV\\[.1cm]  \hline
 {\vrule height 12pt depth 0pt width 0pt} $m_b\simeq \ [m_t \simeq] $ & $4\ [197]$ GeV & $4\ [179]$ GeV&  $4\ [171]$ GeV\\[.1cm]  \hline
 {\vrule height 12pt depth 0pt width 0pt} $\sigma (qq \to tb') \simeq $ & $592$ fb & $8.74$ fb & $0.40$ fb \\[.1cm]  \hline
 {\vrule height 12pt depth 0pt width 0pt} $\sigma (qq \to bb') \simeq $ & $728$ fb & $16.0$ fb & $0.34$ fb\\[.1cm]  \hline
 {\vrule height 12pt depth 0pt width 0pt} $\sigma (qg \to qtb') \simeq $ & $3044$ fb & $20.8$ fb & $0.86$ fb\\[.1cm]  \hline
 {\vrule height 12pt depth 0pt width 0pt} $\sigma (qg \to qbb') \simeq $ & $5668$ fb & $134$ fb & $2.58$ fb\\[.1cm]  \hline
 {\vrule height 12pt depth 0pt width 0pt} $\sigma (bg \to b'Z) \simeq $ & $1392$ fb & $2.88$ fb & $0.01$ fb\\[.1cm]  \hline
 {\vrule height 12pt depth 0pt width 0pt} $\sigma (bg \to b'h) \simeq $ & $2572$ fb & $116$ fb & $1.06$ fb\\[.1cm]  \hline
 {\vrule height 12pt depth 0pt width 0pt} $\sigma (QQ/gg \to bb'Z) \simeq $ & $3074$ fb & $11.6$ fb & $0.04$ fb\\[.1cm]  \hline
 {\vrule height 12pt depth 0pt width 0pt} $\sigma (QQ/gg \to bb'h) \simeq $ & $27000$ fb & $2800$ fb & $26.8$ fb\\[.1cm]  \hline
 {\vrule height 12pt depth 0pt width 0pt} $\sigma (qb \to qb'Z) \simeq $ & $557$ fb & $30.8$ fb & $1.53$ fb\\[.1cm]  \hline
 {\vrule height 12pt depth 0pt width 0pt} $\sigma (qb \to qb'h) \simeq $ & $882$ fb & $67.8$ fb & $1.59$ fb\\[.1cm]  \hline
 {\vrule height 12pt depth 0pt width 0pt} $\sigma (qq/gg \to tb'W) \simeq $ & $453$ fb & $5.84$ fb & $0.05$ fb\\[.1cm]  \hline
 {\vrule height 12pt depth 0pt width 0pt} $\sigma (qb \to qb'W) \simeq $ & $206$ fb & $42.2$ fb & $2.40$ fb\\[.1cm]  \hline
 {\vrule height 12pt depth 0pt width 0pt} $\sigma (QQ/gg \to \bar{b}'b') \simeq $ & $94000$ fb & $271$ fb & $1.23$ fb
\\ \hline
\end{tabular}
\end{center}
\caption{Values (in fb) of the cross sections for the pair and the 
various single $b'$ production reactions at LHC for three sets (A, B, C) of parameters 
with $M_{KK} = 2.7$ TeV, $m_h \simeq 115$ GeV, $g_{Z'} = 1.57$ [see text]. 
For each set, we also give the predicted values of $m_{b'}$,
$m_b$ and $m_t$ (in GeV). $QQ$ denotes either the initial state $qq$ or $bb$.}
\label{SingleCusto_table}
\end{table}

We conclude from the examples of parameter sets considered
in the table that within the present RS framework (reproducing $m_{b,t}$) 
both the single and pair $b'$ productions have promising cross sections
which might allow for their possible detection at LHC, 
if one assumes the typical high luminosity regime expected (${\cal L} \sim 300$ fb$^{-1}$).
The considerable number of events predicted here should still lead to significant signals after including 
the decay analysis, hadronization effects, the {\it Monte Carlo} simulation and effects of detector response.    
Note also that for the extreme situation at $m_{b'} = {\cal O}(1.5)$ TeV, if a $b'$ custodian is to be discovered at LHC
it is mainly via the single production reaction $QQ/gg \to bb'h$ only. Indeed, for this high custodian mass, the pair 
production rate is reduced too much by the phase space suppression.

Note that in comparison, the resonant KK gluon production suffers from typically 
lower cross sections at LHC~\cite{Djouadi:2007eg} (see also 
Ref.~\cite{Agashe:2007ki,Agashe:2008jb,Ledroit:2007ik} for the production of KK EW gauge bosons) 
rendering its observation more tricky.
This is essentially due to the high KK gluon masses ($M_{KK} > {\cal O}(3)$ TeV) 
compared to the lighter $b'$ custodians considered here.

While the dominant single production processes were thought to be only $qg \to qtb'$ and $qg \to qbb'$ 
(partially due to a polarization increase effect), 
we see that for instance the new single production reactions $bg \to b'Z$ and $QQ/gg \to bb'h$ -- 
originally studied here (see also Ref.~\cite{Gopalakrishna:2009xxx}) --
can be of comparable order or even larger than the previously thought dominant ones. The reason is principally 
the possibly pure gluonic initial state for these new processes. The $gg \to tb'W$ reaction, having only a possible pure gluonic initial state, 
is significant at a low $m_{b'}$ value for which the parton density functions are significantly higher for the $gg$ initial state.

Therefore, from a more general point of view, novel reactions such as $bg \to b'h$, $QQ/gg \to bb'Z$ or $QQ/gg \to bb'h$ should be 
now included in the experimental investigations for any heavy quark production predicted by a scenario underlying the SM.
Finally, note also that such new processes -- like the ones drawn in Fig.~(\ref{SingleCusto_graph}) -- 
constitute new channels for the Higgs boson production at LHC
\footnote{In contrast, the corrections arising in the RS model to the usual Higgs production mechanisms 
have already been studied e.g. in Ref.~\cite{Bouchart:2009vq,Djouadi:2007fm}.}.

\section*{Acknowledgements}
The authors thank K.~Agashe and A.~Pukhov for interesting discussions and also 
{\it Les Houches} conveners for organizing this nice Workshop where the present work was started.
SG thanks Brookhaven National
Laboratory for partial financial support to attend the workshop.


\section*{Appendices}
\setcounter{section}{0}
\renewcommand{\thesection}{App.~\Alph{section}}


\section{Goldstone boson equivalence}
\label{GBEq.SEC}

Here we explicitly check the correspondence through the equivalence theorem
between e.g. the $\phi^\pm tb'$ and $W_L^\pm tb'$ couplings in the $R_\xi$ and unitary gauges 
respectively. The correspondence in the neutral sector can be shown analogously. 
  
In the linear realization, the Higgs is written as
\begin{eqnarray}
\Sigma = 
\left ( \begin{array}{cc}  
\Phi_0^* & \phi^+ \\ 
-\phi^- & \Phi_0 
\end{array} \right ) ,
\label{eq:SC_EWSB}
\end{eqnarray} 
with Electroweak symmetry broken by $\left< \Sigma \right> =diag(v,v)$ 
(recall that $v \approx 174$ GeV in our notation) and the three
Goldstone bosons are $\phi^\pm$ and ${\rm Im}{(\Phi^0)} = \phi^0 / \sqrt{2}$.

After reducing the 5$D$ theory to an equivalent 4$D$ theory, we can write the mass matrix in the
$b$-sector as
\begin{eqnarray}
{\cal L}_{4D} \supset - \left ( \begin{array}{cc} \bar{b}_L & \bar{b'}_L  \end{array} \right ) 
\left ( \begin{array}{cc} 
 \lambda^{4D}_b v &   \lambda^{4D}_{b'} v    \\
  0  &  m_{b'} \end{array} \right ) 
\left ( \begin{array}{c}  b_R \\ b'_R \end{array} \right ) + {\rm h.c.} \ ,
\label{eq:SC_matrix-4D}
\end{eqnarray}
with the 4$D$ Yukawa couplings given as
$\lambda^{4D}_b = \frac{\lambda_b}{\pi R_c} f_{Q_L}(\pi R_c) f_{b_R}(\pi R_c)$ and
$\lambda^{4D}_{b'} = \frac{\lambda_t}{\pi R_c} f_{Q_L}(\pi R_c)$ 
$f_{b'_R}(\pi R_c)$
where the $\lambda$'s are the corresponding 5$D$ Yukawa couplings.

In the $\lambda^{4D}_b v \rightarrow 0$ limit, the mass matrix is diagonalized by rotating only the $L$ fields by
$s^L_\theta = -\lambda^{4D}_{b'} v/\sqrt{m_{b'}^2 + (\lambda^{4D}_{b'} v)^2}$ with the heavier mass eigenvalue being 
$m_{b_2} = \sqrt{m_{b'}^2 + (\lambda^{4D}_{b'} v)^2}$. 

Next we turn to showing the correspondence. For simplicity (and since it represents a good approximation) 
we do so in the $\lambda^{4D}_b v \rightarrow 0$ limit.
In order to extract the relevant interaction of the $b'$ quark, we focus on the following two
couplings contained in the first term of Eq.~(\ref{yuk.EQ}):
\begin{eqnarray}
{\cal L} \supset - \lambda^{4D}_{b'} \phi^+ {\bar t}_L b'_R + \lambda^{4D}_t \phi^+ {\bar t}_R b_L  + {\rm h.c.} \ .
\label{eq:SC_terms}
\end{eqnarray} 
Here $\lambda^{4D}_t = \frac{\lambda_t}{\pi R_c} f_{Q_L}(\pi R_c) f_{t_R}(\pi R_c)$.
Writing in terms of the fermion mass eigenstates we obtain
\beq
{\cal L} \supset \phi^+ \left( - \lambda^{4D}_{b'} \bar{t}_L {b_2}_R
+ c^L_\theta \lambda^{4D}_t \bar{t}_R {b_1}_L 
 - s^L_\theta \lambda^{4D}_t \bar{t}_R {b_2}_L \right) + {\rm h.c.} \ . 
\label{Rxicoup.EQ}
\eeq

We will show that we can recover Eq.~(\ref{Rxicoup.EQ}) starting from the
longitudinal $W^\pm_L$ unitary gauge coupling in Eq.~(\ref{b'uni.EQ}).
Including the SM piece, this coupling in unitary gauge in the mass basis is
\beq
{\cal L} \supset \frac{g}{\sqrt{2}} {W_L}^+_\mu \bar{t}_L \gamma^\mu \left( c^L_\theta {b_1}_L - s^L_\theta {b_2}_L \right) + {\rm h.c.} \ ,
\label{Wtbuni.EQ}
\eeq
where we have ignored the $t\leftrightarrow t'$ mixing 
(since $m_{t'} \gg m_{b'}$ for the parameter sets considered) and thus set $c^L_\alpha = 1$.
The Goldstone-boson equivalence theorem implies 
${W_L}^+_\mu \leftrightarrow \partial_\mu \phi^+/m_W = q_\mu \phi^+/m_W$ where $q_\mu$ is the
momentum of the $W$ boson. Using this and momentum conservation $q_\mu = p_{t \ \mu} - p_{b \ \mu}$ 
(where $p_t$ is the momentum out of the vertex and $p_b$ into the vertex) and $m_W = g v/\sqrt{2}$, 
Eq.~(\ref{Wtbuni.EQ}) becomes
\beq
{\cal L} \supset \frac{1}{v} \phi^+ \bar{t}_L 
\left[ c^L_\theta \left( \slashchar{p}_t - \slashchar{p}_{b_1} \right) {b_1}_L 
- s^L_\theta  \left( \slashchar{p}_t - \slashchar{p}_{b_2} \right)  {b_2}_L \right] + {\rm h.c.} \ .
\eeq
Under the assumption that all fermions are on mass-shell and using the equations of 
motion (Dirac equations: $\pslash \psi(p) = m \psi(p)$) we 
recover Eq.~(\ref{Rxicoup.EQ}), thus explicitly showing the correspondence.
One has to recall that if the top quark mixing is neglected, the top mass reads as
$m_t \simeq \lambda^{4D}_t v$.

The contributions to this coupling due to the mixing of the SM $W$ boson with its KK excitations constitute
higher order corrections, in $v/M_{KK}$, which we have already ignored in the present theoretical demonstration 
(but not in the numerical results as mentioned previously). 
Finally, we note that the correspondence between the $\phi^0 bb'$ and $Z_L^0 bb'$ couplings
can be similarly shown.


%% file: Servant/Servant.tex
\chapter{Four top final states }

{\it G.~Servant, M.~Vos, L.~Gauthier and A.-I.~Etienvre}

\begin{abstract}
Of the many interesting final states that may be produced at the LHC, four top production is maybe one of the most spectacular. In this contribution, the sensitivity of this final state to several classes of physics beyond the Standard Model is discussed. The focus is on models where this topology is produced through a heavy resonance. The possibility to reconstruct the top and anti-top quarks in these events is explored.
\end{abstract}

\section{Introduction}
\label{fourtops:introduction}

\begin{figure}[h]
  \begin{center}
    \mbox{
      \begin{tabular}{cc}
	\vspace{-0.6cm}
        \subfigure[]{\includegraphics[angle=0,width=0.2\linewidth]{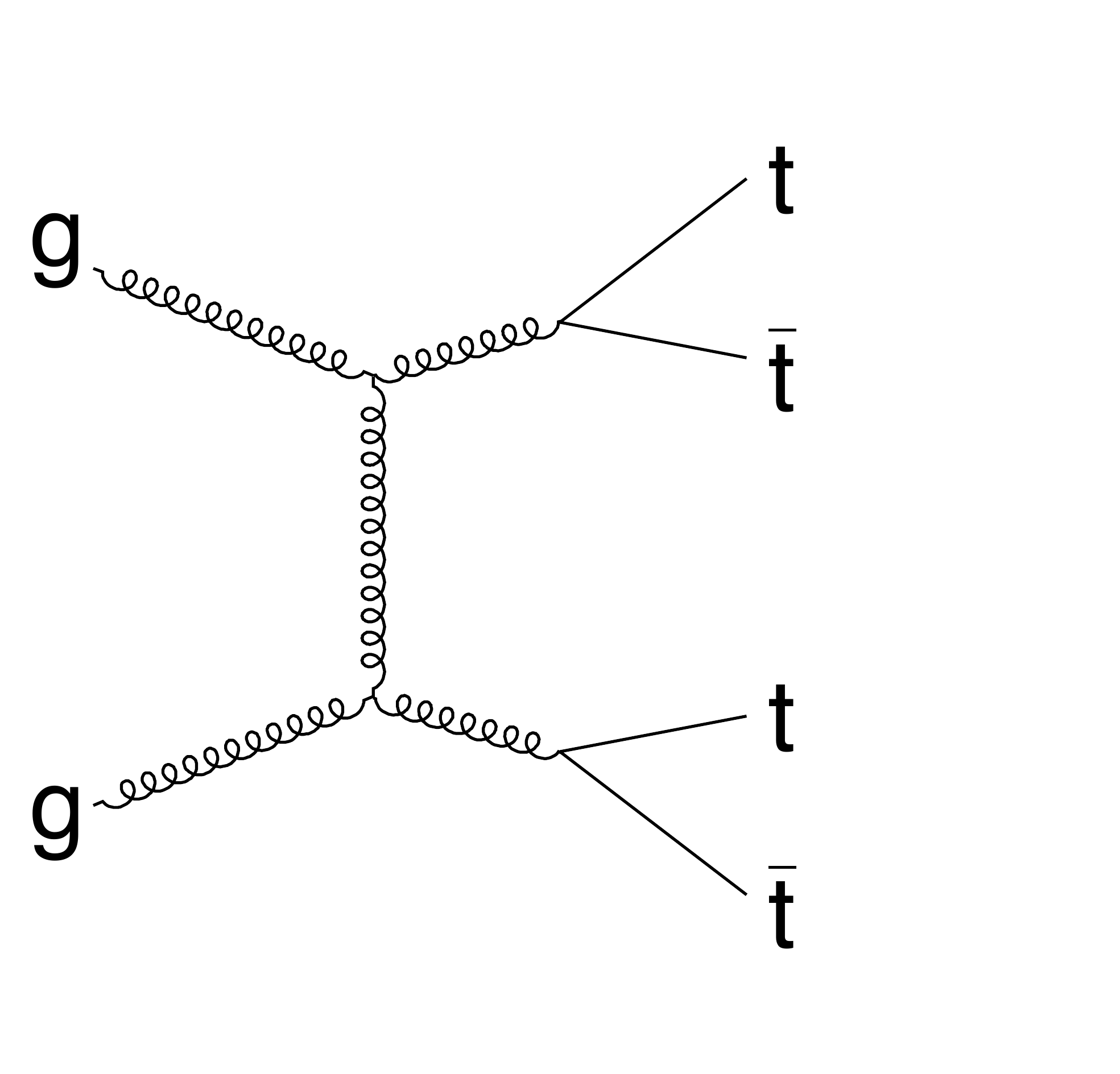}} 
        \subfigure[]{\includegraphics[angle=0,width=0.2\linewidth]{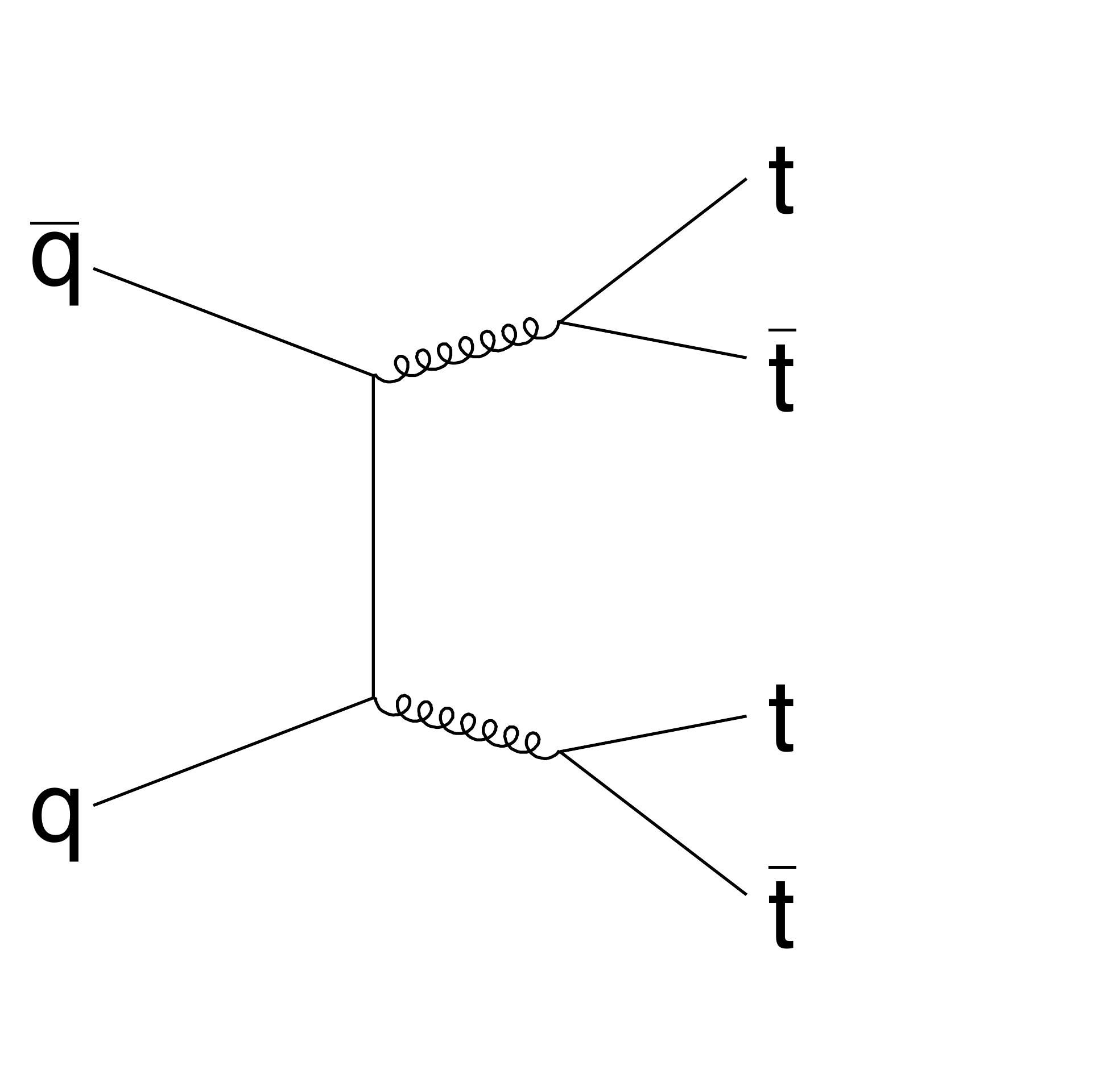}}
        \subfigure[]{\includegraphics[angle=0,width=0.2\linewidth]{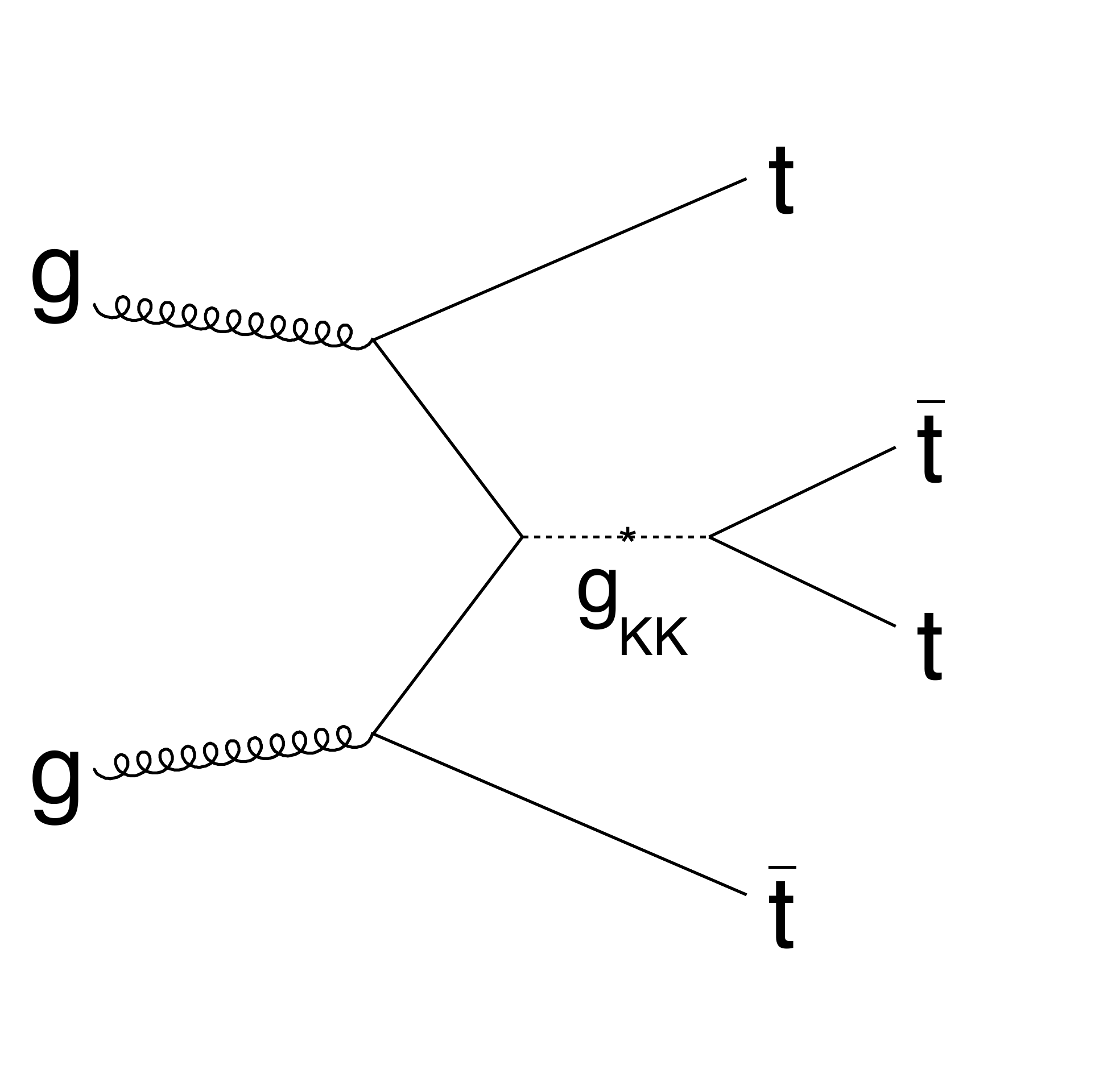}}  
        \subfigure[]{\includegraphics[angle=0,width=0.2\linewidth]{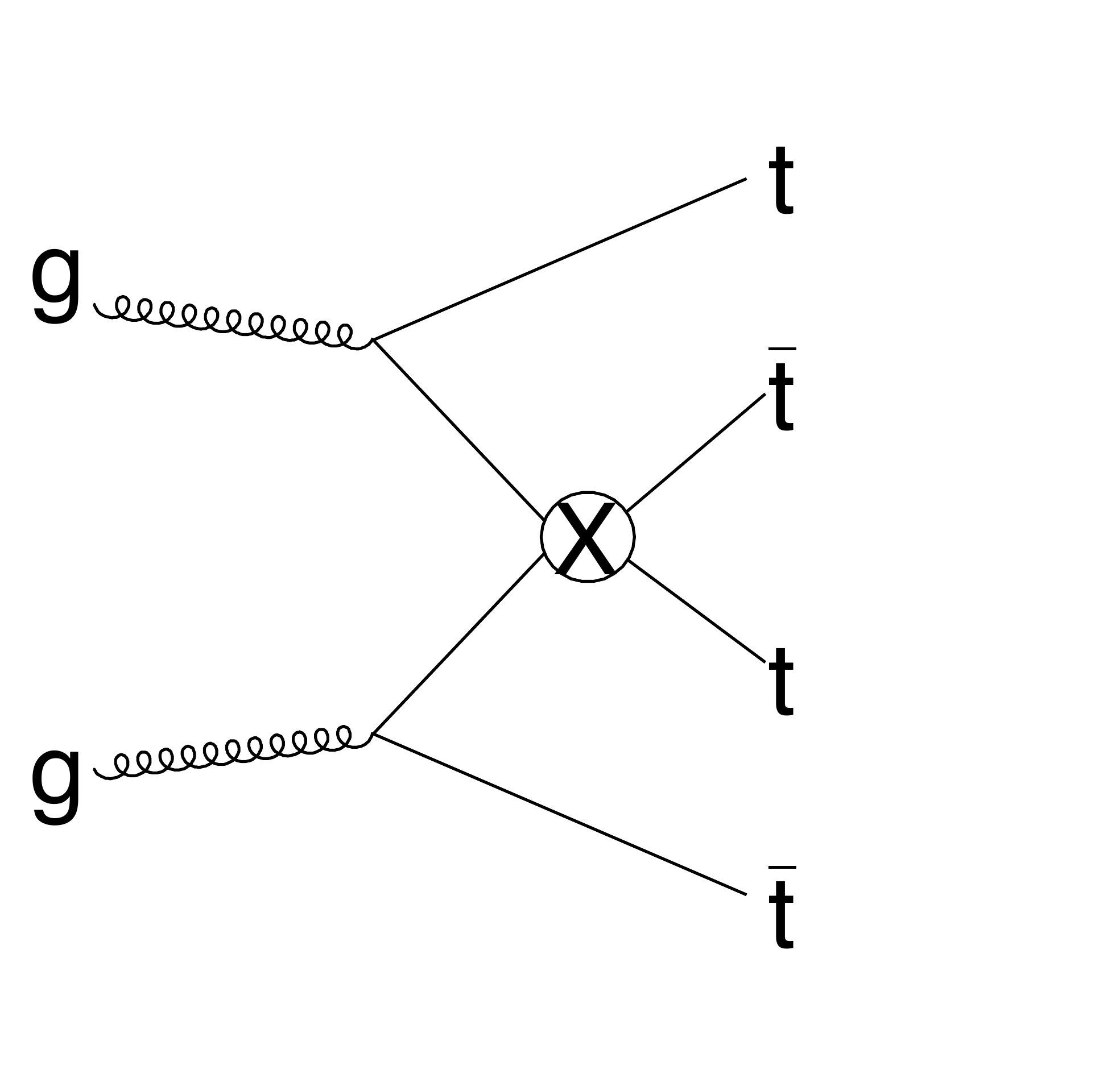}} \\  
      \end{tabular}
    }
  \end{center}
  \caption{(a-b): Two Standard Model diagrams that give rise to the $ t \bar{t} t \bar{t} $ final state. (c-d): Two diagrams involving new physics, that yield to a non-zero event rate even if the new particle does not couple to light quarks. (c) represents s-channel (resonant) $ t\bar{t} $ production. The effective four-top interaction in (d) can result from integrating out a heavy particle.}
  \label{fig:diagrams}
\vspace{0.1cm}
\end{figure}

Four top production occurs in the Standard Model through a large number of diagrams \cite{Barger:1991vn}, two of which are indicated in figure~\ref{fig:diagrams}(a-b). The total $pp\rightarrow t\overline{t} t\overline{t}$ cross-section at 14 TeV  is 7.5 fb in the Standard Model. The production is dominated by gluon-initiated diagrams. 

The interest of this final states lies primarily in its sensitivity to beyond-the-standard-model physics, as recently discussed in ~\cite{tait,tait2,serra}. These authors consider a composite top quark that would give rise to contact interactions like that of figure~\ref{fig:diagrams}(d). The production cross-section through the contact interaction can be as large as several tens of fb.

Another possibility is the production of the $ t \bar{t} t \bar{t} $ final state through an exotic heavy particle. In many extensions of the Standard Model the top quark plays a special role. New particles with a preference for the top quark could yield a sizeable cross-section through processes like that depicted in  figure~\ref{fig:diagrams} (c), where the contact interaction is replaced by a resonance (in this case a Kaluza--Klein gluon). It is interesting to note that this diagram involves only couplings of the new particle to top quarks. In particular, the new particle does not have to couple to light quarks or gluons to be produced at the LHC. In section~\ref{fourtops:resonant} two models giving rise to resonant four top production are discussed in some detail.

A measurement of the four-top production rate would strongly constrain several models. While a complete, detector-level analysis is still missing, several authors~\cite{tait,tait2,serra} have investigated the possibility to isolate this signal. A common aspect of these studies is that the isolation strategy consists in requiring two leptons with the same sign. Thus, processes like $ t \bar{t} +$ jets production, with cross-sections that are several orders of magnitude larger than typical signal cross-sections, are effectively reduced. The power of selecting same-sign dilepton events to study $ttWW$ final states from pair-production of heavy quarks was shown in detail in \cite{Contino:2008hi} and recently applied by CDF to put a strong bound on the mass of fourth generation down-type quarks ($b^{\prime}$) \cite{Aaltonen:2009nr}. The reduction of SM $ t \bar{t} $ production using the same-sign criterion was shown by ATLAS Monte Carlo studies of $t \bar{t} H $ production, with $ H \rightarrow W W^*$ (~\cite{cscnote}, pages 1367-1368).  

Further experimental handles to distinguish the signal are particularly important in the light of the large cross-section of several reducible background processes, like $ t \bar{t} W+$ jets,  $ t \bar{t} WW+$ jets  and $ t \bar{t}+$ jets. Therefore, reconstruction of all or several of the top decays can strengthen the robustness of the analysis considerably. This possibility is particularly interesting in searches for resonant production. Reconstruction of the top quarks allow the reconstruction of the mass of the resonance. The background level can then be normalized in the off-peak region, thus considerably increasing the sensitivity of the search. 

A complete study into the reconstruction of this extremely challenging final state is clearly beyond the scope of this contribution. The results from a first superficial exploration of some ideas is presented in section~\ref{fourtops:reconstruction}.

\section{Resonant production}
\label{fourtops:resonant}

A prototype is based on the Randall-Sundrum (RS) setup where the hierarchy between the Planck and electroweak scales is explained through
warping of an extra dimension. The Standard Model (SM) lives in the bulk \cite{Agashe:2003zs} but the Higgs lives on the IR boundary where the natural scale of physics is $\sim$~TeV.  As a result of the localized Higgs, the zero mode
of the right-handed top quark must also live close to the IR brane, in order to realize the large top mass\footnote{The left-handed top is usually chosen to be further from
the IR brane, in order to mitigate constraints from precision electroweak tests \cite{Agashe:2003zs}.}. As a consequence of the warping, all of the low level Kaluza--Klein (KK) modes have wave functions whose support is concentrated near the IR brane.  Thus, they inevitably couple strongly to $t_R$ and the Higgs.  

\begin{figure}[h]
  \begin{center}
        \includegraphics[width=0.65\linewidth]{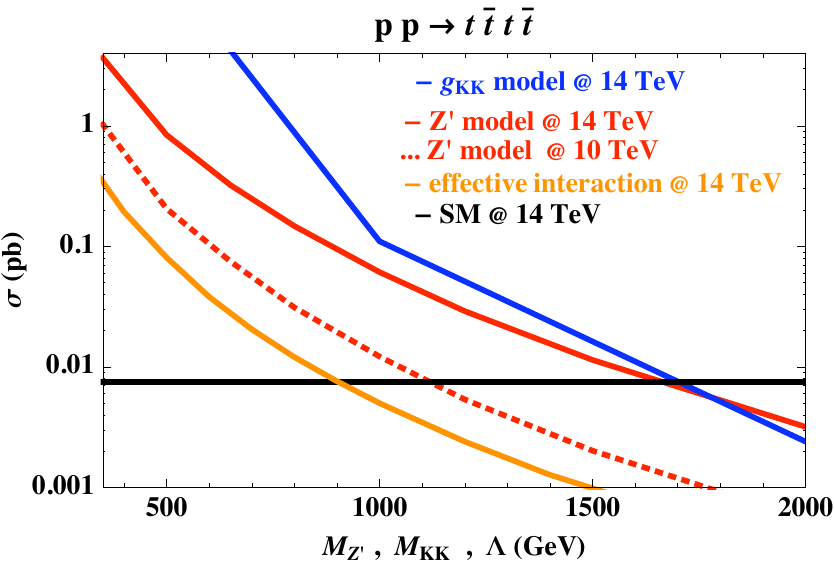}
  \caption{Four top production cross section at the LHC in the different theories discussed in this report. The orange curve refers to the effective 4-fermion interaction $(\overline{t}_R\gamma_{\mu}t_R)(\overline{t}_R\gamma^{\mu}t_R)/\Lambda^2$ leading to fig.~\ref{fig:diagrams}(d).}
  \label{fig:crosssection}
   \end{center}
\end{figure}

The KK gluon~\cite{randall} has a number of features that render it very interesting phenomenologically. First of all, it cannot be revealed by resonance searches in di-lepton final states. As a coloured object it can be produced relatively abundantly (compared to partners of the electro-weak gauge bosons). With the couplings to light quarks of reference~\cite{randall}, $ g_q = g_b^L = -0.2 g_s$, $g^R_b = g^L_t = g_s $ and $ g^R_t = 4 g_s$,  it is still sufficiently narrow ($ \Gamma = 0.15 M$) to yield a resonant signature. Finally, it has a very sizeable branching ratio of 92.5 \% into a top anti-top pair. 
The cross-section for the four-top final state through the KK gluon is represented in figure~\ref{fig:crosssection}. It is noted here that the model of reference~\cite{randall} is experimentally viable for KK gluon masses greater than 2-3 TeV, in which case the 4 top production cross section is well below a fb and LHC prospects are not encouraging.

However, one can easily envision some variation of the RS setup as follows.
Consider a top-philic $Z^\prime$ described by the following lagrangian \cite{Jackson:2009kg}
\begin{eqnarray}
{\cal L}  & = &  {\cal L}_{SM}   - \frac{1}{4} {F}^\prime_{\mu \nu} {F}^{\prime \mu \nu} 
+ M_{{Z}^\prime}^2 {Z}^\prime_{\mu} {Z}^{\prime \mu} 
+ \frac{\chi}{2} {F}^\prime_{\mu \nu} {F}_Y^{\mu \nu} 
+ {g}_t^{Z^\prime} ~ \bar{t} \gamma^\mu P_R {Z}^{\prime}_{\mu}  t 
\label{eq:langrangian}
\end{eqnarray}
where ${F}^{\prime}_{\mu \nu}$ (${F}^{Y}_{\mu \nu}$) is the usual Abelian field strength for the ${Z}^\prime$ (hypercharge boson), 
${g}_t^{Z^\prime}$ is the ${Z}^\prime$ coupling to right-handed 
top quarks\footnote{One can easily include a coupling to the left-handed top (and bottom).
Our choice to ignore such a coupling fits well with typical RS models, balancing the need
for a large top Yukawa interaction with control over corrections to precision electroweak
observables.} (we will take ${g}_t^{Z^\prime}=3$ in our simulations).
The parameter $\chi$ encapsulates the strength of kinetic
mixing between the $Z^\prime$ and SM hypercharge bosons (even if absent in the UV, it is generated in the IR by loops of top quarks).
These extra terms in the lagrangian have a natural connection to Randall--Sundrum theories. 
The $Z^\prime$ represents the lowest KK mode of the 
$U(1)$ contained in $SU(2)_L\times SU(2)_R\times U(1)_{B-L}$.  It typically has mixing with the electroweak
bosons, resulting in strong constraints from precision data.  This will also be
the case when the $Z^\prime$ is a KK mode of the electroweak bosons.  We circumvent these constraints by considering a $Z^\prime$ whose mixing with the 
$Z$ is kinetic.  At large $Z^\prime$ masses this is not operationally different
from the mass-mixing case, but it allows us to consider lower mass $Z^\prime$s which
are not ruled out by precision data.

Through the RS/CFT correspondence \cite{ArkaniHamed:2000ds,Rattazzi:2000hs}, 
the extra-dimensional theory is thought to be dual to an approximately scale-invariant 
theory in which most of the Standard Model is fundamental, but with the 
Higgs and right-handed top largely composite.  
The Higgs couples strongly to composite fields, and the amount of admixture in
a given SM fermion determines its mass \cite{Contino:2003ve}.
In this picture, the
$Z^\prime$ is one of the higher resonances, built out of the same preons as 
$t_R$.  

More generically, in models of partial fermion compositeness, it is natural to expect that only the top quark couples sizably to a new strongly interacting sector.
As a simple example of  a UV completion (see Appendix A of \cite{Jackson:2009kg}) we can treat all SM
 fields (including $t_R$) as uncharged under $U(1)^\prime$.  We include a pair of
fermions $\psi_L$ and $\psi_R$, whose SM gauge quantum numbers are identical to
$t_R$, but with equal charges under $U(1)^\prime$.  
To realize coupling of the $Z^\prime$ to the top quark, we consider the gauge invariant
masses and Yukawa couplings of the top-$\psi$ sector,
\begin{eqnarray}
y H \bar{Q}_3 t_R  + \mu \bar{\psi}_L \psi_R + Y \Phi \bar{\psi}_L t_R
\end{eqnarray}
where $Q_3$ is the 3rd family quark doublet, $H$ is the SM Higgs doublet,
$\Phi$ is the Higgs field
responsible for breaking $U(1)^\prime$, $y$ and $Y$ are dimensionless couplings,
and $\mu$ is a gauge-invariant mass term for $\psi$.  The lightest mass eigenstate is identified as the
top quark. It can have a large coupling to $Z'$ through its $\psi$ component.

\begin{figure}[ht!]
\begin{center}
\includegraphics[width=0.65\textwidth ]{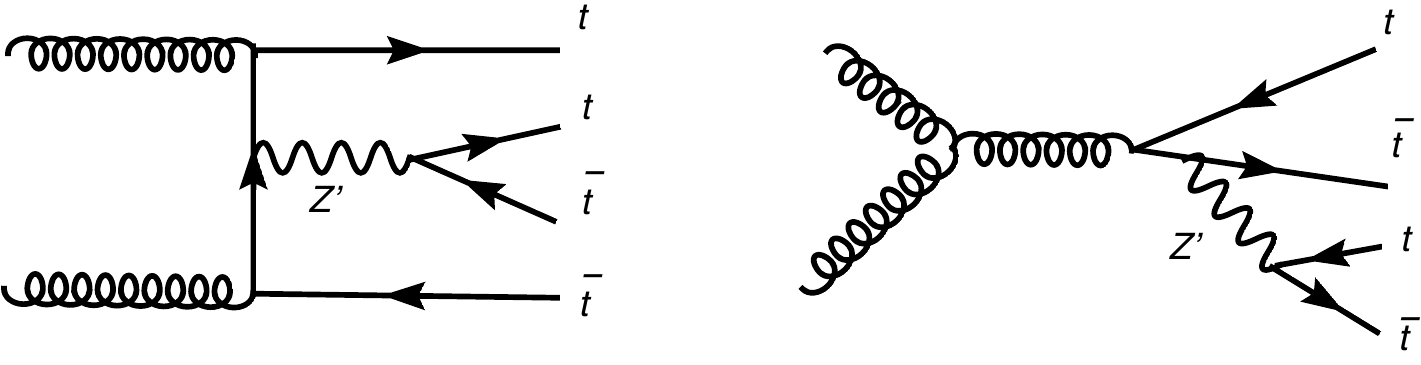}
\caption{\label{fig:diagramZp} \small
Four-top production via $Z^{\prime}$}
\end{center}
\end{figure}

Four-top production arises via the diagrams shown in Fig.~\ref{fig:diagramZp}. Given that the $Z^\prime$ under consideration has suppressed couplings to all SM fields (induced by the kinetic mixing $\chi$) but the top quarks, constraints  are weak and a mass of a few hundreds of GeV is allowed, which can lead to large four-top signals at the LHC. A detailed study is presented in \cite{GauthierServant} and compares with the non-resonant four-top events obtained from the effective four-fermion interaction $(\overline{t}_R\gamma_{\mu}t_R)(\overline{t}_R\gamma^{\mu}t_R)/\Lambda^2$ leading to the diagram \ref{fig:diagrams}(d). The corresponding cross sections  at LHC as a function of the $Z^\prime$ mass and $\Lambda$ are shown in Fig.~\ref{fig:crosssection}.  

\begin{figure}[ht!]
\begin{center}
\includegraphics[width=0.5\textwidth, clip ]{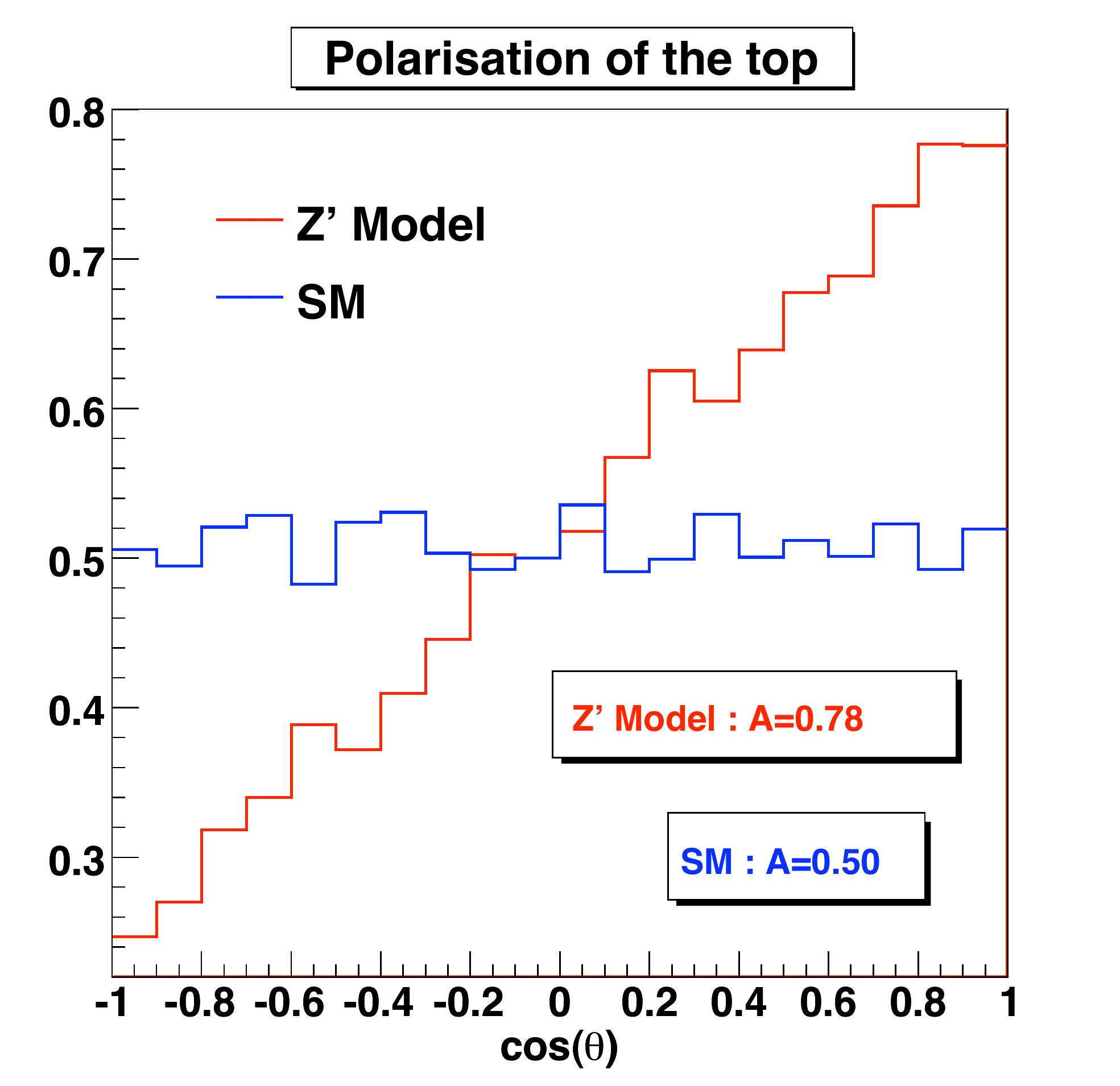}
\caption{\label{fig:polarisation} \small
Distribution of $cos(\theta)$ for the $Z'$ Model with $M_{Z'}=$800 GeV compared to the SM.}
\end{center}
\end{figure}

An interesting way to probe the properties of the top interactions relies on measuring the top polarization. 
The SM four top production being dominated 
by parity invariant QCD processes,  we expect to generate an equal number of left and right-handed
pairs. However, in the new physics models discussed here, there is a strong bias towards RH tops. The angular distribution of the leptons from the top decays enables to analyze the polarisation of the top quarks. The differential cross section can be written as
\begin{equation}
\frac{1}{\sigma}\frac{d\sigma}{d\cos \theta}=\frac{A}{2}(1+\cos \theta) +\frac{1-A}{2}(1-\cos \theta)
\end{equation}
where $\theta$ is the angle between the direction of the lepton in the top rest frame and the direction of the top polarization. The corresponding distribution is illustrated in Fig.~\ref{fig:polarisation}.

\begin{figure}[ht!]
\centering
\begin{minipage}[c]{0.5\textwidth}
\centering
\includegraphics[width=\textwidth]{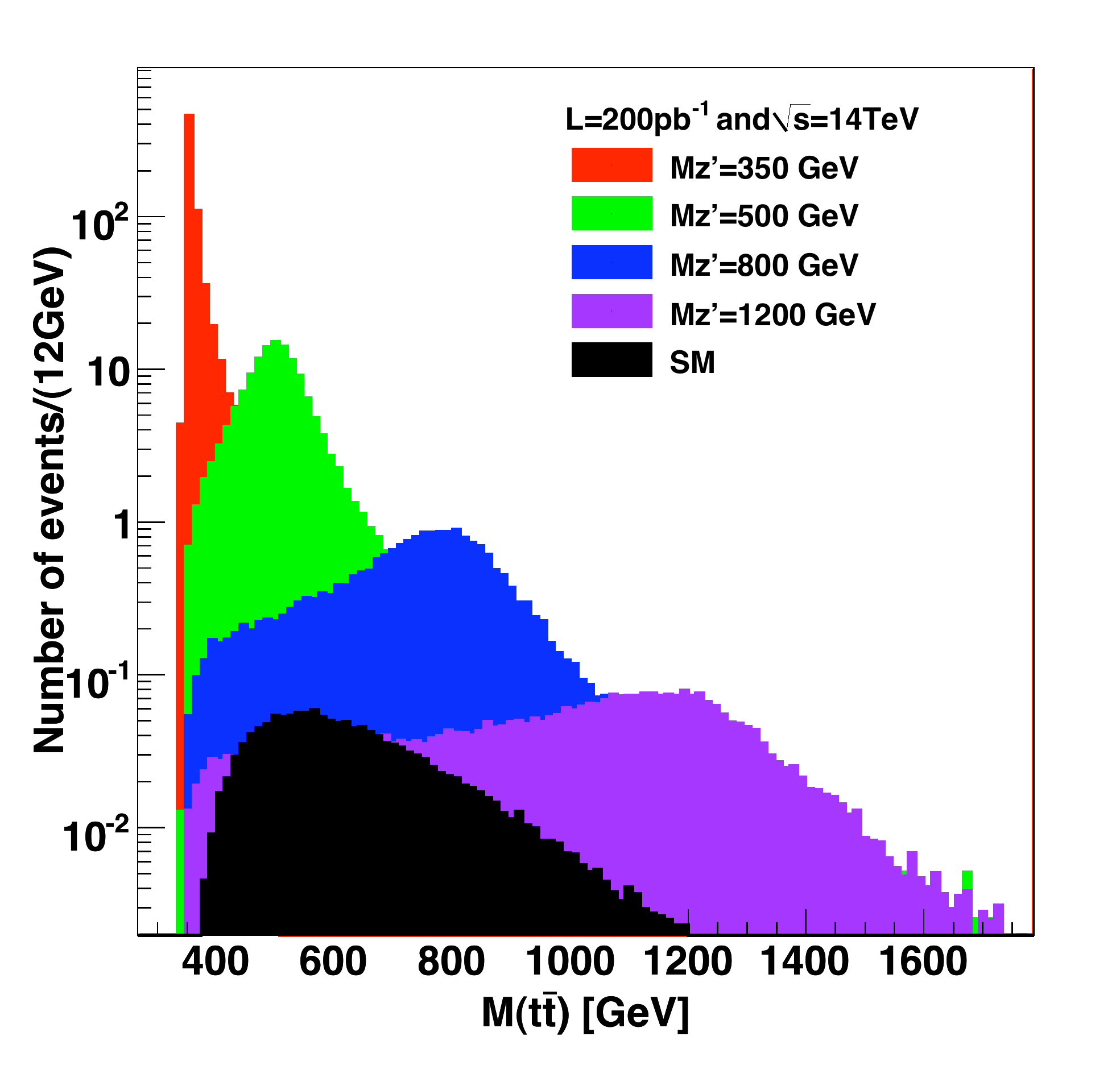}
\end{minipage}%
\begin{minipage}[c]{0.5\textwidth}
\centering
\includegraphics[width=\textwidth]{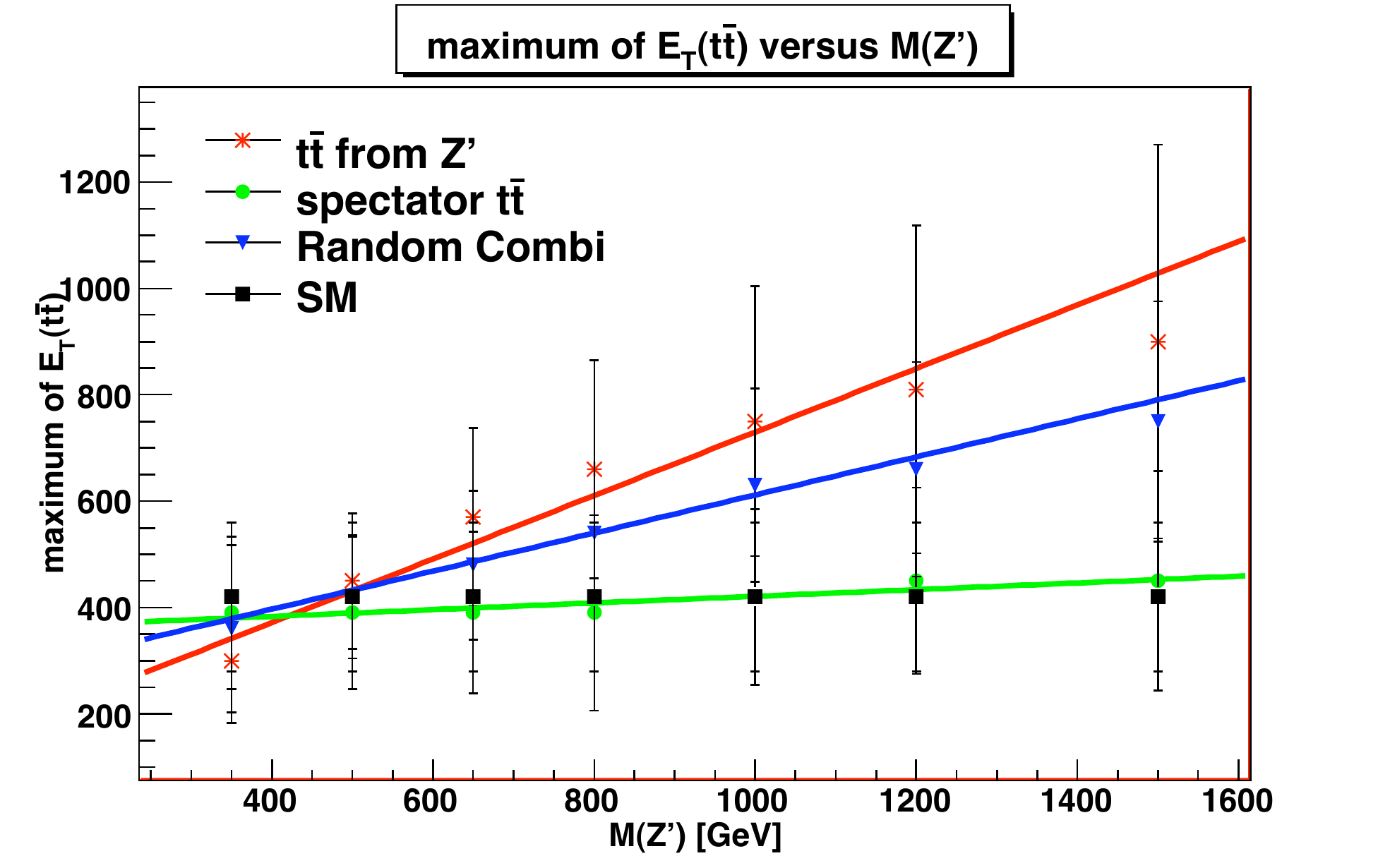}
\end{minipage}%
\caption{\label{fig:Mtt} \small (a)  Invariant mass of the $t\overline{t}$ pair coming from the $Z^{\prime}$
compared with that from the SM four-top events; (b) Position of the maximum of the $E_T$ distribution of the $t\overline{t}$ pair as a function of  $M_{Z^{\prime}}$.}
\end{figure}

\begin{figure}[ht!]
\begin{center}
\includegraphics[width=0.6\textwidth, clip ]{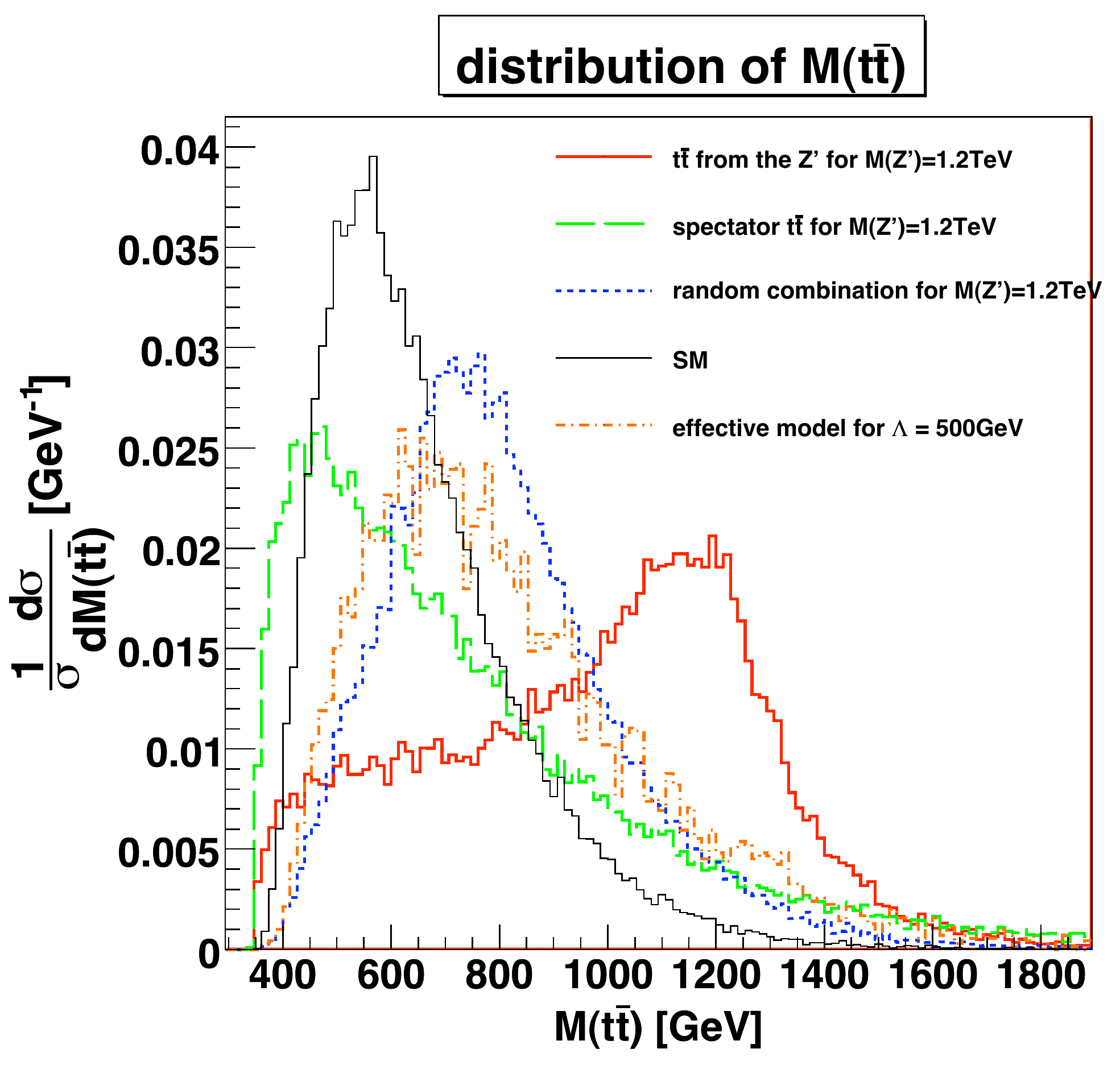}
\caption{\label{fig:Mttsuperpo} \small
Comparison of the different  $t\overline{t}$ invariant mass distributions.}
\end{center}
\end{figure}

In Fig.~\ref{fig:Mtt}, we show the invariant mass $M_{t\overline{t}}$ of the $t\overline{t}$ pair coming from the $Z^{\prime}$ for different $M_{Z^{\prime}}$ masses as well as $M_{t\overline{t}}$ from the SM four-top events. The latter peaks close to 600 GeV. We also display the maximum of the  $t\overline{t}$ pair  transverse energy distribution as a function of $M_{Z^{\prime}}$. Fig.~\ref{fig:Mttsuperpo} compares the  $M_{t\overline{t}}$ distributions of the  $t\overline{t}$ pair emitted by a $Z^{\prime}$ with $M_{Z^{\prime}}=1.2$ TeV,  the spectator $t\overline{t}$ pair, which peaks around 500 GeV and the $t\overline{t}$ pair produced by the effective 4-fermion contact interaction. 
%


\section{Reconstruction}
\label{fourtops:reconstruction} 

Reconstruction of four top events is a challenge to the detector and event reconstruction. The decay of the top quarks gives rise to twelve fermions. To benefit from the same-sign lepton signature two $W$ bosons must decay to lepton-neutrino. The presence of two escaping neutrinos then prevents a complete reconstruction of the twelve momenta. In the most abundantly produced final states, most of the remaining fermions will be quarks, giving rise to a large jet multiplicity. 

\begin{figure}[ht!]
\begin{center}
\includegraphics[width=0.325\textwidth, clip ]{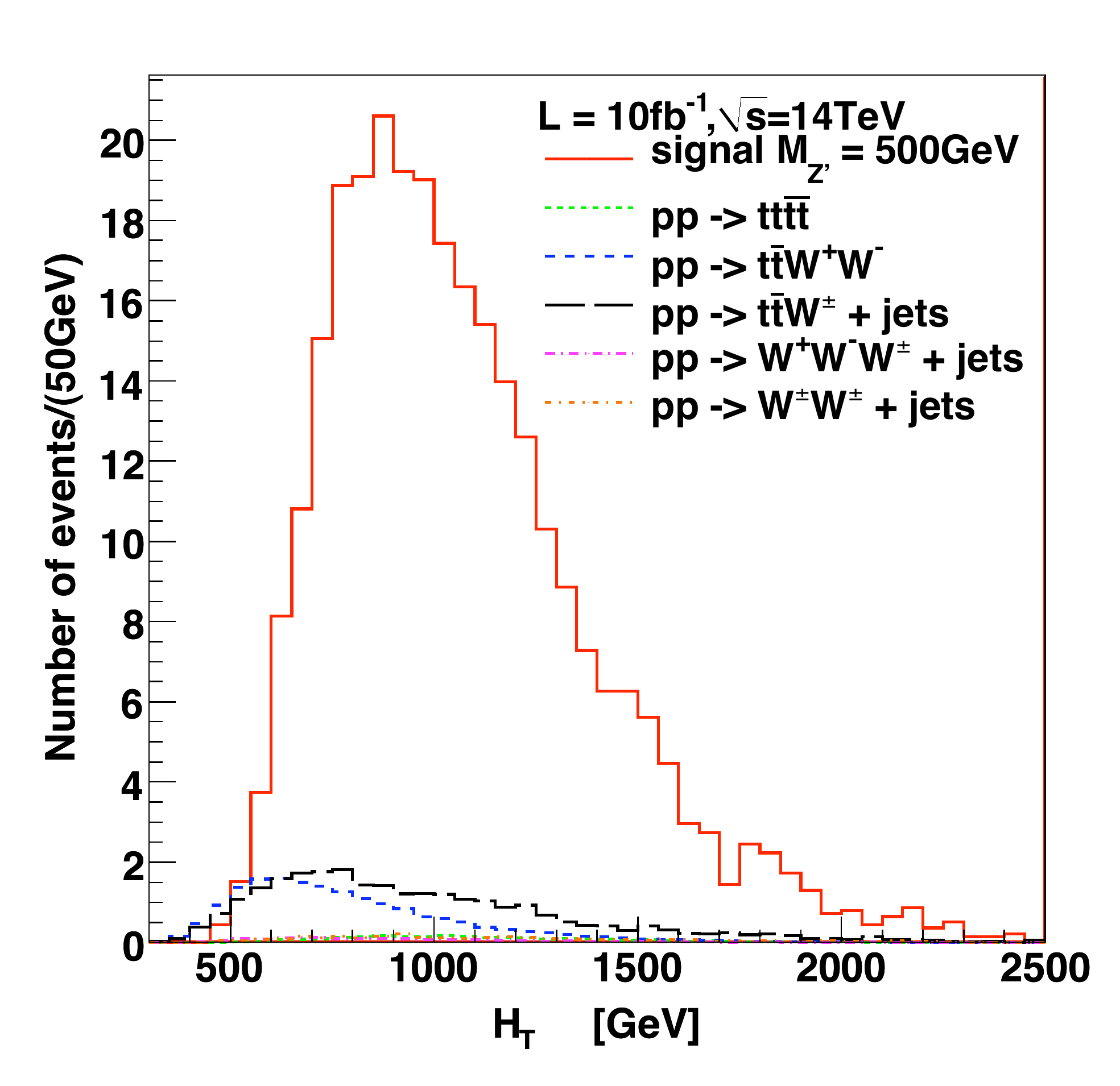}
\includegraphics[width=0.325\textwidth, clip ]{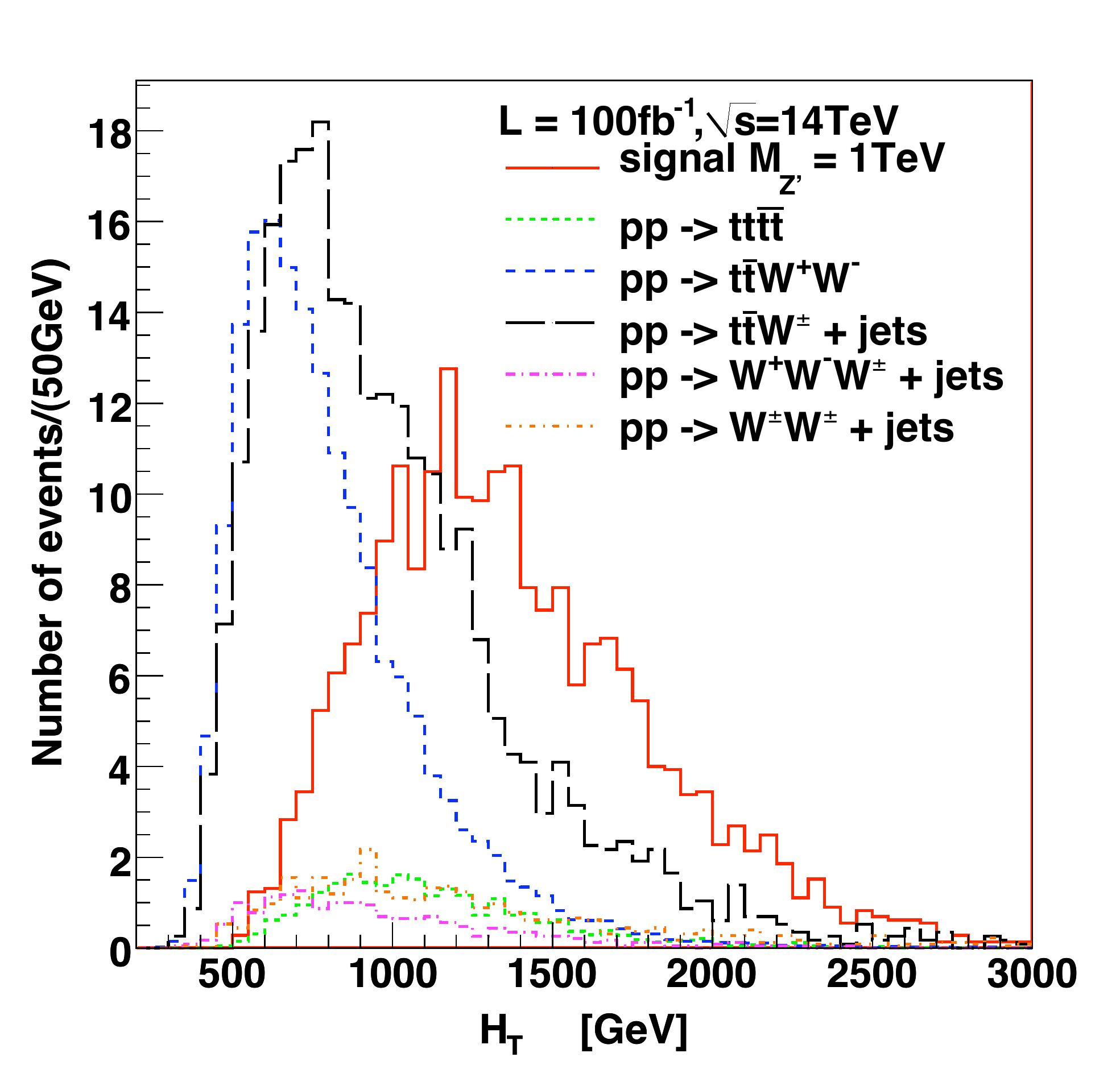}
\includegraphics[width=0.325\textwidth, clip ]{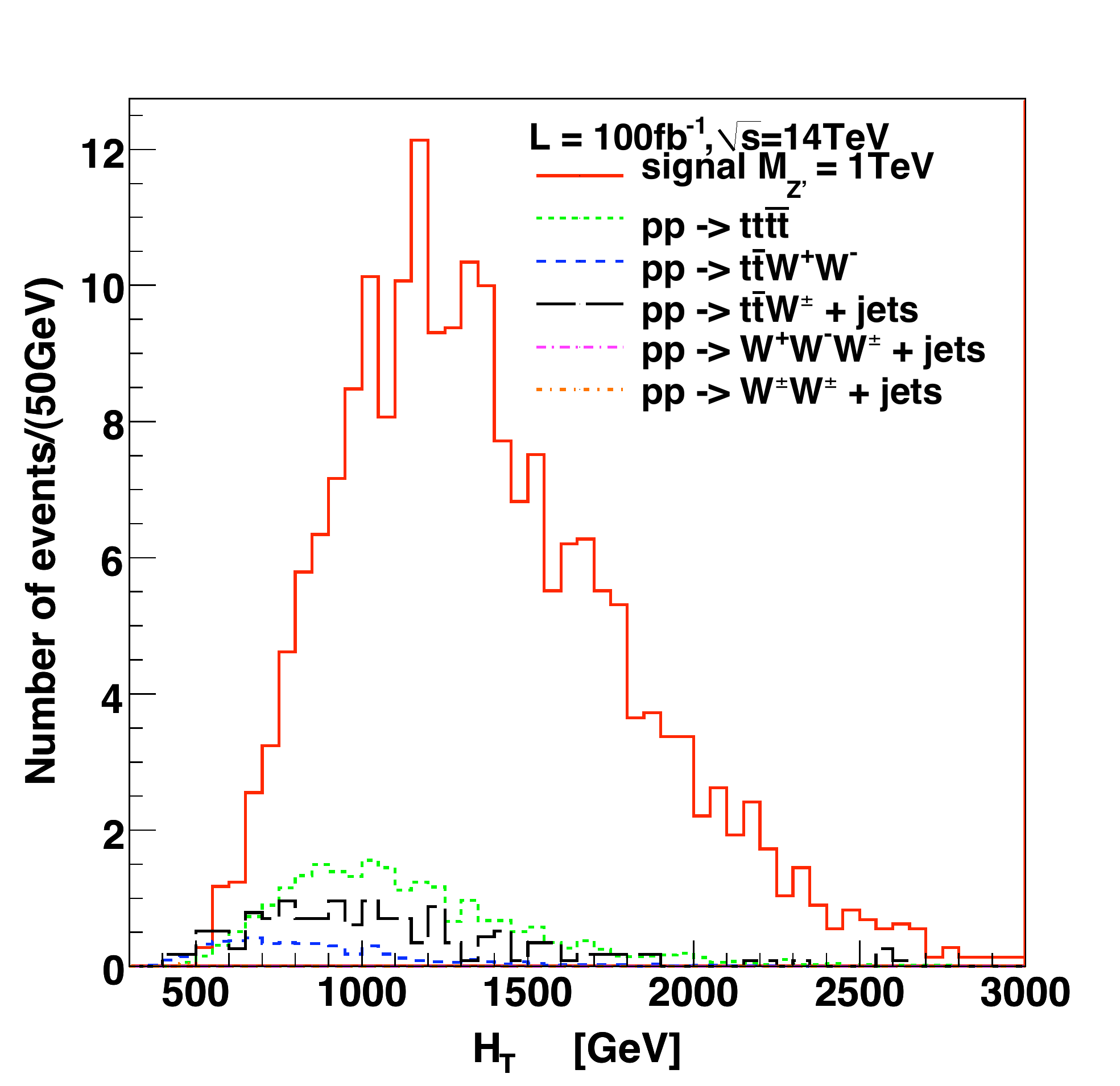}
\caption{ \small Total transverse energy  after demanding  $ n_j \geq 6$,  $p_{T}>30$ GeV (first two plots) and in addition $n_{b-jet}\geq 3$ (third plot).}
\label{fig:HT2bjets}
\end{center}
\end{figure}

The minimal approach to reconstruction merely registers the scalar sum of the transverse energy of all final state objects. The $ H_T $ distribution for a 500 GeV and 1 TeV Z' resonance as described in the previous section are shown in Figure~\ref{fig:HT2bjets}. For sufficiently large resonance mass, i.e. for $ m_{Z'} = $ 1 TeV in the central panel, the signal distribution clearly differs from that of some important (reducible) backgrounds like $ t \bar{t} W^\pm +$ jets and $ t \bar{t} W^+ W^- $.

A further experimental signature of the four-top final states is the large b-jet multiplicity which can be used as a powerful tool to extract the signal even coming from a heavy resonance as shown in Figure~\ref{fig:HT2bjets} and in \cite{GauthierServant}. Reconstruction of (some of) the top quarks in the event can provide additional handles to reduce the background. 

For a complete reconstruction of the $ t \bar{t} t \bar{t} $ final state, one must address the challenge of assigning 12 final state fermions to the four top candidates. Before exploring this very complex final state, we consider the  reconstruction of $ t \bar{t} $ events that is much better understood. 

For the reconstruction of $ t \bar{t} $ pairs the W-mass constraint and a top mass constraint are used to find the correct pairing. This approach is quite successful for $ t \bar{t} $ events and a bit less so for the (simulated) $ t \bar{t} H $ topology. An alternative approach presents itself when one considers the reconstruction of $ t \bar{t} $ events originating in the decay of a heavy resonance. Due to the boost of the top quark, its decay products are collimated in a narrow cone. This top {\em mono-jet} can be identified as such by techniques revealing the jet substructure~\cite{Agashe:2006hk, randall, Baur:2007ck, thalerwang, Almeida:2008yp, Kaplan:2008ie}. Importantly, for sufficiently large resonance mass the decay products of top and anti-top are cleanly separated. A simple assignment based on (geometrical) vicinity is sufficient to find the correct assignment of jets to top candidates. Thus, the ambiguities found in reconstruction of ``tops at rest'' disappear in regime of large top $ p_T $.

\begin{figure}[h]
  \begin{center}
    \mbox{
      \begin{tabular}{cc}
	\vspace{-0.6cm}

        \subfigure[]{\includegraphics[angle=0,width=0.3\linewidth]{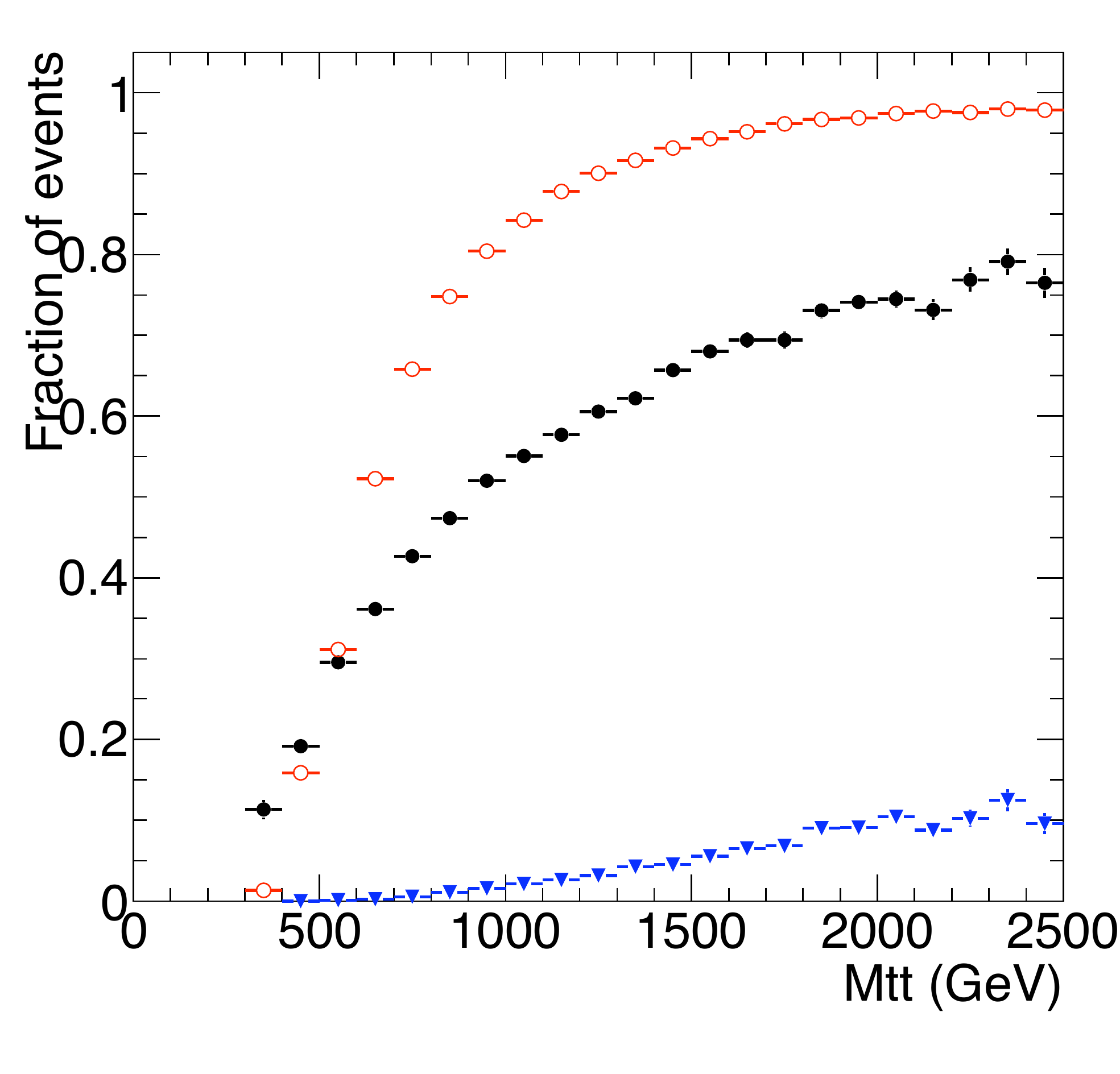}}
        \subfigure[]{\includegraphics[angle=0,width=0.3\linewidth]{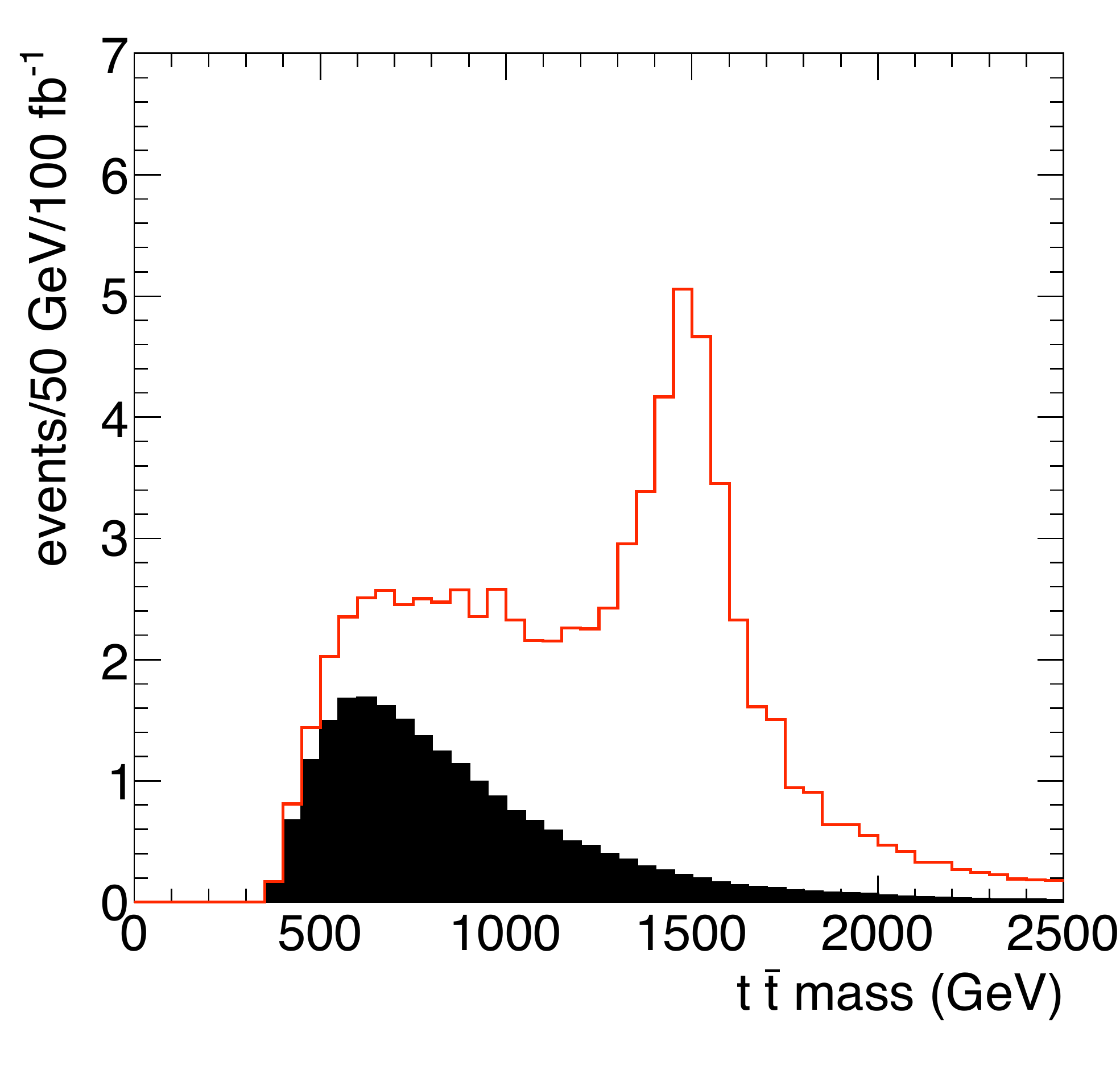}} 
        \subfigure[]{\includegraphics[angle=0,width=0.3\linewidth]{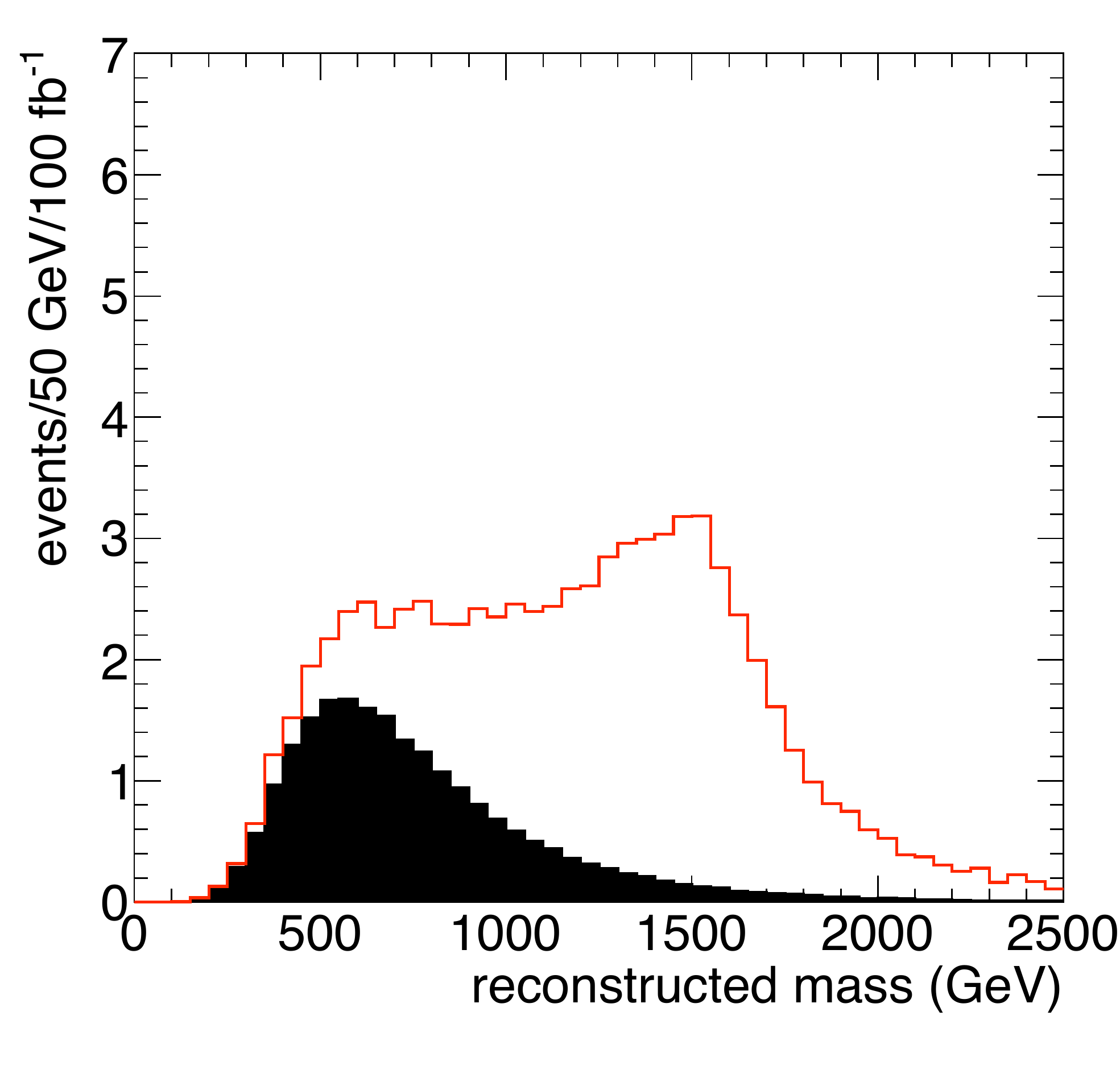}}
\\
      \end{tabular}
    }
  \end{center}
  \caption{The probability as a function of resonance mass that final state fermions are correctly assigned to top and anti-top quarks in $ t \bar{t} $ production (open circles) and $ t \bar{t} t \bar{t} $ production. The filled circles (triangles) indicate the probability to find two (four) correctly paired top quarks. The central panel shows the invariant mass distribution of the two top quarks with highest $p_T $ in SM $ t \bar{t} t \bar{t} $ production (filled histogram) and for production through a 1.5 TeV KK gluon. The rightmost panel shows the invariant mass of the two reconstructed clusters with highest $p_T $.}
  \label{fig:separation}
\vspace{0.1cm}
\end{figure}

To quantify this statement a parton level simulation of $ pp \rightarrow X \rightarrow t \bar{t} $ has been analysed. Lepton+jets events are selected, where one of the W bosons decays to a lepton and a neutrino and the second W boson decay to two jets. The neutrino is discarded and the momenta of the remaining five fermions is presented to the $ k_T $ algorithm~\cite{kt,kt2} for clustering~\footnote{The implementation in FastJet~\cite{fastjet} was used, with E-scheme recombination. The algorithms was used in exclusive mode, forcing it to return exactly two jets. The R-parameter was set to 2.5. For this, somewhat unusual, choice nearly all input objects are clustered into jets (rather than included in the ``beam jets'').}. Clustering is considered correct whenever all decay products from the top (and anti-top) quark are clustered together in a single jet. The result is represented with open circles in the leftmost plot of figure~\ref{fig:separation}. For tops produced at rest the probability of correctly clustering the event is essentially equal to 0. For resonant $ t \bar{t} $ production the probability to find the correct assignment increases rapidly as the resonance mass is increased. The decay products of the top and anti-top quark are collimated more and more in a narrow cone, while the top anti-top are emitted essentially back-to-back. Indeed, for a resonance mass of 1 TeV, the correct assignment is found in nearly eighty percent of events. For a more exhaustive discussion, and results including a complete detector simulation the reader is referred to reference~\cite{atlasttresonance}.

When repeating the exercise for $ t \bar{t} t \bar{t} $ production, the simple clustering has to deal with a much denser topology and is much less successful. As shown with blue triangles in figure~\ref{fig:separation} the probability to find a perfectly clustered event is less than 10 \% over the entire mass range studied here. Of course, the decay of a heavy resonance leads to only one pair of strongly boosted top quarks, while the $ p_T $ of the associated (spectator) top quarks remains relatively small. The third curve (filled circles) in~\ref{fig:separation}  represents the probability that at least two tops out of four are clustered correctly. This probability is quite large even for relatively small resonance mass, reaching approximately 60 \% for a 1 TeV resonance. 

The mass of the resonance is reconstructed as the invariant mass of the two objects with highest $p_T $ in the event. At the parton level this yields good results: the combination of the two top quarks with highest $p_T$ yields the distribution of the central panel of figure~\ref{fig:separation}. The resonance clearly stands out on top of the SM four top production (black). Applying the same criterion to the top quark candidates reconstructed by the clustering algorithm, the distribution in the rightmost panel is obtained. Obviously, the resonant signature is washed out by false combinations and the energy carried away by the escaping neutrinos. Still, the signal and background distributions can clearly be distinguished. 

The additional handle of highly boosted top quarks is found to be quite useful to reduce the combinatoric problem of four top events. Reconstruction of a resonant signature may well be feasible, thus turning the counting experiment into a resonance search. Given the simple-minded nature of this attempt to reconstruct this complex final state this result must be considered as encouragement to develop a more sophisticated approach.

\section{Conclusions}
\label{fourtops:conclusions} 

The four top final state is sensitive to new physics that is relatively unconstrained by precision measurements at LEP or resonance searches at the Tevatron. Examples are models where the top quark is composite, or where a new heavy particle couples strongly (or exclusively) to top quarks.

Reduction of Standard Model processes is achieved primarily through the requirement of two same-sign leptons. The small signal cross-sections (typically 10s of fb) render a counting experiment susceptible to large uncertainties due to large ($ t \bar{t} + $ jets)  backgrounds.

Partial or complete reconstruction of the event enhances the robustness of the measurement. In this contribution we have explored the reconstruction of this complex final state, at the parton level, in the case where a $ t \bar{t} $ pairs originates in the decay of a heavy resonance. The boost of the top quarks is found to greatly reduce the combinatorics involved in assigning the final state fermions. Even with the limited performance of our simple-minded algorithm, a mass peak may be reconstructed for a resonance mass of (or greater than) 1.5 TeV. Reconstruction thus provides a way to distinguish signal and SM background. 

\section*{Acknowledgements}

We thank Emmanuel Bussato and  Javi Serra for discussions.


%% file: Brooijmans/Brooijmans.tex
\chapter{LHC sensitivity to wide Randall--Sundrum gluon excitations}
\label{chapter:KKgluons}

{\it G.~Brooijmans, G.~Moreau and R.K.~Singh}

\begin{abstract}
We apply the results recently obtained by the ATLAS collaboration 
in the reconstruction of high mass $t\bar{t}$ resonances to Kaluza--Klein
excitations of gauge bosons as predicted in Randall--Sundrum models with
fields propagating in the bulk.  The resulting ATLAS sensitivity to 
such signals is determined.
\end{abstract}

\section{Introduction}
Randall--Sundrum (RS) models~\cite{Randall:1999ee} of extra dimensions are 
an attractive approach to dealing with the hierarchy problem (for a short review, see Contribution~\ref{chapter:RS} of these proceedings).  If, furthermore,
standard model fermions  and gauge bosons are allowed to propagate in the bulk,
the model can offer solutions to other major open questions, such as for example the 
existence of dark matter~\cite{Agashe:2004ci,Agashe:2004bm,Belanger:2007dx}.
An interesting way to constrain such models is to hypothesize~\cite{Djouadi:2006rk} that they are
the source of the 
deviation between the standard model prediction~\cite{Djouadi:1989uk} and experimental 
measurement
at LEP and elsewhere~\cite{:2005ema} of the forward-backward 
asymmetry $A^b_{FB}$ in $Z$ boson
decays to bottom quarks.  A study of such a scenario~\cite{Djouadi:2007eg}
\footnote{The production of the Kaluza--Klein excitations of the gluon was also studied in~\cite{Lillie:2007yh,Allanach:2009vz}.}
and more generally the geometrical
mechanism generating a large top quark mass~\cite{Brooijmans:2008se} implies  
that the principal LHC signature is the production of broad, high mass resonances (mainly due to the Kaluza--Klein excitation of the gluon)
decaying primarily to $t\bar{t}$ pairs. Indeed, electroweak precision tests indirectly force the first Kaluza--Klein mass of the gluon 
to be typically larger than $\sim 3$ TeV in a custodially protected framework~\cite{Agashe:2003zs,Bouchart:2008vp,Bouchart:2009vq}.

Due to the collimation of top quark decay products at large top quark 
momentum, the reconstruction of high mass $t\bar{t}$ resonances requires the
development of new experimental approaches~\cite{Agashe:2006hk,Fitzpatrick:2007qr}.  
The ATLAS collaboration recently released a full simulation 
study~\cite{ATL-PHYS-PUB-2009-081} describing the effectiveness of such
new techniques in the reconstruction of narrow $t\bar{t}$ resonances.  In this note,
these results are applied to the four concrete scenarios 
described in Ref.~\cite{Djouadi:2007eg} to 
estimate the integrated luminosity required to exclude these models.

\section{Simulation}

Both signal and $t\bar{t}$ continuum events are generated using 
the {\sc Sherpa}~\cite{Gleisberg:2008ta} event generator for proton-proton
collisions at $\sqrt{s}$ = 14 TeV.  All four scenarios from Ref.~\cite{Djouadi:2007eg}
are considered. 
These scenarios correspond to different localizations of the 
bottom/top quarks along the extra-dimension, all typically reproducing the bottom/top quark masses.  
These different localizations lead to variations in the wave function overlaps between the bottom/top quarks and the 
Kaluza--Klein excitation of the gluon (whose profile is peaked on the so-called TeV-brane), and in turn to
variations in the four-dimensional effective couplings.
The resulting total cross-sections for 2 $< m_{t\bar{t}} <$ 4 TeV
are given in Table~\ref{tab:xs}.

\begin{table}[h]
\centering
\begin{tabular}{|l|c|}
\hline
Signal Model & Cross-section (2 $< m_{t\bar{t}} <$ 4 TeV) (fb) \\
\hline \hline
Standard Model (SM) Only & 365\\  
E1 + SM & 620 \\
E2 + SM & 560 \\
E3 + SM & 615 \\
E4 + SM & 535 \\
\hline
\end{tabular}
\caption{Cross-sections for production of $t\bar{t}$ pairs with 
2 $< m_{t\bar{t}} <$ 4 TeV in the standard model and standard model plus 
the Kaluza--Klein excitations of all the neutral gauge bosons for the model points considered.}
\label{tab:xs}
\end{table}

\section{Experimental reconstruction efficiencies}

The ATLAS collaboration released a study~\cite{ATL-PHYS-PUB-2009-081} of the 
reconstruction efficiency 
of high mass $t\bar{t}$ resonances in the lepton ($e$ and $\mu$) plus jets
channel using splitting scales obtained from reclustering the jets using 
the $k_\perp$ jet clustering algorithm.  In this analysis, fully hadronic top
quark decays were identified using a 
combination of jet mass and the first three $k_\perp$ splitting scales into 
a likelihood variable $y_L$. 
A combination of the fraction 
of visible top mass carried by the lepton~\cite{Thaler:2008ju} and relative
$p_T$ of the lepton w.r.t. the jet were used to tag semileptonic top quark decays.  One
of the main
conclusions of the analysis is that after cuts, the by far dominant background
is the standard model $t\bar{t}$ continuum.  Backgrounds from QCD multijet 
and non-$t\bar{t}$ $W$ boson plus jets are found to be substantially smaller.  
For the study described in this paper, 
the ATLAS operating point chosen is the one with a cut on the hadronic top
likelihood variable $y_L >$ 0.6. The total $t\bar{t}$ selection efficiency
for $e,\mu$ + jets events is parametrized as a function of the hadronically 
decaying top quark's transverse momentum ($p_T^{top}$) with a linear increase from 0\% to 35\% for 
$p_T^{top}$ from 500 to 900 GeV, and a constant value at 35\% for values $p_T^{top} >$
900 GeV.  This selection efficiency is applied at truth level to the generated events.
The $t\bar{t}$ invariant mass is then smeared by 5\% to reflect the resolution 
found in the ATLAS study.

\section{Results}

The semi-frequentist $CLs$ method~\cite{Junk:1999kv} based on a Poisson log-likelihood test 
statistic is used to determine the sensitivity after taking into account branching 
ratios and reconstruction efficiencies.  The invariant mass distribution between 
background-only and signal + background are compared, taking into account flat 
systematic uncertainties on the integrated luminosity (6\%), reconstruction 
efficiency (10\%) and background cross-section (15\%).

The resulting luminosities (for $\sqrt{s} =$ 14 TeV) to exclude the 
model points at 95\% C.L. are given in Table~\ref{tab:limits}.
In all cases these are well below 10 fb$^{-1}$, showing these models
should be accessible in the first few years of LHC running at $\sqrt{s}=$ 
13 or 14 TeV.
\begin{table}[h]
\centering
\begin{tabular}{|l|c|}
\hline
Signal Model & Integrated Luminosity for 95\% C.L. Exclusion (fb$^{-1}$) \\
\hline \hline
E1 + SM & 2.5 \\
E2 + SM & 5.4 \\
E3 + SM & 1.8 \\
E4 + SM & 6.7 \\
\hline
\end{tabular}
\caption{Required integrated LHC luminosities for 95\% C.L. exclusion
of the model points for $\sqrt{s} =$ 14 TeV.}
\label{tab:limits}
\end{table}

\section{Conclusions}

The LHC reach for broad, high mass excitations of the gauge bosons decaying to 
$t\bar{t}$ final states in Randall--Sundrum models has been investigated using
the results of a full simulation study of high $p_T$ top quark reconstruction.
Using the lepton plus jets channel, the LHC experiments should be sensitive to such 
models with integrated luminosities smaller than 10 fb$^{-1}$ collected at 
$\sqrt{s} = 14$ TeV.

\section*{Acknowledgements}
The authors would like to thank the Les Houches workshop organizers
for a very stimulating and enriching workshop.


%% file: Deandrea/resonances.tex
\chapter{Effects of nearby resonances at colliders}

{\it G. Cacciapaglia, A. Deandrea and S. De Curtis}

\begin{abstract}
We describe propagators for particle resonances taking into account the quantum mechanical
interference due to the width of two or more nearby states with common decay channels, 
incorporating the effects arising from the imaginary parts of the one-loop self-energies. 
The interference effect, not usually taken into account in Montecarlo generators, can
modify the cross section or make the more long-lived resonance narrower. We give examples 
of New Physics models for which the effect is sizable for collider physics.
\end{abstract}

\section{Introduction}
In the following we consider a generalisation of the Breit--Wigner
description \cite{Pilaftsis:1997dr} which makes use of a matrix propagator including
non-diagonal width terms in order to describe physical examples in which these effects are relevant.
Indeed for more than one meta-stable state coupled to the same particles, loop effects generate mixings
for the masses as well as mixed contributions for the widths
(imaginary parts). In general a diagonalisation procedure for the
masses (mass eigenstates) will leave non-diagonal terms for the
widths. Usually non-diagonal width terms are discarded. When
two or more resonances are close-by and have common decay channels
such a description is not accurate. The usual Breit--Wigner approximation amounts to sum
the modulus square of the various amplitudes neglecting the
interference terms and this is the usual procedure in Montecarlo generated events. 
When there are common decay channels and the
widths of the unstable particles are of the same order of the mass splitting, the interference terms
may be non-negligible. In the following we shall 
consider models of physics Beyond the Standard Model (BSM) in which new
resonances play a crucial role. Based on these results we suggest that a proper
treatment  should be carefully implemented into Monte Carlo
generators as physical results may be dramatically different from a
naive use of the Breit--Wigner approximation.

\section{The formalism}
We discuss here only the formalism for scalar fields which gives a simpler
overview of the problem without the extra complications of the
gauge and Lorentz structure of the general case. A more detailed analysis can be done also including vector resonances
\cite{Cacciapaglia:2009ic}. 

For a system involving many fields, which do couple to the same intermediate particles,
loops will generate mixings in the masses, but also out-of-diagonal imaginary parts.
In general the real and imaginary parts will not be diagonalisable at the same time. 
The kinetic function is in general a matrix :
\beq
(K_s)_{lk} = (p^2 - m_l^2) \delta_{lk} +  i \Sigma_{lk} (p^2)\,.
\eeq
(We are considering the imaginary part only, the
real one is used to renormalise the masses.) The propagator of the
fields can be defined as the inverse of the matrix:
\beq i (\Delta_s)_{lk}  = i \left( K_s^{-1} \right)_{lk} \,.
\label{delta}\eeq
For simplicity we give here the two-particle case :
\beq \label{eq:twoscalar}
i \Delta_s = \frac{i}{D_s} \left( \begin{array}{cc}
p^2 - m_2^2 + i \Sigma_{22} & -i \Sigma_{12} \\
-i \Sigma_{21} & p^2 - m_1^2 + i \Sigma_{11}
\end{array} \right)\,,
\eeq where \beq D_s = (p^2 - m_1^2 + i \Sigma_{11})(p^2 - m_2^2 + i
\Sigma_{22} ) + \Sigma_{12} \Sigma_{21}\,. \eeq For vanishing
$\Sigma_{12}$ and $\Sigma_{21}$, the propagator is diagonal and  it
reduces to two independent Breit--Wigner propagators with $m_i
\Gamma_i = \Sigma_{ii} (m_i^2)$.

However, the narrow width approximation is not valid if the off-diagonal terms are sizable compared with the mass splitting.
Defining $2 M^2 = m_2^2+m_1^2$
and $2 \delta = m_2^2 - m_1^2$, the poles of the propagator (zeros
of $D_s$) are:
\beq \tilde{m}_{\pm}^2 = M^2 - i \frac{\Sigma_{11} + \Sigma_{22}}{2}
\pm \frac{i}{2} \sqrt{(\Sigma_{22} - \Sigma_{11} + 2 i \delta)^2 + 4
\Sigma_{12} \Sigma_{21}}\,. 
\eeq 
Note that the value of the masses is modified by the presence of the off-diagonal terms due to the
imaginary part of the square root, at the same time the widths are
affected. More importantly, the off-diagonal terms in the propagator
will generate non-negligible interference, which can be in turn
constructive or destructive. 

\section{Numerical examples}
We first study two heavy Higgses where both
the scalars develop a vacuum expectation value (VEV) and therefore couple to the $W$ and $Z$
gauge bosons. This situation is common in supersymmetric models
where two Higgses are required by writing supersymmetric Yukawa
interactions for up and down type fermions, and generic two Higgs
models. The interference between near degenerate Higgses has been
studied in \cite{Ellis:2004fs,Frank:2006yh,Hahn:2007it} focusing in
CP violation effects.

The couplings of the two CP-even Higgses to gauge bosons can be written as
\beq
\lambda_{WWH1} = g\, m_W \cos \alpha\,, &\qquad& \lambda_{WWH2} = g\, m_W \sin \alpha\,, \nonumber\\
\lambda_{ZZH1} = \frac{g\, m_Z}{\cos \theta_W} \cos \alpha\,,
&\qquad& \lambda_{ZZH2} = \frac{g\, m_Z}{\cos \theta_W} \sin
\alpha\,; \label{couplings}\eeq where $\alpha$ is a mixing angle
taking into account the mixing between the two mass eigenstates and
the difference between the two VEVs.

\begin{figure}[tb]
\begin{center}
\includegraphics[width=14cm]{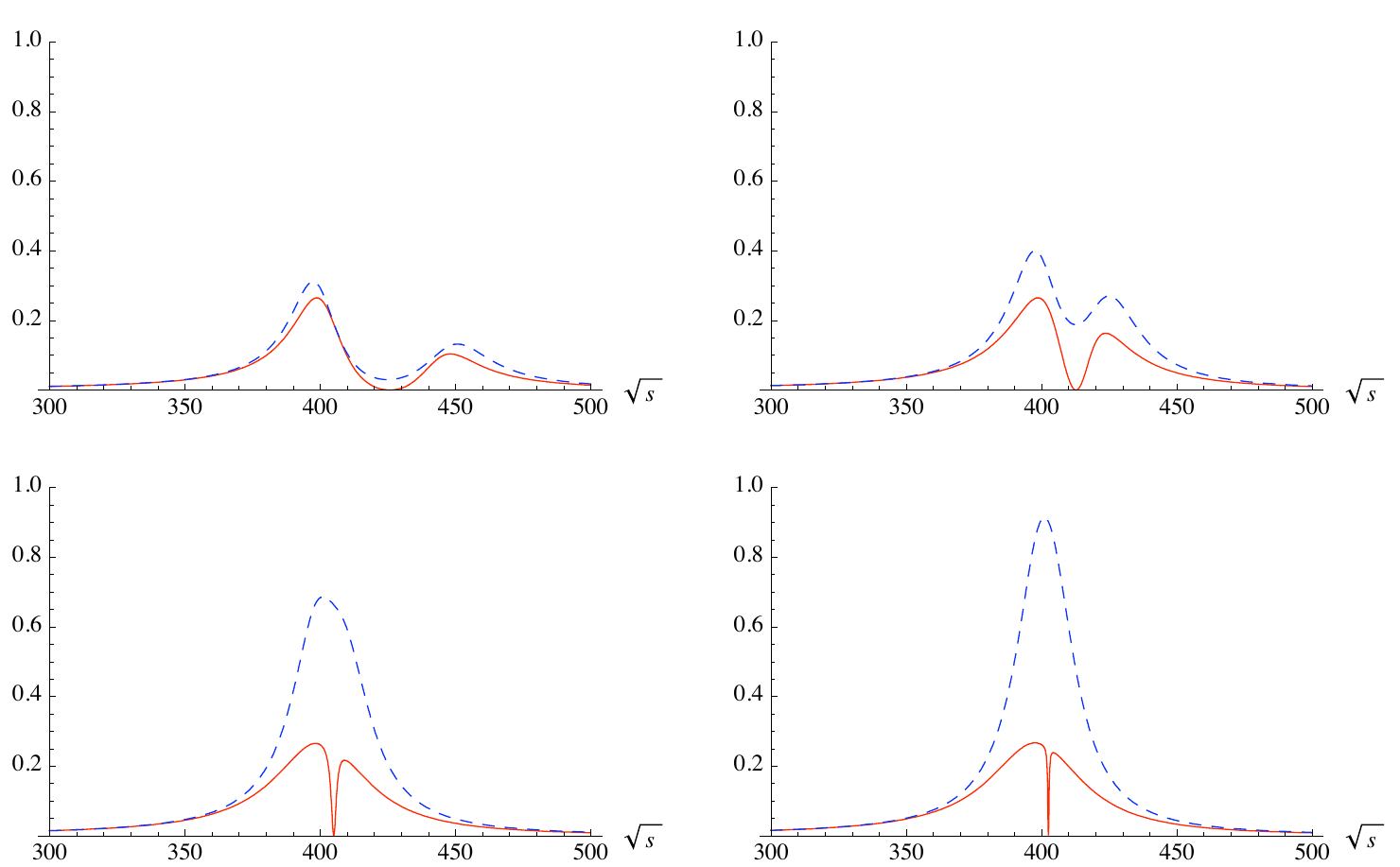}
\end{center}
\caption{\footnotesize Plots of the production cross section (in
arbitrary units) of two nearby Higgses decaying into gauge boson
pairs for the naive Breit--Wigner (blue-dashed) and exact mixing
(red-solid). The mass of the first resonance is fixed to $400$ GeV,
the splitting respectively 50, 25, 10 and 5 GeV and $\alpha=\pi/4$.}
\label{fig:gnomettiHiggs}
\end{figure}

Here we are interested in a generic production cross section of the
two nearby Higgses on the resonances, with decay of the Higgses into
gauge bosons (either $WW$ or $ZZ$). The amplitude of this process is
proportional to the resonant propagator weighted by the couplings
given in eq.(\ref{couplings}). In the case we are considering, the
common decay channels can give off-diagonal terms in
eq.(\ref{eq:twoscalar}) which are sizable compared with the mass
splitting. Here we assume that the coupling to the initial particles are the same :
\beq
\left|(\Delta_s^{11} + \Delta_s^{21}) \cos \alpha + (\Delta_s^{22}+ \Delta_s^{12} ) \sin \alpha \right|^2\,.
\eeq
In Fig.~\ref{fig:gnomettiHiggs}, we plot this quantity in arbitrary units and compare it with the
Breit--Wigner approximation:
 we fix $m_{H1} = 400$ GeV, and vary the splitting from
50 to 5 GeV. For simplicity, in the following we will assume $\alpha
= \pi/4$, so that the two scalars have the same couplings. The exact treatment
of the resonances unveils a destructive interference that  can drastically
reduce the cross section. 

\begin{figure}[tb]
\begin{center}
\includegraphics[width=14cm]{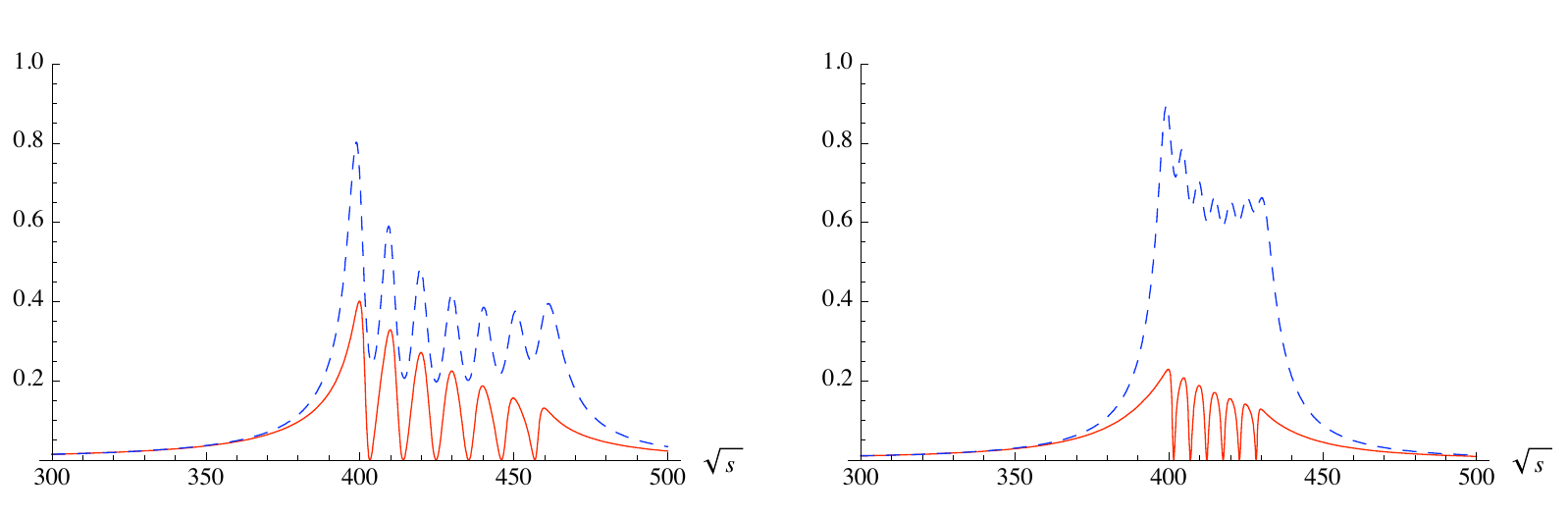}
\end{center}
\caption{\footnotesize Plots of the production cross section (in
arbitrary units) for seven nearby Higgses equally coupled to SM gauge
bosons: the naive Breit--Wigner (blue-dashed) bump reduces to a row of seven dwarfs when the exact mixing
(red-solid) is taken into account. The mass of the first resonance is fixed to $400$ GeV,
the splitting between the six Higgses respectively 10 and 5 GeV.}
\label{fig:gnometti}
\end{figure}

This effect can be even more important for scenarios with a large
number of scalars as predicted in some string models. Our analysis
can be easily extended to an arbitrary number of Higgses. Let's take
for example the couplings to the gauge bosons  to be given by
$g_{\rm SM}/\sqrt{N}$, where $g_{SM}$ is the SM coupling of the
gauge bosons and $N$ is the number of Higgses. In
Fig.~\ref{fig:gnometti} we plot the cross section for seven nearby
Higgses, with the first one at 400 GeV and the others at a distance
of 5 and 10 GeV, the width of each being 6.2 GeV. From the plot it
is clear that the destructive interference reduces the giant
resonance (which is not distinguishable from a single Higgs, once
the experimental smearing is taken  into account) to a bunch of
{\it gnometti} (dwarfs), which will be very hard to detect. The cross
section is in fact reduced by a significant factor with respect to the
naive expectation, and the smearing will wash out the peak
structure. It is intriguing to
compare this analysis with Un-Higgs
models~\cite{Stancato:2008mp,Falkowski:2008yr}, where the Higgs in
indeed a continuum: such behaviour may arise from the superposition
of Kaluza-Klein resonances in extra dimensional realisations or
deconstructed models.

Another striking example involving vector resonances is given by Higgsless models 
\cite{Csaki:2003zu,Cacciapaglia:2004rb}, 
where the first two neutral resonances are nearly degenerate, and they correspond to the first KK excitation
of the $Z$ and of the photon. The masses can be approximated by
\beq m_{Z'}^2 \simeq m_{KK}^2 + 4 m_Z^2\,, \qquad m_{A'}^2 \simeq
m_{KK}^2\,, \eeq so that the mass difference is very small:
\beq
m_{Z'} - m_{A'} \simeq 2 \frac{m_Z^2}{m_{KK}} \sim 16\, {\rm GeV} \cdot \left( \frac{1 {\rm TeV}}{m_{KK}} \right)^2\,.
\eeq
In terms of the parameters of the warped geometry ($R$ is the curvature, $R'$ the position of the Infra-Red brane in covariant coordinates):
\beq
m_{KK} \sim \frac{2.4}{R'}\,, \qquad m_W = \frac{1}{R' \log \frac{R'}{R}}\,;
\eeq
therefore, given the value of the curvature $R$, the KK mass ($R'$) is determined by the $W$ mass.
The determination of the couplings is more involved and we refer to \cite{Cacciapaglia:2009ic} for details.

We consider the following processes: Drell--Yan production and decay
into gauge bosons $W^+ W^-$ (DY), Drell--Yan production and decay
into a pair of leptons (Leptonic) and vector boson fusion production
followed by decay into gauge bosons (VBF).
As in the scalar case, the amplitudes at the resonance are proportional to the propagators weighted by the couplings with the incoming and outcoming particles.

\begin{figure}[tb]
\begin{center}
\includegraphics[width=16cm]{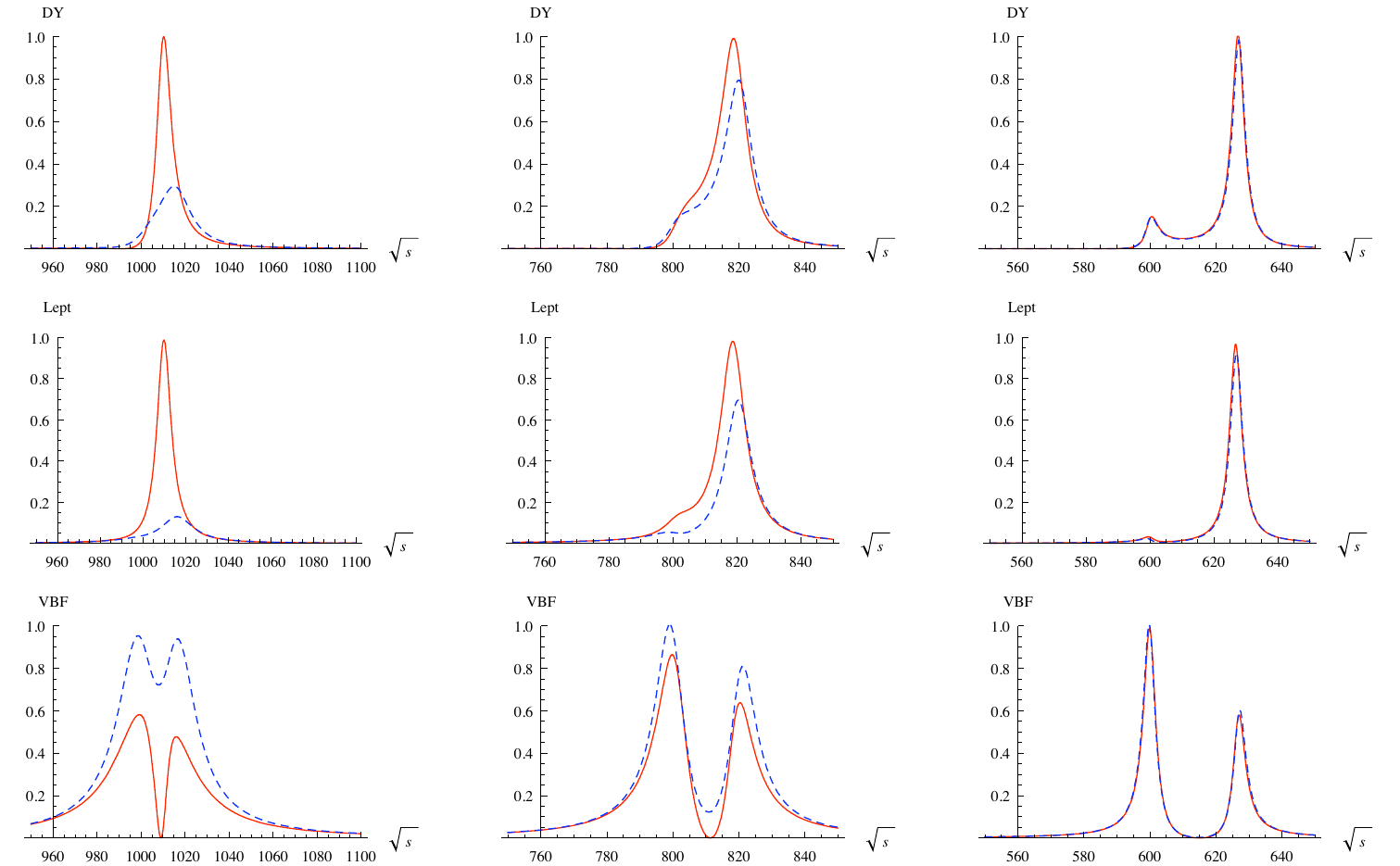}
\end{center}
\caption{\footnotesize Plots of production cross section (in
arbitrary units) of the two low-lying neutral resonances of the
Higgsless model for the naive Breit--Wigner (blue-dashed) and exact
mixing (red-solid). The rows correspond (from top to bottom) to DY,
Leptonic and VBF; the columns (from left to right) correspond to
$m_{KK} = 1000$ GeV, $800$ GeV and $600$ GeV.}
\label{fig:gnomettiHless}
\end{figure}
In  Figure \ref{fig:gnomettiHless} we plot, for illustrative
purposes, the squared matrix element of the three resonant
production channels for $A'$ and $Z'$ as function of $\sqrt{s}$ for
 three different cases: $m_{KK} = 1000$ GeV, $800$ GeV and $600$
GeV. For large masses, the effect of the interference is very
important and it can affect the value of the cross section
significantly. We give in the table the ratio
of the area under the peaks in the figure obtained by the exact
formula and the BW case. This roughly corresponds to the ratio of
the integrated cross sections.
\begin{center}
\begin{tabular}{cccc}
\hline
$M_{KK}=$ & 1000 GeV & 800 GeV & 600 GeV \\
\hline
DY: & 1.6 & 1.15 & 1.02 \\
Lept: & 3.15 & 1.4 & 1.05 \\
VBF: & 0.6 & 0.8 & 0.97 \\
\hline
\end{tabular} \end{center}
In the VBF channel there can be a reduction up to 50\%, while in the
other two channels the interference is constructive and the total
cross section can be enhanced by a factor of 2--3. The interference
is therefore extremely important, especially in the TeV region.
Since this represents the upper bound for Higgsless models, the
interference effects are crucial to determine if the whole
Higgsless parameter space can be probed at the LHC. 

\section{Conclusions}
We have shown that for two or more unstable particles, when there
are common decay  channels and the masses are nearby, the
interference terms may be non-negligible. This kind of scenario is not
uncommon in models of New Physics beyond the Standard Model,
especially in models of dynamical electroweak symmetry breaking or
in extended Higgs sectors. In models with multi-Higgses and in Higgsless models with
near degenerate neutral vector resonances, we showed that
interference induced by the off-diagonal propagators are very
important and they can either suppress or enhance the total cross
sections on resonance depending on the relative sign of the couplings to the initial and final states.
The interference effects can be crucial to study
the phenomenology of such models at the LHC, and to determine its
discovery potential. A proper treatment should be
carefully and systematically implemented into Monte Carlo generators
used to study BSM models.

%% file: Henderson/Henderson.tex




\chapter{An exotic photon cloud trigger for CMS}


{\it C.~Henderson}\\

\begin{abstract}
We propose a novel trigger to be sensitive to a new kind of
beyond-Standard-Model physics: a `photon cloud', consisting of $\sim
200$~GeV of transverse energy emitted through many soft photons
($\lesssim 1$~GeV each).
Such an exotic event could potentially be overlooked by conventional
trigger configurations.
We demonstrate that by considering a simple new variable,
\sumet\ in the electromagnetic calorimeter, which is straight-forward 
to calculate in a high-level trigger, an experiment could be sensitive
to this photon-cloud scenario. 
We estimate rates for this trigger for the expected LHC early
luminosity scenario and show that the proposal is feasible.
This trigger is planned to be implemented in the CMS experiment for
the 2010 LHC running period. 

\end{abstract}



At the Large Hadron Collider (LHC), the eventual goal is to collide
proton bunches at rate up to 40~MHz. However, the rate of events which
can be written to permanent storage is limited to $\sim$ 300~Hz.
Therefore a sophisticated trigger system is required to make an online
selection of the most interesting collision events to be recorded. 

At the Compact Muon Solenoid (CMS) detector~\cite{Bayatian:2006zz,Ball:2007zza}, 
a two-level trigger system is employed. 
At Level~1~\cite{Henderson:cms:trigger_tdr1}, lower resolution information (with
full eta-phi coverage)
from the calorimeter and the muon chambers is 
used to create particle candidates, and specially-programmed firmware
selects events at a rate up to 100 kHz.
A Level~1 accept initiates the complete detector readout, and the full
event is made available to the second trigger stage, the High-Level
Trigger (HLT)~\cite{Henderson:cms:trigger_tdr2}. 
This comprises essentially the full event reconstruction software,
running on a large PC farm. 
Standard particle objects such as jets, photons, muons, electrons,
etc\ldots are all
reconstructed, and form the basis for further event selection.
Generally speaking, events containing high-$E_T$ particles or
combinations of particles are chosen, 
with the goal of selecting 
the most interesting events for permanent storage,
up to a maximum rate of $\sim$ 300~Hz.
The performance of the CMS trigger system in cosmic-ray operations
during 2008 is described in \cite{:2009dq,Collaboration:2009ic}.

The trigger selection is therefore a crucial part of the experiment -
events that do not pass the trigger can never be analysed.
Thus it is of critical importance that we consider all possible types
of collision event, including those arising from exotic new physics
beyond the Standard Model, and ensure that they are not being
unwittingly rejected by the online trigger selection.
Here we consider an unusual type of event topology
that could potentially be missed by conventional trigger
configurations, and propose a novel trigger selection to remedy this.


The unusual event topology that we will consider is a `photon
cloud', which we take to be a large amount of transverse energy 
($\gtrsim 100-200$~GeV) that is emitted as a large number of
soft photons ($\lesssim 1$~GeV each).
An exotic physics scenario which
could potentially produce such a photon cloud is described
in~\cite{Harnik:2008ax}.  
The authors consider an extension of the Standard Model that contains
a new asymptotically free SU(N') gauge force (called QCD') and new
fermions $q'$ which are 
charged under this force (and also carry some SM gauge quantum
numbers, so they can be pair-produced in sufficiently high-energy
collisions). 
For such a gauge field, a wide range of values for the confinement
scale $\Lambda$ can be considered `natural', since
the confinement scale is related to the fundamental gauge coupling
$g_0$ defined at a scale $\mu_0$ by: 
\begin{equation}
  \Lambda = \mu_0 e^{-8\pi^2/bg_0^2}
\end{equation}
where $b$ is the one-loop coefficient of the SU(N') beta function. 

The novelty in this scenario arises when, unlike QCD, one considers
the confinement scale $\Lambda << m_{q'}$, 
where we take $m_{q'}$ to be a few hundred 
GeV or greater (above current observable limits).
This scenario was first discussed in~\cite{Okun:1980kw} and later revived in
the context of the LHC in~\cite{Kang:2008ea}, where the new fermions were
given the name ``quirks''.
The model of~\cite{Harnik:2008ax} specifically considers scalar
quirks (``squirks''), which arise in the context of a folded
supersymmetry scenario~\cite{Burdman:2006tz},
where the superpartners which cancel the ultraviolet divergences
in the Higgs mass due to SM quarks, are charged under the new QCD'
force rather than having normal QCD color.  
Squirks could be pair-produced in LHC
collisions via weak interactions (a Drell--Yan process, or gauge boson
fusion), and their high mass means they will typically be
semi-relativistic. 
The relationship $\Lambda << m_{q'}$ can then result in striking
new phenomenology.  
There will be a QCD' string connecting the squirk-antisquirk
pair, but unlike normal QCD, string fragmentation cannot occur because
the energy density in the string, $\Lambda^2$, is much less than the typical
energy density ($\sim m_{q'}^2$) needed for quirk pair production in a standard
`hadronisation' mechanism. 
Instead, the heavy squirks will continue to separate, until eventually
all their kinetic energy has been transferred into the stretched QCD'
string. 
At this point, the string tension will cause the squirks to start
oscillating, forming a ``squirkonium'' bound state. 
If the squirks are electrically charged, this oscillation will then
radiate photons, with characteristic frequency given by:
\begin{equation}
  \omega \sim \frac{\pi \Lambda^2}{m_{q'}}
\end{equation}
It is possible that a large fraction of the squirkonium energy could
be radiated this way, 
and assuming a squirk mass of 500~GeV, any value of the confinement
scale $\Lambda \lesssim 13$~GeV will result in photons with typical
energy $\lesssim 1$~GeV. 




Such photon cloud events could represent a striking signature of physics
beyond the Standard Model.
The problem is that in such an event, no individual detector region
has very high activity 
and therefore no typical high-energy object would be seen by the trigger.
Triggering on this kind of event therefore requires consideration of
the global properties of the event, not just local regions of
high-activity. 

We propose to introduce a new variable at the trigger level: a sum of
the transverse energy observed in all channels of the electromagnetic
calorimeter (ECAL). 
This is a modification of the standard \sumet\ trigger, which
typically sums contributions from both the electromagnetic and
hadronic calorimeters. 
Not only is this new ECAL-only \sumet\ measure more directly sensitive
to the photon cloud signature, 
it also takes advantage of the reduced background from the
electromagnetic component of minimum bias proton-proton collisions
relative to the hadronic component. 
In addition, by considering only one sub-detector, it is less affected
by detector noise, and it is particularly well suited for the CMS
detector, which has a high-performance electromagnetic
calorimeter~\cite{Henderson:cms:ecal_tdr}. 
These combined benefits allow for a lower trigger threshold,
increasing our sensitivity to potential exotic physics. 






At the High-Level Trigger stage in CMS, the full detector readout is
available and essentially complete 
`offline-like' event reconstruction can be performed. 
Thus it is straight-forward in the software to construct the ECAL-only
\sumet\ variable for event selection at the HLT.
Ideally one would do a similar thing at the Level~1 stage
also. However, unfortunately in this context, the design of the CMS
Level~1 trigger system is such that only the complete calorimeter
\sumet\ can be computed - the electromagnetic and hadronic \sumet\
components are not available separately. Fortunately though, we can
tolerate a higher trigger rate at Level~1, since 
further refinement will be done at the HLT. 
A cut at Level~1 on the complete calorimeter \sumet\ of $\sim 100$~GeV
is expected to be fully efficient for the signal we are considering,
while still producing an acceptable rate. 
For the remainder of this paper, we will discuss only the final
high-level trigger selection. 
For simplicity, we choose to focus only on the ECAL Barrel
sub-detector, which spans the central pseudorapidity region $|\eta| <
1.47$.

Since the HLT is required to be able to accept incoming events at a
rate up to 100 kHz without incurring deadtime, this imposes a limit on
the average total time which can be spent processing each event.
The design goal is that the
average HLT event processing time not exceed 50~ms~\cite{Agostino:2009ic}.
Unlike other region-of-interest based triggers, our proposed new
global trigger 
requires the raw data from all $\sim 60,000$ ECAL barrel channels to be
unpacked; 
it is therefore important to verify that the time spent doing this
will not be prohibitive for HLT operations.
We have studied the CPU-time required for the new selection on minimum
bias Monte Carlo events, and verified that it is well within the
acceptable limits. 
As well as checking the average time, we must also ensure 
that high-occupancy events do not result in unacceptably long
processing times. An excellent opportunity to test this was
provided during the LHC commissioning phase with
so-called `beam splash' events\footnote{As part of the accelerator
commissioning, the beam is fired into the collimators, generating a
large spray of secondary particles that can be seen in the
detector. These events are called `beam splashes'.}
where essentially every calorimeter channel is illuminated by
high-energy particles. 
We studied the time taken for this trigger path on the data
collected during `beam splashes' in November 2009.
The time taken in these extreme high-occupancy conditions forms
an upper bound for the trigger path, and the result is found to be
well within the acceptable range for operations.

The background to this photon cloud signature will come from rare
proton-proton collision events which contain a large number of final-state
photons, either from prompt production or from $\pi^0$ decays. 
We estimate this background using minimum bias events generated by the
{\sc pythia} Monte Carlo event generator~\cite{Sjostrand:2006za}.

\begin{figure}[!htb]
\begin{center}
\includegraphics[width=0.75\textwidth]{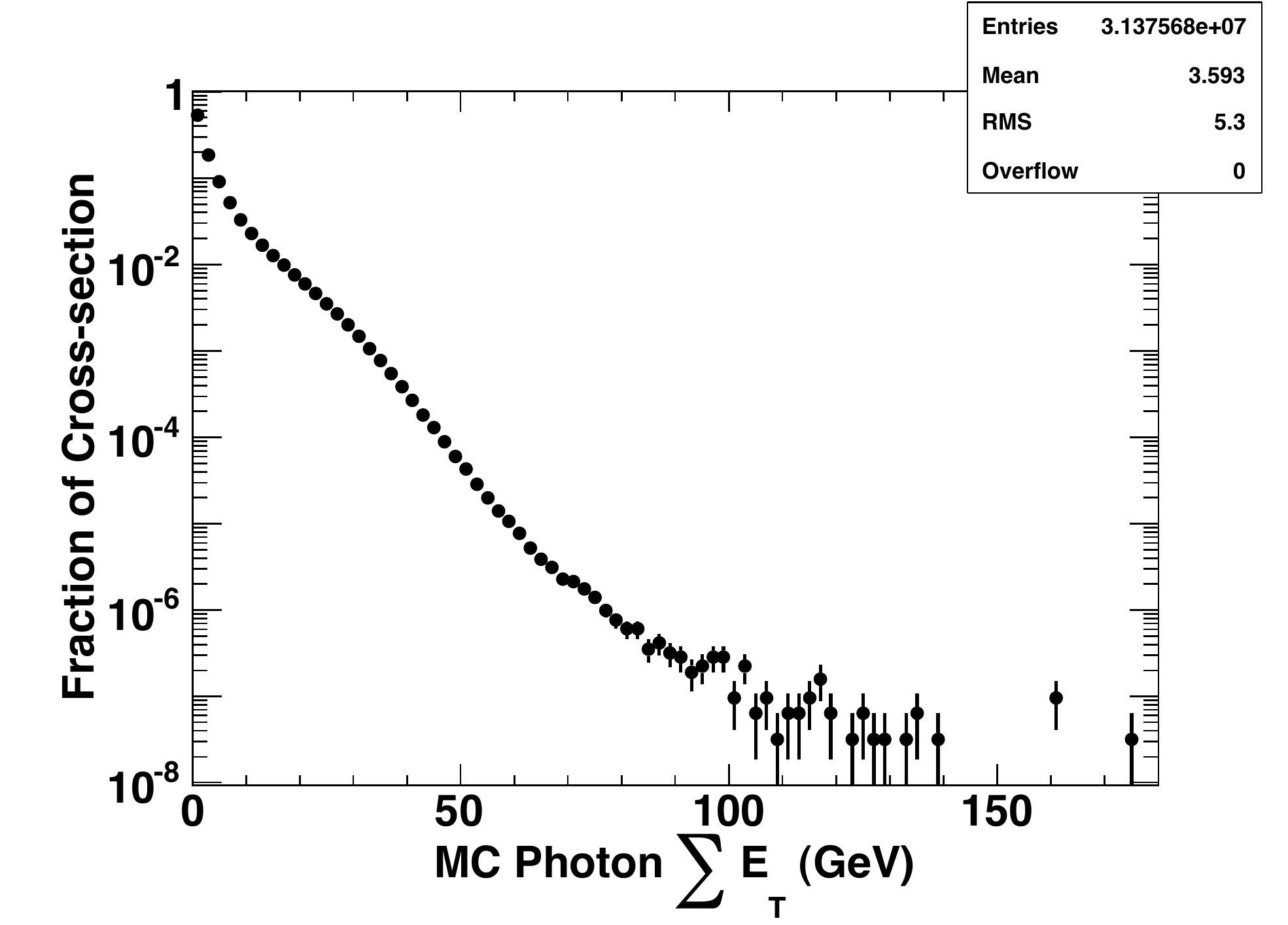}
\end{center}
\caption{Distribution of the \sumet\ of final-state photons in
 {\sc pythia} minimum bias collisions, based on the generator-level
 information. }  
\label{Henderson:fig:minbias_dist1}
\end{figure}

Based on the Monte Carlo generator-level information, the distribution of the
\sumet\ of all final-state photons in the simulated event is shown in 
Figure~\ref{Henderson:fig:minbias_dist1}. 
In combination with the anticipated accelerator luminosity profile,
this will allow us to estimate the expected trigger rate as a function
of selected threshold value. 
Our goal is to choose a trigger threshold that corresponds to a
trigger rate $<1$~Hz (in order to have minimal impact on the limited
trigger bandwidth, and hence the rest of the CMS physics program).

\begin{figure}[!htb]
\begin{center}
\includegraphics[width=0.9\textwidth]{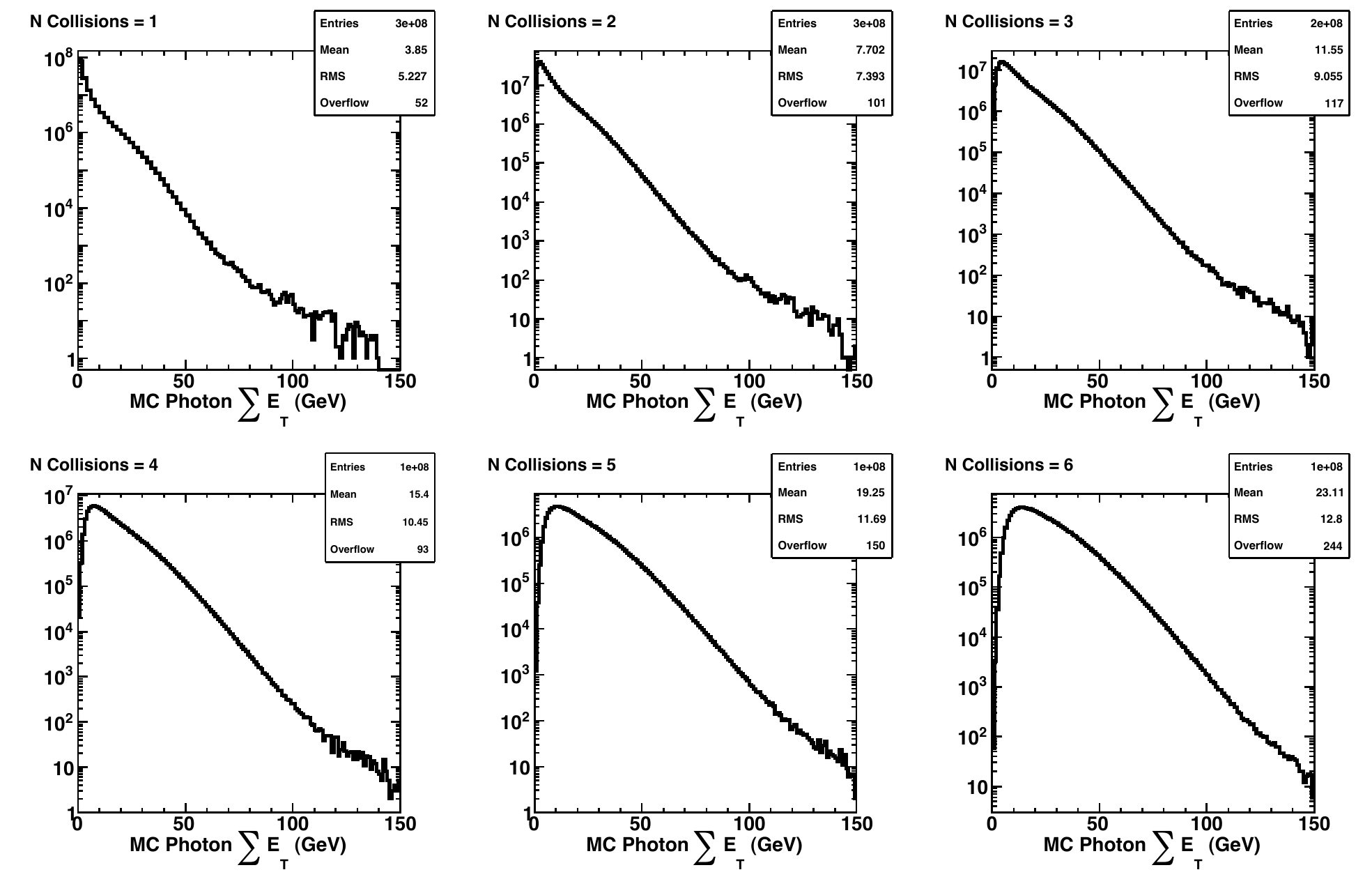}
\end{center}
\caption{Effect of additional collisions per bunch crossing on the
  total photon \sumet\ distribution.}
\label{Henderson:fig:pileup_6colls}
\end{figure}

\begin{figure}[!htb]
\begin{center}
\includegraphics[width=0.8\textwidth]{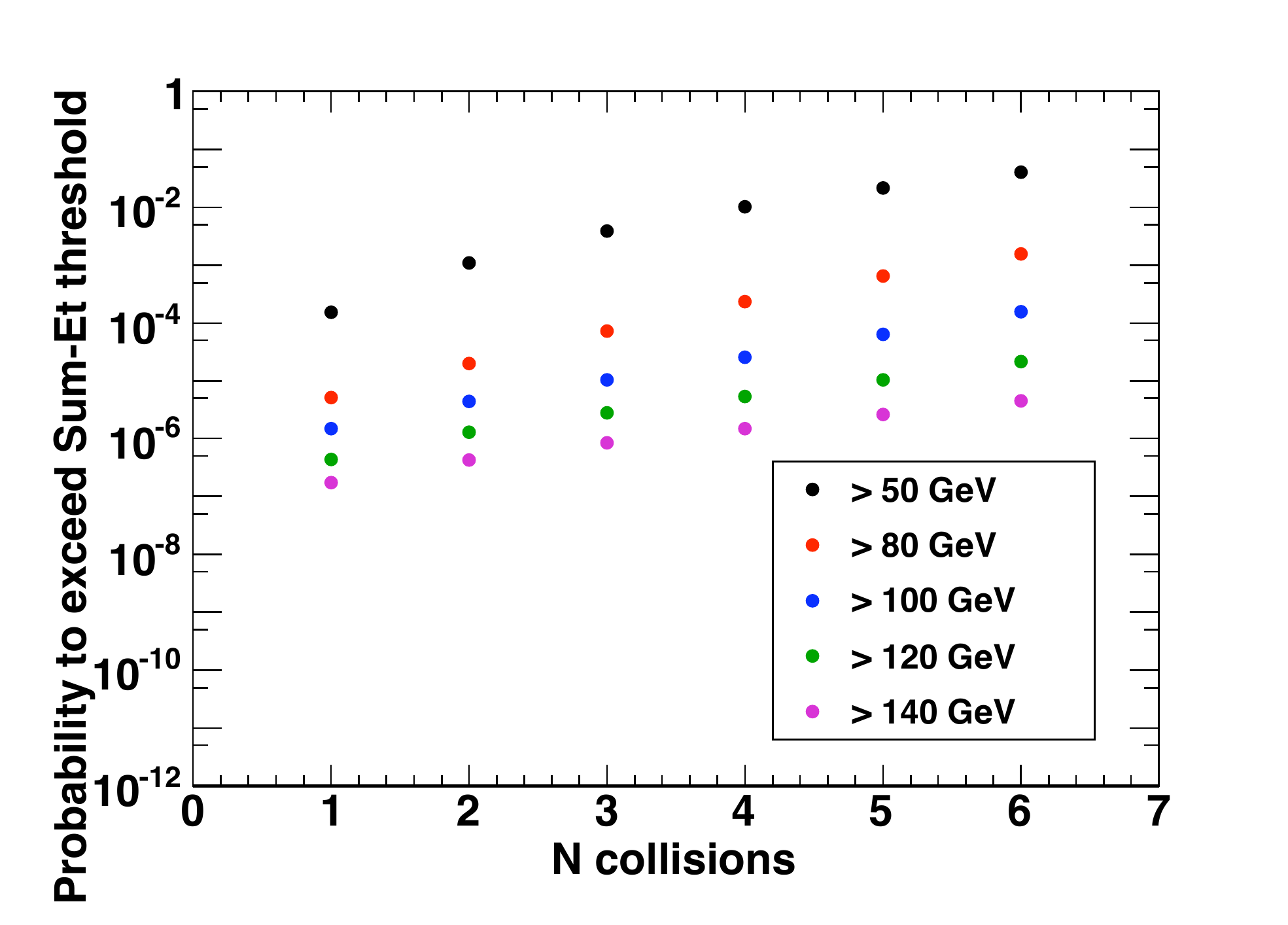}
\end{center}
\caption{Probability to exceed a given threshold value of
  \sumet\ as a function of the number of independent collisions per
  bunch crossing.} 
\label{Henderson:fig:pileup_Vs_ncoll}
\end{figure}

\sumet\ triggers are especially sensitive to the effect of multiple
minimum bias collisions within a single bunch crossing, a phenomenon
known as pileup.
Because the trigger merely sums the total energy deposited in the
calorimeter, it cannot distinguish the separate contributions from the
individual collisions.  
Figure~\ref{Henderson:fig:pileup_6colls} shows how the total photon \sumet\
distribution evolves as a function of the number of separate minimum
bias collisions per bunch crossing. 
To illustrate the point, consider that a total of, say,
100~GeV in the detector can be obtained by two simultaneous
collisions if both generate 50~GeV each, or one generates 60~GeV and the
other 40~GeV, or 70~GeV and 30~GeV, and so on. 
The probability to exceed a given \sumet\ threshold therefore grows
nonlinearly with the number of independent collisions per bunch
crossing, as displayed in Figure~\ref{Henderson:fig:pileup_Vs_ncoll}.

\begin{figure}[!htb]
\begin{center}
\includegraphics[width=0.8\textwidth]{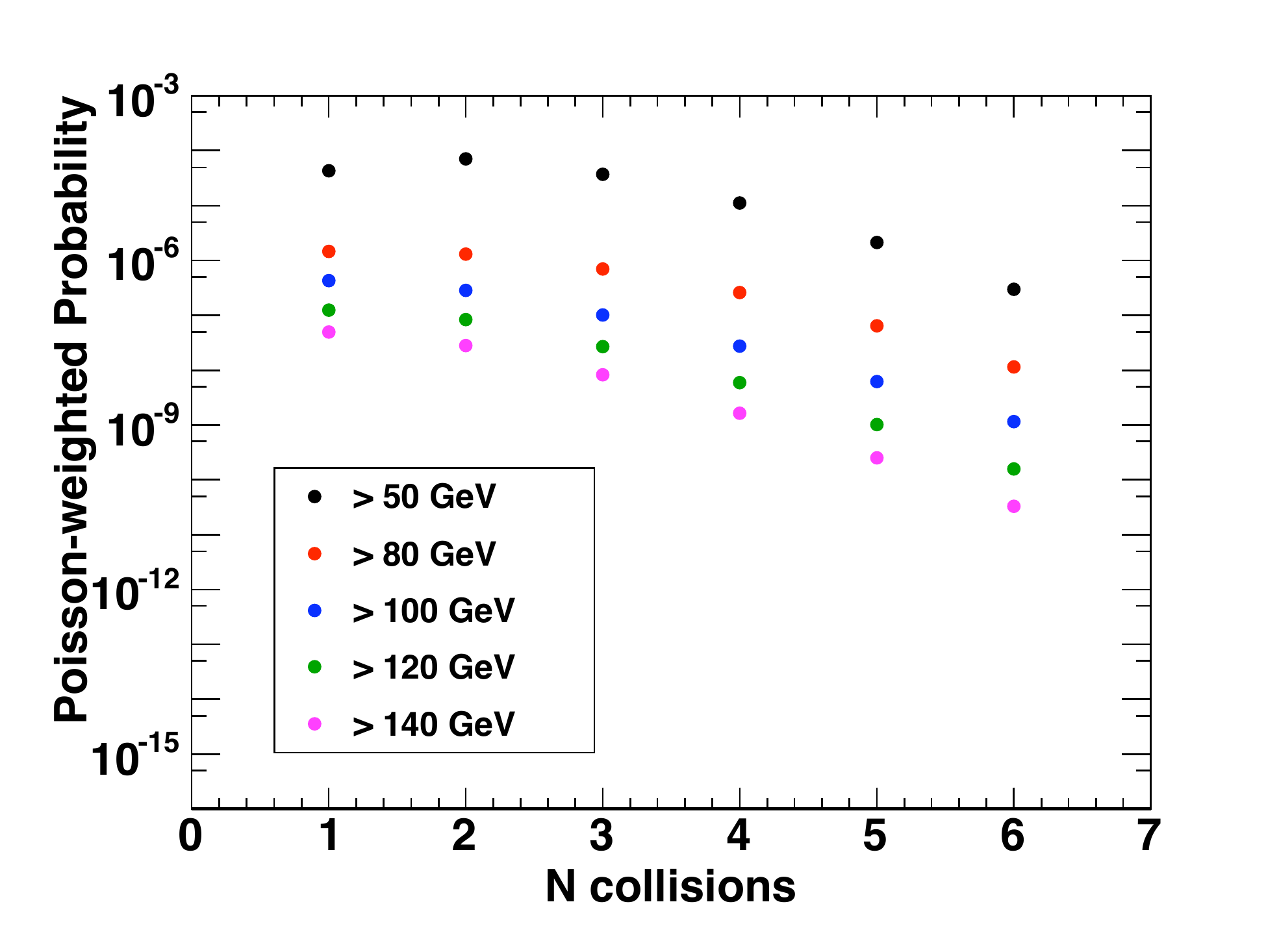}
\end{center}
\caption{Poisson-weighted probability to exceed a given threshold value of
  \sumet\ as a function of the number of independent collisions per
  bunch crossing, assuming $<n> = 0.45$. } 
\label{Henderson:fig:pileup_poisson_weighted04}
\end{figure}

The likelihood of there being $N$ collisions per bunch crossing follows
a Poisson distribution characterised by $<n>$, the mean number of
interactions per crossing. $<n>$ can be determined from the expected
LHC luminosity parameters, in particular the instantaneous luminosity
and the number of colliding bunches. 
A benchmark for the early part of the 2010 running period is for the
LHC to reach an instantaneous luminosity of
$10^{31}$~cm$^{-2}$~s$^{-1}$ with $156 \times 156$ colliding
bunches~\cite{Henderson:lhc_2010_lumi}. 
Taking a proton-proton cross-section of $80$~mb at $7$~TeV, this
corresponds to $<n> \approx 0.45$.  
Figure~\ref{Henderson:fig:pileup_poisson_weighted04} shows the Poisson-weighted
probability to exceed a given trigger threshold, assuming $<n> = 0.45$.  
Fortunately, we find that the
reduced likelihood of having an extra collision more than 
compensates for the increase due to combinatorics, and the observed
rate hence remains under control. 
Estimated trigger rates for the $10^{31}$~cm$^{-2}$~s$^{-1}$
luminosity scenario as a function 
of the chosen threshold value are shown in Table~\ref{Henderson:tbl:rates}.
Note that the values shown here are just from the generator-level
photon \sumet\ --  the actual trigger threshold value that should be used
in the experiment must also account for detector effects such as noise
and resolution, and will therefore be somewhat higher. 
However, it is clear from this table that we can expect to be sensitive to
potential photon clouds with \sumet~$\sim 200$~GeV while maintaining a
trigger rate $<1$~Hz, which was our goal. 

\begin{table}[!htb]
\begin{center}
\begin{tabular}{c|c}
Photon \sumet\ (GeV) &  Trigger Rate (Hz) \\
\hline
50  & 285 \\
80  & 7 \\
100 & 1.5 \\
120 & 0.4 \\
140 & 0.15 \\
\hline
\end{tabular}
\end{center}
\caption{Estimated rates as a function of chosen trigger threshold,
  for LHC $10^{31}$~cm$^{-2}$~s$^{-1}$ startup luminosity scenario.}
\label{Henderson:tbl:rates}
\end{table}


 
In summary, we have considered an unusual potential event topology: a
`photon cloud', consisting of $\sim 200$~GeV of transverse energy
emitted through many soft photons ($\lesssim 1$~GeV each).  
Such an event could arise from the production and decay of `squirks'
in a folded supersymmetry scenario. 
A photon cloud event could be missed by conventional trigger
selections, 
because no individual detector region has particularly high activity. 
We have proposed a new variable, \sumet\ in the electromagnetic
calorimeter, which can be easily computed in an experiment's
high-level trigger and which would be sensitive to this unusual new
physics signature. 
We have estimated the expected rates for this new trigger for the LHC
early luminosity scenario, and shown that the proposal is feasible. 
This trigger is being implemented for the CMS detector and it is
planned to be operational for high-energy LHC collisions in 2010. 

%% file: TomalinStrassler/TomalinStrassler.tex
\chapter[Long-lived exotica production at the LHC/Tevatron]{Models and benchmarks for long-lived exotica production at the LHC/Tevatron}
{\it M.J.~Strassler and I.~Tomalin}

\begin{abstract}
Numerous theories predict that long-lived exotic particles may be
produced at the LHC/Tevatron. These yield highly unusual signatures,
which can prove a major challenge for triggering and event
reconstruction. This note highlights some of these theories, and uses
them to define simple analysis benchmarks, as might be appropriate for
early LHC searches. The note also defines reconstruction/trigger
benchmarks, which are deliberately more difficult to find, designed to
stress the detector and so identify weaknesses to be remedied.
\end{abstract}

\section{Motivation}
\label{TomalinStrassler:LongLived:Motive}

New long-lived particles, with lifetimes such that their decays
commonly occur at distances outside the beam-pipe but inside the
detector volume, do not generally arise in the most popular models of
electroweak symmetry breaking.  Perhaps for this reason, they were
little studied by LHC/Tevatron theorists and experimentalists until
recently.  Only those cases arising in particularly simple models of
gauge-mediated \cite{Dimopoulos:1996vz} supersymmetry, with neutral LSP's decaying
to photon plus gravitino, were covered by early LHC studies.

However, looking across the literature, one finds plenty of models in
which a long-lived particle appears; see for example
\cite{Langacker:1984dc,Martin:2000eq,ArkaniHamed:2004fb}.  Also, it has been emphasized
recently in the literature that long-lived particles arise very
commonly in models with hidden sectors and a mass gap (hidden valleys
\cite{Strassler:2006im}) in which a number of new particles may naturally arise
with a variety of long lifetimes.  (The example of the QCD spectrum,
which has many long-lived hadrons with widely varying decay lengths
and final states, is instructive.)  Moreover, it has also been shown
that finding long-lived particles with current hadron colliders can be
exceptionally difficult, because of challenges in triggering,
reconstruction and detector backgrounds (e.g. secondary interactions.)
This motivates a serious effort to ensure that the collaborations have
a plan to perform searches for long-lived particles and ensure that
the detector hardware and software is used in a way which helps,
rather than hinders, this effort.

Toward this end, it was decided that benchmark models were needed as targets
for the experimental collaborations.  We describe the current status of that
enterprise here.  Our efforts are organized along two different lines, 
with differing goals.  

First, we aim to provide benchmarks for long-lived particle searches
appropriate for early days at the LHC.  The goal in this case is to
provide simple models, with moderately large cross-sections and with
signatures that are relatively uncomplicated.  These models have a
small number of variable parameters on which limits could be placed if
no discovery is made.  We will propose specific possibilities below.

Second, we aim to provide benchmark models that would serve as a
stress-test of the trigger system and reconstruction software.  The goals in
this case are to check whether a challenging signature might cause
problems either for the trigger pathways or the reconstruction
software, or even the methods of data storage.  The underlying concerns
are that a long-lived particle signature might (a) be rejected by the trigger,
(b) cause the event to fail quality control cuts, or (c) be inefficiently 
reconstructed. Even where events are successfully kept, poor reconstruction
can confuse subsequent event skims, so making it difficult to select
a sample of events on which detailed analysis should be performed.
%
The  variety of possible signatures and the 
complexity of the software involved
make it difficult to guess whether the currently designed system is
robust without a test.  We have considered models with signatures that
are in some cases simple, in others exceptionally complex, 
though always realistic.  For
each model we are making available a data set that will serve in such
a test, as well as providing, where possible, the information as to
how to simulate the model without special-purpose software.

\section{Classes of models used in the benchmarks}
\label{TomalinStrassler:LongLived:Models}

In all the models chosen, long-lived particles either arise when a
visible-sector particle decays slowly into a hidden sector, or when a
hidden-sector particle decays slowly to standard model particles.  We
will first discuss the production mechanisms for the long-lived
particle(s). Then we will address the final states emerging in the
decay of the long-lived particle(s).

\subsection{Production}
\label{TomalinStrassler:LongLived:Prod}

The models chosen for the benchmarks produce long-lived particles via
three mechanism: (1) decays of a singly-produced light resonance (such
as a Higgs boson or $Z'$) into a hidden sector with long-lived
particles \cite{Strassler:2006im, Strassler:2006ri}; (2)
R-parity-conserving decays of the lightest super-partner of any
\textit{standard-model} particle (LSP), in the case that the LSP is
heavier than the lightest super-partner in the \textit{hidden} sector
(vLSP) \cite{Dimopoulos:1996vz,Matchev:1999ft,Strassler:2006qa}: and (3) annihilation of
quirks \cite{Okun:1980kw,Strassler:2006im,Kang:2008ea,Juknevich:2009ji}, confined particles
charged under standard model and hidden sector gauge groups, to hidden
valley gluons (also called ``infracolor''.)  In each case, the number
of hidden sector particles produced, and the lifetimes of the
long-lived particles, are strong functions of the essential dynamics
in the hidden sector and of the parameters of the model.  However, the
production rate for a Higgs boson, $Z'$, LSP or other particles occurs
as usual in the standard model or in its appropriate extensions by an
extra $U(1)$ and/or supersymmetry.  For the usual reasons, these rates
may be of order 1 -- 10 pb for light resonances or supersymmetry, and
of order 100 fb -- 1 pb for a $Z'$ or moderately heavy scalar.  (These
numbers are only guidelines and can be exceeded in particular models.)
Quirks are produced through usual standard model pair production, and
to estimate their cross-sections one need merely compare standard
model production rates for standard model fermions of the same mass
and quantum numbers, possibly multiplied by a hidden color factor
(typically a small integer.)


Let us note here that the Higgs scalar that dominantly gives mass to the $W$ and $Z$ 
bosons through its expectation value will dominantly decay to $WW$ and $ZZ$
when these channels becomes kinematically allowed.  For this reason, only a light
Higgs is likely to have exotic decays to long-lived particles with a large branching
fraction (above a percent).  However, any scalars or pseudoscalars
which are not responsible for electroweak symmetry breaking (including for example
the CP-odd $A^0$ boson in two-higgs doublet models such as supersymmetry)
will not (dominantly) decay to gauge bosons and typically will have small widths
to decay to standard model particles.  This makes them susceptible,whatever their
masses might be, to
new couplings to a hidden sector, and thus to new exotic decays.

We should also note that when designing an analysis,
care should be taken to assure that quantum numbers
are properly accounted for in any process
considered.  For example, just as $\rho^0\to\pi^0\pi^0$ is forbidden by Bose
statistics, a process $Z'\to XX$ where $X$ is a real scalar is forbidden.  However
$Z'\to X\bar X$ where $X$ is complex is allowed, as is $\rho^0\to\pi^+\pi^-$.  


\subsection{Decays}
\label{TomalinStrassler:LongLived:Decay}

For simplicity we limit ourselves to scenarios that are both
``natural'' and popular.  It must be emphasized that other scenarios
cannot be excluded on either theoretical or experimental grounds.  All
long-lived particles discussed below are charge- and color-neutral
unless explicitly stated otherwise.  Such particles, if sufficiently
weakly coupled, are poorly constrained by experiment and may be extremely
light, as emphasized for example in
\cite{Dobrescu:2000jt,Strassler:2006im,Bjorken:2009mm}.

New hidden-sector scalar or pseudoscalar particles tend to
decay to the heaviest fermion pair available, due to helicity
suppression and/or coupling proportional to standard model Yukawa
couplings.  For moderate masses these particles will decay dominantly
to $b\bar b$, and perhaps $\tau^+\tau^-$.  An interesting special case
occurs when the mass becomes of order 1 GeV, as for example in the
case of a ``dark axion'' motivated by current anomalies in dark matter
experiments.  In this case, among experimentally detectable decays,
$\mu^+\mu^-$ may be dominant.

New hidden-sector vector or axial-vector particles usually decay to
leptons and quarks in a generation-independent fashion, subject of
course to kinematic constraints.  In this case $\mu^+\mu^-$ and
$e^+e^-$ may be the best channels, though the large rate for dijet
decays may outweigh their relative difficulty.  Again a special case
occurs for masses of order or below 1 GeV.  A vector boson (such as a
``dark photon'' or ``dark gluon'') at this mass, mixing with the
photon, have (like dark axions) been proposed to explain
astrophysics signals from the PAMELA/ATIC and now Fermi-LAT experiments
\cite{ArkaniHamed:2008qp,Pospelov:2008jd}. It would
decay equally to $\mu^+\mu^-$ and $e^+e^-$ (unless below $2m_\mu$.)
The relative branching fractions for decays to leptons versus
$\pi^+\pi^-$ is a complicated but known function of the particle's
mass, determined by electromagnetic couplings and by mixings of the photon
with hadronic resonances \cite{ShihThomasunpub,Abazov:2009hn}.  Additional
results on such particles appear in \cite{Baumgart:2009tn,Cheung:2009su,Bai:2009it}.

New hidden-sector scalar or tensor particles may decay to gauge boson pairs,
including gluon pairs, photon pairs, and (when kinematically allowed) $W^+W^-$
and $ZZ$, and perhaps $Z\gamma$.  In some cases a decay to two Higgs bosons may be permitted.  Branching
fractions of these various final states may vary widely.   

The case of a long-lived neutral LSP (or any analogous particle in
models with KK-parity, T-parity, or other new global symmetries)
offers two different possibilities.  (a) Even without R-parity
violation, the LSP may be an long-lived particle and decay in flight
to a partly visible final state.  Well-known examples include decay to
a gravitino plus a photon, $Z$ or Higgs boson \cite{Matchev:1999ft}.
It is also possible for the decay to go to new hidden sector particles
(as in many supersymmetric hidden valley models \cite{Strassler:2006qa,
ShihThomasunpub,Cheung:2009su}, of which many of the recent dark
matter models \cite{ArkaniHamed:2008qn,Pospelov:2008jd} are examples) in which case their
decay products might appear all at the point of the LSP decay.  (b)
Conversely, it is also possible that the LSP decays promptly into the
hidden sector, producing among other things a long-lived hidden-sector
particle that may decay as described above.

In all cases (except the LSP $\to\gamma +$ gravitino decay mode)
quoted so far, the decay products observed form a resonance ---
generally a new resonance, except in the case of decays of the LSP to
a $Z$ boson (plus a gravitino).  Like the LSP, new hidden sector
particles also might have decays emitting a photon, $Z$ or Higgs and a
second hidden-sector particle \cite{Strassler:2006qa,Juknevich:2009ji}. However, more
complicated decays, where the final state might produce invariant-mass
edges or endpoints rather than resonances, are possible.  For example,
a hidden sector particle, of any spin, might decay to a second while
emitting two standard model fermions $(X\to f\bar f X'$), where the
$f\bar f$ invariant mass is a continuum, with an edge or endpoint.
Another interesting example is the model \cite{Basso:2008iv}, which
predicts $Z'\to\nu_H\bar\nu_H\to (\ell^- W^+)(\ell^+ W^-)$, where
$\nu_H$ is a long-lived, heavy neutrino.  Even more complex decays are
possible, if for example, the decay $X\to f\bar f X'$ is followed by a
prompt decay \cite{Strassler:2006im} such as $X'\to f\bar f$ or $X\to W^+W^-$.  In this case
many standard model particles will be emitted from a single vertex.
The number of possibilities rapidly becomes very large.  Once there is
experience with searching in the simpler scenarios, and it is clearer
how robust the initial analyses are and where the gaps lie between
them, these more complex signatures should be considered, and attempts
made where necessary to check for their presence.

One might ask if there are strong constraints on the masses and lifetimes of the
new particles.  Unfortunately there are not.  In general, in any fixed model, the 
lifetime of a particular particle is often a strongly decreasing function of its mass 
(for roughly the same reason that
the muon lifetime varies inversely with the fifth power of its mass) or of other parameters.  
However, across models there is no correlation between mass and lifetime.  
Even within a model there may be very long-lived particles with large masses 
(just as $B$ mesons live a bit longer than $D$ mesons.)  Thus one ought if 
possible to treat these parameters as independent, 
since otherwise model-based
assumptions will limit the applicability of the results obtained.

\subsection{Multiplicity}
\label{TomalinStrassler:LongLived:Mult}

The decay of a visible sector particle into the hidden sector may lead 
to a final state with any number of long-lived decays, subject only 
to kinematical constraints.  This is partly because of the wide variety of dynamics
that can be present in hidden sectors, affecting  
the intrinsic multiplicities of hidden particles produced, and partly because
hidden sectors may contain several new particles with different lifetimes, some of 
which may decay promptly, others of which may be stable or may decay far outside the 
detector.

For example, models exist in which a $Z'$ may decay into two identical
particles with identical lifetimes (analogous to $\rho\to\pi^+\pi^-$).
In this case one has two back-to-back particles with the same average
lifetime.  But it may also decay to a single long-lived particle and a
promptly-decaying particle (analogous to $a_1\to \pi\rho$) or to a
long-lived particle and an invisibly-decaying (or stable) particle.
At the other extreme, it may decay to particles which have showering
or cascade decays that may lead to a very large number of
hidden-sector particles being produced, possibly producing many
displaced vertices \cite{Strassler:2006im,ATLASHVtrig}.  The same is true if the
LSP in a SUSY model decays promptly to long-lived particles in the
hidden sector; the number of visibly-decaying long-lived hidden-sector
particles produced may vary, and consequently the number in each event
may range from one to a very large number
\cite{Strassler:2006qa,Cheung:2009su}.  By contrast, in SUSY models where the
LSP itself is long-lived \cite{Dimopoulos:1996vz,ArkaniHamed:2004fb} the number of displaced vertices is
generally equal to two (except when one LSP happens to live too long
or too short a time for its displaced decay to be detected.)



\section{Analysis benchmarks}
\label{TomalinStrassler:LongLived:Analysis}

A key requirement for early-data analysis benchmarks is that they
should be sharply defined, theoretically well-motivated, relatively
simple experimentally, and contain a small number of tunable
parameters.  It should be possible to imagine a search strategy (or
small number of strategies) that would make these models appropriate
targets for analyses within the coming few months.

In all the cases listed below, the signatures are simple enough that
it is very easy to implement these models in MadGraph or other similar
event generators.  In particular, all decays in these models are a
sequence of two-body decays.  To obtain precise limits, spin effects 
should be properly included, since angular distributions will affect 
efficiencies. 

\subsection{{\boldmath $Z'$} or {\boldmath $H$ $\to B\bar B \to (f\bar f) (f'\bar f')$}}
\label{TomalinStrassler:LongLived:HBB}

In this benchmark, $B$ is a long-lived, neutral boson and $f$ and
$f'$ are Standard Model fermions.
Experimentally, these events can be identified by reconstructing
the displaced difermion vertices in the Tracker or using
time-of-flight measurements. One can also require that very few
promptly produced hadrons should be flying in a similar direction 
to the displaced fermions. One should trigger on the displaced
fermions, to avoid assumptions about the rest of the event.
Modifications to the trigger are often needed to accomplish this.
Such analyses have been performed at the Tevatron 
\cite{Abazov:2009ik,Abazov:2008zm,Abazov:2006as}.

One should publish the measured the cross-section 
$\sigma[Z'/H\to B\bar B \to (f\bar f) (f'\bar f')]$ 
for individual $f\bar f$ and $f'\bar f'$ species. As explained in 
Sect.~\ref{TomalinStrassler:LongLived:Decay}, $e^+e^-$, $\mu^+\mu^-$ and $q\bar q$
final states should give better sensitivity to spin~1 bosons $B$, 
whilst $\tau^+\tau^-$ or $b\bar b$ should be better for the spin~0 case. 
Searching for mixed
states where one boson decays to leptons and the other to jets
may also be useful, since the leptons will facilitate triggering and
suppress backgrounds.

The cross-section measurements will depend on 3 parameters, which 
can be taken to be the masses $M_{Z'}$ and  $M_B$ and the mean $B$ lifetime 
(or decay length).
Quoting results as a simultaneous function of all 3 variables
is not practical, so one may instead show them as a function
of each in turn. (N.B. For analyses that rely on reconstructing
the displaced vertices with a Tracker, the reconstruction efficiency 
is a strong function of the mean decay length, so it then makes sense
to parametrize the results in terms of it, rather than the lifetime).



The published measurements should be sufficiently complete, to 
allow one to subsequently combine the different channels into
a measurement of $\sigma(Z'/H\to B\bar B)$. This step requires
theoretical assumptions about the branching ratios $B.R.(B\to f\bar f$).
e.g. If $B$ is a pseudo-scalar, a simple model might assume these
to be equal to those of a pseudoscalar Higgs at
the corresponding mass (i.e. no decays to $WW$, $ZZ$).

\subsubsection{More complex Variations on $Z'$ or $H$ $\to B\bar B \to (f\bar f) (f'\bar f')$}
\label{TomalinStrassler:LongLived:HBBvar}

It is worth considering two variations on the previous benchmark, both of which can
be studied with only minor changes in analysis software. They are:
\begin{enumerate}
\item
As explained in Sect.~\ref{TomalinStrassler:LongLived:Prod}, the $B\bar B$ system may originate
from something other than a resonance decay (e.g. a SUSY cascade decay chain).
One could therefore repeat the analysis of the previous section, but
without requiring the presence of a $Z'/H$ peak in the reconstructed $B\bar B$ mass.
This leads to a measurement of $\sigma[B\bar B \to (f\bar f) (f'\bar f')]$,
as a function of 3 principal parameters, which can be taken to be $M_{B\bar B}$, $M_B$ 
and the mean $B$ decay length.
\item
As explained in Sect.~\ref{TomalinStrassler:LongLived:Mult}, the $Z'$ or $H$ may decay to only one boson
that yields a visible, displaced $f\bar f$ vertex. 
One should therefore publish inclusive measurements of $\sigma[B \to f\bar f]$,
as a function of 3 parameters, which can be taken to be the transverse momentum $Pt_{B}$, 
mass $M_B$ and mean decay length of the boson $B$.
\end{enumerate}
In these two variants (especially the second one), the backgrounds will be larger
and harder to control. They may therefore only be practical for leptonic final
states. 

A more complex analysis (but perhaps possible in a well understood detector) is to search for
a pair of long-lived fermions, each decaying to a 3-fermion final state. However, to reduce
the number of free parameters, the early benchmarks described here consider only the special case,
where this occurs via an intermediate 2-body state (fermion plus boson, where the boson
then decays promptly to $f\bar f'$). One suitable benchmark is the model 
\cite{Basso:2008iv}, 
which predicts $Z'\to\nu_H\bar\nu_H$, where $\nu_H$ is a long-lived, heavy neutrino,
followed by $\nu_H\to \ell^- W^+$ or $\nu Z^0$. 

The case where both $\nu_H$ decay
to $\nu Z^0$ is simplest. One can begin by using this channel to make a measurement of 
the cross-section
$\sigma[\nu_H\bar\nu_H \to (\nu Z^0) (\bar\nu Z^0)]$, as a function of 
$M_{\nu_H\bar\nu_H}$, $M_{\nu_H}$ and the mean decay length of $\nu_H$. In this initial
measurement, one should explicitly reconstruct the displaced $Z^0$, but not attempt
to reconstruct the $\nu_H$ or $Z'$. This ensures that the results obtained will also 
constrain other models predicting displaced $Z^0$. (e.g. Long-lived 4$^{th}$ generation 
quarks $b'\to b Z^0$ \cite{Scott:2004wz}).

The case where both $\nu_H$ decay to $\ell^- W^+$ is expected to have a larger cross-section,
so is also worth pursuing. The mixed channel, where one $W$ decays hadronically and the other 
leptonically is perhaps most promising.  Measurements of 
$\sigma[Z'\to\nu_H\bar\nu_H\to (\ell^- W^+)(\ell^+ W^-)]$ could be quoted as function of $M_{Z'}$,
$M_{\nu_H}$ and the mean decay length of $\nu_H$.  (Here one would fully reconstruct the $\nu_H$
and $Z'$).
Ultimately, this benchmark could be further
generalized to consider long-lived fermions decaying to a quark/lepton plus an unknown
boson (charged or neutral) which in turn decays promptly to $f\bar f'$. 

\subsection{SUSY: {\boldmath $\tilde\chi^0_1 \to \tilde D B$}} 
\label{TomalinStrassler:LongLived:SUSY}


In this model, $\tilde D$ is a stable, invisible fermion, assumed to be 
of negligible mass, (e.g. a gravitino). The neutral boson $B$ can be a photon 
(already covered in standard GMSB
searches and ignored here), a $Z$, a Higgs (with a new parameter, the Higgs mass), or a 
new exotic boson (with an unknown mass). This boson is assumed to decay to
difermions $f\bar f$.

The underlying SUSY event can be based on a standard SUSY benchmark \cite{Allanach:2002nj}, 
in which the $\tilde\chi^0_1$ is the LSP.
However, it must be understood that the search strategy, and the parameters
limited by the analysis, will depend strongly upon the particular SUSY
benchmark point chosen. 

A key feature of these events is that, in addition to the fermions from the long-lived exotic,
the rest of the event will contain other hard particles from the SUSY decay chain. One can 
use these additional particles for triggering or background rejection. Doing so
allows one to explore regions of parameter space which would otherwise be 
inaccessible. (e.g. Where the fermions from the long-lived exotic are too soft to be 
triggered upon). However, relying on these additional particles does make the results very dependent
on the particular choice of SUSY benchmark. It is therefore strongly advisable to also quote 
results for the case where these extra particles have not been used.

There are two simple benchmarks based on this model:

\begin{enumerate}
\item
The $\chi^0_1$ decays promptly and the boson $B$ is long-lived. This
leads to an inclusive cross-section measurement of $\sigma[B B \to
(f\bar f) (f'\bar f')]$, as a function of the $B$ boson's mass and
mean decay length. (N.B.  The results obtained will also be influenced
by the $P_t$ of the boson $B$, which depends on the particular SUSY
benchmark chosen). One would probably not need to reconstruct the
$\tilde\chi^0$, except to establish the exact nature of a discovery.

Technically, this benchmark is very similar to the first benchmark
described in Sect.~\ref{TomalinStrassler:LongLived:HBBvar}. Indeed, unless one is using
the rest of the SUSY event for triggering/background rejection, they
are almost identical, and one can query if it is worth studying
both. Doing so does, however, allow one to check if the more crowded environment
of the SUSY event affects the signal selection efficiency.

\item
The $\tilde\chi^0_1$ is long-lived and the boson $B$ decays promptly. In this case, 
one can measure the inclusive cross-section
 $\sigma[\tilde\chi^0_1 \tilde\chi^0_1 \to (\tilde D B)(\tilde D B)\to (\tilde D f\bar f) (\tilde D f'\bar f')]$ 
as a function of the $B$ boson's mass and the $\tilde\chi^0_1$'s mean decay length.
(The results will also depend on the mass and $P_t$ of the $\tilde\chi^0$, 
which depend on the choice of SUSY benchmark).
This is the more experimentally challenging of the two variants, because one can
no longer suppress background by assuming that the
momentum vector of the displaced $f\bar f$ system is collinear with 
direction from the beam spot to the displaced vertex. (However, one could limit oneself
to the special case when $B = Z^0$, which allows one to use the $Z^0$ mass constraint 
to suppress background, and makes the cross-section dependent on one less free parameter). 

Technically, this benchmark is very similar to that based on $\nu_H\to \nu Z^0$.
described in Sect.~\ref{TomalinStrassler:LongLived:HBBvar}. So again, it may not be necessary to
study both. As was the case for that benchmark, more generally applicable results
could be obtained by only explicitely reconstructing the boson $B$, and not
attempting to find the $\chi^0_1$.

\end{enumerate}

\subsection{Charged or Colored Long-Lived Exotics}

In all the above benchmarks, one can consider the possibility of the long-lived exotic being 
charged. The analysis remains similar, although the final states will be subtly different 
(e.g. $l^-\bar\nu$ instead of $l^-l^+$). It may be worth specifically searching not only for
the tracks produced by the daughters of the exotic, but also for the highly ionizing track 
produced by the exotic itself. These should all form a common vertex.
(In the case a signal is seen, this will help clarify its exact nature, whereas if no signal is
seen, it may suppress background). This technique may be particularly helpful for signatures such
as the GMSB SUSY $\tilde\tau^-\to\tau^-\tilde G$, where $\tilde\tau^-$ is long-lived and the
gravitino $\tilde G$ is invisible. The decay point of the $\tilde\tau^-$ can be found by searching 
for the vertex of the $\tilde\tau^-$ and $\tau^-$ tracks.


For colored exotics decaying in flight, there are subtle issues.  (An example 
of such an exotic is the long-lived gluino, predicted in `Split SUSY' theories, where the gluino
is much lighter than the squarks \cite{ArkaniHamed:2004fb}). These exotics may form
neutral or charged exotic hadrons, which may or may not have a track pointing to the decay
vertex.  Moreover, there is some probability of a nearby pion being formed in the hadronization
process (even though gluon radiation is suppressed for a massive particle) and this could give
a nearby soft track that could impact vertex-isolation requirements in an analysis.  On the other
hand, colored particles are generally most often pair-produced (rather than produced in a decay of 
a heavier particle) and therefore they travel at a variety of speeds.  Some fraction of them might
therefore be detectable through the late arrival of their decay products at the ECAL.

\subsection{Very Light Long-Lived Particles}

A special case occurs if the long-lived exotic is very light (less than or of order
1 GeV/$c^2$). In this low mass case, kinematics mean that it can only decay
to pairs of leptons (or pairs of hadrons). These will often have a very
small opening angle, which can make them difficult to resolve
experimentally (and prevent them being selected by isolated lepton
triggers) \cite{Abazov:2009hn}.  Also, they can appear in clusters, further
complicating isolation requirements.

As explained in Sect.~\ref{TomalinStrassler:LongLived:Mult}, there is
theoretical motivation for this scenario, in which the long-lived
particles are known as ``dark photons''. This makes it well worth pursuing.
Since these decay to 
normal matter only as a result of mixing with the photon, their
decay branching ratios are determined by electromagnetic couplings.
After measuring the cross-section of these particles to decay to
individual fermion species, one can therefore subsequently use
this theoretical knowledge to combine the separate channels 
and/or obtain a measurement of the dark photon production cross-section.


\



\


\




\section{Trigger/reconstruction benchmarks}
\label{TomalinStrassler:LongLived:Trig}

A serious concern that has arisen in the study of models with new
long-lived particles involves the behavior of the triggering and
reconstruction software of the LHC (and Tevatron) experiments for such
events.  It is now rather well-appreciated that triggering can be a
serious challenge in certain events with long-lived particles.
Studies of various examples are on-going and there are some public
results \cite{ATLASHVtrig}.  However, only preliminary studies have
been performed of the risks at the step following triggering:
reconstruction and data storage.  The primary reconstruction software,
designed for obvious reasons to look for jets, leptons and photons
emerging from or near the interaction point, may behave unpredictably
when faced with long-lived decays, especially in cases that have not
already been actively studied in the context of gauge-mediated
supersymmetry breaking with photons \cite{Dimopoulos:1996vz}.  While it is
certainly necessary, when looking for long-lived particles, to run
special reconstruction algorithms, it is often not feasible to run
these algorithms on the full data set. It is therefore important that
data set be reduced through some initial event selection to a
manageable size.  (This is particularly true for final states
involving only jets, because of the large QCD background).  If the
primary reconstruction process in some way fails to identify events
with long-lived particles as candidates for specialized
reconstruction, it might turn out to be impossible to collect a large
fraction of the signal into an analysis sample.  Conversely, if the
primary reconstruction software can recognize and flag events with
unusual features that merit inclusion in a long-lived particle
analysis, this may significantly increase the fraction of signal that
can be collected.

After some discussion of this issue, it was generally agreed that a
stress-test of the trigger system and reconstruction software of the experiments is
warranted.  Toward this end, a number of simulated data sets, from a
variety of models with long-lived particles, has been assembled.  Some of the 
models produce simple signatures, while others
produce relatively extreme (though realistic) signals that,
though not especially probable, are appropriate for testing the
behavior of the reconstruction software.  It must be emphasized that
for this reason these more complex
models should {\it not} be viewed as proper
benchmark models for early LHC analyses, and they may turn out not 
to be good benchmark
models for later analyses either.  Moreover, for some models,
the simulation techniques employed in the event generator are crude.
Any serious experimental analysis would deserve more carefully
constructed event generators (which for most of these models are under
construction or consideration.)  However, in all models the events
themselves are consistent with the underlying physical process ---
only the statistical distribution of the events over phase space is
not entirely correct. For this reason, the limitations just described
should not much affect the realism of individual events, and so any
problems observed in the reconstruction of these events should still
allow important lessons regarding the behavior of the software to be
drawn.

Though relatively simple in their signatures, the simple models discussed in 
Sect.~\ref{TomalinStrassler:LongLived:Analysis}
as analysis benchmarks are also  appropriate for the trigger/reconstruction stress-test.
It is already known that certain models with mainly low-energy jets produced by 
long-lived particles ({\it e.g.} $H \to B\bar B$ followed by displaced 
$B\to b\bar b$) can cause trouble for the ATLAS trigger \cite{ATLASHVtrig}. 
Even when triggered, $B\bar B$ final states with decay lengths of order or 
greater than $\sim$20 cm may lack hits in the pixel detectors, which can lead to a 
failure of track reconstruction.   For $B\to$ dijets, a partially successful
reconstruction may be insufficient, because of large QCD backgrounds from
secondary interactions.  

Special problems arise when $B$ is very light (and potentially also for larger masses
if $B$ is sufficiently boosted.)  Even when prompt, 
very light ($\sim$ 1 GeV) dilepton resonances ({\it e.g.} dark photons) with a boost
factor $\gg 1$ can cause
various problems for triggering and reconstruction, including but not limited to
a failure of isolation requirements (with each lepton ruining the isolation
of the other) or because only one of the two close leptons is identified.  A
long lifetime for the new particle compounds these problems.

While there have been some public trigger studies at ATLAS \cite{ATLASHVtrig},
the question of whether these relatively simple signatures 
cause problems for reconstruction at the LHC experiments has so
far been only subject to preliminary studies.
Any issues that arise in these simpler settings will need to be 
addressed before there is any hope of understanding the situation for 
more complex signatures, which we now discuss.

Complex signatures easily arise once the multiplicity of long-lived 
particles exceeds one or two.  The distribution of decay vertices and their daughter particles 
around the detector can be enormously variable and complicated.
Many classes of Hidden Valley models can produce high-multiplicity
final states and/or long-lived particles in some regions of parameter
space.  Consequently, Hidden Valleys serve as a useful set from which to select
examples of physically realistic phenomena that might be especially challenging
for reconstruction software.

As emphasized in \cite{Strassler:2006im}, 
high-multiplicity states may result through a number of mechanisms,
including cascade decays within the hidden sector, parton showering
within the hidden sector, and/or hidden sector hadronization.  For
those hidden sector particles which are forbidden by kinematics and/or
quantum numbers from decaying to final states of purely hidden sector
particles, their decays to standard model particles are relatively
slow, and their lifetimes relatively long.  This arises because of the
weak couplings (through small mixings or irrelevant interactions) of
the hidden sector to the standard model sector.  As for hadrons in
QCD, their lifetimes may be further enhanced by approximately
conserved quantum numbers (analogous to strangeness or CP).
Furthermore, a given hidden sector often has multiple metastable
hidden-sector particles with relatively long lifetimes.  There is
typically, therefore, a wide range of parameters over which a hidden
valley model will produce at least one type of particle that will
generally decay with an observably displaced vertex, often well within
the detector volume.

The models chosen for the reconstruction stress-test draw upon the
same three production mechanisms for v-particles described earlier: 
resonance decay, LSP (or similar) decay, and quirk
annihilation.  (Other mechanisms certainly may arise but these three
suffice to give the wide variety of kinematic distributions needed for
the stress-test.)  Holding the production mechanism fixed, and within
a given class of models, experience has shown that one may usually
vary the v-particle decay chains or showering rates, masses, and
lifetimes as almost independent parameters,\footnote{This is not true
of the most minimal hidden valley models, where certain relations
between these quantities and overall cross-sections often hold.  For
early-LHC {\it analyses}, these more minimal models are more suitable
as benchmarks.  Here, for reconstruction {\it stress-tests}, a more
complex signature, even if it only arises from a non-minimal hidden
valley model, is
sometimes more appropriate.}, subject to relatively weak constraints from
existing data.  These phenomenological parameters may be adjusted so
as to create unusual but nevertheless realistic and plausible final
states which are qualitatively unlike those for which reconstruction
software was designed.

For any hidden valley model, key aspects of its signatures are
determined by the quantum numbers of any metastable v-particles which
produce standard model particles in its decays.  Since v-particles
are always neutral \footnote{This is simply by definition; charged or
colored particles that couple to the hidden sector are not v-particles
but rather ``communicators'' or ``mediators'' between the two sectors,
and are constrained by experiment to be rather heavy.} their masses
are not directly constrained by experiment, so they may be very light.
On the other hand, their neutrality limits their possible final states
to manageable sets.  As discussed earlier,
there are three types of two-body resonances that
commonly arise, assuming the hidden sector does not strongly violate
standard model flavor symmetries: (1) scalar or pseudoscalar v-bosons
that decay to the heaviest fermions available (or to three pions,
etc., if sufficiently light); (2) vector v-bosons that decay in a
generation-democratic way to fermion pairs (or to light lepton and
meson pairs, if sufficiently light) with a ``dark photon'' (a particle
coupling to the standard model only through kinetic mixing with the
photon) as a special case; (3) spin-0 and spin-2 v-bosons that couple
to photon pairs, gluon pairs, and (if kinematically allowed) weak
boson pairs.  Also commonly arising are v-particles (of any spin) that decay to
other v-particles via single emission of a photon, $Z$ or Higgs boson,
or via emission of a fermion pair.  Each one of these types of
particles arises in at least one model used in the stress test, and in
some models several such particles may be present.

Obviously the lifetimes of the v-particles play a key role in
determining the signature.  In models with more than one stable type
of v-particle, the lifetimes of these particles may vary widely,
potentially leading to prompt decays, highly-displaced decays, and/or
missing energy in the same event.  At least one model of this class
arises below.  For many of the models given below, two variants are
presented with different lifetimes for the long-lived particle(s).

A final key determinant of the final states is the multiplicity and
clustering of the v-particles.  This is highly model-dependent and
depends crucially on the details of how particle production in the
hidden sector proceeds.  A general example of a signature with high
multiplicity and complex clustering was discussed in \cite{Strassler:2006im}, where
the visibly decaying particles might decay to dijets \cite{Strassler:2008fv}
or to a mixture of jets and leptons \cite{Han:2007ae}.  Another
is the dark-matter-motivated example of a ``lepton jet'', where we
mean in this context a jet made from {\it more than one} very light
particles which decay (with a branching fraction that is substantial
though possibly not unity) to lepton pairs.

On the website 
http://www.physics.rutgers.edu/$\sim$strassler/LesHouchesModels is presented
the list of models appropriate for the trigger and reconstruction stress
tests.  For each model are provided
\begin{enumerate}
\item A Les Houches Accord event file modelname.lhe with at least 
one thousand events.  The event samples themselves are in the form of
LHE files at parton level; they must still be piped through a showering 
Monte Carlo to account for standard-model showering, decays
and hadronization.  
\item A description of the model (and the simulation technique used) in the file modelname.mdl
\item If the model was generated using a standard Monte Carlo, the appropriate run-card commands will be given in modelname.run
\item If appropriate, an SLHA file that was used in the Monte Carlo generation will be
given in modelname.spc.
\end{enumerate}

For some models the full set of particles in intermediate steps is
provided in the LHE files, but in other cases they are not; this
depends on the simulation method used.  However in all cases the
mother pointers in the LHE file are internally consistent.  Some of
the particles in intermediate steps, and certain stable particles, are
new and have non-standard PDG codes, though since they are charge- and
color-neutral no conflicts or challenges should arise with simulation.

\section{Conclusions}

We have presented some possible analysis benchmarks and for trigger/reconstruction
stress-test benchmarks for long-lived particles.   This work will clearly require revision
after data is acquired and backgrounds to displaced vertices of various types are 
better understood.  
The current benchmarks are available through the website \newline
http://www.physics.rutgers.edu/$\sim$strassler/LesHouchesModels.

There are a number of other topics that were discussed at the Les
Houches workshop that we have not covered here.  These include
interesting but exotic prompt signatures, including four-lepton
(non-$Z$) decays of the Higgs boson, prompt light dilepton resonances,
produced isolated or in clusters; Monte Carlo implementation of
non-abelian hidden sectors in HERWIG and in SHERPA; and the simulation
of the various phenomena associated with (microscopic) quirks.

\section*{Acknowledgements}

The authors would like to thank the University of Washington LHC~group
for hosting the workshop ``Signatures of Long-Lived Exotic Particles
at the LHC'' at which some of this work was initiated, and the DOE
which supported that workshop under Task~TeV of contract
DE-FGO2-96-ER40956.  The work of MJS also was supported under grants
DOE--DE-FG02-96ER40959 and NSF--PHY0904069.

%% file: Morrissey/Morrissey.tex
\chapter{A benchmark SUSY Abelian hidden sector}

{\it D.E.~Morrissey, D.~Poland and K.M. Zurek}

\begin{abstract}
  New and unusual collider signatures can arise if the MSSM couples to
a light hidden sector through gauge kinetic mixing.  In this note we
describe the minimal supersymmetric realization of this scenario.
This model provides a simple benchmark for future LHC collider studies
of light hidden sectors.
\end{abstract}

\section{Introduction}

  Supersymmetry is a well-motivated candidate for new physics beyond
the Standard Model~(SM), and the collider and cosmology signals of the
minimal supersymmetric Standard Model~(MSSM) have been studied
very extensively~\cite{Ball:2007zza,Aad:2009wy}.
However, many new possibilities can arise
if the field content of the MSSM is expanded.
One interesting extension consists of the MSSM coupled to
a new gauged \emph{hidden sector} with characteristic mass scale
near a $\text{GeV}$~\cite{ArkaniHamed:2008qp}.
Models of this type have received attention recently in relation to
potential hints of dark matter, but they are also worthy of study in their
own right since they can produce new and unusual signals at particle
colliders~\cite{ArkaniHamed:2008qp}.

  The largest effect of such a light hidden sector on the phenomenology
at the LHC comes from the fact that, even with exact $R$-parity,
the lightest MSSM superpartner will no longer be stable.
Instead, the LSP of the full theory will lie in the hidden sector.
Supersymmetric cascades initiated by QCD-charged MSSM states will therefore
terminate at the lightest MSSM superpartner, which will subsequently decay
into the hidden sector.  These decays, or subsequent cascades within the hidden
sector itself, can potentially generate highly boosted leptons or jets.

In these proceedings we describe the minimal MSSM extension containing 
a gauged light hidden sector, consisting of a Higgsed Abelian $U(1)_x$ 
gauge group that couples to the MSSM through gauge kinetic mixing.
This simple extension can display a wide variety of collider signals,
and could potentially serve of as benchmark example for collider
studies of light hidden sectors.

\section{Model, parameters, and spectrum}

  The extension of the MSSM that we consider consists of a new
supersymmetric $U(1)_x$ gauge multiplet $(X_{\mu},\,\tilde{X})$,
together with a pair of hidden Higgs fields $H$ and $H^\prime$ 
with charges $x = \pm 1$ under $U(1)_x$,
but singlets under the Standard Model gauge groups.
We take their superpotential to be~\cite{Morrissey:2009ur}
\begin{equation}
W_x  = -\mu'H\,H^\prime.
\end{equation}
We also assume soft supersymmetry-breaking couplings of the form
\begin{equation}
-\mathcal{L} \supset m_H^2|H|^2 + m_{H^\prime}^2|H^\prime|^2
+ \left[-(B\mu)^\prime HH^\prime
+ \frac{1}{2}M_x\tilde{X}\tilde{X} + h.c.\right].
\end{equation}
By redefining these fields, we can take $(B\mu)^\prime$ and $M_x$
to be real and positive with no loss of generality.
For reasons we will discuss below, all dimensionful terms in the
hidden sector are assumed to be on the order of a GeV.

To connect this hidden sector to the MSSM, we introduce a 
kinetic mixing coupling between $U(1)_x$ and hypercharge.
At the supersymmetric level, this is given by
\begin{eqnarray}
\label{Morrissey:mixing}
\mathcal{L} &\supset& \int
d^2\theta\,\left(
\frac{1}{4}B^{\alpha}B_{\alpha}
+\frac{1}{4}X^{\alpha}X_{\alpha}
+\frac{\epsilon}{2}B^{\alpha}X_{\alpha}
\right) + h.c. \\
&\supset& -\frac{1}{4}B_{\mu\nu}B^{\mu\nu} -\frac{1}{4}X_{\mu\nu}X^{\mu\nu}
-\frac{\epsilon}{2}X_{\mu\nu}B^{\mu\nu}
+ i\epsilon\tilde{X}^{\dagger}\bar{\sigma}^{\mu}\partial_{\mu}\tilde{B} + \ldots
\nonumber
\end{eqnarray}
Such a term will be generated radiatively when there are fields
charged under both $U(1)_x$ and
$U(1)_Y$~\cite{Holdom:1985ag,Babu:1996vt,Dienes:1996zr},
\begin{equation}
\Delta\epsilon(\mu) \simeq
\frac{g_x(\mu)g_Y(\mu)}{16\pi^2}\sum_ix_iY_i\;\ln\frac{\Lambda^2}{\mu^2},
\label{Morrissey:epsilonloop}
\end{equation}
where $x_i$ and $Y_i$ denote the charges
of the $i$-th field, $\Lambda$ is the UV cutoff scale, and the log
is cut off below $\mu \simeq m_i$, where $m_i$ is the (supersymmetric)
mass of the $i$-th field.  This leads to values of the kinetic mixing in the
typical range $\epsilon \simeq 10^{-4} - 10^{-2}$.  Conversely, the
kinetic mixing parameter $\epsilon$ can be highly suppressed or
absent if there exist no such bi-fundamentals, or if the
underlying gauge structure consists of a simple group.

  In the present work we concentrate on the case where all the dimensionful
couplings in the hidden $U(1)_x$ sector are on the order of a GeV,
but with the soft supersymmetry breaking (and $\mu$) terms in the MSSM
larger than a few hundred GeV.  This can arise naturally if the
mediator of supersymmetry breaking couples more strongly to the MSSM
than to the hidden sector.  The canonical example we have in mind
is gauge mediation in which the gauge messengers are charged under
the Standard Model gauge groups but not $U(1)_x$.  Supersymmetry
breaking in the MSSM sector will then be communicated to the
hidden sector through gauge kinetic mixing with characteristic size
$m_{soft}^{(x)} \leq \epsilon\,m_{soft}^{(MSSM)}$~\cite{Zurek:2008qg}.
We additionally assume that there are supergravity contributions
to supersymmetry-breaking parameters in all sectors of order 
$m_{3/2} \sim \text{GeV}$.  These contributions will be subleading
in the MSSM sector (and sufficiently small so as to avoid flavor 
mixing problems), but can be very important in the hidden sector.  
Supergravity contributions of this size also provide a natural origin 
for $\mu^{\prime} \sim \text{GeV}$
through the Giudice-Masiero mechanism~\cite{Giudice:1988yz}.

Supersymmetry breaking in the MSSM will also induce an effectively  
supersymmetric contribution to the hidden-sector potential.  
A non-vanishing hypercharge
$D$-term arises in the MSSM when the visible-sector Higgs fields
acquire VEVs with $\tan\beta \neq 1$ (induced by SUSY breaking)
\begin{equation}
\xi_Y = -\frac{g_Y}{2} c_{2\beta} v^2.
\end{equation}
The kinetic mixing operator of Eq.~(\ref{Morrissey:mixing}) then leads to an effective
Fayet-Iliopoulos~~\cite{Fayet:1974jb} term in the hidden-sector $D$-term 
potential~\cite{Dienes:1996zr,Suematsu:2006wh,
Baumgart:2009tn,Cui:2009xq}
\begin{equation}
V_D = \frac{g_x^2}{2}
c_{\epsilon}^2\left(|H|^2 -|H^\prime|^2- \frac{\epsilon}{g_x}\xi_Y
\right)^2,
\end{equation}
with $c_\epsilon = 1/\sqrt{1-\epsilon^2}$.
The hypercharge $D$-term potential retains its usual form.

  Putting together all the contributions,
the hidden sector scalar potential can be written as
\begin{equation}
V = (|\mu'|^2+\tilde m_H^2)|H|^2 + (|\mu|^2+\tilde m_{H'}^2)|H'|^2
- \left[(B\mu)'HH'+(h.c.)\right]
+ \frac{g_x^2}{2}\left(|H|^2-|H'|^2\right)^2,
\end{equation}
where we have defined
\begin{equation}
\tilde m_{H}^2 = m_{H}^2 - \epsilon\,g_x\xi_Y,~~~~~~~~~~
\tilde m_{H^\prime}^2 = m_{H^\prime}^2 + \epsilon\,g_x\xi_Y,
\end{equation}
and we have dropped terms of $O(\epsilon^2)$.  This potential is structurally 
identical to the Higgs potential in the MSSM.
Since we can take $(B\mu)^\prime$ to be real and positive (after
a suitable field redefinition), this potential is minimized with
real and non-negative vacuum expectation values~(VEVs).
It is convenient to write them as
\begin{equation}
\langle H\rangle = \eta\,\sin\alpha,~~~~~~~~
\langle H^\prime\rangle = \eta\,\cos\alpha,
\end{equation}
with $\sin\alpha,\,\cos\alpha \geq 0$.

  Extremizing the potential, we find
\begin{eqnarray}
\sin2\alpha &=& \frac{2(B\mu)^\prime}{2|\mu^\prime|^2
+\tilde{m}_H^2+\tilde{m}_{H^\prime}^2}
\label{Morrissey:alfa}\\
\eta^2 &=& -\frac{|\mu^\prime|^2}{g_x^2}
+\frac{-\tilde{m}_H^2\tan^2\alpha+\tilde{m}_{H^\prime}^2}{g_x^2(\tan^2\alpha-1)}.
\label{Morrissey:eta}
\end{eqnarray}
This solution defines a consistent local minimum provided
$\sin 2\alpha\leq 1$ and $\eta^2\geq 0$.

For non-zero $\eta$, $U(1)_x$ is broken and we obtain a massive gauge 
boson $Z_x$ with mass
\begin{equation}
m_{Z_x} = \sqrt{2}g_x\eta.
\end{equation}
This state will mix with the photon and the $Z$ due to the kinetic coupling,
but the effect on the mass eigenvalues are minuscule for $\epsilon \ll 1$.
Symmetry breaking leads to two physical CP-even scalars and one physical
CP-odd scalar.  The tree-level mass of the CP-odd state $A_x$ is
\begin{equation}
m_{A_x}^2 = \frac{2(B\mu)^\prime}{\sin 2\alpha} = 2|\mu^\prime |^2
+\tilde{m}_H^2+\tilde{m}_{H^\prime}^2.
\label{Morrissey:amass}
\end{equation}
For the two CP-even states, $h_x$ and $H_x$, the mass matrix is given by
\begin{equation}
\mathcal{M}^2_{h_x} = \left(
\begin{array}{cc}
m_{Z_x}^2s_{\alpha}^2+m_{A_x}^2c_{\alpha}^2&-(m_{Z_x}^2+m_{A_x}^2)s_{\alpha}c_{\alpha}
\\
-(m_{Z_x}^2+m_{A_x}^2)s_{\alpha}c_{\alpha}&m_{Z_x}^2c_{\alpha}^2+m_{A_x}^2s_{\alpha}^2
\end{array}
\label{Morrissey:hmass}
\right).
\end{equation}
The remaining states in the hidden sector consist of fermions from the
$U(1)_x$ neutralino (xino) and the hidden higgsinos.  Their mass matrix in
the basis $(\tilde{X},\,\tilde{H},\,\tilde{H}^\prime)$ is
\begin{equation}
\mathcal{M} =
\left(
\begin{array}{ccc}
M_x&m_{Z_x}s_{\alpha}&-m_{Z_x}c_{\alpha}\\
m_{Z_x}s_{\alpha}&0&-\mu^\prime\\
-m_{Z_x}c_{\alpha}&-\mu^\prime&0
\end{array}
\right).
\label{Morrissey:neutmass}
\end{equation}
Note that this mass matrix acquires a zero eigenvalue in the limit
that $M_x \to 0$ and $\sin\alpha \to 0$.  When the only source of supersymmetry
breaking in the hidden sector is residual gauge mediation through
gauge kinetic mixing, one naturally obtains
$M_x^2,\,(B\mu)^\prime\ll m_{Z_x}^2$ and is pushed to this limit.
The resulting light neutralino state can be highly problematic
cosmologically and could potentially overclose the universe.  
It is for this reason we include additional supergravity 
contributions to the soft parameters from the beginning.

  The masses in this sector share a strong structural similarity with
the MSSM.  In particular, we see that the lightest CP-even hidden Higgs
mass is bounded from above at tree level by
\begin{equation}
m_{h_x}^2 \leq m_{Z_x}^2\cos^22\alpha.
\end{equation}
This has important implications for the decay properties of the lighter
Higgs state since there need not be any light hidden fermions in the spectrum.
In particular, the decay of $h_x$ can be very slow if the only channel
available to it involves a loop or a pair of highly off-shell $Z_x$ gauge
bosons.

  For the fermions, the mass matrix has a similar form to the MSSM,
but without a wino state.  Only the hidden higgsinos couple to gauge
bosons.  In the absence of large mixing between the xino and the higgsinos,
the relative mass gap between two mostly higgsino states will be less
than the $Z_x$ mass.  Thus, for $M_x \ll \mu'$ or $M_x \gg \mu'$,
the heaviest neutralino will decay to the lighter state(s) primarily
by emitting a hidden Higgs.  In the latter case, the decay from the
slightly heavier mostly higgsino state to the lighter one will
have to involve an off-shell $Z_x$.  Only when there is a great
deal of mixing between the xino and higgsinos will neutralino
decays through on-shell gauge bosons be relevant~\cite{Morrissey:dmo1}.

  So far we have not considered the effects of gauge kinetic
mixing on the spectrum.  To leading order in $\epsilon$,
the gauge boson kinetic mixing can be removed by shifting
the photon field according to~\cite{Baumgart:2009tn}
\begin{equation}
A_{\mu} \to A_{\mu} - \epsilon X_{\mu}.
\end{equation}
This shift induces a coupling between MSSM fields carrying
electromagnetic charge and the $X_{\mu}$ gauge boson through
the operator
\begin{equation}
-\mathscr{L} \supset -\epsilon\,X_{\mu}J^{\mu}_{em}.
\end{equation}
This coupling allows the decay $Z_x\to f\bar{f}$,
where $f$ is a light Standard Model fermion
with width on the order of
$\Gamma_x \sim \epsilon^2g_x^2\,m_x/12\pi$~\cite{Batell:2009yf}.
Despite its small width, these decay channels will dominate if
there are no channels open kinematically in the hidden sector.
A small mixing with the Standard Model $Z$ is also
induced~\cite{Baumgart:2009tn}.

  Among the neutralinos, the kinetic mixing between the bino
and the xino can be removed most conveniently by shifting
the xino according to
\begin{equation}
\tilde{X}\to \tilde{X}-\epsilon\,\tilde{B}.
\end{equation}
This shift induces a very small mass mixing between the
hidden and MSSM neutralinos on the order of $\epsilon\,m_{Z_x}/M_1$.
More importantly, it leads to a coupling of the Bino to the hidden sector,
\begin{equation}
-\mathscr{L} \supset -\epsilon\,\sqrt{2}g_x\left(
H\;\tilde{H}\tilde{B} - H'\;\tilde{H}\tilde{B} + h.c.\right).
\label{Morrissey:lspdecay}
\end{equation}
On account of these couplings, a would-be MSSM neutralino LSP will
decay to the hidden sector according to
$\tilde{\chi}_1^0 \to \tilde{\chi}_i^x h_x,\,\tilde{\chi}_i^x H_x,\,
\tilde{\chi}_i^x A_x$.  These decays will typically be prompt
for $\epsilon > 10^{-4}$.  In the case of a non-neutralino LSP,
this coupling will also permit its decay, albeit at a lower rate.

  For the purposes of defining a benchmark model, let us point out
that the phenomenology of the model (in the hidden sector) can be
specified by the seven parameters:
\begin{equation}
\left\{g_x,\,\epsilon,\,m_{Z_x},\,m_{A_x},\,\tan\alpha,\,\mu',\,M_x\right\}.
\end{equation}
All these parameters are defined at the low (GeV) scale.
Note that we have implicitly used the minimization conditions to
eliminate $\tilde{m}_H^2$, $\tilde{m}_{H^\prime}^2$, and $(B\mu)^\prime$
in favor of $m_{Z_x}$, $m_{A_x}$, and $\tan\alpha$ which have a more
intuitive interpretation.  We could also have included a soft
mass mixing between the bino and xino, but it is consistent to
neglect such a term provided the origin of $\epsilon$ is
supersymmetric.

\section{Constraints and signatures}

  A light hidden sector of this form can produce striking signals
at the LHC~\cite{ArkaniHamed:2008qp}.
These originate primarily from supersymmetric MSSM cascade
decays.  Instead of terminating at the lightest MSSM superpartner,
these cascades will continue into the hidden sector.
The cascade will continue in
the hidden sector until the lightest superpartner is reached.
If these produce one or more $Z_x$ gauge bosons or $h_x$ hidden Higgses,
a distinctive signature will arise from decays of the $Z_x$  or $h_x$
to Standard Model leptons.  These leptons will typically have a small
invariant mass on account of their origin, but they will also be
highly collimated since their energy scale is set by the mass of the
much heavier MSSM states~\cite{Baumgart:2009tn,Cheung:2009su}.
This construction is a simple and perturbative example of a hidden
valley scenario~\cite{Strassler:2006im,Strassler:2006ri}.

  Before discussing the collider signatures of this scenario, 
let us first outline the bounds on a light hidden sector with a kinetic
mixing to hypercharge.  The most stringent bounds come from
the induced $Z_x$ coupling to SM fermions.  For $m_{Z_x} \sim \text{GeV}$,
and with $Z_x\to f\bar{f}$ the dominant decay mode,
limits from $(g\!-\!2)_{\mu}$ and $B$-physics force $\epsilon < 10^{-3}$
or so~\cite{Pospelov:2008zw,Bjorken:2009mm}.  
Additional constraints can arise if $h_x$ is 
long-lived~\cite{Batell:2009yf,Batell:2009di,Schuster:2009au,Meade:2009mu}.
This arises naturally if $m_{h_x} < m_{Z_x}$ and there are no
lighter $x$-sector fermions, in which case it can only decay through
a loop with $Z_x$'s or through an off-shell pair of $Z_x$'s to
the SM~\cite{Batell:2009yf}.  Constraints on both the $Z_x$ and $h_x$
states are weakened considerably if they decay primarily to the
invisible hidden-sector LSP.

  A wide range of collider signals can arise from the minimal hidden
sector model discussed here.  The dominant mode of production at the
LHC is expected to be through MSSM cascade decays.  These cascades
will proceed to the MSSM LSP, which we assume here to be the lightest
MSSM neutralino.  This LSP will subsequently decay to the hidden 
sector through the interaction of Eq.~(\ref{Morrissey:lspdecay}),
and the decay cascade will continue in the hidden sector down to the LSP.
Along the way, $h_x$, $H_x$, $A_x$ and $Z_x$ states will be emitted,
which can potentially generate new signals.

  Consider first the situation for $\min\{\mu',\,M_x\} < m_{h_x}/2$.
In this case the $Z_x$ and $h_x$ bosons emitted in the decay cascades
will almost always decay to pairs of the lightest neutralino LSP.
The LHC collider signatures are then nearly identical to
those of the MSSM in the absence of the hidden sector:
even though the decay cascades continue, they still only produce
missing energy.

  A more interesting case is $m_{A_x} \gg \mu'\sim M_x > m_{Z_x}$ such that
all the hidden neutralinos are well-mixed with $m_{\chi_1^x} > m_{Z_x}$
and $m_{\chi_3^x}-m_{\chi_1^x} > m_{Z_x}$.  This spectrum allows the
decays $H_x \to Z_xZ_x$ along with $\chi_3^x\to \chi_1^xZ_x$,
while the only decay channel open to the gauge boson is back
to the Standard Model, $Z_x\to f\bar{f}$.  In this situation,
some of the decay cascades will produce new visible signatures
from the $Z_x$ decays to highly boosted Standard Model states.
For $m_{Z_x} < 700\,\text{MeV}$, these decay products will frequently form
\emph{lepton jets} consisting of highly collimated
lepton pairs~\cite{ArkaniHamed:2008qp,Baumgart:2009tn,Cheung:2009su}.
Note that in this case, the $h_x$ state is typically very 
long-lived and escapes the detector~\cite{Batell:2009yf}.
Thus, some of these can also include missing energy.

  We are currently working on implementing this simple benchmark
model into together with a spectrum generator to generate the
necessary SLHA format input data~\cite{Morrissey:dmo1}.


\section{Conclusions}

Light hidden sectors can lead to exciting new signals at the LHC.
We have presented here a simple benchmark Abelian hidden in the context
of supersymmetry that could be useful for future collider studies.

\section*{Acknowledgements}

  We thank the organizers and participants of the
``Physics at Tev Colliders'' program for an excellent workshop,
as well as the \'Ecole de Physique des Houches for the use of
their facilities and the delicious cheese.

%% file: LH_NewPhysics_Report.bbl
\begin{thebibliography}{100}

\bibitem{twitter}
http://twitter.com/CERN/.

\bibitem{Amsler:2008zzb}
Particle Data Group, C. Amsler et~al.,
\newblock Phys. Lett. B667 (2008) 1.

\bibitem{Barr:2010zj}
A.J. Barr and C.G. Lester,
\newblock (2010), arXiv:1004.2732.

\bibitem{Allanach:2002nj}
B.C. Allanach et~al.,
\newblock Eur. Phys. J. C25 (2002) 113, hep-ph/0202233.

\bibitem{Alwall:2008qv}
J. Alwall, S. de~Visscher and F. Maltoni,
\newblock JHEP 02 (2009) 017, hep-ph/0810.5350.

\bibitem{Stelzer:1994ta}
T. Stelzer and W.F. Long,
\newblock Comput. Phys. Commun. 81 (1994) 357, hep-ph/9401258.

\bibitem{Maltoni:2002qb}
F. Maltoni and T. Stelzer,
\newblock JHEP 02 (2003) 027, hep-ph/0208156.

\bibitem{Meade:2007js}
P. Meade and M. Reece,
\newblock (2007), hep-ph/0703031.

\bibitem{pythia}
T. Sjostrand, S. Mrenna and P. Skands,
\newblock JHEP 05 (2006) 026, hep-ph/0603175.

\bibitem{Alwall:2007fs}
J. Alwall et~al.,
\newblock Eur. Phys. J. C53 (2008) 473, hep-ph/0706.2569.

\bibitem{Mangano:2002ea}
M.L. Mangano et~al.,
\newblock JHEP 07 (2003) 001, hep-ph/0206293.

\bibitem{Ovyn:2009tx}
S. Ovyn, X. Rouby and V. Lemaitre,
\newblock (2009), hep-ph/0903.2225.

\bibitem{Belov:2007qg}
S. Belov et~al.,
\newblock Comput. Phys. Commun. 178 (2008) 222, hep-ph/0703287.

\bibitem{Hinchliffe:1996iu}
I. Hinchliffe et~al.,
\newblock Phys. Rev. D55 (1997) 5520, hep-ph/9610544.

\bibitem{Tovey:2000wk}
D.R. Tovey,
\newblock Phys. Lett. B498 (2001) 1, hep-ph/0006276.

\bibitem{Konar:2008ei}
P. Konar, K. Kong and K.T. Matchev,
\newblock JHEP 03 (2009) 085, arXiv:0812.1042.

\bibitem{Barger:1987du}
V.D. Barger, T. Han and R.J.N. Phillips,
\newblock Phys. Rev. D36 (1987) 295.

\bibitem{Chizhov:2006nw}
M.V. Chizhov,
\newblock (2006), hep-ph/0609141.

\bibitem{LesterSummers:1999}
C.G. Lester and D.G. Summers,
\newblock Phys.Lett. B463 (1999) 99, hep-ph/9906349v1.

\bibitem{ChoChoi:2007}
W.S. Cho et~al.,
\newblock Phys. Rev. Lett. 100 (2008) 171801, hep-ph/0709.0288v2.

\bibitem{ChoChoi:2008}
W.S. Cho et~al.,
\newblock JHEP 0802 (2008) 035, hep-ph/0711.4526v2.

\bibitem{BurnsKong:2009}
M. Burns et~al.,
\newblock JHEP 0903 (2009) 143, hep-ph/0810.5576v2.

\bibitem{Bachacou:1999zb}
H. Bachacou, I. Hinchliffe and F.E. Paige,
\newblock Phys. Rev. D62 (2000) 015009, hep-ph/9907518.

\bibitem{:1999fr}
ATLAS Collaboration, A. Airapetian et~al.,
\newblock CERN-LHCC-99-15.

\bibitem{Allanach:2000kt}
B.C. Allanach et~al.,
\newblock JHEP 09 (2000) 004, hep-ph/0007009.

\bibitem{Kawagoe:2004rz}
K. Kawagoe, M.M. Nojiri and G. Polesello,
\newblock Phys. Rev. D71 (2005) 035008, hep-ph/0410160.

\bibitem{Cheng:2007xv}
H.C. Cheng et~al.,
\newblock JHEP 12 (2007) 076, arXiv:0707.0030.

\bibitem{Nojiri:2007pq}
M.M. Nojiri, G. Polesello and D.R. Tovey,
\newblock JHEP 05 (2008) 014, arXiv:0712.2718.

\bibitem{Cheng:2008mg}
H.C. Cheng et~al.,
\newblock Phys. Rev. Lett. 100 (2008) 252001, arXiv:0802.4290.

\bibitem{Cheng:2009fw}
H.C. Cheng et~al.,
\newblock Phys. Rev. D80 (2009) 035020, arXiv:0905.1344.

\bibitem{Webber:2009vm}
B. Webber,
\newblock JHEP 09 (2009) 124, arXiv:0907.5307.

\bibitem{Cheng:2008hk}
H.C. Cheng and Z. Han,
\newblock JHEP 12 (2008) 063, arXiv:0810.5178.

\bibitem{wimpmass}
http://particle.physics.ucdavis.edu/hefti/projects/doku.php?id=wimpmass.

\bibitem{Basso:2008iv}
L. Basso et~al.,
\newblock Phys. Rev. D80 (2009) 055030, arXiv:0812.4313.

\bibitem{oxbridge}
http://www.hep.phy.cam.ac.uk/~lester/mt2/index.html.

\bibitem{Gjelsten:2004ki}
B.K. Gjelsten, D.J. Miller and P. Osland,
\newblock JHEP 12 (2004) 003, hep-ph/0410303.

\bibitem{Wienemann:2008jk}
ATLAS Collaboration, P. Wienemann et~al.,
\newblock AIP Conf. Proc. 1078 (2009) 286, arXiv:0809.2204.

\bibitem{Allanach:2001kg}
B.C. Allanach,
\newblock Comput. Phys. Commun. 143 (2002) 305, hep-ph/0104145.

\bibitem{Porod:2003um}
W. Porod,
\newblock Comput. Phys. Commun. 153 (2003) 275, hep-ph/0301101.

\bibitem{Abe:1991ui}
CDF Collaboration, F. Abe et~al.,
\newblock Phys. Rev. D45 (1992) 1448.

\bibitem{Arnison:1983rp}
UA1 Collaboration, G. Arnison et~al.,
\newblock Phys. Lett. B122 (1983) 103.

\bibitem{Banner:1983jy}
UA2 Collaboration, M. Banner et~al.,
\newblock Phys. Lett. B122 (1983) 476.

\bibitem{Aaltonen:2007ps}
CDF Collaboration, T. Aaltonen et~al.,
\newblock Phys. Rev. D77 (2008) 112001, arXiv:0708.3642.

\bibitem{Lester:1999tx}
C.G. Lester and D.J. Summers,
\newblock Phys. Lett. B463 (1999) 99, hep-ph/9906349.

\bibitem{Barr:2003rg}
A. Barr, C. Lester and P. Stephens,
\newblock J. Phys. G29 (2003) 2343, hep-ph/0304226.

\bibitem{Cho:2008cu}
W.S. Cho et~al.,
\newblock Phys. Rev. D78 (2008) 034019, arXiv:0804.2185.

\bibitem{cdfmt2}
CDF Collaboration, 2009,
\newblock {CDF} note 9769.

\bibitem{Barr:2009mx}
A.J. Barr, B. Gripaios and C.G. Lester,
\newblock (2009), arXiv:0902.4864.

\bibitem{DeRoeck:2009id}
A. De~Roeck et~al.,
\newblock (2009), arXiv:0909.3240.

\bibitem{Nojiri:2003tu}
M.M. Nojiri, G. Polesello and D.R. Tovey,
\newblock (2003), hep-ph/0312317.

\bibitem{gripaios:2007is}
B. Gripaios,
\newblock JHEP 02 (2008) 053, arXiv:0709.2740.

\bibitem{Lester:2007fq}
C. Lester and A. Barr,
\newblock JHEP 12 (2007) 102, arXiv:0708.1028.

\bibitem{Barr:2009jv}
A.J. Barr, B. Gripaios and C.G. Lester,
\newblock JHEP 11 (2009) 096, arXiv:0908.3779.

\bibitem{Kim:2009si}
I.W. Kim,
\newblock (2009), arXiv:0910.1149.

\bibitem{Lester}
C.G. Lester,
\newblock (2009),
\newblock private communication.

\bibitem{Rumsfeld}
D. Rumsfeld,
\newblock (2002),
\newblock US Department of Defense news briefing,
  http://www.defense.gov/transcripts/transcript.aspx?transcriptid=2636.

\bibitem{Cho:2007qv}
W.S. Cho et~al.,
\newblock Phys. Rev. Lett. 100 (2008) 171801, arXiv:0709.0288.

\bibitem{Barr:2007hy}
A.J. Barr, B. Gripaios and C.G. Lester,
\newblock JHEP 02 (2008) 014, arXiv:0711.4008.

\bibitem{Aaltonen:2008dn}
CDF Collaboration, T. Aaltonen et~al.,
\newblock Phys. Rev. D79 (2009) 112002, arXiv:0812.4036.

\bibitem{Abazov:2007ww}
D0 Collaboration, V.M. Abazov et~al.,
\newblock Phys. Lett. B660 (2008) 449, arXiv:0712.3805.

\bibitem{Aaltonen:2008rv}
CDF Collaboration, T. Aaltonen et~al.,
\newblock Phys. Rev. Lett. 102 (2009) 121801, arXiv:0811.2512.

\bibitem{Kilic:2008pm}
C. Kilic, T. Okui and R. Sundrum,
\newblock JHEP 07 (2008) 038, arXiv:0802.2568.

\bibitem{RSW:Essig}
S. Chuang et~al.,
\newblock {Colored Resonances at the Tevatron: Phenomenology and Discovery
  Potential in Multijets}.

\bibitem{Kaplan:2008pt}
D.E. Kaplan and M.D. Schwartz,
\newblock Phys. Rev. Lett. 101 (2008) 022002, arXiv:0804.2477.

\bibitem{Butterworth:2009qa}
J.M. Butterworth et~al.,
\newblock (2009), arXiv:0906.0728.

\bibitem{Butterworth:2002tt}
J.M. Butterworth, B.E. Cox and J.R. Forshaw,
\newblock Phys. Rev. D65 (2002) 096014, hep-ph/0201098.

\bibitem{Butterworth:2007ke}
J.M. Butterworth, J.R. Ellis and A.R. Raklev,
\newblock JHEP 05 (2007) 033, hep-ph/0702150.

\bibitem{Butterworth:2008iy}
J.M. Butterworth et~al.,
\newblock Phys. Rev. Lett. 100 (2008) 242001, arXiv:0802.2470.

\bibitem{Thaler:2008ju}
J. Thaler and L.T. Wang,
\newblock JHEP 07 (2008) 092, arXiv:0806.0023.

\bibitem{Almeida:2008yp}
L.G. Almeida et~al.,
\newblock (2008), arXiv:0807.0234.

\bibitem{Kaplan:2008ie}
D.E. Kaplan et~al.,
\newblock Phys. Rev. Lett. 101 (2008) 142001, arXiv:0806.0848.

\bibitem{Krohn:2009zg}
D. Krohn, J. Thaler and L.T. Wang,
\newblock (2009), arXiv:0903.0392.

\bibitem{Ellis:2009su}
S.D. Ellis, C.K. Vermilion and J.R. Walsh,
\newblock (2009), arXiv:0903.5081.

\bibitem{Plehn:2009rk}
T. Plehn, G.P. Salam and M. Spannowsky,
\newblock (2009), arXiv:0910.5472.

\bibitem{Krohn:2009th}
D. Krohn, J. Thaler and L.T. Wang,
\newblock (2009), arXiv:0912.1342.

\bibitem{Raby:1997bpa}
S. Raby,
\newblock Phys. Lett. B422 (1998) 158, hep-ph/9712254.

\bibitem{Raby:1998xr}
S. Raby and K. Tobe,
\newblock Nucl. Phys. B539 (1999) 3, hep-ph/9807281.

\bibitem{Berger:2008cq}
C.F. Berger et~al.,
\newblock JHEP 02 (2009) 023, arXiv:0812.0980.

\bibitem{Alwall:2008va}
J. Alwall et~al.,
\newblock Phys. Rev. D79 (2009) 015005, arXiv:0809.3264.

\bibitem{Aaltonen:2008pv}
CDF Collaboration, T. Aaltonen et~al.,
\newblock Phys. Rev. Lett. 101 (2008) 251801, arXiv:0808.2446.

\bibitem{Corcella:2000bw}
G. Corcella et~al.,
\newblock JHEP 01 (2001) 010, hep-ph/0011363.

\bibitem{Dreiner:1999qz}
H.K. Dreiner, P. Richardson and M.H. Seymour,
\newblock JHEP 04 (2000) 008, hep-ph/9912407.

\bibitem{Moretti:2002eu}
S. Moretti et~al.,
\newblock JHEP 04 (2002) 028, hep-ph/0204123.

\bibitem{Corcella:2002jc}
G. Corcella et~al.,
\newblock (2002), hep-ph/0210213.

\bibitem{Pumplin:2002vw}
J. Pumplin et~al.,
\newblock JHEP 07 (2002) 012, hep-ph/0201195.

\bibitem{Butterworth:1996zw}
J.M. Butterworth, J.R. Forshaw and M.H. Seymour,
\newblock Z. Phys. C72 (1996) 637, hep-ph/9601371.

\bibitem{Buskulic:1995sw}
ALEPH Collaboration, D. Buskulic et~al.,
\newblock Phys. Lett. B384 (1996) 353.

\bibitem{Abazov:2001yp}
D0 Collaboration, V.M. Abazov et~al.,
\newblock Phys. Rev. D65 (2002) 052008, hep-ex/0108054.

\bibitem{Abbiendi:2003cn}
OPAL Collaboration, G. Abbiendi et~al.,
\newblock Eur. Phys. J. C31 (2003) 307, hep-ex/0301013.

\bibitem{Abbiendi:2004pr}
OPAL Collaboration, G. Abbiendi et~al.,
\newblock Eur. Phys. J. C37 (2004) 25, hep-ex/0404026.

\bibitem{Chekanov:2004kz}
ZEUS Collaboration, S. Chekanov et~al.,
\newblock Nucl. Phys. B700 (2004) 3, hep-ex/0405065.

\bibitem{Acosta:2005ix}
CDF Collaboration, D.E. Acosta et~al.,
\newblock Phys. Rev. D71 (2005) 112002, hep-ex/0505013.

\bibitem{Cacciari:2005hq}
M. Cacciari and G.P. Salam,
\newblock Phys. Lett. B641 (2006) 57, hep-ph/0512210.

\bibitem{RSW:FastJetWeb}
M. Cacciari, G.P. Salam and G. Soyez,
\newblock {FastJet},
\newblock {http://fastjet.fr/}.

\bibitem{Waugh:2006ip}
B.M. Waugh et~al.,
\newblock (2006), hep-ph/0605034.

\bibitem{Catani:1993hr}
S. Catani et~al.,
\newblock Nucl. Phys. B406 (1993) 187.

\bibitem{Ellis:1993tq}
S.D. Ellis and D.E. Soper,
\newblock Phys. Rev. D48 (1993) 3160, hep-ph/9305266.

\bibitem{Abulencia:2007ez}
CDF Collaboration, A. Abulencia et~al.,
\newblock Phys. Rev. D75 (2007) 092006, hep-ex/0701051.

\bibitem{Abulencia:2007iy}
CDF Collaboration, A. Abulencia et~al.,
\newblock Phys. Rev. D75 (2007) 092004.

\bibitem{Higgs:1964ia}
P. Higgs,
\newblock Phys. Lett. 12 (1964) 132.

\bibitem{Higgs:1964pj}
P. Higgs,
\newblock Phys. Rev. Lett. 13 (1964) 508.

\bibitem{Higgs:1966ev}
P. Higgs,
\newblock Phys. Rev. 145 (1966) 1156.

\bibitem{Englert:1964et}
F. Englert and R. Brout,
\newblock Phys. Rev. Lett. 13 (1964) 321.

\bibitem{Guralnik:1964eu}
G.S. Guralnik, C.R. Hagen and T.W.B. Kibble,
\newblock Phys. Rev. Lett. 13 (1964) 585.

\bibitem{Fayet:1974pd}
P. Fayet,
\newblock Nucl. Phys. B90 (1975) 104.

\bibitem{Fayet:1976et}
P. Fayet,
\newblock Phys. Lett. B64 (1976) 159.

\bibitem{Fayet:1977yc}
P. Fayet,
\newblock Phys. Lett. B69 (1977) 489.

\bibitem{Dimopoulos:1981zb}
S. Dimopoulos and H. Georgi,
\newblock Nucl. Phys. B193 (1981) 150.

\bibitem{Sakai:1981gr}
N. Sakai,
\newblock Zeit. Phys. C11 (1981) 153.

\bibitem{Inoue:1982ej}
K. Inoue et~al.,
\newblock Prog. Theor. Phys. 67 (1982) 1889.

\bibitem{Inoue:1982pi}
K. Inoue et~al.,
\newblock Prog. Theor. Phys. 68 (1982) 927.

\bibitem{Inoue:1983pp}
K. Inoue et~al.,
\newblock Prog. Theor. Phys. 71 (1984) 413.

\bibitem{Spira:1997dg}
M. Spira,
\newblock Fortsch. Phys. 46 (1998) 203, hep-ph/9705337.

\bibitem{Djouadi:2005gi}
A. Djouadi,
\newblock Phys. Rept. 457 (2008) 1, hep-ph/0503172.

\bibitem{Djouadi:2005gj}
A. Djouadi,
\newblock Phys. Rept. 459 (2008) 1, hep-ph/0503173.

\bibitem{Graudenz:1992pv}
D. Graudenz, M. Spira and P.M. Zerwas,
\newblock Phys. Rev. Lett. 70 (1993) 1372.

\bibitem{Spira:1993bb}
M. Spira et~al.,
\newblock Phys. Lett. B318 (1993) 347.

\bibitem{Spira:1995rr}
M. Spira et~al.,
\newblock Nucl. Phys. B453 (1995) 17, hep-ph/9504378.

\bibitem{Muhlleitner:2006wx}
M. Muhlleitner and M. Spira,
\newblock Nucl. Phys. B790 (2008) 1, hep-ph/0612254.

\bibitem{Anastasiou:2006hc}
C. Anastasiou et~al.,
\newblock JHEP 01 (2007) 082, hep-ph/0611236.

\bibitem{Aglietti:2006tp}
U. Aglietti et~al.,
\newblock JHEP 01 (2007) 021, hep-ph/0611266.

\bibitem{Bonciani:2007ex}
R. Bonciani, G. Degrassi and A. Vicini,
\newblock JHEP 11 (2007) 095, arXiv:0709.4227.

\bibitem{Kramer:1996iq}
M. Kramer, E. Laenen and M. Spira,
\newblock Nucl. Phys. B511 (1998) 523, hep-ph/9611272.

\bibitem{Djouadi:1991tka}
A. Djouadi, M. Spira and P.M. Zerwas,
\newblock Phys. Lett. B264 (1991) 440.

\bibitem{Dawson:1990zj}
S. Dawson,
\newblock Nucl. Phys. B359 (1991) 283.

\bibitem{Kauffman:1993nv}
R.P. Kauffman and W. Schaffer,
\newblock Phys. Rev. D49 (1994) 551, hep-ph/9305279.

\bibitem{Dawson:1993qf}
S. Dawson and R. Kauffman,
\newblock Phys. Rev. D49 (1994) 2298, hep-ph/9310281.

\bibitem{Dawson:1996xz}
S. Dawson, A. Djouadi and M. Spira,
\newblock Phys. Rev. Lett. 77 (1996) 16, hep-ph/9603423.

\bibitem{Harlander:2002wh}
R.V. Harlander and W.B. Kilgore,
\newblock Phys. Rev. Lett. 88 (2002) 201801, hep-ph/0201206.

\bibitem{Harlander:2002vv}
R.V. Harlander and W.B. Kilgore,
\newblock JHEP 10 (2002) 017, hep-ph/0208096.

\bibitem{Anastasiou:2002yz}
C. Anastasiou and K. Melnikov,
\newblock Nucl. Phys. B646 (2002) 220, hep-ph/0207004.

\bibitem{Anastasiou:2002wq}
C. Anastasiou and K. Melnikov,
\newblock Phys. Rev. D67 (2003) 037501, hep-ph/0208115.

\bibitem{Ravindran:2003um}
V. Ravindran, J. Smith and W.L. van Neerven,
\newblock Nucl. Phys. B665 (2003) 325, hep-ph/0302135.

\bibitem{Harlander:2009bw}
R.V. Harlander and K.J. Ozeren,
\newblock Phys. Lett. B679 (2009) 467, arXiv:0907.2997.

\bibitem{Pak:2009bx}
A. Pak, M. Rogal and M. Steinhauser,
\newblock Phys. Lett. B679 (2009) 473, arXiv:0907.2998.

\bibitem{Harlander:2009mq}
R.V. Harlander and K.J. Ozeren,
\newblock JHEP 11 (2009) 088, arXiv:0909.3420.

\bibitem{Pak:2009dg}
A. Pak, M. Rogal and M. Steinhauser,
\newblock (2009), arXiv:0911.4662.

\bibitem{Harlander:2009my}
R.V. Harlander et~al.,
\newblock (2009), arXiv:0912.2104.

\bibitem{Catani:2003zt}
S. Catani et~al.,
\newblock JHEP 07 (2003) 028, hep-ph/0306211.

\bibitem{Moch:2005ky}
S. Moch and A. Vogt,
\newblock Phys. Lett. B631 (2005) 48, hep-ph/0508265.

\bibitem{Ravindran:2005vv}
V. Ravindran,
\newblock Nucl. Phys. B746 (2006) 58, hep-ph/0512249.

\bibitem{Ravindran:2006cg}
V. Ravindran,
\newblock Nucl. Phys. B752 (2006) 173, hep-ph/0603041.

\bibitem{Harlander:2003bb}
R.V. Harlander and M. Steinhauser,
\newblock Phys. Lett. B574 (2003) 258, hep-ph/0307346.

\bibitem{Harlander:2003kf}
R. Harlander and M. Steinhauser,
\newblock Phys. Rev. D68 (2003) 111701, hep-ph/0308210.

\bibitem{Harlander:2004tp}
R.V. Harlander and M. Steinhauser,
\newblock JHEP 09 (2004) 066, hep-ph/0409010.

\bibitem{Harlander:2005if}
R.V. Harlander and F. Hofmann,
\newblock JHEP 03 (2006) 050, hep-ph/0507041.

\bibitem{Degrassi:2008zj}
G. Degrassi and P. Slavich,
\newblock Nucl. Phys. B805 (2008) 267, arXiv:0806.1495.

\bibitem{Anastasiou:2008rm}
C. Anastasiou, S. Beerli and A. Daleo,
\newblock Phys. Rev. Lett. 100 (2008) 241806, arXiv:0803.3065.

\bibitem{Degrassi:2004mx}
G. Degrassi and F. Maltoni,
\newblock Phys. Lett. B600 (2004) 255, hep-ph/0407249.

\bibitem{Aglietti:2006yd}
U. Aglietti et~al.,
\newblock (2006), hep-ph/0610033.

\bibitem{Actis:2008ug}
S. Actis et~al.,
\newblock Phys. Lett. B670 (2008) 12, arXiv:0809.1301.

\bibitem{Anastasiou:2008tj}
C. Anastasiou, R. Boughezal and F. Petriello,
\newblock JHEP 04 (2009) 003, arXiv:0811.3458.

\bibitem{Georgi:1977gs}
H.M. Georgi et~al.,
\newblock Phys. Rev. Lett. 40 (1978) 692.

\bibitem{Carena:2002qg}
M.S. Carena et~al.,
\newblock Eur. Phys. J. C26 (2003) 601, hep-ph/0202167.

\bibitem{Carena:1999py}
M.S. Carena et~al.,
\newblock Nucl. Phys. B577 (2000) 88, hep-ph/9912516.

\bibitem{Guasch:2003cv}
J. Guasch, P. Hafliger and M. Spira,
\newblock Phys. Rev. D68 (2003) 115001, hep-ph/0305101.

\bibitem{Muhlleitner:2008yw}
M. Muhlleitner, H. Rzehak and M. Spira,
\newblock JHEP 04 (2009) 023, arXiv:0812.3815.

\bibitem{Collins:1978wz}
J.C. Collins, F. Wilczek and A. Zee,
\newblock Phys. Rev. D18 (1978) 242.

\bibitem{Appelquist:1974tg}
T. Appelquist and J. Carazzone,
\newblock Phys. Rev. D11 (1975) 2856.

\bibitem{Brummer:2009ug}
F. Brummer et~al.,
\newblock JHEP 08 (2009) 011, arXiv:0906.2957.

\bibitem{Baer:1995va}
H. Baer et~al.,
\newblock Phys. Rev. D53 (1996) 6241, hep-ph/9512383.

\bibitem{Ball:2007zza}
CMS Collaboration, G.L. Bayatian et~al.,
\newblock J. Phys. G34 (2007) 995.

\bibitem{Aad:2009wy}
ATLAS Collaboration, G. Aad et~al.,
\newblock (2009), arXiv:0901.0512.

\bibitem{Allanach:2006fy}
B.C. Allanach et~al.,
\newblock (2006), hep-ph/0602198.

\bibitem{Allanach:2005kz}
B.C. Allanach and C.G. Lester,
\newblock Phys. Rev. D73 (2006) 015013, hep-ph/0507283.

\bibitem{deAustri:2006pe}
R.R. de~Austri, R. Trotta and L. Roszkowski,
\newblock JHEP 05 (2006) 002, hep-ph/0602028.

\bibitem{Allanach:2007qk}
B.C. Allanach et~al.,
\newblock JHEP 08 (2007) 023, arXiv:0705.0487.

\bibitem{Belanger:2009ti}
G. Belanger et~al.,
\newblock JHEP 11 (2009) 026, arXiv:0906.5048.

\bibitem{Roszkowski:2009ye}
L. Roszkowski, R. Ruiz~de Austri and R. Trotta,
\newblock (2009), arXiv:0907.0594.

\bibitem{Djouadi:2002ze}
A. Djouadi, J.L. Kneur and G. Moultaka,
\newblock Comput. Phys. Commun. 176 (2007) 426, hep-ph/0211331.

\bibitem{Belanger:2004yn}
G. Belanger et~al.,
\newblock Comput. Phys. Commun. 174 (2006) 577, hep-ph/0405253.

\bibitem{Belanger:2006is}
G. Belanger et~al.,
\newblock Comput. Phys. Commun. 176 (2007) 367, hep-ph/0607059.

\bibitem{Lafaye:2004cn}
R. Lafaye, T. Plehn and D. Zerwas,
\newblock (2004), hep-ph/0404282.

\bibitem{Bechtle:2004pc}
P. Bechtle, K. Desch and P. Wienemann,
\newblock Comput. Phys. Commun. 174 (2006) 47, hep-ph/0412012.

\bibitem{AbdusSalam:2009tr}
S.S. AbdusSalam et~al.,
\newblock Phys. Rev. D80 (2009) 035017, arXiv:0906.0957.

\bibitem{Goldberg:1983nd}
H. Goldberg,
\newblock Phys. Rev. Lett. 50 (1983) 1419.

\bibitem{Ellis:1983ew}
J.R. Ellis et~al.,
\newblock Nucl. Phys. B238 (1984) 453.

\bibitem{AbdusSalam:2009qd}
S.S. AbdusSalam et~al.,
\newblock (2009), arXiv:0904.2548.

\bibitem{Kraml:2007pr}
S. Kraml,
\newblock (2007), arXiv:0710.5117.

\bibitem{Pilaftsis:1998dd}
A. Pilaftsis,
\newblock Phys. Lett. B435 (1998) 88, hep-ph/9805373.

\bibitem{Demir:1998dp}
D.A. Demir,
\newblock Phys. Lett. B465 (1999) 177, hep-ph/9809360.

\bibitem{Pilaftsis:1999qt}
A. Pilaftsis and C.E.M. Wagner,
\newblock Nucl. Phys. B553 (1999) 3, hep-ph/9902371.

\bibitem{Falk:1995fk}
T. Falk, K.A. Olive and M. Srednicki,
\newblock Phys. Lett. B354 (1995) 99, hep-ph/9502401.

\bibitem{Gondolo:1999gu}
P. Gondolo and K. Freese,
\newblock JHEP 07 (2002) 052, hep-ph/9908390.

\bibitem{Nihei:2005va}
T. Nihei,
\newblock Phys. Rev. D73 (2006) 035005, hep-ph/0508285.

\bibitem{Gomez:2005nr}
M.E. Gomez et~al.,
\newblock Phys. Rev. D72 (2005) 095008, hep-ph/0506243.

\bibitem{Belanger:2006qa}
G. Belanger et~al.,
\newblock Phys. Rev. D73 (2006) 115007, hep-ph/0604150.

\bibitem{ArkaniHamed:2006mb}
N. Arkani-Hamed, A. Delgado and G.F. Giudice,
\newblock Nucl. Phys. B741 (2006) 108, hep-ph/0601041.

\bibitem{Group:2009qk}
CDF Collaboration, {Tevatron Electroweak Working Group},
\newblock (2009), arXiv:0908.2171.

\bibitem{Belanger:2008sj}
G. Belanger et~al.,
\newblock Comput. Phys. Commun. 180 (2009) 747, arXiv:0803.2360.

\bibitem{Lee:2007gn}
J.S. Lee et~al.,
\newblock Comput. Phys. Commun. 180 (2009) 312, arXiv:0712.2360.

\bibitem{Bechtle:2008jh}
P. Bechtle et~al.,
\newblock Comput. Phys. Commun. 181 (2010) 138, arXiv:0811.4169.

\bibitem{Dunkley:2008ie}
WMAP Collaboration, J. Dunkley et~al.,
\newblock Astrophys. J. Suppl. 180 (2009) 306, arXiv:0803.0586.

\bibitem{Barberio:2008fa}
Heavy Flavor Averaging Group, E. Barberio et~al.,
\newblock (2008), arXiv:0808.1297.

\bibitem{Misiak:2006zs}
M. Misiak et~al.,
\newblock Phys. Rev. Lett. 98 (2007) 022002, hep-ph/0609232.

\bibitem{Aaltonen:2007kv}
CDF Collaboration, T. Aaltonen et~al.,
\newblock Phys. Rev. Lett. 100 (2008) 101802, arXiv:0712.1708.

\bibitem{Regan:2002ta}
B.C. Regan et~al.,
\newblock Phys. Rev. Lett. 88 (2002) 071805.

\bibitem{Romalis:2000mg}
M.V. Romalis, W.C. Griffith and E.N. Fortson,
\newblock Phys. Rev. Lett. 86 (2001) 2505, hep-ex/0012001.

\bibitem{lepsusy}
ALEPH, DELPHI, L3 and OPAL Collaborations, LEP2 SUSY Working Group,
\newblock {\tt http://lepsusy.web.cern.ch/lepsusy/}.

\bibitem{Belanger:inprep}
G. Belanger et~al.,
\newblock in preparation.

\bibitem{Griffith:2009zz}
W.C. Griffith et~al.,
\newblock Phys. Rev. Lett. 102 (2009) 101601.

\bibitem{Grojean:2009fd}
C. Grojean,
\newblock (2009), arXiv:0910.4976.

\bibitem{Giudice:2007fh}
G.F. Giudice et~al.,
\newblock JHEP 06 (2007) 045, hep-ph/0703164.

\bibitem{Contino:2010mh}
R. Contino et~al.,
\newblock (2010), arXiv:1002.1011.

\bibitem{Dimopoulos:1981xc}
S. Dimopoulos and J. Preskill,
\newblock Nucl. Phys. B199 (1982) 206.

\bibitem{Banks:1984gj}
T. Banks,
\newblock Nucl. Phys. B243 (1984) 125.

\bibitem{Kaplan:1983fs}
D. Kaplan and H. Georgi,
\newblock Phys. Lett. B136 (1984) 183.

\bibitem{Kaplan:1983sm}
D. Kaplan, H. Georgi and S. Dimopoulos,
\newblock Phys. Lett. B136 (1984) 187.

\bibitem{Georgi:1984ef}
H. Georgi, D. Kaplan and P. Galison,
\newblock Phys. Lett. B143 (1984) 152.

\bibitem{Georgi:1984af}
H. Georgi and D. Kaplan,
\newblock Phys. Lett. B145 (1984) 216.

\bibitem{Dugan:1984hq}
M. Dugan, H. Georgi and D. Kaplan,
\newblock Nucl. Phys. B254 (1985) 299.

\bibitem{Contino:2003ve}
R. Contino, Y. Nomura and A. Pomarol,
\newblock Nucl. Phys. B671 (2003) 148, hep-ph/0306259.

\bibitem{Agashe:2004rs}
K. Agashe, R. Contino and A. Pomarol,
\newblock Nucl. Phys. B719 (2005) 165, hep-ph/0412089.

\bibitem{Contino:2006qr}
R. Contino, L. Da~Rold and A. Pomarol,
\newblock Phys. Rev. D75 (2007) 055014, hep-ph/0612048.

\bibitem{Barate:2003sz}
LEP Working Group for Higgs boson searches, R. Barate et~al.,
\newblock Phys. Lett. B565 (2003) 61, hep-ex/0306033.

\bibitem{Schael:2006cr}
ALEPH Collaboration, S. Schael et~al.,
\newblock Eur. Phys. J. C47 (2006) 547, hep-ex/0602042.

\bibitem{LEPHWG}
LEP Higgs Working Group,
\newblock (2002), LHWG Note 2002-02.

\bibitem{Aaltonen:2010yv}
CDF and D0 Collaborations, T. Aaltonen et~al.,
\newblock Phys. Rev. Lett. 104 (2010) 061802, arXiv:1001.4162.

\bibitem{tautaulim}
CDF and D0 Collaborations, TEVNPH Working Group,
\newblock FERMILAB-PUB-09-394-E , CDF Note 9888, DO Note 5980-CONF.

\bibitem{Peskin:1991sw}
M.E. Peskin and T. Takeuchi,
\newblock Phys. Rev. D46 (1992) 381.

\bibitem{Djouadi:1997yw}
A. Djouadi, J. Kalinowski and M. Spira,
\newblock Comput. Phys. Commun. 108 (1998) 56, hep-ph/9704448.

\bibitem{Cahn:1983ip}
R.N. Cahn and S. Dawson,
\newblock Phys. Lett. B136 (1984) 196.

\bibitem{Hikasa:1985ee}
K. Hikasa,
\newblock Phys. Lett. B164 (1985) 385.

\bibitem{Altarelli:1987ue}
G. Altarelli, B. Mele and F. Pitolli,
\newblock Nucl. Phys. B287 (1987) 205.

\bibitem{Han:1992hr}
T. Han, G. Valencia and S. Willenbrock,
\newblock Phys. Rev. Lett. 69 (1992) 3274, hep-ph/9206246.

\bibitem{Glashow:1978ab}
S.L. Glashow, D.V. Nanopoulos and A. Yildiz,
\newblock Phys. Rev. D18 (1978) 1724.

\bibitem{Kunszt:1991xk}
Z. Kunszt, Z. Trocsanyi and W.J. Stirling,
\newblock Phys. Lett. B271 (1991) 247.

\bibitem{Han:1991ia}
T. Han and S. Willenbrock,
\newblock Phys. Lett. B273 (1991) 167.

\bibitem{Raitio:1978pt}
R. Raitio and W.W. Wada,
\newblock Phys. Rev. D19 (1979) 941.

\bibitem{Ng:1983jm}
J.N. Ng and P. Zakarauskas,
\newblock Phys. Rev. D29 (1984) 876.

\bibitem{Kunszt:1984ri}
Z. Kunszt,
\newblock Nucl. Phys. B247 (1984) 339.

\bibitem{Gunion:1991kg}
J.F. Gunion,
\newblock Phys. Lett. B261 (1991) 510.

\bibitem{Marciano:1991qq}
W.J. Marciano and F.E. Paige,
\newblock Phys. Rev. Lett. 66 (1991) 2433.

\bibitem{Beenakker:2001rj}
W. Beenakker et~al.,
\newblock Phys. Rev. Lett. 87 (2001) 201805, hep-ph/0107081.

\bibitem{Beenakker:2002nc}
W. Beenakker et~al.,
\newblock Nucl. Phys. B653 (2003) 151, hep-ph/0211352.

\bibitem{Dawson:2002tg}
S. Dawson et~al.,
\newblock  D67 (2003) 071503, hep-ph/0211438.

\bibitem{Spira:1995mt}
M. Spira,
\newblock (1995), hep-ph/9510347.

\bibitem{Spira:Prog}
http://people.web.psi.ch/spira/proglist.html.

\bibitem{Espinosa:2010vn}
J.R. Espinosa, C. Grojean and M. Muhlleitner,
\newblock (2010), arXiv:1003.3251.

\bibitem{Weinberg:1979bn}
S. Weinberg,
\newblock Phys. Rev. D19 (1979) 1277.

\bibitem{Susskind:1978ms}
L. Susskind,
\newblock Phys. Rev. D20 (1979) 2619.

\bibitem{Lane:2002wv}
K. Lane,
\newblock (2002), hep-ph/0202255.

\bibitem{Hill:2002ap}
C.T. Hill and E.H. Simmons,
\newblock Physics Reports 381 (2003) 235, hep-ph/0203079.

\bibitem{Eichten:1979ah}
E. Eichten and K.D. Lane,
\newblock Phys. Lett. B90 (1980) 125.

\bibitem{Holdom:1981rm}
B. Holdom,
\newblock Phys. Rev. D24 (1981) 1441.

\bibitem{Appelquist:1986an}
T.W. Appelquist, D. Karabali and L.C.R. Wijewardhana,
\newblock Phys. Rev. Lett. 57 (1986) 957.

\bibitem{Yamawaki:1986zg}
K. Yamawaki, M. Bando and K.i. Matumoto,
\newblock Phys. Rev. Lett. 56 (1986) 1335.

\bibitem{Akiba:1986rr}
T. Akiba and T. Yanagida,
\newblock Phys. Lett. B169 (1986) 432.

\bibitem{Cohen:1988sq}
A.G. Cohen and H. Georgi,
\newblock Nucl. Phys. B314 (1989) 7.

\bibitem{Hill:1994hp}
C.T. Hill,
\newblock Phys. Lett. B345 (1995) 483, hep-ph/9411426.

\bibitem{Peskin:1990zt}
M.E. Peskin and T. Takeuchi,
\newblock Phys. Rev. Lett. 65 (1990) 964.

\bibitem{Golden:1990ig}
M. Golden and L. Randall,
\newblock Nucl. Phys. B361 (1991) 3.

\bibitem{Holdom:1990tc}
B. Holdom and J. Terning,
\newblock Phys. Lett. B247 (1990) 88.

\bibitem{Altarelli:1991fk}
G. Altarelli, R. Barbieri and S. Jadach,
\newblock Nucl. Phys. B369 (1992) 3.

\bibitem{Lane:1993wz}
K.D. Lane,
\newblock (1993), hep-ph/9401324.

\bibitem{Lane:1994pg}
K.D. Lane,
\newblock (1994), hep-ph/9409304.

\bibitem{Lane:1991qh}
K.D. Lane and M.V. Ramana,
\newblock Phys. Rev. D44 (1991) 2678.

\bibitem{Appelquist:1997fp}
T. Appelquist, J. Terning and L.C.R. Wijewardhana,
\newblock Phys. Rev. Lett. 79 (1997) 2767, hep-ph/9706238.

\bibitem{Lane:1989ej}
K.D. Lane and E. Eichten,
\newblock Phys. Lett. B222 (1989) 274.

\bibitem{Dietrich:2005wk}
D.D. Dietrich, F. Sannino and K. Tuominen,
\newblock Phys. Rev. D73 (2006) 037701, hep-ph/0510217.

\bibitem{Eichten:1996dx}
E. Eichten and K.D. Lane,
\newblock Phys. Lett. B388 (1996) 803, hep-ph/9607213.

\bibitem{Eichten:1997yq}
E. Eichten, K.D. Lane and J. Womersley,
\newblock Phys. Lett. B405 (1997) 305, hep-ph/9704455.

\bibitem{Lane:1999uh}
K.D. Lane,
\newblock Phys. Rev. D60 (1999) 075007, hep-ph/9903369.

\bibitem{Lane:2002sm}
K. Lane and S. Mrenna,
\newblock Phys. Rev. D67 (2003) 115011, hep-ph/0210299.

\bibitem{Eichten:2007sx}
E. Eichten and K. Lane,
\newblock Phys. Lett. B669 (2008) 235, arXiv:0706.2339.

\bibitem{Abdallah:2001ft}
DELPHI Collaboration, J. Abdallah et~al.,
\newblock Eur. Phys. J. C22 (2001) 17, hep-ex/0110056.

\bibitem{Schael:2004tq}
ALEPH Collaboration, S. Schael et~al.,
\newblock Phys. Lett. B614 (2005) 7.

\bibitem{Abazov:2006iq}
D0 Collaboration, V.M. Abazov et~al.,
\newblock Phys. Rev. Lett. 98 (2007) 221801, hep-ex/0612013.

\bibitem{Abazov:2009eu}
D0 Collaboration, V.M. Abazov et~al.,
\newblock (2009), arXiv:0912.0715.

\bibitem{Aaltonen:2009jb}
CDF Collaboration, T. Aaltonen et~al.,
\newblock (2009), arXiv:0912.2059.

\bibitem{Sjostrand:2006za}
T. Sjostrand, S. Mrenna and P. Skands,
\newblock JHEP 05 (2006) 026, hep-ph/0603175.

\bibitem{Brooijmans:2008se}
G.H. Brooijmans et~al.,
\newblock (2008), arXiv:0802.3715.

\bibitem{LSTCPGS}
J. Conway et~al.,
\newblock {PGS -- Pretty Good Simulator},
\newblock http://is.gd/bYsbQ.

\bibitem{Lane:2009ct}
K. Lane and A. Martin,
\newblock Phys. Rev. D80 (2009) 115001, arXiv:0907.3737.

\bibitem{LSTCCMS}
CMS Collaboration,
\newblock CMS PAS EXO-09-007, 2009.

\bibitem{LSTCbose}
T. Bose,
\newblock CMS Note, CMS CR-2008/004, 2008.

\bibitem{LSTCcmsZPrime}
CMS Collaboration, C. Collaboration,
\newblock CMS PAS EXO-09-006, 2009.

\bibitem{LSTCcmsZPrime14TeV}
CMS Collaboration, C. Collaboration,
\newblock CMS PAS EXO-08-001, 2008.

\bibitem{:2007sb}
CDF Collaboration, T. Aaltonen et~al.,
\newblock Phys. Rev. Lett. 99 (2007) 171802, arXiv:0707.2524.

\bibitem{Bando:1984ej}
M. Bando et~al.,
\newblock Phys. Rev. Lett. 54 (1985) 1215.

\bibitem{Bando:1987br}
M. Bando, T. Kugo and K. Yamawaki,
\newblock Phys. Rept. 164 (1988) 217.

\bibitem{Wess:1971yu}
J. Wess and B. Zumino,
\newblock Phys. Lett. B37 (1971) 95.

\bibitem{Witten:1983tw}
E. Witten,
\newblock Nucl. Phys. B223 (1983) 422.

\bibitem{Harvey:2007ca}
J.A. Harvey, C.T. Hill and R.J. Hill,
\newblock Phys. Rev. D77 (2008) 085017, arXiv:0712.1230.

\bibitem{Kawarabayashi:1966kd}
K. Kawarabayashi and M. Suzuki,
\newblock Phys. Rev. Lett. 16 (1966) 255.

\bibitem{Riazuddin:1966sw}
Riazuddin and Fayyazuddin,
\newblock Phys. Rev. 147 (1966) 1071.

\bibitem{Georgi:1989xy}
H. Georgi,
\newblock Nucl. Phys. B331 (1990) 311.

\bibitem{Quigg:2009gg}
C. Quigg,
\newblock (2009) 311, 0908.3660.

\bibitem{Randall:1999ee}
L. Randall and R. Sundrum,
\newblock Phys. Rev. Lett. 83 (1999) 3370, hep-ph/9905221.

\bibitem{Davoudiasl:1999tf}
H. Davoudiasl, J.L. Hewett and T.G. Rizzo,
\newblock Phys. Lett. B473 (2000) 43, hep-ph/9911262.

\bibitem{Pomarol:1999ad}
A. Pomarol,
\newblock Phys. Lett. B486 (2000) 153, hep-ph/9911294.

\bibitem{Chang:1999nh}
S. Chang et~al.,
\newblock Phys. Rev. D62 (2000) 084025, hep-ph/9912498.

\bibitem{Grossman:1999ra}
Y. Grossman and M. Neubert,
\newblock Phys. Lett. B474 (2000) 361, hep-ph/9912408.

\bibitem{Gherghetta:2000qt}
T. Gherghetta and A. Pomarol,
\newblock Nucl. Phys. B586 (2000) 141, hep-ph/0003129.

\bibitem{Davoudiasl:2009cd}
H. Davoudiasl et~al.,
\newblock (2009), arXiv:0908.1968.

\bibitem{Agashe:2005vg}
K. Agashe, R. Contino and R. Sundrum,
\newblock Phys. Rev. Lett. 95 (2005) 171804, hep-ph/0502222.

\bibitem{Agashe:2004ci}
K. Agashe and G. Servant,
\newblock Phys. Rev. Lett. 93 (2004) 231805, hep-ph/0403143.

\bibitem{Agashe:2004bm}
K. Agashe and G. Servant,
\newblock JCAP 0502 (2005) 002, hep-ph/0411254.

\bibitem{Goldberger:1999uk}
W.D. Goldberger and M.B. Wise,
\newblock Phys. Rev. Lett. 83 (1999) 4922, hep-ph/9907447.

\bibitem{Garriga:2002vf}
J. Garriga and A. Pomarol,
\newblock Phys. Lett. B560 (2003) 91, hep-th/0212227.

\bibitem{Maldacena:1997re}
J.M. Maldacena,
\newblock Adv. Theor. Math. Phys. 2 (1998) 231, hep-th/9711200.

\bibitem{Gubser:1998bc}
S.S. Gubser, I.R. Klebanov and A.M. Polyakov,
\newblock Phys. Lett. B428 (1998) 105, hep-th/9802109.

\bibitem{Witten:1998qj}
E. Witten,
\newblock Adv. Theor. Math. Phys. 2 (1998) 253, hep-th/9802150.

\bibitem{ArkaniHamed:2000ds}
N. Arkani-Hamed, M. Porrati and L. Randall,
\newblock JHEP 08 (2001) 017, hep-th/0012148.

\bibitem{Rattazzi:2000hs}
R. Rattazzi and A. Zaffaroni,
\newblock JHEP 04 (2001) 021, hep-th/0012248.

\bibitem{Huber:2003tu}
S.J. Huber,
\newblock Nucl. Phys. B666 (2003) 269, hep-ph/0303183.

\bibitem{Agashe:2004cp}
K. Agashe, G. Perez and A. Soni,
\newblock Phys. Rev. D71 (2005) 016002, hep-ph/0408134.

\bibitem{Csaki:2008zd}
C. Csaki, A. Falkowski and A. Weiler,
\newblock JHEP 09 (2008) 008, arXiv:0804.1954.

\bibitem{Blanke:2008zb}
M. Blanke et~al.,
\newblock JHEP 03 (2009) 001, arXiv:0809.1073.

\bibitem{Bauer:2009cf}
M. Bauer et~al.,
\newblock (2009), arXiv:0912.1625.

\bibitem{Agashe:2009tu}
K. Agashe,
\newblock Phys. Rev. D80 (2009) 115020, arXiv:0902.2400.

\bibitem{Gedalia:2009ws}
O. Gedalia, G. Isidori and G. Perez,
\newblock Phys. Lett. B682 (2009) 200, arXiv:0905.3264.

\bibitem{Fitzpatrick:2007sa}
A.L. Fitzpatrick, G. Perez and L. Randall,
\newblock (2007), arXiv:0710.1869.

\bibitem{Chen:2008qg}
M.C. Chen and H.B. Yu,
\newblock Phys. Lett. B672 (2009) 253, arXiv:0804.2503.

\bibitem{Perez:2008ee}
G. Perez and L. Randall,
\newblock JHEP 01 (2009) 077, arXiv:0805.4652.

\bibitem{Csaki:2008qq}
C. Csaki et~al.,
\newblock JHEP 10 (2008) 055, arXiv:0806.0356.

\bibitem{Santiago:2008vq}
J. Santiago,
\newblock JHEP 12 (2008) 046, arXiv:0806.1230.

\bibitem{Csaki:2008eh}
C. Csaki, A. Falkowski and A. Weiler,
\newblock Phys. Rev. D80 (2009) 016001, arXiv:0806.3757.

\bibitem{Csaki:2009wc}
C. Csaki et~al.,
\newblock (2009), arXiv:0907.0474.

\bibitem{Chen:2009hr}
M.C. Chen, K.T. Mahanthappa and F. Yu,
\newblock (2009), arXiv:0909.5472.

\bibitem{Agashe:2003zs}
K. Agashe et~al.,
\newblock JHEP 08 (2003) 050, hep-ph/0308036.

\bibitem{Agashe:2006at}
K. Agashe et~al.,
\newblock Phys. Lett. B641 (2006) 62, hep-ph/0605341.

\bibitem{Carena:2006bn}
M.S. Carena et~al.,
\newblock Nucl. Phys. B759 (2006) 202, hep-ph/0607106.

\bibitem{Carena:2007ua}
M.S. Carena et~al.,
\newblock Phys. Rev. D76 (2007) 035006, hep-ph/0701055.

\bibitem{Giudice:2000av}
G.F. Giudice, R. Rattazzi and J.D. Wells,
\newblock Nucl. Phys. B595 (2001) 250, hep-ph/0002178.

\bibitem{Rizzo:2002pq}
T.G. Rizzo,
\newblock JHEP 06 (2002) 056, hep-ph/0205242.

\bibitem{Toharia:2008tm}
M. Toharia,
\newblock Phys. Rev. D79 (2009) 015009, arXiv:0809.5245.

\bibitem{Csaki:2007ns}
C. Csaki, J. Hubisz and S.J. Lee,
\newblock Phys. Rev. D76 (2007) 125015, arXiv:0705.3844.

\bibitem{Callan:1970ze}
C.G. Callan, Jr., S.R. Coleman and R. Jackiw,
\newblock Ann. Phys. 59 (1970) 42.

\bibitem{Almeida:2008tp}
L.G. Almeida et~al.,
\newblock Phys. Rev. D79 (2009) 074012, arXiv:0810.0934.

\bibitem{Skiba:2007fw}
W. Skiba and D. Tucker-Smith,
\newblock Phys. Rev. D75 (2007) 115010, hep-ph/0701247.

\bibitem{Holdom:2007nw}
B. Holdom,
\newblock JHEP 03 (2007) 063, hep-ph/0702037.

\bibitem{Holdom:2007ap}
B. Holdom,
\newblock JHEP 08 (2007) 069, arXiv:0705.1736.

\bibitem{Seymour:1993mx}
M.H. Seymour,
\newblock Z. Phys. C62 (1994) 127.

\bibitem{Butterworth:2008sd}
J.M. Butterworth et~al.,
\newblock AIP Conf. Proc. 1078 (2009) 189, arXiv:0809.2530.

\bibitem{Ellis:2009me}
S.D. Ellis, C.K. Vermilion and J.R. Walsh,
\newblock (2009), arXiv:0912.0033.

\bibitem{Benchekroun:2001je}
D. Benchekroun, C. Driouichi and A. Hoummada,
\newblock Eur. Phys. J. direct C3 (2001) N3.

\bibitem{Davoudiasl:2007wf}
H. Davoudiasl, T.G. Rizzo and A. Soni,
\newblock Phys. Rev. D77 (2008) 036001, arXiv:0710.2078.

\bibitem{Agashe:2006hk}
K. Agashe et~al.,
\newblock Phys. Rev. D77 (2008) 015003, hep-ph/0612015.

\bibitem{Lillie:2007ve}
B. Lillie, J. Shu and T.M.P. Tait,
\newblock Phys. Rev. D76 (2007) 115016, arXiv:0706.3960.

\bibitem{Djouadi:2007eg}
A. Djouadi, G. Moreau and R.K. Singh,
\newblock Nucl. Phys. B797 (2008) 1, arXiv:0706.4191.

\bibitem{Guchait:2007jd}
M. Guchait, F. Mahmoudi and K. Sridhar,
\newblock Phys. Lett. B666 (2008) 347, arXiv:0710.2234.

\bibitem{Baur:2007ck}
U. Baur and L.H. Orr,
\newblock Phys. Rev. D76 (2007) 094012, arXiv:0707.2066.

\bibitem{Baur:2008uv}
U. Baur and L.H. Orr,
\newblock Phys. Rev. D77 (2008) 114001, arXiv:0803.1160.

\bibitem{Lillie:2007yh}
B. Lillie, L. Randall and L.T. Wang,
\newblock JHEP 09 (2007) 074, hep-ph/0701166.

\bibitem{Fitzpatrick:2007qr}
A.L. Fitzpatrick et~al.,
\newblock JHEP 09 (2007) 013, hep-ph/0701150.

\bibitem{Agashe:2007zd}
K. Agashe et~al.,
\newblock Phys. Rev. D76 (2007) 036006, hep-ph/0701186.

\bibitem{Antipin:2007pi}
O. Antipin, D. Atwood and A. Soni,
\newblock Phys. Lett. B666 (2008) 155, arXiv:0711.3175.

\bibitem{Antipin:2008hj}
O. Antipin and A. Soni,
\newblock JHEP 10 (2008) 018, arXiv:0806.3427.

\bibitem{Agashe:2008jb}
K. Agashe et~al.,
\newblock Phys. Rev. D80 (2009) 075007, arXiv:0810.1497.

\bibitem{Agashe:2007ki}
K. Agashe et~al.,
\newblock Phys. Rev. D76 (2007) 115015, arXiv:0709.0007.

\bibitem{Dennis:2007tv}
C. Dennis et~al.,
\newblock (2007), hep-ph/0701158.

\bibitem{Contino:2008hi}
R. Contino and G. Servant,
\newblock JHEP 06 (2008) 026, arXiv:0801.1679.

\bibitem{Mrazek:2009yu}
J. Mrazek and A. Wulzer,
\newblock (2009), arXiv:0909.3977.

\bibitem{delAguila:2008iz}
F. del Aguila et~al.,
\newblock Eur. Phys. J. C57 (2008) 183, arXiv:0801.1800.

\bibitem{Aaltonen:2009nr}
CDF Collaboration, T. Aaltonen et~al.,
\newblock (2009), arXiv:0912.1057.

\bibitem{Carena:2007tn}
M. Carena et~al.,
\newblock Phys. Rev. D77 (2008) 076003, arXiv:0712.0095.

\bibitem{:1999fq}
ATLAS~Collaboration,
\newblock {Detector and physics performance technical design report.
  Vol. 1},
\newblock CERN-LHCC-99-14 (1999).

\bibitem{Bayatian:2006zz}
CMS Collaboration, G.L. Bayatian et~al.,
\newblock CERN-LHCC-2006-001.

\bibitem{PGS}
J. Conway,  (2006),
\newblock {PGS 4: Pretty Good Simulation of high energy collisions},
\newblock
  http://www.physics.ucdavis.edu/~conway/research/software/pgs/pgs4-general.htm.

\bibitem{:2007dz}
CDF Collaboration, T. Aaltonen et~al.,
\newblock Phys. Rev. Lett. 100 (2008) 231801, arXiv:0709.0705.

\bibitem{Abazov:2008ny}
D0 Collaboration, V.M. Abazov et~al.,
\newblock Phys. Lett. B668 (2008) 98, arXiv:0804.3664.

\bibitem{Aaltonen:2009iz}
CDF Collaboration, T. Aaltonen et~al.,
\newblock Phys. Rev. Lett. 102 (2009) 222003, arXiv:0903.2850.

\bibitem{mcatnlo}
S. Frixione and B.R. Webber,
\newblock JHEP 06 (2002) 029, hep-ph/0204244.

\bibitem{mcatnlo2}
S. Frixione, P. Nason and B.R. Webber,
\newblock JHEP 08 (2003) 007, hep-ph/0305252.

\bibitem{cms_ttbar_resonance_semi}
CMS Collaboration,
\newblock CMS Physics Analysis Summary EXO-09-008 (2009).

\bibitem{cms_ttbar_resonance_allhad}
CMS Collaboration, 
\newblock CMS Physics Analysis Summary EXO-09-002 (2009).

\bibitem{atlas_ttbar_resonance}
ATLAS Collaboration, 
\newblock to be published as ATLAS-PHYS-PUB .

\bibitem{Brooijmans:2008zz}
G. Brooijmans,
\newblock ATL-PHYS-CONF-2008-008.

\bibitem{brooijmans2}
B. Chapleau and G. Brooijmans,
\newblock ATL-PHYS-INT-2009-037(ATLAS collaboration only).

\bibitem{Vos:2008zzb}
M. Vos,
\newblock ATL-PHYS-CONF-2008-016.

\bibitem{Agashe:2009di}
K. Agashe and R. Contino,
\newblock Phys. Rev. D80 (2009) 075016, arXiv:0906.1542.

\bibitem{Azatov:2009na}
A. Azatov, M. Toharia and L. Zhu,
\newblock Phys. Rev. D80 (2009) 035016, arXiv:0906.1990.

\bibitem{Azatov:2008vm}
A. Azatov, M. Toharia and L. Zhu,
\newblock Phys. Rev. D80 (2009) 031701, arXiv:0812.2489.

\bibitem{AguilarSaavedra:2004wm}
J.A. Aguilar-Saavedra,
\newblock Acta Phys. Polon. B35 (2004) 2695, hep-ph/0409342.

\bibitem{AguilarSaavedra:2000aj}
J.A. Aguilar-Saavedra and G.C. Branco,
\newblock Phys. Lett. B495 (2000) 347, hep-ph/0004190.

\bibitem{Bejar:2003em}
S. Bejar, J. Guasch and J. Sola,
\newblock Nucl. Phys. B675 (2003) 270, hep-ph/0307144.

\bibitem{Han:2000jz}
T. Han and D. Marfatia,
\newblock Phys. Rev. Lett. 86 (2001) 1442, hep-ph/0008141.

\bibitem{Skands:2003cj}
P.Z. Skands et~al.,
\newblock JHEP 07 (2004) 036, hep-ph/0311123.

\bibitem{Lehmacher:2008hs}
M. Lehmacher,
\newblock (2008), arXiv:0809.4896.

\bibitem{mistaggcharm}
Alagoez, E. et~al.,
\newblock http://unizh.web.cern.ch/unizh/Activities/cms.htm.

\bibitem{Fogli:2005cq}
G.L. Fogli et~al.,
\newblock Prog. Part. Nucl. Phys. 57 (2006) 742, hep-ph/0506083.

\bibitem{Fogli:2006yq}
G.L. Fogli et~al.,
\newblock Phys. Rev. D75 (2007) 053001, hep-ph/0608060.

\bibitem{Carena:2004xs}
M.S. Carena et~al.,
\newblock Phys. Rev. D70 (2004) 093009, hep-ph/0408098.

\bibitem{Cacciapaglia:2006pk}
G. Cacciapaglia et~al.,
\newblock Phys. Rev. D74 (2006) 033011, hep-ph/0604111.

\bibitem{Aaltonen:2008vx}
CDF Collaboration, T. Aaltonen et~al.,
\newblock Phys. Rev. Lett. 102 (2009) 031801, arXiv:0810.2059.

\bibitem{Aaltonen:2008ah}
CDF Collaboration, T. Aaltonen et~al.,
\newblock Phys. Rev. Lett. 102 (2009) 091805, arXiv:0811.0053.

\bibitem{Buchmuller:1991ce}
W. Buchmuller, C. Greub and P. Minkowski,
\newblock Phys. Lett. B267 (1991) 395.

\bibitem{Khalil:2006yi}
S. Khalil,
\newblock J. Phys. G35 (2008) 055001, hep-ph/0611205.

\bibitem{BL_master_thesis}
L. Basso,
\newblock Master's thesis, {Universit\`a degli Studi di Padova}, 2007.

\bibitem{Minkowski:1977sc}
P. Minkowski,
\newblock Phys. Lett. B67 (1977) 421.

\bibitem{VanNieuwenhuizen:1979hm}
P. Van~Nieuwenhuizen and D.Z. Freedman,
\newblock Amsterdam, Netherlands: North-holland (1979) 341p.

\bibitem{Yanagida:1979as}
T. Yanagida,
\newblock In Proceedings of the Workshop on the Baryon Number of the Universe
  and Unified Theories, Tsukuba, Japan, 13-14 Feb 1979.

\bibitem{S.L.Glashow}
S.L. Glashow, in \emph{Quarks and Leptons}, eds. M.L\`evy et al. (Plenum, New
  York $1980$), p.~$707$.

\bibitem{Mohapatra:1979ia}
R.N. Mohapatra and G. Senjanovic,
\newblock Phys. Rev. Lett. 44 (1980) 912.

\bibitem{Huitu:2008gf}
K. Huitu et~al.,
\newblock Phys. Rev. Lett. 101 (2008) 181802, arXiv:0803.2799.

\bibitem{Langacker:1991pg}
P. Langacker and M.x. Luo,
\newblock Phys. Rev. D45 (1992) 278.

\bibitem{Rizzo:1996ce}
T.G. Rizzo,
\newblock (1996), hep-ph/9612440.

\bibitem{Contino:2008xg}
R. Contino,
\newblock Nuovo Cim. 123B (2008) 511, arXiv:0804.3195.

\bibitem{Gulov:2009tn}
A.V. Gulov and V.V. Skalozub,
\newblock (2009), arXiv:0905.2596.

\bibitem{Erler:2009jh}
J. Erler et~al.,
\newblock JHEP 08 (2009) 017, arXiv:0906.2435.

\bibitem{indico_chamonix}
http://indico.cern.ch/conferenceDisplay.py?confId=83135.

\bibitem{B-L_observ}
L. Basso et~al.,
\newblock In progress.

\bibitem{Basso:2010pe}
L. Basso et~al.,
\newblock (2010), arXiv:1002.3586.

\bibitem{Basso:2010jt}
L. Basso et~al.,
\newblock (2010), arXiv:1002.1939.

\bibitem{Basso:2010jm}
L. Basso, S. Moretti and G.M. Pruna,
\newblock (2010), arXiv:1004.3039.

\bibitem{Pukhov:2004ca}
A. Pukhov,
\newblock (2004), hep-ph/0412191.

\bibitem{Semenov:1996es}
A.V. Semenov,
\newblock (1996), hep-ph/9608488.

\bibitem{Balka:1987ty}
CDF Collaboration, L. Balka et~al.,
\newblock Nucl. Instrum. Meth. A267 (1988) 272.

\bibitem{Bityukov:2000tt}
S.I. Bityukov and N.V. Krasnikov,
\newblock Nucl. Instrum. Meth. A452 (2000) 518.

\bibitem{Basso:2009hf}
L. Basso et~al.,
\newblock JHEP 10 (2009) 006, arXiv:0903.4777.

\bibitem{CTEQ_website}
http://durpdg.dur.ac.uk/hepdata/pdf.html.

\bibitem{Salvioni:2009mt}
E. Salvioni, G. Villadoro and F. Zwirner,
\newblock JHEP 11 (2009) 068, arXiv:0909.1320.

\bibitem{Emam:2008zz}
W. Emam and P. Mine,
\newblock Erratum-ibid. G36 (2009) 129701.

\bibitem{deSandes:2008yx}
H. de~Sandes and R. Rosenfeld,
\newblock J. Phys. G36 (2009) 085001, arXiv:0811.0984.

\bibitem{Burdman:2008qh}
G. Burdman et~al.,
\newblock Phys. Rev. D79 (2009) 075026, arXiv:0812.0368.

\bibitem{AguilarSaavedra:2005pv}
J.A. Aguilar-Saavedra,
\newblock Phys. Lett. B625 (2005) 234, hep-ph/0506187.

\bibitem{AguilarSaavedra:2009es}
J.A. Aguilar-Saavedra,
\newblock JHEP 11 (2009) 030, arXiv:0907.3155.

\bibitem{Atre:2008iu}
A. Atre et~al.,
\newblock Phys. Rev. D79 (2009) 054018, arXiv:0806.3966.

\bibitem{Azuelos:2004dm}
G. Azuelos et~al.,
\newblock Eur. Phys. J. C39S2 (2005) 13, hep-ph/0402037.

\bibitem{Yue:2009cq}
C.X. Yue, H.D. Yang and W. Ma,
\newblock Nucl. Phys. B818 (2009) 1, arXiv:0903.3720.

\bibitem{Berger:2009qy}
E.L. Berger and Q.H. Cao,
\newblock (2009), arXiv:0909.3555.

\bibitem{Campbell:2009gj}
J.M. Campbell et~al.,
\newblock JHEP 10 (2009) 042, arXiv:0907.3933.

\bibitem{Bouchart:2009vq}
C. Bouchart and G. Moreau,
\newblock Phys. Rev. D80 (2009) 095022, arXiv:0909.4812.

\bibitem{Gopalakrishna:2009xxx}
S. Gopalakrishna, G. Moreau and R.K. Singh,
\newblock Work in progress .

\bibitem{Bouchart:2008vp}
C. Bouchart and G. Moreau,
\newblock Nucl. Phys. B810 (2009) 66, arXiv:0807.4461.

\bibitem{Djouadi:2006rk}
A. Djouadi, G. Moreau and F. Richard,
\newblock Nucl. Phys. B773 (2007) 43, hep-ph/0610173.

\bibitem{Djouadi:2009nb}
A. Djouadi et~al.,
\newblock (2009), arXiv:0906.0604.

\bibitem{Agashe:2008uz}
K. Agashe, A. Azatov and L. Zhu,
\newblock Phys. Rev. D79 (2009) 056006, arXiv:0810.1016.

\bibitem{Ledroit:2007ik}
F. Ledroit, G. Moreau and J. Morel,
\newblock JHEP 09 (2007) 071, hep-ph/0703262.

\bibitem{Djouadi:2007fm}
A. Djouadi and G. Moreau,
\newblock Phys. Lett. B660 (2008) 67, arXiv:0707.3800.

\bibitem{Barger:1991vn}
V.D. Barger, A.L. Stange and R.J.N. Phillips,
\newblock Phys. Rev. D44 (1991) 1987.

\bibitem{tait}
B. Lillie, J. Shu and T.M.P. Tait,
\newblock JHEP 04 (2008) 087, arXiv:0712.3057.

\bibitem{tait2}
K. Kumar, T.M.P. Tait and R. Vega-Morales,
\newblock JHEP 05 (2009) 022, arXiv:0901.3808.

\bibitem{serra}
A. Pomarol and J. Serra,
\newblock Phys. Rev. D78 (2008) 074026, arXiv:0806.3247.

\bibitem{cscnote}
ATLAS Collaboration, G. Aad et~al.,
\newblock (2009), arXiv:0901.0512.

\bibitem{randall}
B. Lillie, L. Randall and L.T. Wang,
\newblock JHEP 09 (2007) 074, hep-ph/0701166.

\bibitem{Jackson:2009kg}
C.B. Jackson et~al.,
\newblock JCAP 1004 (2010) 004, arXiv:0912.0004.

\bibitem{GauthierServant}
L. Gauthier and G. Servant,
\newblock {In Preparation}.

\bibitem{thalerwang}
J. Thaler and L.T. Wang,
\newblock JHEP 07 (2008) 092, arXiv:0806.0023.

\bibitem{kt}
S. Catani et~al.,
\newblock Nucl. Phys. B406 (1993) 187.

\bibitem{kt2}
S.D. Ellis and D.E. Soper,
\newblock Phys. Rev. D48 (1993) 3160, hep-ph/9305266.

\bibitem{fastjet}
M. Cacciari and G.P. Salam,
\newblock Phys. Lett. B641 (2006) 57, hep-ph/0512210.

\bibitem{atlasttresonance}
ATLAS~Collaboration,
\newblock ATLAS-PHYS-PUB in preparation.

\bibitem{Belanger:2007dx}
G. Belanger, A. Pukhov and G. Servant,
\newblock JCAP 0801 (2008) 009, arXiv:0706.0526.

\bibitem{Djouadi:1989uk}
A. Djouadi, J.H. Kuhn and P.M. Zerwas,
\newblock Z. Phys. C46 (1990) 411.

\bibitem{:2005ema}
ALEPH Collaboration, 
\newblock Phys. Rept. 427 (2006) 257, hep-ex/0509008.

\bibitem{Allanach:2009vz}
B.C. Allanach et~al.,
\newblock (2009), arXiv:0910.1350.

\bibitem{ATL-PHYS-PUB-2009-081}
ATLAS Collaboration,
\newblock CERN preprint ATL-PHYS-PUB-2009-081. ATL-COM-PHYS-2009-255 (2009).

\bibitem{Gleisberg:2008ta}
T. Gleisberg et~al.,
\newblock JHEP 02 (2009) 007, arXiv:0811.4622.

\bibitem{Junk:1999kv}
T. Junk,
\newblock Nucl. Instrum. Meth. A434 (1999) 435.

\bibitem{Pilaftsis:1997dr}
A. Pilaftsis,
\newblock Nucl. Phys. B504 (1997) 61, hep-ph/9702393.

\bibitem{Cacciapaglia:2009ic}
G. Cacciapaglia, A. Deandrea and S. De~Curtis,
\newblock Phys. Lett. B682 (2009) 43, arXiv:0906.3417.

\bibitem{Ellis:2004fs}
J.R. Ellis, J.S. Lee and A. Pilaftsis,
\newblock Phys. Rev. D70 (2004) 075010, hep-ph/0404167.

\bibitem{Frank:2006yh}
M. Frank et~al.,
\newblock JHEP 02 (2007) 047, hep-ph/0611326.

\bibitem{Hahn:2007it}
T. Hahn et~al.,
\newblock (2007), arXiv:0709.1907.

\bibitem{Stancato:2008mp}
D. Stancato and J. Terning,
\newblock JHEP 11 (2009) 101, arXiv:0807.3961.

\bibitem{Falkowski:2008yr}
A. Falkowski and M. Perez-Victoria,
\newblock Phys. Rev. D79 (2009) 035005, arXiv:0810.4940.

\bibitem{Csaki:2003zu}
C. Csaki et~al.,
\newblock Phys. Rev. Lett. 92 (2004) 101802, hep-ph/0308038.

\bibitem{Cacciapaglia:2004rb}
G. Cacciapaglia et~al.,
\newblock Phys. Rev. D71 (2005) 035015, hep-ph/0409126.

\bibitem{Henderson:cms:trigger_tdr1}
{CMS Trigger and Data Acquisition Project Technical Design Report Vol. I: The
  Trigger Systems}, 2000,
\newblock http://cmsdoc.cern.ch/cms/TDR/TRIGGER-public/trigger.html.

\bibitem{Henderson:cms:trigger_tdr2}
{CMS Trigger and Data Acquisition Project Technical Design Report Vol. II: Data
  Acquisition and High Level Trigger}, 2002,
\newblock http://cmsdoc.cern.ch/cms/TDR/DAQ/daq.html.

\bibitem{:2009dq}
CMS Collaboration, S. Chatrchyan et~al.,
\newblock (2009), arXiv:0911.5422.

\bibitem{Collaboration:2009ic}
CMS Collaboration,
\newblock (2009), arXiv:0911.4889.

\bibitem{Harnik:2008ax}
R. Harnik and T. Wizansky,
\newblock arXiv:0810.3948.

\bibitem{Okun:1980kw}
L.B. Okun,
\newblock JETP Lett. 31 (1980) 144.

\bibitem{Kang:2008ea}
J. Kang and M.A. Luty,
\newblock JHEP 11 (2009) 065, arXiv:0805.4642.

\bibitem{Burdman:2006tz}
G. Burdman et~al.,
\newblock JHEP 02 (2007) 009, hep-ph/0609152.

\bibitem{Henderson:cms:ecal_tdr}
{CMS ECAL Technical Design Report}, 1997,
\newblock http://cmsdoc.cern.ch/cms/TDR/ECAL/ecal.html.

\bibitem{Agostino:2009ic}
L. Agostino et~al.,
\newblock JINST 4 (2009) P10005, arXiv:0908.1065.

\bibitem{Henderson:lhc_2010_lumi}
{LHC 2009--2010 luminosity performance},
\newblock
  http://lhc-commissioning.web.cern.ch/lhc-commissioning/luminosity/09-10-lumi-estimate.htm.

\bibitem{Dimopoulos:1996vz}
S. Dimopoulos et~al.,
\newblock Phys. Rev. Lett. 76 (1996) 3494, hep-ph/9601367.

\bibitem{Langacker:1984dc}
P. Langacker, R.W. Robinett and J.L. Rosner,
\newblock Phys. Rev. D30 (1984) 1470.

\bibitem{Martin:2000eq}
S.P. Martin,
\newblock Phys. Rev. D62 (2000) 095008, hep-ph/0005116.

\bibitem{ArkaniHamed:2004fb}
N. Arkani-Hamed and S. Dimopoulos,
\newblock JHEP 06 (2005) 073, hep-th/0405159.

\bibitem{Strassler:2006im}
M.J. Strassler and K.M. Zurek,
\newblock Phys. Lett. B651 (2007) 374, hep-ph/0604261.

\bibitem{Strassler:2006ri}
M.J. Strassler and K.M. Zurek,
\newblock Phys. Lett. B661 (2008) 263, hep-ph/0605193.

\bibitem{Matchev:1999ft}
K.T. Matchev and S.D. Thomas,
\newblock Phys. Rev. D62 (2000) 077702, hep-ph/9908482.

\bibitem{Strassler:2006qa}
M.J. Strassler,
\newblock (2006), hep-ph/0607160.

\bibitem{Juknevich:2009ji}
J.E. Juknevich, D. Melnikov and M.J. Strassler,
\newblock JHEP 07 (2009) 055, arXiv:0903.0883.

\bibitem{Dobrescu:2000jt}
B.A. Dobrescu, G.L. Landsberg and K.T. Matchev,
\newblock Phys. Rev. D63 (2001) 075003, hep-ph/0005308.

\bibitem{Bjorken:2009mm}
J.D. Bjorken et~al.,
\newblock Phys. Rev. D80 (2009) 075018, arXiv:0906.0580.

\bibitem{ArkaniHamed:2008qp}
N. Arkani-Hamed and N. Weiner,
\newblock JHEP 12 (2008) 104, arXiv:0810.0714.

\bibitem{Pospelov:2008jd}
M. Pospelov and A. Ritz,
\newblock Phys. Lett. B671 (2009) 391, arXiv:0810.1502.

\bibitem{ShihThomasunpub}
D. Shih and S. Thomas,
\newblock unpublished.

\bibitem{Abazov:2009hn}
D0 Collaboration, V.M. Abazov et~al.,
\newblock Phys. Rev. Lett. 103 (2009) 081802, arXiv:0905.1478.

\bibitem{Baumgart:2009tn}
M. Baumgart et~al.,
\newblock JHEP 04 (2009) 014, arXiv:0901.0283.

\bibitem{Cheung:2009su}
C. Cheung et~al.,
\newblock (2009), arXiv:0909.0290.

\bibitem{Bai:2009it}
Y. Bai and Z. Han,
\newblock Phys. Rev. Lett. 103 (2009) 051801, arXiv:0902.0006.

\bibitem{ArkaniHamed:2008qn}
N. Arkani-Hamed et~al.,
\newblock Phys. Rev. D79 (2009) 015014, arXiv:0810.0713.

\bibitem{ATLASHVtrig}
ATLAS Collaboration, Y. Bai et~al.,
\newblock ATL-PHYS-PUB-2009-082.

\bibitem{Abazov:2009ik}
D0 Collaboration, V.M. Abazov et~al.,
\newblock Phys. Rev. Lett. 103 (2009) 071801, arXiv:0906.1787.

\bibitem{Abazov:2008zm}
D0 Collaboration, V.M. Abazov et~al.,
\newblock Phys. Rev. Lett. 101 (2008) 111802, arXiv:0806.2223.

\bibitem{Abazov:2006as}
D0 Collaboration, V.M. Abazov et~al.,
\newblock Phys. Rev. Lett. 97 (2006) 161802, hep-ex/0607028.

\bibitem{Scott:2004wz}
CDF Collaboration, A.L. Scott,
\newblock Int. J. Mod. Phys. A20 (2005) 3263, hep-ex/0410019.

\bibitem{Strassler:2008fv}
M.J. Strassler,
\newblock (2008), arXiv:0806.2385.

\bibitem{Han:2007ae}
T. Han et~al.,
\newblock JHEP 07 (2008) 008, arXiv:0712.2041.

\bibitem{Morrissey:2009ur}
D.E. Morrissey, D. Poland and K.M. Zurek,
\newblock JHEP 07 (2009) 050, arXiv:0904.2567.

\bibitem{Holdom:1985ag}
B. Holdom,
\newblock Phys. Lett. B166 (1986) 196.

\bibitem{Babu:1996vt}
K.S. Babu, C.F. Kolda and J. March-Russell,
\newblock Phys. Rev. D54 (1996) 4635, hep-ph/9603212.

\bibitem{Dienes:1996zr}
K.R. Dienes, C.F. Kolda and J. March-Russell,
\newblock Nucl. Phys. B492 (1997) 104, hep-ph/9610479.

\bibitem{Zurek:2008qg}
K.M. Zurek,
\newblock Phys. Rev. D79 (2009) 115002, arXiv:0811.4429.

\bibitem{Giudice:1988yz}
G.F. Giudice and A. Masiero,
\newblock Phys. Lett. B206 (1988) 480.

\bibitem{Fayet:1974jb}
P. Fayet and J. Iliopoulos,
\newblock Phys. Lett. B51 (1974) 461.

\bibitem{Suematsu:2006wh}
D. Suematsu,
\newblock JHEP 11 (2006) 029, hep-ph/0606125.

\bibitem{Cui:2009xq}
Y. Cui et~al.,
\newblock JHEP 05 (2009) 076, arXiv:0901.0557.

\bibitem{Morrissey:dmo1}
D.E. Morrissey,
\newblock in preparation .

\bibitem{Batell:2009yf}
B. Batell, M. Pospelov and A. Ritz,
\newblock Phys. Rev. D79 (2009) 115008, arXiv:0903.0363.

\bibitem{Pospelov:2008zw}
M. Pospelov,
\newblock Phys. Rev. D80 (2009) 095002, arXiv:0811.1030.

\bibitem{Batell:2009di}
B. Batell, M. Pospelov and A. Ritz,
\newblock Phys. Rev. D80 (2009) 095024, arXiv:0906.5614.

\bibitem{Schuster:2009au}
P. Schuster, N. Toro and I. Yavin,
\newblock (2009), arXiv:0910.1602.

\bibitem{Meade:2009mu}
P. Meade et~al.,
\newblock (2009), arXiv:0910.4160.

\end{thebibliography}
